\documentclass[a4paper,11pt]{article}
\pdfoutput=1 % if your are submitting a pdflatex (\textit{i.e.} if you have
\usepackage{jheppub} % for details on the use of the package, please
\usepackage[T1]{fontenc} % if needed

\newif\ifsol\solfalse %false if solutions should not appear, true to have solution appearing
\numberwithin{equation}{subsection}

\usepackage{quoting}
\usepackage{ragged2e}
\usepackage{tensor}
\usepackage{booktabs}
\usepackage{enumerate}
\usepackage[most]{tcolorbox}
\tcbset{colback=cyan!5!white, colframe=black!50!black, 
        highlight math style= {enhanced, %<-- needed for the ’remember’ options
            colframe=black,colback=black!20!white,boxsep=0pt}
        }
        
\tikzset{
    partial ellipse/.style args={#1:#2:#3}{
        insert path={+ (#1:#3) arc (#1:#2:#3)}
    }
}

\newenvironment{RCText}[1][2em]
  {\begin{quoting}[leftmargin=#1,rightmargin=#1]\RaggedRight}
  {\end{quoting}}

\usepackage{fontawesome}
\usepackage[normalem]{ulem}
\usepackage{slashed}
\usepackage{bbold}
\usepackage[multiple]{footmisc}

\newtheorem{theorem}{Theorem}[section]
\newtheorem{exercise}[theorem]{Exercise}

\newtheorem{remark}[theorem]{Remark}

\usepackage{import}
\usepackage{multicol}
\usepackage{utfsym}

\usepackage{multirow}

\usepackage{slashed}
\newcommand{\B}{{\scriptscriptstyle\text{B}}}
\newcommand{\F}{{\scriptscriptstyle\text{F}}}
\renewcommand{\L}{{\scriptscriptstyle\text{L}}}
\newcommand{\R}{{\scriptscriptstyle\text{R}}}
\newcommand{\z}{{\scriptscriptstyle\text{o}}}
\newcommand{\psu}{\mathfrak{psu}}
\newcommand{\su}{\mathfrak{su}}

\newcommand{\Ql}{\gen{Q}_\L}
\newcommand{\Sl}{\gen{S}_\L}
\renewcommand{\sl}{\gen{s}_\L}

\newcommand{\Qr}{\gen{Q}_\R}
\newcommand{\sr}{\gen{s}_\R}
\newcommand{\Sr}{\gen{S}_\R}

\newcommand{\ql}{\gen{q}_\L}
\newcommand{\qr}{\gen{q}_\R}

\newcommand{\de}{\text{d}}

\newcommand{\com}[2]{[#1,#2]}
\newcommand{\anticom}[2]{\{#1,#2\}}
\newcommand\gen[1]{\mathbf #1}
\newcommand{\extder}{\mathrm{d}}
\newcommand{\alg}[1]{\mathfrak{#1}}

\usepackage{tikz}

\newcommand{\ket}[1]{| #1 \rangle}
\newcommand{\bra}[1]{\langle #1 |}

\newcommand{\cir}{R}

\newcommand{\tr}{\text{tr}}

\newcommand{\lr}{\left(}
\newcommand{\rr}{\right)}
\newcommand{\ls}{\left[}
\newcommand{\rs}{\right]}
\newcommand{\lc}{\left\{}
\newcommand{\rc}{\right\}}

\newcommand{\p}{\partial}

\newcommand{\be}{\begin{equation}}
\newcommand{\ee}{\end{equation}}

\newcommand{\CF}{\mathcal{F}}

\newcommand{\CL}{\mathcal{L}}

\newcommand{\RR}{\mathbb{R}}
\newcommand{\CC}{\mathbb{C}}

\newcommand{\pd}{\partial}
\newcommand{\vphi}{\varphi}
\newcommand{\inv}{^{-1}}

\newcommand{\qiq}{\quad\implies\quad}

\title{\boldmath Exact approaches on the string worldsheet }

\author[a]{Saskia Demulder,}
\author[b]{Sibylle Driezen,}
\author[c]{Bob Knighton,}
\author[d]{Gerben Oling,}
\author[e]{Ana L. Retore,}
\author[f]{Fiona K.~Seibold,}
\author[g,h,1]{Alessandro Sfondrini,\note{MATRIX Simons Fellow.}}
\author[i]{Ziqi Yan}

\affiliation[a]{Ben Gurion University of the Negev,\\ David Ben Gurion Blvd 1,  84105 Be'er Sheva, Israel}
\affiliation[b]{Institut f\"ur Theoretische Physik, ETH Z\"urich,\\ Wolfgang-Pauli-Stra\ss e 27, Z\"urich 8093, Switzerland}
\affiliation[c]{Department of Applied Mathematics and Theoretical Physics, University of Cambridge,\\
Cambridge CB3 0WA, United Kingdom}
\affiliation[d]{School of Mathematics and Maxwell Institute for Mathematical Sciences,
  University of Edinburgh,\\
  Peter Guthrie Tait Road, Edinburgh EH9 3FD, UK
}
\affiliation[e]{Department of Mathematical Sciences, Durham University, Durham DH1 3LE, U.K.}
\affiliation[f]{Blackett Laboratory, Imperial College, London SW7 2AZ, U.K.}
\affiliation[g]{Dipartimento di Fisica e Astronomia, Universit\`{a} degli Studi di Padova,\\ via Marzolo 8, 35131 Padova, Italy}
\affiliation[h]{Istituto Nazionale di Fisica Nucleare, Sezione di Padova,\\ via Marzolo 8, 35131 Padova, Italy}
\affiliation[i]{Nordita, KTH Royal Institute of Technology and Stockholm University\\ Hannes Alfv\'ens v\"{a}g 12, SE-106 91 Stockholm, Sweden}

\emailAdd{demulder@post.bgu.ac.il}
\emailAdd{sdriezen@phys.ethz.ch}
\emailAdd{rik23@cam.ac.uk}
\emailAdd{gerben.oling@ed.ac.uk}
\emailAdd{ana.retore@durham.ac.uk}
\emailAdd{f.seibold21@imperial.ac.uk}
\emailAdd{alessandro.sfondrini@unipd.it}
\emailAdd{ziqi.yan@su.se}

\abstract{
We review different exact approaches to string theory. In the context of the Green-Schwarz superstring, we discuss the action in curved backgrounds and its supercoset formulation, with particular attention to superstring backgrounds of the $AdS_3$ type supported by both Ramond-Ramond and Neveu-Schwarz-Neveu-Schwarz fluxes. This is the basis for the discussion of classical integrability, of worldsheet-scattering factorisation in the uniform lightcone gauge, and eventually  of the string spectrum through the mirror thermodynamic Bethe ansatz, which for $AdS_3$ backgrounds was only derived and analysed very recently. We then illustrate some aspects of the Ramond-Neveu-Schwarz string, and introduce the formalism of Berkovits-Vafa-Witten, which has seen very recent applications to $AdS_3$ physics, which we also briefly review. Finally, we present the relation between M-theory in the discrete lightcone quantisation and decoupling limits of string theory that exhibit non-relativistic behaviors,
highlighting the connection with integrable $T\bar{T}$ deformations, as well as the relation between spin-matrix theory and  Landau-Lifshitz models.

This review is based on lectures given at the Young Researchers Integrability School and Workshop 2022 ``Taming the string worldsheet'' at NORDITA, Stockholm.
}

\begin{document} 
\maketitle
\flushbottom

\section{Introduction}
String theory has occupied a central role in high-energy theoretical physics for decades. Initially constructed as a potential model to describe strong interactions from the late 1960s, it became a candidate for a quantum theory of gravity, or even for a ``theory of everything'' in the subsequent decades. In the process, a theory (in fact, more than one such theory) of supersymmetric strings, or \textit{superstrings}, was  formulated as a way to incorporate fermions into the formalism.
More recently, string theory has been instrumental in providing a concrete realisation of the holographic principle~\cite{tHooft:1993dmi,Susskind:1994vu} in terms of the celebrated AdS/CFT correspondence~\cite{Maldacena:1997re,Gubser:1998bc,Witten:1998qj}. Even though it is unclear whether string theory can describe the universe as we see it --- with (quantum) gravity, all the fundamental particles of the Standard Model, inflation, and so on --- we can confidently state that string theory can help us understand the general features that a consistent theory of quantum gravity must have.

Here we will just consider a theory of closed superstrings, which is the one most directly relevant to quantum gravity. The surface swept by the string is the \textit{worldsheet}, the generalisation of the worldline of a particle. The string can split and merge, making ``loops''. The ``loop order'' of this expansion is the \textit{genus} of the string worldsheet, and for this reason we talk of a \textit{genus expansion} of string observables, whose expansion parameter is the \textit{string coupling constant}~$g_s$. More often than not, one considers a perturbative expansion in the genus. For our purposes here, we can be even more modest and consider the case of free strings propagating on a fixed geometry, \textit{i.e.}~$g_s=0$.

Even with such simplifications, understanding string theory is far from easy. First of all, the ``target'' geometry on which superstrings can consistently propagate (\textit{i.e.}, without running into anomalies in the quantum theory) cannot be arbitrary: it has to be a ten-dimensional solution of the supergravity equations, described not only by the metric, but by additional fields such as the dilaton, the Kalb-Ramond field, and by the so-called Ramond-Ramond (RR) fluxes (which are important for coupling the fermions to the geometry). Secondly, even if $g_s=0$ there are many parameters in this theory. One important parameter is the string tension~$T$ (which can be treated as a dimensionless quantity by rescaling it by a typical length scale in the target space). When the tension is very large, the strings shrink to a point; in this limit, supergravity captures well the features of the strings. Vice versa, when the tension is small, the strings are ``floppy'' and probe the geometry very differently from how a probe particle would. Very often, depending on the details of the target space, there may be more parameters that influence the string dynamics.

In this limit of free strings, solving string theory means solving the two-dimensional quantum field theory on the string worldsheet. Thankfully, this theory  has some special features: it is invariant under two-dimensional diffeomorphism and Weyl symmetry. In the Ramond-Neveu-Schwarz (RNS) approach, one supplements these invariances with \textit{worldsheet supersymmetry}. Then, it is possible to quantise the theory in a covariant way by using the Becchi-Rouet-Stora-Tyutin (BRST) approach. A big drawback of this approach, however, is that it becomes very unwieldy in the presence of RR fluxes: the worldsheet CFT is non-local, and the matter and ghosts sectors become intertwined~\cite{Berenstein:1999jq, Berenstein:1999ip, Cho:2018nfn}.
An alternative approach is to write an action which is supersymmetric in target-space; this is the Green-Schwarz (GS) approach. In this case, the theory is local but in general it is not known how to quantise it covariantly. One instead can use a lightcone gauge, which, however, in general yields a non-conformal interacting two-dimensional QFT in finite-volume --- this is hardly a tractable theory!

Because of these limitations, it is difficult to study superstring theory \textit{even at $g_s=0$}, on most backgrounds. A few notable exceptions are
\begin{itemize}
    \item Flat space; this is the best-studied string background and it can be considered in detail both in the RNS formalism (in fact, one can even go beyond the $g_s=0$ setup which we have been discussing here) and in the GS formalism; in that case, the light-cone gauge-fixed theory is free (with an appropriate choice of the gauge), see \textit{e.g.}~\cite{Green:1987sp,Green:2012pqa, Polchinski:1998rq,Polchinski:1998rr}.
    \item pp-wave geometries~\cite{Blau:2001ne,Metsaev:2001bj}, \textit{even in the presence of RR fluxes} (which here are constant). Here the preferred approach is the GS formalism because, with a suitable gauge-fixing, again we find free theories~\cite{Metsaev:2002re}.
    \item Certain curved geometries without RR fluxes; an important example in this context is the $AdS_3\times S^3\times T^4$ background~\cite{Giveon:1998ns}. In the RNS formalism this yields a supersymmetric WZW model, which can be solved in terms of the free action of Ka\v{c}-Moody algebras of $\mathfrak{sl}(2,\mathbb{R})$ and $\mathfrak{su}(2)$ up to suitably physical-state constraints~\cite{Maldacena:2000hw}.
\end{itemize}
However, there are many important examples which cannot be solved in such a way. An important case is the maximally supersymmetric Anti-De Sitter geometry, \textit{i.e.}~$AdS_5\times S^5$; free strings here are dual~\cite{Maldacena:1997re} to the four-dimensional $SU(N_c)$ $\mathcal{N}=4$ supersymmetric Yang-Mills theory in the 't Hooft limit $N_c\to\infty$~\cite{tHooft:1973alw}, and the string tension~$T$ is related to the 't Hooft coupling. In this sense, it is quite \textit{unsurprising} that we might struggle to solve such a string theory, as this should be as hard as solving a large-$N_c$ non-abelian Yang-Mills theory in four dimensions --- a far cry from the free (or almost-free) theories discussed in the bullet points above!  

For the specific case of $AdS_5\times S^5$ strings and $\mathcal{N}=4$ SYM, a remarkable fact provides an alternative way to solve the planar limit of the theory --- the model is \textit{integrable}, meaning that it possess infinitely many symmetries which constrain its dynamics. This was first famously observed on the $\mathcal{N}=4$ SYM side of the duality~\cite{Minahan:2002ve}, but it is perhaps easiest to understand from the string worldsheet~\cite{Bena:2003wd}. After all, the model on the string worldsheet is a 2D field theory, and integrability for such theories have been studied since the 1970s. In this context, integrability provides a way to make sense of the lightcone gauge-fixed theory non-perturbatively: By demanding that integrability is preserved by the scattering matrix, we can fix it (almost) uniquely, and use Bethe ansatz techniques to derive the spectrum. The worldsheet-integrability approach to $AdS_5\times S^5$ strings is by now well-established, and well reviewed: see~\cite{Arutyunov:2009ga} for a review of worldsheet integrability specifically, and~\cite{Beisert:2010jr} for a more general review of AdS$_5$/CFT$_4$ integrability.

Integrability is not the only way, at least in principle, to incorporate the Ramond-Ramond flux on the worldsheet.  The hybrid formalism~\cite{Berkovits:1999im} is one way to modify the worldsheet CFT approach in such a way to include RR flux. Since its inception, it was clear that this would be especially suitable for backgrounds such as $AdS_3\times S^3\times T^4$. As we mentioned, that background can be realised as a level-$k$ supersymmetric WZW model (without any RR flux, but with a Kalb-Ramond $B$-field). The RR flux can be turned on by a continuous modulus (in perturbative string theory) starting from this simple CFT. The hybrid formalism saw a recent revival in the context of AdS$_3$/CFT$_2$. However, at least for now, it has not yet provided a solution for the spectrum in presence of RR and Neveu-Schwarz-Neveu-Schwarz (NSNS) fluxes related to the Kalb-Ramond field; instead, it has been extremely valuable in better understanding the $k=1$ model~\cite{Eberhardt:2018ouy}.

The $AdS_3\times S^3\times T^4$ background, and more generally AdS$_3$/CFT$_2$, can also be studied by integrability. In fact, quite remarkably, these backgrounds are integrable for arbitrary combinations of RR and NSNS fluxes~\cite{Cagnazzo:2012se}. While the case of pure-NSNS background can be studied by integrability in lightcone gauge~\cite{Dei:2018mfl}, its integrable structure is perhaps ``too simple'', and very close to that of a free theory. This is also the case for flat-space strings~\cite{Dubovsky:2012wk,Baggio:2018gct}, and indeed both cases can be thought of as a sort of ``$T\bar{T}\,$'' deformation~\cite{Zamolodchikov:2004ce, Smirnov:2016lqw, Cavaglia:2016oda} of a free model. Conversely, the integrable structure of $AdS_3\times S^3\times T^4$ is particularly intriguing at the pure-RR point, where it bears some similarities (and a few crucial differences) to the famous case of $AdS_5\times S^5$, both at the classical~\cite{Babichenko:2009dk} and quantum~\cite{Borsato:2012ud} levels. In a way, we have two competing pictures. On the one hand, the worldsheet-integrability one, which can give the spectrum of pure-RR backgrounds (as well as the pure-NSNS backgrounds, but somewhat trivially), and then requires introducing $k=1,2,3,\dots$ to incorporate the NSNS flux;  the integrability construction for mixed-flux backgrounds has been partially worked out at the quantum level in terms of a factorised S~matrix~\cite{Lloyd:2014bsa}, but it has not yet been completed and the equations describing the exact mixed-flux spectrum are currently unknown. On the other hand, we have the hybrid formalism, which works very well for pure-NSNS backgrounds, but then requires turning on the (continuous) RR flux; currently, it is not clear how to do this in a way that allows to solve for the spectrum of generic states, not to mention other observables. Either way, the setup with both RR and NSNS fluxes is currently \textit{terra incognita}, and a current focus of research.

In view of all this, a large part of this review is devoted to review these approaches to the string worldsheet with particular reference to AdS$_3$/CFT$_2$. While the integrability approach has been reviewed in~\cite{Sfondrini:2014via} relatively recently, a number of new results have been found since then, as we will see. By contrast, our presentation of the hybrid approach is as far as we know the first recent review of the subject, and we hope that it will be helpful to those who would like to enter the field. The last two chapters introduce a related, but more ambitious, topic --- understanding M~theory, at least in some decoupling limits where it is related to corners of string theory that exhibit non-relativistic behaviors. 
Intriguingly, this appears to be closely connected to some of the ideas encountered when discussing the string theory, such as the uniform lightcone gauge~\cite{Arutyunov:2005hd} and the aforementioned $T\bar{T}$ deformations.

\paragraph{Note to the ArXiv submission.}
This review was elaborated based on the lecture notes of the Young Researchers Integrability School and Workshop ``Taming the string worldsheet'', held at NORDITA (Stockholm) in the period 23--29 October, 2022. As you will see, we have decided to leave into the text some ``exercises'' --- simple computations that interested readers should be able to complete themselves.
Each of the subsequent sections has been originally drafted by a single author, with particular expertise on the subject. However, through a number of subsequent revisions, we have tried to craft an overall coherent narrative and keep our discussion pedagogical and our notation uniform. The text has come a long way from the original lecture notes used in the school, in terms of overall length as well as, importantly, refinement.
Given the breadth and sheer length of the review, however, we anticipate that there might be further room for improvement, and we would be grateful for any comments --- from typos to conceptual points.

\newpage
\section{General aspects of Green-Schwarz superstrings} \label{s:sibylle}

%\textit{Current author: Sibylle Driezen. } For questions, comments, typo's, \textit{etc.}, feel free to write to \href{mailto:sib.driezen@gmail.com}{sib.driezen@gmail.com}.

%%%% PART I (Sibylle)
%\subsection{Setting and motivation}
As understood from the main introduction, 
strings propagating in curved backgrounds are notoriously hard to quantise and  describe non-perturbatively  at the level of the worldsheet. The worldsheet theory is in fact a two-dimensional interacting field theory with  infinitely many interaction terms and coupling constants. Nevertheless, their strengths are all measured by a dimensionless quantity characterised by the tension of the string, and this allows us to still make some interesting statements. In particular,  at the lowest order of the weakly-coupled regime (corresponding to large tension, \textit{i.e.}~point-like strings), consistency requirements demand that the background fields satisfy the supergravity equations of motion---a generalisation of ordinary Einstein gravity. Higher orders in perturbation theory as well as non-perturbative statements are, however, quite hard to probe. 

However, when the string moves in flat space, the string action is quadratic. This fact allows for the quantization of the theory using two commonly employed canonical methods. The first goes under the name of covariant quantisation, in which part of the worldsheet gauge symmetries are fixed such that the theory becomes free, though constrained, and conformally invariant. Quantisation is then done by common techniques for CFTs subjected to constraints. In the second method, the constrained system is avoided by identifying the full  physical phase space prior to the act of quantisation. In doing so one fixes all worldsheet gauge symmetries but one destroys manifest Lorentz covariance. This method goes under the name of lightcone quantisation. Both methods  of course yield the same results but they can offer different viewpoints on the matter under question.

The generalisation of both methods to strings in curved space each come with their own issues. For covariant quantisation one would need to know the (interacting) worldsheet CFT description at the exact level in the string tension. Each different supergravity solution in fact corresponds to a different worldsheet CFT.\footnote{Knowing the supergravity solution relates to the worldsheet CFT at lowest order. The fully-fledged exact CFT corresponds however to the complete quantum-corrected background fields.  } Good toy models to probe such theories are based on Wess-Zumino-Witten CFTs. On the other hand, lightcone quantisation will highly depend on the particular coordinate system used for the curved spacetime, as well as the particular gauge chosen. Moreover, the  resulting gauge-fixed theory will  become highly non-linear, making in general only perturbative statements possible. 

Introducing superstrings makes a clear distinction in the favourable method to use. The Ramond-Neveu-Schwarz (RNS) superstring, where fermions are introduced on the worldsheet on the same footing as worldsheet bosons, is typically quantised using covariant quantisation. The Green-Schwarz (GS) superstring, where fermions are introduced through manifest target space supersymmetry, is typically quantised using lightcone quantisation, since  additional phase-space constraints arising in this formalism are not understood in a covariant way. The main advantage of the GS superstring, however, is that they can be described in generic curved spaces---supergravity theories where all fields are turned on---while this is not the case for RNS superstrings.\footnote{In particular, in the presence of Ramond-Ramond (RR) background fields the RNS worldsheet model becomes nonlocal, or in any case features an intricate and interacting system of ghost fields that seems hard to decouple.}
While worldsheet CFT techniques are lost, however, many interesting classes of non-trivial GS superstrings (e.g.~those present in the AdS/CFT correspondence) instead enjoy the property of worldsheet integrability. Such theories are characterised by an infinite set of conserved charges and correspondingly unique mathematical structures, which can provide a large number of  techniques facilitating the computation of observables (chiefly, the spectrum) of the quantum theory exactly in the string tension. 

This section gives a general introduction to the Green-Schwarz formulation of superstring theory. We start in section \ref{s:sib:bosonic} with a review on the bosonic string, laying-out conventions and notation, and highlighting the important facts for the sections to come. We will discuss how to fix the gauge that sets the stage to lightcone quantisation, both in flat and in curved space. In section \ref{sec:Sib:prereq:GS} we then move on to the Green-Schwarz superstring. We will put particular importance on the introduction of a new  worldsheet gauge symmetry in this formalism, known as $\kappa$-symmetry, that similarly should be fixed for lightcone quantisation. Again, we will distinguish the flat- and curved-space GS superstring. After the general description, we will consider in section \ref{sec:Sib:supercoset_constr_GS}  a particular class of GS strings propagating in geometries realised as supercosets. This includes e.g.~$AdS_n\times S^n$ backgrounds relevant for the AdS/CFT correspondence. Compared to the generic curved GS string, they can be realised using a slightly different construction which takes advantage of the geometrical supercoset structure. In this language, classical worldsheet integrability materialises quite elegantly, as we will show in section \ref{sec:sib:integrability}. We end this section with some concluding remarks in \ref{sec:sib:concl}. 
References to the original literature, as well as books and  lecture notes, will be given along the way. 

The subsequent section \ref{s:saskia} will then treat the GS formalism for specific supergravity backgrounds, \textit{i.e.}~$AdS_3\times S^3$ backgrounds, which are ideally suited to highlight important but subtle differences between the general GS and the supercoset GS decription. It will particularly illustrate the need to have a hybrid description between the RNS and GS formalism, which will be discussed in section \ref{s:bob}, in order to progress in its exact worldsheet description in general. In sections \ref{s:fiona} and \ref{ana:sec}, however,  the quantum level will rather be explored in the GS formalism using worldsheet integrability. 

\subsection{Bosonic strings} \label{s:sib:bosonic}
Before we introduce the Green-Schwarz superstring,  let us first briefly review  the non-linear sigma-model describing bosonic strings propagating in  flat and curved backgrounds.\footnote{For good notes and reviews see e.g.~\cite{Tong:2009np,Callan:1989nz,Green:1987sp}.} This is a two-dimensional (respectively free and interacting) field theory defined on the surface $\Sigma$, called the worldsheet,  swept out by the string.  We will  parametrise $\Sigma$ by a time $\tau$ and spatial $\sigma$ coordinate as  $\sigma^\alpha = (\tau, \sigma)$, $\alpha=0,1$. Throughout this section, we will only consider the classical theory of closed strings with 
\begin{equation}
    \sigma \in (0, R) \ , \qquad \sigma \sim \sigma + R \ ,
\end{equation}
with the exception of a few comments regarding the quantum level. We will focus in particular on the classical global and local symmetries of the action, as well as fixing the gauge.

\subsubsection{In flat space} 

\noindent {\bf Action and equations of motion} --- When the bosonic string propagates in a $D$-dimensional flat Minkowski spacetime ${\cal M}$, its dynamics is encoded by the  Polyakov action defined on $\Sigma$
\begin{equation}\label{eq:polyakov}
    S_\text{P}[X,h] = - \frac{T}{2} \int_\Sigma \de^2\sigma \sqrt{-h} h^{\alpha\beta} \partial_\alpha X^\mu \eta_{\mu\nu} \partial_\beta X^\nu , \qquad \de^2\sigma =\de\tau\de\sigma \ ,
\end{equation}
where $T=(2\pi\alpha')^{-1}$ is the tension of the string, $\alpha'$ the worldsheet coupling parameter,  $\sigma^\alpha = (\tau,\sigma)$ the worldsheet coordinates, $h_{\alpha\beta}$  the  worldsheet metric, $\eta_{\mu\nu}$, with $\mu,\nu=0,\ldots, D-1$,  the  Minkowski metric and  $X^\mu$ the  coordinates on ${\cal M}$. 
Intuitively, the tension $T$ measures the energy per unit length stored by the string. As a result, for large $T$ (small $\alpha'$) the string behaves effectively point-like and the theory enters the semi-classical (supergravity) approximation. On the other hand, when $T$ is nearly vanishing, the string becomes very large and stringy effects dominate.
The coordinates $X^\mu(\sigma) \equiv X^\mu(\tau,\sigma) $  can be interpreted both as free  scalar fields  on $\Sigma$ coupled to two-dimensional gravity as well as  the embedding coordinates of the string in ${\cal M}$. Having these two interpretations is the defining feature of a non-linear  sigma-model, \textit{i.e.}~a field theory where the fields are maps $X^\mu(\sigma)$ from a base space, here $\Sigma$, to a target space, here ${\cal M}$.  We assume that the string is not winding around one of the target space directions. 
%\FS{This may be confusing, I presume here you write $\sigma$ for the vector $\sigma=(\tau, \sigma)$... I would not use it as a vector (but now I realise that it makes notation easier for the rest of the section --- then maybe bold for a vector? or keep the index $\sigma^\alpha$?), and also earlier define $\extder^2 \sigma=\extder \tau \extder\sigma$}

The equations of motion for the two dynamical fields $X^\mu$ and $h_{\alpha\beta}$ are
\begin{align}
    \delta X^\mu : \qquad & \partial_\alpha \left( \sqrt{-h} h^{\alpha\beta} \partial_\beta X^\mu \right) =0 \ , \\
    \delta h^{\alpha\beta} : \qquad & T_{\alpha\beta} = \partial_\alpha X^\mu \eta_{\mu\nu} \partial_\beta X^\nu - \frac{1}{2} h_{\alpha\beta} h^{\gamma\delta} \partial_\gamma X^\mu \eta_{\mu\nu} \partial_\delta X^\nu = 0 \ , \label{eq:virasoro-flat}
\end{align}
where $T_{\alpha\beta}$ is the worldsheet energy-momentum tensor. The latter are known as the Virasoro constraints for the flat space string. Integrating  the solution for $h^{\alpha\beta}$  out of the Polyakov action $S_{\text P}[X,h]$ gives the Nambu-Goto action 
\begin{equation}
    S_{\text{NG}}[X] = - T \int \de^2\sigma \sqrt{- \mathrm{det} \left(\partial_\alpha X^\mu \eta_{\mu\nu} \partial_\beta X^\nu \right) } \ ,
\end{equation}
which is proportional to the area of the worldsheet (and, therefore, is the natural generalisation of particles to strings).\\

\noindent {\bf Symmetries} --- The Polyakov action enjoys a number of symmetries. The worldsheet theory is invariant under two gauge symmetries: worldsheet diffeomorphisms $\sigma^\alpha \rightarrow \tilde{\sigma}^\alpha =\tilde{\sigma}^\alpha(\sigma)$
\begin{equation} \label{eq:ws-diffeo}
     X^\mu (\sigma) \rightarrow \tilde{X}^\mu(\tilde{\sigma}) = X^\mu(\sigma) \ , \qquad h_{\alpha\beta}(\sigma) \rightarrow \tilde{h}_{\alpha\beta}(\tilde{\sigma}) = \frac{\partial\sigma^\gamma}{\partial\tilde{\sigma}^\alpha} \frac{\partial\sigma^\delta}{\partial\tilde{\sigma}^\beta} h_{\gamma\delta}(\sigma) \ ,
\end{equation}
and Weyl rescalings 
\begin{equation} \label{eq:weyl-rescaling}
    X^\mu (\sigma) \rightarrow X^\mu(\sigma) \ , \qquad h_{\alpha\beta}(\sigma) \rightarrow \Omega(\sigma) h_{\alpha\beta}  (\sigma) \ , 
\end{equation}
for some local scalar factor $\Omega$.
Furthermore, because we are in flat space, the theory has a global symmetry which corresponds to Poincar\'e invariance in target space
\begin{equation}
    X^\mu \rightarrow \Lambda^\mu{}_\nu X^\nu + c^\nu \ .
\end{equation}

\noindent {\bf Gauge fixing} --- With the purpose of quantisation in mind, it is useful to identify the physical degrees of freedom and phase-space of the worldsheet theory. To completely do this, one would need to fix the gauge symmetries and solve the Virasoro constraints. A first useful gauge choice is the so-called conformal gauge, which fixes part of the worldsheet diffeomorphisms in $\tau$ and $\sigma$ as well as the Weyl rescalings such that the worldsheet metric is flat:
\begin{equation} \label{eq:conformal-gauge}
    \sqrt{-h} h^{\alpha\beta} = \eta^{\alpha\beta} = \mathrm{diag} (-1,1)^{\alpha\beta} \ .
\end{equation}
The Polyakov action then simplifies to the action of free scalars governed by a  wave equation $\partial^\alpha \partial_\alpha X^\mu = 0$ subjected to the Virasoro constraints \eqref{eq:virasoro-flat}. The former can be solved by
\begin{equation}
    X^\mu (\tau, \sigma) = X^\mu_L (\sigma^+) + X^\mu_R(\sigma^-) \ , \qquad \sigma^\pm = \frac{1}{2} (\tau \pm \sigma) \ ,
\end{equation}
for some arbitrary periodic functions $X^\mu_L$ and $X^\mu_R$ describing left- and right-moving waves respectively. They can generally be expanded in Fourier modes, $\alpha_{L,n}^\mu$ and $\alpha_{R,n}^\mu$, as 
\begin{equation}
    X^\mu_{L/R} (\sigma^\pm) = \frac{1}{2} x^\mu +  p^\mu \sigma^\pm +  \sum_{n\neq 0 } \alpha_{L/R,n}^\mu e^{-i n \sigma^\pm} \ ,
\end{equation}
where the constants $x^\mu$ and $p^\mu$ correspond to the position and momentum of the center of mass of the string respectively.

After imposing conformal gauge, there are still some residual gauge symmetries: diffeomorphisms that can be compensated by a Weyl transformation, which change the worldsheet coordinates only holomorphically (\textit{i.e.}~by $\sigma^\pm \rightarrow \tilde{\sigma}^\pm (\sigma^\pm)$). These are precisely conformal symmetries on $\Sigma$, and thus the action $S_{\text P} [X, h=\eta]$ describes a conformal field theory (CFT) of free bosonic fields $X^\mu$. Quantisation of such free CFTs, which preserves the conformal gauge symmetries, is well understood and is known as covariant quantisation.\footnote{At the worldsheet quantum level, this theory in general suffers from a gauge anomaly, known as the Weyl anomaly, which can be cancelled by taking $D=26$ free bosonic fields, or by considering any other matter CFT with central charge  $c_{\text{\tiny matter}}=26$. 
%These matter CFTs can for instance be a theory of interacting bosons  describing strings in curved spacetimes, such as Wess-Zumino-Witten models. 
\label{f:weyl-anomaly}}  More details  can be found in~section \ref{s:bob}.

Instead of covariant quantisation, one can also quantise the worldsheet theory by adapting a further gauge choice of the residual holomorphic diffeomorphisms. A very convenient gauge to do this is the so-called  lightcone gauge. Introducing the lightcone coordinates
\begin{equation}\label{eq:lightcone-coordinates}
    X^\pm = \frac{X^0 \pm X^{D-1}}{\sqrt{2}} \ ,
\end{equation}
the lightcone gauge of the holomorphic symmetries fixes the solutions $X^+_L(\sigma^+)$ and $X^+_R(\sigma^-)$ such that   the spacetime coordinate $X^+$ is identified with worldsheet time $\tau$ as
\begin{equation} \label{eq:lightcone-gauge}
    X^+  = x^+ + p^+ \tau \ .
\end{equation}
The advantage of working in  this gauge is that   the field $X^-$, and consequently $p^-$, is completely determined (up to the integration constant $x^-$) by the Virasoro constraints: they can be written   in terms of $p^+ = - p_-$ and the remaining fields $X^1, \ldots, X^{D-2}$, which will be the only physical degrees of freedom left, and are referred to as the transverse coordinates. As they can be split in left and right movers, there are in total $2\times (D-2)$ of them. Note that  $p^-=-p_+$  can be thought of as the lightcone Hamiltonian generating shifts in $X^+$ and thus $\tau$. The Virasoro constraints are then partially solved up to one constraint, known as the level-matching condition, which is a mass-shell constraint that can be tracked to demanding periodicity of the fields $X^\mu(\tau,\sigma)$ in $\sigma$.
After quantisation, the mass-shell constraint rather straightforwardly allows one to determine the target space spectrum. However, the  price to pay for  using lightcone gauge is that  Lorentz covariance is explicitly broken. Enforcing Lorentz covariance of the quantum theory (which is also done by taking $D=26$ as in footnote \ref{f:weyl-anomaly}) gives a spectrum that besides  massive excitations (and a tachyonic excitation)  includes  three fundamental massless excitations: a spin two graviton, the antisymmetric B-field, and the dilaton
\begin{equation} \label{eq:G-B-Dil}
    G_{\mu\nu}(X) \ , \qquad B_{\mu\nu}(X) \ , \qquad \Phi(X) \ ,
\end{equation}
respectively. We will not go into further details, but the  reader can e.g.~consult \cite{Tong:2009np} and \cite{Green:1987sp}.

\subsubsection{In curved backgrounds} 

One can now consider a collection of the massless excitations \eqref{eq:G-B-Dil} of the flat space closed string to produce a curved background in which  another (probe) closed string can propagate. In particular, one constructs in this way a curved target space geometry with a curved metric $G_{\mu\nu}(X)$, which is  built out of quantised graviton states, an antisymmetric B-field $B_{\mu\nu}(X)$, and a dilaton $\Phi(X)$.

\bigskip
\noindent {\bf Action and equations of motion} ---  The corresponding string action is
\begin{equation} \label{eq:S-G-B}
\begin{aligned}
    S[X,h] = -\frac{T}{2} \int_\Sigma \de^2\sigma &\left(\sqrt{-h}h^{\alpha\beta} \partial_\alpha X^\mu G_{\mu\nu}(X) \partial_\beta X^\nu + \epsilon^{\alpha\beta} \partial_\alpha X^\mu B_{\mu\nu}(X) \partial_\beta X^\nu \right. \\ \ \ &\left. + \alpha' \sqrt{-h} \mathcal{R} \Phi(X) \right) \ ,
\end{aligned}
\end{equation}
where $\epsilon^{\alpha\beta}$ is the Levi-Civita symbol with convention $\epsilon^{\tau\sigma}=1$ and $\mathcal{R}$ is the worldsheet Ricci scalar.
From now on we will ignore the dilaton contribution as it can be understood as a higher-order worldsheet quantum effect (in $\alpha'$).

The equations of motions now are
\begin{align}
    \delta X^\mu : \qquad & \partial_\alpha \left( \sqrt{-h} h^{\alpha\beta} \partial_\beta X^\mu \right) + \Gamma^{(-\alpha\beta)\mu}_{\nu\rho}\partial_\alpha X^\nu \partial_\beta X^\rho =0 \ , \label{eq:eom-curved} \\
    \delta h^{\alpha\beta} : \qquad & T_{\alpha\beta} = \partial_\alpha X^\mu G_{\mu\nu} \partial_\beta X^\nu - \frac{1}{2} h_{\alpha\beta} h^{\gamma\delta} \partial_\gamma X^\mu G_{\mu\nu} \partial_\delta X^\nu = 0 \ , \label{eq:virasoro-curved}
\end{align}
where $\Gamma^{(-\alpha\beta)\mu}_{\nu\rho}=\sqrt{-h} h^{\alpha\beta}\Gamma^\mu_{\nu\rho}-\frac{1}{2} \epsilon^{\alpha\beta} H^\mu_{\nu\rho}$ with $\Gamma^\mu_{\nu\rho}$  the usual Christoffel symbol and $H=\mathrm{d}B$ the so-called torsion three-form. The latter equations \eqref{eq:virasoro-curved} are the Virasoro constraints for the curved space string.\footnote{If we would not ignore the dilaton term, the energy-momentum tensor $T_{\alpha\beta}$ would receive a contribution from $\Phi$ which breaks classical Weyl invariance. Weyl invariance can, however, be restored by quantum loop contributions from worldsheet interactions. \label{f:weyl-anomaly2}}

\vspace{.2cm}

\begin{centering}
\begin{tcolorbox}
\begin{exercise}
Derive the equations of motion for $X^\mu(\tau, \sigma)$ from the action \eqref{eq:S-G-B} with vanishing dilaton term. Note that this is a generalisation of the geodesic equation for point particles $X^\mu(\tau)$ in curved spaces.
\end{exercise}
\end{tcolorbox}
\end{centering}

\vspace{.2cm}

\noindent {\bf Symmetries} --- The action and equations of motion are invariant under target space diffeomorphisms $X^\mu \rightarrow \tilde{X}^\mu=\tilde{X}^\mu (X)$, transforming the metric and B-field as 
\begin{equation} \label{eq:G-B-diffeo}
    \tilde{G}_{\mu\nu}(\tilde{X}) = \tilde{\partial}_\mu X^\rho\tilde{\partial}_\nu X^\lambda G_{\rho \lambda}(X) , \qquad \tilde{B}_{\mu\nu}(\tilde{X}) = \tilde{\partial}_\mu X^\rho\tilde{\partial}_\nu X^\lambda B_{\rho \lambda}(X) \ ,
\end{equation}
as well as B-field gauge transformations, which in form notation reads
\begin{equation} \label{eq:b-gauge-transf}
    B \rightarrow B+ \mathrm{d}\Lambda ,
\end{equation}
for some arbitrary  one-form $\Lambda (X)$. The gauge-invariant field strength is  the torsion three-form $H=\mathrm{d}B$.  However, the target space diffeomorphisms are only worldsheet symmetries  when they correspond to isometries of target space, \textit{i.e.}~when $\tilde{G}_{\mu\nu}(\tilde{X})=G_{\mu\nu}(\tilde{X})$ and $\tilde{H}_{\mu\nu\rho}(\tilde{X})=H_{\mu\nu\rho}(\tilde{X})$ (otherwise they are not a true transformation of the system, but rather a reformulation). In that case, they are in fact \textit{global} symmetries from the point of view of the worldsheet.  Consider the infinitesimal diffeomorphism
\begin{equation}\label{eq:inf-diffeo}
    \delta X^\mu = \xi^{A} k_{A}^\mu (X) \ ,
\end{equation}
where $\xi^{A}$  are a collection of infinitesimal parameters, indexed by $A$,  of the coordinate transformation generated by the vector fields $k_{A}^\mu(X)$. The isometry condition holds to ${\cal O}(\xi^2)$ if the $k_{A}^\mu (X)$ are Killing vector fields, \textit{i.e.}~if
\begin{equation}
    L_{k_{A}} G_{\mu\nu} = L_{k_{A}} H_{\mu\nu\rho} = 0 \ ,
\end{equation}
where $L_{k}$ denotes the usual Lie derivative with respect to the vector $k$.

\vspace{.2cm}

\begin{centering}
\begin{tcolorbox}
\begin{exercise}
Show that, up to total derivative terms, the transformation \eqref{eq:inf-diffeo} is  a global symmetry of  the action \eqref{eq:S-G-B} (again let $\Phi(X)=0$)  when $L_{k_{A}} G_{\mu\nu} =0$ and $L_{k_{A}} B_{\mu\nu} =(\mathrm{d}\omega)_{\mu\nu}$ for some arbitrary one-form $\omega(X)$. 
\end{exercise}
\end{tcolorbox}
\end{centering}

\vspace{.2cm}

\noindent The corresponding Noether currents ${\cal J}^{\alpha}_{A}$ can be found by promoting $\xi^A$ to a local parameter, $\xi^{A} = \xi^{A}(\tau, \sigma)$,  varying the action \eqref{eq:G-B-Dil} under \eqref{eq:inf-diffeo}, and then reading ${\cal J}^{\alpha}_{A}$ off from terms  proportional to $\partial_\alpha\xi^{A}$ as 
\begin{equation}
\delta S = -T \int_\Sigma \de^2\sigma \   \partial_\alpha \xi^{A} {\cal J}^{\alpha}_{A}\ .
\end{equation}
One finds 
\begin{equation}
   {\cal J}^{\alpha}_{A} = k_{A}^\mu\left(\sqrt{-h} h^{\alpha\beta}  G_{\mu\nu} + \epsilon^{\alpha\beta} B_{\mu\nu}\right) \partial_\beta X^\nu - \epsilon^{\alpha\beta} \omega_\mu \partial_\beta X^\mu \ ,
\end{equation}
which satisfies the conservation equation
\begin{equation}
    \partial_\alpha {\cal J}^{\alpha}_{A} = 0 \ ,
\end{equation}
upon using the equations of motion and Killing equations. 
The associated Noether charges then are
\begin{equation}
    Q_{A} =- T \int^{\cir}_0 \de \sigma \ {\cal J}^{\tau}_{A} \ .
\end{equation}

\vspace{.2cm}

\begin{centering}
\begin{tcolorbox}
\begin{exercise}
Show that $\partial_\alpha {\cal J}^{\alpha}_{A} = 0$ upon using  the equations of motions and  the property that $k_{A}^\mu$ are Killing vectors. 
\end{exercise}
\end{tcolorbox}
\end{centering}

\vspace{.2cm}

Recall that Killing vectors fields close under the Lie bracket and thus form a Lie algebra $\mathfrak{g}$ corresponding to the isometries of the target manifold. Classically, the Noether charges close under the Poisson brackets into the same Lie algebra. The index $A$ thus corresponds to a Lie algebra index $A=1, \ldots , \mathrm{dim}\,\mathfrak{g}$.

\bigskip
\noindent {\bf Gauge fixing} --- Besides the \textit{target space} gauge transformations \eqref{eq:G-B-diffeo} and \eqref{eq:b-gauge-transf}, the worldsheet theory is still invariant under the  worldsheet diffeomorphism \eqref{eq:ws-diffeo} and Weyl rescaling \eqref{eq:weyl-rescaling} gauge symmetries. One can therefore still fix the conformal gauge as in \eqref{eq:conformal-gauge}, but the difference now is that $S[X,h=\eta]$ does not reduce to a free field theory of bosons, but rather to a non-trivial interacting theory known as the non-linear sigma-model.\footnote{Strictly speaking, the term non-linear sigma-model refers to the theory $S[X,\eta]$ with only non-trivial metric coupling $G$. However, it is also used in the string community to describe propagation of strings in all backgrounds fields. The meaning should be clear from the context.}
%In fact, there are an infinite number of coupling parameters as can be observed by Taylor expanding the action around a classical solution $X^\mu(\sigma) = X^\mu_{\text{\tiny cl.}} + \sqrt{\alpha'} X^\mu_{\text{\tiny fl.}}(\sigma)$. Each derivative of the metric and B-field can be thought of as coupling constants for interactions in the fluctuations, and they all come acompanied with a factor $\sqrt{\alpha'}$. The worldsheet perturbation series is thus an $\alpha'$-expansion with $\alpha'$ the loop counting parameter.
In conformal gauge and lightcone coordinates $\sigma^\pm =\frac{1}{2}(\tau \pm \sigma)$ the action then reads 
\begin{equation}
    S[X] =  \frac{T}{2} \int_\Sigma \de^2\sigma \  \partial_+ X^\mu \left(G_{\mu\nu}(X) + B_{\mu\nu}(X) \right) \partial_- X^\nu \ ,
\end{equation}
supplemented by the Virasoro constraints \eqref{eq:virasoro-curved} 
\begin{equation}
    \partial_\pm X^\mu G_{\mu\nu}(X) \partial_\pm X^\nu = 0 \ .
\end{equation}
As opposed to the flat space case, the  equations of motion of $X^\mu$ \eqref{eq:eom-curved}  in conformal gauge are  in general  not easily solvable without restricting to a certain class of solutions\footnote{A trivial ``class'' of solutions is the vacuum $X^\mu = 0$.} or making further assumptions about the target space geometry. The quantum level is further complicated by  non-trivial interaction terms which can be obtained by expanding the action around a constant classical solution $\bar{X}^\mu$ with small fluctuations $\hat{X}^\mu(\tau, \sigma)$ as $X^\mu(\tau, \sigma) = \bar{X}^\mu + \sqrt{\alpha'} \hat{X}^\mu(\tau, \sigma)$. Each derivative of the metric and B-field can then be thought of as coupling constants for interactions in the fluctuations, and they all come acompanied with a factor $\sqrt{\alpha'}$. The worldsheet perturbation series is thus an $\alpha'$-expansion with $\alpha'$ the loop counting parameter. Because of the (local) residual conformal symmetry, this theory should be an \textit{exact} interacting boson CFT, without any gauge anomaly. For curved backgrounds, the Weyl anomaly cancellation conditions not only requires $c_{\text{\tiny matter}}=26$, but cancelling additional loop contributions  will furthermore severely constrain the target space geometry. In particular, to first order in $\alpha'$ (in which the dilaton term is relevant, see also footnote \ref{f:weyl-anomaly}) the background fields $G$, $B$, and $\Phi$ must satisfy
\begin{align}
    {\cal R}_{\mu\nu} - \frac{1}{4} H_{\mu\nu}^2 + 2 \nabla_\mu \nabla_\nu \Phi &=0 \ , \\
    \nabla^\rho H_{\rho\mu\nu} - 2 \nabla^\rho\Phi H_{\rho\mu\nu} &=0 \ ,
\end{align}
with ${\cal R}_{\mu\nu}$  the target space Ricci tensor. These equations are known as the type II bosonic supergravity equations in the absence of Ramond-Ramond fields, justifying the name ``supergravity approximation'' for the region of small $\alpha'$ (large $T$). Interestingly, the one-loop Weyl anomaly coefficients  can be related to  the one-loop renormalisation group  $\beta$-functions obtained from a worldsheet perturbative expansion, which of course  should also vanish  to preserve worldsheet conformal invariance at the quantum level. 
For more details see \cite{Callan:1989nz}. For generic curved backgrounds, quantisation at higher orders is much less understood. In particular,  covariant quantisation is generally not available when the exact worldsheet CFT is unknown. An important exception are string models where the worldsheet CFT is a  Wess-Zumino-Witten theory, in which the target space corresponds to a group manifold. As we mentioned earlier, more information can  be found in section \ref{s:bob}.

 Under some mild assumptions, one can instead perform a type of lightcone quantisation  by fixing all of the  gauge symmetries, in a similar but different way as in the flat space case. In particular, this can be done conveniently  when the curved background has at least two commuting isometries, one time-like and one space-like,  that  are realised as shifts of the  target space coordinates $X^0$ and $X^{D-1}$ respectively.\footnote{For example, backgrounds of the type $AdS_n\times S^n$ have this property, where the time-like $X^0$ can parametrise the global $AdS$ time and the space-like $X^{D-1}$  can parametrise an angle of the sphere.} In that case, one can consider a simple but non-trivial class of classical pointlike string solutions  that are at most linear in the worldsheet time and are of the form
 \begin{equation} \label{eq:linear-pointlike-string}
     \bar{X}^0 = a^0 \tau ,   \qquad \bar{X}^{D-1} = a^{D-1} \tau , \qquad \bar{X}^i = 0, \qquad i = \{1, \ldots, D-2\} \ ,
 \end{equation}
 with $a^0$ and $a^{D-1}$ real non-zero constants. We assume an adapted coordinate system in which the region defined by \eqref{eq:linear-pointlike-string} is not pathological. Assuming furthermore a gauge of Weyl rescalings in which $\sqrt{-h}h^{\tau \tau}$ is constant,  the equations of motion \eqref{eq:eom-curved} are then indeed trivially solved for all $X^\mu$, since the Christoffel symbol on this pointlike solution vanishes. Introducing $a^\mu = (a^0, 0, \ldots, 0, a^{D-1})$, we can write the Virasoro constraints as
 \begin{equation}\label{eq:virasoro-pointlike}
     a^\mu G_{\mu\nu} (\bar{X}) a^\nu = 0 \ ,
 \end{equation}
 with $G_{\mu\nu} (\bar{X})$ the metric evaluated on \eqref{eq:linear-pointlike-string}.
 For simplicity, let us from now on consider $a^0=a^{D-1}=1$, $G_{00}(\bar{X}) = - G_{D-1,D-1}(\bar{X})=-1$, and $G_{0,D-1}(\bar{X}) = G_{0i}(\bar{X}) = G_{D-1,i}(\bar{X}) =0 $ for all $i=1,\ldots, D-1$. Note that this solves the Virasoro constraints \eqref{eq:virasoro-pointlike}.  As a non-trivial exercise, the reader can drop these assumptions if they wish. 
 On this solution, the conjugate momenta
 \begin{equation}
     P_\mu (X) = \frac{\delta S}{\delta \partial_\tau X^\mu} = - T \sqrt{-h} h^{\tau \beta} \partial_\beta X^\nu G_{\mu\nu}(X) - T \partial_\sigma X^\nu B_{\mu\nu}(X) \ ,
 \end{equation}
 will now read 
 \begin{equation}
     \bar{P}_+ = 0, \qquad  \bar{P}_{-} = \sqrt{2} T \sqrt{-h} h^{\tau \tau} , \qquad \bar{P}_i = 0 \ ,
 \end{equation}
 in which we recall the definition of the lightcone coordinates \eqref{eq:lightcone-coordinates}.
 The lightcone quantisation procedure now starts by gauge-fixing the worldsheet diffeomorphisms  through adopting the so-called \textit{uniform} lightcone gauge, which eliminates the fluctuations
 \begin{equation} \label{eq:sib-ulcg}
     \hat{X}^+ = 0 , \qquad \hat{P}_- = 0 \ ,
 \end{equation}
 and thus the momentum $P_-$ will be distributed \textit{uniformly} along the string. See also \cite{Arutyunov:2009ga} for a review or the recent \cite{Borsato:2023oru}.
 Because of the vanishing vacuum values of the remaining coordinates $\bar{X}^{-},\bar{X}^{i}$ and momenta $\bar{P}_{+}, \bar{P}_{i}$, the only physical remaining degrees of freedom are the fluctuations $\hat{X}^{-}, \hat{X}^{i}$ and $\hat{P}_{+},\hat{P}_{i}$. In the literature, the hatted notation is therefore usually dropped.   
As in the flat space case, the lightcone gauge thus reduces the number of physical bosonic degrees of freedom to $2(D-2)$, at the price of losing  Lorentz covariance. After solving the Virasoro constraints for the gauge-fixed theory, one can show that again $P_+$ can be identified with the worldsheet lightcone Hamiltonian and that it can be completely expressed in terms of the transverse fields ${X}^i$ and $P_i$ only. For consistency, there will again be a type of level-matching condition that can be tracked to demanding periodicity of the fields. For more details, we refer the reader to section \ref{s:strings-lightcone-gauge}, which starts with a more general set-up of uniform lightcone quantisation and discusses in detail how to extract the gauge-fixed worldsheet Hamiltonian. 
 Nevertheless, let us just mention already here that lightcone quantisation for strings in curved backgrounds remains non-trivial, essentially because the lightcone  Hamiltonian will be non-linear, which makes   only perturbative quantisation  possible. Rather than the Hamiltonian, however, the vital object to focus on in order to possibly obtain all-loop results for the spectrum is the worldsheet S-matrix. This is indeed the topic of  section \ref{s:fiona}.

\vspace{.2cm}

 \textit{Remark: lightcone gauge in flat space vs.~curved space} --- Note that the lightcone gauge for the flat space string is a combination of the conformal gauge \eqref{eq:conformal-gauge} and the gauge \eqref{eq:lightcone-gauge}, while for the curved space string the conformal gauge is not adopted, and worldsheet diffeomorphism and Weyl rescalings are fixed rather differently. In fact, for a generic curved background, in lightcone gauge the worldsheet metric~$h_{\alpha\beta}$ must be determined by the Virasoro constraints, and in general $h_{\alpha\beta}\neq\eta_{\alpha\beta}$.
 Another  crucial difference is that in curved space there is no general solution to the equations of motion, and therefore the gauge choice will depend on the class of classical solutions one is interested in. The subsequent procedure of quantisation will therefore   be vacua-dependent. In the flat space case, on the other hand, 
 one can study quantisation universally on top of the completely general classical solution of the free wave equation.

 \vspace{.2cm}

 We are now ready to introduce the Green-Schwarz superstring in the next section. Already in the flat space case, quantisation of this theory is only understood by adopting a type of lightcone gauge --- here covariant quantisation is not available because quantising the Green-Schwarz superstring without fixing a gauge is non-trivial due to additional phase-space constraints for which it is not entirely understood how to treat them without losing  manifest Lorentz invariance (for more remarks, see e.g.~§5.4 of \cite{Green:1987sp}).

\subsection{Green-Schwarz superstrings}\label{sec:Sib:prereq:GS}

\subsubsection{In flat space}\label{sec:Sib:prereq:flatGS}

In this section we will introduce fermions in the flat space Polyakov string by means of the Green-Schwarz (GS) formalism. An important difference with the Ramond-Neveu-Schwarz (RNS) formalism treated in section \ref{s:bob} is that, instead of introducing  worldsheet fermions and worldsheet  supersymmetry, for the GS string fermions and supersymmetry are manifested in target space. When adopting a lightcone gauge, however, one can show that the physical RNS and GS superstring are equivalent \cite{Green:1987sp}. As is well known, demanding Lorentz invariance of the  quantum theory of these superstrings  singles out the critical dimension $D=10$. Instead of  starting with the introduction of the classical GS superstring in general dimensions, and proving that we must have $D=10$ due to consistency with the quantum theory, we will already assume $D=10$ from the start (however, for the first-principles route see e.g.~\cite{Green:1987sp}). Furthermore, we will introduce a maximal number of supersymmetries in  10-dimensional Minkowski space, as then the less-supersymmetric cases can be readily obtained.

 Let us first recall that spinors in $D=10$ can, in general, take $2^{D/2}=32$ \textit{complex} values, but the most fundamental (irreducible) representation are the Majorana-Weyl spinors, which can have $16$ \textit{real} independent values.\footnote{For more details, see e.g.~chapter 3 of \cite{Freedman:2012zz}.} Given that we consider theories of particles with spin greater than $2$ as unphysical, the maximal amount of supercharges (fermionic generators of the super-Poincar\'e algebra) is $32$ in $D=10$ and thus the maximal number of supersymmetries we can introduce is ${\cal N}=2$. 

\bigskip
\noindent {\bf Action and symmetries} --- The starting point to obtain the  Green-Schwarz superstring action in flat spacetime is to supersymmetrise the bosonic Polyakov  action \eqref{eq:polyakov} as follows\footnote{This supersymmetrisation procedure follows from doing the analogue as for the supersymmetric pointlike particle, where one shifts $\partial_\tau X^\mu$ with $-i \bar{\theta}^I \Gamma^\mu \partial_\tau \theta^I$ in the worldline action, see §5.1.1 of \cite{Green:1987sp}.}
\begin{equation} \label{eq:flat-GS-part1}
    S_1[X,\theta,h] = - \frac{T}{2} \int \de^2\sigma \sqrt{-h} h^{\alpha\beta} \Pi^\mu_\alpha \eta_{\mu\nu} \Pi^\nu_{\beta} \ ,
\end{equation}
where $\mu,\nu=0,\ldots , 9$ and
\begin{equation}
    \Pi^\mu_\alpha = \partial_\alpha X^\mu  - i \bar{\theta}^I \Gamma^\mu \partial_\alpha \theta^J \delta_{IJ} \ .
\end{equation}
Here we introduced two $D=10$ Majorana-Weyl (MW) spinors $\theta^I$, where   $I=1,2$ indicates the number of target space supersymmetries,\footnote{The ${\cal N}=1$ or ${\cal N}=0$ string can be obtained by setting one or both of the $\theta$'s to zero, respectively.} and $\bar{\theta}= {\theta}^\dag \Gamma^0={\theta}^T \Gamma^0 $. Note that we will always suppress spinor indices. At the level of the worldsheet, however, the components of these spinors are anti-commuting \textit{scalar} fields. The objects $\Gamma^\mu$ are the $D=10$ $32\times 32$  gamma-matrices  satisfying the Clifford algebra
\begin{equation}
    \{ \Gamma^\mu , \Gamma^\nu \} = 2 \eta^{\mu\nu}\mathbb{1}_{32} , \qquad \eta^{\mu\nu} = \mathrm{diag}(-1, +1, \ldots, +1)^{\mu\nu} \ .
\end{equation}
Its hermiticity properties are $(\Gamma^\mu)^\dag = \Gamma^0 \Gamma^\mu \Gamma^0$ and thus $\Gamma^\mu$ is unitary.
An identity for MW spinors $\psi_i$ in $D=10$ that will be useful in the following is
\begin{equation}
   \bar{\psi}_1 \Gamma^\mu \psi_2 = -\bar{\psi}_2 \Gamma^\mu \psi_1 \ . 
\end{equation}
%The Majorana-Weyl condition implies that, for any spinor $\psi$, \bar{\psi} = \psi^t \Gamma^0 \\

The action \eqref{eq:flat-GS-part1} is still invariant under worldsheet diffeomorphisms \eqref{eq:ws-diffeo} and Weyl rescalings \eqref{eq:weyl-rescaling} (incl.~transforming $\theta^I$ as $\theta^I(\sigma)\rightarrow\tilde{\theta}^I(\tilde{\sigma}) = \theta^I(\sigma)$ and $\theta^I(\sigma)\rightarrow \theta^I(\sigma)$  respectively). In addition, the action also has global symmetries capturing bosonic target space Poincar\'e invariance (as in the bosonic case) and invariance under global supersymmetry transformations acting  as
\begin{equation} \label{eq:global-susy-part1}
    \delta_\epsilon\theta^I = \epsilon^I, \qquad \delta_\epsilon X^\mu = i \bar{\epsilon}^I \Gamma^\mu \theta^J \delta_{IJ} , \qquad \delta_\epsilon h^{\alpha\beta} = 0 ,
\end{equation}
where $\epsilon^I$ are constant MW spinors of the same chirality as $\theta^I$. Under \eqref{eq:global-susy-part1} one simply finds that $\delta_\epsilon\Pi^\mu_\alpha=0$ and so the action is indeed invariant under ${\cal N}=2$ supersymmetry in $D=10$. However, as we have noted above, the theory  \eqref{eq:flat-GS-part1} has $2\times 16$ real target space spinors (16 for each $I$), while after lightcone gauge only $2\times 8$ physical bosonic degrees of freedom (left- and right-moving) would remain. 
Therefore, the physical target space theory cannot be supersymmetric, unless there is a fermionic gauge symmetry which reduces the number of spinors to $2\times 8$. In fact, such a symmetry is also present in the superparticle case \cite{Siegel:1983hh} and goes under the name of $\kappa$-symmetry. One can thus expect it to be generalised to superstrings too \cite{Green:1987sp}, and it can in particular be obtained by adding a second term $S_2$ to the action $S_1$ \eqref{eq:flat-GS-part1} to find a supersymmetric superstring action  $S=S_1+S_2$ with\footnote{Interestingly, $S_2$ can not be constructed for ${\cal N}>2$.}
\begin{equation} \label{eq:flat-GS-part2}
    S_2[X,\theta] = -\frac{T}{2} \int \de^2\sigma \ \epsilon^{\alpha\beta} \left(2i \partial_\alpha X^\mu (\bar{\theta}^1 \Gamma_\mu \partial_\beta \theta^1 - \bar{\theta}^2 \Gamma_\mu \partial_\beta\theta^2) -2 \bar{\theta}^1 \Gamma^\mu \partial_\alpha\theta^1 \bar{\theta}^2 \Gamma_\mu \partial_\beta \theta^2 \right) \ .
\end{equation}
Before describing the $\kappa$-gauge symmetry let us first note some important facts. Firstly, also  $S_2$ on its own is invariant under worldsheet diffeomorphisms, Weyl rescalings, global Poincar\'e symmetries, and the ${\cal N}=2$ global supersymmetries \eqref{eq:global-susy-part1}.

\begin{centering}
\begin{tcolorbox}
\begin{exercise}
Verify that $S_2$ \eqref{eq:flat-GS-part2} is invariant under \eqref{eq:global-susy-part1} up to total derivative terms.  One can use the following identity  for 10-dimensional MW spinors $\psi_i$
\begin{equation} \label{eq:identity-triple-mw-10d}
    \bar{\epsilon} \Gamma_\mu \psi_{[1} \bar{\psi}_2 \Gamma^\mu \psi_{3]}=0 \ . 
\end{equation}
\end{exercise}
{\footnotesize
A proof of this identity can be found in §4.A of~\cite{Green:1987sp}. It is interesting to note that the classical superstring theory only exists in $D=3,4,6,10$ where the identity~\eqref{eq:identity-triple-mw-10d} holds for Majorana, Majorana or Weyl, Weyl and MW respectively. This is in contrast to the bosonic string theory which classically can exist in any spacetime dimension. Of course, as we have mentioned above, the quantum theory singles out $D=10$ ($D=26$ in the bosonic case).}
\end{tcolorbox}
\end{centering}

\noindent Secondly, note that  the action $S_2$ does not depend on the worldsheet metric $h^{\alpha\beta}$. Therefore, it does  not contribute to the Virasoro constraints, which read
\begin{equation} \label{eq:virasoro-gs-flat}
    T_{\alpha\beta} = \Pi_\alpha^\mu \eta_{\mu\nu} \Pi_\beta^\nu - \frac{1}{2} h_{\alpha\beta} h^{\gamma \delta} \Pi_\gamma^\mu \eta_{\mu\nu} \Pi_\delta^\nu = 0 \ .
\end{equation}

\textit{Remark} --- The action $S=S_1+S_2$ for the Green-Schwarz superstring in flat space can more systematically be obtained by generalising the 10-dimensional Minkoswki target space of the 2-dimensional sigma-model to a 10-dimensional flat ${\cal N}=2$ superspace parametrised by bosonic coordinates $X^\mu$ and two anticommuting spinor coordinates $\theta^I$. The term $S_2$ is also known as the Wess-Zumino term (see \cite{Henneaux:1984mh}) and is constructed  from an exact three-form. It only survives for strings, not for particles, due to their extended nature. 

\bigskip
\noindent {\bf $\kappa$-symmetry} --- We will now show that the total action $S=S_1+S_2$ is invariant under a local fermionic symmetry (while its individual terms are not), which will reduce the number of spinorial degrees of freedom to $2\times 8$. The infinitesimal transformation parameter is a local $D=10$ MW spinor $\kappa^I_\alpha (\sigma)$ which carries a worldsheet vector index $\alpha$, and is of the same chirality as $\theta^I$.  In addition, the parameters $\kappa^I_\alpha (\sigma)$ are restricted to be ``anti-self-dual'' for $I=1$ and ``self-dual'' for $I=2$, which corresponds to the  two irreducible representations of the two-dimensional Lorentz group  achieved by the worldsheet projectors
\begin{equation} \label{eq:worldsheet-projectors}
\begin{gathered}
    P_\pm^{\alpha\beta} = \frac{1}{2} \left( h^{\alpha\beta} \pm \frac{\epsilon^{\alpha\beta}}{\sqrt{-h}} \right) \ , 
\end{gathered}
\end{equation}
which satisfy
\begin{equation} \label{eq:ws-projectors-id1}
   P_+^{\alpha\beta} + P_-^{\alpha\beta}=h^{\alpha\beta}, \qquad  P_\pm^{\alpha\gamma} h_{\gamma\delta} P_\pm^{\delta\beta} = P_\pm^{\alpha\beta}, \qquad P_\pm^{\alpha\gamma} h_{\gamma\delta} P_\mp^{\delta\beta} = 0 \ .
\end{equation}
In particular, 
\begin{equation}
\begin{aligned}
    \kappa^{1\alpha} &= (P_+^{\alpha\beta} + P_-^{\alpha\beta}) \kappa^1_\beta = P_-^{\alpha\beta} \kappa^1_\beta , \\
    \kappa^{2\alpha} &= (P_+^{\alpha\beta} + P_-^{\alpha\beta}) \kappa^2_\beta = P_+^{\alpha\beta} \kappa^2_\beta , 
\end{aligned}
\end{equation}
In other words, using the notation $A_\pm^\alpha = P_\pm^{\alpha\beta} A_\beta$ for  worldsheet vectors $A_\alpha$, the parameters are restricted by  $\kappa_+^{1\alpha}=\kappa_-^{2\alpha}=0$.\footnote{At the quantum level, the $I=1$ spinors will therefore describe right-moving modes while $I=2$ will describe left-moving modes. Note of course  that $A_\pm^\alpha = P_\pm^{\alpha\beta} A_{\pm \beta}$ while  $P_\pm^{\alpha\beta} A_{\mp\beta} = 0$. } 
The $\kappa$-symmetry transformation now reads \cite{Green:1987sp}
\begin{equation} \label{eq:kappatransformation}
    \delta_\kappa \theta^I = 2 i\Gamma_\mu \Pi^\mu_{\alpha}  \kappa^{I\alpha}  , \qquad \delta_\kappa X^\mu = i \bar{\theta}^I \Gamma^\mu \delta_\kappa \theta^J \delta_{IJ} \ .
\end{equation}

\textit{Remark} --- $\kappa$-symmetry is rather peculiar: it involves a fermionic symmetry with a  transformation parameter which is local in the worldsheet coordinates and which caries both a worldsheet vector index and a spinorial target space index,  whilst the theory itself does not involve any worldsheet spinors. From the point of view of the worldsheet, the $\kappa$-transformation rather transforms all objects as (worldsheet) vectors.

\begin{centering}
\begin{tcolorbox}
\begin{exercise}
Show the relations \eqref{eq:ws-projectors-id1} and the properties
\begin{equation} \label{eq:ws-projectors-id2}
    P_\pm^{\alpha\gamma} P_\pm^{\beta\delta}= P_\pm^{\alpha\delta} P_\pm^{\beta\gamma} , \qquad P_+^{\alpha\beta} = P_-^{\beta\alpha} \ .
\end{equation}
Hint: use  $\epsilon^{\alpha\beta}\epsilon^{\gamma\delta} = -h (h^{\alpha\delta}h^{\beta\gamma} - h^{\alpha\gamma}h^{\beta\delta})$ and $h^{\alpha\beta} \epsilon^{\gamma\delta}-\epsilon^{\alpha\beta} h^{\gamma\delta} = h^{\alpha\gamma} \epsilon^{\beta\delta} - \epsilon^{\alpha\gamma} h^{\beta\delta}$.

In the notation $A_\pm^\alpha = P_\pm^{\alpha\beta} A_\beta$ observe that $A_{\pm\tau}$ and $A_{\pm \sigma}$ are not independent. In conformal gauge in fact $A_{\pm\tau} =\mp A_{\pm \sigma}$.
\end{exercise}
\end{tcolorbox}
\end{centering}

\textit{Remark} --- {Using the worldsheet projectors, the Virasoro constraints  \eqref{eq:virasoro-gs-flat} can be rewritten as 
\begin{equation} \label{eq:virasoro-gs-flat2}
    \Pi^\mu_{\pm \alpha} \eta_{\mu\nu} \Pi^\nu_{\pm \beta} = 0 \ .
\end{equation}
To show this, observe that the energy-momentum tensor $T_{\alpha\beta}$ has four independent components under the 2d-Lorentz group, namely $T^{\alpha\beta}_{\pm\pm} = P_\pm^{\alpha\gamma} P_\pm^{\beta\delta} T_{\gamma\delta}$ and $T^{\alpha\beta}_{\pm\mp} = P_\pm^{\alpha\gamma} P_\mp^{\beta\delta} T_{\gamma\delta}$. Then, after some algebra, one finds $T^{\alpha\beta}_{\pm\pm}=\Pi^\mu_{\pm \alpha} \eta_{\mu\nu} \Pi^\nu_{\pm \beta}$ while $T^{\alpha\beta}_{\pm\mp} = 0$ identically. It will be convenient to use that $\epsilon^{\alpha\beta}\epsilon^{\gamma\delta} = -h (h^{\alpha\delta}h^{\beta\gamma} - h^{\alpha\gamma}h^{\beta\delta})$ and $h^{\alpha\beta} \epsilon^{\gamma\delta}-\epsilon^{\alpha\beta} h^{\gamma\delta} = h^{\alpha\gamma} \epsilon^{\beta\delta} - \epsilon^{\alpha\gamma} h^{\beta\delta}$.
}

\begin{centering}
\begin{tcolorbox}
\begin{exercise}
    Show that $\Pi_\alpha^\mu$ transforms under \eqref{eq:kappatransformation} as 
    \begin{equation}
        \delta_\kappa \Pi_\alpha^\mu = 2i \partial_\alpha \bar{\theta}^I \Gamma^\mu \delta_\kappa \theta^J \delta_{IJ}.
    \end{equation}
\end{exercise}
\end{tcolorbox}
\end{centering}

To show that the full action $S=S_1+S_2$ is invariant under \eqref{eq:kappatransformation} we must derive the variation of $\delta ( \sqrt{-h}h^{\alpha\beta})$ accordingly. This can be done by varying
%To start, let us consider the variation of 
the terms $S_1$ \eqref{eq:flat-GS-part1} and $S_2$ \eqref{eq:flat-GS-part2} up to quadratic order in fermions. We can write
\begin{equation}
    \delta S_1 \ : \ \delta_\kappa \left( \sqrt{-h}h^{\alpha\beta}\right) \Pi_\alpha^\mu \eta_{\mu\nu} \Pi_\beta^\nu + 4i \sqrt{-h}h^{\alpha\beta} \Pi_{\alpha}^\mu  \partial_\beta \bar{\theta}^I \Gamma_\mu \partial_\kappa \theta^J \delta_{IJ}  \ ,
\end{equation}
and
\begin{equation}
\begin{aligned}
     \delta S_2 \ : \ &2i \epsilon^{\alpha\beta} \Pi_\alpha^\mu  \delta_\kappa\left(\bar{\theta}^1 \Gamma_\mu \partial_\beta \theta^1   \right)- (1\leftrightarrow 2) +{\cal O}(\theta,\delta_\kappa\theta)^4 \ , \\
     &=2i \epsilon^{\alpha\beta} \Pi_\alpha^\mu  \left(\delta_\kappa\bar{\theta}^1 \Gamma_\mu \partial_\beta \theta^1 + \bar{\theta}^1 \Gamma_\mu \partial_\beta \delta_\kappa\theta^1  \right)- (1\leftrightarrow 2)  +{\cal O}(\theta,\delta_\kappa\theta)^4 \ , \\
     &=2i \epsilon^{\alpha\beta} \Pi_\alpha^\mu  \left(\partial_\beta (\bar{\theta}^1 \Gamma_\mu  \delta_\kappa\theta^1) - 2 \partial_\beta \bar{\theta}^1 \Gamma_\mu  \delta_\kappa\theta^1 \right) - (1\leftrightarrow 2)   +{\cal O}(\theta,\delta_\kappa\theta)^4 \ .
\end{aligned}
\end{equation}
We can now combine both variations conveniently as
\begin{equation}
    \begin{aligned}
        \delta S \ : \ &\delta_\kappa \left( \sqrt{-h}h^{\alpha\beta}\right) \Pi_\alpha^\mu \eta_{\mu\nu} \Pi_\beta^\nu + 2i \epsilon^{\alpha\beta} \Pi_\alpha^\mu \partial_\beta \left(\bar{\theta}^1 \Gamma_\mu  \delta_\kappa\theta^1-(1\leftrightarrow 2) \right)  \\
        &+8i \sqrt{-h}  \Pi_\alpha^\mu P_-^{\alpha\beta} \partial_\beta \bar{\theta}^1 \Gamma_\mu \delta_\kappa\theta^1 + 8i \sqrt{-h}  \Pi_\alpha^\mu P_+^{\alpha\beta} \partial_\beta \bar{\theta}^2 \Gamma_\mu \delta_\kappa\theta^2 +{\cal O}(\theta,\delta_\kappa\theta)^4 \ , \\
        &=\delta_\kappa \left( \sqrt{-h}h^{\alpha\beta}\right) \Pi_\alpha^\mu \eta_{\mu\nu} \Pi_\beta^\nu + 2i \epsilon^{\alpha\beta} \partial_\alpha X^\mu \partial_\beta \left(\bar{\theta}^1 \Gamma_\mu  \delta_\kappa\theta^1-(1\leftrightarrow 2) \right)  \\
        &+8i \sqrt{-h}  \Pi_\alpha^\mu P_-^{\alpha\beta} \partial_\beta \bar{\theta}^1 \Gamma_\mu \delta_\kappa\theta^1 + 8i \sqrt{-h}  \Pi_\alpha^\mu P_+^{\alpha\beta}\partial_\beta \bar{\theta}^2 \Gamma_\mu \delta_\kappa\theta^2 +{\cal O}(\theta,\delta_\kappa\theta)^4 \ .
    \end{aligned}
\end{equation}
Doing  partial integration and using $\epsilon^{\alpha\beta} \partial_\alpha\partial_\beta = 0$ on the second term of the first line we see that it is simply a total derivative term that can be dropped. As for the second line, let us now use the expression for $\delta\theta^I$ as well as the (anti)-self-duality conditions on $\kappa^I_\alpha$. We get
\begin{equation}
    \begin{aligned}
        8i \sqrt{-h}  \Pi_\alpha^\mu P_-^{\alpha\beta} \partial_\beta \bar{\theta}^1 \Gamma_\mu \delta_\kappa\theta^1 &= -16 \sqrt{-h} \Pi_\alpha^\mu P_-^{\alpha\beta} \partial_\beta \bar{\theta}^1 \Gamma_\mu \Gamma_\nu \Pi_\gamma^\nu P_-^{\gamma\delta} \kappa^1_\delta \ ,\\
        &= - 8 \sqrt{-h} \Pi_\alpha^\mu P_-^{\alpha\beta} \partial_\beta \bar{\theta}^1 \{\Gamma_\mu \Gamma_\nu \} \Pi_\gamma^\nu P_-^{\gamma\delta} \kappa^1_\delta \ , \\
        &= - 16 \sqrt{-h} \Pi_\alpha^\mu P_-^{\alpha\beta} \partial_\beta \bar{\theta}^1 \eta_{\mu\nu} \Pi_\gamma^\nu P_-^{\gamma\delta} \kappa^1_\delta \ , \\
        &= - 16 \sqrt{-h} \Pi_\alpha^\mu P_-^{\alpha\beta} \partial_\beta \bar{\theta}^1 \eta_{\mu\nu} \Pi_\gamma^\nu  \kappa^{1\gamma} \ ,
    \end{aligned}
\end{equation}
where we have used \eqref{eq:ws-projectors-id2}, and similarly
\begin{equation}
    \begin{aligned}
        8i \sqrt{-h}  \Pi_\alpha^\mu P_+^{\alpha\beta}\partial_\beta \bar{\theta}^2 \Gamma_\mu \delta_\kappa\theta^2 = - 16 \sqrt{-h} \Pi_\alpha^\mu P_+^{\alpha\beta} \partial_\beta \bar{\theta}^2 \eta_{\mu\nu} \Pi_\gamma^\nu  \kappa^{2\gamma} \ .
    \end{aligned}
\end{equation}
Concluding, the variation of the full action $S$ vanishes under \eqref{eq:kappatransformation} to quadratic order in fermions when\footnote{Note that the right-hand-side of \eqref{eq:kappatransformation-metric} is symmetric and unimodular (traceless), as it should.}
\begin{equation} \label{eq:kappatransformation-metric}
    \delta_\kappa \left( \sqrt{-h}h^{\alpha\beta}\right) = 16 \sqrt{-h} \left( P_-^{\alpha\gamma} \partial_\gamma \bar{\theta}^1 \kappa^{1\beta} + P_+^{\alpha\gamma} \partial_\gamma \bar{\theta}^2 \kappa^{2\beta} \right) \ .
\end{equation}
In fact, the full action $S$ can be shown to be invariant under \eqref{eq:kappatransformation} and \eqref{eq:kappatransformation-metric} to all orders in fermions by using again the identity \eqref{eq:identity-triple-mw-10d}. One may do this as an exercise. 

\bigskip
\noindent {\bf On-shell rank of $\kappa$-symmetry} --- The final important step is to show that the $\kappa$-gauge symmetry indeed reduces the number of physical spinorial degrees of freedom of $\theta^I$ to $2\times 8$. For this we must derive its on-shell rank, \textit{i.e.}~the number of degrees of freedom $\kappa$-transformations can fix once the Virasoro constraints (which reduce the bosonic degrees of freedom) are satisfied. First, we can note that in general $\kappa^I_\alpha$ would have $2\times 2\times 16 = 64$ components in total. However, because of the (anti-)self-duality conditions $\kappa_+^{1\alpha} = \kappa_-^{2\alpha} = 0$ this number is already reduced to $32$ components. We will now show that only half of these can fix components of $\theta^1$ and $\theta^2$. Let us rewrite
\begin{equation}
    \begin{aligned}
        \delta_\kappa \theta^1 &= 2i \Gamma_\mu \Pi_{+\alpha}^\mu  \kappa_-^{1\alpha} \ , \qquad
        \delta_\kappa \theta^2 = 2i \Gamma_\mu \Pi_{-\alpha}^\mu  \kappa_+^{1\alpha} \ ,
    \end{aligned}
\end{equation}
where we have used the properties \eqref{eq:ws-projectors-id1} and \eqref{eq:ws-projectors-id2}, and the (anti-)self-duality conditions. Then, define the operators
\begin{equation}
    O_{\pm\alpha} =  \Gamma_\mu \Pi_{\pm\alpha}^\mu \ ,
\end{equation}
so that $\delta_\kappa\theta^1 = 2 i O_{+\alpha} \kappa^{1\alpha}_-$ and $\delta_\kappa\theta^2 = 2 i O_{-\alpha} \kappa^{2\alpha}_+$. Squaring these operators in their spinor indices and using again \eqref{eq:ws-projectors-id2} we find 
\begin{equation}
    O_{\pm\alpha} O_{\pm\beta} = \Gamma_\mu \Gamma_\nu \Pi_{\pm\alpha}^\mu \Pi_{\pm\beta}^\nu = \frac{1}{2} \{ \Gamma_\mu , \Gamma_\nu \} \Pi_{\pm\alpha}^\mu \Pi_{\pm\beta}^\nu  = \eta_{\mu\nu} \Pi_{\pm\alpha}^\mu \Pi_{\pm\beta}^\nu \mathbb{1} \ ,
\end{equation}
and thus they square to zero on-shell, upon the Virasoro constraints \eqref{eq:virasoro-gs-flat2}, which means that they are both at most half-rank (\textit{i.e.}~rank 16). Generically, however, their sum $O_{\pm\alpha}+O_{\mp\alpha}=\Gamma_\mu \Pi_\alpha^\mu$ is not constrained and thus of full-rank. Altogether, this means that on-shell $O_{+\alpha}$ and $O_{-\alpha}$ are each of rank $16$. Therefore, the independent components of $\delta_\kappa\theta^1$ and $\delta_\kappa\theta^2$ are only half of those of $\kappa^{1\alpha}_-$ and $\kappa^{2\alpha}_+$ respectively. Concluding, the rank of $\kappa$-symmetry is $16$ and thus it is precisely the called-for fermionic gauge symmetry which reduces the physical degrees of freedom of $\theta^I$ from $2\times 16$ to $2\times 8$. The flat target space of the Green-Schwarz physical worldsheet action thus is indeed ${\cal N}=2$ supersymmetric. 

\bigskip
%\SIB{Equations of motion? Commutator of two $\kappa$ transformations?}\\
\noindent {\bf Equations of motion} --- The equations of motion of the action $S=S_1+S_2$ are highly non-linear and read\footnote{The latter two equations are obtained by using the equations of motion \eqref{eq:GS-flat-EOMX} from $X^\mu$ as well as the Fierz identities for MW-spinors in ten dimensions (see e.g.~eq.~(7.4.51) of \cite{Green:1987sp} combined with table 3.1 of \cite{Freedman:2012zz}).} 
\begin{alignat}{2}
    \delta X^\mu &: \qquad && \partial_\alpha \left(\sqrt{-h} (h^{\alpha\beta} \partial_\beta X^\mu -2i P_-^{\alpha\beta} \bar{\theta}^1 \Gamma^\mu \partial_\beta \theta^1 -2i P_+^{\alpha\beta} \bar{\theta}^2 \Gamma^\mu \partial_\beta \theta^2) \right) = 0 \ , \label{eq:GS-flat-EOMX} \\
    \delta \theta^1 &: \qquad && P_-^{\alpha\beta} \Pi_\alpha^\mu \eta_{\mu\nu} \Gamma^\nu \partial_\beta \theta^1=0 \ ,\\
    \delta \theta^2 &: \qquad && P_+^{\alpha\beta} \Pi_\alpha^\mu \eta_{\mu\nu} \Gamma^\nu \partial_\beta \theta^2 = 0 \ ,
\end{alignat}
which should be supplemented by the Virasoro constraints \eqref{eq:virasoro-gs-flat2} obtained from $\delta h^{\alpha\beta}$. 

 \bigskip
\noindent {\bf Gauge fixing} --- As we mentioned at the end of the previous section, with  quantisation  in mind, it is paramount to fix the lightcone gauge symmetries of the Green-Schwarz superstring, as covariant quantisation of this theory is extremely non-trivial and in fact not well understood.\footnote{This is due to the existence of non-trivial phase-space constraints arising from the fact that the momenta $P_\theta^I$, conjugate to the coordinates $\theta^I$, are not independent, cf.~§5.4 of \cite{Green:1987sp}.} Luckily, lightcone quantisation works nicely and straightforwardly both in flat and curved spaces. As this procedure depends on the precise form of the equations of motion, it is instructive to see in detail how this is done in the flat setting.   In this case it is again appropriate to partially fix the worldsheet diffeomorphisms and Weyl rescalings   by the conformal gauge \eqref{eq:conformal-gauge}. Opposed to the bosonic case, however, the equations of motions will not yet reduce to free wave equations. To achieve this one needs to fix the residual bosonic gauge symmetry as well as the fermionic $\kappa$-gauge symmetry. To do so, let us first introduce lightcone gamma-matrices
\begin{equation}
    \Gamma^\pm = \frac{1}{\sqrt{2}} \left( \Gamma^0 + \Gamma^{D-1} \right) \ .
\end{equation}
Because of the Clifford algebra relations, they satisfy
\begin{equation}
    (\Gamma^+)^2=(\Gamma^-)^2=0, \qquad \{ \Gamma^+ , \Gamma^- \}=2 \ .
\end{equation}
Hence, as their sum and difference are non-singular, $\Gamma^\pm$ are half-rank. As in general $\mathrm{rank} (AB) \leq \mathrm{min} (\mathrm{rank} A , \mathrm{rank} B)$, when also $\Gamma^+ O_{\pm\alpha}$ is half-rank then $\kappa$-symmetry transformations can be used to fix the gauge
\begin{equation} \label{eq:kappa-lightcone-gauge}
    \Gamma^+ \theta^1 = \Gamma^+ \theta^2 = 0 \ ,
\end{equation}
which enforces half of the components of $\theta^I$ to be zero. In this case, we see from the equations of motion \eqref{eq:GS-flat-EOMX} that the equations for $X^+$ and $X^i$ become  free wave equations. As in the bosonic case, cf.~\eqref{eq:lightcone-gauge}, the residual holomorphic symmetries can then be used to impose
\begin{equation} \label{eq:lightcone-gauge-2}
    X^+ = x^+ + p^+ \tau \ .
\end{equation}
This is entirely consistent since in this gauge
\begin{equation}
    \Gamma^+ O_{\pm\alpha} = \Gamma^+ \Gamma_\mu \Pi^\mu_{\pm\alpha} \propto \Gamma^+ \Gamma_+ p^+ + {\cal O}(X^i , \theta^I) , \qquad i=1,\ldots,D-2 \ ,
\end{equation}
and $\Gamma^+\Gamma^-$ is indeed half-rank.
We refer to \eqref{eq:kappa-lightcone-gauge} and \eqref{eq:lightcone-gauge-2} as the lightcone gauge for the flat GS superstring.  The Virasoro constraints then again determine the field $X^-$ completely in terms of  $p^+, X^i$ and the physical components of $\theta^I$.  In addition, the  equations for  $\theta^I$ linearise as well and can be solved explicitly. After lightcone gauge, the remaining manifest global symmetry of the ten-dimensional $SO(1,9)$ Lorentz invariance is  an $SO(8)$ rotation symmetry of the $X^i$ fields. Interestingly, it is the triality property of $SO(8)$ that can be used to show that the gauge-fixed GS superstring is equivalent to the RNS superstring after the GSO projection. For more details on the latter we refer again to section~\ref{s:bob}.

\begin{centering}
\begin{tcolorbox}
\begin{exercise}
 Show that the expression (without summation over $I$)
 \begin{equation}
     \bar{\theta}^I \Gamma^\mu \theta^I  \ ,
 \end{equation}
 vanishes for $\mu=+,i$ but not for $\mu=-$.
\end{exercise}
\end{tcolorbox}
\end{centering}

An alternative way to  see the $SO(8)$ symmetry is to fix the gauges in the GS sigma-model action directly. For details on how to do this we refer to   section \ref{s:fiona} and \cite{Hoare:LN}, where it is done for a curved background: you can follow the same procedure, but simply set $G_{\mu\nu} = \eta_{\mu\nu}$ and $B_{\mu\nu}=0$ to simplify calculations and obtain expressions valid for the flat GS superstring. Recall, however, that in the curved setting one  typically adopts the \textit{uniform} lightcone gauge. The resulting gauge-fixed action  in lightcone coordinates is of the form
\begin{equation}
    S_{\mathrm{\tiny GS-fixed}} \sim \int \de^2\sigma \, \partial_+ X^i \partial_- X_i - 2i \bar{\theta}^1 \Gamma^- \partial_+ \theta^1 +2i \bar{\theta}^2 \Gamma^- \partial_- \theta^2 \ .
\end{equation}
The theory thus indeed describes $2\times 8$ left-moving and $2\times 8$ right-moving bosons and fermions. The global supersymmetry transformations that preserve the $\kappa$-gauge (possibly using compensating $\kappa$-transformations) transpire into an ${\cal N}=(8,8)$ worldsheet supersymmetry.

\bigskip
\noindent{\bf Relation to type II supergravity} --- The Green-Schwarz superstring in ten-dimensional Minkowski space describes a $D=10$ type II supergravity solution with following bosonic field content
\begin{equation} \label{eq:sugra-flatspace}
    G_{\mu\nu} = \eta_{\mu\nu} \ , \qquad \Phi = \Phi_0 \ , \qquad B = F_{(n)}=0 \ ,
\end{equation}
where $\Phi_0$ is constant and $F_{(n)}$ denotes the $n$-form RR field strengths of type IIA/B ($n=0,2,4$ for IIA, $n=1,3,5$ for IIB). When the spinors $\theta^1$ and $\theta^2$ are of the same chirality  the  supergravity is of type IIB, while opposite chirality gives  type IIA. Defining $\Gamma_{11} = \Gamma_0 \Gamma_1 \ldots \Gamma_9$, which satisfies $\Gamma_{11}^2=1$ and $\mathrm{Tr}(\Gamma_{11}) = 0$, spinors of different chirality can be distinguished as
\begin{equation}
    \Gamma_{11} \psi = \pm \psi \ .
\end{equation}
The global supersymmetry transformations \eqref{eq:global-susy-part1} correspond to the superisometries of the background, \textit{i.e.}~the constant transformation parameter $\epsilon^I$ satisfies the Killing spinor equation for the gravitino, while the Killing spinor equation for the dilatino is automatically satisfied. 

In general, $D=10$ type IIA/B supergravity can admit a bosonic field content with a curved metric $G$ and with the forms $B$ and $F_{(n)}$ non-constant and non-vanishing. They are constrained to satisfy the type IIA/B supergravity equations of motion, which follow from the  type IIA/B bosonic actions 
\begin{align}\label{sec:sib:prereqs:effective_actions}
    S_{\text{IIA}} ={}& \int d^{10}X \sqrt{-G} \left( e^{-2\Phi} ({\cal R} + 4 (\nabla \Phi)^2 - \frac{1}{2} H^2) -\frac{1}{2} (F_{(2)}^2  + F_{(4)}^2 ) \right) \nonumber\\ &-\frac{1}{2} \int B\wedge d C_3 \wedge dC_3 \ , \\
    S_{\text{IIB}} ={}& \int d^{10}X \sqrt{-G} \left( e^{-2\Phi} ({\cal R} + 4 (\nabla \Phi)^2 - \frac{1}{2} H^2) -\frac{1}{2} (F_{(1)}^2 + F_{(3)}^2 +\frac{1}{2} F_{(5)}^2 ) \right) \nonumber\\ &-\frac{1}{2} \int C_{(4)} \wedge H \wedge F_{(3)} \ ,
\end{align}
where $\cal R$ is the target space Ricci scalar, and $C_{(n)}$ are the gauge potentials of the RR field strengths $F_{(n)}$  as
\begin{gather}
    F_{(2)} = d C_{(1)} \ , \quad F_{(4)} = d C_{(3)} - H \wedge C_{(1)} \ , \\
    F_{(1)} = d C_{(0)} \ , \quad F_{(3)} = d C_{(2)} - H\wedge C_{(0)} \ , \quad F_{(5)} = d\left(C_{(4)} + \frac{1}{2} B\wedge C_{(2)} \right) - H\wedge C_{(2)} \ . \nonumber
\end{gather}
Bosonic isometries of such backgrounds are characterised by Killing vectors $k$ that satisfy $L_k G =L_k H = L_k F_{(n)} = 0$ whilst superisometries are characterised by Killing spinors $\epsilon$ that satisfy the Killing spinor equations obtained from varying the fermionic (dilatino and gravitino) fields. 
The main scope of this review is to study  strings that can give  rise to these more generic curved backgrounds.

\subsubsection{In curved backgrounds} \label{s:GS-curved}
One of the main advantages of the Green-Schwarz formalism is that it can perfectly describe semi-classical strings in any background with a curved metric, B-field, dilaton, and Ramond-Ramond (RR) fields.\footnote{In  the RNS formalism, where supersymmetry is introduced on the worldsheet, the RR fields originate from fermionic creation operators  and are realised by  vertex operators that are non-local in the worldsheet fields. Therefore RR fields   cannot be easily coupled to the worldsheet metric in the standard way.} As for the GS superstring in flat space, in curved backgrounds the GS action can   be derived by constructing a two-dimensional non-linear sigma-model for which the target space  is a ten-dimensional  ${\cal N}=2$ superspace with non-trivial curvature and torsion (cf.~the remark after eq.~\eqref{eq:virasoro-gs-flat}). By construction, this theory is   invariant under ${\cal N}=2$ global supersymmetry. We will not delve into many details of the (derivation of) the superspace GS action, as we will quickly move to its supercoset formulation  in the next section, but we will simply give a bit of its taste here.  In the Nambu-Goto formulation, the superspace action is of the form 
\begin{equation}
    S_{\text{\tiny SNG}} = -T \int \de^2\sigma \sqrt{-\mathrm{det}{\cal G}_{\alpha\beta}({\cal Z})} - \int_\Sigma {\hat B}({\cal Z}) \ ,
\end{equation}
with   ${\cal Z}^{ M} = (X^\mu , \theta^I)$ the superspace coordinates indexed by ${ M} = (\mu,I)$,   ${\hat B}({\cal Z})$ the pull-back of a superspace two-form and ${\cal G}_{\alpha\beta}({\cal Z})$ the induced metric
\begin{equation}
    {\cal G}_{\alpha\beta}({\cal Z})  = E_\alpha{}^{ A} ({\cal Z}) \eta_{{ A}{ B}} E_\beta{}^{ B} ({\cal Z}) ,
\end{equation}
with $A$ the flat  index for the coordinates $X^\mu$, $\eta_{AB}$ the flat metric, and $E_\alpha{}^{ A}({\cal Z})$ are components of the   generic superspace vielbeins  
\begin{equation}\label{eq:superspacevielbeinscomps}
    E_\alpha{}^{\cal A} ({\cal Z}) = \partial_\alpha {\cal Z}^{ M} E_{ M}{}^{\cal A} ({\cal Z}) . 
\end{equation}
with ${\cal A}=(A,a I)$  the flat  indices corresponding to the  superspace indices ${ M} = (\mu,I)$. 
%Note that the components ${\cal A}=A$ of E_\alpha{}^{\cal A} ({\cal Z})   depend on \textit{all} superspace coordinates.
By generalising the flat space case \cite{Henneaux:1984mh}, this action must be invariant under the following $\kappa$-symmetry transformations \cite{Grisaru:1985fv}
\begin{equation} \label{eq:kappa-gs-superspace}
    \delta_\kappa {\cal Z}^{ M} E_{ M}{}^A = 0 , \qquad \delta_\kappa {\cal Z}^{ M} E_{ M}{}^{a I} = \frac{1}{2} (1+\Gamma)^{aI}{}_{bJ}\kappa^{bJ} ,
\end{equation}
where  the operator $\Gamma$ is defined for type IIB as~\footnote{We will explain how to obtain the type IIA expressions  around eq.~\eqref{eq:toIIA}.}
\begin{equation}
    \Gamma =\frac{1}{2 \sqrt{-{\mathcal G}}}  \epsilon^{\alpha\beta} E_\alpha{}^A E_{\beta}{}^B \Gamma_{AB} \sigma^3 \ .
\end{equation}
 Since this $\Gamma$ also satisfies $\mathrm{Tr}(\Gamma)=0$ and $\Gamma^2=1$, the operator $(1+\Gamma)$ is of half-rank and thus $\kappa$-transformations  remove half the fermionic degrees of freedom. This and the question of the completeness of the $\kappa$ gauge-fix will be discussed in more detail in section \ref{s:saskia-complete-kg}, illustrated for the particular case of holographic $AdS_3$-backgrounds.
Recall that the coupling of a two-form $\hat{B}$  is necessary for  $\kappa$-symmetry, \textit{i.e.}~the non-linear sigma-model must be extended by a type of Wess-Zumino (WZ) term constructed from an exact three-form of the ten-dimensional  superspace and which generalises the flat space WZ term \cite{Grisaru:1985fv}. The curvature,  torsion and the WZ form should satisfy Bianchi identities and $\kappa$-symmetry constraints.\\

\textit{Remark} --- An interesting fact is that for a long time it was conjectured that, for  $D=10$ ${\mathcal N}=2$ curved superspaces, the requirement of  the curved $\kappa$-symmetry version imposes the background to satisfy the constraints of  $D=10$ supergravity~\cite{Grisaru:1985fv,Shapiro:1986yy}. However, only recently  it was understood that the $D=10$ supergravity constraints are sufficient but not necessary: in general, the background is required to solve the so-called ``modified'' supergravity equations~\cite{Wulff:2016tju}. \\

In explicit component form, \textit{i.e.}~in terms of the individual fields $X^\mu(\sigma)$ and $\theta^I(\sigma)$, the curved GS action is however very convoluted and obtaining its expression requires expanding the super-objects in fermions and solving the Bianchi identities and $\kappa$-symmetry constraints order by order in fermions. In fact, as opposed to flat space, in general it does not terminate at quartic order in the fermions anymore.  To give an idea, we present here its explicit form to quadratic order in fermions, which captures all the bosonic type II supergravity background fields generically. The action to zeroeth order in fermions is that  of the low-energy bosonic  string
\begin{equation}  \label{eq:GS-curved-zeroeth}
    S_{\mathrm{GS}}^{(0)}[X,h] = -\frac{T}{2} \int_\Sigma \de^2\sigma \left(\sqrt{-h}h^{\alpha\beta} \partial_\alpha X^\mu G_{\mu\nu}(X) \partial_\beta X^\nu + \epsilon^{\alpha\beta} \partial_\alpha X^\mu B_{\mu\nu}(X) \partial_\beta X^\nu  \right) \ ,
\end{equation}
where the metric is obtained from $G_{\mu\nu} = e_\mu{}^A \eta_{AB} e_\nu{}^B$
with  $e_\mu{}^A = E_\mu{}^A (X, \theta=0)$  the bosonic vielbein and the B-field comes from the zeroeth order  component of the superspace two-form $\hat{B}$. At  quadratic order the action reads \cite{Tseytlin:1996hs,Cvetic:1999zs} (see also \cite{Wulff:2013kga}) for type IIB
\begin{equation} \label{eq:GS-curved-quadratic}
    S_{\mathrm{GS}}^{(2)}[X,h,\theta] = T \int_\Sigma d^2 \sigma \ i \bar{\theta}^I \left(\sqrt{-h}h^{\alpha\beta} \delta_{I}^{J} - \epsilon^{\alpha\beta} (\sigma_3)_{I}^{J} \right) e_\alpha{}^A \Gamma_A {\cal D}_{\beta JK} \theta^K \ ,
\end{equation}
where  $(\sigma_i)_{I}^{J}$ are simply Pauli matrices in the $I,J=1,2$ indices,  $e_\alpha{}^A = \partial_\alpha X^\mu e_\mu{}^A$,  $\Gamma^A = \Gamma^\mu e_\mu{}^A$, and ${\cal D}_{\alpha IJ}$ is the operator
\begin{equation}
\begin{aligned}
    {\cal D}_{\alpha}^{IJ} ={}& \delta^{IJ} \left( \partial_\alpha + \frac{1}{4} \omega_\alpha{}^{AB} \Gamma_{AB} \right) + \frac{1}{8} \sigma_3^{IJ} e_\alpha{}^A H_{ABC} \Gamma^{BC} - \frac{1}{8} e^{\Phi} e_\alpha{}^A {\cal S}^{IJ} \Gamma_A  \ ,
\end{aligned}
\end{equation}
where the higher-rank gamma-matrices are defined as\footnote{We use the convention $\Gamma^{[AB]} = \frac{1}{2} \left(\Gamma^A \Gamma^B - \Gamma^B \Gamma^A\right)$.} $\Gamma^{AB \ldots C} = \Gamma^{[A} \Gamma^B \ldots \Gamma^{C]}$, $H_{ABC}$ is the field strength $H=dB$ in flat indices, 
and $\omega_\alpha{}^{AB} = \partial_\alpha X^\mu \omega_\mu{}^{AB}$ with $\omega_\mu{}^{AB}$ the spin-connection
\begin{equation}
    \omega_\mu{}^{AB} = e^{\nu[A|} \left( \partial_\mu e_\nu{}^{|B]} - \partial_\nu e_\mu{}^{|B]} + e^{\rho|B]} e_{\mu}{}^C \partial_\rho e_{\nu C} \right) \ .
\end{equation}
Finally, the operator ${\cal S}^{IJ}$ encodes the RR field strengts. For type IIB it reads
\begin{equation}\label{eq:RR-bispinor-IIB}
     {\cal S}^{IJ} = \epsilon^{IJ} \Gamma^B F_B^{(1)}  +\frac{1}{3!} \sigma_1^{IJ} \Gamma^{BCD}F^{(3)}_{BCD} + \frac{1}{2.5!}\epsilon^{IJ} \Gamma^{BCDEF} F^{(5)}_{BCDEF}  \ .
\end{equation}
To quadratic order in fermions $\kappa$-symmetry acts  as
\begin{equation} \label{eq:kappa-symmetry}
\begin{gathered}
    \delta_\kappa \theta^I = 2i \Gamma_\mu \partial_\alpha X^\mu \kappa^{I\alpha} + {\cal O}(\theta^2), \qquad \delta_\kappa X^\mu = i \bar{\theta}^I \Gamma^\mu \delta_\kappa \theta^J \delta_{IJ} + {\cal O}(\theta^3) \ , \\
    \delta_\kappa \left(\sqrt{-h}h^{\alpha\beta} \right) = 16 \sqrt{-h}  P_{I}^{J\alpha\gamma} P_{J}^{K\beta\delta} \bar{\kappa}^I_\gamma {\cal D}_{\delta KL} \theta^L \ ,
\end{gathered}
\end{equation}
where $P_{I}^{J\alpha\gamma} \equiv \frac{1}{2} \left( h^{\alpha\beta} \delta_{I}^{J} - \frac{\epsilon^{\alpha\beta}}{\sqrt{-h}} (\sigma_3)_{I}^{J} \right)$. How to appropriately fix all of the gauge symmetries in order to proceed with perturbative quantisation of the theory will be discussed  for an arbitrary  type IIB background with two commuting isometries in section \ref{s:strings-lightcone-gauge}.\\

 The expressions for the type IIA case are obtained by merging the two MW spinors into one 32-component Major spinor $\Theta$, 
%by replacing the gamma-matrices with the 32-dimensional ones, 
by replacing
\begin{equation} \label{eq:toIIA}
    (\sigma_3)_{IJ} \rightarrow \Gamma_{11} = \Gamma_0 \Gamma_1 \ldots \Gamma_9 \ ,
\end{equation}
and finally by replacing
\begin{equation} \label{eq:IIA_fluxes}
    {\cal S}^{IJ} \rightarrow \frac{1}{2}\sigma_3^{IJ} \Gamma^{BC} F_{BC}^{(2)}  + \frac{1}{4!} \delta^{IJ}\Gamma^{BCDE} F_{BCDE}^{(4)} \ .
\end{equation}
For more details see e.g.~\cite{Wulff:2013kga}.
%\SIB{For vanishing gravitino and dilatino}

\begin{centering}
\begin{tcolorbox}
\begin{exercise}
    At quadratic order in $\theta$, show that the type IIA and type IIB action $S_{\mathrm GS} = S_{\mathrm GS}^{(0)} + S_{\mathrm GS}^{(2)}$ reduces to the Green-Schwarz action $S=S_1+S_2$ in flat space for the field content \eqref{eq:sugra-flatspace}. 
\end{exercise}
\end{tcolorbox}
\end{centering}

\vspace{.4cm}

As we have mentioned, higher orders in fermions can in principle be obtained by starting from the formal superspace GS action of \cite{Grisaru:1985fv}, expanding in fields, and solving $\kappa$-constraints and Bianchi identities order by order.
This procedure was in particular used in \cite{Wulff:2013kga} to obtain the quartic order in the case of a type IIA/B supergravity background with  vanishing gravitino and dilatino. 
For generic backgrounds, however, this   quickly becomes  highly complex to do and discussing it in more detail falls out of the scope of this review.  However, for a large class of interesting backgrounds one can actually fix a $\kappa$-gauge  such that the full GS action to quartic order  is exact \cite{Wulff:2013kga}. This includes e.g.~$AdS_5\times S^5$, $AdS_4\times \mathbb{C}P^3$, $AdS_3\times S^3\times T^4$ and $AdS_3\times S^3\times S^3\times S^1$. \\

In the remaining of this section, we will focus on a special class of backgrounds, including the ones just mentioned, whose geometry and superisometries can be combined into a supercoset. As in the superspace formulation, superstrings propagating in supercoset backgrounds are described by  sigma-models that are all-order in fermions. They are particularly interesting  because they in addition enjoy a classical integrable structure on the worldsheet. 
In general however, the supercoset sigma-models correspond to the curved GS action only after a certain $\kappa$-gauge is fixed in the latter. Importantly, some configurations of the string may not be compatible with this particular $\kappa$-gauge. Such subtleties arise e.g.~in $AdS_3\times S^3\times M_4$, among others, and  will be discussed  in detail  in section \ref{s:saskia-gf}. The canonical exception is the type IIB $AdS_5\times S^5$ background.

\subsection{Supercoset construction of Green-Schwarz superstrings}\label{sec:Sib:supercoset_constr_GS}

Although the Green-Schwarz worldsheet action with a generic target space is formally known to all orders in fermions in terms of its superspace expression, obtaining its explicit form is highly non-trivial. The superspace expression can thus be hard to work with for explicit purposes. As we have alluded to in the previous section, for some special backgrounds with an underlying algebraic structure, there is a simple and convenient approach to constructing the action, known as the supercoset construction. As we will explain, when the target space is furthermore a semi-symmetric space, its form will be highly constrained, and it will be very simple to show that the theory is classically integrable.

Historically, the supercoset construction of the GS action for semi-symmetric spaces started with the  work of Henneaux and Mezincescu \cite{Henneaux:1984mh} who understood that the flat space GS action can be viewed as a Wess-Zumino type sigma-model action into the  space 
\begin{equation} 
    \frac{ISO(1,9 \ | \ {\cal N}=2 )}{SO(1,9)} ,
\end{equation}
where $ISO(1,9 \ | \ {\cal N}=2 )$ is the ${\cal N}=2$ super-Poincar\'e group in ten dimensions and $SO(1,9)$ its Lorentz subgroup.
The Wess-Zumino term obtained from a closed three-form of this space  captures the term $S_2[X,\theta]$ \eqref{eq:flat-GS-part2} quartic in fermions.
The bosonic part of this supercoset is of course $ISO(1,9)/SO(1,9)$ which represents  Minkowski space.\footnote{There is a general theorem that any manifold ${\cal M}$ with a transitive symmetry group action $G$, here $G=ISO(1,9)$ the Poincar\'e group, is isomorphic to the coset of $G$  by it stability group $H_p$ of any point $p\in {\cal M}$. It is very easy to verify that here the stability group is indeed the Lorentz group. 
We will come back to this later but for more details, a proof, and examples, on how to understand certain manifolds as coset spaces $G/G^{(0)} = \{ g \sim g h | g\in G, h\in G^{(0)} \}$ we refer the reader to section 2 of \cite{Zarembo:2017muf}.}  It was later realised by Metsaev and Tseytlin in \cite{Metsaev:1998it} that the work  \cite{Henneaux:1984mh} can be generalised to other semi-symmetric spaces  with curved target space metrics such as, in particular, $AdS_5 \times S^5$ which can be realised as
\begin{equation}
    AdS_5 \times S^5 + \text{fermions} \cong \frac{PSU(2,2|4)}{SO(1,4)\times SO(5)} \ .
\end{equation}
The bosonic subgroup of the Lie supergroup $PSU(2,2|4)$ is $SO(2,4)\times SO(6)$ and indeed $AdS_5$ and $S^5$ corresponds to the cosets $SO(2,4)/SO(1,4)$ and $SO(6)/SO(5)$ respectively.  Crucial in the supercoset construction is the existence of a $\mathbb{Z}_4$ symmetry in the algebra of target space superisometries. In this section, in particular section \ref{sec:SSSM_action}, we will show how one can then easily construct a minimal sigma-model action on any supercoset with this property. We will call this  the canonical semi-symmetric space sigma-model (SSSSM). The construction will not be the most general, however, and in fact we  will  see in section \ref{sec:CZ-WZ} that it can be generalised to accommodate for more generic background fields.

\subsubsection{Lie superalgebras and semi-symmetric spaces}

Let us first introduce some important concepts of Lie superalgebras $\mathfrak{g}$ and  semi-symmetric spaces relevant for superstring theory. A more detailed introduction of Lie superalgebras can be found in \cite{KAC19778}. 

\bigskip
\noindent {\bf Lie superalgebras and their $\mathbb{Z}_2$ grading} --- A Lie superalgebra $\mathfrak{g}$ is a generalisation of an ordinary Lie algebra as it is generated by both bosonic and fermionic generators (supercharges). We will denote them respectively by $T_{A_{[0]}}$ and $T_{A_{[1]}}$ such that $\mathfrak{g}=\mathrm{span}(T_A) = \mathrm{span}(T_{A_{[0]}}, T_{A_{[1]}})$.

Let us first consider the general linear superalgebra $\mathfrak{g}=\mathfrak{gl}(n|m)$.  An element $M$ of $\mathfrak{gl}(n|m)$ can be represented  as an $(n+m)\times (n+m)$ supermatrix acting on $\mathbb{C}^{n|m}$, \textit{i.e.}~the vector space ordered first by  $n$ Grassmann-even components followed by $m$ Grassmann-odd components, by
\begin{equation}
    M_{n+m, n+m} = \left(\begin{array}{c|c} a_{n,n} & \theta_{n,m} \\\hline \psi_{m,n} & b_{m,m} \end{array}\right) . 
\end{equation}
To preserve the Grassmann structure $a,b$ and $\theta,\psi$ are complex matrices whose elements are Grassmann-even (bosonic) and Grassmann-odd (fermionic) respectively. Hence we are actually working with the Grassmann enveloping superalgebra
of $\mathfrak{g}$, \textit{i.e.} 
\begin{equation}
    M =  M^{A_{[0]}} T_{A_{[0]}}+M^{A_{[1]}} T_{A_{[1]}}  ,
\end{equation}
with $M^{A_{[0]}}$ Grassmann-even, $M^{A_{[1]}}$ Grassmann-odd, and $T_A$ ordinary numerical matrices. \\

\noindent The Lie superalgebra is equipped with
\begin{itemize}
    \item a $\mathbb{Z}_2$ grading $|A| \equiv |T_A|$ called the degree of $T_A$.  It distinguishes even ($T_A=T_{A_{[0]}}$) and odd ($T_A=T_{A_{[1]}}$) elements by assigning $|A|=0$ and $|A|=1$ respectively.
    On the Grassmann enveloping superalgebra the $\mathbb{Z}_2$ grading can be realised through an involutive automorphism $\upsilon$ by the following element of $\mathfrak{gl}(n|m)$
\begin{equation} \label{eq:hypercharge}
    \Upsilon = \left(\begin{array}{c|c}
    \mathbb{1}_n & 0 \\ \hline 0 & -\mathbb{1}_m
    \end{array}\right) \ .
\end{equation}
It acts as
\begin{equation} \label{eq:Z2automorphism}
    \upsilon (M) = \Upsilon M \Upsilon^{-1} =\left(\begin{array}{c|c} a & -\theta \\ \hline -\psi & b \end{array}\right) , \qquad \upsilon^2 (M) = M \ ,
\end{equation}
and thus indeed defines a $\mathbb{Z}_2$ characterising the parity of the supermatrices: writing $\upsilon (M) = (-)^F M$, we call the supermatrix  even when $F=0$ and  odd when $F=1$.
\item a graded Lie superbracket $[\cdot, \cdot \}$ satisfying bilinearity, super skew-symmetry
\begin{equation}
    [T_A , T_B\} = T_A T_B - (-)^{|A||B|} T_B T_A = - (-)^{|A||B|} [T_B,T_A\} ,
\end{equation}
and the super Jacobi identity
\begin{equation}
    (-1)^{|A||C|} [T_A,[T_B,T_C\}\} + (-1)^{|B||A|} [T_B,[T_C,T_A\}\} + (-1)^{|C||B|} [T_C,[T_A,T_B\}\} = 0 .
\end{equation}
%Note that $|[T_A,T_B]| = |A|+|B| \ \mathrm{ mod} 2$.
 On the Grassmann enveloping superalgebra we then have (for $i,j=0,1$)
\begin{equation}
    [M,N] = M^{A_{[i]}} N^{A_{[j]}} [T_{A_{[i]}} , T_{A_{[j]}} \} = -[N,M] \ ,
\end{equation}
\textit{i.e.}~its Lie bracket is antisymmetric as usual.
\item a bilinear form $\kappa (T_A, T_B)$ realised through the supertrace
\begin{equation}\label{eq:supertrace}
    \kappa(T_A, T_B) = \mathrm{STr} (T_A T_B),   
\end{equation}
which satisfies
\begin{equation}
    \mathrm{STr}(T_A T_B) = (-1)^{|A||B|} \mathrm{STr}(T_B  T_A) , \qquad \mathrm{STr}([T_A , T_B\}) = 0 \ .
\end{equation}
It is defined as 
\begin{equation}
    \mathrm{STr}M = \mathrm{STr}\left(\begin{array}{c|c} a & \theta \\\hline \psi & b \end{array}\right) = \mathrm{Tr}a - \mathrm{Tr} b ,
\end{equation}
and  on the Grassmann enveloping superalgebra it satisfies the properties
\begin{equation} \label{eq:supertrace-cyclicity-adinvariant}
   \mathrm{STr}(MN) = \mathrm{STr}(NM) , \qquad \mathrm{STr}(M[P,N]) + \mathrm{STr}([P,M]N) = 0 \ .
\end{equation}
The supertrace thus defines an ad-invariant bilinear form on the Grassmann enveloping superalgebra.\footnote{Note that the usual trace would define an ad-invariant form for an ordinary Lie algebra but not for a (Grassmann enveloping) superalgebra. }
\end{itemize}

\vspace{.4cm}

\begin{centering}
\begin{tcolorbox}
\begin{exercise}
    Show that $\mathrm{STr}([T_A , T_B\}) = 0$ and $\mathrm{STr}([M,N]) = 0$. Verify that $\mathrm{Tr}([M,N])\neq 0$.
\end{exercise}
\end{tcolorbox}
\end{centering}

\vspace{.4cm}

The $\mathfrak{gl}(n|m)$ is not a simple Lie algebra: as $\mathrm{STr}([M,N])=0$ the subspace of supertraceless supermatrices, called the special linear superalgebra $\mathfrak{sl}(n|m)$, is a sub-superalgebra that forms an ideal of $\mathfrak{gl}(n|m)$.
%\footnote{An ideal $\mathfrak{i}$ is a Lie subalgebra of $\mathfrak{g}$ that satisfies $[\mathfrak{g}, \mathfrak{i}]\subset \mathfrak{i}$.} 
The $\mathfrak{sl}(n|m)$ itself is simple only if $n\neq m$. When $m=n$ the identity matrix is  a central element of  $\mathfrak{sl}(n|n)$. In that case, to obtain a simple algebra one must consider $\mathfrak{psl}(n|n) = \mathfrak{sl}(n|n)/\mathbb{1}_{2n}$, \textit{i.e.}~elements of $\mathfrak{sl}(n|n)$ that differ by a supermatrix proportional to $\mathbb{1}_{2n}$ must be identified.\footnote{Interestingly, the defining representation of $\mathfrak{sl}(n|n)$ in terms of supermatrices is not a representation for    $\mathfrak{psl}(n|n)$. This is due to possible non-trivial Grassmann-odd terms which cause the fact that for $M_1,M_2\in \mathfrak{psl}(n|n)$ it is not generally true that also $[M_1, M_2] \in \mathfrak{psl}(n|n)$. Therefore, one usually works with $\mathfrak{sl}(n|n)$ and implements the quotient using a gauge symmetry. 
} \\

Let us now consider real forms of $\mathfrak{(p)sl}(n|m)$ defined as the unitary superalgebras $\mathfrak{(p)su}(p,q|r,s)$ which is a real sub-superalgebra whose elements satisfy the reality condition
\begin{equation}
    M^\star = -M \ ,
\end{equation}
defined by
\begin{equation} \label{eq:psurealitycondition}
    M^\star = H M^\dag H^{-1} , \qquad H = \mathrm{diag}(\mathbb{1}_p , -\mathbb{1}_q | \mathbb{1}_r , -\mathbb{1}_s) \ ,
\end{equation}
where $M^\dag=(M^\ast)^t$ denotes the usual hermitian conjugation of matrices. Note that
\begin{equation}
    (M^\star)^\star = M , \qquad (MN)^\star = N^\star M^\star , \qquad (c M)^\star = c^\ast M^\star ,
\end{equation}
for $c$ a Grassmann number, for which we take the convention 
\begin{equation}
    (c_1 c_2)^\ast = c_2^\ast c_1^\ast = (-)^{|c_1| |c_2|} c_1^\ast c_2^\ast \ ,
\end{equation}
which guarantees $(MN)^\dag = N^\dag M^\dag$.

Another important class of superalgebras are the orthosymplectic superalgebras denoted by $\mathfrak{osp}(n|m)$, which are the super-counterparts of ordinary orthogonal Lie algebras. Instead of transposition, however, they are defined through \textit{supertransposition}  which acts as
\begin{equation}
    \left(\begin{array}{c|c}
    a & \theta \\ \hline \psi & b
    \end{array}\right)^{\mathrm{st}} = \left(\begin{array}{c|c}
    a^t & - \psi^t \\ \hline \theta^t & b^t
    \end{array}\right) ,
\end{equation}
and satisfies\footnote{Note that for general supermatrices $(MN)^t \neq N^t M^t$ and therefore conditions such as $M^t=-M$ (the defining relation of ordinary orthogonal Lie algebras) would not preserve commutation relations.}
\begin{equation} \label{eq:propst}
    (MN)^{\mathrm{st}} = N^{\mathrm{st}} M^{\mathrm{st}} , \qquad (M^{\mathrm{st}})^{\mathrm{st}} = \upsilon (M) \ .
\end{equation}
Elements of the orthosymplectic superalgebra $\mathfrak{osp}(n|2m)$ satisfy
\begin{equation} \label{eq:def-st-J2m}
    M^{\mathrm{st}} = - \Sigma M \Sigma^{-1} , \qquad \Sigma = \left(\begin{array}{c|c}
    \mathbb{1}_n & 0 \\ \hline 0 & J_{2m}
    \end{array}\right) , \qquad J_{2m} = \left(\begin{array}{c c}
    0 & -\mathbb{1}_m \\ \mathbb{1}_m & 0 
    \end{array}\right) \ .
\end{equation}

\noindent {\bf Semi-symmetric spaces and $\mathbb{Z}_4$ grading} --- Other interesting and important superalgebras are those whose $\mathbb{Z}_2$ grading defined above actually sits in a larger $\mathbb{Z}_4$ grading. This means that $\mathfrak{g}$ admits an automorphism $\Omega : \mathfrak{g} \rightarrow \mathfrak{g}$ that squares to the $\mathbb{Z}_2$ automorphism \eqref{eq:Z2automorphism} defining parity
\begin{equation}
    \Omega^2(M) = \upsilon (M) ,
\end{equation}
and which has order four
\begin{equation}
    \Omega^4 (M) = M \ .
\end{equation}
It therefore has four eigenvalues, $\pm 1$, $\pm i$, and decomposes $\mathfrak{g}$ into a direct sum of graded eigenspaces as
\begin{equation} \label{eq:Z4grading}
    \mathfrak{g} = \mathfrak{g}^{(0)} \oplus \mathfrak{g}^{(1)} \oplus \mathfrak{g}^{(2)} \oplus \mathfrak{g}^{(3)} ,
\end{equation}
defined by
\begin{equation}
    \Omega (\mathfrak{g}^{(k)} )= i^k \mathfrak{g}^{(k)}  \ .
\end{equation}
Since $\Omega$ is an automorphism $\Omega[\mathfrak{g},\mathfrak{g}] = [\Omega(\mathfrak{g}), \Omega(\mathfrak{g})]$,  the graded eigenspaces satisfy
\begin{equation} \label{eq:ssscommrel}
    [\mathfrak{g}^{(k)} , \mathfrak{g}^{(l)}] \subset \mathfrak{g}^{(k+l \ \mathrm{mod} \ 4)}
\end{equation}
for $k,l = 0,1,2,3$. Hence, $\mathfrak{g}^{(0)}$ forms a subalgebra of $\mathfrak{g}$ and $\mathfrak{g}^{(k\neq 0)}$ are representations of $\mathfrak{g}^{(0)}$. This decomposition is also called a semi-symmetric space decomposition, and either the existence of the $\mathbb{Z}_4$ automorphism $\Omega$ or  the property \eqref{eq:ssscommrel} of the  commutation relations  can be seen as its defining feature. We can introduce projectors $P^{(k)} : \mathfrak{g} \rightarrow \mathfrak{g}^{(k)}$ for every $k=0,1,2,3$ realised as
\begin{equation} \label{eq:Z4projector}
    \begin{aligned}
        P^{(k)} &= \frac{1}{4} (\mathbb{1}+i^{2k}\Omega^2 + i^{3k}\Omega + i^k\Omega^3) , 
    \end{aligned}
\end{equation}
and which satisfy
\begin{equation}
    \mathbb{1}=P^{(0)}+P^{(1)}+P^{(2)}+P^{(3)} , \qquad P^{(k)} P^{(l\neq k)} = 0 , \qquad P^{(k)} P^{(k)} = P^{(k)} \ . 
\end{equation}
We will denote $M^{(k)}\equiv P^{(k)}M$.

\vspace{.4cm}

\begin{centering}
\begin{tcolorbox}
\begin{exercise}
Show that $\mathfrak{g}^{(0)}$ and $\mathfrak{g}^{(2)}$ span even (bosonic) supermatrices, while $\mathfrak{g}^{(1)}$ and $\mathfrak{g}^{(3)}$ span odd (fermionic) supermatrices. 
\end{exercise}
\end{tcolorbox}
\end{centering}

\vspace{.4cm}

\textit{Remark} ---  Equivalently, the $\mathbb{Z}_2$ automorphism $\upsilon$ defined in \eqref{eq:Z2automorphism} decomposes the superalgebra as
\begin{equation}
    \mathfrak{g} = \mathfrak{g}^{[0]} \oplus \mathfrak{g}^{[1]} ,
\end{equation}
with $\mathfrak{g}^{[k]}$, $k=0,1$ defined as\footnote{The generators $T_{A_{[0]}}$ and $T_{A_{[1]}}$ of $\mathfrak{g}$ that we defined in the beginning of this section thus span $\mathfrak{g}^{[0]}$ and $\mathfrak{g}^{[1]}$ respectively.}
\begin{equation}
    \upsilon ( \mathfrak{g}^{[k]} ) = (-)^k \mathfrak{g}^{[k]} ,
\end{equation}
and satisfying
\begin{equation} \label{eq:sscommrel}
    [\mathfrak{g}^{[k]} , \mathfrak{g}^{[l]}] \subset \mathfrak{g}^{[k+l \ \mathrm{mod} \ 2]} .
\end{equation}
Hence $\mathfrak{g}^{[0]}$ is a subalgebra.
Such a decomposition is called a symmetric space decomposition, whose defining feature is the existence of a $\mathbb{Z}_2$ or, equivalently, the commutation relations \eqref{eq:sscommrel}.
%(in particular $[\mathfrak{g}^{[1]},\mathfrak{g}^{[1]}]\subset \mathfrak{g}^{[0]}$). 
Note that if $\Omega$ defined above exists, then 
\begin{equation}
    \mathfrak{g}^{[0]} = \mathfrak{g}^{(0)}\oplus\mathfrak{g}^{(2)} \ , \qquad \mathfrak{g}^{[1]} = \mathfrak{g}^{(1)}\oplus\mathfrak{g}^{(3)} \ ,
\end{equation}
and thus both $\mathfrak{g}^{(0)}$ and $\mathfrak{g}^{(0)}\oplus\mathfrak{g}^{(2)}$ are subalgebras of $\mathfrak{g}$. Let us finally mention that also ordinary Lie algebras can have the algebraic structure of a symmetric space---and they are in fact quite interesting---but, in contrast to Lie superalgebras, this property is special and not guaranteed (similary as the $\mathbb{Z}_4$ for Lie superalgebras is not guaranteed).

\vspace{.4cm}

 \begin{centering}
\begin{tcolorbox}
\begin{exercise}
    When restricting to bosonic supermatrices $\mathfrak{g}^{[0]}$, observe that $\Omega$ also defines a symmetric space. 
\end{exercise}
\end{tcolorbox}
\end{centering}

\vspace{.4cm}

Let us now denote the generators of $\mathfrak{g}$ respecting the $\mathbb{Z}_4$ decomposition as $T_{A_{(k)}}$, in the sense that $T_{A_{(k)}} \in \mathfrak{g}^{(k)}$ for $k=0,1,2,3$. Then, the supermatrix decomposes as $M=M^{A_{(i)}} T_{A_{(i)}}$ with $M^{A_{(0)}},M^{A_{(2)}}$ Grassmann even, $T_{A_{(0)}}, T_{A_{(2)}}$ bosonic generators, $M^{A_{(1)}},M^{A_{(3)}}$ Grassmann odd, and $T_{A_{(1)}}, T_{A_{(3)}}$ fermionic supercharges. The ad-invariant bilinear form can then be denoted as
\begin{equation}
    \kappa_{A_{(k)}B_{(l)}} \equiv \kappa (T_{A_{(k)}}, T_{B_{(l)}} )=\mathrm{STr} (T_{A_{(k)}} T_{B_{(l)}} ) \ .
\end{equation}

\vspace{.4cm}

 \begin{centering}
\begin{tcolorbox}[breakable]
\begin{exercise}
Show that the supertrace is $\Omega$-invariant, meaning
that it respects the $\mathbb{Z}_4$ grading
\begin{equation} \label{eq:STrZ4invariant}
    \mathrm{STr} (\Omega (M) \Omega (N) )= \mathrm{STr} (MN) \,. 
\end{equation}
You can use that  $\Omega$ defines a representation for $\mathfrak{g}$, and thus the vector space $\mathbb{C}^{n|m}$ on which $\mathfrak{g}$ acts must also decompose in the $\mathbb{Z}_4$ grading. If we then span $\mathbb{C}^{n|m}$ by basis elements in the order $e^{(0)}_1, \ldots , e^{(0)}_{\dim\mathfrak{g}^{(0)}}, \allowbreak e^{(2)}_1, \ldots , e^{(2)}_{\dim\mathfrak{g}^{(2)}},e^{(1)}_1, \ldots , e^{(1)}_{\dim\mathfrak{g}^{(1)}},e^{(3)}_1, \ldots , e^{(3)}_{\dim\mathfrak{g}^{(3)}}$ then in order to preserve the grading the supermatrix $M$ must decompose as
\begin{equation}
    M = \left( \begin{array}{c c | c c}
    a_{(0)(0)} & a_{(0)(2)} & \theta_{(0)(3)} & \theta_{(0)(1)} \\
    a_{(2)(0)} & a_{(2)(2)} & \theta_{(1)(0)} & \theta_{(3)(0)} \\ \hline
    \psi_{(0)(1)} & \psi_{(0)(3)} & b_{(0)(0)} & b_{(0)(2)} \\
    \psi_{(3)(0)} & \psi_{(1)(0)} & b_{(2)(0)} & b_{(2)(2)}
    \end{array}\right) ,
\end{equation}
with $a_{(k)(l)},\theta_{(k)(l)}, \ldots \in \mathfrak{g}^{(k+l \ \mathrm{mod} \ 4)}$.\\

Show that \eqref{eq:STrZ4invariant} now implies that 
\begin{equation} \label{eq:supertrace-property}
    \kappa_{A_{(k)}B_{(l)}} = 0 \quad \text{when} \quad k+l \neq 0 \ \mathrm{mod} \ 4 \ .
\end{equation}
Hence the only non-vanishing components are $\kappa_{A_{(0)}B_{(0)}}$, $\kappa_{A_{(2)}B_{(2)}}$, and $\kappa_{A_{(1)}B_{(3)}}= - \kappa_{B_{(3)}A_{(1)}}$.
\end{exercise}
\end{tcolorbox}
\end{centering}

\vspace{.4cm}

\textit{Remark} --- The above exercise shows that the decomposition of $\mathfrak{g}$ into its $\mathbb{Z}_4$ graded eigenspaces is \textit{not} an orthogonal decomposition with respect to the supertrace. \\

Assuming that the Lie supergroup $G$ corresponding to the Lie superalgebra $\mathfrak{g}$ is connected, the $\mathbb{Z}_4$ algebra automorphism $\Omega$ extends to a $\mathbb{Z}_4$ group automorphism $\omega :G \rightarrow G$ as $\omega (g) = \omega (e^{ X}) = e^{ \Omega(X)} $ for $g=e^{ X}\in G$ and $X\in \mathfrak{g}$. The subalgebra $\mathfrak{g}^{(0)}$ invariant under $\Omega$ now defines a bosonic subgroup $ G^{(0)}$ which is the set of fixed points of $\omega$. This algebraic structure in turn defines a semi-symmetric space $G/G^{(0)}$  which is a manifold with a transitive supergroup action $G$ of which the stability group of the manifold is defined by the $\mathbb{Z}_4$, \textit{i.e.}~$G^{(0)}$. Side note: when the group automorphism defining $G^{(0)}$ is a $\mathbb{Z}_2$, then this structure defines a symmetric space. \\

\textit{Remark} --- Note that semi-symmetric and symmetric spaces are \textit{special} cases of homogeneous spaces. Homogeneous spaces  are simply manifolds with a transitive $G$ action and a stability group $G^{(0)}$. They are isomorphic to the (super)coset $G/G^{(0)}$ (for more detail see \cite{Zarembo:2017muf}). Only when the stability group $G^{(0)}$ arises as the invariant of a $\mathbb{Z}_2$ ($\mathbb{Z}_4$) automorphism,  then the homogeneous space is a (semi-)symmetric space. In the string literature, (semi-)symmetric  spaces are sometimes also referred to as (super)coset spaces although---as is hopefully clear from this paragraph---that is an abuse of language. \\ 

Before introducing the sigma-model action to the semi-symmetric space $G/G^{(0)}$, let us illustrate some of the abstract concepts introduced in this section with some examples.\\

\noindent {\bf Example: $S^n$} ---  Let us first warm-up using an example without fermionic generators. Take the unit $n$-dimensional sphere $S^n$ embedded in $\mathbb{R}^{n+1}$ by
\begin{equation} \label{eq:embedding-sphere}
    X_I X^I = 1 ,
\end{equation}
with $I = 1, \ldots, n+1$. $G=SO(n+1)$ is a  rotational symmetry group that leaves $S^n$ invariant and which acts transitively. To determine the stability group $G^{(0)}$, consider a particular point on $S^n$ such as, e.g., the north pole. It is not hard to convince oneself that $G^{(0)}=SO(n)$. Thus
\begin{equation}
    S^n \cong \frac{SO(n+1)}{SO(n)} \ ,
\end{equation}
is a homogeneous space. In fact, it is also a symmetric space. Consider for simplicity $S^2 = SO(3)/SO(2)$. The commutation relations of $\mathrm{Lie}(SO(3))$ and the bilinear form are
\begin{equation}
    [T_1, T_2] =  T_3 \ , \qquad \kappa (T_A, T_B) = -2 \delta_{AB} \ .
\end{equation}
We can define a $\mathbb{Z}_2$ automorphism
\begin{equation}
    \sigma (T_3) = T_3 \ , \qquad \sigma (T_{1,2} ) = -T_{1,2} \  ,
\end{equation}
which in turn defines the stability group $SO(2)$ generated by $T_3$. Indeed the commutation relations \eqref{eq:sscommrel} for $\mathfrak{g}^{[0]} = \mathrm{span}(T_3)$ and $\mathfrak{g}^{[1]} = \mathrm{span}(T_1, T_2)$ are satisfied. Let us remark that also $AdS_n \cong SO(2,n-1)/SO(1,n-1)$ geometries arise as symmetric spaces.   \\

\noindent {\bf Example: $\mathfrak{psu}(2,2|4)$} --- For GS superstrings relevant for AdS/CFT, the  canonical example  of a superalgebra with a $\mathbb{Z}_4$ grading is $\mathfrak{psu}(2,2|4)$, which is related to the maximally supersymmetric $AdS_5\times S^5$ background. The reality condition \eqref{eq:psurealitycondition} here implies that 
\begin{equation}
     a = - h a^\dag h , \qquad b= -b^\dag, \qquad \theta = - h \psi^\dag , \qquad h = \mathrm{diag}(\mathbb{1}_2 , - \mathbb{1}_2) \ , 
\end{equation}
and thus the matrices $a$ and $b$ span $\mathfrak{u}(2,2)$ and $\mathfrak{u}(4)$ respectively. 
Hence the bosonic subalgebra $\mathfrak{g}^{[0]}$ of $\mathfrak{psu}(2,2|4)$  is 
\begin{equation}
    \mathfrak{su}(2,2) \oplus \mathfrak{su}(4) \ .
\end{equation}
Note that the central element $\mathbb{1}_8$ of $\mathfrak{su}(2,2|4)$, which we quotient away in $\mathfrak{psu}(2,2|4)$,   is bosonic and the only element with non-vanishing trace.

A nice realisation of the $\mathbb{Z}_4$ grading of $\mathfrak{psu}(2,2|4)$ is\footnote{More generally this is  a $\mathbb{Z}_4$ automorphism of $\mathfrak{gl}(4|4)$. }
\begin{equation} \label{eq:z4-ads5-s5}
    \Omega (M) = - {\cal K} M^{\mathrm{st}} {\cal K}^{-1} , \qquad {\cal K} = \mathrm{diag}\left(J_2, J_2 | J_2 , J_2\right) ,
\end{equation}
with $J_2$ defined in \eqref{eq:def-st-J2m}. 
Note, however, that the $\mathbb{Z}_4$ grading is not unique: any automorphism $\hat{\Omega}$ related to $\Omega$ with a similarity transformation also has order four and  would define a different grading of $\mathfrak{psu}(2,2|4)$. An example here is $\hat{\Omega}(M)=-M^{\mathrm{st}}$.

\vspace{.4cm}

\begin{centering}
\begin{tcolorbox}[breakable]
\begin{exercise}
Show that $\Omega (M) \in \mathfrak{psl}(2,2|4)$ when $M\in \mathfrak{psl}(2,2|4)$. In particular, 
\begin{equation}
    \mathrm{STr} (\Omega (M)) = 0,  \quad \text{and} \quad \mathrm{Tr} (\Omega (M)) = 0 \ .
\end{equation}
Observe that 
\begin{equation}
    \Omega(M_1 M_2) = - \Omega(M_2) \Omega(M_1) ,
\end{equation}
and thus $\Omega$ is indeed an automorphism $\Omega[M_1, M_2] = [\Omega(M_1), \Omega(M_2)]$. 
\end{exercise}
\end{tcolorbox}
\end{centering}
Let us point out that $(M^\mathrm{st})^\dag \neq (M^\dag)^{\mathrm{st}}$
and in particular
\begin{equation}
\begin{alignedat}{2}
    \Omega (M)^\dag &= \Omega (M^\dag) \qquad &&M \in \mathfrak{g}^{[0]} , \\
    \Omega (M)^\dag &=- \Omega (M^\dag) \qquad &&M \in \mathfrak{g}^{[1]} ,
\end{alignedat}
\end{equation}
or equivalently $\Omega(M)^\dag = \upsilon (\Omega(M^\dag))$.
This highlights a subtlety about possible $\mathbb{Z}_4$ gradings: strictly speaking on the full algebra $\Omega$ does not respect the real form \eqref{eq:psurealitycondition}.  Nevertheless, this issue is easily circumvented as each projection $M^{(k)}$ does take values in $\mathfrak{psu}(2,2|4)$.  We can show this by using \eqref{eq:Z4projector} and \eqref{eq:psurealitycondition} which implies
\begin{equation}
    M^{(k)\dag} = -\frac{1}{4} H^{-1}\left( M + i^{2k} \Omega^2(M) +i^k \upsilon\Omega(M) +i^{3k} \upsilon\Omega^3(M)  \right) H \ .
\end{equation}
Note that this follows in particular because $[H,\Upsilon]=[H,{\cal K}]=[{\cal K},\Upsilon]=0$. Now using $\upsilon = \Omega^2$ and $\Omega^4=1$ we have
\begin{equation}
    M^{(k)\dag} = - H^{-1}M^{(k)} H \ ,
\end{equation}
and thus  indeed the components $M^{(k)}$ belong to $\mathfrak{psu}(2,2|4)$ for any $k=0,1,2,3$. For explicitness, let us write down each of the projections separately. We have
\begin{equation}
    \begin{alignedat}{2}
        M^{(0)} &= \frac{1}{2} \left(\begin{array}{c|c}
             a- K a^t K^{-1}&   \\\hline
             & b- K b^t K^{-1}
        \end{array} \right) , \quad
        &&M^{(2)} = \frac{1}{2} \left(\begin{array}{c|c}
             a+ K a^t K^{-1}&   \\\hline
             & b+ K b^t K^{-1}
        \end{array} \right), \\
        M^{(1)} &= \frac{1}{2} \left(\begin{array}{c|c}
             & \theta- i K \psi^t K^{-1}  \\\hline
             \psi+ i K \theta^t K^{-1} & 
        \end{array} \right),\quad
        &&M^{(3)} = \frac{1}{2} \left(\begin{array}{c|c}
             & \theta+ i K \psi^t K^{-1}  \\\hline
             \psi- i K \theta^t K^{-1} & 
        \end{array} \right) ,
    \end{alignedat}
\end{equation}
where $K= \mathrm{diag}(J_2, J_2)$.  Using an explicit matrix realisation of $\mathfrak{psu}(2,2|4)$ (see e.g.~section 1.1. of \cite{Arutyunov:2009ga}), one can show that the  Lie algebra $\mathfrak{g}^{(0)}$ invariant under the $\mathbb{Z}_4$ is the subalgebra
\begin{equation}
    \mathfrak{so}(1,4) \oplus \mathfrak{so}(5) \subset \mathfrak{g}^{[0]} \ ,
\end{equation}
and that with $\mathfrak{g}^{(2)} \subset \mathfrak{g}^{[0]}$ this defines another symmetric space decomposition,  that is $[\mathfrak{g}^{(2)},\mathfrak{g}^{(2)}]\subset \mathfrak{g}^{(0)}$, which is of course different from the parity $\mathbb{Z}_2$. The bosonic subalgebra of $\mathfrak{psu}(2,2|4)$ thus describes the  isometries of a symmetric  space which is precisely   $AdS_5 \times S^5$ 
\begin{equation}
      AdS_5 \times S^5 \cong \frac{SU(2,2)}{SO(1,4)} \times \frac{SU(4)}{SO(5)} \cong \frac{SO(2,4)}{SO(1,4)} \times \frac{SO(6)}{SO(5)} \ .
\end{equation}
Including the fermionic generators  one will cover  the superisometries of the maximally supersymmetric $AdS_5 \times S^5$ superspace
\begin{equation}
    AdS_5 \times S^5 + \text{fermions} \cong  \frac{PSU(2,2|4)}{SO(1,4)\times SO(5)} \ .
\end{equation}

\noindent {\bf Example: direct sums of Lie supergroups and $AdS_3\times S^3$} --- A special case of semi-symmetric spaces are the so-called permutation supercosets \cite{Babichenko:2009dk,Cagnazzo:2012se} for which the symmetry group  is the direct product of two simple supergroups $\tilde{G}= {G} \times {G}$. In fact, at the algebra level, any direct sum of two superalgebras admits a $\mathbb{Z}_4$ defined as a permutation of the factors by
\begin{equation}
    \Omega = \begin{pmatrix}
        0 & \mathbb{1} \\
        (-)^F & 0
    \end{pmatrix} \ .
\end{equation}
The invariant subalgebra is $\tilde{\mathfrak{g}}^{(0)} = ({\mathfrak{g}}^{[0]}\oplus {\mathfrak{g}}^{[0]})_{\text{\tiny diag}}$, \textit{i.e.}~$X\in \tilde{\mathfrak{g}}^{(0)}$ is of the form $X=(\xi, \xi)$ with $\xi\in {\mathfrak{g}}^{[0]}$. The semi-symmetric space is then ${G} \times {G}/(G^B\times G^B)_{\text{\tiny diag}}$  and thus its bosonic section is just the group manifold  $G^B = \exp \mathfrak{g}^{[0]}$. This is in particular the case for the  $AdS_3\times S^3$ semi-symmetric spaces, for which ${G}=PSU(1,1|2)$. Their bosonic subgroup is $SU(1,1)\times SU(2)$ corresponding to the $AdS_3\times S^3$ manifold. The fermionic elements  on the other hand come from the odd generators of $\mathfrak{psu}(1,1|2)\oplus \mathfrak{psu}(1,1|2)$, of which there are 16 in total. This example will be discussed in much more detail in section \ref{sec:saskia:AdS3coset}.

%\noindent {\bf Example: $AdS_3\times S^3$} --- We will give  many details about this interesting example in the second part of these lectures, see section \ref{sec:saskia:AdS3coset}.

\subsubsection{Action of the canonical semi-symmetric space sigma-model}\label{sec:SSSM_action}

The supercoset construction of the Green-Schwarz superstring  starts from the main requirement that the target space of the sigma-model action is a semi-symmetric space  $G/G^{(0)}$ with the key existence of a $\mathbb{Z}_4$ \cite{Berkovits:1999zq}. This builds in from the very beginning the property of invariance under  global transformations of a supergroup G (\textit{i.e.}~supersymmetries) on the field configurations. We assume from now on that the associated Lie superalgebra is $\mathfrak{g}=\mathfrak{su}(p,q|r,s)$.

The supercoset construction  is most elegantly achieved by using the formalism of Maurer-Cartan forms of Lie superalgebras.
Consider a supergroup-valued element $g \in G$ to which we associate the field configuration of the sigma-model action as a map from the worldsheet $\Sigma$ to the space $G$
\begin{equation}
    g  \ : \ \Sigma \rightarrow G \ : \ \sigma \mapsto g(\sigma) \ .
\end{equation}
The group element thus plays the analogous role of the spacetime coordinates $X^\mu : \Sigma \rightarrow {\cal M} : \sigma \mapsto X^\mu(\sigma)$.
From $g$ we can construct the  superalgebra-valued (left-invariant) Maurer-Cartan form $J$
\begin{equation} \label{eq:mc-form}
    J = g^{-1} d g \  \in \  \mathfrak{g} \otimes \Omega^1(G) , 
\end{equation}
which we can expand as $J=J^A T_A$ in the Lie superalgebra, with $J^A\in \Omega^1(G)$ one-forms. This one-form satisfies identically the famous Maurer-Cartan equation 
\begin{equation} \label{eq:mc-identity-1}
    dJ + J\wedge J = 0 \ .
\end{equation}
Importantly, $J$ is clearly invariant under global $G$ transformations from the left
\begin{equation} \label{eq:left-G-action}
    g(\sigma) \rightarrow g_L g(\sigma) , 
\end{equation}
with $g_L \in G$ constant on the worldsheet.
To require however that the target space is $G/G^{(0)}$, the degrees of freedom of the field configurations must actually be supercoset representatives $[g]$. In particular, since  $G^{(0)}$ is the stability group of the manifold,   points in $G$ that are related by   transformations with elements in $G^{(0)}$ must be identified, \textit{i.e.}
\begin{equation}
    g \sim g h , \qquad [g] = [gh] ,
\end{equation}
with $h\in G^{(0)}$.\footnote{More generally, the left-acting $G$ transformation \eqref{eq:left-G-action} can thus act as $g\rightarrow g_0 g h$ with $h\in G^{(0)}$ a compensating local $G^{(0)}$ transformation. } This physical equivalence  can be achieved by realising the right action of $G^{(0)}$ as a worldsheet gauge symmetry 
\begin{equation}\label{eq:H-gauge-invariance}
    g(\sigma) \rightarrow g(\sigma) h(\sigma) , \qquad h(\sigma) \in G^{(0)} . 
\end{equation}
Under this transformation, the left-invariant Maurer-Cartan form $J$ transforms as
\begin{equation}
    J \rightarrow h^{-1} J h + h^{-1} dh .
\end{equation}
Decomposing $J$ under the $\mathbb{Z}_4$ grading as in \eqref{eq:Z4grading} 
\begin{equation}
    J = J^{(0)} + J^{(1)} + J^{(2)} + J^{(3)} ,
\end{equation}
each projection $J^{(k)} = P^{(k)}J$ transforms under the gauge transformation as
\begin{equation} \label{eq:H-action-graded-J}
    J^{(0)} \rightarrow h^{-1}J^{(0)}h + h^{-1}dh, \qquad J^{(1,2,3)} \rightarrow h^{-1} J^{(1,2,3)} h \ ,
\end{equation}
because, since $h\in G^{(0)}$, $h^{-1}dh\in \mathfrak{g}^{(0)}$. Hence,  $J^{(0)}$ transforms as a gauge field,  while $J^{(1,2,3)}$ transform with a similarity transformation. \\

\noindent The simplest way to built the semi-symmetric space sigma-model (SSSSM)  in terms of the fields $[g]\in G/G^{(0)}$, and which
\begin{itemize}
    \item is invariant under the global superisometry group $G$ acting from the left as in \eqref{eq:left-G-action},
    \item is gauge-invariant the local bosonic subgroup $G^{(0)}\subset G$ acting from the right as in \eqref{eq:H-gauge-invariance}, with $G^{(0)}$ the invariant of a $\mathbb{Z}_4$,
    \item has a bosonic truncation  as in \eqref{eq:GS-curved-zeroeth} with a target space $G^B/G^{(0)}$, where $G^B$ is the bosonic subgroup of $G$ generated by $\mathfrak{g}^{[0]}=\mathfrak{g}^{(0)}\oplus \mathfrak{g}^{(2)}$,
    \item and would reduce to the Green-Schwarz superstring in Minkoswki space in an appropriate flat space limit,
\end{itemize}
is to pull-back the objects $J^{(1,2,3)}$ to the worldsheet and pair them accordingly using the ad-invariant bilinear form of the Lie superalgebra, \textit{i.e.}~the supertrace \eqref{eq:supertrace}.\footnote{For a more first-principle construction of the action \eqref{eq:SSSSM1} we refer to  \cite{Metsaev:1998it,Berkovits:1999zq}.} Firstly, the pull-back is obtained by introducing local coordinates $Z^{ M}=(X^\mu, \theta^I)$ parametrising the target superspace $G/G^{(0)}$ which picks one representative per orbit  as $[g(Z^{ M})]$.
%\footnote{Recall from section \ref{s:GS-curved} that the local coordinates $Z^{ M}=(X^\mu, \theta^I)$ capture both the (bosonic) spacetime coordinates $X^\mu$ as well as the fermionic superspace coordinates $\theta^I$.} 
We can then expand the Maurer-Cartan form  as $J = J_{ M} dZ^{ M} = J_\alpha \de \sigma^\alpha$ using the pull-back map $Z^{ M}(\sigma^\alpha)$ between worldsheet and target space, \textit{i.e.}~simply $d Z^{ M} = \partial_\alpha Z^{ M} \de \sigma^\alpha$. Secondly, recalling that the supertrace respects the $\mathbb{Z}_4$ grading and thus satisfies the property \eqref{eq:supertrace-property}, a natural pairing of the objects $J^{(1,2,3)}$ in the sigma-model action is as follows
\begin{equation} \label{eq:SSSSM1}
    S_{\text{\tiny SSSSM}} = -\frac{T}{2} \int \de^2\sigma\ \mathrm{STr} \left(\sqrt{-h}h^{\alpha\beta} J_{\alpha}^{(2)} J_{\beta}^{(2)} - \varkappa \epsilon^{\alpha\beta} J_{\alpha}^{(1)}J_{\beta}^{(3)} \right) ,
\end{equation}
where $\varkappa$ is a yet undetermined but real constant, which  will be fixed by demanding $\kappa$-symmetry.  This action is clearly invariant under the global left-acting $G$ transformations and the  local right-acting $G^{(0)}$ transformations as well as worldsheet diffeomorphisms and Weyl rescalings.\footnote{Note that the action would not have been gauge-invariant under right-acting $G^{(0)}$ transformations if instead of the supertrace we would have used the standard trace.} Note however that this is not necessarily the most generic ansatz for a pairing of $J^{(1,2,3)}$, and we will come back to this point later. For now, we will call \eqref{eq:SSSSM1} the \textit{canonical} SSSSM. Generalisation of this action will be discussed in section \ref{sec:saskia:AdS3coset}. 
%\SIB{put correct ref?}
However, let us remark here that in the case of $\mathfrak{psu}(2,2|4)$ giving rise to the supersymmetric $AdS_5\times S^5$ background, it was shown in \cite{Metsaev:1998it} that  \eqref{eq:SSSSM1} for $\varkappa=1$ is the \textit{unique} action which satisfies the above requirements together with local $\kappa$-invariance.

\begin{centering}
\begin{tcolorbox}
\begin{exercise}
    Show that $\varkappa$ must be real in order for the Lagrangian to be real when $\mathfrak{g} = \mathfrak{su}(p,q|r,s)$.
\end{exercise}
\end{tcolorbox}
\end{centering}

\textit{Remark} --- Using the pull-back,  the Maurer-Cartan identity \eqref{eq:mc-identity-1} reads
\begin{equation} \label{eq:mc-identity-2}
    \partial_\alpha J_\beta - \partial_\beta J_{\alpha} + [J_\alpha , J_\beta] = 0 \ .
\end{equation}

\begin{centering}
\begin{tcolorbox}
\begin{exercise}\label{ex:mc-projection}
    Project the Maurer-Cartan identity \eqref{eq:mc-identity-2} on each of the $\mathbb{Z}_4$ graded components.
\end{exercise}
\end{tcolorbox}
\end{centering}

\noindent {\bf The exact WZ-term} --- The second term of the action \eqref{eq:SSSSM1} should be thought of as the analogue of the term \eqref{eq:flat-GS-part2} quartic in fermions and is commonly also referred to as the exact Wess-Zumino term. On the supercoset it  ascends  from a closed three-form which respects the $\mathbb{Z}_4$ grading. In form language it reads 
\begin{equation}\label{eq:exact_form_tweede_term}
    \Theta_3 = \mathrm{STr} \left(J^{(2)} \wedge J^{(3)} \wedge J^{(3)} - J^{(2)} \wedge J^{(1)} \wedge J^{(1)} \right) ,
\end{equation}
and in the sigma-model action it is integrated over a three-cycle ${\cal B}$ whose boundary is the two-dimensional worldsheet $\partial {\cal B} = \Sigma$.

\begin{centering}
\begin{tcolorbox}
\begin{exercise}
    Show that $\Theta_3$ is closed, $d\Theta_3=0$ upon the Jacobi identity. Recall that for a $q$-form $\xi$ and an $r$-form $\omega$ one has $d(\xi\wedge\omega) = d\xi \wedge\omega + (-)^q \xi \wedge d \omega$.
\end{exercise}
\end{tcolorbox}
\end{centering}

\noindent In general, closed three-forms of manifolds are not necessarily exact, this depends on whether or not the third cohomology group is trivial. Due to the $\mathbb{Z}_4$ grading, however, this is in fact the case for $\Theta_3$ 
\begin{equation} \label{eq:theta3-is-exact}
    \Theta_3 = \frac{1}{2} d \ \mathrm{STr} \left( J^{(1)} \wedge J^{(3)} \right) ,
\end{equation}
and consequently, using Stokes' theorem, the Wess-Zumino term can be reduced to a local integral over the two-dimensional worldsheet $\Sigma$ giving rise precisely to the latter term of \eqref{eq:SSSSM1}. Generically, for GS superstrings, finding such exact three-forms is essential for $\kappa$-symmetry and here relies completely on the $\mathbb{Z}_4$ (semi-symmetric space) structure \cite{Berkovits:1999zq}.

\begin{centering}
\begin{tcolorbox}
\begin{exercise}
    Show \eqref{eq:theta3-is-exact}.
\end{exercise}
\end{tcolorbox}
\end{centering}

\noindent {\bf Reducing to a GS-type action} --- To compare \eqref{eq:SSSSM1} with the generic Green-Schwarz action for curved spaces given in section \ref{s:GS-curved}  (and in particular to find the bosonic truncation, \textit{i.e.}~setting fermions to zero), let us parametrise the group element $g$ as
\begin{equation} \label{eq:g-parametrisation}
    g = g_B e^\theta , \qquad g_B \in G^B = \exp \left( \mathfrak{g}^{(0)}\oplus \mathfrak{g}^{(2)}\right), \qquad \theta\in \mathfrak{g}^{(1)}\oplus \mathfrak{g}^{(3)} ,
\end{equation}
and expand the Maurer-Cartan form \eqref{eq:mc-form} to quadratic order in fermions
\begin{equation} \label{eq:exp-mc-form}
    J = J_B + {\cal D}_B\theta  -\frac{1}{2} [\theta , {\cal D}_B\theta]  +{\cal O}( \theta^3) \ ,
\end{equation}
where $J_B=g_B^{-1}d g_B$ and ${\cal D}_B = d + \mathrm{ad}_{J_B}$. Projecting onto each of the $\mathbb{Z}_4$ graded components then gives up to quadratic order
\begin{equation} \label{eq:mc-to-quadraticfermions}
\begin{aligned}
    J^{(0)} &=J_B^{(0)}-\frac{1}{2} [\theta , {\cal D}_B\theta]^{(0)} , \qquad 
    J^{(1)} ={\cal D}_B\theta^{(1)},\\
    J^{(2)} &=J_B^{(2)}-\frac{1}{2} [\theta , {\cal D}_B\theta]^{(2)} , \qquad
    J^{(3)} ={\cal D}_B\theta^{(3)}  .
\end{aligned}
\end{equation}
In the bosonic truncation, we see that $J^{(1,3)}=0$ and thus   only the first term of \eqref{eq:SSSSM1} survives. It reproduces the  metric coupling of \eqref{eq:GS-curved-zeroeth}, after expanding $J_{B\alpha}= J_{B\mu}^A  \partial_\alpha X^\mu T_A $, as
\begin{equation}
    S^{(0)}_{\text{\tiny SSSSM}} 
    %= -\frac{T}{2}\int \de^2\sigma \mathrm{STr} \left(\sqrt{-h}h^{\alpha\beta} J_{\alpha}^{(2)} J_{\beta}^{(2)} \right)
    = -\frac{T}{2}\int \de^2\sigma \ \sqrt{-h}h^{\alpha\beta} \partial_\alpha X^\mu G_{\mu\nu}(X) \partial_\beta X^\nu ,
\end{equation}
with 
\begin{equation}\label{eq:metric_form_SSSSM}
    G_{\mu\nu}(X) = J_{B\mu}^{A_{(2)}} \kappa_{A_{(2)}B_{(2)}} J_{B\nu}^{B_{(2)}} \ .
\end{equation}

\begin{centering}
  \begin{tcolorbox}
    \begin{exercise}\label{ex:2-sphere-metric} 
    Let us turn off all the fermions and consider
    the $S^2$ example discussed in the previous section (starting from \eqref{eq:embedding-sphere}). Derive its metric by taking the (bosonic) parametrisation
      \begin{equation}
          g_B = e^{\phi T_1} e^{\theta T_2}e^{\xi T_3} .
      \end{equation}
      \end{exercise}
      \noindent \textbf{Solution}: The right-acting gauge symmetry can be used to remove $\xi$. You should find $J_B^{(2)} = \cos \theta d\phi T_1+d\theta T_2$ and
      \begin{equation}
          ds^2 \propto \cos^2\theta d\phi^2 + d\theta^2 \ .
        \end{equation}
  \end{tcolorbox}
\end{centering}

\noindent This kinetic term is precisely  the symmetric-space sigma-model (SSSM) action on $G^B/G^{(0)}$.
There is not a similar way, however, to reproduce the $B$-field coupling after bosonic truncation, as rather trivially terms such as 
\begin{equation}
    \epsilon^{\alpha\beta} \mathrm{STr} \left( J_\alpha^{(2)} J_\beta^{(2)} \right) \ ,
\end{equation}
vanish. This means that actions of the type  \eqref{eq:SSSSM1} which pairs only two of the  one-forms $J^{(1,2,3)}$ will not give rise to supergravity backgrounds with non-trivial NSNS fluxes. However, as we will discuss in section \ref{s:saskia}, $B$-field couplings can arise from considering non-trivial Wess-Zumino terms. This is rather important to cover the most general $AdS_3$ supergravity backgrounds, in particular those with mixed NSNS and RR fluxes.

Let us also point out that  terms such as
\begin{equation}
    \sqrt{-h}h^{\alpha\beta} \mathrm{STr} \left( J_\alpha^{(1)} J_\beta^{(3)} \right)
\end{equation}
are in principle allowed under the requirements of global $G$- and local $G^{(0)}$-invariance. However,  they  are ruled out by the observation that, after substituting \eqref{eq:mc-to-quadraticfermions}, they give rise to purely quadratic fermionic terms  which are not present in the standard GS superstring action, cf.~eqs.~\eqref{eq:flat-GS-part2} or \eqref{eq:GS-curved-quadratic}. 
%(there, the quadratic fermionic terms  are always accompanied with at least one bosonic field).

For completeness, let us  substitute \eqref{eq:mc-to-quadraticfermions} in the action \eqref{eq:SSSSM1} up to quartic order in fermions, giving
\begin{equation} \label{eq:exp-S-SSSSM}
    S_{\text{\tiny SSSSM}} = -\frac{T}{2}\int \de^2\sigma \ \mathrm{STr}\left(\sqrt{-h}h^{\alpha\beta} (J^{(2)}_{B\alpha}J^{(2)}_{B\beta} - J^{(2)}_{B\alpha} [\theta, {\cal D}_B \theta]_\beta^{(2)} ) -\varkappa\epsilon^{\alpha\beta} ({\cal D}_B\theta)_{\alpha}^{(1)}({\cal D}_B\theta)_\beta^{(3)} \right)  \ .
\end{equation}
Note that we have a term purely quadratic in fermions, but here it is a total derivative
\begin{equation}
    \int \de^2\sigma \ \mathrm{STr} ( \epsilon^{\alpha\beta} \partial_\alpha \theta^{(1)} \partial_\beta \theta^{(3)}) = \int \de^2\sigma \ \mathrm{STr} \partial_\alpha( \epsilon^{\alpha\beta}  \theta^{(1)} \partial_\beta \theta^{(3)}) \ .
\end{equation}
The remaining terms have the structure of the type II Green-Schwarz action in curved spaces. More precisely, however, depending on the particular supercoset this may be the case only after (partially) fixing the $\kappa$-symmetry gauge in the GS action. E.g.~the $\kappa$-gauge is not needed to obtain the supercoset for $AdS_5\times S^5$ \cite{Metsaev:1998it}, it should be partially fixed to obtain the supercoset for $AdS_4\times CP^3$ \cite{Arutyunov:2008if,Gomis:2008jt}, and it should be completely fixed to obtain the supercosets for $AdS_3\times S^3$ \cite{Babichenko:2009dk}.
The flat space limit is obtained by expanding around $g=1$ and can be shown to give the Green-Schwarz superstring in Minkowski space \cite{Metsaev:1998it}.\\
%\SIB{be more presice on flat spadce?}

\noindent {\bf Equations of motion} --- Our next goal is to write down the equations of motion of the SSSSM \eqref{eq:SSSSM1}. Using the $\mathbb{Z}_4$ invariance and the cyclicity of the supertrace, \textit{i.e.}~\eqref{eq:STrZ4invariant} and \eqref{eq:supertrace-cyclicity-adinvariant}, we can write the variation of the action for $\delta g$ as
\begin{equation}
    \delta S_{\text{\tiny SSSSM}} = -T \int \de^2\sigma \ \mathrm{STr}\left( \delta J_{\alpha} ( \sqrt{-h}h^{\alpha\beta} J^{(2)}_{\beta} + \frac{\varkappa}{2} \epsilon^{\alpha\beta}( J^{(1)}_{\beta} -  J^{(3)}_{\beta}) )\right) , 
\end{equation}
where, in terms of the actual field configurations, 
\begin{equation}
    \delta J_\alpha = \delta (g^{-1} \partial_\alpha g ) = - g^{-1} \delta g J_\alpha + g^{-1} \partial_\alpha (\delta g) ,
\end{equation}
and thus with 
\begin{equation}
    \Lambda^\alpha \equiv \sqrt{-h}h^{\alpha\beta} J^{(2)}_{\beta} + \frac{\varkappa}{2} \epsilon^{\alpha\beta} (J^{(1)}_{\beta} -J^{(3)}_{\beta}) ,
\end{equation}
we obtain after partial integration
\begin{equation}
    \delta S_{\text{\tiny SSSSM}} = T \int \de^2\sigma \ \mathrm{STr}\left( g^{-1} \delta g ([J_\alpha ,\Lambda^\alpha ]  + \partial_\alpha \Lambda^\alpha )\right) , 
\end{equation}
up to total derivative terms that vanish for periodic strings. Then, if we assume that the  superalgebra $\mathfrak{g}$ is such that the bilinear form  is non-degenerate, the   superstring equations of motion  read
\begin{equation} \label{eq:SSSSM1-eom}
    \partial_\alpha \Lambda^\alpha + [J_\alpha , \Lambda^\alpha] = 0 \ .
\end{equation}

\begin{centering}
\begin{tcolorbox}
\begin{exercise}
    Project the equations of motion \eqref{eq:SSSSM1-eom} on each of the $\mathbb{Z}_4$ graded components to show that the only non-trivial equations are
    \begin{equation} \label{eq:ssssm1-eoms}
        \begin{gathered}
        {\partial_\alpha (\sqrt{-h} h^{\alpha\beta} J_\beta^{(2)} ) + \sqrt{-h} h^{\alpha\beta} [J_\alpha^{(0)}, J_\beta^{(2)}] +\frac{\varkappa}{2} \epsilon^{\alpha\beta} \left(  [J_\alpha^{(1)} , J_\beta^{(1)}]- [J_\alpha^{(3)} , J_\beta^{(3)}] \right) = 0 ,} \\
        \sqrt{-h} h^{\alpha\beta}[J_\alpha^{(3)}, J_\beta^{(2)}] - \varkappa\epsilon^{\alpha\beta} [J_\alpha^{(2)}, J_\beta^{(3)}] =0, \\
        \sqrt{-h} h^{\alpha\beta}[J_\alpha^{(1)}, J_\beta^{(2)}] + \varkappa\epsilon^{\alpha\beta} [J_\alpha^{(2)}, J_\beta^{(1)}] =0 .
        \end{gathered}
    \end{equation}
\end{exercise}
\end{tcolorbox}
\end{centering}

\noindent The equations of motion for the  worldsheet metric $h^{\alpha\beta}$ corresponds, as usual, to the vanishing of the worldsheet energy-momentum tensor
\begin{equation}\label{eq:virasoro-ssssm1}
T_{\alpha\beta} = \mathrm{STr} \left( J_\alpha^{(2)} J_\beta^{(2)} \right) -\frac{1}{2} h_{\alpha\beta} h^{\gamma\delta} \mathrm{STr} \left( J_\gamma^{(2)} J_\delta^{(2)} \right) = 0 \ ,
\end{equation}
\textit{i.e.}~the Virasoro constraints,  cast in the  superalgebraic language. \\

\noindent{\bf Conservation law of the superisometry group} --- By Noether's theorem, the equations of motion \eqref{eq:SSSSM1-eom} are related to the conservation laws of global symmetries. Indeed, \eqref{eq:SSSSM1-eom} can be recast as 
\begin{equation}
    g^{-1} \partial_\alpha \left( g \Lambda^\alpha g^{-1} \right) g = 0
\end{equation}
and thus as a conservation equation for the current
\begin{equation}
    {\cal J}^\alpha \equiv g \Lambda^\alpha g^{-1} \ .
\end{equation}
This is precisely the Noether current for the left-acting global $G$ symmetry \eqref{eq:left-G-action}.

\begin{centering}
\begin{tcolorbox}
\begin{exercise}
    Apply Noether's theorem to show the above statement. Use that for a left-acting $G$ symmetry, the infinitesimal transformation of \eqref{eq:left-G-action} can be written as $\delta g = X^A T_A g$ when we paramterise $g_L = e^{X^A T_A}\in \mathfrak{g}$.  
\end{exercise}
\end{tcolorbox}
\end{centering}

\noindent  The associated Noether charges generating this symmetry are 
\begin{equation} \label{eq:noether-global-susy}
    Q = -T \int^{2\pi}_0 \de \sigma {\cal J}^\tau = -T \int^{2\pi}_0 \de \sigma \ g \left( \sqrt{-h} h^{\tau \alpha} J_{\alpha}^{(2)} + \frac{\varkappa}{2} (J_\sigma^{(1)} - J_\sigma^{(3)}) \right) g^{-1} \ .
\end{equation}
Its projections onto an element of $\mathfrak{g}$ with non-degenerate bilinear form is $Q_A = \mathrm{STr}(Q T_A)$.

\subsubsection{Kappa-symmetry}\label{sec:Sibylle:supercoset:kappa}
As we know from section \ref{sec:Sib:prereq:flatGS}, having a global superisometry group  is not enough to have   target space supersymmetry of the on-shell spectrum and thus a Green-Schwarz theory. The sigma-model action must have a  fermionic gauge symmetry in order to ensure the correct counting of bosonic and fermionic degrees of freedom. In this section, we discuss this local fermionic $\kappa$-symmetry  in the supercoset language for the canonical SSSSM action \eqref{eq:SSSSM1}.

Appropriate  $\kappa$-symmetry transformations can be obtained as  particular local right-actions on $g$ that depend only on fermionic parameters $\xi \in \mathfrak{g}^{(1)}\oplus \mathfrak{g}^{(3)}$ as \cite{McArthur:1999dy}\footnote{This action can be understood as an enlargement of the local right transformations by the bosonic stabilizer $G^{(0)}$  to a  subgroup including fermionic generators. }
\begin{equation} \label{eq:xi-kappa-trans}
    g \rightarrow g e^{\xi} , \qquad g^{-1}\delta g = \xi=\xi^{(1)} + \xi^{(3)} .
\end{equation} 
In this case one has $\delta Z^{{ M}} J_{{ M}}^{(2)}=0$ as for the generic superspace expression \eqref{eq:kappa-gs-superspace}.
Furthermore, this transformation clearly  commutes with the actions from the left and thus fixing a $\kappa$-gauge will be consistent with  global supersymmetry.
For arbitrary $\xi$, however, the action \eqref{eq:SSSSM1} will not be invariant under \eqref{eq:xi-kappa-trans}. First, as we know  from the flat space case, also the worldsheet metric will need to transform. Second, further conditions on $\xi$ are required to guarantee that $\delta S_{\text{\tiny SSSSM}}=0$. Lastly, the normalisation $\varkappa\in \mathbb{R}$ between the kinetic and Wess-Zumino  term must be fixed appropriately. An appropriate ansatz for the local $\kappa$-symmetry parameters $\xi$ is 
\begin{equation}
    \begin{gathered}
    \xi^{(1)} =  \left\{ J^{(2)}_{\alpha + },  \kappa^{(1)\alpha}_- \right\} , \qquad \xi^{(3)} =  \left\{ J^{(2)}_{\alpha - } ,\kappa^{(3)\alpha}_+ \right\} , 
    \end{gathered}
\end{equation}
where $\kappa^{(1)\alpha}_+=\kappa^{(3)\alpha}_-=0$ and we recall   the notation $A_\pm^\alpha = P_\pm^{\alpha\beta} A_\beta$ with $P_\pm^{\alpha\beta}$ defined in \eqref{eq:worldsheet-projectors}.
The parameters have the correct grading $\xi^{(i)}\in \mathfrak{g}^{(i)}$ when the $\mathbb{Z}_4$ acts on products of matrices as $\Omega (MN) = \pm \Omega(N)\Omega(M)$. This is the case in all relevant gradings in the literature. In particular, note that  $\Omega$ is then readily an automorphism. To compensate the transformation of $g^{-1}\delta g$, the variation of the worldsheet metric can be found when the parameter $\varkappa$ is fixed to $\pm 1$. Interestingly, as we will see in section \ref{s:lax-ssssm},  the same values for $\varkappa$ are required for classical integrability. After some algebra, making   use of the identities for the worldsheet projectors, the variation of the metric can then be written as\footnote{For more clues on how to proof this we refer to \cite{Arutyunov:2009ga}, however one should be careful with changes in conventions (in particular $J\rightarrow -J$,  $\epsilon^{\alpha\beta}\rightarrow - \epsilon^{\alpha\beta}$ and  $P_\pm^{\alpha\beta} \rightarrow (-h)^{-1/2} P_\mp^{\alpha\beta}$).}
\begin{equation}\label{eq:kappa_var_metric}
    \delta \left( \sqrt{-h}h^{\alpha\beta} \right) \kappa_{A_{(2)} B_{(2)}} = -8 \sqrt{-h}  \mathrm{STr}\left( T_{A_{(2)}} (  [J_-^{(1)\alpha}, \kappa_-^{(1)\beta}] +  [J_+^{(3)\alpha}, \kappa_+^{(3)\beta}])T_{B_{(2)}} \right) \ .
\end{equation}
 More useful expressions in terms of the Maurer-Cartan currents can be obtained once a particular superalgebra is chosen, see e.g.~\cite{Grigoriev:2007bu,Arutyunov:2008if}, which consider $\mathfrak{psl}(n|n)$ and $\mathfrak{osp}(2,2|6)$ respectively. Instead, in the remaining of this section we will derive a generic formula to count the rank of $\kappa$-symmetry of the canonical SSSSM for generic semi-symmetric $G/G^{(0)}$. \\
%\begin{equation}
%\begin{gathered}
%    \delta {\phi}^{\tilde{M}} J_{\tilde{M}}^{(2)}= 0, \qquad \delta {\phi}^{\tilde{M}} J_{\tilde{M}}^{(I)}=\xi^{(I)} = 2 J_\alpha^{(2)}  \kappa^{\alpha I} , \\
%    \delta ( \sqrt{-h}h^{\alpha\beta})= -16 i \sqrt{-h} \left(\overline{J}_{-}^{(1)\alpha} \kappa^{(1)\beta} +\overline{J}_{+}^{(3)\alpha} \kappa^{(3)\beta} \right)
%\end{gathered}
%\end{equation}
%for $(I)=(1),(3)$ (to be compared with $I=1,2$ of section \ref{s:gs-flat}) and $\kappa_+^{(1)\alpha} =\kappa_-^{(3)\alpha}= 0$. Invariance of the action then fixes $\varkappa=\pm 1$ Although we use here the notation for the $\mathbb{Z}_4$ grading, it is not clear the $\epsilon^{(1,3)} \in \mathfrak{g}^{(1,3)}$. This of course depends on the automorphism $\Omega$ used. }
%
%$\varkappa$ is fixed to $+ 1$ and the transformations read\footnote{Equivalently, one can fix $\varkappa =-1$ up to minor changes on the transformation rules. }
%\begin{equation}
%    \begin{gathered}
%    \xi^{(1)} = i \left\{ J^{(2)}_{\alpha + },  \kappa^{(1)\alpha}_- \right\} , \qquad \xi^{(3)} = i \left\{ J^{(2)}_{\alpha - } ,\kappa^{(3)\alpha}_+ \right\} , \\
%    \delta ( \sqrt{-h} h^{\alpha\beta} ) = 
%    \end{gathered}
%\end{equation}
%
%
%\SIB{add motivation why its useful to first study the rank}

\noindent {\bf On-shell rank of $\kappa$-symmetry} ---  We will analyse the presence of $\kappa$-symmetry, as well as its on-shell rank, by doing a semi-classical analysis of fermionic fluctuations around a bosonic background solution following \cite{Zarembo:2010sg} (see also \cite{Vicedo:2010qd}).

In particular, consider again the parametrisation \eqref{eq:g-parametrisation} of the group element $g$. We will now take $g_B = \bar{g}_B(\sigma)$ as a bosonic background solution around which we will  analyse  fermionic fluctuations $\theta = \epsilon \hat{\theta}$ with $\epsilon$ a small parameter. Then \eqref{eq:exp-mc-form} becomes
\begin{equation} 
    J = \bar{J} + \epsilon \bar{{\cal D}} \hat{\theta} - \frac{\epsilon^2}{2} [\hat{\theta}, \bar{{\cal D}} \hat{\theta}] + {\cal O}(\epsilon^3) , 
\end{equation}
where $\bar{J} = \bar{g}_B^{-1} d \bar{g}_B$ and $\bar{\cal D} = d + \mathrm{ad}_{\bar{J}}$ satisfies the classical equations of motion \eqref{eq:ssssm1-eoms}, which can be written in this case as
\begin{equation} \label{eq:bg-eom}
    \bar{\nabla}_\alpha \bar{J}^{\alpha(2)} = 0 ,
\end{equation}
where 
\begin{equation}
    \bar{\nabla}_\alpha A^\beta = \bar{D}_\alpha A^\beta  + \Gamma_{\alpha\gamma}^\beta A^\gamma , \qquad \bar{D} = d + \mathrm{ad}_{\bar{J}^{(0)}}  \ ,
\end{equation}
and $\Gamma_{\alpha\gamma}^\beta$ is the usual Christoffel symbol. 
In addition, we have the Maurer-Cartan identities for $\bar{J}$ projected on $\mathfrak{g}^{(0)}$ and $\mathfrak{g}^{(2)}$. They can be written respectively as
\begin{equation} \label{eq:mc-0-2-identities}
\begin{gathered}
    \bar{F}_{\alpha\beta} + \left[ \bar{J}_\alpha^{(2)}, \bar{J}_\beta^{(2)} \right]=0 , \qquad
\bar{D}_\alpha {\bar J}_\beta^{(2)} - \bar{D}_\beta {\bar J}_\alpha^{(2)} =0   .
\end{gathered}
\end{equation}

\begin{centering}
\begin{tcolorbox}
\begin{exercise}
    Show that the background equations of motion \eqref{eq:bg-eom} and the  identities \eqref{eq:mc-0-2-identities} imply that
    \begin{equation}
        \begin{aligned}
            &[\bar{D}_\alpha , \sqrt{-h}h^{\alpha\beta} \mathrm{ad}_{\bar{J}_\beta^{(2)}}] = 0 , \\
            &\epsilon^{\alpha\beta} \bar{D}_\alpha \bar{D}_\beta = -\epsilon^{\alpha\beta} \mathrm{ad}_{\bar{J}_\alpha^{(2)}}\mathrm{ad}_{\bar{J}_\beta^{(2)}} , \\
            &\epsilon^{\alpha\beta} [\bar{D}_\alpha , \mathrm{ad}_{{\bar{J}_\beta^{(2)}}} ] = 0, 
            \end{aligned}
    \end{equation}
\end{exercise}
\end{tcolorbox}
\end{centering}

\noindent Using the above identities, as well as the properties \eqref{eq:ssscommrel}, \eqref{eq:supertrace-cyclicity-adinvariant}, and those related to the worldsheet projectors \eqref{eq:worldsheet-projectors},\footnote{Particularly important are the orthogonality properties of the worldsheet projectors.}  the expansion of the action \eqref{eq:exp-S-SSSSM} around $\bar{g}_B$  can be written for $\varkappa=1$ as
\begin{equation} \label{eq:sc-f-ssssm}
\begin{aligned}
    S_{\text{\tiny SSSSM}} = \bar{S} - \epsilon^2 T \int \de^2\sigma \ \sqrt{-h}h^{\alpha\beta} \mathrm{STr} &\left( \hat{\theta}^{(1)} \bar{\nabla}_{\alpha -} [\bar{J}^{(2)}_{\alpha +},\hat{\theta}^{(1)}] + \hat{\theta}^{(3)} \bar{\nabla}_{\alpha +} [\bar{J}^{(2)}_{\alpha -},\hat{\theta}^{(3)}] \right. \\&\qquad \left.  -2 [\bar{J}^{(2)}_{\alpha +},\hat{\theta}^{(1)}] [\bar{J}^{(2)}_{\beta -},\hat{\theta^{(3)}}] \right)  + {\cal O}(\epsilon^3) ,
\end{aligned}
\end{equation}
with $\bar{S} = -\frac{T}{2}\int \de^2\sigma \ \sqrt{-h}h^{\alpha\beta} \mathrm{STr} (\bar{J}^{(2)}_\alpha \bar{J}^{(2)}_\beta)$,  subjected to the background Virasoro constraints \eqref{eq:virasoro-ssssm1}. Using the worldsheet projectors, the latter can be written similarly as in \eqref{eq:virasoro-gs-flat2} but read now
\begin{equation}
    \mathrm{STr} \left( \bar{J}^{(2)}_{\alpha\pm}\bar{J}^{(2)}_{\beta\pm} \right) = 0 \ .
\end{equation}
This shows that the currents $\bar{J}^{(2)}_{\alpha\pm}$ are null.

We thus see that the semi-classical action \eqref{eq:sc-f-ssssm} for the fermionic fluctuations  would degenerate when  the currents $\bar{J}^{(2)}_{\alpha +}$ ($\bar{J}^{(2)}_{\alpha -}$) have a vanishing commutator in some of the $\mathfrak{g}^{(0)}$ ($\mathfrak{g}^{(3)}$) directions:
%depends on the fermionic fluctuations $\hat{\theta}^{(1)}$ and $\hat{\theta}^{(3)}$ only in the combinations $[\bar{J}^{(2)}_{\alpha +},\hat{\theta}^{(1)}]$ and $[\bar{J}^{(2)}_{\alpha -},\hat{\theta}^{(3)}]$. The Lagrangian for the fluctuation would thus degenerate when  the currents $\bar{J}^{(2)}_{\alpha +}$ ($\bar{J}^{(2)}_{\alpha -}$) have a vanishing commutator in some of the $\mathfrak{g}^{(0)}$ ($\mathfrak{g}^{(3)}$) directions. 
the corresponding fermionic fluctuations in $\hat{\theta}^{(1)}$ ($\hat{\theta}^{(3)}$) decouple and  do not contribute to the semi-classical dynamics. This is precisely the physical consequence of $\kappa$-symmetry, a gauge symmetry which renders some fermionic degrees of freedom unphysical. The simplest way to fix the $\kappa$-gauge is to set those components of $\hat{\theta}^{(1)}$ ($\hat{\theta}^{(3)}$) to zero that do not contribute to the semi-classical action. 
%They are precisely the components proportional to the directions of $\mathfrak{g}^{(0)}$ ($\mathfrak{g}^{(3)}$) that commute with the  (sufficiently generic) background currents $\bar{J}^{(2)}_{\alpha +}$ ($\bar{J}^{(2)}_{\alpha -}$). 
The on-shell rank of the $\kappa$-symmetry is thus
\begin{equation}
    N_+ =  \mathrm{dim} \ \mathrm{ker} \ \mathrm{ad}_{\bar{J}^{(2)}_{\alpha +}} \vert_{\mathrm{g}^{(1)}}  , \qquad N_- =  \mathrm{dim} \ \mathrm{ker} \ \mathrm{ad}_{\bar{J}^{(2)}_{\alpha -}} \vert_{\mathrm{g}^{(3)}} \ ,
\end{equation}
and  highly depends on the superalgebra $\mathfrak{g}$ under consideration. For a systematic analysis of their values see \cite{Zarembo:2010sg}. Note that  the numbers $N_\pm $ must be  independent of the background solution $\bar{g}_B$: they must be determined only through the structure constants of $\mathfrak{g}$ and should not depend on the particular evaluation of the background currents $\bar{J}^{(2)}_{\alpha\pm}$. In other words, we consider $\bar{J}^{(2)}_{\alpha\pm}$ to be sufficiently generic null elements of $\mathfrak{g}^{(2)}$. However, for some special (singular) solutions the $\kappa$-rank can become larger. 

%This concludes our discussion of $\kappa$-symmetry for the canonical semi-symmetric space sigma-model \eqref{eq:SSSSM1}.

\subsubsection{Conformal GS sigma-models on semi-symmetric spaces}
%\SIB{Recap of https://arxiv.org/pdf/1003.0465.pdf}
Certain superalgebras have the interesting property that their quadratic Casimir $c_2 (G)$, defined as $\mathrm{STr}_{\mathrm{adj}} (MN) = c_2(G) \mathrm{STr}(MN)$, vanishes. For the SSSSM, the dependence of its one-loop worldsheet $\beta$-function on the supergroup $G$ is in fact only through the number $c_2(G)$, which simply determines the rate of the flow (see e.g.~\cite{Zarembo:2010sg,Zarembo:2017muf}).  In other words,  a  coupling constant $\lambda$
(obtained from the Taylor expansion of one of the components of e.g.~the curved metric $G_{\mu\nu}$) would run as
\begin{equation}\label{eq:beta_fct_GS}
    \beta (\lambda) = \frac{d}{d\log\mu} \lambda(\mu) = \alpha' c_2(G) f(\lambda) + {\cal O}(\alpha'^2) \ ,
\end{equation}
with $\mu$ the momentum cut-off scale and $f(\lambda)$ some function of $\lambda$. If $c_2(G)=0$, the coupling does not run and thus the worldsheet sigma-model is conformal to ${\cal O}(\alpha')$. This is the case for $\mathfrak{psl}(n|n)$, $\mathfrak{osp}(2n+2|2n)$, and $\mathfrak{d}(2,1;\alpha)$ \cite{KAC19778,Kac:1977em,Frappat:1996pb} (see also \cite{Bershadsky:1999hk,Berkovits:1999im} for an exact result for $\mathfrak{g} = \mathfrak{psu}(n|n)$). For more details, we refer to \cite{Zarembo:2010sg} which also analyses the correct counting of the central charge due to fixing conformal and $\kappa$-gauge. An important conclusion of that work is that the list of SSSSMs of the type \eqref{eq:SSSSM1} giving rise to consistent string backgrounds up to one-loop in $\alpha'$ is not very long (see §5 of \cite{Zarembo:2010sg}). Important examples are $PSU(2,2|4)/SO(1,4)\times SO(5)$ for $AdS_5\times S^5$, $OSp(6|4)/U(3)\times SO(3,1)$ for $AdS_4\times CP^3$, $PSU(1,1|2)\times PSU(1,1|2)/SU(1,1)\times SU(2)$ for $AdS_3\times S^3$ and $D(2,1;\alpha)\times D(2,1;\alpha)/SO(4)\times SL(2,R)$ for $AdS_3\times S^3\times S^3$, of which the latter two will be discussed in section \ref{s:saskia}. 

\subsection{Classical  integrability}\label{sec:sib:integrability}
In this section, we will show that the canonical SSSSM \eqref{eq:SSSSM1} is classically integrable. In general, for field theories, the precise definition of classical integrability is  subtle but commonly given in terms of the existence of an infinite tower of conserved charges. This property severely constrains the dynamics and essentially provides a large toolkit of mathematical techniques to solve the theory exactly.
We will first introduce the general principles of classical integrablity for two-dimensional field theories in terms of the so-called Lax integrability and then  show how the action \eqref{eq:SSSSM1} falls under this umbrella. For simplicity, we will do this first for a simpler cousin of the SSSSM, the Principal Chiral Model (PCM), which has a trivial $G^{(0)}$, and generalise the proof later to the canonical SSSSM.

\subsubsection{Preamble: classical Lax integrability in two-dimensional field theories}

A finite-dimensional classical system is classically integrable when \textit{(i)} the number of independent conserved charges $Q_i$ equals the number of degrees of freedom $n$, and  \textit{(ii)} these charges are all in involution (Poisson commute), \textit{i.e.}~$\{Q_i, Q_j\}_{\text{\tiny P.B.}} =0$ for all $i,j=1,\ldots, n$. A convenient way to formulate this is in terms of Lax pairs and $r$-matrices who have to represent the equations of motion and the Poisson brackets of the theory in a particular way. For more details, we refer e.g.~to the lecture notes \cite{Driezen:2021cpd} of one of the  authors or the book \cite{Babelon:2003qtg}. 
Similarly, for two-dimensional field theories, which have an infinite number of degrees of freedom,   Lax integrability is understood as the possibility of recasting the equations of motion in a certain way: In this case,  as a zero-curvature or flatness condition of a one-form ${\cal L}(z)$ (called the Lax connection) which besides the field configurations must depend on a free parameter $z\in \mathbb{C}$ (called the spectral parameter).\footnote{In what follows, we will suppress the dependence on the field configurations.} This is also known as the ``zero-curvature formulation'' or ``zero-curvature Lax representation''. 
In form language, this condition reads
\begin{equation} \label{eq:zero-curvature-form}
    d{\cal L}(z) + {\cal L}(z)\wedge {\cal L}(z) = 0 \ ,
\end{equation}
and it should hold $\forall z \in \mathbb{C}$. 
In terms of coordinates $\sigma^\alpha = (\tau, \sigma)$ on the two-dimensional spacetime $\Sigma$ (for us the worldsheet, as before) we have ${\cal L}(z) = {\cal L}_\tau (z) \de\tau + {\cal L}_\sigma (z) \de \sigma$ and then \eqref{eq:zero-curvature-form} simply becomes
\begin{equation}\label{eq:zero-curvature-coordinates}
\partial_\tau {\cal L}_\sigma (z) - \partial_\sigma {\cal L}_\tau (z) + [{\cal L}_\tau (z),{\cal L}_\sigma (z)] = 0 , \qquad \forall z\in\mathbb{C} \ .
\end{equation}
 The ${\cal L}_\alpha (z)$  typically take values in a finite-dimensional matrix representation of a  non-abelian Lie algebra $\mathfrak{g}$ (or more precisely, due to the freedom in $z$, in the loop algebra $\mathfrak{g}\otimes \mathbb{C}$). \\

An important remark is that the Lax connection is not unique. Besides depending on the  representation, 
%Firstly, the dimension of the matrices ${\cal L}_\alpha(z)$ can be different for different zero-curvature representations. Secondly, and more importantly,
${\cal L}(z)$ is defined up to local ``gauge'' transformations acting as
\begin{equation}\label{eq:lax-gauge-transformation}
    {\cal L} (z) \rightarrow {\cal L}^g(z) = g {\cal L}(z) g^{-1} - d g g^{-1} \ ,
\end{equation}
which leave the zero-curvature condition \eqref{eq:zero-curvature-form} invariant. The objects $g$ are matrix elements of the same dimension as ${\cal L} (z)$ but can be completely arbitrary; they can depend on the dynamical variables as well as  the spectral parameter $z$. \\

\noindent{\bf Conserved charges} --- The freedom in the parameter $z$ in the zero-curvature condition is crucial and in fact allows us to construct an infinite tower of conserved charges. To show this, let us first note that \eqref{eq:zero-curvature-coordinates} corresponds to a compatibility condition of the following auxiliary linear system\footnote{This can be seen by taking the derivative to $\tau$ of the first equation, and substracting it from the derivative to $\sigma$ of the second equation, which implies a consistency condition given precisely by \eqref{eq:zero-curvature-coordinates}.}
\begin{equation}
    (\partial_\sigma + {\cal L}_\sigma (z) ) \Psi (z) = 0, \qquad (\partial_\tau + {\cal L}_\tau (z)) \Psi = 0 ,
\end{equation}
where $\Psi$ is sometimes called the ``wave-function''. 
Fixing now the initial condition
\begin{equation}
\Psi (0,0 ;z) = 1\,,  
\end{equation}
the solution to the linear problem is obtained by parallel transportation from the origin to a point $(\tau, \sigma)$ along an arbitrary path $\gamma$ with the connection ${\cal L}(z)$. In other words,
\begin{equation}
    \Psi(z) = \Psi(\tau, \sigma ; z) = \overleftarrow{P \exp}  \left( - \int_\gamma {\cal L}(z) \right) \ ,
\end{equation}
which is  well-defined (it does not depend on the chosen path $\gamma$)
due to the zero-curvature  property \eqref{eq:zero-curvature-form} of the Lax connection. 
Here $ \overleftarrow{P \exp}$ is  the path ordered exponential defined on a fixed time slice  as
\begin{equation}
\begin{aligned}
 \overleftarrow{P \exp} \left( \int_0^\sigma \de \sigma' \, A(\sigma')  \right)  &= \sum_{k = 0}^\infty \frac{1}{k!} \int^\sigma_0 \cdots \int^\sigma_0   \overleftarrow{P } \{ A(\sigma_1') \cdots A(\sigma_k') \} \, \de \sigma_1' \cdots \de \sigma_k' \\
 &=  \sum_{k = 0}^\infty \int^\sigma_0 \de \sigma_k' \int^{\sigma_k'}_0 \de \sigma_{k-1}'  \cdots \int^{\sigma_2'}_0 \de \sigma_1'  \ A(\sigma_k') \cdots A(\sigma_1')   \, . 
\end{aligned}
\end{equation}
An infinite set of conserved charges can now be obtained by considering a path at a fixed time slice to define the \textit{transport matrix} $\Omega(b,a;z) \equiv \Psi (\tau , b ; z) \Psi (\tau , a ; z)^{-1}$, \textit{i.e.}
\begin{equation} \label{eq:monodromy}
\Omega(b,a; z ) = \overleftarrow{P \exp} \left(- \int^b_a \mathrm{d}\sigma\; \mathcal{L}_\sigma ( z) \right)\, .
\end{equation}

\begin{centering}
\begin{tcolorbox}
\begin{exercise}
Write out the first three terms of the the expansion of $\Omega(b,a;z)$. Note that in general this is a highly non-local expression.
\end{exercise}  
\end{tcolorbox}
\end{centering}

\noindent The transport matrix satisfies the following  properties
\begin{align}
\delta \Omega(b,a ;z) &= - \int^b_a \mathrm{d}\sigma\, \Omega(b,\sigma;z) \delta \mathcal{L}_\sigma (\tau, \sigma ; z) \Omega(\sigma, a;z)\, ,\\
\partial_\sigma \Omega(\sigma,a ;z) &= - \mathcal{L}_\sigma (\tau,\sigma ;z) \Omega(\sigma, a;z)\, , \label{eq:Transport2}\\
\partial_\sigma \Omega(b,\sigma ;z) &= \Omega(b, \sigma ;z)  \mathcal{L}_\sigma (\tau,\sigma ;z)\, , \label{eq:Transport3}\\
\Omega(a,a ;z ) &=1\,  \label{eq:Transport4},
\end{align}
Using \eqref{eq:zero-curvature-coordinates} together with the above properties, one can  show that
\begin{equation}\label{eq:MonToTime}
\begin{aligned}
\partial_\tau \Omega(b,a ;z) 
%&=- \int^b_a \mathrm{d}\sigma\, T(b,\sigma;z) \partial_\tau \mathcal{L}_\sigma (\tau, \sigma ; z) T(\sigma, a;z)\, ,\\
%&=- \int^b_a \mathrm{d}\sigma\, T(b,\sigma;z)\left( \partial_\sigma \mathcal{L}_\tau (\tau, \sigma ; z)  - [\mathcal{L}_\tau (\tau, \sigma ; z) , \mathcal{L}_\sigma (\tau, \sigma ; z)]  \right)T(\sigma, a;z)\, ,\\
%&= - \int^b_a \de \sigma \partial_\sigma \left[ \overleftarrow{P \exp} \left( \int^b_s ds {\cal L}_\sigma (s, t;z) \right) \mathcal{L}_\tau (\tau, \sigma ; z)  \overleftarrow{P \exp} \left( \int^s_a ds  {\cal L}_\sigma (s, t;z)  \right)  \right] , \\
&= \Omega(b,a ;z) \mathcal{L}_\tau (\tau, a;z)  - \mathcal{L}_\tau (\tau, b ;z) \Omega(b,a;z) \,  .
\end{aligned}
\end{equation}

\begin{centering}
\begin{tcolorbox}
\begin{exercise}
Show  \eqref{eq:MonToTime}.
\end{exercise}  
\end{tcolorbox}
\end{centering}

\noindent If we still assume that $\Sigma$ has the topology of a cylinder and could thus describe the closed string worldsheet, then ${\cal L}(\tau, \sigma ;z) = {\cal L}(\tau, \sigma + R ;z)$ and
\begin{equation}
    \partial_\tau \Omega (R , 0 ;z) = [\Omega (R , 0;z) , {\cal L}_\tau (\tau, 0 ; z)] \ .
\end{equation}
Therefore, defining the \textit{monodromy matrix} $\Omega(z) \equiv \Omega(R , 0;z)$,  its trace as well as the trace of its powers is conserved, \textit{i.e.}
\begin{equation} \label{eq:monodromy-conservation}
    \partial_\tau \mathrm{Tr} \Omega (z)^n = 0 , \qquad \forall n\in \mathbb{N}, \qquad \forall z\in\mathbb{C} \ .
\end{equation}
Taylor expanding $\mathrm{Tr}\Omega(z)$ around any suitable value $z_\star$ in the complex plane in which the monodromy is analytic thus produces an infinite set of conserved charges. These charges can be both local and non-local.\footnote{In fact, changing the value of $z_\star$, one can obtain several of those infinite sets of charges, possibly with different properties.} Another  way of looking at this is that \eqref{eq:monodromy-conservation} implies that the eigenvalues $\lambda(z)$ of $\Omega(z)$, defined by the characteristic equation
\begin{equation} \label{eq:characteristic-equation}
    \Gamma  = \mathrm{det} \left(\Omega(z) - \lambda(z) \mathbb{1} \right) = 0 ,
\end{equation}
are conserved. This is also apparent from the fact that under the gauge transformation \eqref{eq:lax-gauge-transformation} the transport matrix transforms as
\begin{equation}
    \Omega(b,a;z) \rightarrow \Omega^g(b,a; z) = g(\tau, b) \Omega(b,a;z) g^{-1}(\tau, a) ,
\end{equation}
which can be used to diagonalise $\Omega(z)$.
Hence, equivalently, Taylor expanding $\lambda(z)$ produces an infinite tower of  conserved charges.
The spectral properties of the theory are thus encoded in the spectral properties of $\Omega(z)$. Therefore, the characteristic equation \eqref{eq:characteristic-equation}, which defines an algebraic curve in $z\in\mathbb{C}$, is also called the classical spectral curve (CSC). In an $N\times N$ matrix representation for ${\cal L}(z)$, the CSC is a polynomial equation in $\lambda(z)$ of degree $N$, and thus gives rise to an $N$-sheeted Riemann surface. The sheets may degenerate at certain branch points in $z$, whose number and location depends on the degree and form of the polynomial $\Omega(z)$ in $z$, which in turn depends on the particular form of  ${\cal L}_\sigma (z)$ on thus the particular solution of the equations of motion. From a different point of view, this data can be used to classify families of such solutions (see e.g.~\cite{Dorey:2006zj,Vicedo:2008ryn}).\\

\textit{Remark: local charges} --- An interesting set of conserved charges is obtained by expanding the monodromy matrix around poles of the Lax connection. One can show that in the vicinity of each pole $z_k$ of ${\cal L}(z)$ one can perform a gauge transformation that diagonalises the Lax connection. Consequently, the path-ordered exponential of the gauge-transformed monodromy matrix becomes a normal exponential and thus the Taylor expansion of its eigenvalues will exhibit an infinite set of \textit{local} conserved charges. This generic procedure is also known as abelianization (see e.g.~\cite{Babelon:2003qtg}). In the CSC, on the other hand, local charges only appear in the leading order of the expansion of $\lambda(z)$ around a value $z_\star$ in which ${\cal L}_\sigma(z_\star)$ or ${\cal L}^g_\sigma(z_\star)$ vanishes (see also comments in the explicit examples discussed later).  \\

\noindent {\bf Comments on involution of charges} --- While the zero-curvature formulation only ensures us that we have infinite towers of conserved charges, and does not ensure anything about their involution, possessing the structures
\begin{equation}
\{ {\cal L}(z) , \ \Psi (z) , \ T(z) \} \ ,
\end{equation}
already  opens-up interesting and well-known classical integrable techniques (such as e.g.~the CSC, also known as the finite-gap integration technique). However, at the quantum level the knowledge of the Poisson bracket algebra and involutivity of the charges plays a crucial role: it connects to factorisation  of scattering into $2\rightarrow 2 $ elastic scattering processes and underlies  Bethe Ans\"atze techniques. 
Unfortunately, we do not have the space-time to discuss the necessary Poisson-bracket structure here. Instead, let us e.g.~refer to §3.3 of \cite{Driezen:2021cpd} in general and, as for the discussion on the Poisson brackets for the canonical SSSSM representing $AdS_5\times S^5$,  to Ch.II.3 §2.3 of \cite{Beisert:2010jr}. For scattering matrices and Bethe Ans\"atze techniques, we refer to the coming  sections \ref{s:fiona} and \ref{ana:sec} respectively.

\subsubsection{Lax formulation of the Principal Chiral Model}

Before deriving the Lax formulation of the  semi-symmetric space sigma-model, we warm up with a simpler model: the Principal Chiral Model (PCM).
This is a non-linear sigma-model to a bosonic Lie group manifold $G$, whose action can be written as
\begin{equation}
    S_{\text{\tiny PCM}} = -\frac{T}{2}\int_\Sigma \de^2\sigma \sqrt{-h}h^{\alpha\beta}\mathrm{Tr} \left( J_\alpha J_\beta \right) , 
\end{equation}
with $J=g^{-1}dg$ still defined as before through a Lie group element $g$ that is a map from the worldsheet $\Sigma$ to $G$. Note that the PCM can be seen as a simpler version of the SSSSM \eqref{eq:SSSSM1} for which fermionic fields have been set to zero, and for which the subgroup $G^{(0)}$ is trivial.

The PCM action enjoys a number of interesting properties. It has a global $G_L\times G_R$ invariance, which acts independently as
\begin{equation}
    g \rightarrow g_L g , \qquad g \rightarrow g g_R ,
\end{equation}
for $g_L, g_R \in G$ constant elements on the worldsheet. The conservation laws of the corresponding Noether currents (${\cal J}_L^\alpha = \sqrt{-h} h^{\alpha\beta} g J_\beta g^{-1}$ and ${\cal J}_R^\alpha=\sqrt{-h} h^{\alpha\beta}J_\beta$ respectively) coincide with the equations of motion for $g$, which read 
\begin{equation} \label{eq:eom-pcm}
    \partial_\alpha \left( \sqrt{-h} h^{\alpha\beta} J_\beta \right) =0.
\end{equation}
In form language, this can be rewritten simply as 
\begin{equation} \label{eq:eom-pcm-form}
    d\star J = 0 ,
\end{equation}
where we define the action of the Hodge star on one-forms living on a two-dimensional space $\Sigma$ as
\begin{equation}
    \star J = J_\alpha \star \de \sigma^\alpha =  J_\alpha\epsilon^{\alpha\beta} \gamma_{\beta\gamma}   \de \sigma^\gamma \ ,  
\end{equation}
and we introduced $\gamma^{\alpha\beta} \equiv \sqrt{-h} h^{\alpha\beta}$ and its inverse $\gamma_{\alpha\beta} \equiv (-h)^{-1/2} h_{\alpha\beta}$.
The equations of motion for $h^{\alpha\beta}$ (the Virasoro constraints) can be written as usual
\begin{equation}
    \mathrm{Tr} (J_{\alpha\pm} J_{\beta\pm } ) = 0 ,
\end{equation}
using  the worldsheet projectors $\eqref{eq:worldsheet-projectors}$.

\begin{centering}
\begin{tcolorbox}
\begin{exercise} \label{ex:hodge-star}
    Show the equivalence between \eqref{eq:eom-pcm} and \eqref{eq:eom-pcm-form}. Show that the Hodge-star action on one-forms satisfies $\star^2 J = J$ and $\star J_1 \wedge J_2 = - J_1\wedge \star J_2$.
\end{exercise} 
{\footnotesize Hint: recall the identity $\epsilon^{\alpha\beta}\epsilon^{\gamma\delta} = -h (h^{\alpha\delta}h^{\beta\gamma} - h^{\alpha\gamma}h^{\beta\delta})$.}
\end{tcolorbox}
\end{centering}

An important observation to obtain the Lax formulation of the PCM is that, using the Maurer-Cartan identity \eqref{eq:mc-identity-1}, we can relax the identification of the current $J$ with the Maurer-Cartan one-form formulated in terms of the field $g$. That is to say, rather than viewing $g$ as the fundamental field satisfying a single second-order differential equation \eqref{eq:eom-pcm-form}, we can view $J$ as the fundamental field whose dynamics is captured by two first-order equations, namely the conservation law \eqref{eq:eom-pcm-form} and the Maurer-Cartan identity \eqref{eq:mc-identity-1}. To travel between the two pictures one can use the pure-gauge condition for the flat current, \textit{i.e.}~the identification $J=g^{-1}dg$. Proving Lax integrability is now very convenient in the $J$-picture. Indeed, the existence of a conserved current that is also flat immediately guarantees  a Lax formulation.
%\footnote{See e.g.~also the lecture notes \cite{Hoare:2021dix}, where conserved flat currents are identified, and correspondingly the Lax formulation, for a large class of integrable deformations of sigma-models.} 
Let us show explicitly why this is. Recall that we need a Lax connection ${\cal L}(z)$ that on-shell has zero curvature  for every value of an arbitrary parameter $z\in\mathbb{C}$, \textit{i.e.}~it must satisfy \eqref{eq:zero-curvature-form}. To obtain this, let us combine the two equations \eqref{eq:eom-pcm-form} and \eqref{eq:mc-identity-1} for $J$ by considering the object
\begin{equation}
    {\cal L}(\alpha, \beta) = \alpha J + \beta \star J ,
\end{equation}
where $\alpha, \beta \in \mathbb{C}$ are constants that we will fix such that \eqref{eq:zero-curvature-form} is satisfied on-shell. This system can not be fully determined: there should be at least one redundancy in the parameters  indicating the existence of a free  parameter $z$. Using \eqref{eq:eom-pcm-form} and \eqref{eq:mc-identity-1} we find
\begin{equation}
\mathrm{d} {\cal L}(\alpha, \beta ) + {\cal L}(\alpha, \beta ) \wedge {\cal L}(\alpha, \beta ) = \left( \alpha^2 - \beta^2 -\alpha \right) J \wedge J .
\end{equation}
Hence, we can indeed  solve \eqref{eq:zero-curvature-form} with a single redundancy (we have one equation,  $\alpha^2 - \beta^2 -\alpha \overset{!}{=} 0$, for two variables). Taking $\alpha = \frac{1}{1-z^2}$ and $\beta = \frac{z}{1-z^2}$  the Lax connection  reads 
\begin{equation} \label{eq:lax-pcm}
{\cal L}(z) = \frac{J + z \star J}{1- z^2} \ .
\end{equation}
Its zero-curvature condition \eqref{eq:zero-curvature-form} now implies the equations \eqref{eq:eom-pcm-form} and \eqref{eq:mc-identity-1} for any value $z \in \mathbb{C}$. 

Before generalising this to the SSSSM, let us show the interesting relation between the Lax connection and the generators of global Noether symmetries. In conformal gauge \eqref{eq:conformal-gauge}, the spatial component of the Lax reads
\begin{equation}
    {\cal L}_\sigma (z) = \frac{J_\sigma + z J_\tau}{1-z^2} \ .
\end{equation}
Around $z\sim \infty$, ${\cal L}_\sigma (z)$ thus behaves as ${\cal L}_\sigma (z)= - J_\tau/z + {\cal O}(z^{-2})$.
 Taylor expanding the monodromy matrix around $z\sim \infty$ thus gives
\begin{equation}
    \Omega(z) = 1 + z^{-1} Q_R + {\cal O}(z^{-2}), \qquad Q_R = \int^{R}_0 \de \sigma J_\tau ,
\end{equation}
where $Q_R$ is precisely the (local) Noether charge corresponding to the global $G_R$ symmetry. Similarly, the first non-trivial term obtained by expanding the monodromy matrix  gauge transformed by the field $g$ around the value $z\sim 0$ gives the (local) Noether charge $Q_L = \int^{R}_0 \de \sigma   g J_\tau g^{-1}$ corresponding to the global $G_L$ symmetry. In both cases, note that  the spatial component of the Lax (resp.~its gauge transformation by $g$) vanishes at $z\sim \infty$ (resp.~$z\sim 0$) and thus its expansion starts only at the next order. Leading order terms of the monodromy (resp.~its gauge transformation) then do not receive contributions from nested integrals and thus are local. 
Higher order terms will of course still give rise to non-local conserved charges, which interestingly are hidden from a first naive analysis of the PCM action principle. Further details and references can again be found in \cite{Driezen:2021cpd}.

\subsubsection{Lax formulation of the semi-symmetric space sigma-model} \label{s:lax-ssssm}
To find the Lax formulation of the canonical SSSSM \eqref{eq:SSSSM1}, we employ a similar strategy as for the PCM, \textit{i.e.}~relaxing the identification of $J$ with $g$ and combining the equations of motion \eqref{eq:ssssm1-eoms} with the Maurer-Cartan identity. First, let us rewrite \eqref{eq:ssssm1-eoms} in form language. One can show that the equations are equivalent to
\begin{equation}\label{eq:ssssm1-eoms-form}
    \begin{aligned}
        & d\star J^{(2)} +  J^{(0)}\wedge\star J^{(2)}+ \star J^{(2)} \wedge J^{(0)} - \varkappa J^{(1)} \wedge J^{(1)} + \varkappa J^{(3)} \wedge J^{(3)} = 0 \ , \\
        &J^{(3)} \wedge \star J^{(2)} +\star J^{(2)} \wedge J^{(3)}   + \varkappa J^{(2)} \wedge J^{(3)}+\varkappa J^{(3)} \wedge J^{(2)} = 0 \ , \\
        &J^{(1)} \wedge \star J^{(2)}+\star J^{(2)} \wedge \star J^{(1)} - \varkappa J^{(2)} \wedge J^{(1)}-\varkappa J^{(1)} \wedge J^{(2)} = 0 \ ,
    \end{aligned}
\end{equation}
while the projections of the Maurer-Cartan identities on the graded eigenspaces read\footnote{This is the solution to exercise \ref{ex:mc-projection}.}
\begin{equation} \label{eq:mc-identity-2-projections}
    \begin{aligned}
        & dJ^{(0)} + J^{(0)}\wedge J^{(0)} + J^{(2)}\wedge J^{(2)} +J^{(1)}\wedge J^{(3)}+J^{(3)}\wedge J^{(1)} = 0 \ , \\
        & dJ^{(1)} + J^{(0)}\wedge J^{(1)} + J^{(1)}\wedge J^{(0)} +J^{(2)}\wedge J^{(3)}+J^{(3)}\wedge J^{(2)} = 0 \ , \\
        & dJ^{(2)} + J^{(0)}\wedge J^{(2)} + J^{(2)}\wedge J^{(0)} +J^{(1)}\wedge J^{(1)}+J^{(3)}\wedge J^{(3)} = 0 \ , \\
        & dJ^{(3)} + J^{(0)}\wedge J^{(3)} + J^{(3)}\wedge J^{(0)} +J^{(1)}\wedge J^{(2)}+J^{(2)}\wedge J^{(1)} = 0 \ .
    \end{aligned}
\end{equation}
Now, noticing that the first equation of \eqref{eq:ssssm1-eoms-form} is an equation for $d\star J^{(2)}$, while the others are algebraic, and the projections of the Maurer-Cartan identity are all equations for $dJ^{(i)}$, $i=0,\ldots , 3$, we propose the following ansatz for the Lax connection 
\begin{equation}
    {\cal L}(\alpha_0,\alpha_1,\alpha_2,\alpha_3,\alpha_4) = \alpha_0 J^{(0)} +\alpha_1 J^{(2)} + \alpha_2 \star J^{(2)} + \alpha_3 J^{(1)} + \alpha_4 J^{(3)} \ ,
\end{equation}
with $\alpha_i$, $i=0,1,2,3,4$ free parameters yet to be determined by demanding on-shell flatness of ${\cal L}$. Imposing \eqref{eq:zero-curvature-form} and projecting onto each of the $\mathbb{Z}_4$-eigenspaces $\mathfrak{g}^{(i)}$, we find using \eqref{eq:ssssm1-eoms-form} and \eqref{eq:mc-identity-2-projections}, as well as the properties of the Hodge-star of exercise \ref{ex:hodge-star}, that
\begin{itemize}
    \item on $\mathfrak{g}^{(0)}$, 
    \begin{equation}
    \begin{aligned}
        &(\alpha_0^2-\alpha_0) J^{(0)}\wedge J^{(0)}  + (\alpha_1^2-\alpha_2^2-\alpha_0)J^{(2)}\wedge J^{(2)} \\& +(\alpha_3\alpha_4 - \alpha_0) \left(J^{(1)}\wedge J^{(3)} +J^{(3)}\wedge J^{(1)}  \right) = 0 \ ,
    \end{aligned}
    \end{equation}
    which implies (excluding the trivial possibility $\alpha_0=0$),
    \begin{equation} \label{eq:flat-lax-cond-0}
        \alpha_0 =1, \qquad \alpha_1^2-\alpha_2^2=1, \qquad \alpha_3\alpha_4=1 \ .
    \end{equation}
    \item on $\mathfrak{g}^{(1)}$,
    \begin{equation}
     \begin{aligned}
        &(\alpha_0\alpha_3 - \alpha_3)\left(J^{(0)}\wedge J^{(1)} + J^{(1)}\wedge J^{(0)} \right) + \\ &(\alpha_1\alpha_4 - \varkappa \alpha_2\alpha_4 - \alpha_3)\left(J^{(2)}\wedge J^{(3)} + J^{(3)}\wedge J^{(2)} \right) = 0
         \end{aligned}
    \end{equation}
    which, with the above, implies
    \begin{equation} \label{eq:flat-lax-cond-1}
        \frac{\alpha_1\alpha_4 - \alpha_3}{\alpha_2\alpha_4} = \varkappa \ .
    \end{equation}
    \item similarly, on $\mathfrak{g}^{(2)}$, one will find the conditions
    \begin{equation} \label{eq:flat-lax-cond-2}
        \frac{\alpha_3^2-\alpha_1}{\alpha_2} = -\varkappa , \qquad \frac{\alpha_4^2-\alpha_1}{\alpha_2} =\varkappa \ .
    \end{equation}
    \item and finally on $\mathfrak{g}^{(3)}$, the conditions
    \begin{equation}\label{eq:flat-lax-cond-3}
        \frac{\alpha_1\alpha_3 - \alpha_4}{\alpha_2\alpha_3} = -\varkappa \ .
    \end{equation}
\end{itemize}
This means we have 7 conditions for 6 variables (including $\varkappa$ in the counting). However, luckily, by close inspection one will find that there is a degree 2 redundancy. 
First, one will see that adding \eqref{eq:flat-lax-cond-2}  gives the same condition as adding \eqref{eq:flat-lax-cond-1} and \eqref{eq:flat-lax-cond-3}, \textit{i.e.}
\begin{equation}
    \alpha_3^2 + \frac{1}{\alpha_3^2} = 2\alpha_1 \ .
\end{equation}
Secondly,  one can rewrite the second equation of \eqref{eq:flat-lax-cond-0} for $\alpha_3\neq 0 $ (equivalently $\alpha_4\neq 0$) as
\begin{equation}
    \left(\frac{1}{\alpha_3^4}-1 \right)^2 \left(\frac{1}{\varkappa^2 - 1}\right) = 0 \ .
\end{equation}
Thus, for $\varkappa = \pm 1$, $\alpha_3$ will remain free. Note that the other conditions fix $\alpha_{1,2,4}$ in terms of $\alpha_3$. Thus, if on the other hand $\varkappa^2 \neq  1$, then $\alpha_3$, and by extension all other variables, would be fixed so that instead of having a one-parameter family of flat Lax connections ${\cal L}(z)$, we would have a trivially flat current.  Concluding, the theory is classically integrable only for  $\varkappa = \pm 1$.
Interestingly, this is precisely the same condition as the one ensuring $\kappa$-symmetry.  It would be very interesting to understand if there is a deeper connection between integrability and $\kappa$-symmetry. 

In summary, taking $\varkappa=1$ and writing $\alpha_3 \equiv z$, the Lax connection thus becomes \cite{Bena:2003wd}
\begin{equation} \label{eq:lax-ssssm}
    {\cal L}(z) =  J^{(0)} + z J^{(1)} +\frac{1}{2} \left(z^2+z^{-2} \right) J^{(2)} + \frac{1}{2} \left(z^2-z^{-2} \right) \star J^{(2)}  + z^{-1} J^{(3)} ,
\end{equation}
whose zero-curvature condition \eqref{eq:zero-curvature-form} implies the equations \eqref{eq:ssssm1-eoms-form} and \eqref{eq:mc-identity-2-projections} for any value of $z\in \mathbb{C}$. This demonstrates the classical integrability of the canonical SSSSM \eqref{eq:SSSSM1}. Note, however, that the equations of motion for $h^{\alpha\beta}$ (and thus the Virasoro constraints) \eqref{eq:virasoro-ssssm1} do not immediately follow from \eqref{eq:zero-curvature-form}. \\ 

\textit{Remark} --- As mentioned before, the Lax connection takes values in $\mathfrak{g}\otimes \mathbb{C}$, and thus for $\mathfrak{g}=\mathfrak{su}(p,q|r,s)$, ${\cal L}(z)\in \mathfrak{sl}(p+q|r+s)$. In the case of $AdS_5\times S^5$ this is $\mathfrak{psl}(4|4)$. For this example, the Lax connection behaves under  the $\mathbb{Z}_4$ automorphism $\Omega$ \eqref{eq:z4-ads5-s5} as $\Omega({\cal L}(z)) = {\cal L}(i z)$.
Another useful parametrisation often used is $z=\sqrt{\frac{1+x}{1-x}}$ in which case the $\mathbb{Z}_4$ acts as $\Omega({\cal L}(x)) = {\cal L}(1/x)$. The eigenvalues $\lambda(z)$ of the monodromy matrix must respect this symmetry.\\

\noindent {\bf Transformation of the Lax  under local symmetries} --- The superstring sigma-model \eqref{eq:SSSSM1} exhibits a number of local symmetries, namely Weyl rescalings, worldsheet diffeomorphism invariance,  right-acting $G^{(0)}$ transformations, and $\kappa$-symmetry. Here, we will briefly discuss the behaviour of the Lax connection \eqref{eq:lax-ssssm} under them. 

First, under Weyl rescalings, the Lax connection of course transforms trivially, as $\gamma^{\alpha\beta}=\sqrt{-h}h^{\alpha\beta}$ stays invariant under \eqref{eq:weyl-rescaling}. Secondly, under an infinitesimal  worldsheet diffeomorphisms $\sigma^\alpha \rightarrow \tilde{\sigma}^\alpha= \sigma^\alpha + f^\alpha(\sigma)$ we would have ${\cal L} (\sigma) = {\cal L}_\alpha (\sigma) \de \sigma^\alpha = {\cal L}_\alpha (\tilde \sigma) d\tilde{\sigma}^\alpha$ and thus $\delta {\cal L}_\alpha = \partial_\beta {\cal L}_\alpha f^\beta + {\cal L}_\beta \partial_\alpha f^\beta$. Using the zero-curvature condition, this becomes simply an infinitesimal gauge transformation \eqref{eq:lax-gauge-transformation} with the parameter $\delta g =- f^\beta {\cal L}_\beta$. The eigenvalues $\lambda(z)$ of the monodromy matrix will thus be invariant under worldsheet diffeomorphisms. 

Thirdly, under the right-acting $G^{(0)}$ transformations \eqref{eq:H-gauge-invariance}, or equivalently \eqref{eq:H-action-graded-J}, the Lax connection simply transforms as a gauge transformation \eqref{eq:lax-gauge-transformation} with the parameter $g=h(\sigma)^{-1}$. The eigenvalues $\lambda(z)$ of the monodromy matrix will thus be invariant under the right-acting gauge transformations by $G^{(0)}$. 

Finally, in \cite{Grigoriev:2007bu,Arutyunov:2008if,Arutyunov:2009ga} it was shown that for $\mathfrak{psl}(n|n)$ and $\mathfrak{osp}(6|2,2)$ the eigenvalues of the monodromy will also be invariant under $\kappa$-symmetries if and only if the Virasoro constraints are satisfied. Hence, although the Virasoro constraints do not directly follow from the zero-curvature of the Lax, they are required such that the conserved eigenvalues $\lambda(z)$ are invariant under all the local gauge symmetries.\\

\noindent {\bf Relation to  global supersymmetries} --- Recall that in the PCM case, the expansion of the monodromy around  values $z_\star$ where ${\cal L}_\sigma(z_\star)$ or ${\cal L}^g_\sigma(z_\star)$ vanishes holds at the first non-trivial  order a local Noether charge associated to global symmetries. The same happens for the SSSSM. As there is no value $z_\star$ where the Lax \eqref{eq:lax-ssssm} vanishes, let us do a gauge transformation by the superfield $g$ of ${\cal L}_\sigma(z)$. Assuming for simplicity conformal gauge, it reads
\begin{equation}
    {\cal L}_\sigma^g (z) = (z-1) a_\sigma^{(1)} + \frac{1}{2} (z^{-2} - z^2) a_\tau^{(2)} + \frac{1}{2} (z^{-2}+z^2 - 2) a_\sigma^{(2)} + (z^{-1}-1)a_\sigma^{(3)} ,
\end{equation}
where $a^{(i)} = g J^{(i)} g^{-1}$.\footnote{Notice that despite the notation, $a^{(i)}$ does not necessarily belong to $\mathfrak{g}^{(i)}$.} Clearly ${\cal L}^g_\sigma (z=1)=0$. The expansion of ${\cal L}^g_\sigma(z)$ around $z =1+\epsilon$ is then
\begin{equation}
    {\cal L}^g_\sigma (z) = \epsilon \left( a_\sigma^{(1)}-a_\sigma^{(3)} - 2 a_\tau^{(2)} \right) + {\cal O}(\epsilon^2) ,
\end{equation}
and expanding the gauge transformed monodromy thus gives
\begin{equation}
    \Omega^g(z) = 1 + \epsilon \int^{R}_0 \de \sigma \left( 2 a_\tau^{(2)} -a_\sigma^{(1)}+a_\sigma^{(3)} \right) +{\cal O}(\epsilon^2).
\end{equation}
Here we see, in conformal gauge, precisely the local Noether charge of \eqref{eq:noether-global-susy} associated to the global $G$ symmetry appearing (for $\varkappa=1$). Again, higher-order terms in $\epsilon$ will contain non-local, hidden, charges.

\subsubsection{Comments on the integrability of the GS superspace \texorpdfstring{$\sigma$}{sigma}-model}

Although the supercoset structure is quite nice to work with---in particular the Lax connection can be easily derived, something that is in general not obvious at all---one of its main problems is that it is not entirely the right language to study backgrounds that are not maximally supersymmetric in $D=10$ (\textit{i.e.}~not flat space, nor $AdS_5\times S^5$, nor its limits). In particular,  they correspond to a GS action only after a certain $\kappa$-gauge is fixed, and this gauge may not be compatible with all string solutions. The existence of a flat Lax should of course not depend on which solution is taken and therefore, in such cases, it is desirable to have a Lax connection of the curved GS action before fixing the $\kappa$-gauge. This problem was addressed for backgrounds of the form $AdS \times S \times S \times T$ 
%arising from near horizon limits of intersecting brane constructions 
in \cite{Wulff:2014kja} (for earlier work on
$AdS_4\times CP^3$,  $AdS_3\times S^3 \times S^3 \times S^1$, and $AdS_2 \times S^2\times T^6$, see the references in \cite{Wulff:2014kja}). Along the same lines as above, the construction uses the components of the Noether current of the superisometries to build the Lax  and  its flatness was then shown  up to quadratic order in fermions. 

%\SIB{ 1207.5531, https://arxiv.org/pdf/1402.3122.pdf and sec4.4. 1111.4197}

\subsection{Summary and concluding remarks} \label{sec:sib:concl}
In this section, we reviewed the Green-Schwarz superstring formalism  in flat and curved space, as well as its supercoset formulation and the presence of classical integrability. While staying at the classical level, we  discussed in these various scenarios the presence of worldsheet gauge symmetries and how they should be fixed in order to proceed with lightcone quantisation. Let us here just briefly summarise and conclude with  a few main messages.
%Keeping the conclusions short, let us just summarise with a few take-home messages.

Although the GS formalism in principle is ideally suited to describe superstrings in generic curved supergravity backgrounds, the practical story can be quite complicated. The generic superspace GS action is only explicitly known to quadratic order in fermions (to quartic order under some assumptions), since obtaining higher orders involves solving certain equations whose complexity grows quickly at each order. Furthermore, even at the bosonic level,  the lightcone gauge of the curved string results in a highly non-linear theory which generically can only be quantised perturbatively.

However, a large class of interesting backgrounds realised as supercoset geometries circumvent some of these issues. They can be realised using an alternative GS description, known as the supercoset construction, which in principle can be expanded to any order in fermions without the need to solve additional constraints and equations.
%\footnote{For many of these supercoset backgrounds, the action moreover terminates after quartic order. }
One of the main advantages of  the supercoset language is that the classical integrability of these theories manifests itself rather naturally, which should aid the quantum level significantly. It is also the natural language to study large classes of integrable deformations of supergravity backgrounds (see e.g.~\cite{Hoare:2021dix} and references therein).

Nevertheless, in some cases the supercoset GS string is encapsulated by the generic GS formalism only in certain gauges.
%\footnote{The canonical exception is the type IIB $AdS_5\times S^5$ background.} 
Some relevant classical string configurations may not be compatible with that particular gauge choice, and this can introduce important subtleties for lightcone quantisation, which is both gauge- and solution-dependent. These subtleties will be illustrated in quite some detail in the upcoming section \ref{s:saskia} by means of the GS superstring propagating in $AdS_3\times S^3$ backgrounds.

%%%% END OF PART I SIBYLLE

%%%% PART II SASKIA
\newpage
\section{Green-Schwarz superstring in holographic \texorpdfstring{$\text{AdS}_3$}{AdS3} backgrounds} \label{s:saskia}

%\noindent
%\textit{Current author: Saskia Demulder.}

%\noindent
%For  comments, questions, typos,... feel free to send a mail to \href{mailto:saskia.demulder@gmail.com}{saskia.demulder@gmail.com}.

%%%%%%%%%%%%%%%%%%%%%%%%%%%%%%%%%%%%%
%%%%%%%         Intro         %%%%%%%
%%%%%%%%%%%%%%%%%%%%%%%%%%%%%%%%%%%%%
%\subsection{Introduction}
In this section and in the following ones, we will apply integrability and conformal field theory techniques to explore holographic $AdS_3$-backgrounds. In the present section, by exploiting the large (super)symmetry structure of the $AdS_3\times S^3\times M_4$ superstring, we will review its supercoset realisation together with the corresponding Lax connection, which warrants the integrability of the supercoset action.
As it will become clear, even if the supercoset action can help unravel many properties and observables of superstrings propagating in holographic $AdS_3$-backgrounds, it is incomplete; the issue is that it does not correctly reproduce the $T^4$ part of the geometry, as well as some of the fermions. 

In section \ref{sec:saskia:AdS3/CFT2props} we will first review the main features of $\text{AdS}_3/\text{CFT}_2$ by highlighting some of its unique properties as compared to other holographic backgrounds. Having set the stage, we will discuss how the corresponding holographic backgrounds are generated from the back-reaction of the corresponding D-brane configurations. This will provide the necessary insights to unravel the symmetry properties of the $AdS_3\times S^3$ backgrounds. Section \ref{sec:saskia:AdS3coset} will treat the supercoset realisation of these backgrounds and their relation to the GS superstring action. The massless modes which are particular to the spectrum of $AdS_3\times S^3 \times M_4$ backgrounds will be discussed in section \ref{sec:saskia:spectrumplanewave}. Supergravity solutions for these backgrounds are generically supported by a mixture of NSNS- and RR-fluxes. To address this, section \ref{sec:CZ-WZ} will review how a WZ-term can be introduced in the supercoset action, while preserving, amongst other properties, integrability of the supercoset action. In section \ref{s:saskia-gf} we will temper the successes of the supercoset action, reviewing how the required gauge-fixing puts severe restrictions on the string configurations the supercoset action can describe. Finally, we will close this section with a summary and discussion in section \ref{sec:saskia:summary}.

Finally let us add a caveat and an apology. Classical integrability of the $AdS_3\times S^3$-backgrounds forms a very rich and wide subject of past and present research. As a result, having both in mind pedagogy and conciseness, several more advanced aspects of the integrable $AdS_3\times S^3$-string could not be covered in this section. When possible, references to these subjects are however provided. 

%%%%%%%%%%%%%%%%%%%%%%%%%%%%%%%%%%%%%
%%%%%%% Facts about AdS3/CFT2 %%%%%%%
%%%%%%%%%%%%%%%%%%%%%%%%%%%%%%%%%%%%%
\subsection{\texorpdfstring{$\text{AdS}_3/\text{CFT}_2$}{AdS3/CFT2} holography: facts and peculiarities}\label{sec:saskia:AdS3/CFT2props}
Although less well-known and less well-understood than the celebrated AdS$_5$/CFT$_4$ holographic correspondence, the AdS$_3$/CFT$_2$ correspondence shows some remarkable and challenging features. The goal of this section is to provide a bird's eye view of the fundamental peculiarities and challenges presented by AdS$_3$/CFT$_2$ holography.

%%%%%%%%%%%%%%%%%%%%%%%%%%%%%%%%%%%%%
\subsubsection{\texorpdfstring{$\text{AdS}_3/\text{CFT}_2$}{AdS3/CFT2}: ``Less is more''} 
AdS$_3$/CFT$_2$ holography is in many regards ``less'' than AdS$_5$/CFT$_4$: it is lower dimensional, the maximal supersymmetric solutions have only half the number of supersymmetries, the underlying symmetry groups are smaller\footnote{By ``smaller'' we mean here that all $AdS_3$-holographic with 16 supercharges have, besides a curved part, also some flat directions. The latter are described by abelian group factors appended to the non-abelian groups, completing the background to a 10 dimensional string background. This is not the case of  $AdS_5\times S^5$ or $AdS_4\times \mathbb CP^4$, which are already 10 dimensional backgrounds.}, etc.  Oftentimes however, when looking at systems in lower dimensions or with less symmetries, peculiar things can happen:
\begin{itemize}
	\item Gravity in $AdS_3$ has no propagating graviton, but is nonetheless non-trivial. $AdS_3$ gravity admits black-hole solutions \cite{Banados:1992wn,Strominger:1996sh} which behave in many ways like their higher-dimensional counterparts as they e.g.~satisfy the Bekenstein-Hawking area law. 
	\item In addition, $AdS_3$ gravity enjoys an infinite dimensional algebra of asymptotic symmetries. This is simply a reflection of the fact that, on the dual side (and in contrast to higher dimensional CFTs) two dimensional CFTs have an infinite dimensional conformal symmetry group.\footnote{This might at first look somewhat problematic since the gravity side only has a \textit{finite} dimensional group of symmetry, \textit{i.e.}~$SO(2,2)$. Brown and Henneaux \cite{Brown:1986nw} realised however that the finite group on the gravity side only accounts for \textit{globally} defined generators of all symmetry in $AdS_3$ gravity. What was missing to match the infinite number of generators on the CFT side are asymptotic symmetry generators of $AdS_3$ spacetime. These asymptotic symmetries are gauge transformations leaving the field configurations at the boundary invariant and as such do not have globally defined generators. Note that this result by Brown and Henneaux predates Maldacena's celebrated conjecture \cite{Maldacena:1997re}.}
	\item Another peculiar property is that, in the lightcone gauge, the string spectrum for $AdS_3$-backgrounds contains  modes with different masses, including massless ones. This deos not happen for higher dimensional $AdS$-backgrounds, whose spectrum exclusively contains massive excitations only. The presence of massless modes is a consequence of the existence of flat~directions completing the curved part of the geometry of (maximal supersymmetric) $AdS_3$ holographic backgrounds. Massless modes present a significant challenge towards unravelling the integrable structure of superstrings propagating in $AdS_3$-backgrounds. Some of these issues and their resolutions in the context of the $AdS_3\times S^3\times T^4$ S-matrix will be explored in section~\ref{s:fiona}. 
	%\item As we will come to see, the supercoset approach to prove the integrability of superstrings in holographic $AdS_3$-backgrounds has many advantages, it also has severe drawbacks. These limitations called for an alternative strategy which was realised in the so-called hybrid formalism which can be applied whenever the background is supported by RR-flux. The hybrid formalism  will be the topic of section \ref{s:bob}.  
 %%% AS: I commented out the above point, I do not think it makes sense.
	\item As we will discuss at length in the main text,  $AdS_3$-backgrounds, in contrast to  $AdS_5$- or  $AdS_4$-backgrounds, need not only be supported by pure RR-flux but also by a mixture of NSNS- and RR-fluxes. The relative contribution of RR- and NSNS-fluxes to the $AdS_3$-background is controlled by a parameter $q$. This can be schematically summarised by the diagram

\begin{center}
\tikzset{every picture/.style={line width=0.75pt}} %set default line width to 0.75pt        
\begin{tikzpicture}[x=0.75pt,y=0.75pt,yscale=-1,xscale=1]
\draw    (218,60) -- (354.64,60.2) ;
\draw [shift={(354.64,60.2)}, rotate = 0.08] [color={rgb, 255:red, 0; green, 0; blue, 0 }  ][fill={rgb, 255:red, 0; green, 0; blue, 0 }  ][line width=0.75]      (0, 0) circle [x radius= 3.35, y radius= 3.35]   ;
\draw [shift={(218,60)}, rotate = 0.08] [color={rgb, 255:red, 0; green, 0; blue, 0 }  ][fill={rgb, 255:red, 0; green, 0; blue, 0 }  ][line width=0.75]      (0, 0) circle [x radius= 3.35, y radius= 3.35]   ;
%Shape: Brace [id:dp5368072566720683] 
\draw   (355.48,48.83) .. controls (355.48,44.16) and (353.15,41.83) .. (348.48,41.83) -- (297.79,41.83) .. controls (291.12,41.83) and (287.79,39.5) .. (287.79,34.83) .. controls (287.79,39.5) and (284.46,41.83) .. (277.79,41.83)(280.79,41.83) -- (225.48,41.83) .. controls (220.81,41.83) and (218.48,44.16) .. (218.48,48.83) ;
%Curve Lines [id:da6416546413792393] 
\draw    (354.64,60.2) .. controls (367.93,72.03) and (373.07,65.13) .. (384.07,55.71) ;
\draw [shift={(385.48,54.52)}, rotate = 140.19] [color={rgb, 255:red, 0; green, 0; blue, 0 }  ][line width=0.75]    (10.93,-4.9) .. controls (6.95,-2.3) and (3.31,-0.67) .. (0,0) .. controls (3.31,0.67) and (6.95,2.3) .. (10.93,4.9)   ;

% Text Node
\draw (194,73.4) node [anchor=north west][inner sep=0.75pt]    {$q=0$};
% Text Node
\draw (337,73.4) node [anchor=north west][inner sep=0.75pt]    {$q=1$};
% Text Node
\draw (381,35) node [anchor=north west][inner sep=0.75pt]   [align=left] {WZW point};
% Text Node
\draw (251,13) node [anchor=north west][inner sep=0.75pt]   [align=left] {integrability};
\end{tikzpicture}
\end{center}
\vspace{-5pt}
\end{itemize}
Let us discuss each portion of this diagram:
\vspace{-5pt}
\paragraph{\boldmath The $q=1$ point:} For this value of the parameter $q$, the $AdS_3$-background is supported by pure NSNS three-form flux. When a pure NSNS supergravity solution is available, the worldsheet sigma model can be described within the RNS formalism and admits extended chiral symmetry \cite{Maldacena:2000hw}. In this particular case, the worldsheet CFT is a supersymmetric WZW model, and we may exploit the full power of chiral algebras and their representation theory to solve the system \cite{Elitzur:1998mm,Giveon:1998ns,Kutasov:1999xu,Maldacena:2000hw}.

\paragraph{\boldmath The $q=0$ point:} This is the pure RR-flux background. In contrast to the pure NSNS solution, in the pure RR case the worldsheet CFT is nonlocal. This is problematic when one tries to quantise it, and indeed RR-backgrounds are infamously hard to quantise. Despite the apparent intractability of pure RR backgrounds, these backgrounds are, as we will see in section \ref{sec:saskia:AdS3coset}, nonetheless integrable by virtue of the existence of a supercoset realisation \cite{Babichenko:2009dk} (see also the earlier section \ref{s:lax-ssssm}).

\paragraph{\boldmath Arbitrary value of $0<q<1$ :} For generic value of the parameter $q$, the background is supported by a mixture of NSNS and RR fluxes. In section \ref{sec:CZ-WZ}, we will discuss how the supercoset realisation can be modified to accommodate for both NSNS and RR fluxes to support the geometry. Although the new term spoils the $\mathbb Z_4$-symmetry, the system as a whole remains integrable. Understanding mixed flux backgrounds is especially intriguing as it can offer a potential bridge between the knowledge made available at the conformal point $q=1$ via CFT techniques and results obtained using integrability tools at generic values of~$q$.

%%%%%%%%%%%%%%%%%%%%%%%%%%%%%%%%%%%%%
\subsubsection{\texorpdfstring{$AdS_3 \times S^3\times M_4$}{AdS3xS3xM4} string backgrounds}\label{saskia:sec:AdS3S3M4_bckgrds}
Holographic backgrounds can be obtained as the near-horizon geometries curved by stacks of branes in a weak-coupling limit. In the following we will discuss how the holographic correspondence between gravity in $AdS_3\times S^3\times M_4$ and certain superconformal field theories in two dimensions can be inferred by a particular configuration of (intersecting) branes. For more details we refer the reader to the textbooks \cite{Blumenhagen:2013fgp,Ammon:2015wua}.

The branes will be placed in type IIB string theory in a $\mathbb R^{1,4}\times S^1\times M_4$ background, where $M_4$ is a compact manifold which describes the internal degrees of freedom. What the manifold $M_4$ can be taken to be is fixed by requiring that the total background admits a maximal number of supersymmetries. It turns out that $AdS_3$ backgrounds can preserve at most 16 of the 32 supersymmetries of type IIA/B supergravity solutions. There are three possible backgrounds preserving 16 real supercharges that fit in the AdS$_3$/CFT$_2$ correspondence: $M_4=S^3\times S^1,T^4$ or $K3$. The latter is CY$_2=K3$ and is the unique non-trivial compact Calabi-Yau manifold in two complex dimensions. Since $K3$ can be seen as an orbifold limit of $T^4$, many of the results for the $K3$ background can be derived from that of $T^4$ since most integrability tools can be likewise applied to orbifolds, orientifolds or deformations of dual pairs, see e.g.~\cite{Zoubos:2010kh} for an introduction in the context of the $AdS_5$/CFT$_4$ correspondence.

In this review we will restrict the discussion to $M_4=S^3\times S^1$ or $T^4$.  Each of these backgrounds is realised as the near-horizon of a different D-brane set-up, which we will now briefly summarise. We will first consider a brane configuration leading to type IIB solutions supported by pure RR-flux. Later we will comment on how backgrounds supported by NSNS-flux and also how type IIA solutions can be obtained by applying string dualities.
%%%%%%%%%%%%%%%%
\subsubsection*{D-brane construction}

\begin{figure}
    \begin{minipage}{0.30\linewidth} 
\tikzset{every picture/.style={line width=0.75pt}} %set default line width to 0.75pt        
\begin{tikzpicture}[x=0.75pt,y=0.75pt,yscale=-1,xscale=1]
%uncomment if require: \path (0,300); %set diagram left start at 0, and has height of 300

%Straight Lines [id:da40146947967555713] 
\draw    (72.71,25.86) -- (107.94,13.83) -- (108.6,110.73) -- (73.37,127.83) -- cycle ;
%Straight Lines [id:da5067881011690214] 
\draw    (130.78,77.16) -- (159.48,62.59) -- (108.27,62.28) -- (73.04,76.84) -- cycle ;
%Straight Lines [id:da019091681675087346] 
\draw    (73.37,61.96) -- (55.59,62.28) -- (22.48,77.16) -- (73.04,76.84) ;
%Straight Lines [id:da47369881861654506] 
\draw [line width=1.5]    (108.27,62.28) -- (73.04,76.84) ;

%Curve Lines [id:da2165272055299241] 
\draw    (34.48,90.83) .. controls (60.96,86.65) and (43.48,81.16) .. (61.48,69.16) ;
%Curve Lines [id:da35040120136986264] 
\draw    (96.27,67.28) .. controls (102.48,58.83) and (96.48,40.83) .. (122.48,40.83) ;
%Curve Lines [id:da16271321409836237] 
\draw    (94.48,97.83) .. controls (120.96,93.65) and (113.48,101.83) .. (128.48,104.83) ;

% Text Node
\draw (14,84) node [anchor=north west][inner sep=0.75pt]  [font=\small] [align=left] {{\footnotesize D5'}};
% Text Node
\draw (130,97) node [anchor=north west][inner sep=0.75pt]  [font=\small] [align=left] {{\footnotesize D5}};
% Text Node
\draw (125,32) node [anchor=north west][inner sep=0.75pt]  [font=\small] [align=left] {{\footnotesize D1}};
\end{tikzpicture}  

\end{minipage}\hfill
\begin{minipage}{0.70\linewidth}
  \begin{tabular}{lcccccccccc}
    & \multicolumn{6}{c}{$\mathbb R^{1,5}$}& \multicolumn{4}{c}{$M_4$}\\
	& \multicolumn{6}{c}{\downbracefill}& \multicolumn{4}{c}{\downbracefill} \\
    & 0 & 1 & 2 & 3 & 4 & 5 & 6 & 7 & 8 & 9 \\ \hline
    & $t$ & $x_1$ & $x_2$ & $x_3$ & $x_4$ & $x_5$ &   &   &  &  \\ \hline
    $N_1$ D1-branes  & $-$ & $-$  & {$\sim$} & {$\sim$} & {$\sim$} & {$\sim$} &  {\boldmath$\cdot$} & {\boldmath$\cdot$} &  {\boldmath$\cdot$} &  {\boldmath$\cdot$} \\
  $N_5$ D5-branes  & $-$ &  $-$ & {\boldmath$\cdot$} & {\boldmath$\cdot$} & {\boldmath$\cdot$} & {\boldmath$\cdot$} & $-$ & $-$ & $-$ & $-$  \\
   $N_5'$ D5-branes  & $-$ &  $-$ & $-$ & $-$ & $-$ & $-$ & {\boldmath$\cdot$} & {\boldmath$\cdot$} &  {\boldmath$\cdot$} & {\boldmath$\cdot$} 
  \end{tabular}
 \end{minipage} 
 
 \caption[something]{The D1-D5-D5' brane configuration. A dash $-$ means that the brane is extended in that direction, while a dot {\boldmath$\cdot$} means that the brane is perpendicular to the direction, correspondingly to Neumann and Dirichlet boundary conditions respectively. For completeness, the tilde $\sim$ indicates that the brane can be smeared or delocalised in that direction. From the diagram it is apparent that the original $SO(1,9)$ symmetry is broken down to a $SO(4)_{2345}\times SO(4)_{6789}$ symmetry, where the subscript indicates the relevant directions. Note however that the last factor, $SO(4)_{6789}$ remains only unbroken at low energy, e.g.~at the supergravity regime, when the compactified manifold in these directions is small. At larger energies the four dimensional manifold $T^4$ further breaks down the symmetry group to $SO(4)_{6789}\rightarrow U(1)^4$. Including the last row (adding an additional set of D5 branes) realises the $AdS_3\times S^3\times S^3\times S^1$-background.}\label{table:D1D5_system}
\end{figure}

Considering first the $AdS_3\times S^3\times T^4$ background and its type IIB solution, the associated D-brane configuration \cite{Strominger:1996sh,deBoer:1999gea} is realised by a stack of $N_5$ D5-branes intersecting $N_1$ D1-branes as indicated in the two first rows of table \ref{table:D1D5_system}. Note that in the table we added the possibility to have the D1-branes smeared or delocalised in the 2345-direction. The idea is to first construct a periodic array of $D$-branes in question along a direction transverse to its location. Subsequently taking the continuum limit, the background admits an isometry in that direction by virtue of losing an explicit dependence in the harmonic functions. Compactifying in that direction one can then apply a T-duality. For more details and applications see \cite{Johnson:2003glb,Gauntlett:1997cv,Mohaupt:2000gc}.\label{saskia:ftnt:smearing}

Turning to the D-branes, they bend the original flat space $\mathbb R^{1,5}$ and the resulting curved geometry is described by a metric of the form
\begin{align}\label{eq:metric_D1/D5}
	\mathrm d s^2=(H_1H_5)^{-1/2}\mathrm d s_{\mathbb R^{1,1}}^2+(H_1H_5)^{1/2}\mathrm d s_{\mathbb R^{4}}^2+\left(\frac{H_1}{H_5}\right)^{1/2}\mathrm ds^2_{M_4}\,,
\end{align}
where we have broken down the $\mathbb R^{1,5}$-plane into two pieces $(\mathbb R^{1,1},\mathbb R^{4})$.  The functions $H_{1/5}$ represents the gravitational back-reaction of the large number of D1 and D5-branes.  If we set $r^2=\sum_{i=6}^9 x^i x^i$, then $H_1(r)$ and $H_5(r)$ are harmonic functions given by the expressions\footnote{There is a subtlety here, one would expect naively that the $H_1$ function would depend on the distance perpendicular to their location. In particular, besides the already $6-9$-directions also $2-5$. The reason is that the D1-brane is in fact not localised but `smeared' or `delocalised' along these directions. The dependence then results from intersecting the two D-brane stacks and applying the harmonic function rule, which prescribes how when intersecting branes, the composite solution will be described by products of powers of the harmonic functions $H_i$. See \cite{Tseytlin:1996bh,Gauntlett:1997cv,Peet:2000hn}.}%thesis van Pol
\begin{align}
	H_1=1+\frac{Q_1}{r^2}\,,\quad H_5=1+\frac{Q_5}{r^2}\,,\qquad \,\mathrm{where}\quad Q_1= (2\pi)^4 g_s N_1 (\alpha')^3/ V_4\,,\; Q_5=g_sN_5\alpha'\,,
\end{align}
where $g_s$ is the string coupling constant, $V_4$ is the volume of the internal manifold and $N_{1/5}$ denotes the number of D1/D5 branes, respectively.
Note that, as expected, at a large distance away from the stack of branes, the metric \eqref{eq:metric_D1/D5} is simply flat ten dimensional space.
The metric in \eqref{eq:metric_D1/D5} has to be supported by an RR three-form flux $F^{(3)}$ to solve the type IIB supergravity solution together with a dilaton field $\Phi$ given by 
\begin{align}\label{eq:remainingfields_IIB}
	H^{(3)}&=2 Q_1  e^{-2\Phi}\star_6\mathrm{Vol}({S^3})+2 Q_5\mathrm{Vol}({S^3})\,,\quad \text{and}\quad e^{-2\Phi}=\frac{H_5}{H_1} \,,
\end{align} 
where $H^{(3)}$ is the three-form flux, $\mathrm{Vol}({S^3})$ denotes the unit volume form on the three-sphere $S^3$ and the Hodge star $\star_6$ is with respect to coordinates in the $\mathbb R^{1,5}$-plane.

% 0612201
To realise the $AdS_3\times S^3\times S^3\times S^1$ background we need an additional $D5$ brane, that is a $D1/D5/D5'$ system.  The brane configuration is summarised in table \ref{table:D1D5_system}, now including the last row. The derivation of the background metric and fields is very similar and we refer the reader to \cite{deBoer:1999gea,Gukov:2004ym} for more details.

%%%%%%%%%%%%%%%%
\subsubsection*{Near-horizon geometry and supergravity}
To enter the supergravity regime, we need to demand that the string length is much smaller than the string scale (corresponding to the ``point-like limit'') and by turning off all quantum fluctuations, this corresponds to taking
\begin{align}
    g_sN_1\,, g_sN_5\gg 1\,,\quad \text{and}\quad N_1,N_5\gg 1\,.
\end{align}
The near-horizon limit \cite{Maldacena:1998bw} means taking $\alpha'\rightarrow 0$ but with the following quantities fixed
\begin{align}\label{eq:near-horizon-limit}
    \frac{r}{\alpha'}=\text{fixed}\,,\quad v_4\equiv \frac{\mathrm{Vol}(M_4)}{(2\pi)^4\alpha'{}^2}=\text{fixed}\,,\quad \frac{(2\pi)^2\alpha'g_S}{\sqrt{\mathrm{Vol}(M_4)}}=\mathrm{fixed}\,.
\end{align}
In this limit the harmonic functions $H_{1/5}$ just become 1 and the metric \eqref{eq:metric_D1/D5} becomes
\begin{align}
    \mathrm d s^2=\frac{r^2}{\alpha'Q_5}(-\mathrm d x_0^2+\mathrm d x_1^2)+ \frac{\alpha'Q_5}{r^2}\mathrm dr^2+\alpha'Q_5\mathrm d\Omega_3^2+\mathrm ds_{T^4}\,.
\end{align}
Rewriting the above metric by changing the coordinates $r=R^2/u$ where $R^2=Q_5\alpha'$, leads to the metric for $AdS_3\times S^3\times T^4$ given by
\begin{align}
    \mathrm d s^2=R\left(\mathrm d s_{AdS_3}^2+\mathrm d s_{S^3}^2\right)+\mathrm ds_{T^4}^2\,,
\end{align}
here we have identified the metric for the three-dimensional sphere and the three-dimensional AdS-space in Poincar\'e coordinates
\begin{align}
	\mathrm d s_{AdS_3}^2= u^{-2}(-\mathrm d x_0^2+\mathrm d x_1^2+\mathrm d u^2)\,.
\end{align}
For the remaining fields supporting the type IIB supergravity solution in \eqref{eq:remainingfields_IIB}, we see that in the near-horizon limit, the D1/D5-systems curved the space into $AdS_3\times S^3\times T^4$ threaded by RR-three form flux and with dilaton field
\begin{align}
    F_{(3)}=2Q_5(\mathrm{Vol}(AdS_3)+\mathrm{Vol}(S^3))\,,\quad e^{-2\Phi}=Q_5/Q_1\,.
\end{align}

\noindent
In turn, the D1/D5/D5' brane set-up leads to a near-horizon geometry with metric
\begin{align}\label{eq:metricAdS3_S3_S3_S1cpt}
	\mathrm d s^2=R_{AdS}^2\mathrm d s^2_{AdS_3}+R_+^2\mathrm d s_{S_+^3}+R_-^2\mathrm d s_{S_-^3}+\mathrm d s^2_{S^1}\,,
\end{align}
supported by the RR three-form flux 
\begin{align}
	F_{(3)}&=2Q_5\left(\mathrm{Vol}(AdS_3)+ \,\mathrm{Vol}(S^3_+)+ \,\mathrm{Vol}(S^3_-)\right)
\end{align} 
and the dilaton is again related to the relative number of D1 and D5 branes $e^{-2\Phi}=Q_5/Q_1$.
	
Demanding the supergravity equations to be satisfied imposes additional relations between the moduli of the geometry. The $AdS_3\times S^3\times S^3\times S^1$ background preserves 16 supersymmetries provided the radii of the two three-spheres, denoted by $R_\pm$, and the AdS radius $l$ are related by a triangle relation \cite{Gauntlett:1998fz}
\begin{align}\label{eq:triangle_S3S3}
	\frac{1}{R^2_+}+\frac{1}{R^2_-}=\frac{1}{R_{AdS}^2}\,,
\end{align}
with $R_{AdS}$ the radius of the $AdS$-space.
Often this identity is parametrised by an angle $\varphi$ or the parameter $\alpha$ as follows
\begin{align}\label{eq:alpha_to_phi}
	\frac{R_{ AdS}^2}{R_+^2}=\cos^2\varphi=\alpha\,,\quad \frac{R_{AdS}^2}{R_-^2}=\sin^2\varphi\,.
\end{align}
Anticipating what is to come, the value $\cos^2\varphi=\alpha$ will precisely match the parameter $\alpha$ in the superisometry algebra of the $AdS_3\times S^3\times S^3\times S^1$ background appearing in the exceptional Lie superalgebra $\mathfrak{d}(2,1;\alpha)$ underlying its superisometries.  
There are two limiting cases for values of the parameter $\alpha$. When $\alpha \rightarrow 1$, the radius of one of the three-spheres blows up and the other is traded for flat space by being compactified on $T^3$. This limit thus leads to $AdS_3\times S^3\times T^4$. In this case the symmetry algebra degenerates to $\mathfrak{d}(2,1;0)^2=\mathfrak{psu}(1,1|2)^2 $. Another special limit is when $\alpha=1/2$ at which point the two three-spheres are described by equal radii and the symmetry algebra becomes $\mathfrak{osp}(4,2)$.

%%%%%%%%%%%%%%%%
\subsubsection*{T- and S-dual set-ups}
Applying a U-duality transformation to the D1/D5 set-up, and the corresponding type IIB solutions, leads to new set-up of D-branes. These solutions will in general solve supergravity equations of a different supergravity type and, correspondingly with different fluxes, as we now summarise:
\begin{itemize}
    \item  By performing an S-duality,\footnote{Here we restrict to type IIB backgrounds, since S-duality acts only on type IIB supergravity. S-duality has as distinctive properties that it swaps the sign of the dilaton, in effect inverting the string couple constant $g_S$ to $1/g_s$.} the RR-three form flux is traded for an NSNS-three form flux. Or in terms of branes, the D1/D5 system becomes an NS5/F1 brane system, sourcing the corresponding fluxes. The duality transformation trades the geometry and fields in \eqref{eq:metric_D1/D5} and \eqref{eq:remainingfields_IIB}, 
   where the three-form flux $F^{(3)}$ is exchanged for the Kalb-Ramond three-form flux $H_{(3)}$. Note that S-duality maps a strong to weak coupled regime, and vice-versa, since it inverts the string coupling  $g_s$ .
     The resulting geometry takes the form
\begin{gather}
\begin{aligned}
	\mathrm d s^2 &=(H_1H_5)^{-1/2}(\mathrm -d x_0^2+\mathrm d x_1^2)+(H_1H_5)^{1/2}\mathrm d x^i\mathrm d x^j +(H_1H_5)^{1/2}\mathrm d s_{T_4}^2\,,\\
	H_{(3)}&=2 Q_5 \mathrm d \mathrm{Vol}(S^3)+2Q_1e^{-2\Phi}\star_6\mathrm d \mathrm{Vol}(S^3)\,,\\
	 e^{-2\Phi}&=H_5/H_1\,,
\end{aligned} 
\end{gather}
where the $(t,x_5)$-direction spans the worldvolume of the F1-branes and the same $(t,x_5)$-direction along with the $T^4$-directions span the NS5-brane worldvolume. $H_{(3)}$ is the type IIB supergravity three form flux, $\mathrm d \mathrm{Vol}(M)$ stands for the volume element on the manifold $M$ and $\star_6$ is the Hodge star in six dimensions. $H_1$ and $H_5$ are harmonic functions depending on the {radial coordinate $r=\sqrt{\sum_i x_i^2}$} given by the expressions
\begin{align}
	&\quad H_n(r)= 1+Q_n/r^2\,,\quad \text{for}\quad n=1,5\,,
	\\
	Q_1&= (2\pi)^4g_s{\alpha'}^3N_1/\mathrm{Vol}(M_4)\,,\quad Q_5=g_s{\alpha'}N_5\,.
\end{align}
In general, we can conclude that an $AdS_3\times S^3\times M_4$ type II superstring can be realised by a F1/NS5/(NS5')-brane system, or equivalently after duality transformation by a D1/D5(/D5')-brane system. Such a system of F1/NS5/D1/D5 branes then give combined rise to a geometry supported by a mixture of the NSNS- and RR-fluxes, which in the case of the $AdS_3\times S^3\times S^3\times S^1$ reads
\begin{gather}
\begin{aligned}\label{eq:fluxesAdS3_S3_S3_S1}
	F_{(3)}&=2\hat q\left(\mathrm{Vol}(AdS_3)+\cos \phi \,\mathrm{Vol}(S^3_+)+\sin \phi \,\mathrm{Vol}(S^3_-)\right)\,,\\
	H&=2 q\left(\mathrm{Vol}(AdS_3)+\cos \phi \,\mathrm{Vol}(S^3_+)+\sin \phi \,\mathrm{Vol}(S^3_-)\right)\,,
\end{aligned}
\end{gather}
sourced respectively by the D1/D5 and F1/NS5 brane systems.
Since S-duality maps NSNS into RR fluxes, the whole D-brane system is invariant under S-duality. In section \ref{sec:CZ-WZ}, we will see that the parameters $q$ and $\hat q$ are constraint to be related via
\begin{align}
	q^2+\hat q^2=1\,.
\end{align} 	
	In particular, we obtain the single parameter interpolation that was anticipated in the introduction, yielding a one-parameter family of (integrable) supersymmetric backgrounds. When $q=0$, the geometry is purely supported by RR three-form flux, while at $q=1$ we are in presence of a pure NSNS solution. 
	\item Performing a T-duality along one of the (compactified) $M_4$ directions maps the type IIB to a type IIA background. Remember that for a T-duality acting on an single direction, a Neumann boundary condition is turned into a Dirichlet boundary condition, and vice-versa. Under a T-duality transformation with respect to, e.g.,  a direction transversal to the D$p$-brane, we thus obtain a D$(p-1)$-brane, where the transverse direction has been swapped for a direction along the world-volume of the D-brane. 	
\end{itemize}

	\begin{centering}
	\begin{tcolorbox}
	\begin{exercise}
	    Find the succession of T-duality transformation that, when applied to the D1/D5 set-up in table \ref{table:D1D5_system} (that is forget the last row of D5'-branes), turns the D5-brane into the D1, and vice-versa.
	\end{exercise}  
	\end{tcolorbox}	
	\end{centering}
The radii of the three-sphere $S^3$ and $AdS_3$ are essentially related to the string tension that is in turn determined by the level of the NSNS-flux $k$ and the RR coupling $h$ by
\begin{equation}\label{eq:string-tension}
T=\frac{R^2}{2\pi \alpha'}=\sqrt{h^2 +\frac{k^2}{4\pi^2}}\,,
\end{equation}
where $h$ is a continuous and $k$ is quantised. We will return to this equation and its relation to the parameters $q$ and $\hat q$ introduced above in section \ref{sec:saskia:wzwterm}. As mentioned in the introduction and explained in section \ref{s:sib:bosonic} for large values of the string tension $T\gg 1$ one enters the supergravity approximation. 

%%%%%%%
\subsubsection*{Global symmetries and symmetry algebras}
The bosonic isometry group for the curved part of the $AdS_3\times S^3\times T^4$  supergravity background is described by two copies of the algebra $\mathfrak{su}(1,1)\oplus \mathfrak{su}(2)$. This is not a coincidence. $AdS_3$ backgrounds are dual to two dimensional conformal field theories. This factorised structure of the global symmetry group underlying $AdS_3$/CFT$_2$ simply reflects the two sectors of the dual CFT$_2$: each copy of two the non-compact subalgebras $\mathfrak{su}(1,1)$ accounts for the left- and right-movers. Combining the two copies and the four $\alg{u}(1)$s coming from the torus directions we have the isomorphism
\begin{align}\label{eq:bosonic_isom_AdS3_S3_T4}
[\mathfrak{su}(1,1)\oplus \mathfrak{su}(2)]_\L\oplus [\mathfrak{su}(1,1)\oplus \mathfrak{su}(2)]_\R\oplus \mathfrak{u}(1)^4\cong \mathfrak{so}(2,2)\oplus \mathfrak{so}(4)\oplus \mathfrak{u}(1)^4\,.
\end{align}
The first factor on the right-hand side is isomorphic to the (global) conformal symmetry group in two dimensions, while the second factor accounts holographically for its R-symmetry group. We will now discuss both factors in algebraic detail.

\begin{centering}
		\begin{tcolorbox}
	\begin{exercise}%1505.06767
Convince yourself of the above isomorphisms.
\end{exercise}	
\end{tcolorbox}	
\end{centering}
\noindent
Writing $\gen{L}_0, \gen{L}_\pm$ for the generators of  $\mathfrak{su}(1|1)=\mathfrak{sl}(2)$ and $\gen{J}_3,\gen{J}_\pm$ for the generators of $\mathfrak{su}(2)$, the corresponding commutation relations are 
\begin{gather}
\begin{aligned}\label{eq:bos_comm_rel_psu112}
    [\gen{L}_0,\gen{L}_\pm]=\mp \gen{L}_\pm\,,&\quad [\gen{L}_+,\gen{L}_-]=2\gen{L}_0\,,\\
    [\gen{J}_3,\gen{J}_\pm]=\pm \gen{J}_\pm\,,&\quad [\gen{J}_+,\gen{J}_-]=2\gen{J}_3\,.
\end{aligned}
\end{gather}
The four-torus $T^4$ of the geometry admits in addition a local rotation symmetry, which we denote by $\mathfrak{so}(4)_{T^4}$. The presence of this symmetry group can be directly inferred from the D1-D5 system in table \ref{table:D1D5_system}. The brane configuration is localised at a point in the four spatial dimensions and the system is invariant under the associated $SO(4)$ rotational group. Holographically, this symmetry group can be identified with $SO(4)\cong SU(2)_L\times SU(2)_R$, where the $SU(2)$'s are the left- and right moving R-symmetry groups. Although this is not part of the isometry group, when decomposed as $\mathfrak{so}(4)_{T^4}=\mathfrak{su}(2)\oplus \mathfrak{su}(2)$, this local symmetry will be crucial for unraveling the structure of the S-matrix, as will be the topic of section \ref{s:fiona}.

Similarly, the bosonic isometric group for the $AdS_3\times S^3\times S^3\times S^1$ supergravity background has the bosonic isometry algebra
\begin{align}\label{eq:bosonic_isom_AdS3_S3_S3_S1}
   \mathfrak{so}(2,2)\oplus \mathfrak{so}(4)\oplus \mathfrak{so}(4)\oplus \mathfrak{u}(1) \cong [\mathfrak{su}(1,1)\oplus \mathfrak{su}(2)\oplus \mathfrak{su}(2)]^2_{\L/\R}\oplus \mathfrak{u}(1)\,.
\end{align}
\noindent
Provided that the radius moduli satisfy the respective equalities in eqs. \eqref{eq:triangle_S3S3}, we also have that he backgrounds  $AdS_3\times S^3\times M_4$ with $M_4= S^3\times S^1$ or $T^4$ have 16 (real) supersymmetries. Combining the algebra of the corresponding supercharges together with the bosonic isometry groups leads to the Lie superalgebras
\begin{align}
    AdS_3\times S^3\times T^4&: \; \mathfrak{psu}(1,1|2)\oplus \mathfrak{psu}(1,1|2)\,,\\    AdS_3\times S^3\times S^3\times S^1&:\; \mathfrak{d}(2,1;\alpha)\oplus \mathfrak{d}(2,1;\alpha)\,.
\end{align}
For later use, let us spell out the generators and super-commutation relations of the $\mathfrak{psu}(1,1|2)\oplus \mathfrak{psu}(1,1|2)$ superisometry algebra governing superstrings propagating in an $AdS_3\times S^3\times T^4$-background. The superalgebra $\mathfrak{psu}(1,1|2)$ possesses, besides the six bosonic generators $(\gen{L}_0,\gen{L}_+,\gen{L}_-,\gen{J}_3,\gen{J}_+,\gen{J}_-)$ (with commutation relation given by eq. \eqref{eq:bos_comm_rel_psu112}), also eight fermionic generators $\gen{Q}_{a\alpha A}$, where $a,\alpha,A\in \{\pm\}$ with non-vanishing fermionic and mixed commutation relations 
\begin{gather} \label{eq:symAdS3}
	\begin{aligned}
		\{\gen{Q}_{\pm + A},\gen{Q}_{\pm + B}\}=\pm\epsilon_{AB}\gen{L}_{\pm} \,,&\qquad \{\gen{Q}_{+\pm A},\gen{Q}_{-\pm B}\}=\mp\epsilon_{AB}\gen{J}_{\pm}\,,\\
		\{\gen{Q}_{+\pm A},\gen{Q}_{-\mp B}\}&=\epsilon_{AB}\big(-\gen{L}_0\pm \gen{J}_3\big)\,,\\
    [\gen{J}_3,\gen{Q}_{a\pm A}]=\pm\frac{1}{2}\gen{Q}_{a\pm A}\,,&\qquad
    [\gen{J}_\pm,\gen{Q}_{a\mp A}]=\gen{Q}_{a\pm A}\,,\\
    [\gen{L}_0,\gen{Q}_{\pm\alpha A}]=\pm\frac{1}{2}{\gen{Q}}_{\pm\alpha A}\,,&\qquad [\gen{L}_\pm,\gen{Q}_{\mp\alpha A}]={\gen{Q}}_{\pm\alpha A}\,,
	\end{aligned}
\end{gather}
where $\epsilon_{AB}$ denotes the anti-symmetric $\epsilon$-symbols with $\epsilon_{+-} = 1$.

The superalgebra $\mathfrak{d}(2,1;\alpha)$ is in turn\footnote{Note that in the following equations,  the parameter $\alpha$ in the algebra $\mathfrak{d}(2,1;\alpha)$ should not be confused with the fermionic index $\alpha$.} spanned by nine bosonic generators $\gen{T}_{A'}$ for $A'=0,\dots, 8$ and eight fermionic generators $\gen{Q}_{\alpha'}$ for $\alpha'=1,\dots,8$. Since we have two copies, this tallies to 16 fermionic and 16 bosonic generators. The bosonic subalgebra of $\mathfrak{d}(2,1;\alpha)$ coincides with three commuting copies of $\mathfrak{sl}(2)$
\begin{align}\label{eq:comm_rel_bos_d12a}
	[\gen{T}_{A'},\gen{T}_{B'}]=\varepsilon_{A'B'C'}\gen{T}^{C'}\,,\quad \text{for}\quad A',B',C'=0,\dots,8\,,
\end{align}
with $\varepsilon_{ABC}$ an antisymmetric tensor with non-zero entries $\varepsilon_{012}=\varepsilon_{345}=\varepsilon_{678}=1$. The fermionic and mixed commutation relations are given by
\begin{gather}
	\begin{aligned}
	\,[\gen{T}_{A'},\gen{Q}_{\alpha'}]&=-(-i)^{A'}\tfrac{i}{2}\gen{Q}_{\beta'}(\tilde\gamma_{A'})^{\beta'}{}_{\alpha'}\,,\\
	\{\gen{Q}_{\alpha'},\gen{Q}_{\beta'}\}&= (\tilde C\gamma^a)_{\alpha'\beta'}\gen{T}_a+i\cos ^2\phi (\tilde C\tilde \gamma^{\hat a})_{\alpha'\beta'}\gen{T}_{\hat a}+i\sin^2\phi (\tilde C\tilde \gamma^{a'})_{\alpha'\beta'}\gen{T}_{a'}\,,\label{eq:comm_rel_ferm_d12a}
\end{aligned}
\end{gather}
where $(-i)^a=-i$, $(-i)^{\hat a}=1=(-i)^{a'}$ and the eight-by-eight matrices $\tilde \gamma$ and $\tilde C$ can, using the index notation in eq. \eqref{eq:AdS3_S3_S3_S1_indices} , be realised by
\begin{align}
	\tilde \gamma^a&=\rho^a\otimes \mathbb 1\otimes \mathbb 1\,,\quad \rho^a=(i\sigma^2,\sigma^1,\sigma^3)\\
	\tilde \gamma^{\hat a}&= \mathbb 1\otimes \rho^{\hat a}\otimes\mathbb 1\,,\quad \rho^{\hat a}=(\sigma^1,\sigma^2,\sigma^3)\\
	\tilde\gamma^{a'}&=\mathbb 1\otimes \mathbb 1 \otimes \rho^{a'}\,,\quad\rho^{a'}=(\sigma^1,\sigma^2,\sigma^3)\\
	\tilde C&=\sigma^2\otimes \sigma^2\otimes \sigma^2\,.
\end{align} 
We see that in the commutation relations in eq. \eqref{eq:comm_rel_ferm_d12a}, the free parameter $\alpha$ characterising the algebra makes it appearance in the mixed commutation relation.

In what follows, we will thus use the following choices of labels for the coordinates of the coset space
\begin{align}\label{eq:AdS3_S3_S3_S1_indices}
	\underbrace{AdS_3}_{a=0,\dots,2}\times\underbrace{S^3}_{\hat a=3,4,5}\times \underbrace{S^3}_{a'=6,7,8}\times S^1\,.
\end{align}
The backgrounds $AdS_3\times S^3\times S^3\times S^1$ and $AdS_3\times S^3\times T^4$ are related by the limits detailed in section \ref{saskia:sec:AdS3S3M4_bckgrds}, now via their superisommetry groups as well. To see this, remember that the commutation relations for a single copy $\mathfrak{d}(2,1;\alpha)$ of the superisometry group for the  $AdS_3\times S^3\times S^3\times S^1$ background as given in eqs. \eqref{eq:comm_rel_bos_d12a} and \eqref{eq:comm_rel_ferm_d12a}. When one takes the limit $\phi\rightarrow 0$ or $\alpha\rightarrow 1$, the radius of the one of the spheres and the $AdS$ radius coincide, while the second sphere `blows up'. Indeed, consider the generators of the left sphere $\gen{T}_{\hat a}$ and introduce the sphere radius $R_-$ by rescaling the generators $\gen{T}_{\hat a}\mapsto R_-\gen{T}_{\hat a}$. According to eq. \eqref{eq:alpha_to_phi}, we need to take $R_-$ to infinity in the commutation relations of $\mathfrak{d}(2,1;\alpha)$ above, leading to the new (anti-)commutation relations
\begin{align}
	[\gen{T}_{\hat a},\gen{T}_{\hat b}]&=0\,,\\
	\{Q_{\alpha'},Q_{\beta'}\}&= (\tilde C\gamma^a)_{\alpha'\beta'}\gen{T}_a+i (\tilde C\tilde \gamma^{\hat a})_{\alpha'\beta'}\gen{T}_{\hat a}\,.
\end{align}
These can been recognised as the commutation relations for the superalgebra  $\mathfrak{psu}(1,1|2)$ (together with three commuting generators $S_{\hat a}$). In fact, this is a basic property of the superalgebra $\mathfrak{d}(2,1;\alpha)$ when taking $\alpha\rightarrow 1$, see e.g. \cite{Frappat:1996pb}.

%%%%%%%%%%%%%%%%
\subsubsection*{Dual CFTs} 
According to the AdS/CFT conjecture, superstring theory on $AdS_3\times S^3\times M_4$ is expected to be holographically dual to a two-dimensional superconformal field theory. 
The holographic dual to the superstring on $AdS_3\times S^3\times M_4$-backgrounds remains however largely mysterious. 
If we investigate the duality starting from the string side, it is relatively easy to work out the description of pure-NSNS backgrounds as WZW models, and this can give us some insight on their holographic duals.
In presence of both NSNS and RR flux this is much harder, and we know very little about the dual models. Some intuition about their properties may be obtained by general AdS/CFT arguments and by the fact that they should be, in a suitable sense, marginal deformations of those appearing at the WZW points. Furthermore, very recently, integrability provided some quantitative insight on some of these models.

Let us start by briefly reviewing the AdS/CFT intuition. When the length scale describing the manifold $M_4$ is small compared to the $S^1$ part of the $\mathbb{R}^{1,4}\times S^1\times T^4$ geometry where we placed the D-branes, the low-energy dynamics is described by a $1+1$ dimensional supersymmetric gauge theory with $U(N_1)\times U(N_5)$ gauge group. In the IR this gauge theory flows to the candidate dual CFT. This CFT should have $\mathcal N= (4,4)$ superconformal symmetry and the central charge equal to $c=6N_1 N_5$, see e.g. \cite{deBoer:1999gea,Gukov:2004ym}.
The $\mathfrak{psu}(1,1|2)^2$ algebra which we discussed above is the global part of the infinite-dimensional $\mathcal N= (4,4)$ algebra, just like $\su(1,1)^2$ is the global part of the Virasoro algebra of a an ordinary bosonic CFT.
Similar considerations also hold for the D1-D5-D5' construction, but they yield instead the so-called \textit{large} $\mathcal N= (4,4)$ algebra, whose global part is $\mathfrak{d}(2,1;\alpha)^2$.

The case involving the compact space $T^4$ has always been the most studied and the best understood. Since the beginning of the study of AdS/CFT, it was conjectured that the dual CFT of this set-up should be a symmetric product orbifold CFT based on~$T^4$~\cite{Dijkgraaf:1998gf,deBoer:1998us,deBoer:1999gea,Maldacena:1999bp,Larsen:1999uk,Seiberg:1999xz,Argurio:2000tb,Gukov:2004ym}.
This means considering a free supersymmetric $\mathcal{N}=(4,4)$ theory of on~$T^4$, taking its $N$-fold tensor product, and imposing that the theory is invariant under the action of the symmetric group~$S_N$.%
\footnote{Such orbifold CFTs are well studied in string theory, see e.g.~\cite{Arutyunov:1997gt,Lunin:2000yv}.} 
The central charge of the model is then $c=6N$, and in order to compare this with perturbative string theory we are interested in the $N\gg1$ limit. Indeed, the permutation orbifold structure is important in order to reproduce not only the central charge, but also other generic features of a good holographic CFT, such as its density of states~\cite{Belin:2014fna}. Moreover, some robust features of this holographic duality, such as the protected spectrum, can be matched against that of the symmetric-product CFT. Similar arguments suggest that the symmetric-product orbifold of K3 and $S^3\times S^1$ should appear as duals of the more general holographic setups involving such compact spaces.
All this leaves an important open question: where, in the vast parameter space of \textit{e.g.} $AdS_3\times S^3\times T^4$ strings should the symmetric-product orbifold of $T^4$ sit?

We can loosely compare this situation with the well-known duality between $AdS_5\times S^5$ strings and $SU(N_c)$ $\mathcal{N}=4$ SYM. The analogue of the symmetric-product orbifold point is where the dual CFT is free and planar. This would be $N_c\gg1$, with the 't Hooft coupling~$\lambda=0$. In terms of string theory, this should be the limit where strings have vanishingly small tension.
Things are more complicated for the case of $AdS_3\times S^3 \times T^4$, because we have more parameters than $N_c$ and $\lambda$ --- in particular, the string tension is sourced by both RR and NSNS field strengths.

The answer to this question came from studying quantitatively the spectrum of non-protected states in string theory, starting from the worldsheet. The first indications by studying the supersymmetric WZW models at the lowest value of the level~$k=1$, without any RR flux~\cite{Giribet:2018ada,Gaberdiel:2018rqv}. Because of some subtleties with the WZW formalism at this particular value of the level (in order to decouple the fermions, one formally ends up with a $\su(2)_{-1}$ Ka\v{c}-Moody algebra, where the level is negative), a better description is provided by the hybrid formalism, which we present in section~\ref{s:bob}; this allowed a detailed check of the duality~\cite{Eberhardt:2018ouy}.
Already when considering the case of no RR flux and larger level $k=2,3,4,\dots$, the construction of the holographic dual is more involved, see~\cite{Eberhardt:2021vsx}.

But how to turn on the RR flux? In principle, this can be done in conformal perturbation theory by singling out the correct marginal deformation in the spectrum of the underformed symmetric-product orbifold CFT.
There have been multiple efforts in this direction. In~\cite{Pakman:2009mi}, this idea was pursued to the end of constructing an integrable spin-chain of the type of the Minahan-Zarembo one~\cite{Minahan:2002ve}, which however here turns out to be substantially more involved.
In~\cite{David:2010yg}, the algebraic structure behind integrability was studied for the purpose of building an S~matrix in a spirit of the one of Beisert~\cite{Beisert:2005tm}; however, the precise identification of the representations of the integrable symmetries was not entirely correct (this was corrected in~\cite{Borsato:2012ud,Borsato:2014hja}). More recently, the study of the integrable symmetries from the symmetric-product orbifold was revisited and a  first-order perturbation theory computation was done in~\cite{Gaberdiel:2023lco}; as explained in~\cite{Frolov:2023pjw} this reproduces the weak-RR-flux expansion of the integrable symmetries and S~matrix which were previously bootstrapped in~\cite{Lloyd:2014bsa} and which we will discuss in section~\ref{s:fiona}.
This perturbative approach is sure to produce important data to check any worldsheet-based prediction. It should however be noted that the conformal perturbation theory has its limitations; at higher orders,  the explicit computations are fairly involved, see see for instance~\cite{Apolo:2022fya,Hughes:2023fot} and references therein, as well as~\cite{Pakman:2009zz} for a different (diagrammatic) approach. With the current ``technology'', it seems unlikely that we will be able to obtain fifth-order corrections to the energies as it was (remarkably) the case for $AdS_5\times S^5$, see~\cite{Eden:2012fe}.

One may approach this question from the worldsheet. Unfortunately, the hybrid formalism also seems to become unwieldy in this case, as we shall review below.
The case of RR flux (or mixed RR/NSNS flux) is precisely where integrability is expected to be helpful. In fact, thanks to the construction of the mirror TBA which we will review in section~\ref{ana:sec}, it was possible to quantitatively study the tensionless limit of the $k=0$ case, where all the tension is sourced by the RR fluxes~\cite{Brollo:2023pkl} (which play a role very similar to the 't~Hooft coupling~$\lambda$).
At precisely zero tension, infinitely many states are degenerate with each other, but they lift starting at linear order in the tension. Interestingly, at this order the linear contribution comes from the $T^4$ modes and their superpartners. The energy spectrum is not, however, that of a symmetric-product orbifold CFT; in fact, the dynamics even at leading order in the tension is that of an interacting theory of magnons, whose precise dynamics is not yet understood (at least, not in terms of a dual Hamiltonian)~\cite{Baglioni:2023zsf}.
This goes to show that, in perturbative string theory, the ``tensionless'' limit can be drastically different depending on the fluxes used to realise it.

%%%%%%%%%%%%%%%%%%%%%%%%%%%%%%%%%%%%%
%% Equivalence coset and sector GS %%
%%%%%%%%%%%%%%%%%%%%%%%%%%%%%%%%%%%%%
\subsection{Supercoset action for \texorpdfstring{$AdS_3 \times S^3\times M_4$}{AdS3xS3xM4}}\label{sec:saskia:AdS3coset}
In this section we will show how $AdS_3\times S^3\times M_4$-backgrounds supported by pure RR-flux, for $M_4=T^4$ and $M_4=S^3\times S^1$, can be realised as semisymmetric supercoset sigma-models (SSSM). The coset will admit a $\mathbb Z_4$-grading warranting, as reviewed in section \ref{s:lax-ssssm}, the existence of a Lax connection and the classical integrability of the action. A generic D-brane configuration sourcing $AdS_3$ holographic backgrounds leads however to supergravity solutions supported not only by RR fluxes but also by NSNS fluxes. In particular, the semisymmetric supercoset action presented in section \ref{sec:SSSM_action} cannot be the final answer. In order to describe the mixture of RR and NSNS fluxes sustaining the $AdS_3$-backgrounds, one has to be able to add a Wess-Zumino-term modelling the NSNS fluxes. Remarkably, supercosets underlying $AdS_3\times S^3\times S^3\times S^1$ and $AdS_3\times S^3\times T^4$ are part of a special family of supercosets for which it is possible to add a WZ-term. We will see that, within the supercoset formulation, the equations of motion for the superstring propagating in $AdS_3\times S^3\times M_4$ admit a Lax representation, establishing the classical integrability of these backgrounds even after the inclusion of NSNS fluxes. Along the way we will however gradually come to realise that the supercoset formulation suffers from a  serious drawback: it is not a complete description of the Green-Schwarz superstring.

%%%%%%%%%%%%%%%%%%%%%%%%%%%%%%%%%%%%%
%%%%%%%%%%%%%%%%%%%%%%%%%%%%%%%%%%%%%
\subsubsection{Supercoset and integrability}
Section \ref{s:lax-ssssm} showed that, provided the string background can be written as a (super)coset manifold and admits $\mathbb Z_4$-invariant supercoset action, the corresponding superstring equations of motion are automatically classically integrable.  In this section we will apply the supercoset construction to the particular case of $AdS_3\times S^3\times M_4$ backgrounds.

%%%%%%%%%%%%%%%%
\subsubsection*{Permutation cosets}
Looking back at the discussion of the global symmetry structure of $AdS_3$ backgrounds, the observant reader might have noticed that in both cases the symmetry group is of the form $G=H\times H$. As was already discussed earlier, holographic $AdS_3$-backgrounds that are homogeneous space, will automatically have a supercoset realisation of the form $H\times H/H_0$, where the stabeliser subgroup $H_0$ is bosonic and has to be a fixed by the $\mathbb Z_4$ automorphism.

The crucial point now is that this type of cosets, called ``permutation cosets'' (for reasons which become clear soon) can naturally be endowed with a $\mathbb Z_4$ structure. This can be made explicit by constructing the $\mathbb Z_4$ automorphism as a (super-)matrix $\Omega$ acting on elements $(X_L,X_R)$ of the direct sum algebra $\mathfrak h\oplus \mathfrak h$
\begin{align}\label{eq:AdS3:supercoset_autorm}
	\Omega=\begin{pmatrix}
		0&\mathrm{id}\\ (-1)^F &0
	\end{pmatrix}\,,
\end{align}
where $(-1)^F$ is the fermionic parity operator.  
The operator $\Omega$ effectively permutes the two factors in the Lie algebra $G$ (hitting one of the factors with  fermion parity operator $(-1)^F$ on the way), hence the name ``permutation supercosets''. One can easily check that the fourth power is the identity, $\Omega^4=1$ and preserves the (anti-)commutative relations of $\mathfrak h\oplus\mathfrak h$.

\begin{centering}
\begin{tcolorbox}
	\begin{exercise}
Check by direct computation that the $\mathbb Z_4$-decomposition of the symmetry algebra $\mathfrak g=\mathfrak g^{(0)}\oplus \mathfrak g^{(1)}\oplus \mathfrak g^{(2)}\oplus \mathfrak g^{(3)}$ according to the grading defined by $\Omega$, \textit{i.e.}~$\Omega(\mathfrak h^{(k)})=i^k\mathfrak h^{(k)}$.
%\noindent
%Solution:
%\begin{align}
%    \mathfrak h^{(0)}&=\{(X,X)\mid X\in \mathfrak h_\mathrm{bos}\}\\
%    \mathfrak h^{(1)}&=\{(X,iX)\mid X\in \mathfrak h_\mathrm{ferm}\}\\
%    \mathfrak h^{(2)}&=\{(X,-X)\mid X\in \mathfrak h_\mathrm{bos}\}\\
%    \mathfrak h^{(3)}&=\{(X,-iX)\mid X\in \mathfrak h_\mathrm{ferm}\}
%\end{align}
\end{exercise}
\end{tcolorbox}
\end{centering}
\noindent
In the previous exercise, you should have identified the subalgebras 
\begin{align}
   \mathfrak h^{(0)}&=\{(X,X)\mid X\in \mathfrak h_\mathrm{bos}\}\,,\\
   \mathfrak h^{(1)}&=\{(X,iX)\mid X\in \mathfrak h_\mathrm{ferm}\}\,,\\
   \mathfrak h^{(2)}&=\{(X,-X)\mid X\in \mathfrak h_\mathrm{bos}\}\,,\\
   \mathfrak h^{(3)}&=\{(X,-iX)\mid X\in \mathfrak h_\mathrm{ferm}\}\,.
\end{align}
In particular, the bosonic subalgebra $\mathfrak h _0$ corresponds to the diagonal subalgebra of $\mathfrak h\oplus \mathfrak h$ with element in the Grassmann-even or bosonic part of the algebra:
\begin{align}
		 \mathfrak{h}_0=\{(X,X)\mid X\in \mathfrak h_\mathrm{bos}\}\,.
\end{align}
Note that the bosonic part of this coset space, that is $H_\mathrm{bos}\times H_\mathrm{bos}/ H_0$ is the bosonic subgroup $H_\mathrm{bos}$ itself, as it should be.

%%%%%%%%%%%%%%%%
\subsubsection*{$AdS_3\times S^3(\times S^3)$-supercosets}
With these elements in hand, together with the global symmetry groups discussed in the previous section, the supercoset space is easily identified. The supercoset sigma model for the six-dimensional part $AdS_3\times S^3$ is constructed starting from the superisometry group $G$ with background bosonic part $G_\mathrm{bos}$ and subgroup $H_0\subset G_\mathrm{bos}$ such that $G_\mathrm{bos}/H_0$ is $AdS_3\times S^3$
\begin{align}\label{eq:coset_AdS3_S3}
	\frac{G}{H_0}=\frac{PSU(1,1|2)\times PSU(1,1|2)}{SU(1,1)\times SU(2)}=AdS_3\times S^3+\text{ fermions}
\end{align}
with the bosonic part of the coset being 
\begin{align}
	G=\frac{[SU(1,1)\times SU(2)]^2}{SU(1,1)\times SU(2)}=\frac{SO(2,2)}{SO(1,2)}\times \frac{SO(4)}{SO(3)}\,.
\end{align}
Adding an additional $S^3$-factor to the $AdS_3\times S^3$-geometry enhances the supercoset to a one-parameter family of supercosets, where the parameter $\alpha$ captures the relative sizes between the two $S^3$ (see eq. \eqref{eq:alpha_to_phi}). That is, the sigma-model for $AdS_3\times S^3\times S^3$ is realised by the coset based on an exceptional Lie superalgebra
\begin{align}\label{eq:coset_AdS3_S3_S3}
\frac{G}{H_0}=\frac{D(2,1;\alpha)\times D(2,1;\alpha)}{SL(2)\times SU(2)\times SU(2)}=AdS_3\times S^3\times S^3 + \text{fermions}\,.
\end{align}
Falling short of being ten dimensional backgrounds, neither of these two cosets can be proper superstring backgrounds, and we know that we have to append flat directions $T^4$ or $S^1$, respectively.

For the supercoset space in equations \eqref{eq:coset_AdS3_S3} and \eqref{eq:coset_AdS3_S3_S3} respectively, we can now construct the corresponding SSSSM action given in eq. \eqref{eq:SSSSM1}. After taking their appropriate bosonic group element  $g_B$ parametrising $SU(1,1)\times SU(2)$ one can derive the metric \eqref{eq:metric_form_SSSSM}, as e.g. in exercise \ref{ex:2-sphere-metric} by using the commutation relations (or faster: with Mathematica).
 
%%%%%%%%%%%%%%%%%
%\subsubsection*{The addition of flat directions }
The two supercoset spaces in \eqref{eq:coset_AdS3_S3} and \eqref{eq:coset_AdS3_S3_S3} only account for a six and nine dimensional spacetime, respectively. The flat directions, \textit{i.e.}~the remaining $T^4$, respectively $S^1$, factor expected from the near-horizon limit of the D-brane set-up that complete the supergravity backgrounds have to be added `by hand' \cite{Berkovits:1999im,Babichenko:2009dk}
\begin{align}
	\frac{D(2,1;\alpha)\times D(2,1;\alpha)}{SL(2)\times SU(2)\times SU(2)}\times U(1)\quad \text{and}\quad \frac{PSU(1,1|2)\times PSU(1,1|2)}{SL(2)\times SU(2)}\times U(1)^4\,.
\end{align}

%%%%%%%%%%%%%%%%
\subsubsection*{Supersymmetry}
It remains to be checked that the supercosets display the expected number of fermionic degrees of freedom to account for the sixteen supersymmetries present in the corresponding supergravity solutions. To do so, we have to distinguish between the two $AdS_3\times S^3$-backgrounds.

When $M_4=S^3\times S^1$, the coset space $AdS_3\times S^3\times S^3$ is described by the exceptional algebra $\mathfrak{d}(2,1;\alpha)$ and its supercoset space has 16 fermionic generators. Computing the corresponding kappa-symmetry, see e.g. \cite{Zarembo:2010sg}, one finds that its rank is zero and the space is already fully gauge fixed with respect to the fermionic gauge invariance and the number of fermionic degrees of freedom remains 16.

When $M_4=T^4$, the story is a little more subtle.  The six-dimensional supercoset space 
\begin{align}
    AdS_3\times S^3\cong PSU(1,1|2)\times PSU(1,1|2)/SU(1,1)\times SU(2)
\end{align} 
describes 16 fermionic degrees of freedom, but this time has kappa-symmetry rank that is non-zero. After fixing the kappa-symmetry gauge symmetry we are left with only eight physical degrees of freedom. Although this is the correct number for a six-dimensional GS action, we expect twice as many in a 10-dimensional GS action when the flat $T^4$-direction is added.  At first, this seems to pose somewhat of a conundrum as, in the conformal gauge, the additional $T^4$ decouples the equation of motion from the curved supercoset part of the geometry. Key to the solution \cite{Babichenko:2009dk}, is that the total metric transforms non-trivially under kappa-symmetry, see eq. \eqref{eq:kappa_var_metric}, coupling effectively the fermionic variations on both portion of the geometry. Indeed, the additional extra bosons coming from the $T^4$ invalidate the original  kappa-symmetry gauge fixing that gauged away eight fermionic degrees of freedom. As a result, the sixteen original fermionic degrees of freedom are effectively reinstated in the coset plus $T^4$ background.

%%%%%%%%%%%%%%
\subsubsection*{Lax connection and integrability }
From the above we can thus conclude that both the $AdS_3\times S^3\times S^3\times S^1$ and $AdS_3\times S^3\times T^4$ backgrounds admit a supercoset realisation with a $\mathbb Z_4$ automorphism $\Omega$ given in eq. \eqref{eq:AdS3:supercoset_autorm}. One can thus follow the steps spelled out in section \ref{sec:SSSM_action}: using the $\mathbb Z_4$ automorphism to decompose the zero-curvature current one-form $J$ as given in eq. \eqref{eq:lax-ssssm}, ensuring the integrability of the supercoset action. Note, that for the matter of classical integrability or the existence of a Lax connection, the flat directions do not play any role.

%%%%%%%%%%%%%%
\subsubsection{Matching to quadratic gauge-fixed GS action}\label{sec:equiv_w_gfGS}
The attentive reader might worry whether the $AdS_3\times S^3$-supercoset action we constructed above actually coincides with the Green-Schwarz superstring action for the same backgrounds. This is indeed not  obviously the case. In the supercoset action, the flat-directions $T^4$ or $S^1$ are by construction, but up to Virasoro constraints, decoupled from the curved or coset-part, \textit{i.e.}~$AdS_3\times S^3$ or $AdS_3\times S^3\times S^3$. From this observation we have to conclude that, to have a chance to coincide with the supercoset action, the Green-Schwarz action should be in a particular gauge. This gauge has to be chosen such as to decouple the flat directions from the rest of the geometry. The apparent issue is that for the Green-Schwarz action this is not the case. The GS action  eq. \eqref{eq:GS-curved-quadratic} contains a kinetic term  which couples the $M_4$ directions with the worldsheet fermions via terms of the form
\begin{align}
\bar\theta^Ie_\alpha{}^A\Gamma_A\partial_J\theta^K\,.
\end{align}
Generically, it is simply not consistent to have any bosonic directions decoupling from the rest since all Dirac matrices in the kinetic term are non-zero\footnote{This is in stark contrast to the hybrid formalism which will be the topic of section \ref{s:bob}. There the four-torus $T^4$ can be added independently since it is completely orthogonal to the remaining non-linear part of the action.}. Remember however that the GS action has to first be kappa-gauge fixed, halving the initial number of fermionic degrees of freedom. This opens the possibility, by picking a wisely chosen gauge-fixing, to decouple the fermions which couple to the bosons, leading to the decoupled flat directions.
It was shown in \cite{Babichenko:2009dk} that there indeed exists a kappa-gauge fixing for the (quadratic) GS action for which the flat direction decouple. For the demonstration of how the two actions coincide in this gauge we refer the reader to the original paper \cite{Babichenko:2009dk}.  The gauged fixed Green-Schwarz action up to quadratic order in fermions can then be shown to coincide with the supercoset action on $AdS_3\times S^3(\times S^3)$. We will postpone spelling out this particular kappa-gauge fixing to section \ref{sec:coset_kappa_gauge_fix}. Although clearing out on obstacle,  we will also argue in that same section that this gauge fixing choice will pose severe problems when it comes to capturing all the fundamental string excitations in $AdS_3$ holographic backgrounds.

%%%%%%%%%%%%%%%%%%%%%%%%%%%%%%%%%%%%%
%%  Mass spectrum plane-wave limit %%
%%%%%%%%%%%%%%%%%%%%%%%%%%%%%%%%%%%%%
\subsection{Mass spectrum in the plane-wave/BMN limit}\label{sec:saskia:spectrumplanewave}

One of the distinguishing features of $AdS_3$-backgrounds is the presence of massless modes in the lightcone gauge. As mentioned already in the introduction, these formed initially a significant challenge for the implementation of several integrability techniques, which in recent years have been the subject of much attention. In this section, we will see how these massless modes can be detected in the Penrose or BMN limit within the supercoset formalism. The  massless modes will be revisited within the S-matrix formalism in section \ref{s:fiona}.
%%%%%%%%%%%%%%%%%%%%%%%%%%%%%%%%%%%%%
%%%%%%%%%%%%%%%%%%%%%%%%%%%%%%%%%%%%%
\subsubsection{Plane-wave/BMN limit and plane-wave backgrounds}
The hurried reader can safely skip to the next subsection, as the content of this section is not vital to what is to follow. Since the plane-wave/BMN limit is a useful tool when it comes to studying the integrability of holographic backgrounds, we briefly review it here. For a more comprehensive treatment and a more complete list of references, we point the reader towards \cite{Sadri:2003pr,Nastase:2017cxp}  or \cite{Puletti:2010ge} in the context of integrability.

The Green-Schwarz action on generic background is interacting, and thus  hard to quantise. Fortunately, in a certain limit, the so-called Penrose or plane-wave limit, the gauge-fixed GS action becomes free.  On the gauge side of the holographic duality, this limit is called the BMN limit. The name BMN stems from a paper by Berenstein, Maldacena and Nastase \cite{Berenstein:2002jq}, see also \cite{Gubser:2002tv}. There, ``BMN'' established a precise connection between the spectrum of superstring theory on a pp-wave background and a particular class of operators in the gauge theory, the so-called BMN operators. Note that the terms BMN limit and plane-wave limit are often used interchangeably.

A pp-wave geometry is defined as a Lorentzian manifold that admits a covariantly constant null vector field $k$, the direction in which the wave moves: \textit{i.e.}~$\nabla k=0$. In Brinkmann coordinates the generic form of such metric, when taking $k=\partial_v$ is 
\begin{align}
\label{eq:ppwavelineel}
	\mathrm d s^2=-2\mathrm d u\mathrm d v -f(u,x,y)\mathrm d u^2+ \mathrm d x^2 + \mathrm d y^2\,,
\end{align}
where $f$ is a smooth function. The existence of a covariantly constant null Killing vector field $k$ guarantees that this family of backgrounds are in fact $\alpha'$-exact supergravity solutions. When $f$ is just a function of $u$, the metric is that of a plane-wave. This is equivalent to demanding for the existence of a \textit{globally} defined covariantly constant null Killing vector field. 

It was Penrose's insight \cite{penrose1976any} that, close to a null geodesic, any Lorentzian metric looks like the plane-wave metric~\eqref{eq:ppwavelineel}. Later this observation was extended to supergravity backgrounds in ten or eleven dimensional Lorentzian space-time by \cite{Gueven:2000ru}. More precisely, given a null geodesic one can always pick a set of adapted coordinates such that the metric takes on the form
\begin{align}
	\mathrm d s^2 =R^2\left(-2\mathrm d u\mathrm d\tilde v+\mathrm d \tilde v (\mathrm d \tilde v+A_I(u,\tilde v,\tilde x^I)\mathrm d \tilde x^I)+g_{JK}(u,\tilde v, \tilde x^I)\mathrm d\tilde x^J\mathrm d\tilde x^K\right)\,,
\end{align}
where $R$ is a constant, which will enable us to zoom in into the geodesic's path. As one can see, the coordinate $u$ parametrises the null geodesic, whilst $\tilde v$ measures the distance between these geodesics. The remaining coordinates are denoted by $\tilde x^I$. Taking $R$ to infinity, together with the rescalings 
\begin{align}
	\tilde v=v/R^2\,,\quad \tilde x^I=x^I/R\,,
\end{align}
keeping $u,v$ and $x^I$ fixed, effectively zooming in into a region infinitesimally close to the light-like geodesic. The metric in this limit becomes
\begin{align}
	\mathrm d s^2=-2\mathrm d u \mathrm d v+g_{IJ}(u)\mathrm d x^I \mathrm d x^J\,.
\end{align}
Note that the metric $g_{IJ}$ is now only a function of the coordinates.

 Finally, another crucial property of plane-wave solutions is that one can show that they preserve at least half of the maximal number of supersymmetries \cite{Cvetic:2002si}. For the $AdS_5\times S^5$-background, which is maximally supersymmetric, the Penrose limit yields a maximally supersymmetric plane-wave solution \cite{Blau:2001ne}. Actually more is true, the Penrose limit in general does not break any supersymmetry, and in some cases even creates new ones \cite{Itzhaki:2002kh}.

One can include leading quantum corrections to the string energies to the BMN limit. Starting from the lightcone gauge fixed superstring action, one can perform an expansion in higher powers of the fields. This expansion, which is referred to as the near-BMN expansion, can be seen as a perturbation away from the pp-wave background. A detailed and pedagogical account of this limit can be found in \cite{Puletti:2010ge}. This limit will also play a major role in section \ref{s:fiona}.

%%%%%%%%%%%%%%%%%%%%%%%%%%%%%%%%%%%%%
%%%%%%%%%%%%%%%%%%%%%%%%%%%%%%%%%%%%%
\subsubsection{On masses, no masses and other problems}
The fundamental bosonic excitations\footnote{... and by supersymmetry for each bosonic state, we have a corresponding fermionic states.} of the $AdS_3\times S^3\times S^3\times S^1$ superstring in the lightcone gauge are critical to understand why the $AdS_3$-backgrounds are simultaneously a challenge and opportunity to explore integrable systems.
Unlike the $AdS_5\times S^4$ and $AdS^4\times \mathbb CP^4$ backgrounds, the string spectrum of $AdS_3\times S^3\times M_4$ admits modes with different masses and even massless modes. 

The modes of different masses appearing on the $AdS_3\times S^3\times M_4$-spectra were studied using the near-BMN limit in \cite{Gomis:2002qi,Hikida:2002in,Gomis:2002qi,Sommovigo:2003kd,Babichenko:2009dk}.
The masses of the string spectrum can be derived by studying the plane-wave limit of the theory in the lightcone gauge where the metric then becomes~\cite{penrose1976any}
\begin{align}\label{eq:plane_wave_metric_AdS3xS3xS3xS1}
    \mathrm d s^2=-4\mathrm d x^+ \mathrm d x^-+\sum_{i=0}^8 (m_i)^2x_i^2(\mathrm d x^+)^2 +\sum_{i=1}^8\mathrm d x^2_i\,,
\end{align}
which is the pp-wave metric. Its derivation for the specific case of $AdS_3\times S^3\times S^3\times S^1$ can be found in some detail in~\cite{Dei:2018yth}.
This backgrounds has both light and heavy modes: the $AdS_3\times S^3\times S^3\times S^1$ features four type of excitations\footnote{The plane-wave limit of the $AdS_3\times S^3\times S^3\times S^1$ is in fact a one-parameter family of metrics (on top of the parameter $\alpha$). For the sake of simplicity we will neglect this parameter. For more details see \cite{Lloyd:2013wza}.}
\begin{align}
    (m_{1,2})^2=1,\quad (m_{3,4})^2=\alpha,\quad (m_{5,6})^2=1-\alpha,\quad  m_{7,8}=0.
\end{align}
For each mass(less) state above there are two  corresponding excitations (in the CFT picture this reflects the fact that we should have one for each chirality state). We see that for the $AdS_3\times S^3\times S^3\times S^1$, there are six massive states and two massless states. In the massive modes there are `light' ($m^2=\alpha,1-\alpha$) and `heavy' ($m^2=1$) states. One can identify the heavy states, by deriving the finite gap equations \cite{Lloyd:2013wza}, to be composite states of the two light states $m^2=\alpha$ and $m^2=1-\alpha$. The first massless state comes from the $S^1$ factor whilst the second one is a linear combination of the two equatorial directions in the two $S^3$-spheres: $\psi$ and $\vartheta_\perp=\tan \varphi\,\vartheta_++\cot \varphi\,\vartheta_-$. The massless mode along the two three-spheres is often called the coset (massless) boson whilst the one along the $S^1$ is called the non-coset (massless) boson. Taking the $\varphi\rightarrow 0$, the $S^3\times S^1$ part of the geometry becomes flat, like $T^4$, and two new massless modes appear. All massless modes are non-coset modes.

The presence of both fundamental states of different masses and massless modes form a serious challenge to the integrability program. This can already be seen at the level of the supercoset action: the coset formulation has the distinct advantage of making integrability manifest, it fails to capture these massless modes. To see this, let us first for concreteness consider the $AdS_3\times S^3\times S^3\times S^1$. It is clear from how the flat direction $S^1$ had to be added by hand in the supercoset action, that the massless mode associated to that direction will be missing\footnote{Remember that the coset action is only equivalent to the GS superstring action when the $S^1$ part of the geometry is completely decoupled from the coset part. The $S^1$ only interacts with the coset through the Virasoro constraint.}. The absence of the second type of massless mode can be traced back to how the Virasoro constraints are imposed on the $S^3\times S^3$ part of the geometry. We follow here mostly the original paper \cite{Rahmfeld:1998zn,Metsaev:2000mv} which constructed the supercoset of the $AdS_3$-backgrounds. Later, in \cite{Babichenko:2009dk}, it was realised that this formulation overimposed the Virasoro constraints, and that at the level of the algebraic curves a weaker condition can be imposed, that effectively enables one to see the massless modes.

Remember now that the superstring Green-Schwarz action correctly describes the physical fermionic degrees of freedom only upon fixing a kappa-symmetry. %{\color{red}In exercise \ref{ex:kappa-zero-curv}, you will show that although the Virasoro constraints have to be imposed independently of the Lax connection}, demanding the Lax connection to keep satisfying the zero curvature condition under kappa-transformation is equivalent to imposing the Virasoro constraints.
 The supercoset sigma-model is equivalent to a (subsector) of the kappa-gauge fixed Green-Schwarz action and thus subsumes enforced Virasoro constraints. {\color{black} Unfortunately, this condition is too strong and sets the massless mode to zero.}
 To see this, we go to conformal gauge where the Virasoro constraint is equivalent to the vanishing of the total energy-momentum tensor. Restricted to the $S^3\times S^3$ part of the geometry it splits into a contribution for each sphere
\begin{align}
    T^{S^3_+}_{\pm\pm}+ T^{S^3_-}_{\pm\pm}=0\,,
\end{align}
using the notation in eq. \eqref{eq:triangle_S3S3}. However, since the equations of motion factorise for each three-sphere, the two terms vanish independently:  $T^{S^3_{\pm}}_{\pm\pm}=0$. Equivalently (after fixing the residual symmetries of the action in the conformal gauge) the two stress tensors can be fixed to the same but opposite constant, \textit{i.e.}~$T^{S^3_{\pm}}_{\pm\pm}=\pm \mathrm{cst}$. This choice effectively kills any degree of freedom that would have a momentum component along the two spheres, and in particular the second type of the massless modes.

%%%%%%%%%%%%%%%%%%%%%%%%%%%%%%%%%%%%%
%%%%%%%%%%%%%%%%%%%%%%%%%%%%%%%%%%%%%
\subsection{The CZ-WZ-term and integrability with mixed fluxes}\label{sec:CZ-WZ}
In the previous sections, we have seen that the Green-Schwarz action on sufficiently symmetric spaces which are supported by RR-fluxes can be written in terms of a supercoset action. When that action admits a $\mathbb Z_4$ symmetry, the supercoset action becomes a semisymmetric space sigma-model and is automatically integrable.  Supergravity solutions for $AdS_3\times S^3\times M_4$ backgrounds on the other hand admit in the most general case a mixture of RR- and NSNS-fluxes. This in turn raises the question to whether the NSNS flux $H$ can be embedded in the supercoset formulation as well and, if so, whether the resulting action remains (classically) integrable. Tackling the former question first: remember that NSNS fluxes are described by a Wess-Zumino type term, that is a contribution to the action of the form
\begin{align}
	S_{\mathrm{WZ}}\propto \int_B\kappa(J\stackrel{\wedge}{,} [J\stackrel{\wedge}{,} J])\propto \int_{\bar g(B)} H\,,
\end{align}
where $B$ is a three-dimensional surface with boundary the two-dimensional string worldsheet $\partial B=\Sigma$, where $\kappa$ is a non-degenerate bilinear form and $J=g^{-1}\mathrm dg$ is a Maurer-Cartan form characterised by the extension $\bar g$ to $B$ of the coordinate element $g:\Sigma\rightarrow M$ of the manifold. The proportionality symbol $\propto$ indicates that (at least for the moment) we take no heed of proportionality factors. To distinguish this WZ-term to the one featuring in the Green-Schwarz action of section \ref{sec:Sib:prereq:flatGS}, we will refer to the WZ-term capturing the NSNS content of $AdS_3$-backgrounds as the Cagnazzo-Zarembo-WZ-term (shortened to CZ-WZ-term).

When introducing a WZ-term to any action one has to tread carefully. Indeed, the WZ-term is not an integral over the worldsheet $\Sigma$ but over a three-dimensional surface with boundary $\Sigma$.  One has thus to make sure that the total action still makes sense: its variation should ``localise'' to an integral over  $\Sigma$ for the equations of motion to be local. That is, the variation of the WZ term has to be a total derivative.
What will save the day here, is that the coset space realising the $AdS_3$-backgrounds is of a very special type: they are permutation supercosets. 

Adding a new term to the action, a second impeding hurdle when considering WZ-terms is that the $\mathbb Z_4$ symmetry might (and as we will see in this case, will) be broken. Without the $\mathbb Z_4$-symmetry at hand, the integrability of the system as a whole is no longer guaranteed by a canonical Lax connection. Fortunately, although the $\mathbb Z_4$ symmetry is broken, a Lax connection can be constructed guaranteeing that the system remains classically integrable.

%%%%%%%%%%%%%%
\subsubsection{The Wess-Zumino-term}\label{sec:saskia:wzwterm}  	
After these many words of caution, we are now in a position to introduce the form of the Wess-Zumino term capturing the NSNS-flux. To this effect, remember that the $\mathbb Z_4$-automorphism acts on the supercurrent by $\Omega(J_a^{(n)})=i^nJ_a^{(n)}$. We can thus write the supercurrent in terms of four terms, distinguished by their $\mathbb Z_4$ grading
\begin{align}
	J_a=g^{-1}\partial_ag=J^{(0)}_a+J^{(1)}_a+J^{(2)}_a+J^{(3)}_a\,.
\end{align}
In terms of the $\mathbb Z_4$-components of the supercurrent the action takes on the form
\begin{align}
	S&=\frac{1}{2}\int \mathrm d^2x \;\mathrm{Str}\left(\sqrt{h}h^{ab}J^{(2)}_a J^{(2)}_b+\hat q \varepsilon^{ab}J^{(1)}_a J^{(3)}_b\right)\\
	&=\frac{1}{2}\int \mathrm{Str}(J^{(2)}\wedge\star J^{(2)}+ \hat q J^{(1)}\wedge J^{(3)})\,.
\end{align}
Guided by the fact that $J_2$ is the only current that remains after reducing the supercoset action to its bosonic part, one may be tempted to posit the following combination as a first attempt for a WZ-term
\begin{align}\label{eq:WZ-term_AdS3_naive}
    S_\mathrm{WZ}^\mathrm{naive}=\frac{2}{3}\int_B\mathrm{Str}\,J_2\wedge J_2\wedge J_2\,.
\end{align}
Albeit simple, this guess in wrong. Indeed, as you will show in exercise \ref{ex:SWZnaivedoesnotlocalise}, it turns out that the variation of $S_\mathrm{WZ}^\mathrm{naive}$  does not `localise'.

\begin{centering}
\begin{tcolorbox}
\begin{exercise}
Compute the variation of $ S_\mathrm{WZ}^\mathrm{naive}$ and argue that it is not a total derivative term.
\end{exercise}\label{ex:SWZnaivedoesnotlocalise}
\end{tcolorbox}	
\end{centering}
\noindent
The authors in \cite{Cagnazzo:2012se} showed that when completing the naive Ansatz for the WZ-term in eq. \eqref{eq:WZ-term_AdS3_naive} with two additional contributions, the variations becomes a total derivative. The WZ-term yielding a meaningful contribution to the equations of motion takes on the form
\begin{equation}\label{eq:WZ-term_AdS3_tot}
	S_{\mathrm{WZ}\text{-}\mathrm{CZ}}=\hat q\int_B\mathrm d^3y\,\varepsilon^{abc}\mathrm{Str}\left( \frac{2}{3}J^{(2)}_aJ^{(2)}_bJ^{(2)}_c+J^{(1)}_aJ^{(3)}_bJ^{(2)}_c+J^{(3)}_aJ^{(1)}_bJ^{(2)}_c\right)\,.
\end{equation}
Here $\hat q$ is at the moment an arbitrary constant, which, as we will see shortly, will be constrained by either kappa-symmetry or integrability of the total action. 

\begin{centering}
\begin{tcolorbox}
	\begin{exercise}
Show that the variation of the WZ-term in eq. \eqref{eq:WZ-term_AdS3_tot} is
	\begin{gather}
\begin{aligned}
	\delta S_\mathrm{WZ}&=\int_\Sigma \mathrm{Str}\left(\xi_2 (2J^{(2)}\wedge J^{(2)}+ J^{(1)}\wedge J^{(3)}+J^{(3)} \wedge J^{(2)}) \right.\\
	&\qquad\qquad \quad \left. + \xi_1(J^{(2)}\wedge J^{(3)}+J^{(3)}\wedge J^{(2)}) +\xi_3(J^{(2)}\wedge J^{(1)}+J^{(1)}\wedge J^{(2)}) \right)\,,
\end{aligned}
\end{gather}
which is indeed a total derivative.
 \end{exercise}
{\footnotesize
As pointed out in \cite{Cagnazzo:2009zh}, this computation is nearly completely similar to that in \cite{Berkovits:1999zq}, from which the reader is invited to take inspiration from.
}
 \end{tcolorbox}
\end{centering}
\noindent
In conclusion, the action $S+S_\mathrm{CZ-WZ}$ describes an $AdS_3$-background featuring both RR- and NSNS-fluxes and the total action becomes
\begin{gather}\label{eq:CZ-action}
\begin{aligned}
	S_\mathrm{tot}&=\frac{1}{2}\int_\Sigma\mathrm{Str}(J^{(2)}\wedge\star J^{(2)}+\hat q J^{(1)}\wedge J^{(3)})\\
	&\quad+q\int_B\mathrm{Str}\left(\frac{2}{3}J^{(2)}\wedge J^{(2)}\wedge J^{(2)}+J^{(1)}\wedge J^{(3)}\wedge J^{(2)}+J^{(3)}\wedge J^{(1)}\wedge J^{(2)} \right)\,,
\end{aligned}
\end{gather}
where we remark for later use that the parameter $q$ can be identified with the WZ-level $k$ via the relation $k=2\pi T q$ with $T$ the string tension.

Having written down the action, let us reinstate the string tension explicitly to understand how to access the supergravity regime. Working out the relation between the level $k$ and the parameters $q$ and $\hat q$ leads us to the string tension already mentioned in eq. \eqref{eq:string-tension} and the different terms of the action read
\begin{align}
S_\mathrm{bos}=\frac{T}{2}S+\frac{k}{4\pi}S_\mathrm{WZ}\,.
\end{align}
where the first action is the sigma-model action and the second is the WZ action we just constructed.

\noindent
 Having introduced a new term to the system we have been studying so far, all `nice' properties that were discussed in the first part of this course may be potentially lost. Fortunately, we will see in the rest of this section that this is not the case and we will, even after adding the  CZ-WZ-term, see that  the properties of integrability, kappa-symmetry and conformal invariance are preserved.

%%%%%%%%%%%%%%
\subsubsection{\texorpdfstring{$\mathbb Z_4$}{Z4}-symmetry and integrability}  	
The first crucial observation is that in this new action \eqref{eq:CZ-action}, the $\mathbb Z_4$ symmetry is now explicitly broken since the WZ-term has grading two. With no $\mathbb Z_4$-invariance at hand, the construction discussed in section \ref{s:lax-ssssm} for the conventional Lax connection of the SSSSM-action can no longer be applied.  

	\begin{centering}
	\begin{tcolorbox}
	\begin{exercise}
\noindent
That the WZ term has grading zero (and not two) is in fact crucial. Imagine we would  have a coset based on simple supergroup instead of a permutation coset, the WZ term would be identically zero. Why is that and why isn't that a problem for permutation cosets? %Zarembo p8
\end{exercise}
	\end{tcolorbox}
	\end{centering}	
\noindent
In spite of the absence of an $\mathbb Z_4$-invariance, it was shown in \cite{Cagnazzo:2012se} that the action with WZ-term remains integrable. To see this consider the following general Ansatz for a Lax connection
\begin{align}\label{eq:lax_Ansatz}
	L=J^{(0)}+\alpha_1J^{(2)}+\alpha_2\star J^{(2)}+\beta_1J^{(1)}+\beta_2J^{(3)}\,.
\end{align}
For the system to be integrable with a Lax connection of this form, the matrix $L$ should have a zero-curvature condition, \textit{i.e.}
\begin{align}
	\mathrm d L+L\wedge L=0\,,
\end{align}
that is equivalent to the equations of motion, given the currents satisfy the (graded) Maurer-Cartan equations.

\begin{centering}
\begin{tcolorbox}
	\begin{exercise}
\noindent 
The flatness condition for the Ansatz $L$ in eq. \eqref{eq:lax_Ansatz} has to be equivalent to the equations of motion (and taking into account the Maurer-Cartan equations). It turns out \cite{Cagnazzo:2009zh} (although you are most welcome to check to as well) that this is true provided that the coefficients in the Ansatz satisfy
\begin{align}
	-\alpha_1+\hat{q} \alpha_2+\beta_1^2&=0\,,\qquad\qquad\qquad\;\,  2q\alpha_2-1+\alpha_1^2-\alpha_2^2=0\,,\\
	  -\alpha_1-\hat{q} \alpha_2+\beta_2^2&=0\,,\qquad-\beta_1+\alpha_1\beta_2+q\alpha_2\beta_1-\hat{q}\alpha_2\beta_2=0\\
     q\alpha_2-1+\beta_1\beta_2&=0\,,\qquad  -\beta_1+\alpha_1\beta_2+q\alpha_2\beta_1-\hat{q}\alpha_2\beta_2=0\,.
\end{align} 	
This algebraic system admits a solution under the condition that the parameters $\hat{q}$ and $q$ are not independent. Derive that relation and note that the system then becomes underconstrained. Solve for $\alpha_2$, $\beta_1$ and $\beta_2$. This yields then the form for the Lax connection for the (permutation) coset action with WZ term where $\alpha_1$ plays the role of spectral parameter.
\end{exercise}
\end{tcolorbox}
\end{centering}
\noindent
The relation between the parameters featured in the action you should obtained in the last exercise is the, by now familiar, relation
\begin{align}
	q^2+\hat{q}^2=1\,,
\end{align}
and, from the same exercise follows the one-parameter set of solutions
\begin{align}
	\alpha_2&=q\pm \sqrt{-1+\alpha_1^2+q^2}\,,\quad \beta_i=\pm\sqrt{\alpha_1+(-1)^i\hat q\alpha_2}\quad (i=1,2)\,.
\end{align}
The only remaining non-determined parameter, $\alpha_1$, becomes the spectral parameter of the Lax connection. For completeness, let us write down the Lax connection, 
\begin{align}
	L&=J^{(0)}+\hat q\frac{\boldsymbol{\mathrm{x}}^2+1}{\boldsymbol{\mathrm{x}}^2-1}J^{(2)}+\left(q-\frac{2\hat q \boldsymbol{\mathrm{x}}}{\boldsymbol{\mathrm{x}}^2-1}\right)\star J^{(2)}\\
	&\qquad +\left(\boldsymbol{\mathrm{x}} +\frac{\hat q}{1-q}\right)\sqrt{\frac{\hat q(1-q)}{\boldsymbol{\mathrm{x}}^2-1}}J^{(1)}+\left(\boldsymbol{\mathrm{x}}-\frac{\hat q}{1+q}\right)\sqrt{\frac{\hat q(1+q)}{\boldsymbol{\mathrm{x}}^2-1}}J^{(3)}\,,
\end{align} 
which is obtained from the Ansatz for the Lax connection \eqref{eq:lax_Ansatz} after parametrising the only remaining free coefficient as 
\begin{align}
	\alpha_1=\hat q\frac{\boldsymbol{\mathrm{x}}^2+1}{\boldsymbol{\mathrm{x}}^2-1}\,,
\end{align}
and where $\star$ is the 10-dimensional Hodge star.

Although the property of integrability survives, the lack of $\mathbb Z_4$ invariance in the new action poses a severe problem for the application of many integrability techniques. Fortunately, it was quickly realised in \cite{Babichenko:2014yaa} that the $\mathbb Z_4$ symmetry can be reinstated. The trick consists of introducing the grading-two matrix\footnote{Yet another way around the absence of $\mathbb Z_4$-invariance in the action \eqref{eq:CZ-action}, can be circumvented by redefining the `supertrace' for the WZ term \cite{Hoare:2013pma}.}
\begin{align}
	W=\begin{pmatrix}
		1&0\\ 0&-1
	\end{pmatrix}\,,
\end{align} 
as an overall factor in the WZ-term
\begin{gather}\label{eq:GS_w_WZterm}
\begin{aligned}
	S&=\frac{1}{2}\int_\Sigma\mathrm{Str}(J^{(2)}\wedge\star J^{(2)}+\hat{q} J^{(1)}\wedge J^{(3)})\\
	&\quad+q\int_B\mathrm{Str}\,W\left(\frac{2}{3}J^{(2)}\wedge J^{(2)}\wedge J^{(2)}+J^{(1)}\wedge J^{(3)}\wedge J^{(2)}+J^{(3)}\wedge J^{(1)}\wedge J^{(2)} \right)
\end{aligned}
\end{gather}
the last term has now grading 4 and the action is again $\mathbb Z_4$ invariant. The matrix $W$  is a non-dynamical field and the action of the $\mathbb Z_4$ automorphism on $W$ generates a non-physical symmetry which effectively flips the sign of the coupling parameter $q\rightarrow -q$. This new action \eqref{eq:GS_w_WZterm}, again admits a Lax representation provided the very same identity, $q^2=1-\hat{q}^2$, is satisfied. 

Let us conclude by mentioning that the additional CZ-WZ term does not spoil conformality of the original coset model. As was shown in \cite{Cagnazzo:2012se}, the one-loop beta-function for the coset model $+$ CZ-WZ-term differs by an overall power of the parameter $\hat{q}$ compared to the original beta-function give in eq. \eqref{eq:beta_fct_GS}:
\begin{align}
    \beta(g)_\mathrm{CZ}= \hat{q}^2 \beta(g)+\mathcal{O}(\alpha'{}^2)\,.
\end{align}
Note here that when $\hat q=0$, \textit{i.e.}~when we have a pure NSNS-background, we obtain, as expected, a fixed point of the group renormalisation flow. For all other values $\hat q\neq 0$, the beta-function is proportional, as in  \eqref{eq:beta_fct_GS}, to the  Killing form of the superalgebra. 

 %%%%%%%%%%%%%%%%%%%%%%%%%%%%%%%%%%%%%
%% Type II GS action in AdS3 space %%
%%%%%%%%%%%%%%%%%%%%%%%%%%%%%%%%%%%%%
\subsection{Gauge fixings and yet another problem} \label{s:saskia-gf}
In this section we will finally return to the problem mentioned at the end of section \ref{sec:equiv_w_gfGS}.
Remember that the coset action can only be equivalent to the Green-Schwarz action (up to quadratic order in the fermions) after imposing a kappa-gauge fixing and for which the flat directions decouple. Picking this gauge comes however at a hefty price: this particular gauge fixing turns out not to be compatible with some motions of the string. In fact, as we will shortly see, this choice is incompatible with superstrings whose motion is restricted to the $AdS_3$ subspace of the $AdS_3\times S^3\times S^3\times S^1$ geometry (or, even worse, the $AdS_3\times S^3$ subspace when considering the $AdS_3\times S^3\times T^4$ background). Often these problematic solutions in the supercoset formalism are referred to in the literature as  \textit{singular}. Finally let us comment that, in this gauge the action contains a kinetic term for the massless fermions that has no quadratic contribution \cite{Babichenko:2009dk}. In absence of a conventional kinetic term, there is no straightforward way to derive the Poisson brackets and thus quantise the action. In addition the coset description, due to the kappa-symmetry gauge fixing, is incomplete and cannot describe all string configurations. This issue will be the topic of section \ref{sec:saskia:supercoset_incomplete}. As a result an analysis of the quantised spectrum necessitates working directly with the Green-Schwarz action in the BMN lightcone kappa gauge \cite{Borsato:2014hja,Lloyd:2014bsa}.

%%%%%%%%%%%%%
\subsubsection{Intermezzo: the superspace Green-Schwarz superstring action}
In order to fulfil this program, we first need to review the necessary superspace notation and technology.
As already reviewed in section \ref{sec:Sib:supercoset_constr_GS}, the complete GS action can be compactly written in superspace variables. Assuming for simplicity that the background is a supergravity solution with vanishing background fermionic fields (\textit{i.e.}\ the gravitino and dilatino), the Green-Schwarz superstring action for a general type II supergravity background takes on the form \cite{Grisaru:1985fv,Wulff:2013kga}
\begin{align}\label{eq:GS-superspace-action}
	S=-T\int_\Sigma\left(\frac{1}{2}\star  E^A E^B\eta_{AB}- B\right)\,,
\end{align}
where we have the same notation as in section \ref{s:sibylle}, see eq.~\eqref{eq:superspacevielbeinscomps}, \textit{i.e.}~$  E^A$ are the vector supervielbeins pullbacked to the worldsheet, $B$ is the (pullbacked) NSNS two-form potential and $T$ is the string tension. The worldsheet Hodge star $\star$ is taken with respect to the worldsheet metric. Note that the supervielbeins $ E^A(\mathcal Z)$ and $ B(\mathcal Z)$ are superfields and thus depend both on the ten bosonic coordinates $X^\mu$  and as well as, in the case of a type IIA solution, on the 32 fermionic coordinates $\theta$ of the type IIA superspace. The type IIA spinor can be written as $\theta_{IIA}=\theta^1_{IIA}+\theta^2_{IIA}$ where 
\begin{align}
	\theta^1_{IIA}=\begin{pmatrix}
		\hat \theta^1_{IIA} \\ 0
	\end{pmatrix}\,,\qquad \theta^2_{IIA}=\begin{pmatrix}
		 0 \\ \hat \theta^2_{IIA}
	\end{pmatrix}\,,
\end{align}
and the components $\hat \theta^i_{IIA}$ are 16-components Majorana spinors of opposite chirality. Performing a T-duality with respect to any of the flat compact directions $v$ of the geometry, maps the type IIA to a type IIB solutions with two 16-component Majorana-Weyl worldsheet spinors $\theta^1_{IIB}=\theta^1_{IIA}$ and $\theta^2_{IIB}=\Gamma_v\theta^2_{IIA}$ \cite{Cvetic:1999zs}.

Taken altogether we have the coordinates $\mathcal Z^{M}=(X^\mu,\theta^I)$. The conventions for the indices are as follows $\mu=0,\dots,9$ are the 10-dimensional spacetime indices, the 32 Grassmann-odd coordinates $\theta^I$ for $I=0,\dots,32$, and the worldsheet coordinates are denoted by $\xi^\alpha=(\tau,\sigma)$. With this notation the pullback of the supervielbein on the worldsheet is 
\begin{align}
 E^A(X,\theta)=(\partial_\alpha X^\mu E_\mu{}^A(X,\theta)+\partial_\alpha\theta^I  E_I{}^A(X,\theta))\mathrm d \xi^\alpha\,.
\end{align}

\noindent
Since type IIA superspace features 32 fermionic coordinates, the Green-Schwarz action has an expansion (by expanding the superfields) in even powers of the fermions $\theta$ up to the 32$^\mathrm{nd}$ order in principle, though for practical applications and sufficiently supersymmetric backgrounds only the second or fourth order is needed. As of date though the GS action is only known up to quartic order \cite{Wulff:2013kga} in the fermions. Note that already at this order in fermions the action is rather unwieldy.

We have seen in section \ref{sec:Sibylle:supercoset:kappa} that the Green-Schwarz superstring action remains invariant under local fermionic transformations forming the so-called kappa-symmetry. In the superspace formulation this symmetry is parameterised by a 32-component spinor $\kappa(\xi)$ and the target space coordinates $\mathcal Z^{ M}=(X^\mu,\theta^I)$ transform as
\begin{gather}
\begin{aligned}\label{eq:kappa_var_superfields} {}
	\delta_\kappa \mathcal Z^{ M}  E_{ M}{}^{aI} &=\frac{1}{2}(1+\Gamma)^{aI}{}_{bJ} \,\kappa^{bJ}\,,\quad 
	\delta_\kappa \mathcal  Z^{ M}  E_{ M}{}^A =0\,,
\end{aligned}
\end{gather}
introduced earlier in \eqref{eq:kappa-gs-superspace}. Remember that here  $( E_{ M}{}^A, E_{ M}{}^{aI})$ are the background supervielbeins and $\tfrac{1}{2}(1+\Gamma)^{aI}{}_{bJ}$ is a spinor projection matrix where the matrix $\Gamma$ is explicitly given by 
\begin{align}\label{eq:matrix_Gamma_Superspace}
	\Gamma=\frac{1}{2\sqrt{-\mathcal G}} \varepsilon^{\alpha\beta} E_\alpha{}^A E_\beta{}^B\Gamma_{AB}\Gamma_{11}\,,\quad \Gamma^2=1\,,
\end{align}
which involves the determinant of the induced metric on the worldsheet $\mathcal{G}_{\alpha\beta}= E_\alpha{}^A E_\beta{}^B\eta_{AB}$ and note that the projector only involves the bosonic supervielbeins. We encountered this projector already in eq. \eqref{eq:kappa-gs-superspace}  but for type IIB superspace, where $\Gamma_{11}$ is replaced for the Pauli matrix $\sigma^3$ (but see also \eqref{eq:toIIA}). For later use, one can check that the variations of the components under kappa-symmetry to linear order in $\theta$ from eq. \eqref{eq:kappa_var_superfields}: 
\begin{align}\label{eq:kappa_var_lin_order}
	\delta_\kappa X^\mu E_\mu=\bar\theta(1+\Gamma)\kappa\,,\quad \delta_\kappa\theta=\frac{1}{2}(1+\Gamma)\kappa\,.
\end{align}
As we will soon see, the critical observation to make when fixing a kappa-gauge is that the matrix $\Gamma$ appearing in the kappa-symmetry variations depends on the pullback of the vielbeins $ E^A$. This implies that which gauge-fixing one should pick depends on where the string moves through the background.

%%%%%%%%%%%%%
\subsubsection{Completeness of kappa-gauge fixings} \label{s:saskia-complete-kg}
% notation 1306.6918 and 1204.4742
Before introducing the kappa-gauge fixing that enables one to identify the (quadratic) Green-Schwarz superstring action with the supercoset sigma-model introduced in sections \ref{sec:Sib:prereq:GS} and \ref{sec:Sib:supercoset_constr_GS}, we need to make some general considerations about the completeness of a given choice of kappa-gauge fixing. Without loss of generality we will discuss the gauge fixing in type IIA supergravity. Remember from section \ref{sec:Sib:prereq:GS}, that in the case of type IIA supergravity solutions, the two Majorana-Weyl spinors featured in the GS action can be described as one single 32-component Majorana spinor $\theta$.  Assume that the gauge fixing making this selection is captured by a certain $32\times 32$-dimensional matrix $M$:
\begin{align}\label{eq:kappa_gauge_fix_M}
	M\theta=0\,,
\end{align}
which thus kills at most 16 of the fermionic fields in $\theta$. We use the notation $\theta=(\vartheta,v)$, where $\vartheta$ denote the fermions corresponding to the unbroken supersymmetries and $v$ to those broken by $M$. 

When fixing a kappa-gauge, counting how many fermionic degrees of freedom remain after enforcing the gauge fixing determined by $M$ is not sufficient. One should also ensure that the requirement for the gauge fixing condition to be invariant under infinitesimal kappa-symmetry variations is equivalent to putting all kappa-symmetry parameters to zero. If this is true, the gauge fixing is called complete. To see how this constraints the, so-far, unspecified gauge fixing matrix $M$: vary the gauge condition \eqref{eq:kappa_gauge_fix_M} according to the linearised kappa-transformation rule in eq. \eqref{eq:kappa_var_lin_order}
\begin{align}\label{eq:comm_kappa_gauge_fix}
	0=\frac{1}{2}M(1+\Gamma)\kappa=\frac{1}{4}[M,\Gamma](1+\Gamma)\kappa\,.
\end{align}

\begin{centering}
\begin{tcolorbox}
\begin{exercise}
\noindent
Show that the second equality in eq. \eqref{eq:comm_kappa_gauge_fix} by first showing that 
\begin{align}
	\frac{1}{2}M(1+\Gamma)\kappa=\frac{1}{2}(1+\Gamma)M(1+\Gamma)\kappa +\frac{1}{2}[M,\Gamma](1+\Gamma)\kappa\,,
\end{align}
and by arguing that the first term vanishes. 
\end{exercise}	
\end{tcolorbox}
\end{centering}
\noindent
We conclude from eq. \eqref{eq:comm_kappa_gauge_fix},
that for the gauge fixing to be complete, the commutator $[M,\Gamma]$, when restricted to the subspace generated by the physical fermionic degrees of freedom, has to be invertible\footnote{If $M$ defines the projection into an $n$-dimensional space, where for $AdS_3$-space we have that $n\leq 16$, this condition means that $\mathrm{rank}[M,\Gamma]\geq n/2$.}. When this requirement is not satisfied, the matrix $M$ has a non-trivial kernel. The corresponding gauge fixing choice then can potentially set physical fermionic degrees of freedom to zero or keep unphysical fermionic degrees of freedom unfixed.

%%%%%%%%%%%%%
\subsubsection{The coset kappa-gauge for \texorpdfstring{$AdS_3 \times S^3$}{AdS3xS3}-backgrounds}\label{sec:coset_kappa_gauge_fix}

Looking back at the construction of the supercoset action in section \ref{sec:saskia:AdS3coset}, we also see that the specific choice of kappa-gauge fixing should be chosen such as to describe the superstring as a supercoset sigma-model. In particular, the gauge-fixing has to project out the 16 non-coset fermions
\begin{align}
	v=M\theta =0\,.
\end{align}
To get some insight on what form $M$ could take, a clever trick is to consider the supersymmetry variation of the dilatino\footnote{As per usual, here $012$ are the $AdS_3$ direction, $345$ and $678$ accounts for the two three-spheres and 9 is the $S^1$-factor. See also the D-brane set-up of table \ref{table:D1D5_system}.}
	\begin{align}\label{eq:dilaton_SUSY_var}
		\delta\gamma=\Gamma^A\slashed{F}\Gamma_A\epsilon=8\Gamma^{012}(1-\mathcal P)\epsilon\,,
	\end{align}
	where $\mathcal P$ is a projector whose form and properties you will unravel in exercise \ref{ex:four-form_projec_trick} below and 
	\begin{align}
		\slashed{F}=e^{\phi}\left(-\frac{1}{2}\Gamma^{AB}\Gamma_{11}F_{AB}+\frac{1}{4!}\Gamma^{ABCD}F_{ABCD}\right)\,.
	\end{align} 
	 Together with the fact that the four-form RR-flux supporting the type II solutions on $AdS_3\times S^3\times S^3\times S^1$, the expression for the dilaton variation can be rewritten as
	\begin{align}
	\slashed{F}=4\Gamma^{012} \Gamma^9(1-\mathcal P)\,.
\end{align}
Remember that the four-form flux supporting a IIA supergravity is of the form \eqref{eq:IIA_fluxes} and takes on the explicit form 
\begin{align}\label{eq:IIA_F4_flux_S3S1}
    F_{(4)}=\frac{e^{-\phi}}{3}\left( E^a E^b E^c\varepsilon_{ABC}+\cos\varphi\,  E^{\hat a} E^{\hat b} E^{\hat c}\varepsilon_{\hat a\hat b\hat c}+\sin\varphi\,  E^{ a'} E^{ b'} E^{\hat c}\varepsilon_{ a' b' c'} \right) E^9
\end{align}
where as before $\phi$ is the dilaton and the $AdS_3$ radius has been fixed to one and $\varphi$ is the parameter parametrising the ratio of the $S^3$ radii in eq. \eqref{eq:triangle_S3S3}. The ten indices in $A$ have been partitioned over the different components of the geometry as in eq. \eqref{eq:AdS3_S3_S3_S1_indices}.

From the supersymmetry variation of the dilatino in \eqref{eq:dilaton_SUSY_var}, we see that a total of 16 supersymmetry parameters are preserved by the background. That this, they verify $\epsilon = \mathcal P\epsilon$ such that  $\delta \lambda=0$. Taken altogether, we are led to the following choice of kappa gauge fixing \cite{Rughoonauth:2012qd}	
	\begin{align}
		\vartheta=\mathcal P\theta\,,\quad v=(1-\mathcal P)\theta\,,
	\end{align}
	where the sixteen $\vartheta$ are the coset fermions (corresponding to the preserved supersymmetries) and $v$ are the non-coset fermions (\textit{i.e.}~the broken supersymmetries). This is precisely the gauge fixing used in \cite{Sundin:2012gc} to reduce the GS superstring action (to quadratic order in fermions) and to show its equivalence to the supercoset action presented in section \ref{sec:saskia:AdS3coset}.

	\begin{centering}
\begin{tcolorbox}
\begin{exercise}\label{ex:four-form_projec_trick}
From the GS action in \eqref{eq:GS-superspace-action}, we saw that the RR fields couple to the other fields via the matrix $\slashed{F}$ which in type II takes on the form
\begin{align}
	\slashed{F}=-\frac{1}{2} E^{\phi}\Gamma^{AB}\Gamma_{11}F_{AB}+\frac{1}{4!}\Gamma^{ABCD}F_{ABCD}\,.
\end{align}
where $A=0,\dots, 9$ is the index running over the 10-dimensional background and $\phi$ is the dilaton field. Show that the four-form flux field $F_4$ in eq. \eqref{eq:IIA_F4_flux_S3S1} can be rewritten into the form
\begin{align}
	\slashed{F}=4\gamma_\ast \Gamma^9(1-\mathcal P)\,,
\end{align}
where $\gamma_\ast=\Gamma^{012}$ (note that $\gamma_\ast^2=1$) and $\mathcal P$ is a projection matrix given by
\begin{align}\label{eq:kappa_gauge_P}
	\mathcal P=\frac{1}{2}(1+\cos\varphi \,\Gamma^{012}\Gamma^{345}+\sin\varphi\, \Gamma^{012}\Gamma^{678})\,.
\end{align}
Check that this is a projector.
\end{exercise}	
\end{tcolorbox}
\end{centering}
\noindent

%%%%%%%%%%%%%
\subsubsection{The supercoset action is not enough}\label{sec:saskia:supercoset_incomplete}
We still need to check whether the gauge fixing is compatible with \textit{any} motion of the strings through the $AdS_3\times S^3\times S^3\times S^1$. Assume that the superstring only propagates in the $AdS_3$ subspace of the background, using the notation in \eqref{eq:AdS3_S3_S3_S1_indices} that means that most of the components of the supervielbein $ E^A$ vanish: $ E_I{}^{\hat a}= E_I{}^{a'}= E_I{}^{9}=0$. Since the matrix $\Gamma$ \eqref{eq:matrix_Gamma_Superspace} controlling the kappa-symmetry variation as given in eq. \eqref{eq:kappa_var_superfields} directly depends on the supervielbein and consequenly simplifies to the expression
\begin{align}
	\Gamma \propto \epsilon^{\alpha\beta} E_\alpha{}^a   E_\beta{}^b \Gamma_{ab}\Gamma_{11}\,,
\end{align}
where $a=0,1,2$ runs over the $AdS_3$ directions only. Since the projector $\mathcal P$ from exercise \ref{ex:four-form_projec_trick} has terms either proportional to $\Gamma^{012}$ or to the identity $\mathbb 1$ {\color{black}and taking into account that $\{\Gamma_a,\Gamma_\bullet\}=0$ where $\bullet=\hat a,a'$ or $9$}, we have to conclude that the projector $\mathcal P$ commutes with $\Gamma$. The commutator $[M,\Gamma]$ is non-invertible and the consistency condition in eq. \eqref{eq:comm_kappa_gauge_fix} is violated. When the string moves entirely in $AdS$-part of the geometry, the total number of kappa-symmetries in the sigma-model is effectively increased. Indeed, in this singular situation, the projector $\mathcal P$ commutes with the kappa-symmetry projector, and effectively singles out a number of non-coset or broken supersymmetries which cannot be eliminated. We can but conclude that the supercoset model in this case only captures a \textit{subsector} of the kappa-symmetry gauge-fixed Green-Schwarz superstring \cite{Rughoonauth:2012qd}. As a result, in order to study strings whose motion is restricted to the $AdS_3\subset AdS_3\times S^3\times T^4$-region which lies outside of the description of the supercoset action, one has to resort directly to the superstring Green-Schwarz action. The situation worsens for the  $AdS_3\times S^3\times T^4$-background. In that case the projector is given by $\mathcal P$ in eq. \eqref{eq:kappa_gauge_P} but taking $\varphi\rightarrow 0$. There the motion of the superstring in the whole $AdS_3\times S^3$-subspace is incompatible with the gauge fixing. 

The fact that the supercoset action cannot describe all string motion is clear drawback, and raised the question to the integrability of the $AdS_3$-superstring for any subsectors of its geometry. This issue prompted efforts toward finding a Lax connection directly for the GS superstring (to quadratic order in the fermions) rather than for the (incomplete) supercoset construction. This hope was realised in \cite{Sundin:2012gc,Murugan:2012mf}, where the Lax connection was constructed explicitly without any kappa-gauge fixing, warranting the integrability of superstrings in $AdS_3\times S^3\times M_4$ backgrounds. Although a Lax connection has up until now not been constructed for higher order in fermions of the GS action, mostly due to the quickly rising level of complexity at each new order, essentially the same approach is expected to be applicable without any obstruction.

This phenomenon of the coset description not describing all possible string configurations is not restricted to $AdS_3\times S^3$ superstrings sigma-model. One can show in a very similar way that no gauge-fixing of the $AdS_4\times \mathbb{C}P^3$ superstrings sigma-model can describe all possible string motions while at the same time gauge fixing all unphysical fermionic degrees of freedom and missing physical degrees of freedom \cite{Grassi:2009yj,Cagnazzo:2009zh,Sorokin:2011rr}.

%%%%%%%%%%%%%%%%%%%%%%%%%%%%%%%%%%%%%%%%%%%%%%%%%%%%%%
%%%%%%%%%%%%%%%%%%%%%%%%%%%%%%%%%%%%%%%%%%%%%%%%%%%%%%
\subsection{Summary and concluding remarks}\label{sec:saskia:summary}
In this section we reviewed the integrability of the Green-Schwarz superstring action propagating in $AdS_3\times S^3$-backgrounds using the supercoset action, both with and without NSNS three-form flux.
The supercoset description, although appealing in its simplicity when compared to the full-fledged GS action, is unfortunately incomplete. In establishing this fact, we identified the kappa-gauge as the main culprit, which is however necessary to warrant the matching of the supercoset action with the GS action. In this gauge, strings propagating  only in a subsector of geometry involving the $AdS$-part can simply not be described using the coset action.  Fortunately, this issue can be remedied by working directly with the GS action. In fact, the statement of classical integrability can be proven in the GS formalism without having to rely on the supercoset formulation.  This lays the basis for the study of $AdS_3$ superstring backgrounds by integrability which we will undertake in the next two sections.
%Rather than being merely a failure of the formalism, the supercoset approach and its shortcomings heralded the key properties and challenges of superstrings on  AdS3, which distinguish it from other holographic backgrounds. In the sections to come, we will trail research spanning from the beginning of the last decade till today. By harnessing several integrability and CFT techniques, we will be able to systematically unravel many of the challenges and questions raised in this section.
%
Having established classical integrability of the system, we are naturally led to wonder whether integrability is preserved at the quantum level. A related question is that of the puzzling spectrum of both massive and massless particles the $AdS_3\times S^3$ confronted us with. These points will be addressed in section \ref{s:fiona} by studying the worldsheet S-matrix of $AdS_3 \times S^3\times T^4$ superstring.  We will return to the spectrum of the $AdS_3 \times S^3$-superstring and its massless modes, in section \ref{ana:sec}, where it will be treated with the Betze Ansatz. It is also worth noting that one may employ a different approach altogether, which does not rely on the GS string. The hybrid approach, which is discussed at length in section~\ref{s:bob}, does allow to treat in principle $AdS_3$ superstrings supported by  RR and NSNS fluxes in a unified way. However, it is practically very difficult to do so away from the WZW points (\textit{i.e.}, when the RR flux is not zero).

Although much has been recently understood concerning the propagation of strings in $AdS_3\times S^3\times T^4$-backgrounds, their full solution (either through integrability or by other means) and the complete characterisation of their holographic duals is still full of mysteries and challenges, whose solution is bound to advance our understanding of string theory, holography, CFT, and integrable models. %Their study form a very active field of research providing a guiding example towards expanding the integrability toolbox as well as steadily solidifying our understanding of the AdS/CFT correspondence. 

\newpage

%%%%%%%%%%%%%%%%%%%%%%%%%%%%%%%%%%%%%%%%%%
%%%%%%%%%%%%%%%%%%%%%%%%%%%%%%%%%%%%%%%%%%
%%%%%%%%%%%%%%%%%%%%%%%%%%%%%%%%%%%%%%%%%%
\newpage
\section{The worldsheet S matrix} \label{s:fiona}
% \textit{Current author: Fiona Seibold. } For questions, comments, typo's, \textit{etc.},~on this section, feel free to write to \href{mailto:fiona.seibold@gmail.com}{fiona.seibold@gmail.com}.
%\subsection{Introduction}
In section~\ref{s:sibylle} we discussed the Green-Schwarz formulation of superstring theory in curved spacetime. Starting from this action one would like to compute physical observables such as the string spectrum: what are the allowed energies for closed string states?
For integrable theories, an important stepping stone to arrive at this result is to obtain the worldsheet S-matrix, describing the scattering of excitations on the two-dimensional worldsheet of the string. Due to the presence of a large amount of conserved charges, the structure of this S-matrix is heavily constrained, see~\cite{Dorey:1996gd} for a review on integrable S-matrices. In particular, the set of incoming momenta and outgoing momenta must be the same for any scattering process, and a $n \rightarrow n$ scattering event decomposes into a sequence of $2 \rightarrow 2$ scattering events. Because such a decomposition can be performed in different ways, consistency then requires the two-body S-matrix to satisfy the quantum Yang-Baxter equation. Computing the S-matrix is therefore also useful to check that the classical integrability discussed in the previous sections survives quantisation. In the next section~\ref{ana:sec} we will then see how to compute physical quantities starting from this two-body S-matrix.

The Green-Schwarz formulation of superstring theory, in contrast to the RNS formalism or the hybrid model, is particularly well-suited to obtain the worldsheet S-matrix for strings in curved space-times. The first step consists in fixing the gauge redundancy (reparametrisation invariance, Weyl rescalings, kappa-symmetry) of the action. A convenient way to fix these is through the uniform lightcone gauge, which we already briefly encountered in section~\ref{s:sibylle} and makes it possible to analyse strings in curved spaces~\cite{Arutyunov:2005hd}. This then leads to a two-dimensional non-relativistic field theory. For closed strings, the two-dimensional worldsheet is topologically a cylinder. If the radius of that cylinder is small then there is no notion of asymptotically ``in'' and ``out'' states. The worldsheet S-matrix  is defined in the ``decompactification'' limit, when the radius of the cylinder becomes so big that the cylinder can be traded for a plane and asymptotic states are well-defined.

There are two ways to obtain the worldsheet S-matrix. One is through perturbation theory, by considering an expansion around the classical solution used for the lightcone gauge fixing --- this is tantamount to expanding around the pp-wave limit considered in section~\ref{s:saskia}. 
Another possibility is through the integrable ``bootstrap''. In that case, one uses the symmetries of the lightcone gauge fixed theory, as well as the assumption of quantum integrability, to completely fix the worldsheet S-matrix. For strings on $AdS_3 \times S^3 \times T^4$ the perturbative S-matrix was first obtained in the pure-RR case~\cite{Rughoonauth:2012qd,Sundin:2012gc,Sundin:2013ypa} and then for the mixed-flux theory~\cite{Hoare:2013pma}. In parallel, a proposal for the pure-RR exact S-matrix was formulated in~\cite{Borsato:2012ud}. This analysis was then extended to include the massless modes~\cite{Borsato:2014exa,Borsato:2014hja} and the mixed-flux case~\cite{Lloyd:2014bsa}.

This section is organised as follows. First we provide a general discussion of strings in lightcone gauge. We then specialise to strings propagating in $AdS_3 \times S^3 \times T^4$ and sketch the construction of the perturbative and exact S matrices.

\subsection{Strings in lightcone gauge} \label{s:strings-lightcone-gauge}
In this section we explain how, starting from the Green-Schwarz (GS) action, one can compute the worldsheet S-matrix perturbatively in the string tension. The discussion essentially follows the review~\cite{Arutyunov:2009ga}, though we will work directly in terms of the GS action, rather than its supercoset formulation.

Let us start by recalling the form of the GS non-linear sigma model action to quadratic order in the fermions, first encountered in \ref{s:GS-curved},
\begin{equation}
S = - \frac{T}{2} \int \extder \tau \extder \sigma \left[ \left(\gamma^{\alpha \beta} \hat{G}_{\mu \nu} + \epsilon^{\alpha \beta} \hat{B}_{\mu \nu} \right) \partial_\alpha X^\mu \partial_\beta X^\nu + \mathcal L_{kin} \right] \,,
\end{equation}
where we have combined the bosonic part of the action with the ``mass'' terms of the fermions, 
\begin{align}
\hat{G}_{\mu \nu} &= G_{\mu \nu} - \frac{i}{4} \bar{\theta}_I \Gamma_{(\mu} \slashed{\omega}_{\nu)} \delta^{IJ} \theta_J + \frac{i}{8} \bar{\theta}_I \Gamma_{(\mu} H_{\nu) \rho \sigma} \Gamma^{\rho \sigma} \sigma_3^{IJ} \theta_J + \frac{i}{8} \bar{\theta}_I \Gamma_{(\mu} \mathcal S^{IJ} \Gamma_{\nu)} \delta^{IJ} \theta_J \,, \\
\hat{B}_{\mu \nu} &= B_{\mu \nu} - \frac{i}{4} \bar{\theta}_I \Gamma_{[\mu}  \slashed{\omega}_{\nu]} \sigma_3^{IJ} \theta_J + \frac{i}{8} \bar{\theta}_I  \Gamma_{[\mu} H_{\nu] \rho \sigma} \Gamma^{\rho \sigma}  \delta^{IJ}\theta_J + \frac{i}{8} \bar{\theta}_I  \Gamma_{[\mu} \mathcal S^{IJ} \Gamma_{\nu]} \sigma_3^{IJ} \theta_J \,,
\end{align}
while the last piece $\mathcal L_{kin}$ contains the kinetic terms for the fermions,
\begin{equation}
\mathcal L_{kin} =   i \gamma^{\alpha \beta} \partial_\alpha X^\mu \bar{\theta}_I \Gamma_\mu  \delta^{IJ} \partial_\beta\theta_J + i \epsilon^{\alpha \beta} \partial_\alpha X^\mu \bar{\theta}_I \Gamma_\mu   \sigma_3^{IJ}\partial_\beta\theta_J \,.
\end{equation}
The two-dimensional base manifold is parametrised by $\tau \in (- \infty, + \infty)$ and $\sigma \in (0,\cir)$. Here we are considering closed strings, so that the base manifold is an infinitely-long cylinder with circumference $\cir$. Then, $\gamma^{\alpha \beta} = \sqrt{-h} h^{\alpha \beta}$ is the Weyl-invariant metric on the two-dimensional worldsheet, and $\epsilon^{\alpha \beta}$ is the antisymmetric Levi-Civita symbol with the convention $\epsilon^{\tau \sigma} = - \epsilon^{\sigma \tau} = \epsilon_{\sigma \tau} = -\epsilon_{\tau \sigma} = +1 $. $T$ is an overall constant (the string tension). The quantities $X^\mu$ with $\mu=1,\dots, \dim(\mathcal M)$ can be seen either as fields $X^\mu(\tau, \sigma)$ on the two-dimensional worldsheet, or as coordinates of the target space $\mathcal M$. Then, $G_{\mu \nu}$ and $B_{\mu \nu}$ denote respectively the metric and antisymmetric B-field (Kalb-Ramond field), characterising the geometry of the target space $\mathcal M$. In what follows we have in mind target-space geometries of the form $AdS_n \times S^n \times \mathcal X$ where $\dim \mathcal X=10-2n$, for instance $AdS_5 \times S^5$ or  $AdS_3 \times S^3 \times T^4$. These are particularly relevant in the context of the AdS/CFT correspondence.

As explained in the previous sections, the GS action has redundancies. It is invariant under worldsheet reparametrisations, Weyl rescalings as well as fermionic kappa symmetry.
To analyse the physical degrees of freedom it is important to remove these redundancies by choosing a specific gauge. There are several possible gauges, and the best choice often depends on the goal to be achieved. Here we shall use the  so-called ``uniform lightcone gauge'', introduced in~\cite{Arutyunov:2005hd}. As we will see, in that gauge the worldsheet Hamiltonian is related to the target space energy, which makes it particularly convenient to analyse the spectral problem, and the size~$R$ of the worldsheet is related to a physical charge, making it easier to take the decompactification limit.  This bosonic gauge-fixing should then be supplemented by a compatible kappa gauge on the fermions.

Working in the Hamiltonian formalism, there are two steps to write down the gauge-fixed action. The first step is to go to the first order formalism, introducing conjugate momenta. The next step is to impose the gauge fixing to this first order action.

\subsubsection{First order form}

The momentum conjugate to the bosonic coordinate $X^\mu$ is defined as
\begin{equation}
P_\mu = \frac{\delta S }{\delta (\partial_\tau X^\mu)} = -T \left(\gamma^{\tau \beta} \hat{G}_{\mu \nu} + \epsilon^{\tau \beta} \hat{B}_{\mu \nu} \right) \partial_\beta X^\nu - \frac{T}{2} i \bar{\theta}_I (\gamma^{\tau \beta} \delta^{IJ} + \epsilon^{\tau \beta} \sigma_3^{IJ} ) \Gamma_\mu \partial_\beta \theta_J\,.
\end{equation}
Solving in terms of $\partial_\tau X^\mu$ and replacing in the action gives, up to quadratic order in fermions
\begin{equation} \label{eq:1st}
\begin{aligned}
S &= \int \extder \tau \extder \sigma \Big(P_\mu \dot{X}^\mu + \frac{i}{2} P_\mu \bar{\theta}_I \Gamma^\mu \dot{\theta}_I + \frac{T}{2} i \acute{X}^\mu (\sigma_3^{IJ} G_{\mu \nu} - \delta^{IJ} B_{\mu \nu}) \bar{\theta}_I \Gamma^\nu \dot{\theta}_J \\
&\qquad \qquad \qquad + \frac{\gamma^{\tau \sigma}}{\gamma^{\tau \tau}} C_1 + \frac{1}{2 T \gamma^{\tau \tau}} C_2\Big)\,,
\end{aligned}
\end{equation}
with
\begin{align}
C_1 =&\, P_\mu \acute{X}^\mu + \frac{i}{2} P_\mu \bar{\theta}_I \Gamma^\mu \acute{\theta}_I + \frac{T}{2} i \sigma_3^{IJ} \acute{X}^\mu \bar{\theta}_I \Gamma_\mu \acute{\theta}_J -\frac{T}{2} i B_{\mu \nu} \acute{X}^\mu \bar{\theta}_I \Gamma^\nu \acute{\theta}_I \,, \\
C_2 =&\, \hat{G}^{\mu\nu}P_\mu P_\nu+T^2\hat{G}_{\mu\nu} \acute{X}^\mu\acute{X}^\nu+2 T \hat{G}^{\mu\nu} \hat{B}_{\nu\kappa}P_\mu \acute{X}^\kappa+T^2\hat{G}^{\mu\nu}\hat{B}_{\mu\kappa}\hat{B}_{\nu\lambda}\acute{X}^\kappa\acute{X}^\lambda \\
 &\quad + i T^2 \acute{X}^\mu \bar{\theta}_I \Gamma_\mu \acute{\theta}_I + i T \sigma_3^{IJ} P_\mu \bar{\theta}_I \Gamma^\mu \acute{\theta}_J + i T^2 B_{\mu \nu} \acute{X}^\nu \sigma_3^{IJ} \bar{\theta}_I \Gamma^\mu \acute{\theta}_J \,.\nonumber
\end{align}
We introduced the shorthand notation $\dot{X}^\mu = \partial_\tau X^\mu$ and $\acute{X}^\mu = \partial_\sigma X^\mu$ (similarly for fermions). We also used the fact that the Weyl-invariant metric on the worldsheet satisfies $\det \gamma=-1$, so that we can eliminate $\gamma^{\sigma \sigma}$ through the relation 
\begin{equation}
    \gamma^{\sigma \sigma} = {[(\gamma^{\tau \sigma})^2-1]}/{\gamma^{\tau \tau}}\,.
\end{equation}
The action \eqref{eq:1st} is no longer manifestly invariant under worldsheet reparametrisation. Remember however that we have the Virasoro constraints, and the equations of motion for the (unphysical) degrees of freedom associated to the worldsheet metric are simply $C_1=C_2=0$. 

The idea is to pick the global $AdS$ time $t$ (time-like) and an angle $\varphi$ (space-like) parametrising one of the big circles of $S^n$.
The choice of the space-like coordinate is made in such a way as to preserve as much supersymmetry as possible.~\footnote{Other choices are possible, see for instance~\cite{Frolov:2019nrr,Frolov:2019xzi,Borsato:2023oru}, which may (albeit non-trivially) lead to an invariance under larger superalgebras~\cite{Borsato:2023oru}.} 
 The remaining coordinates are the transverse fields and we write $X=(t, \varphi, X^1, \dots X^8)$. We introduce the (target-space) lightcone coordinates and their conjugate momenta
\begin{align}
X^+ &= (1-a) t + a \varphi \,, &\qquad X^- &= - t+\varphi \,, \\
P_+ &= P_t + P_\varphi \,, &\qquad P_- &= - a P_t +(1-a) P_\varphi\,,
\end{align}
where we include a gauge parameter $0 \leq a \leq 1$. The inverse relations are
\begin{align}
    t &= X^+ - a X^- \,, &\qquad \varphi &= X^+ +(1-a) X^- \,, \\
    P_t &= (1-a) P_+ - P_- \,, &\qquad P_\varphi &= a P_+ + P_- \,.
\end{align}
We consider backgrounds that depend on $t$ and $\varphi$ only through their derivatives (with respect to $\tau$ and $\sigma$), so that translations in these two coordinates are manifest symmetries of the action;  $t$ and $\varphi$ then parametrise two $\alg{u}(1)$ isometric directions, with corresponding conserved (Noether) charges given by the space-time energy $\gen{E}$ and the angular momentum $\gen{J}$ respectively, 
\begin{equation} \label{eq:Noether1}
\gen{E} = - \int_{0}^{\cir} \extder \sigma \, P_t \,, \qquad \gen{J} = + \int_{0}^{\cir} \extder \sigma \, P_\varphi \,,
\end{equation}
or, in terms of the target space lightcone momenta,
\begin{equation} \label{eq:Noether2}
\gen{P}_+ = \int_{0}^{\cir} \extder \sigma \, P_+ = \gen{J}- \gen{E} \,, \qquad  
\gen{P}_- = \int_{0}^{\cir} \extder \sigma \, P_- = (1-a)\gen{J}+ a \gen{E} \,.
\end{equation}
We can see that the cases $a=0$ and $a=1/2$ are special; in the former, $\gen{P}_-=\gen{J}$ is simply the (quantised) angular momentum along~$\varphi$, while in the latter we get a more ``symmetrical'' setup (which turns out to simplify some computations).

\subsubsection{Uniform lightcone gauge}

The uniform lightcone gauge consists in eliminating the fluctuations of $X^+$ and $P_-$ as in \eqref{eq:sib-ulcg}, which fixes here
\begin{equation} \label{eq:lc}
X^+ = \tau+ 2 \pi m a \frac{\sigma}{\cir} \,, \qquad P_- =1 \,,
\end{equation}
where we introduce  the integer winding number $m$ along the circle parametrized by $\varphi$, so that $\varphi(\cir) - \varphi(0)= 2 \pi m$.
Because $P_-=1$ the lightcone momentum is spread uniformly along the string, hence the name of ``uniform lightcone gauge''. Moreover, we have that
\begin{equation}
\label{eq:worldsheetlength}
    \gen{P}_- =\int_{0}^{R} d\sigma\, P_-= \cir \,,
\end{equation}
 and therefore after lightcone gauge fixing the worldsheet size is fixed in terms of the charges of the state, namely its energy $\gen E$ and angular momentum $\gen J$ through the relation \eqref{eq:Noether2}. 

After lightcone gauge fixing the bosonic part of the first order action \eqref{eq:1st} becomes
\begin{equation}
S = \int \extder \tau \extder \sigma \left(\dot{X}^- + P_+  + P_j \dot{X}^j + \frac{\gamma^{\tau \sigma}}{\gamma^{\tau \tau}} C_1^{g.f.} + \frac{1}{2 T \gamma^{\tau \tau}} C_2^{g.f.}\right)\,.
\end{equation}
The first term is a total derivative, and assuming fall-off conditions at $\tau = - \infty$ and $\tau = +\infty$ it drops out. We are now left with imposing the lightcone Virasoro constraints. The first constraint reads
\begin{equation} \label{eq:C1gf}
C_1^{g.f.}=0 \quad \Rightarrow \quad \acute{X}^- = - P_j \acute{X}^j - \frac{2 \pi m a}{R} P_+ \,,
\end{equation}
and plugging into the second constraint gives
\begin{equation} \label{eq:C2gf}
C_2^{g.f.} =0 \quad \Rightarrow \quad  P_+ = P_+(X^j,P_j) \,.
\end{equation}  
Finally, the action takes the form
\begin{equation} \label{eq:lcaction}
S = \int \extder \tau \extder \sigma \left(  P_j \dot{X}^j - H \right) \,, \qquad H=-P_+(X^j,P_j) \,.
\end{equation}
The quantity $H$ then naturally takes the interpretation of the lightcone Hamiltonian (density). A consequence of this and the relation in \eqref{eq:Noether2} is that the worldsheet Hamiltonian is related to the string target-space energy and angular momentum through
\begin{equation}
\label{eq:lcHamiltonian}
\mathbf{H} = \int_{0}^{\cir} \extder \sigma H = \mathbf{E} - \mathbf{J} \,.
\end{equation} 
It is important to note that this Hamiltonian usually takes a rather involved form, featuring square roots. For instance, assuming zero winding ($m=0$) and a metric and B-field of the schematic form
\begin{equation} \label{eq:GBsimp}
    G_{\mu \nu} = 
    \begin{pmatrix} 
    G_{++} & G_{+-} & 0 \\
    G_{-+} & G_{--} & 0 \\
    0 & 0 & G_{ij} 
    \end{pmatrix} \,, \qquad B_{\mu \nu} = 
    \begin{pmatrix} 
    0 & 0 & 0 \\
    0 & 0 & 0 \\
    0 & 0 & B_{ij}
    \end{pmatrix} \,,
\end{equation}
the (bosonic) Hamiltonian density is
\begin{equation} \label{eq:hden}
    H = \frac{G^{+-} +\sqrt{(G^{+-})^2 - G^{++} (G^{--}+T^2 G_{--} P_i P_j \acute{X}^i \acute{X}^j + \mathcal H})}{G^{++}} \,,
\end{equation}
where
\begin{equation}
    \mathcal H = G^{ij}P_i P_j + T^2 G_{ij} \acute{X}^i \acute{X}^j + 2 T G^{ij} B_{ik} P_i \acute{X}^k + T^2 G^{ij} B_{ik} B_{jl} \acute{X}^k \acute{X}^l \,.
\end{equation}
Note that because $C_2^{g.f.}=0$ is a quadratic equation for $P_+$, there are two possible solutions. One needs to choose the sign in such a way as to obtain a positive Hamiltonian. Later we will expand this Hamiltonian in powers of the transverse fields (or equivalently in the large tension expansion), and choosing the correct branch usually ensures that the expansion starts at quadratic order in the fields.
\begin{remark}
It is possible to rewrite the lightcone gauge condition \eqref{eq:lc} as well as the lightcone gauge fixed action \eqref{eq:lcaction} without having to resort to the conjugate momenta. For this we recall that winding and momentum interchage under T-duality transformation. We can then dualise the model in $X^-$, and calling the dual coordinate $\tilde{X}^-$, the condition $P_-=1$ becomes $\tilde{X}^- =\sigma$. These two ways of imposing the lightcone gauge should of course be equivalent, but this way of thinking makes it possible to work with the Lagrangian formulation rather than the Hamiltonian one. Such a formulation has been carefully worked out in \cite{Arutyunov:2014jfa}, also including fermions.
\end{remark}

At this point let us mention that \eqref{eq:C1gf} fixed the spatial derivative $\acute{X}^-$. Consistency then requires to impose
\begin{equation}
X^-(\cir) - X^-(0) = \int_{0}^{\cir} \extder \sigma \, \acute{X}^-(\sigma) = \int_{0}^{\cir} \extder \sigma \, \left(- P_j \acute{X}^j - \frac{2 \pi m a}{R} P_+ \right)= 2 \pi m \,.
\end{equation}
This is the ``level-matching'' condition. The worldsheet momentum, namely the charge $\gen{p}$ associated to translations along the $\sigma$ direction of the worldsheet, is precisely given by this combination. Consequently, physical states obey 
\begin{equation}
    \gen{p} |\text{phys}\rangle= 2 \pi m |\text{phys}\rangle\,,\qquad m\in\mathbb{Z} \,.
\end{equation}

\subsubsection{Kappa gauge}
When discussing the GS action we have seen that there is another redundancy in its description, realised as a local ``kappa-symmetry'' on the fermionic fields. To be left with only physical fields we also need to choose a particular kappa gauge, which will set some of the fermions to zero. Of course, this kappa-gauge should be compatible with the bosonic uniform lightcone gauge chosen above. We recall that the kappa-symmetry involves a variation of the bosonic coordinates $X^\mu$, the fermions $\theta^I$ as well as the worldsheet metric, see \eqref{eq:kappa-symmetry}, which we rewrite here for convenience:
\begin{equation} 
\begin{gathered}
    \delta_\kappa \theta^I = 2i \Gamma_\mu \partial_\alpha X^\mu \kappa^{I\alpha} + {\cal O}(\theta^2), \qquad \delta_\kappa X^\mu = i \bar{\theta}^I \Gamma^\mu \delta_\kappa \theta^J \delta_{IJ} + {\cal O}(\theta^3) , \\
    \delta_\kappa \gamma^{\alpha\beta} = 16 \sqrt{-h}  P_{I}^{J\alpha\gamma} P_{J}^{K\beta\delta} \bar{\kappa}^I_\gamma {\cal D}_{\delta KL} \theta^L \ .
\end{gathered}
\end{equation}
% Now, the bosonic lightcone gauge \eqref{eq:lc} fixes $X^+=\tau + a m \sigma$ and hence compatibility with kappa-symmetry requires that $\delta_\kappa X^+=0$. This in turns imposes a constraint on the fermions, namely
% \begin{equation}
% \Gamma^+ \theta =0\,, \qquad \Gamma^+ = \frac{1}{\sqrt{2}} (\Gamma^0 + \Gamma^1),
% \end{equation}
% where $\Gamma^0$ and $\Gamma^1$ are the (tangent-space) Gamma-matrices associated with the coordinates $t$ and $\varphi$. 
A convenient gauge is then (recall that in our convention the coordinates used for lightcone gauge fixing are $X^0 = t$ and $X^1 =\varphi$)
\begin{equation}
    G^+ \theta = \bar{\theta} G^+ =0 \,, \qquad G^\pm = \frac{1}{2}(\Gamma^0 \pm \Gamma^1)\,.
\end{equation}
The Gamma-matrices with flat lightcone indices satisfy the identities
\begin{equation}
    (G^\pm)^2 =0\,, \qquad G^+ G^- + G^- G^+ = -1\,, \qquad G^+ + G^-=\Gamma^0\,.
\end{equation}
Moreover, by using the second identity, it follows that
\begin{equation}
    \bar{\theta} \Gamma_{j_1} \dots \Gamma_{j_n} \theta =0\,,
\end{equation}
where $\Gamma_j$ denotes a Gamma-matrix in the transverse directions. Again assuming a metric and B-field of the form \eqref{eq:GBsimp}, the action after lightcone gauge fixing takes the form (dropping total derivative terms and assuming zero winding)
\begin{equation} \begin{aligned}
    S = &\int \extder \tau \extder \sigma \Big( P_+ + P_j \dot{X}^j + \frac{i}{2} \bar{\theta}^I \left( \delta^{IJ}(P_+ \Gamma^+ +  \Gamma^-) + T \sigma_3^{IJ} \acute{X}^-  \Gamma_- \right)\dot{\theta}_I \\
    &\qquad \qquad + \frac{\gamma^{\tau \sigma}}{\gamma^{\tau \tau}} C_1^{g.f.} + \frac{1}{2 T \gamma^{\tau \tau}} C_2^{g.f.} \Big)\,.
    \end{aligned}
\end{equation}
In the above, 
\begin{equation} \begin{aligned}
    \Gamma^+ &= (1-a) \Gamma^t + a \Gamma^\varphi\,, &\qquad \Gamma^- &= \Gamma^\varphi - \Gamma^t\,, \\
    \Gamma_- &= -a \Gamma_t +(1-a) \Gamma_{\varphi}\,, &\qquad \Gamma_+ &= \Gamma_t + \Gamma_\varphi\,.
    \end{aligned}
\end{equation}
As in the bosonic case the Hamiltonian density is related to the lightcone momentum, $H = -P_+(X^j,P_j, \theta^I)$ upon solving the Virasoro constraints. Its explicit expression will now also include fermionic fields.

\subsubsection{Perturbative expansion}

The lightcone gauge-fixed theory is a 2d (non-relativistic) theory defined on a cylinder of size $\cir=  \gen{P_-}$, and its  time-evolution if governed by the Hamiltonian $\gen H$. In order to analyse this theory we want to be able to define asymptotic ``in'' and ``out'' states and diagonalise $\gen{H}$ in this basis. 
The first step towards this goal is to take the decompactification limit, in which the cylinder decompactifies to a plane. This corresponding to sending $\cir \rightarrow \infty$. In that limit the lightcone momentum $\gen{ P_-}$ becomes infinite. However, the Hamiltonian should remain finite.  This means that both $\gen J$ and $\gen E$ become infinite, but their difference $\gen J - \gen E$ stays finite. 
%Furthermore, the periodic boundary conditions should be replaced by fall-off conditions at infinity.
Then, we perturbatively expand the Hamiltonian around the classical solution \eqref{eq:lc}.  To do this we rescale the worldsheet coordinate $\sigma \rightarrow T \sigma$, and the bosonic and fermionic fields as
\begin{equation}
X^j \rightarrow \frac{1}{\sqrt{T}} X^j\,, \qquad P_j \rightarrow \frac{1}{\sqrt{T}} P_j\,, \qquad \theta^I \rightarrow \frac{1}{\sqrt{T}} \theta^I\,.
\end{equation}
The worldsheet Hamiltonian can then be naturally expanded in powers of the transverse fields (or equivalenty, in inverse power of the string tension)
\begin{equation}
\gen H= \gen{H^{(2)}} + \frac{1}{\sqrt{T}} \gen{H}^{(3)} + \frac{1}{T} \gen{H}^{(4)} + \dots
\end{equation}
The order $\gen{H}^{(n)}$ contains $n$ powers of the transverse fields $X^j$ (or derivative/momentum thereof) and its explicit expression depends on the theory that we consider. Henceforth we will focus on the $AdS_3 \times S^3 \times T^4$ superstring.

\subsection{Application to \texorpdfstring{$AdS_3 \times S^3 \times T^4$}{AdS3xS3xT4}}

In global coordinates, the metric of $AdS_3 \times S^3 \times T^4$ is given by
\begin{equation}
    \extder s^2 = -(1+\rho^2) \extder t^2 + \frac{\extder \rho^2}{1+\rho^2} + \rho^2 \extder \psi^2  + (1-r^2) \extder \varphi^2 +\frac{\extder r^2}{1-r^2} + r^2 \extder \phi^2 + \extder x^i \extder x^i\,,
\end{equation}
with $i=6,7,8,9$.
Note that the radius of $AdS_3$ and $S^3$ has been reabsorbed into the coordinates, this is why it does not appear explicitly in the metric.
On top of that, there is also the B-field and its corresponding H-flux, given by
\begin{equation} \begin{aligned}
    B_{(2)} &= -q\left( \rho^2 \extder t \wedge \extder \psi + r^2 \extder \varphi \wedge \extder \phi\right)\,, \\
    H_{(3)} &= \extder B_{(2)} = 2q \left( \rho \, \extder t \wedge \extder \rho \wedge \extder \psi + r \, \extder \varphi \wedge \extder r \wedge \extder \phi\right)\,,
    \end{aligned}
\end{equation}
with $0\leq q \leq 1$ a free parameter of the mixed flux theory. The $q=1$ case corresponds to the pure NSNS point, and one can check that the metric and H-flux alone solve the supergravity equations of motion with constant dilaton $\Phi$. For generic $q$ on the other hand, to satisfy the supergravity equations of motion the above NSNS sector needs to be supplemented with RR fluxes (again with constant dilaton), and the easiest way to do that is to add a 3-form RR flux 
\begin{equation}
    F_{(3)} = 2\sqrt{1-q^2} \left( \rho \, \extder t \wedge \extder \rho \wedge \extder \psi + r \, \extder \varphi \wedge \extder r\wedge \extder \phi\right)\,.
\end{equation}

In the curved part of the background there are four $\alg{u}(1)$ isometries, realised as shifts in $t, \varphi, \psi$ and $\phi$. The associated Noether charges are respectively the target space energy, target space angular momentum, spin in $AdS$ and spin in the sphere. We will use $t$ and $\varphi$ for the lightcone gauge fixing. Then, there are also four $\alg{u}(1)$ isometries coming from the $T^4$, simply realised as shifts in $x^i$ with $i=6,7,8,9$. Performing the field redefinition
\begin{equation} \begin{aligned}
    \rho &= \frac{\sqrt{(X_1)^2 + (X_2)^2}}{1-\frac{1}{4}((X_1)^2+(X_2)^2)}\,, &\qquad \psi &= \arctan \left( \frac{X_2}{X_1}\right)\,, \\
    r &= \frac{\sqrt{(X_3)^2 + (X_4)^2}}{1+\frac{1}{4}((X_3)^2+(X_4)^2)}\,, &\qquad \phi &= \arctan \left( \frac{X_4}{X_3}\right)\,,
    \end{aligned}
\end{equation}
the metric becomes
\begin{equation} \begin{aligned}
    \extder s^2 &= - \left( \frac{1+\frac{1}{4}(X_1^2+X_2^2)}{1-\frac{1}{4}(X_1^2+X_2)}\right)^2 \extder t^2 + \left( \frac{1}{1 - \frac{1}{4} (X_1^2+X_2^2)}\right)^2 \left( \extder X_1^2 + \extder X_2^2 \right) \\ 
     &\qquad + \left( \frac{1-\frac{1}{4}(X_3^2+X_4^2)}{1+\frac{1}{4}(X_3^2+X_4^2)}\right)^2 \extder \varphi^2 + \left( \frac{1}{1 + \frac{1}{4}(X_3^2+X_4^2)}\right)^2 \left( \extder X_3^2 + \extder X_4^2 \right) + \extder x^i \extder x^i\,,
    \end{aligned}
\end{equation}
while the B-field is
\begin{equation}
    B_{(2)} = q \left( \frac{X_1 \extder X_2 - X_2 \extder X_1}{(1-\frac{1}{4}(X_1^2+X_2^2))^2} \wedge \extder t + \frac{X_3 \extder X_4 - X_4 \extder X_3}{(1+\frac{1}{4}(X_3^2+X_4^2))^2} \wedge  \extder \varphi  \right)\,.
\end{equation}
This change of coordinate makes it easier to construct states with well defined eigenvalues under the charges associated to the four $\alg{u}(1)$ isometries. This will be important to make the link with the fundamental excitations scattered by the S-matrix.

% There are four $\alg{u}(1)$ isometries realised as translation along $t, \psi, \varphi$ and $\phi$ that give rise to four conserved charges. These are the target space energy $\mathcal E = - \int \extder \sigma P_t = \gen{L}_\L + \gen{L}_\R$, the target space angular momentum  $\mathcal J = \int \extder \sigma P_\varphi = \gen{J}_\L + \gen{J}_\R$, the AdS spin $\mathcal J_\psi = \int \extder \sigma P_\psi = \gen{L}_\L - \gen{L}_\R$ as well as the spin on the sphere, $\mathcal J_\phi = \int \extder \sigma P_\phi = \gen{J}_\L - \gen{J}_\R$. We define the four linear combinations
% \begin{align}
%     \gen{H} &= \gen{L}_\L + \gen{L}_\R - \gen{J}_\L - \gen{J}_\R\,, \\
%     \gen{M} &= \gen{L}_\L - \gen{L}_\R - \gen{J}_\L + \gen{J}_\R\,, \\
%     \gen{B} &= \gen{L}_\L - \gen{L}_\R + \gen{J}_\L - \gen{J}_\R\,, \\
%     \gen{R} &= \frac{1}{2} \left( \gen{L}_\L + \gen{L}_\R + \gen{J}_\L + \gen{J}_\R \right)\,.
% \end{align}
% From the general discussion about lightcone gauge fixing we deduce that $\gen{H}$ is the worldsheet Hamiltonian, while $\gen{R}=2 r$ in the $a=1/2$ gauge. $\gen{M}$ has an interpretation in terms of a worldsheet angular momentum, while $\gen{B}$ generates an outer-automorphism of the lightcone gauge fixed symmetry algebra. In particular,
% \begin{equation}
%     \com{\gen{B}}{\gen{Q}_\L{}^1}
% \end{equation}

\paragraph{Fields and conjugate momenta.}
We define the complex linear combinations
\begin{equation}
    Z = \frac{X_1+i X_2}{\sqrt{2}}\,, \quad \bar{Z} = \frac{X_1-i X_2}{\sqrt{2}}\,, \quad Y = \frac{X_3 + i X_4}{\sqrt{2}}\,, \quad \bar{Y} = \frac{X_3 - i X_4}{\sqrt{2}}\,.
\end{equation}
These have conjugate momenta given by
\begin{equation}
    P_Z =  \frac{P_1 - i P_2}{\sqrt{2}}\,, \qquad P_{\bar{Z}} =  \frac{P_1 + i P_2}{\sqrt{2}}\,, \qquad P_Y =  \frac{P_3 - i P_4}{\sqrt{2}}\,, \qquad P_{\bar{Y}} =  \frac{P_3 + i P_4}{\sqrt{2}}\,.
\end{equation}
Similarly, for the torus directions we introduce
\begin{equation}
    X^{11} = \frac{x_6 + i x_7}{\sqrt{2}}\,, \quad X^{22} = \frac{x_6 - i x_7}{\sqrt{2}}\,, \quad X^{12} = \frac{x_8 + i x_9}{\sqrt{2}}\,, \quad X^{21} = -\frac{x_8 - i x_9}{\sqrt{2}}\,,
\end{equation}
with conjugate momenta
\begin{equation}
    P_{\dot{a} a} =  \epsilon_{\dot{a} \dot{b}} \epsilon_{ab} \dot{X}^{\dot{b} b}\,.
\end{equation}

We then fix the lightcone gauge as in \eqref{eq:lc}, with $m=0$. Notice that this classical solution used for lightcone gauge fixing does not involve any winding in the torus direction and hence we formally consider the theory on $\mathbb{R}^4$ instead of $T^4$.

% They satisfy the canonical Poisson structure (taking at equal worldsheet time $\tau$)
% \begin{equation} \begin{aligned}
%     \anticom{Z(\sigma)}{P_{\bar{Z}}(\sigma')}_{P.B.} &=  \anticom{\bar{Z}(\sigma)}{P_Z(\sigma')}_{P.B.} = i \delta(\sigma - \sigma')\,, \\
%     \anticom{Y(\sigma)}{P_{\bar{Y}}(\sigma')}_{P.B.} &=  \anticom{\bar{Y}(\sigma)}{P_Y(\sigma')}_{P.B.} = i \delta(\sigma - \sigma')\,, \\
%     \anticom{X^{\dot{a} a}(\sigma)}{P_{\dot{b} b}(\sigma')}_{P.B.} &=   i \delta^{\dot{a}}_{\dot{b}} \delta^a_b \delta(\sigma - \sigma')\,.
%     \end{aligned}
% \end{equation}
\paragraph{Quadratic Hamiltonian.} Applying the procedure explained in the previous section leads to the quadratic Hamiltonian (we include the fermions for completeness)
\begin{equation} \begin{aligned}
    H^{(2)} &= P_{Z} P_{\bar{Z}} + \acute{Z} \acute{\bar{Z}} + Z \bar{Z} - i q (Z \acute{\bar{Z}} - \bar{Z}\acute{Z}) \\
    &\qquad + P_{Y} P_{\bar{Y}} +  \acute{Y} \acute{\bar{Y}} +  Y \bar{Y} - i q (Y \acute{\bar{Y}} - \bar{Y}\acute{Y}) \\
    &\qquad  + i \zeta_\R^* \acute{\zeta}_\R + q \zeta_\R^* \zeta_\R - i \zeta_\L^* \acute{\zeta}_\L - q \zeta_\L^* \zeta_\L + \sqrt{1-q^2} (\zeta_\R^* \zeta_\L + \zeta_\L^* \zeta_\R)\\
    &\qquad  + i \eta_\R^* \acute{\eta}_\R + q \eta_\R^* \eta_\R - i \eta_\L^* \acute{\eta}_\L - q \eta_\L^* \eta_\L + \sqrt{1-q^2} (\eta_\R^* \eta_\L + \eta_\L^* \eta_\R)\\
    &\qquad + P_{\dot{a} a} P^{\dot{a} a} +  \acute{X}_{\dot{a} a} \acute{X}^{\dot{a} a} + i \chi_{+ a}^* \acute{\chi}_+^a - i \chi_{- a}^* \acute{\chi}_-^a\,.
    \end{aligned}
\end{equation} 
This is the Hamiltonian describing two  $AdS_3$ massive bosons $Z$ and $\bar{Z}$, two $S^3$ massive bosons $Y$ and $\bar{Y}$, as well as four massive fermions $\zeta_{\L, \R}$ and $\eta_{\L, \R}$ (recall that for fermions the fields and their complex conjugate are not independent). They all have unit mass. Then we also have four massless bosons $X^{\dot{a} a}$ coming from the torus direction, as well as four massless fermions $\chi^a_\pm$ for $a=1,2$ (as required by supersymmetry). The fields are free, but the presence of NSNS flux for $q \neq 0$ introduces a parity-breaking term (a term that changes sign under $\sigma \rightarrow - \sigma$).

Notice that the presence of the factor of $i$ in the Hamiltonian is not problematic and is in fact required to have reality of $H^{(2)}$. Indeed, the fields $Z,Y$ are complex and the linear combination appearing in the bracket of the first two lines is purely imaginary. Therefore to have a real Hamiltonian these terms need to be supplemented with a factor of $i$. Had we written the Hamiltonian in terms of the real fields $X^j$ this term would simplify have been $q (X_2 \acute{X}_1 - X_1 \acute{X}_2)$, which is manifestly real. Similarly for the fermions, the factor of $i$ is required from their reality condition.

\paragraph{Equations of motion and oscillator representation.}
The massive bosonic fields satisfy the equations of motion
\begin{equation} \begin{aligned}
    (\Box + 1) Z &= -2 i q \acute{Z}\,, \qquad (\Box + 1) Y = -2 i q \acute{Y}\,, \\
    (\Box + 1) \bar{Z} &= +2 i q \acute{\bar{Z}}\,, \qquad (\Box + 1) \bar{Y} = +2 i q \acute{\bar{Y}}\,,
    \end{aligned}
\end{equation}
with $\Box = \partial_\tau^2 - \partial_\sigma^2$.
There is a slight modification from the usual Klein-Gordon equation for free bosonic fields due to the  parity-breaking term related to the presence of NSNS flux. The solutions are still of plane-wave form, for instance
\begin{align}
    Z(\sigma) &= \frac{1}{\sqrt{2 \pi}}\int  \extder p \,  \left(e^{-i \omega \tau + i p \sigma} \frac{a_Z(p)}{\sqrt{2 \omega(p)}} + e^{i \bar{\omega} \tau - i p \sigma} \frac{a^\dagger_{\bar{Z}} (p)}{\sqrt{2 \bar{\omega}(p)}} \right)\,, \\
    \bar{Z}(\sigma) &= \frac{1}{\sqrt{2 \pi}} \int  \extder p \,  \left(e^{-i \bar{\omega} \tau + i p \sigma} \frac{a_{\bar{Z}}(p)}{\sqrt{2 \bar{\omega}(p)}} + e^{i \omega \tau - i p \sigma} \frac{a^\dagger_{Z} (p)}{\sqrt{2 \omega(p)}} \right)\,,
\end{align}
where $a_Z(p)$, $a^\dagger_Z(p)$ and $a_{\bar{Z}}(p)$, $a^\dagger_{\bar{Z}}(p)$ satisfy the canonical commutation relations,
\begin{equation}
    \com{a_Z(p)}{ a_{Z}^\dagger(p')} = \com{a_{\bar{Z}}(p)}{ a_{\bar{Z}}^\dagger(p')}  = \delta(p-p')\,.
\end{equation}
These can  be used to generate the excitations
\begin{equation}
    \ket{Z(p)} = a_Z^\dagger(p) \ket{0}\,,  \qquad \ket{\bar{Z}(p)} = a_{\bar{Z}}^\dagger(p) \ket{0}\,.
\end{equation}
One has similar expressions for the fields $Y(\sigma)$ and $\bar{Y}(\sigma)$. 

The massless bosonic fields satisfy
\begin{equation}
    \Box X^{\dot{a} a} =0\,,
\end{equation}
whose plane-wave solutions are simply
\begin{align}
    X^{\dot{a} a}(\sigma) &= \frac{1}{\sqrt{2 \pi}} \int  \extder p \,  \left(e^{-i \omega_\circ \tau + i p \sigma} \frac{a^{\dot{a} a}(p)}{\sqrt{2 \omega_\circ(p)}} + e^{i \omega_\circ \tau - i p \sigma} \frac{a^\dagger_{\dot{a} a} (p)}{\sqrt{2 \omega_\circ(p)}} \right).
\end{align}

\begin{tcolorbox}
  \begin{exercise}
    Show that one has the dispersion relations
    \begin{equation}
        \omega= \sqrt{1- 2 q p + p^2}\,, \qquad \bar{\omega}= \sqrt{1 + 2 q p + p^2}\,, \qquad \omega_\circ = \sqrt{p^2}\,.
    \end{equation}
  \end{exercise}
\end{tcolorbox}
\ifsol
\begin{tcolorbox}
  \textbf{Solution:} When going from position space to momentum space one has, for a plane wave of the form $Z(\sigma) \sim \int \extder p \, e^{-i \omega \tau + i p \sigma} a_Z(p)$
  \begin{equation}
      \Box Z (\sigma)\rightarrow (- \omega^2 + p^2) a_Z(p), \qquad \acute{Z}(\sigma) \rightarrow - i p \, a_Z(p)
  \end{equation}
\end{tcolorbox}
\fi

The massive fermionic fields satisfy the equations of motion
\begin{equation} \begin{aligned}
    (\partial_- + i q) \zeta_{\R} + i \sqrt{1-q^2} \zeta_\L &=0\,, &\qquad (\partial_+ - i q) \zeta_\L + i \sqrt{1-q^2} \zeta_{\R} &=0\,,  \\
     (\partial_- + i q) \eta_{\R} + i \sqrt{1-q^2} \eta_\L &=0\,, &\qquad (\partial_+ - i q) \eta_\L + i \sqrt{1-q^2} \eta_{\R} &=0\,,
    \end{aligned}
\end{equation}
with $\partial_\pm = \partial_\tau \pm \partial_\sigma$.
These are Dirac equations, modified by the presence of the parity breaking term. Notice that combining the two equations gives
\begin{equation} \begin{aligned}
   (\Box +1) \zeta_{\L, \R} &= -2 i q \, \acute{\zeta}_{\L,\R} \,, &\qquad (\Box+1) \eta_{\L,\R}  &= -2 i q \, \acute{\eta}_{\L,\R}\,, \\
   \end{aligned}
\end{equation}
which is the same equation as for the massive bosonic fields $Z$ and $Y$. 
The solution of the above equations are of plane-wave form, featuring coefficients. For instance (one has similar expression for the fields $\eta_{\L,\R}$ since they satisfy the same equations of motion),
\begin{align}
        \zeta_\L(\sigma) &= \frac{1}{\sqrt{2 \pi}} \int  \extder p \, \left(e^{-i \omega \tau + i p \sigma} f_\L(p) a_\zeta(p) + e^{i \bar{\omega} \tau - i p \sigma} g_\L(p) a^\dagger_{\bar{\zeta}} (p) \right)\,, \\
        \zeta_\L^*(\sigma) &= \frac{1}{\sqrt{2 \pi}} \int  \extder p \, \left(e^{-i \bar{\omega} \tau + i p \sigma} g^*_\L(p) a_{\bar{\zeta}}(p) + e^{i \omega \tau - i p \sigma} f^*_\L(p) a^\dagger_{\zeta} (p) \right)\,, \\
        \zeta_\R(\sigma) &= \frac{1}{\sqrt{2 \pi}} \int  \extder p \, \left(e^{-i \omega \tau + i p \sigma} f_\R(p) a_\zeta(p) + e^{i \bar{\omega} \tau - i p \sigma} g_\R(p) a^\dagger_{\bar{\zeta}} (p) \right)\,, \\
        \zeta^*_\R(\sigma) &= \frac{1}{\sqrt{2 \pi}} \int  \extder p \, \left(e^{-i \bar{\omega} \tau + i p \sigma} g^*_\R(p) a_{\bar{\zeta}}(p) + e^{i \omega \tau - i p \sigma} f^*_\R(p) a^\dagger_{\zeta} (p) \right)\,.
\end{align}
The various functions are fixed by requiring that the equations of motion are satisfied, and that the ladder operators $a_\zeta(p)$,$a^\dagger_\zeta(p)$ and $a_{\bar{\zeta}}(p)$,$a^\dagger_{\bar{\zeta}}(p)$ satisfy canonical commutation relations. The creation operators can be used to define the excitations
\begin{equation}
    \ket{\zeta(p)} = a_{\zeta}^\dagger(p) \ket{0}\,, \qquad \ket{\bar{\zeta}(p)} = a_{\bar{\zeta}}^\dagger(p) \ket{0}\,.
\end{equation}
\begin{tcolorbox}
  \begin{exercise}
    Show that this fixes 
\begin{align}
    \omega &= \sqrt{1 - 2 q p + p^2}\,, & f_\L(p) &= \frac{1}{\sqrt{2 \omega}} \frac{\sqrt{1-q^2}}{\sqrt{\omega-p+q}}\,, & f_\R(p) &= \frac{1}{\sqrt{2 \omega}} \sqrt{\omega-p+q}\,, \\
    \bar{\omega} &= \sqrt{1 + 2 q p + p^2}\,, & g_\L(p) &= \frac{1}{\sqrt{2 \bar{\omega}}} \frac{\sqrt{1-q^2}}{\sqrt{\bar{\omega}+p+q}}\,, & g_\R(p) &= \frac{1}{\sqrt{2 \bar{\omega}}} \sqrt{\bar{\omega}+p+q}\,.
\end{align}
Check that these functions satisfy
\begin{equation}
    f_\L(p)^2 + f_\R(p)^2= g_\L(p)^2 + g_\R(p)^2=1\,.
\end{equation}
  \end{exercise}
\end{tcolorbox}
\ifsol
\begin{tcolorbox}[breakable, enhanced]
  \textbf{Solution:} The equations of motion in momentum space read
  \begin{equation} \begin{aligned}
      (-\omega + p - q) f_\L(p) + i \sqrt{1-q^2} f_\R(p) &=0\,, \\
      (-\omega - p + q) f_\R(p) + i \sqrt{1-q^2} f_\L(p) &=0\,, \\
      (+\bar{\omega} - p - q) g_\L(p) + i \sqrt{1-q^2} g_\R(p) &=0\,, \\
      (+\bar{\omega} + p + q) g_\R(p) + i \sqrt{1-q^2} g_\L(p) &=0\,.
      \end{aligned}
  \end{equation}
  Combining the first two equations as well as the last two equations immediately leads to the desired dispersion relations. We also want that the Hamiltonian takes the canonical form $ H = \int \extder p \, \sum_{\mathcal X} \omega_{\mathcal X} a^\dagger_{\mathcal X} a_{\mathcal X}$ where the sum is over all excitations. Plugging the oscillator representation into the Hamiltonian gives rise to the two equations
  \begin{equation} \begin{aligned}
  \omega &= (-p + q) |f_\R(p)|^2  + (p-q) |f_\L(p)|^2 + \sqrt{1-q^2} (f_\R^*(p) f_\L(p) + f_\L^*(p) f_\R(p))\,, \\
  \bar{\omega} &= (p + q) |g_\R(p)|^2  + (-p-q) |g_\L(p)|^2 + \sqrt{1-q^2} (g_\R^*(p) g_\L(p) + g_\L^*(p) g_\R(p))\,.
  \end{aligned}
  \end{equation}
  In combination with the equations of motion written previously these give the desired expressions for the functions $f_{\L,\R}(p)$ and $g_{\L,\R}(p)$.
\end{tcolorbox}
\fi

Using the ladder operators the Hamiltonian becomes diagonal,
\begin{equation}
    \gen H^{(2)} =  \int \extder p \sum_{\mathcal X} \omega_{\mathcal X} \,a^\dagger_{\mathcal X}(p) a_{\mathcal X}(p)\,,
\end{equation}
where $\mathcal X$ denotes all types of particles. The dispersion relation reads
\begin{equation} \label{eq:mu-perturb}
    \omega_{\mathcal X} = \sqrt{\mu^2 + 2 \mu q p + p^2}\,, \qquad 
    \mu = \left\{\begin{aligned}
        -&1 \text{ for } \mathcal X \in \{Z,Y, \zeta, \eta\}\,, \\
        +&1 \text{ for } \mathcal X \in \{\bar{Z},\bar{Y}, \bar{\zeta}, \bar{\eta} \}\,, \\
        &0 \text{ for }  \mathcal{X} \in \{ X^{\dot{a} a}, \chi_+^a, \chi_-^a\}\,.
         \end{aligned} \right.
\end{equation}
Furthermore, also the worldsheet momentum operator becomes diagonal, with
\begin{equation}
    \gen{p} = \int \extder p  \sum_{\mathcal X} p \, a^\dagger_{\mathcal X}(p) a_{\mathcal X}(p)\,.
\end{equation}

\paragraph{Higher order terms and perturbative S matrix.}

The cubic Hamiltonian vanishes, $\gen H^{(3)} =0$, while the expression for the quartic Hamiltonian $\gen H^{(4)}$, encoding the first non-trivial interactions, is already too involved to be written down explicitly here. For an integrable theory it takes the generic form
\begin{equation}
    \gen H^{(4)} = \dots + \int \extder p_1 \extder p_2 T_{ij}^{kl}(p_1,p_2) a_k^\dagger(p_1) a_l^\dagger(p_2) a^i(p_1) a^j(p_2) + \dots
\end{equation}
and allows one to identify the tree level S matrix 
\begin{equation}
    \gen{S} = 1 + \frac{i}{T} \gen{T}\,, \qquad \gen{T} \ket{\mathcal X_i \mathcal X_j} = T_{ij}^{kl} \ket{\mathcal X_k \mathcal X_l}\,.
\end{equation}
With the notation $(Z_+,Y_+,\zeta_+,\eta_+) = (Z,Y,\zeta,\eta)$ and $(Z_-,Y_-,\zeta_-,\eta_-)=(\bar{Z},\bar{Y},\bar{\zeta},\bar{\eta})$  one finds~\cite{Hoare:2013pma} the following tree-level scattering among massive particles: 

\medskip \noindent
Boson-Boson:
\begin{equation} \begin{aligned}
&\begin{aligned}
    \gen T \ket{Z_\pm Z_\pm} &= (- l_1 + c) \ket{Z_\pm Z_\pm}\,, &\, \gen T \ket{Z_\pm Z_\mp} &= \left(- l_2  + c \right) \ket{Z_\pm Z_\mp} - l_4 \ket{\zeta_\pm \zeta_\mp} - l_4 \ket{\mathcal \eta_\pm \mathcal \eta_\mp}\,, \\
    \gen T \ket{Y_\pm Y_\pm} &= (+l_1 +c) \ket{Y_\pm Y_\pm}\,, &\, \gen T \ket{Y_\pm Y_\mp} &= (+l_2 + c) \ket{Y_\pm Y^\mp} + l_4 \ket{\zeta_\pm \zeta_\mp} + l_4 \ket{\mathcal \eta_\pm \mathcal \eta_\mp}\,,
    \end{aligned}\\
&\begin{aligned}
    \gen T \ket{Z_\pm Y_\pm} &= (-l_3+c) \ket{Z_\pm Y_\pm} +l_5 \ket{\zeta_\pm \eta_\pm} - l_5 \ket{\mathcal \eta_\pm \mathcal \zeta_\pm}\,, &\, \gen T \ket{Z_\pm Y_\mp} &= (-l_3+ c) \ket{Z_\pm Y_\mp}\,, \\
    \gen T \ket{Y_\pm Z_\pm} &= (+l_3+c) \ket{Y^\pm Z^\pm} +l_5 \ket{\zeta_\pm \eta_\pm} -l_5 \ket{\mathcal \eta_\pm \mathcal \zeta_\pm}\,, &\,
    \gen T \ket{Y_\pm Z_\mp} &= (+l_3+c) \ket{Y_\pm Z_\mp}\,, 
    \end{aligned}
    \end{aligned}
\end{equation}
Fermion-Fermion:
\begin{equation} \begin{aligned}
    \gen T \ket{\zeta_\pm \zeta_\pm} &= c \ket{\zeta_\pm \zeta_\pm}\,, &\qquad  \gen T \ket{\zeta_\pm \zeta_\mp} &= - l_4 \ket{Z_\pm Z_\mp} + l_4 \ket{Y_\pm Y_\mp}\,, \\
    \gen T \ket{\eta_\pm \eta_\pm} &= c \ket{\eta_\pm \eta_\pm}\,, &\qquad \gen T \ket{\eta_\pm \eta_\mp} &= - l_4\ket{Z_\pm Z_\mp} + l_4  \ket{Y_\pm Y_\mp}\,,\\
    \gen T \ket{\zeta_\pm \eta_\pm} &= +l_5\ket{Z_\pm Y_\pm} +l_5 \ket{Y_\pm Z_\pm} \,, &\qquad \gen T \ket{\zeta_\pm \eta_\mp} &= c \ket{\zeta_\pm \eta_\mp}\,, \\
    \gen T \ket{\eta_\pm \zeta_\pm} &= -l_5 \ket{Z_\pm Y_\pm} -l_5\ket{Y_\pm Z_\pm}\,, &\qquad
    \gen T \ket{\eta_\pm \zeta_\mp} &= c \ket{\eta_\pm \zeta_\mp}\,, 
    \end{aligned}
\end{equation}
Boson-Fermion:
\begin{equation} \begin{aligned}
    \gen T \ket{Y_\pm \zeta_\pm} &= (l_6 + c) \ket{Y_\pm \zeta_\pm} - l_5 \ket{\zeta_\pm Y_\pm}\,, &\qquad  \gen T \ket{Y_\pm \zeta_\mp} &= (l_7 + c) \ket{Y_\pm \zeta_\mp} + l_4 \ket{\eta_\pm Z_\mp}\,, \\
    \gen T \ket{\zeta_\pm Y_\pm} &= (l_8 + c) \ket{\zeta_\pm Y_\pm} - l_5 \ket{Y_\pm \zeta_\pm}\,, &\qquad  \gen T \ket{\zeta_\pm Y_\mp} &= (l_9 + c) \ket{\zeta_\pm Y_\mp} - l_4 \ket{Z_\pm \eta_\mp}\,, \\
    \gen T \ket{Y_\pm \eta_\pm} &= (l_6 + c) \ket{Y_\pm \eta_\pm} - l_5 \ket{\eta_\pm Y_\pm}\,, &\qquad  \gen T \ket{Y_\pm \eta_\mp} &= (l_7 + c) \ket{Y_\pm \eta_\mp} - l_4 \ket{\zeta_\pm Z_\mp}\,, \\
    \gen T \ket{\eta_\pm Y_\pm} &= (l_8 + c) \ket{\eta_\pm Y_\pm} - l_5 \ket{Y_\pm \eta_\pm}\,, &\qquad  \gen T \ket{\eta_\pm Y_\mp} &= (l_9 + c) \ket{\eta_\pm Y_\mp} + l_4 \ket{Z_\pm \eta_\mp}\,, \\%
    \gen T \ket{Z_\pm \zeta_\pm} &= (-l_6 + c) \ket{Z_\pm \zeta_\pm} + l_5 \ket{\zeta_\pm Z_\pm}\,, &\qquad  \gen T \ket{Z_\pm \zeta_\mp} &= (-l_7 + c) \ket{Z_\pm \zeta_\mp} + l_4 \ket{\eta_\pm Y_\mp}\,, \\
    \gen T \ket{\zeta_\pm Z_\pm} &= (-l_8 + c) \ket{\zeta_\pm Z_\pm} + l_5 \ket{Z_\pm \zeta_\pm}\,, &\qquad  \gen T \ket{\zeta_\pm Z_\mp} &= (-l_9 + c) \ket{\zeta_\pm Z_\mp} - l_4 \ket{Y_\pm \eta_\mp}\,, \\
    \gen T \ket{Z_\pm \eta_\pm} &= (-l_6 + c) \ket{Z_\pm \eta_\pm} + l_5 \ket{\eta_\pm Z_\pm}\,, &\qquad  \gen T \ket{Z_\pm \eta_\mp} &= (-l_7 + c) \ket{Z_\pm \eta_\mp} - l_4 \ket{\zeta_\pm Y_\mp}\,, \\
    \gen T \ket{\eta_\pm Z_\pm} &= (-l_8 + c) \ket{\eta_\pm Z_\pm} + l_5 \ket{Z_\pm \eta_\pm}\,, &\qquad  \gen T \ket{\eta_\pm Z_\mp} &= (-l_9 + c) \ket{\eta_\pm Z_\mp} + l_4 \ket{Y_\pm \eta_\mp}\,.
    \end{aligned}
\end{equation}
The coefficients are given by
\begin{equation} \begin{aligned}
    l_1 &= \frac{1}{2} \frac{p_1+p_2}{p_1-p_2} (p_1 \omega_2 + p_2 \omega_1)\,, \quad l_2 = \frac{1}{2} \frac{p_1-p_2}{p_1+p_2}(p_1 \omega_2 + p_2 \omega_1)\,, \quad l_3 = -\frac{1}{2} (p_1 \omega_2 + p_2 \omega_1)\,, \\
    l_4 &= - \frac{p_1 p_2}{2(p_1+p_2)} \left( \sqrt{(\omega_1+p_1+\mu_1)(\omega_2+p_2+\mu_2)} - \sqrt{(\omega_1-p_1-\mu_1)(\omega_2-p_2-\mu_2)} \right)\,, \\
    l_5 &= - \frac{p_1 p_2}{2(p_1-p_2)} \left( \sqrt{(\omega_1+p_1+\mu_1)(\omega_2+p_2+\mu_2)} + \sqrt{(\omega_1-p_1-\mu_1)(\omega_2-p_2-\mu_2)} \right)\,, \\
    l_6 &= \frac{1}{2}(l_1 + l_3)\,, \qquad l_7 = \frac{1}{2}(l_2 +l_3)\,, \qquad l_8 = \frac{1}{2}(l_1-l_3)\,,\qquad l_9 = \frac{1}{2}(l_2-l_3)\,, \\
    \end{aligned}
\end{equation}
and
\begin{equation} 
c = -\big(a-\frac{1}{2}\big) (p_1 \omega_2 - p_2 \omega_1)\,.
\end{equation}
Some comments are in order. First of all, in terms of notation we use
\begin{equation}
    \ket{\mathcal X \mathcal Y} \equiv \ket{\mathcal X(p_1) \mathcal Y(p_2)}\,.
\end{equation}
The excitations' energies $\omega_1$ and $\omega_2$ should be replaced by the dispersion relation corresponding to the scattered excitations, namely $\mathcal X$ (with momentum $p_1$) and $\mathcal Y$ (with momentum $p_2$). A similar statement holds for $\mu_1$ and $\mu_2$ which should be replaced according to \eqref{eq:mu-perturb}. Then, when considering only bosonic fields, the tree-level S-matrix becomes diagonal. This is specific to the bosonic truncation and to $AdS_3 \times S^3 \times T^4$, a similar calculation for (bosonic) $AdS_5 \times S^5$ gives non-diagonal elements. Because the starting point for the perturbative calculation was the gauge-fixed theory up to quadratic order in fermions, this perturbative S-matrix does not capture processes involving four fermions, of the type Fermion+Fermion $\rightarrow$ Fermion+Fermion. The classical Yang-Baxter equation, 
\begin{equation}
    \com{\gen T_{23}}{\gen T_{13}} + \com{\gen T_{23}}{\gen T_{12}} + \com{\gen T_{13}}{\gen T_{12}}=0\,,
\end{equation}
indicating that the theory is classically integrable, is obeyed up to processes involving four fermions. Finally, the tree-level S-matrix simplifies in the $a=\frac{1}{2}$ gauge, and choosing other values for $a$ results in a shift of the tree-level S-matrix that is proportional to the identity. This is a generic statement that is also true at higher-loop order, and comes from the fact that changing the $a$-gauge corresponds to performing a so-called $T\bar{T}$-deformation of the lightcone gauge-fixed theory. Such a deformation dresses the exact S-matrix (to all loop orders) by a CDD factor~\cite{Castillejo:1955ed}. 
In principle one could go on and compute higher-loop contributions to $\gen{S}$, but this turns out to be difficult. In the next section we discuss how to find the exact S matrix (to all-loop order) using the symmetries of the lightcone gauge fixed model, together with some results from this perturbative calculation.

\subsection{Symmetry algebra and representations}

We recall that the bosonic isometries of $AdS_3\times S^3$ are given by $\alg{so}(2,2)\oplus\alg{so}(4)$, and that they decompose into a direct sum~$\alg{sl}(2)_{\L}\oplus\alg{sl}(2)_{\R}\oplus \alg{su}(2)_{\L}\oplus\alg{su}(2)_{\R}$, where we used the ``left'' and ``right'' labels $\L$ and $\R$ to distinguish the two copies of the various algebras. We will call the Cartan elements of these four algebras $\gen{L}_{\L} \equiv -(\gen{L}_{0})_\L$, $\gen{L}_{\R} \equiv -(\dot{\gen{L}}_{0})_\R$, $\gen{J}_{\L} \equiv +(\gen{J}_{3})_\L$ and $\gen{J}_{\R}= +(\dot{\gen{J}}_{3})_\R$ respectively.
Furthermore, the energy and  the angular momentum are
\begin{equation}
    \gen{E}% = - \int \extder \sigma P_t
    = \gen{L}_\L + \gen{L}_\R,\qquad
    \gen{J} %=\int \extder \sigma P_\varphi
    = \gen{J}_\L + \gen{J}_\R.
\end{equation}
The remaining combinations $\gen{L}_\L - \gen{L}_\R$ and $\gen{J}_\L - \gen{J}_\R$ have the interpretation of the spin in AdS~\footnote{The two $\alg{sl}(2)$ algebras are non-compact but when considering their embedding into $\alg{so}(2,2)$ it turns out that this particular linear combination is compact and hence the eigenvalues will be quantitised, motivating the interpretation as a spin. } and in the sphere, respectively. In terms of the string NLSM, they correspond to the conserved charges associated to shifts in $\psi$ and $\phi$.

It follows from the relation \eqref{eq:lcHamiltonian} that the worldsheet Hamiltonian  is
\begin{equation}
\gen{H} = (\gen{L}_\L + \gen{L}_\R) - (\gen{J}_\L + \gen{J}_\R)\,.
\end{equation}
From now on we will work in the $a=1/2$ gauge. The charge $\gen{P}_-$ which measures the circumference of the worldsheet, see eq.~\eqref{eq:worldsheetlength}, is then given by
\begin{equation} 
    \gen{P}_- = \frac{1}{2} \left(\gen{L}_\L + \gen{L}_\R + \gen{J}_\L + \gen{J}_\R\right)  \equiv \gen{R}\,.
\end{equation}

The classical solution which was used to gauge-fix the theory, see \eqref{eq:lc}, which identifies the vacuum of the gauge-fixed theory~$|0\rangle_{\cir}$ for a given size $\cir$, does not have any spin in AdS or in the sphere (it is invariant under shifts of $\psi$ and $\phi$). Moreover, the charges under $\gen{E}$ and $\gen{J}$ of the classical solution are precisely equal (in the $a=1/2$ gauge). In other words
\begin{equation}
    \gen{L}_\L|0\rangle_\cir =\gen{L}_\R|0\rangle_\cir =\gen{J}_\L|0\rangle_\cir =\gen{J}_\R|0\rangle_\cir = \frac{\cir}{2}\, |0\rangle_\cir\,,  
\end{equation}
where to obtain the last equality we have used that
\begin{equation}
    \gen{R}\,|0\rangle_\cir = \cir \, |0\rangle_\cir\,.
\end{equation}
%Moreover, $r\in\mathbb{Z}/2$ because the eigenvalues of~$\gen{J}_\L$ and~$\gen{J}_\R$ are quantised as usual for unitary representations of $\alg{su}(2)$. 

One can then define two related but not identical symmetry algebras:
\begin{enumerate}
    \item The algebra of charges which commute with the lightcone Hamiltonian~$\gen{H}$,
    \item The algebra of charges which annihilate the ground state~$|0\rangle_\cir$.
\end{enumerate}
The first algebra represents the symmetries of the gauge-fixed theory. The second algebra is important to study the excitations over the vacuum. In fact, if $\gen{X}$ is any generator of the second algebra, and $|\Psi_O\rangle_\cir=\gen{O}^\dagger|0\rangle_\cir$ is some excited state of the theory, we have that
\begin{equation}
    \gen{X}\,|\Psi_O\rangle_\cir=\gen{X}\,\gen{O}^\dagger\,|0\rangle_\cir 
    =
    \big[\gen{X},\,\gen{O}^\dagger\big]\,|0\rangle_\cir +
    \gen{O}^\dagger\,\gen{X}\,|0\rangle_\cir =\big[\gen{X},\,\gen{O}^\dagger\big]\,|0\rangle_\cir\,,
\end{equation}
so that $\gen{O}^\dagger$ itself must transform in a representation of the second algebra. 

Notice that the two algebras are different: in particular, $\gen{R}$ commutes with~$\gen{H}$ (because they are both a linear combination of Cartan generators), but it does not annihilate the vacuum. As it turns out, this is the  only difference between the two algebras (but it is an important one). In what follows we will be interested in the decompactification limit $\gen R \rightarrow \infty$, in which the two algebras become essentially equivalent.

% At quadratic level we used the fact that the theory is essentially free to use ladder operators to quantise the theory, so that $\gen{O} \sim \gen{a}_\mathcal{X}$. In general it is difficult, or impossible, to quantise interacting field theories, of which our lightcone gauge fixed model is an example. To make progress we use integrability and assume that the theory retains this property after lightcone gauge fixing and at the quantum level. 
% For integrable models the worldsheet S matrix satisfies a set of properties, in particular factorisation, which allows us to define a Zamolodchikov-Fadeev algebra. An all-loop excitation can be written as $\gen{A}^\dagger \ket{0}$ for an abstract element $\gen{A}$ satisfying the commutation relation
% \begin{equation}
%     \gen{A}_i^\dagger(p_1) \gen{A}_j^\dagger(p_2) = S_{ij}^{kl}(p_1,p_2) \gen{A}_l^\dagger(p_2) \gen{A}_k^\dagger(p_1)\,.
% \end{equation}

\subsubsection{On-shell symmetry of the lightcone gauge fixed theory}
To find the on-shell symmetry algebra of the lightcone gauge-fixed theory we need to find the generators of the isometry algebra $\alg{psu}(1,1|2)_\L \oplus \alg{psu}(1,1|2)_\R$ that commute with the worldsheet Hamiltonian $\gen{H}$. For this purpose it is convenient to decompose
\begin{equation}
\gen{H} = \gen{H}_\L + \gen{H}_\R\,, \qquad \gen{H}_\L = \gen{L}_\L - \gen{J}_\L\,, \qquad \gen{H}_\R = \gen{L}_\R - \gen{J}_\R\,.
\end{equation}
We know that the generators of $\alg{psu}(1,1|2)_\L$ commute with the ones of $\alg{psu}(1,1|2)_\R$. So we just need to focus on one copy, let's say $\alg{psu}(1,1|2)_\L$ and ask which generators commute with $\gen{H}_\L$. The commutation relations of the generators of $\alg{psu}(1,1|2)$ can be found in \eqref{eq:symAdS3}.

 Obviously, the generators $\gen{L}_\L$ and $\gen{J}_\L$ commute with $\gen{H}_\L$. On the other hand, it is easy to see that this cannot be the case for $\gen{L}_\pm$ and $\gen{J}_\pm$. Let us now consider the supercharges. From the commutation relations 
 \begin{equation}
 \com{\gen{L}_\L}{\gen{Q}_{\pm \alpha A}}= \mp \gen{Q}_{\pm \alpha A}\,, \qquad \com{\gen{J}_\L}{\gen{Q}_{\pm \alpha A}}= \pm \gen{Q}_{\pm \alpha A}\,,
 \end{equation} 
 it follows that all the fermionic generators that have two opposite first two indices will commute with $\gen{H}_\L$. We denote those with
\begin{equation}
\gen{Q}_\L{}^1 = + \gen{Q}_{-+2}\,, \qquad \gen{Q}_\L{}^2 = - \gen{Q}_{-+1}\,, \qquad \gen{S}_{\L 1} = + \gen{Q}_{+-1}\,, \qquad \gen{S}_{\L 2} = + \gen{Q}_{+-2}\,.
\end{equation}
The commutation relations for these generators read
\begin{equation} \label{eq:comL}
\anticom{\gen{Q}_\L{}^A}{\gen{S}_{\L B}} = \delta^A_B \gen{H}_\L\,.
\end{equation}
Furthermore, we also inherit the reality conditions
\begin{equation}
\gen{H}_\L^\dagger = \gen{H}_\L\,, \qquad (\gen{Q}_\L{}^A)^\dagger = \gen{S}_{\L A}\,, \qquad (\gen{S}_{\L A})^\dagger = \gen{Q}_\L{}^A\,,
\end{equation}
which imposes the following BPS bound on the algebra
\begin{equation}
    \gen{H}_\L \geq 0\,.
\end{equation}
\begin{centering}
\begin{tcolorbox}
\begin{exercise}
  Show that we have this BPS bound.
\end{exercise}
\end{tcolorbox}
\end{centering}
\ifsol
\begin{centering}
\begin{tcolorbox}[breakable, enhanced]
\textbf{Solution:} Let us take an arbitrary state $\ket{\psi}$ and compute the overlap (no sum over the index $a$)
\begin{equation}
    \bra{\psi} \anticom{\gen{Q}_\L{}^A}{\gen{S}_{\L A}} \ket{\psi} = \left\{  \begin{aligned}
    &\bra{\psi} \gen{H}_\L \ket{\psi}\,, \\
    &\bra{\psi} \anticom{\gen{Q}_\L{}^A}{(\gen{Q}_{\L}{}^A)^\dagger} \ket{\psi} = || (\gen{Q}_\L{}^A)^\dagger \ket{\psi}||^2 + ||\gen{Q}_\L{}^A \ket{\psi}||^2 \geq 0\,.
    \end{aligned} \right.
\end{equation}
    For the equality of the first line we have used the commutation relation. For the equality of the second line we have used the reality condition. This shows that $\gen{H}_\L$ is positive-definite.
\end{tcolorbox}
\end{centering}
\fi

% From all this we deduce that we are in presence of two copies of the $\alg{su}(1|1)_\L$ algebra (the index $\L$ is inherited from $\alg{psu}(1,1|2)_\L$). {\color{purple}{Is this $\alg{psu}(1|1) \oplus \alg{u}(1)$?}} 

The discussion in the ``right'' sector goes through the same way. We define the generators
\begin{equation}
\gen{Q}_{\R 1} = + \bar{\gen{Q}}_{-\dot{+}1}\,, \qquad \gen{Q}_{\R 2} = + \bar{\gen{Q}}_{-\dot{+}2}\,, \qquad \gen{S}_\R{}^1 = - \bar{\gen{Q}}_{+\dot{-}2}\,, \qquad \gen{S}_\R{}^2 = \bar{\gen{Q}}_{+\dot{-}1}\,,
\end{equation} 
which obey the two decoupled commutation relations
\begin{equation} \label{eq:comR}
\anticom{\gen{Q}_{\R A}}{\gen{S}_\R{}^B} = \delta_A^B \gen{H}_\R\,.
\end{equation}
From the reality conditions
\begin{equation}
\gen{H}_\R^\dagger = \gen{H}_\R\,, \qquad (\gen{Q}_\R{}^A)^\dagger = \gen{S}_{\R A}\,, \qquad (\gen{S}_{\R A})^\dagger = \gen{Q}_\R{}^A\,.
\end{equation}
we deduce another BPS bound
\begin{equation}
    \gen{H}_\R \geq 0\,.
\end{equation}

% The commutation relations \eqref{eq:comL} and \eqref{eq:comR} can be written
% \begin{equation}
%     \anticom{\gen{Q}_{\L}{}^a}{\gen{S}_{\L b}} = \frac{1}{2}\delta^a_b \left(\gen{H}+\gen{M} \right)\,, \qquad \anticom{\gen{Q}_{\R a}}{\gen{S}_{\R}{}^b} = \frac{1}{2} \delta^b_a \left( \gen{H}-\gen{M}\right)\,,
% \end{equation}
% were we recall that the total Hamiltonian is
% \begin{equation}
%     \gen{H} = \gen{H}_\L + \gen{H}_\R\,,
% \end{equation}
% and we found it convenient to also introduce the angular momentum
% \begin{equation}
%     \gen{M} = \gen{H}_\L - \gen{H}_\R = \gen{L}_\L + \gen{L}_\R - \gen{J}_\L - \gen{J}_\R\,.
% \end{equation}

%This shows that there are also two copies of the $\alg{su}(1|1)_\R$ algebra.

Then, we also know that all the Cartan elements should commute with $\gen{H}$. Those can be an arbitrary linear combination of left and right generators. In particular, on top of $\gen{H}$ and $\gen{R}$ we also define 
\begin{equation}
     \gen{M} = \gen{L}_\L - \gen{L}_\R - \gen{J}_\L + \gen{J}_\R\,, \qquad \gen{B} = \gen{L}_\L - \gen{L}_\R + \gen{J}_\L - \gen{J}_\R\,,
\end{equation}
which also commute with $\gen{H}$. Notice that while the fermionic charges appearing in \eqref{eq:comL} and \eqref{eq:comR} commute with $\gen{H}$ and $\gen{M}$, this is not the case for $\gen{R}$ and $\gen{B}$. In fact we have
\begin{equation} \begin{aligned}
    \com{\gen{R}}{ \gen{Q}_{\L}{}^A} &= + \frac{1}{2} \gen{Q}_\L{}^A\,, \qquad \com{\gen{R}}{\gen{S}_{\L A}} = - \frac{1}{2} \gen{S}_{\L A}\,, \\ \com{\gen{R}}{\gen{Q}_{\R A}} &= + \frac{1}{2} \gen{Q}_{\R A}\,, \qquad \com{\gen{R}}{\gen{S}_{\R}{}^A} = - \frac{1}{2} \gen{S}_{\R}{}^A\,,
    \end{aligned}
\end{equation}
as well as
\begin{equation} \begin{aligned}
    \com{\gen{B}}{ \gen{Q}_{\L}{}^A} &= +  \gen{Q}_\L{}^A\,, \qquad \com{\gen{B}}{\gen{S}_{\L A}} = -  \gen{S}_{\L A}\,, \\ \com{\gen{B}}{\gen{Q}_{\R A}} &= - \gen{Q}_{\R A}\,, \qquad \com{\gen{B}}{\gen{S}_{\R}{}^A} = + \gen{S}_{\R}{}^A\,,
    \end{aligned}
\end{equation}
so that both $\gen{R}$ and $\gen{B}$ define automorphisms of the symmetry algebra.

Finally, all the generators of $\alg{su}(2)_\bullet \oplus \alg{su}(2)_\circ$ are preserved by the lightcone gauge fixing.
The fermionic generators $\gen{S}_{\L a},\gen{Q}_{\R a}$ and $\gen{S}_\L{}^a,\gen{Q}_{\R}{}^a$ transform respectively in the fundamental and anti-fundamental representations of $\alg{su}(2)_\bullet$, for instance
\begin{equation}
\com{(\gen{J}_\bullet)_a{}^b}{\gen{S}_{\L c}} = \delta^b_c \gen{S}_{\L a} - \frac{1}{2} \delta_a^b \gen{S}_{\L c}\,, \qquad \com{(\gen{J}_\bullet)_a{}^b}{\gen{Q}_\L{}^c} = - \delta^c_a \gen{Q}_\L{}^b + \frac{1}{2} \delta_a^b \gen{Q}_\L{}^c\,,
\end{equation} 
while all the generators commute with $\alg{su}(2)_\circ$.

\begin{tcolorbox}
\begin{exercise}
  Find the algebra of charges which annihilate the ground state $\ket{0}_r$. You should find the same algebra, but without $\gen{R}$.
\end{exercise}
\end{tcolorbox}
\ifsol
\begin{tcolorbox}[breakable, enhanced]
\textbf{Solution:} From the BPS bound on the left $\alg{psu}(1,1|2)$ copy it follows that it must be
\begin{equation}
    \gen{Q}_\L{}^A \ket{0}_r =0\,, \qquad \gen{S}_{\L A} \ket{0}_r =0\,.
\end{equation}
Similarly, from the BPS bound on the right $\alg{psu}(1,1|2)$ we have that
\begin{equation}
    \gen{Q}_{\R A} \ket{0}_r =0\,, \qquad \gen{S}_{\L}{}^A \ket{0}_r =0\,.
\end{equation}
It is also clear that $\gen{H}$, $\gen{M}$ and $\gen{B}$ all annihilate the vacuum. However we have that $\gen{R} \ket{0}_r = r \ket{0}_r$. All the other generators do not annihilate the ground state. For instance, we have that
\begin{equation}
    \gen{L}_+ \ket{0}_r =2 \com{\gen{L}_0}{\gen{L}_+} \ket{0}_r = 
\end{equation}
\end{tcolorbox}
\fi

\subsubsection{Off-shell symmetry algebra}
The vacuum is a physical state (it carries zero momentum). This is why the algebra we find is a subalgebra of the original $\alg{psu}(1,1|2)_\L \oplus \alg{psu}(1,1|2)_\R$ isometry algebra. To construct the S matrix, we want to relax this level-matching condition, so that we can also scatter particles that are off-shell. Why do we want this? Let us consider the scattering of three particles with respective momenta $p_1,p_2,p_3$. The initial state should be physical, so that we impose the level matching condition $p_1+p_2+p_3=0$ (assuming no winding). Then,  from integrability we know that this factorises into a sequence of $2 \rightarrow 2$ scattering events, and the thing we are computing is really this 2-body S matrix. Such an S matrix will have, for instance, incoming particles with momenta $p_1$ and $p_2$, while $p_3$ is left alone. But we do not have the level matching condition when restricting to particles 1 and 2 only, in general we have $p_1 + p_2 \neq 0$. Therefore, while for the $n \rightarrow n$ scattering event the in and out states are physical, this is not the case for the individual $2 \rightarrow 2$ scattering events.

The algebra constructed in the previous section will get modified when considering excitations above the vacuum that do not satisfy the level-matching condition. An algebra is quite a rigid object and there are not many ways in which it can be deformed to accommodate for a non-vanishing $\gen{p} \neq0$. Guided from a perturbative calculation, one finds that the symmetry algebra gets extended by two central elements coupling the left and right sectors,
\begin{align} 
\label{eq:com1}
    \anticom{\gen{Q}_{\L}{}^A}{\gen{S}_{\L B}} &= \frac{1}{2}\delta^A_B \left(\gen{H}+\gen{M} \right)\,, &\qquad \anticom{\gen{Q}_{\R A}}{\gen{S}_{\R}{}^B} &= \frac{1}{2} \delta^B_A \left( \gen{H}-\gen{M}\right)\,, \\
    \label{eq:com2}
    \anticom{\gen{Q}_{\L}{}^A}{\gen{Q}_{\R B}} &= \delta^A_B \gen{C}\,, &\qquad  \anticom{\gen{S}_{\L A}}{\gen{S}_{\R}{}^B} &= \delta^B_A \bar{\gen{C}}\,.
\end{align}
Moreover, we have the reality conditions
\begin{equation}
    (\gen{Q}_\L{}^A)^\dagger = \gen{S}_{\L A}\,, \qquad (\gen{Q}_{\R A})^\dagger = \gen{S}_\R{}^A\,, \qquad \gen{H} \geq 0\,, \qquad \gen{C}^\dagger = \bar{\gen{C}}\,,
\end{equation}
where the last two relations are a consequence of the reality conditions imposed on the fermionic generators (the first two equations). 
\eqref{eq:com1} and \eqref{eq:com2} are the commutation relations of the algebra
\begin{equation} \label{eq:A}
    \mathcal A = (\alg{su}(1|1)_\L \oplus \alg{su}(1|1)_\R)_{c.e.}^{\oplus 2}\,,
\end{equation}
where $c.e.$ stands for ``central extension''. The newly introduced generator $\gen{C}$ and its hermitian conjugate $\gen{C}^\dagger$ indeed commute with all the other generators of the algebra and is hence central. For completeness, let us note that
\begin{equation}
    \com{\gen{B}}{\gen{C}} = 0\,, \qquad \com{\gen{B}}{\gen{C}^\dagger}=0\,,
\end{equation}
while
\begin{equation}
    \com{\gen{R}}{\gen{C}} = + \gen{C}\,, \qquad \com{\gen{R}}{\gen{C}^\dagger}=- \gen{C}^\dagger\,. 
\end{equation}
From these last two equations we see that, had we also included $\gen{R}$ in the algebra, then $\gen{C}$ would no longer have been central.

The commutation relations \eqref{eq:com1} are the same both on-shell and off-shell. The commutators of \eqref{eq:com2} however are not. 
In fact, from a perturbative calculation one can argue that it must be (while this result comes from a perturbative calculation, it can be argued that these expression should be valid to all loop order)
\begin{equation} \label{eq:central1}
    \gen{M} = \mu + \frac{k}{2 \pi} \gen{p} \,,  \qquad \gen{C} = \frac{i h}{2} (e^{i \gen{p}}-1)\,, \qquad \gen{C}^\dagger = - \frac{i h}{2} (e^{-i\gen{p}} -1)\,,
\end{equation}
with $\gen{p}$ the worldsheet momentum operator, $k \in \mathbb N$ the WZ level and $h \geq 0$ is related to the amount of RR flux (there is a non-trivial relation between $h$, the tension $T$ appearing in the classical action and the WZ level). Finally,
\begin{equation}
\mu \in \mathbb{Z}\,,    
\end{equation}
is a free parameter characterising different representations. 
It is then easy to see that $\gen{C}$ and $\gen{C}^\dagger$ are only present off-shell. On a physical state we have 
\begin{equation}
    \gen{p} \ket{\text{phys}} = 2 \pi m \ket{\text{phys}}\,, \quad \gen{M} \ket{\text{phys}} = (\mu + k m) \ket{\text{phys}}\,, \quad \gen{C} \ket{\text{phys}} = \gen{C}^\dagger \ket{\text{phys}} = 0\,.
\end{equation}

\begin{centering}
\begin{tcolorbox}
\begin{exercise}
  Show that we indeed have $\mu \in \mathbb{Z}$.
\end{exercise}
\end{tcolorbox}
\end{centering}
\ifsol
\begin{centering}
\begin{tcolorbox}[breakable, enhanced]
\textbf{Solution:}  We recall that, on physical states,
\begin{equation}
    \gen{M} \ket{\text{phys}} = (\gen{L}_\L - \gen{L}_\R - \gen{J}_\L + \gen{J}_\R) \ket{\text{phys}} = (\mu + k m) \ket{\text{phys}}\,,
\end{equation}
with $k \in \mathbb{N}$ and $m \in \mathbb{Z}$. 
The generators $\gen{J}_\L$ and $\gen{J}_\R$ are the Cartans of two compact $\alg{su}(2)$ algebras. The eigenvalues of $\gen{J}_\L - \gen{J}_\R$ are thus quantised, with integer spins for bosons and half-integer spin for fermions. The other two generators $\gen{L}_\L$ and $\gen{L}_\R$ are Cartans of a non-compact algebra $\alg{sl}(2)$, but considering their embedding into $\alg{so}(2,2)$ it turns out that the particular linear combination $\gen{L}_\L - \gen{L}_\R$ is associated to a compact direction (this is why it is associated to the AdS spin). Again, it gives rise to an integer spin for bosons and half integer for fermions. For bosons, the difference of two integers is always an integer. For fermions, the difference of two half-integers is always an integer. Therefore, we conclude that for physical states we have $\mu+ k m \in \mathbb{Z}$ and hence $\mu \in \mathbb{Z}$. 
\end{tcolorbox}
\end{centering}
\fi

\subsubsection{One-particle representations}\label{sec:one-particle}

The algebra $\mathcal A$ has two types of representations: long representations, which are sixteen-dimensional, and short representations, which are four-dimensional. From a perturbative calculation we observe that the transverse excitations transform in short representations (they arrange themselves into four-dimensional spaces that share the same value of $\mu$), obeying the shortening condition
\begin{equation} \label{eq:shortening}
    \gen{H}^2 = \gen{M}^2 + 4 \gen{C}^\dagger \gen{C}\,.
\end{equation}
Since we do not expect the dimensionality of the representations to change when going to all-loop order, we will always consider particles transforming in short representations.
Using the explicit expression for the central elements $\gen{M}$ and $\gen{C}$ in \eqref{eq:central1}, together with the fact that $\gen{H} \geq 0$ to choose the branch of the square root, we deduce that it must be
\begin{equation}
    \gen{H} = \sqrt{\left(\mu + \frac{k}{2 \pi} p\right)^2+ 4 h^2 \sin^2\left(\frac{p}{2}\right)}\,.
\end{equation}
This is the all-loop dispersion relation. It is non-relativistic and shares some similarities with a lattice-like dispersion relation.

How can we make the link with the dispersion relation found in the perturbative calculation? First, we need to know that to leading order we have that $T \sim h$. Then, recall that in the perturbative calculation we rescaled the coordinate $\sigma \rightarrow \sigma/T$. The worldsheet momentum being associated with shifts in $\sigma$, to make the link with the perturbative calculation we need to rescale $p \rightarrow p/T \sim p/h$. We then take the large tension limit $h \rightarrow \infty$. In doing so, the Hamiltonian becomes
\begin{equation}
    \gen{H}^{(2)} = \sqrt{\mu^2+2 \mu q p + p^2}\,,
\end{equation}
where the amount of NSNS flux $q=\frac{k}{2 \pi T}$. This is precisely what was found at quadratic order. From this we also deduce that $|\mu|$ has the natural interpretation of a mass. In fact this allows us to identify the following representations:

\paragraph{Left representation.}
The states $\ket{Z_\L(p)},\ket{Y_\L(p)},\ket{\Psi_\L^1(p)}, \ket{\Psi_\L^2(p)}$ are characterised by $\mu=+1$. They fall into a four-dimensional irreducible representation of the symmetry algebra that we shall call ``left'' and denote with $\varrho_\L$. The name ``left'' comes from the fact that for $p=0$ we have $\gen M=1$ and $\gen H=1$ and hence $\gen H_\R \equiv \gen H- \gen M=0$. Using the commutation relation it follows that the  supercharges act as (we only report the non-vanishing elements)
\begin{equation} \label{eq:repleft}
    \begin{aligned}
        &(\Ql)^{A}|Y_\L(p)\rangle&=&\,a_{\L}(p)\,|\Psi_\L^A(p)\rangle\,,\quad
        &&(\Ql)^{A}|\Psi_\L^B(p)\rangle&=&\,\varepsilon^{AB}a_{\L}(p)\,|Z_\L(p)\rangle\,,\\
        &(\Sl)_{A}|\Psi_\L^B(p)\rangle&=&\,\delta_A{}^B\,a_{\L}^*(p)\,|Y_\L(p)\rangle\,,\quad
        &&(\Sl)_{A}|Z_\L(p)\rangle&=&-\varepsilon_{AB}\,a_{\L}^*(p)\,|\Psi_\L^{B}(p)\rangle\,,\\
        &(\Sr)^{A}|Y_\L(p)\rangle&=&\,b_{\L}^*(p)\,|\Psi_\L^A(p)\rangle\,,\quad
        &&(\Sr)^{A}|\Psi_\L^B(p)\rangle&=&\,\varepsilon^{AB}b^*_{\L}(p)\,|Z_\L(p)\rangle\,,\\
        &(\Qr)_{A}|\Psi_\L^B(p)\rangle&=&\,\delta_A{}^B\,b_{\L}(p)\,|Y_\L(p)\rangle\,,\quad
        &&(\Qr)_{A}|Z_\L(p)\rangle&=&-\varepsilon_{AB}\,b_{\L}(p)\,|\Psi_\L^{B}(p)\rangle\,.
    \end{aligned}
\end{equation}
In our convention $\varepsilon^{12}=-\varepsilon_{12}=+1$. Moreover, $\ket{\Psi_\L^1}$ and $\ket{\Psi_\L^2}$ transform as a doublet under $\alg{su}(2)_\bullet$, while all states are neutral under $\alg{su}(2)_\circ$. This can be schematically summarised as (we only mention the lowering operators for readability)
\begin{equation}
\begin{aligned}
\begin{tikzpicture}
\node (Y) at (0,2) {$|Y_\L(p)\rangle$};
\node (psi1) at (-2.5,0) {$|\Psi_\L^1(p)\rangle$};
\node (psi2) at (2.5,0) {$|\Psi_\L^2(p)\rangle$};
\node (Z) at (0,-2) {$|Z_\L(p)\rangle$};
\draw[->, thick, color=black!50!cyan] (Y) -- (psi1) node[pos=0.3,left] {$\Ql^1,\Sr^1$ \phantom{x}};
\draw[->, thick, color=black!50!cyan] (Y) -- (psi2) node[pos=0.3,right] { \phantom{x} $\Ql^2,\Sr^2$};
\draw[->, thick, color=black!50!cyan] (psi1) -- (Z) node[pos=0.7,left] {$\Ql^2,\Sr^2$  \phantom{x}};
\draw[->, thick, color=black!50!cyan] (psi2) -- (Z) node[pos=0.7,right] { \phantom{x} $\Ql^1,\Sr^1$};

\draw[dotted, <->, thick] (psi1) -- (psi2) node[pos=0.5,below] {$\gen{J}_\bullet$};

\end{tikzpicture}
\end{aligned}
\end{equation}

\begin{centering}
\begin{tcolorbox}
\begin{exercise}
  Convince yourself that the representation is indeed of the form \eqref{eq:repleft}.
\end{exercise}
\end{tcolorbox}
\end{centering}
The free functions $a_\L(p)$ and $b_\L(p)$ are the representation parameters. From the commutation relations \eqref{eq:com1} and \eqref{eq:com2} it follows that
\begin{equation} \label{eq:ableft} \begin{aligned}
    \gen H = |a_\L(p)|^2 + |b_\L(p)|^2\,, \qquad \gen M = |a_\L(p)|^2 - |b_\L(p)|^2\,, \qquad \gen C = a_\L(p) b_\L(p)\,,
    \end{aligned}
\end{equation}
and hence the shortening condition is automatically satisfied. On the other hand, using \eqref{eq:central1} it must be that
\begin{equation} \label{eq:valL}
    \gen H = \sqrt{\left(1+\frac{k}{2 \pi}\right)^2 + 4 h^2 \sin^2 \left( \frac{p}{2} \right)}\,, \qquad \gen M = 1+ \frac{k}{2 \pi}p\,, \qquad \gen C = \frac{ih}{2} \left( e^{ip} -1 \right)\,.
\end{equation}
This gives two equations on $a_\L(p)$ and $b_\L(p)$ (one equation is automatically satisfied due to the shortening condition), and hence the representation parameters are completely fixed.

Note that comparision with the perturbative calculation motivates the identification
\begin{equation}
    \ket{Z_\L(p)} = \ket{\bar{Z}(p)}\,, \quad \ket{Y_\L(p)} = \ket{\bar{Y}(p)}\,, \quad \ket{\Psi^1_\L(p)} = \ket{\bar{\zeta}(p)}\,, \quad \ket{\Psi_\L^2(p)} =\ket{\bar{\eta}(p)}\,.
\end{equation}
% \begin{center}
% \begin{tikzpicture}
%   \draw[->] (-3, 0) -- (4.2, 0) node[right] {$p$};
%   \draw[->] (0, -3) -- (0, 4.2) node[above] {$E(p)$};
%   \draw[scale=0.5, domain=-4:4, smooth, variable=\x, blue] plot ({\x}, {sqrt(1+4*(0.1)*(0.1)*sin(deg{\x/2))^2)});
% \end{tikzpicture}
% \end{center}

% It will be useful to introduce new types of functions, the \textit{Zhukovski variables} $x_\L^+(p)$ and $x_\L^-(p)$, and write
% \begin{equation}
%     a_\L(p) = \eta_{\L}(p)\,, \qquad b_\L(p) = - \frac{e^{-ip/2}}{x_{\L}^-(p)} \eta_\L(p)\,, \qquad \eta_{\L}(p) = e^{ip/4} \sqrt{\frac{i h}{2} (x_\L^-(p) - x_{\L}^+(p))}\,.
% \end{equation}
% The consistency conditions then become
% \begin{equation}
%     H = \frac{i h}{2} \left( x_\L^- + \frac{1}{x^+_\L} - x_\L^+ + \frac{1}{x_\L^-} \right)\,, \qquad \frac{x_\L^+}{x_\L^-} = e^{i p}\,.
% \end{equation}

\paragraph{Right representation.} Then, the states $\ket{Z_\R(p)}, \ket{Y_\R(p)}, \ket{\Psi_\R^1(p)}, \ket{\Psi_\R^2(p)}$ have $\mu=-1$ and fall into another four-dimensional irreducible representation of the symmetry algebra that we will call ``right'' and denote by $\varrho_\R$. 
The name ``right'' comes from the fact that for $p=0$ we have $\gen M=-1$ and $\gen H=+1$ and hence $\gen H_\L \equiv \gen H+ \gen M=0$.
The supercharges act as
\begin{equation}
    \begin{aligned}
        &(\Ql)^{A}|Z_\R(p)\rangle&=&\,b_{\R}(p)\,|\Psi_\R^A(p)\rangle\,,\qquad
        &&(\Ql)^{A}|\Psi_\R^B(p)\rangle&=&-\varepsilon^{AB}b_{\R}(p)\,|Y_\R(p)\rangle\,,\\
        &(\Sl)_{A}|\Psi_\R^B(p)\rangle&=&\,\delta_A{}^B\,b_{\R}^*(p)\,|Z_\R(p)\rangle\,,\qquad
        &&(\Sl)_{A}|Y_\R(p)\rangle&=&\,\varepsilon_{AB}\,b_{\R}^*(p)\,|\Psi_\R^{B}(p)\rangle\,,\\
        &(\Sr)^{A}|Y_\R(p)\rangle&=&\,a_{\R}^*(p)\,|\Psi_\R^A(p)\rangle\,,\qquad
        &&(\Sr)^{A}|\Psi_\R^B(p)\rangle&=&-\varepsilon^{AB}a^*_{\R}(p)\,|Z_\R(p)\rangle\,,\\
        &(\Qr)_{A}|\Psi_\R^B(p)\rangle&=&\,\delta_A{}^B\,a_{\R}(p)\,|Y_\R(p)\rangle\,,\qquad
        &&(\Qr)_{A}|Z_\R(p)\rangle&=&\,\varepsilon_{AB}\,a_{\R}(p)\,|\Psi_\R^{B}(p)\rangle\,.
    \end{aligned}
\end{equation}
Again, $\ket{\Psi_\R^1}$ and $\ket{\Psi_\R^2}$ transform as a doublet under $\alg{su}(2)_\bullet$, while all states are neutral under $\alg{su}(2)_\circ$.
The representation takes the schematic form
\begin{equation}
\begin{aligned}
\begin{tikzpicture}
\node (Y) at (0,2) {$|Z_\R(p)\rangle$};
\node (psi1) at (-2.5,0) {$|\Psi_\R^1(p)\rangle$};
\node (psi2) at (2.5,0) {$|\Psi_\R^2(p)\rangle$};
\node (Z) at (0,-2) {$|Y_\R(p)\rangle$};
\draw[->, thick, color=black!50!cyan] (Y) -- (psi1) node[pos=0.3,left] {$\Ql^1,\Sr^1$ \phantom{x}};
\draw[->, thick, color=black!50!cyan] (Y) -- (psi2) node[pos=0.3,right] {\phantom{x} $\Ql^2,\Sr^2$};
\draw[->, thick, color=black!50!cyan] (psi1) -- (Z) node[pos=0.7,left] {$\Ql^2,\Sr^2$ \phantom{x}};
\draw[->, thick, color=black!50!cyan] (psi2) -- (Z) node[pos=0.7,right] {\phantom{x} $\Ql^1,\Sr^1$};

\draw[dotted,<->, thick] (psi1) -- (psi2) node[pos=0.5,below] {$\gen{J}_\bullet$};

\end{tikzpicture}
\end{aligned}
\end{equation}
The representation parameters obey
\begin{equation} \label{eq:abright} \begin{aligned}
    \gen H = |a_\R(p)|^2 + |b_\R(p)|^2\,, \qquad \gen M = -|a_\R(p)|^2 + |b_\R(p)|^2\,, \qquad \gen C = a_\R(p) b_\R(p)\,.
    \end{aligned}
\end{equation}
The are fixed by requiring that
\begin{equation} \label{eq:valR}
    \gen H = \sqrt{\left(-1+\frac{k}{2 \pi}\right)^2 + 4 h^2 \sin^2 \left( \frac{p}{2} \right)}\,, \qquad \gen M = -1+ \frac{k}{2 \pi}p\,, \qquad \gen C = \frac{ih}{2} \left( e^{ip} -1 \right)\,.
\end{equation}
To make the link with the perturbative calculation one needs to make the identification
\begin{equation}
    \ket{Z_\R(p)} = \ket{Z(p)}\,, \quad \ket{Y_\R(p)} = \ket{Y(p)}\,, \quad \ket{\Psi^1_\R(p)} = \ket{\zeta(p)}\,, \quad \ket{\Psi_\R^2(p)} =\ket{\eta(p)}\,.
\end{equation}

\paragraph{Left-Right symmetry.}
Looking at the two above diagrams it immediately follows that there is a discrete left-right symmetry when swapping the particles
\begin{equation}
    |Y_\L\rangle\leftrightarrow |Y_\R\rangle\,,
    \qquad
    |\Psi_\L^A\rangle \leftrightarrow |\Psi_\R^A\rangle\,,
    \qquad
    |Z_\L\rangle\leftrightarrow |Z_\R\rangle\,.
\end{equation}

\paragraph{Massless representations.} Finally, the bosons $T^{\dot{a} a}$ and fermions $\chi^{\dot{a}}$, $\bar{\chi}^{\dot{a}}$ have $\mu=0$. These arrange themselves into two four-dimensional irreducible representations, transforming as a doublet under $\alg{su}(2)_\circ$. 
The generators act as
\begin{equation}
    \begin{aligned}
        &(\Ql)^{A}|\chi^{\dot{A}}(p)\rangle &=&\,a_{\z}(p)\,|T^{\dot{A}A}(p)\rangle\,,\qquad
        &&(\Ql)^{A}|T^{\dot{A}B}(p)\rangle &=&\,\varepsilon^{AB}a_{\z}(p)\,|\tilde{\chi}^{\dot{A}}(p)\rangle\,,\\
        &(\Sl)_{A}|T^{\dot{A}B}(p)\rangle &=&\,\delta_A{}^B\,a_{\z}^*(p)\,|\chi^{\dot{A}}(p)\rangle\,,\qquad
        &&(\Sl)_{A}|\tilde{\chi}^{\dot{A}}(p)\rangle &=&-\varepsilon_{AB}\,a_{\z}^*(p)\,|T^{\dot{A}B}(p)\rangle\,,\\
        &(\Sr)^{A}|\chi^{\dot{A}}(p)\rangle&=&\,b_{\z}^*(p)\,|T^{\dot{A}A}(p)\rangle\,,\qquad
        &&(\Sr)^{A}|T^{\dot{A}B}(p)\rangle&=&\,\varepsilon^{AB}b^*_{\z}(p)\,|\tilde{\chi}^{\dot{A}}(p)\rangle\,,\\
        &(\Qr)_{A}|T^{\dot{A}B}(p)\rangle&=&\,\delta_A{}^B\,b_{\z}(p)\,|\chi^{\dot{A}}(p)\rangle\,,\qquad
        &&(\Qr)_{A}|\tilde{\chi}(p)\rangle&=&-\varepsilon_{AB}\,b_{\z}(p)\,|T^{\dot{A}B}(p)\rangle\,.
    \end{aligned}
\end{equation}
This is depicted as
\begin{equation}
\begin{aligned}
\begin{tikzpicture} 

\def\xl{-3}
\def\xr{3}
\def\hl{-0.5}
\def\hr{0.5}

\node (Yl) at (\xl,\hl+2) {$|\chi^{\dot{1}}(p)\rangle$};
\node (psi1l) at (\xl-2.5,\hl) {$|T^{\dot{1}1}(p)\rangle$};
\node (psi2l) at (\xl+2.5,\hl) {$|T^{\dot{1}2}(p)\rangle$};
\node (Zl) at (\xl,\hl-2) {$|\tilde{\chi}^{\dot{1}}(p)\rangle$};
\draw[->, thick, color=black!50!cyan] (Yl) -- (psi1l) node[pos=0.3,left] {$\Ql^1,\Sr^1$ \phantom{x}};
\draw[->, thick, color=black!50!cyan] (Yl) -- (psi2l) node[pos=0.3,right] {\phantom{x} $\Ql^2,\Sr^2$};
\draw[->, thick, color=black!50!cyan] (psi1l) -- (Zl) node[pos=0.7,left] {$\Ql^2,\Sr^2 $ \phantom{x}};
\draw[->, thick, color=black!50!cyan] (psi2l) -- (Zl) node[pos=0.7,right] {\phantom{x} $\Ql^1,\Sr^1$};

\node (Yr) at (\xr,\hr+2) {$|\chi^{\dot{2}}(p)\rangle$};
\node (psi1r) at (\xr-2.5,\hr) {$|T^{\dot{2}1}(p)\rangle$};
\node (psi2r) at (\xr+2.5,\hr) {$|T^{\dot{2}2}(p)\rangle$};
\node (Zr) at (\xr,\hr-2) {$|\tilde{\chi}^{\dot{2}}(p)\rangle$};
\draw[->, thick, color=black!50!cyan] (Yr) -- (psi1r) node[pos=0.3,left] {$\Ql^1,\Sr^1$ \phantom{x}};
\draw[->, thick, color=black!50!cyan] (Yr) -- (psi2r) node[pos=0.3,right] {\phantom{x} $\Ql^2,\Sr^2$};
\draw[->, thick, color=black!50!cyan] (psi1r) -- (Zr) node[pos=0.7,left] {$\Ql^2,\Sr^2 $ \phantom{x}};
\draw[->, thick, color=black!50!cyan] (psi2r) -- (Zr) node[pos=0.7,right] {\phantom{x} $\Ql^1,\Sr^1$};

\draw[dashed, thick] (Yl) -- (Yr) node[pos=0.3,left] {};
\draw[dashed, thick] (Zl) -- (Zr) node[pos=0.6,below] {$\gen{J}_\circ$};
\draw[dashed, thick] (psi1l) -- (psi1r) node[pos=0.7,left] {};
\draw[dashed, thick] (psi2l) -- (psi2r) node[pos=0.7,right] {};

\draw[dotted, <->,thick] (psi1l) -- (psi2l) node[pos=0.5,below] {$\gen{J}_\bullet$};
\draw[dotted, <->,thick] (psi1r) -- (psi2r) node[pos=0.5,above] {$\gen{J}_\bullet$};

\end{tikzpicture}
\end{aligned}
\end{equation}

The representation parameters are such that
\begin{equation} \label{eq:abmassless}\begin{aligned}
    \gen H = |a_\circ(p)|^2 + |b_\circ(p)|^2\,, \qquad \gen M = |a_\circ(p)|^2 - |b_\circ(p)|^2\,, \qquad \gen C = a_\circ(p) b_\circ(p)\,.
    \end{aligned}
\end{equation}

Notice that for $k=0$ (pure RR case), the dispersion relation becomes periodic in $p$, with period $ 2 \pi$, so that we can restrict to the domain $- \pi \leq p \leq + \pi$. The massless dispersion relation then reads
\begin{equation}
    \gen H = \left| 2 h \sin \left(\frac{p}{2} \right)\right| = \left\{
    \begin{aligned}
        &+ 2 h \sin\left( \frac{p}{2}\right)\,, &\qquad 0 \leq &p \leq \pi\,, \\
        &- 2 h \sin\left( \frac{p}{2}\right)\,, &\qquad -\pi \leq &p \leq 0\,.
    \end{aligned}
    \right.
\end{equation}
Depending on the momentum we will have two types of massless particles: chiral (positive momentum, first line above) or anti-chiral (negative momentum, second line above). These two different branches are also present in the more general mixed flux case, since for small $p$ we have that
\begin{equation}
    \gen H \sim |p| \,\sqrt{\left(\frac{k}{2 \pi} \right)^2 + h^2} \,, \qquad p \ll 1\,.
\end{equation}

\paragraph{Representation coefficients.} The constraints \eqref{eq:ableft}, \eqref{eq:abright} and \eqref{eq:abmassless} are best solved using the \textit{Zhukovski variables} $x^+_*(p)$ and $x^-_*(p)$ (we introduce a set of Zhukovski variables for each sector),
\begin{equation}
\label{eq:xpm}
\begin{aligned}
    x^{\pm}_{\L,p} &=& \frac{e^{\pm i p /2}}{2h\,\sin \big(\tfrac{p}{2}\big)} \Bigg(\big(1+\tfrac{k}{2\pi}p\big)+\sqrt{\big(1+\tfrac{k}{2\pi}p\big)^2+4h^2\sin^2\big(\tfrac{p}{2}\big)}\Bigg)\,,\\
    x^{\pm}_{\R,p} &=& \frac{e^{\pm i p /2}}{2h\,\sin \big(\tfrac{p}{2}\big)} \Bigg(\big(1-\tfrac{k}{2\pi}p\big)+\sqrt{\big(1-\tfrac{k}{2\pi}p\big)^2+4h^2\sin^2\big(\tfrac{p}{2}\big)}\Bigg)\,,\\
    x^{\pm}_{\z,p} &=& \frac{e^{\pm i p /2}}{2h\,\sin \big(\tfrac{p}{2}\big)} \Bigg(\big(0+\tfrac{k}{2\pi}p\big)+\sqrt{\big(0+\tfrac{k}{2\pi}p\big)^2+4h^2\sin^2\big(\tfrac{p}{2}\big)}\Bigg)\,.
\end{aligned}
\end{equation}
These satisfy
\begin{equation}
\label{eq:xpm-shortening}
\begin{aligned}
x^+_{\L,p}+\frac{1}{x^+_{\L,p}}-x^-_{\L,p}-\frac{1}{x^-_{\L,p}}&=\frac{2i\,\big(1+\tfrac{k}{2\pi}p\big)}{h}\,,\\
x^+_{\R,p}+\frac{1}{x^+_{\R,p}}-x^-_{\R,p}-\frac{1}{x^-_{\R,p}}&=\frac{2i\,\big(1-\tfrac{k}{2\pi}p\big)}{h}\,,\\
x^+_{\z,p}+\frac{1}{x^+_{\z,p}}-x^-_{\z,p}-\frac{1}{x^-_{\z,p}}&=\frac{2i\,\big(0+\tfrac{k}{2\pi}p\big)}{h}\,,
\end{aligned}
\end{equation}
as well as
\begin{equation}
  \frac{x^+_{*,p}}{x^-_{*,p}} = e^{i p}\,, \qquad  x^+_{*,p}-\frac{1}{x^+_{*,p}}-x^-_{*,p}+\frac{1}{x^-_{*,p}}=\frac{2i\,\gen H}{h}\,.
  \label{eq:energymomentum}
\end{equation}
Defining
\begin{equation}
\label{eq:abparam}
\begin{aligned}
&a_\L=\eta_{\L,p}\,,\ 
&&b_\L=-\frac{e^{-i p/2}}{x^-_{\L,p}}\eta_{\L,p}\,,
\quad
&&a^*_\L=e^{-ip/2}\eta_{\L,p}\,,\ 
&&b^*_\L=-\frac{1}{x^{+}_{\L,p}}\eta_{\L,p}\,,\\
&b_\R=\eta_{\R,p}\,,\ 
&&a_\R=-\frac{e^{-i p/2}}{x^-_{\R,p}}\eta_{\R,p}\,,
\quad
&&b^*_\R=e^{-ip/2}\eta_{\R,p}\,,\ 
&&a^*_\R=-\frac{1}{x^{+}_{\R,p}}\eta_{\R,p}\,,\\
&a_\z=\eta_{\z,p}\,,\ 
&&b_\z=-\frac{e^{-i p/2}}{x^-_{\z,p}}\eta_{\z,p}\,,
\quad
&&a^*_\z=e^{-ip/2}\eta_{\z,p}\,,\ 
&&b^*_\z=-\frac{1}{x^{+}_{\z,p}}\eta_{\z,p}\,,
\end{aligned}
\end{equation}
with
\begin{equation}
\label{eq:etaparameter}
    \eta_{*,p}=e^{ip/4}\sqrt{\frac{ih}{2}(x^-_{*,p}-x^+_{*,p})}\,,
\end{equation}
all the relations \eqref{eq:ableft}, \eqref{eq:abright} and \eqref{eq:abmassless} become satisfied.

\subsubsection{Two particle representations}

The S matrix is an operator $S : \mathcal V(p_1,m_1) \otimes \mathcal V(p_2,m_2) \rightarrow  \mathcal V(p_1,m_1) \otimes \mathcal V(p_2, m_2)$, which should ``commute'' (this will be made more precise in the next section) with the symmetry generators. Hence, we need to understand how the generators act on a two-particle representation, given by the tensor product of two one-particle representations. Assume that we have a bosonic generator $\gen{J}$ acting on a one-particle state as $\gen{J} \ket{\Phi} = \ket{\Phi'}$. Then a two-particle representation can be constructed using the trivial coproduct,
\begin{equation} \label{eq:copJtriv}
\text{(trivial)}\qquad
\gen{J}_{12} = \Delta(\gen{J}) = \gen{J} \otimes 1 + 1 \otimes \gen{J}\,,
\end{equation}
so that $\gen{J}_{12} \ket{\Phi_1 \Phi_2} = \ket{\Phi'_1 \Phi_2} + \ket{\Phi_1 \Phi'_2}$. 
For fermionic generators $\gen{Q}$ we need to take into account the minus sign arising when passing one fermion through the other and the trivial coproduct reads
\begin{equation}
\text{(trivial)}\qquad
\gen{Q}_{12} = \Delta(\gen{Q}) = \gen{Q} \otimes 1 + \Sigma \otimes \gen{Q}\,,
\end{equation}
with $\Sigma$ the diagonal matrix that acts with $+1$ on bosons and $-1$ on fermions.~\footnote{The tensor product on the other hand is the standard one. Alternatively, it is possible to redefine the tensor product so that it picks up a sign when acting on fermions, and not introducing the additional $\Sigma$ operator.} For a generic fermionic operator $\gen{Q}$ we have that $\Sigma \gen{Q} \Sigma = - \gen{Q}$. 
However, it turns out that due to the presence of the central extension, the coproduct needs to be modified. This can be seen looking at the action of $\gen{C}$ on a two-particle state. On one hand, we expect that
\begin{equation}
\text{(trivial)}\qquad
\gen{C}_{12} \ket{\Phi_1 \Phi_2} = \frac{i h}{2} \left( e^{i \gen{P}_{12}}-1\right)\ket{\Phi_1 \Phi_2} = \frac{i h}{2} \left( e^{i (p_1+p_2)}-1\right)\ket{\Phi_1 \Phi_2}\,,
\end{equation}
because $\gen{P}_{12}$ measures the total momentum of the two-particle state.
On the other hand, assuming a trivial coproduct of the form \eqref{eq:copJtriv} leads to
\begin{equation}
\text{(trivial)}\qquad\gen{C}_{12} \ket{\Phi_1 \Phi_2} = \frac{i h}{2} \left( (e^{i p_1}-1) + (e^{i p_2}-1)  \right)\ket{\Phi_1 \Phi_2}\,,
\end{equation}
which is not what we want.
In order to reproduce the expected result, one needs to add a braiding factor into the coproduct. One can show that either of the following choices gives the correct result:
\begin{equation}
\Delta(\gen{C}) = \gen{C} \otimes 1 + \gen{U}^2 \otimes \gen{C}\,, \qquad \Delta(\gen{C}) = \gen{C} \otimes \gen{U}^2 + 1 \otimes \gen{C}\,, \qquad \gen{U} = e^{\tfrac{i}{2} \gen{p}}\,.
\end{equation}

To have compatibility with the commutation relations, also the coproduct of fermionic charges need to include the braiding factor $\gen{U}$. Using the first choice in the above, we get
\begin{equation} \begin{aligned}
\Delta(\gen{H}) &= \gen{H} \otimes 1 + 1 \otimes \gen{H}\,, \\
\Delta(\gen{Q}) &= \gen{Q} \otimes 1 + \Sigma \gen{U} \otimes \gen{Q}\,, \\
\Delta(\gen{S}) &= \gen{S} \otimes 1 + \Sigma \gen{U}^{-1} \otimes \gen{S}\,, \\
\Delta(\gen{C}) &= \gen{C} \otimes 1 + \gen{U}^2 \otimes \gen{C}\,, \\
\Delta(\gen{U}) &= \gen{U} \otimes \gen{U}\,.
\end{aligned}
\end{equation}
The second choice will eventually lead to the same S matrix.

\begin{tcolorbox}
  \begin{exercise}
    A singlet state $\ket{1}$ is a two-particle state annihilated by all the generators, $\Delta(\gen{J}) \ket{1} =0$ for all $\gen{J}$ in $\mathcal A$. Find such a singlet state in the massive sector.
  \end{exercise}
\end{tcolorbox}

\subsubsection{Factorisation}

\paragraph{Of the algebra.}

Within $\mathcal A$ we observe the presence of two copies of the subalgebra 
\begin{equation}
\mathcal B = \left[\alg{su}(1|1)_\L \oplus \alg{su}(1|1)_\R\right]_{c.e.}\,,
\end{equation}
sharing the same central elements.%
\footnote{A similar factorisation happens in the case of $AdS_5 \times S^5$. There, the symmetry breaking pattern is given by $
\alg{psu}(2,2|4) \  \rightarrow \  \mathcal A = \alg{su}(2|2)_{c.e.}^{\oplus 2}$,
and hence we can identify $\mathcal B = \alg{su}(2|2)_{c.e.}$.} 
 This algebra has four fermionic generators as well as four central elements,
\begin{equation}
\anticom{\gen{q}_\L}{\gen{s}_\L} = \gen{h}_\L\,, \qquad \anticom{\gen{q}_\R}{\gen{s}_\R} = \gen{h}_\R\,, \qquad \anticom{\gen{q}_\L}{\gen{q}_\R} = \gen{c}\,, \qquad \anticom{\gen{s}_\L}{\gen{s}_\R} = \bar{\gen{c}}\,.
\end{equation}
 We construct the generators
\begin{equation} \begin{aligned}
\gen{H}_\L{}^1 &= \gen{h}_\L \otimes 1\,, &\qquad \gen{H}_\L{}^2 &= 1 \otimes \gen{h}_\L\,, &\qquad \gen{C}^1 &= \gen{c} \otimes 1\,, &\qquad \gen{C}^2 &= 1 \otimes \gen{c}\,, \\
\gen{H}_\R{}^1 &= \gen{h}_\R \otimes 1\,, &\qquad \gen{H}_\R{}^2 &= 1 \otimes \gen{h}_\R\,, &\qquad \bar{\gen{C}}^1 &= \bar{\gen{c}} \otimes 1\,, &\qquad \bar{\gen{C}}^2 &= 1 \otimes \bar{\gen{c}}\,, 
\end{aligned}
\end{equation}
as well as
\begin{equation} \begin{aligned}
\gen{Q}_\L{}^1 &= \gen{q}_\L \otimes 1\,, &\qquad \gen{S}_{\L 1} &= \gen{s}_\L \otimes 1\,,
 &\qquad \gen{Q}_\L{}^2 &= \Sigma \otimes \gen{q}_\L\,, &\qquad \gen{S}_{\L 2} &= \Sigma \otimes \gen{s}_\L\,, \\ 
\gen{Q}_{\R 1} &= \gen{q}_\R \otimes 1\,, &\qquad \gen{S}_\R{}^1 &= \gen{s}_\R \otimes 1\,, &\qquad \gen{Q}_{\R 2} &= \Sigma \otimes \gen{q}_\R\,, &\qquad \gen{S}_\R{}^2 &= \Sigma \otimes \gen{s}_\R\,.
\end{aligned}
\end{equation}
%Indices are raised and lowered with the antisymmetric Levi-Civita symbol $\epsilon^{12} = -\epsilon_{12}=+1$. 
 The central elements should be the same in the $1$ and $2$ copies, which imposes
\begin{equation}
\gen{H}_\L{}^{1} = \gen{H}_\L{}^2 = \gen{H}_\L\,, \qquad \gen{H}_\R{}^{1} = \gen{H}_\R{}^2 = \gen{H}_\R\,, \qquad \gen{C}^1 = \gen{C}^2 = \gen{C}\,, \qquad \bar{\gen{C}}^1 = \bar{\gen{C}}^2 = \bar{\gen{C}}\,.
\end{equation}
The generators defined in this way precisely obey the commutation relations \eqref{eq:com1} and \eqref{eq:com2}.

\paragraph{Of the representations.}
The factorised structure carries over at the level of the representation. Short representations of $\mathcal B$ are two-dimensional and we define the following four representations
\begin{equation}
\rho_\L=(\phi_\L^\B | \varphi_\L^\F)\,, \qquad \rho_\R=(\phi_\R^\F | \varphi_\R^\B)\,, \qquad \rho_\z=(\phi_\z^\B | \varphi_\z^\F)\,, \qquad \rho_\z'=(\phi_\z^\F | \varphi_\z^\B)\,.
\end{equation}
In our notation the first component is always the heighest weight state (denoted by $\phi_*^*$) while the second component is always the lowest weight state (denoted by $\varphi_*^*$). Depending on the representation, these are bosonic (denoted by the upper index $^\B$) or fermionic (denoted by the upper index $^\F$). The representations are all of the same form
\begin{equation}
\begin{aligned}
    &\rho_\L: &\quad &\ql\, |\phi_\L^\B\rangle = a_\L\,|\varphi_\L^\F\rangle,\ 
    &&\sl\, |\varphi_\L^\F\rangle = a^*_\L\,|\phi_\L^\B\rangle,\ 
    &&\sr\, |\phi_\L^\B\rangle = b^*_\L\,|\varphi_\L^\F\rangle,\ 
    &&\qr\, |\varphi_\L^\F\rangle = b_\L\,|\phi_\L^\B\rangle\,,\\[0.2cm]
    &\rho_\R: &\quad &\ql\, |\phi_\R^\F\rangle = a_\R\,|\varphi_\R^\B\rangle,\ 
    &&\sl\, |\varphi_\R^\B\rangle = a^*_\R\,|\phi_\R^\F\rangle,\ 
    &&\sr\, |\phi_\R^\F\rangle = b^*_\R\,|\varphi_\R^\B\rangle,\ 
    &&\qr\, |\varphi_\R^\B\rangle = b_\R\,|\phi_\R^\F\rangle\,,\\[0.2cm]
    &\rho_\circ: &\quad &\ql\, |\phi_\z^\B\rangle = a_\z\,|\varphi_\z^\F\rangle,\ 
    &&\sl\, |\varphi_\z^\F\rangle = a^*_\z\,|\phi_\z^\B\rangle,\ 
    &&\sr\, |\phi_\z^\B\rangle = b^*_\z\,|\varphi_\z^\F\rangle,\ 
    &&\qr\, |\varphi_\z^\F\rangle = b_\z\,|\phi_\z^\B\rangle\,,\\[0.2cm]
    &\rho_\circ': &\quad &\ql\, |\phi_\z^\F\rangle = a_\z\,|\varphi_\z^\B\rangle,\ 
    &&\sl\, |\varphi_\z^\B\rangle = a^*_\z\,|\phi_\z^\F\rangle,\ 
    &&\sr\, |\phi_\z^\F\rangle = b^*_\z\,|\varphi_\z^\B\rangle,\ 
    &&\qr\, |\varphi_\z^\B\rangle = b_\z\,|\phi_\z^\F\rangle\,,
    \label{eq:irreducibleRepresentations}
\end{aligned}    
\end{equation}
with only the value of the representation coefficients  differing for each representation.
The representations of $\mathcal A$ are then constructed from tensor products of representations of $\mathcal B$,
\begin{equation}
    \varrho_\L = \rho_\L \otimes \rho_\L\,, \qquad \varrho_\R = \rho_\R \otimes \rho_\R\,, \qquad \varrho_\circ^{\dot{a}} = (\rho_\circ \otimes \rho_\circ') \oplus (\rho_\circ' \otimes \rho_\circ)\,.
\end{equation}
In particular we have that
\begin{equation} \begin{aligned}
    \varrho_\L &=\{Y_\L=\phi_\L^\B \otimes \phi_\L^\B,\, \Psi_\L^1 = \varphi_\L^\F \otimes \phi_\L^\B,\, \Psi_\L^2 = \phi_\L^\B \otimes \varphi_\L^\F,\, Z_\L = \varphi_\L^\F \otimes \varphi_\L^\F \}\,, \\
    \varrho_\R &=\{Z_\R=\phi_\R^\F \otimes \phi_\R^\F,\, \Psi_\R^1 = \varphi_\R^\B \otimes \phi_\R^\F,\, \Psi_\R^2 = \phi_\R^\F \otimes \varphi_\R^\B,\, Y_\R = \varphi_\R^\B \otimes \varphi_\R^\B \}\,.
    \end{aligned}
\end{equation}

\subsection{The worldsheet S matrix}

Now that we have a good understanding of the symmetry algebra of the lightcone gauge fixed theory as well as the representations in which the excitations transform, we can bootstrap the worldsheet S matrix.  As usual for an integrable theory, a $n$-body scattering event factorises into a sequence of 2-body scattering events and the crucial building block is therefore the two-body worldsheet S matrix. The latter should commute with the symmetries of the lightcone gauge fixed theory,
\begin{equation} \label{eq:Smat-eq}
P_g \Delta(\gen J) \gen{S} = \gen S \Delta(\gen J)\,, \qquad \forall \, \gen{J} \in \mathcal A\,,
\end{equation}
where $P_g$ denotes the graded permutation operator acting e.g.~as
\begin{equation}
    P_g \ket{\phi^\B_p \phi^\B_q} = \ket{\phi^\B_q \phi^\B_p}\,, \quad P_g \ket{\phi^\B_p \phi^\F_q} = \ket{\phi^\F_q \phi^\B_p}\,, \quad P_g \ket{\phi^\F_p \phi^\B_q} = \ket{\phi^\B_q \phi^\F_p}\,, \quad P_g \ket{\phi^\F_p \phi^\F_q} = -\ket{\phi^\F_q \phi^\F_p}\,.
\end{equation}
Solving these equations will in fact completely fix the S-matrix, up to scalar factors.

\subsubsection{Factorised S matrix}

First we consider the ``factorised'' S matrix, which governs the scattering of two representations of $\mathcal B$.  Because we have 4 different short representations (left, right and two massless), the S matrix naturally arranges into 16 different blocks. 4 blocks belong to the ``massive'' sector (when both representations scattered are massive, these are left-left, left-right, right-left and right-right), 4 blocks belong to the ``massless'' sector (when both representations scattered are massless), and the remaining 8 blocks are of mixed mass type. It is important to note that the equation \eqref{eq:Smat-eq} only fixes the S-matrix up to a prefactor in each block (so that we are left with 16 different scalar prefactors). Below, when writing the S-matrix elements, we will choose a particular normalisation in each block.

\paragraph{Left-left scattering.}
First of all let us consider the case in which both representations are of the ``Left'' type. The scattering matrix takes the form
\begin{equation}
\begin{aligned}
    S|\phi_{\L,p}^\B\phi_{\L,q}^\B\rangle &= A^{\L\L}_{pq}\,|\phi_{\L,p}^\B\phi_{\L,q}^\B\rangle,&\quad
    S|\phi_{\L,p}^\B\varphi_{\L,q}^\F\rangle&= B^{\L\L}_{pq}|\phi_{\L,p}^\B\varphi_{\L,q}^\F\rangle + C^{\L\L}_{pq}|\varphi_{\L,p}^\F\phi_{\L,q}^\B\rangle,\\
    S|\varphi_{\L,p}^\F\varphi_{\L,q}^\F\rangle &= F^{\L\L}_{pq}\,|\varphi_{\L,p}^\F\varphi_{\L,q}^\F\rangle,&\quad
    S|\varphi_{\L,p}^\F\phi_{\L,q}^\B\rangle&= D^{\L\L}_{pq}|\varphi_{\L,p}^\F\phi_{\L,q}^\B\rangle + E^{\L\L}_{pq}|\phi_{\L,p}^\B\varphi_{\L,q}^\F\rangle,
\end{aligned}
\end{equation}
with matrix elements 
\begin{equation}
\label{eq:SLLexplicit}
\begin{aligned}
    &A_{pq}^{\L\L}=1\,,\qquad
    &&B_{pq}^{\L\L}=e^{-\frac{i}{2}p}\frac{x^+_{\L,p} - x^+_{\L,q}}{x^-_{\L,p} - x^+_{\L,q}}\,,\\
    &C_{pq}^{\L\L}=e^{-\frac{i}{2}p}e^{+\frac{i}{2}q}\frac{x^-_{\L,q} - x^+_{\L,q}}{x^-_{\L,p} - x^+_{\L,q}}\frac{\eta_{\L,p}}{\eta_{\L,q}}\,,\qquad
    &&D_{pq}^{\L\L}=e^{+\frac{i}{2}q}\frac{x^-_{\L,p} - x^-_{\L,q}}{x^-_{\L,p} - x^+_{\L,q}}\,,\\
    &E_{pq}^{\L\L}=C_{pq}^{\L\L}\,,\qquad
    &&F_{pq}^{\L\L}= e^{-\frac{i}{2}p}e^{+\frac{i}{2}q} \frac{x^+_{\L,p} - x^-_{\L,q}}{x^-_{\L,p} - x^+_{\L,q}}\,.
\end{aligned}
\end{equation}
The coefficients are written in terms of the Zhukhovski variables $x^\pm_{\L,p}$ and are valid for any value of the mass $\mu>0$. In particular, we can scatter particles of the same mass, but also of different masses, as long as both of them have positive mass. Because the equation \eqref{eq:Smat-eq} fixes the S-matrix elements up to an overall factor, it is possible to multiply all the matrix elements in \eqref{eq:SLLexplicit} by an overall prefactor $\Sigma^{\L \L}_{pq}$.

\paragraph{Right-right scattering.}
Then, we consider the scattering of two ``Right'' which takes the same schematic form,
\begin{equation}
\begin{aligned}
    S|\varphi_{\R,p}^\B\varphi_{\R,q}^\B\rangle &= A^{\R\R}_{pq}\,|\varphi_{\R,p}^\B\varphi_{\R,q}^\B\rangle,&\quad
    S|\varphi_{\R,p}^\B\phi_{\R,q}^\F\rangle&= B^{\R\R}_{pq}|\varphi_{\R,p}^\B\phi_{\R,q}^\F\rangle + C^{\R\R}_{pq}|\phi_{\R,p}^\F\varphi_{\R,q}^\B\rangle,\\
    S|\phi_{\R,p}^\F\phi_{\R,q}^\F\rangle &= F^{\R\R}_{pq}\,|\phi_{\R,p}^\F\phi_{\R,q}^\F\rangle,&\quad
    S|\phi_{\R,p}^\F\varphi_{\R,q}^\B\rangle&= D^{\R\R}_{pq}|\phi_{\R,p}^\F\varphi_{\R,q}^\B\rangle + E^{\R\R}_{pq}|\varphi_{\R,p}^\B\phi_{\R,q}^\F\rangle.
\end{aligned}
\end{equation}
The S-matrix elements simply obtained from the previous ones through replacing ``Left'' with ``Right'' Zhukovski variables,
\begin{equation}
\begin{aligned}
    &A_{pq}^{\R\R}=1\,,\qquad
    &&B_{pq}^{\R\R}=e^{-\frac{i}{2}p}\frac{x^+_{\R,p} - x^+_{\R,q}}{x^-_{\R,p} - x^+_{\R,q}}\,,\\
    &C_{pq}^{\R\R}=e^{-\frac{i}{2}p}e^{+\frac{i}{2}q}\frac{x^-_{\R,q} - x^+_{\R,q}}{x^-_{\R,p} - x^+_{\R,q}}\frac{\eta_{\R,p}}{\eta_{\R,q}}\,,\qquad
    &&D_{pq}^{\R\R}=e^{+\frac{i}{2}q}\frac{x^-_{\R,p} - x^-_{\R,q}}{x^-_{\R,p} - x^+_{\R,q}}\,,\\
    &E_{pq}^{\R\R}=C_{pq}^{\R\R}\,,\qquad
    &&F_{pq}^{\R\R}= e^{-\frac{i}{2}p}e^{+\frac{i}{2}q} \frac{x^+_{\R,p} - x^-_{\R,q}}{x^-_{\R,p} - x^+_{\R,q}}\,.
\end{aligned}
\end{equation}
These are fixed up to an overall prefactor $\Sigma_{pq}^{\R\R}$. Again, the above formuli hold for the scattering of two representations with arbitrary values of $\mu<0$.

\paragraph{Left-right scattering.}
Here we have
\begin{equation}
\begin{aligned}
    S|\phi_{\L,p}^\B\varphi_{\R,q}^\B\rangle&= A^{\L\R}_{pq}|\phi_{\L,p}^\B\varphi_{\R,q}^\B\rangle + B^{\L\R}_{pq}|\varphi_{\L,p}^\F\phi_{\R,q}^\F\rangle,
    &\quad
    S|\phi_{\L,p}^\B\phi_{\R,q}^\F\rangle &= C^{\L\R}_{pq}\,|\phi_{\L,p}^\B\phi_{\R,q}^\F\rangle,\\
    S|\varphi_{\L,p}^\F\phi_{\R,q}^\F\rangle&= E^{\L\R}_{pq}|\varphi_{\L,p}^\F\phi_{\R,q}^\F\rangle + F^{\L\R}_{pq}|\phi_{\L,p}^\B\varphi_{\R,q}^\B\rangle,&\quad
    S|\varphi_{\L,p}^\F\varphi_{\R,q}^\B\rangle &= D^{\L\R}_{pq}\,|\varphi_{\L,p}^\F\varphi_{\R,q}^\B\rangle,
\end{aligned}
\end{equation}
with
\begin{equation}
   \begin{aligned}
    &A_{pq}^{\L\R}= e^{-\frac{i}{2}p}\frac{1-x^+_{\L,p} x^-_{\R,q}}{1-x^-_{\L,p} x^-_{\R,q}}\,,\qquad
    &&B_{pq}^{\L\R}=e^{-\frac{i}{2}p}e^{-\frac{i}{2}q} \frac{2 i}{h} \frac{\eta_{\L,p}\eta_{\R,q}}{1-x^-_{\L,p} x^-_{\R,q}}\,,\\
    &C_{pq}^{\L\R}=1\,,\qquad
    &&D_{pq}^{\L\R}=e^{-\frac{i}{2}p}e^{-\frac{i}{2}q}\frac{1- x^+_{\L,p} x^+_{\R,q}}{1-x^-_{\L,p} x^-_{\R,q}}\,,\\
    &E_{pq}^{\L\R}=  e^{-\frac{i}{2}q}\frac{1-x^-_{\L,p} x^+_{\R,q}}{1-x^-_{\L,p} x^-_{\R,q}}\,,\qquad
    &&F_{pq}^{\L\R}=B_{pq}^{\L\R}\,.
\end{aligned}
\end{equation}

\paragraph{Right-left scattering.}

The right-left S~matrix reads
\begin{equation}
\begin{aligned}
    S|\varphi_{\R,p}^\B\phi_{\L,q}^\B\rangle&= A^{\R\L}_{pq}|\varphi_{\R,p}^\B\phi_{\L,q}^\B\rangle + B^{\R\L}_{pq}|\phi_{\R,p}^\F\varphi_{\L,q}^\F\rangle,
    &\quad
    S|\varphi_{\R,p}^\B\varphi_{\L,q}^\F\rangle &= C^{\R\L}_{pq}\,|\varphi_{\R,p}^\B\varphi_{\L,q}^\F\rangle,\\
    S|\phi_{\R,p}^\F\varphi_{\L,q}^\F\rangle&= E^{\R\L}_{pq}|\phi_{\R,p}^\F\varphi_{\L,q}^\F\rangle + F^{\R\L}_{pq}|\varphi_{\R,p}^\B\phi_{\L,q}^\B\rangle,&\quad
    S|\phi_{\R,p}^\F\phi_{\L,q}^\B\rangle &= D^{\R\L}_{pq}\,|\phi_{\R,p}^\F\phi_{\L,q}^\B\rangle,
\end{aligned}
\end{equation}
with
\begin{equation}
   \begin{aligned}
     &A_{pq}^{\R\L}=e^{+\frac{i}{2}q}\frac{1-x^+_{\R,p} x^-_{\L,q}}{1-x^+_{\R,p} x^+_{\L,q}}\,,\qquad
    &&B_{pq}^{\R\L}= \frac{2 i}{h} \frac{\eta_{\R,p}\eta_{\L,q}}{1-x^+_{\R,p} x^+_{\L,q}}\,,\\
    &C_{pq}^{\R\L}= e^{+\frac{i}{2}p}e^{+\frac{i}{2}q}\frac{1-x^-_{\R,p} x^-_{\L,q}}{1-x^+_{\R,p} x^+_{\L,q}}\,,\qquad
    &&D_{pq}^{\R\L}=1\,,\\
    &E_{pq}^{\R\L}= e^{+\frac{i}{2}p}\frac{1-x^-_{\R,p} x^+_{\L,q}}{1-x^+_{\R,p} x^+_{\L,q}}\,,\qquad
    &&F_{pq}^{\R\L}=B_{pq}^{\R\L}\,.
\end{aligned}
\end{equation}

\paragraph{Massless sector.}

For the massless S matrix, when both particles are in the $\rho_\z$ representation, 
\begin{equation}
\begin{aligned}
    S|\phi_{\z,p}^\B\phi_{\z,q}^\B\rangle &= A^{\z\z}_{pq}\,|\phi_{\z,p}^\B\phi_{\z,q}^\B\rangle,&\quad
    S|\phi_{\z,p}^\B\varphi_{\z,q}^\F\rangle&= B^{\z\z}_{pq}|\phi_{\z,p}^\B\varphi_{\z,q}^\F\rangle + C^{\z\z}_{pq}|\varphi_{\z,p}^\F\phi_{\z,q}^\B\rangle,\\
    S|\varphi_{\z,p}^\F\varphi_{\z,q}^\F\rangle &= F^{\z\z}_{pq}\,|\varphi_{\z,p}^\F\varphi_{\z,q}^\F\rangle,&\quad
    S|\varphi_{\z,p}^\F\phi_{\z,q}^\B\rangle&= D^{\z\z}_{pq}|\varphi_{\z,p}^\F\phi_{\z,q}^\B\rangle + E^{\z\z}_{pq}|\phi_{\z,p}^\B\varphi_{\z,q}^\F\rangle.
\end{aligned}
\end{equation}
The S-matrix elements are obtained from the left-left S-matrix elements through the $m \rightarrow 0$ limit. This can simply be achieved by replacing $x^\pm_\L$ by $x^\pm_\z$ in \eqref{eq:SLLexplicit}.
When both particles are in the $\rho_\z'$ representation we have, instead
\begin{equation}
\begin{aligned}
    S|\varphi_{\z,p}^\B\varphi_{\z,q}^\B\rangle &= F^{\z\z}_{pq}\,|\varphi_{\z,p}^\B\varphi_{\z,q}^\B\rangle,&\quad
    S|\varphi_{\z,p}^\B\phi_{\z,q}^\F\rangle&= D^{\z\z}_{pq}|\varphi_{\z,p}^\B\phi_{\z,q}^\F\rangle - E^{\z\z}_{pq}|\phi_{\z,p}^\F\varphi_{\z,q}^\B\rangle,\\
    S|\phi_{\z,p}^\F\phi_{\z,q}^\F\rangle &= A^{\z\z}_{pq}\,|\phi_{\z,p}^\F\phi_{\z,q}^\F\rangle,&\quad
    S|\phi_{\z,p}^\F\varphi_{\z,q}^\B\rangle&= B^{\z\z}_{pq}|\phi_{\z,p}^\F\varphi_{\z,q}^\B\rangle - C^{\z\z}_{pq}|\varphi_{\z,p}^\B\phi_{\z,q}^\F\rangle.
\end{aligned}
\end{equation}
Similarly, in the mixed case we have
\begin{equation}
\label{eq:Slefttildeleft}
\begin{aligned}
    S|\phi_{\z,p}^\B\varphi_{\z,q}^\B\rangle&= B^{\z\z}_{pq}|\phi_{\z,p}^\B\varphi_{\z,q}^\B\rangle +C^{\z\z}_{pq}|\varphi_{\z,p}^\F\phi_{\z,q}^\F\rangle,
    &\quad
    S|\phi_{\z,p}^\B\phi_{\z,q}^\F\rangle &= A^{\z\z}_{pq}\,|\phi_{\z,p}^\B\phi_{\z,q}^\F\rangle,\\
    S|\varphi_{\z,p}^\F\phi_{\z,q}^\F\rangle&= D^{\z\z}_{pq}|\varphi_{\z,p}^\F\phi_{\z,q}^\F\rangle + E^{\z\z}_{pq}|\phi_{\z,p}^\B\varphi_{\z,q}^\B\rangle,&\quad
    S|\varphi_{\z,p}^\F\varphi_{\z,q}^\B\rangle &= F^{\z\z}_{pq}\,|\varphi_{\z,p}^\F\varphi_{\z,q}^\B\rangle,
\end{aligned}
\end{equation}
and finally
\begin{equation}
\begin{aligned}
    S|\varphi_{\z,p}^\B\phi_{\z,q}^\B\rangle&= D^{\z\z}_{pq}|\varphi_{\z,p}^\B\phi_{\z,q}^\B\rangle - E^{\z\z}_{pq}|\phi_{\z,p}^\F\varphi_{\z,q}^\F\rangle,
    &\quad
    S|\varphi_{\z,p}^\B\varphi_{\z,q}^\F\rangle &= F^{\z\z}_{pq}\,|\varphi_{\z,p}^\B\varphi_{\z,q}^\F\rangle,\\
    S|\phi_{\z,p}^\F\varphi_{\z,q}^\F\rangle&= B^{\z\z}_{pq}|\phi_{\z,p}^\F\varphi_{\z,q}^\F\rangle - C^{\z\z}_{pq}|\varphi_{\z,p}^\B\phi_{\z,q}^\B\rangle,&\quad
    S|\phi_{\z,p}^\F\phi_{\z,q}^\B\rangle &= A^{\z\z}_{pq}\,|\phi_{\z,p}^\F\phi_{\z,q}^\B\rangle.
\end{aligned}
\end{equation}

\paragraph{Mixed-mass sector.}

Finally, in the mixed mass sector we have
\begin{equation}
\begin{aligned}
    S|\phi_{\L,p}^\B\varphi_{\z,q}^\B\rangle&= B^{\L\z}_{pq}|\phi_{\z,p}^\B\varphi_{\z,q}^\B\rangle +C^{\L\z}_{pq}|\varphi_{\z,p}^\F\phi_{\z,q}^\F\rangle,
    &\quad
    S|\phi_{\L,p}^\B\phi_{\z,q}^\F\rangle &= A^{\L\z}_{pq}\,|\phi_{\L,p}^\B\phi_{\z,q}^\F\rangle,\\
    S|\varphi_{\L,p}^\F\phi_{\z,q}^\F\rangle&= D^{\L\z}_{pq}|\varphi_{\L,p}^\F\phi_{\z,q}^\F\rangle + E^{\L\z}_{pq}|\phi_{\L,p}^\B\varphi_{\z,q}^\B\rangle,&\quad
    S|\varphi_{\L,p}^\F\varphi_{\z,q}^\B\rangle &= F^{\L\z}_{pq}\,|\varphi_{\L,p}^\F\varphi_{\z,q}^\B\rangle,
\end{aligned}
\end{equation}
and 
\begin{equation}
\begin{aligned}
    S|\varphi_{\z,p}^\B\phi_{\L,q}^\B\rangle&= D^{\z\L}_{pq}|\varphi_{\z,p}^\B\phi_{\L,q}^\B\rangle - E^{\z\L}_{pq}|\phi_{\z,p}^\F\varphi_{\L,q}^\F\rangle,
    &\quad
    S|\varphi_{\z,p}^\B\varphi_{\L,q}^\F\rangle &= F^{\z\L}_{pq}\,|\varphi_{\z,p}^\B\varphi_{\L,q}^\F\rangle,\\
    S|\phi_{\z,p}^\F\varphi_{\L,q}^\F\rangle&= B^{\z\L}_{pq}|\phi_{\z,p}^\F\varphi_{\L,q}^\F\rangle - C^{\z\L}_{pq}|\varphi_{\z,p}^\B\phi_{\L,q}^\B\rangle,&\quad
    S|\phi_{\z,p}^\F\phi_{\L,q}^\B\rangle &= A^{\z\L}_{pq}\,|\phi_{\z,p}^\F\phi_{\L,q}^\B\rangle.
\end{aligned}
\end{equation}
as well as
\begin{equation}
\begin{aligned}
    S|\varphi_{\R,p}^\B\varphi_{\z,q}^\B\rangle&=
    C^{\R\z}_{pq}\,|\varphi_{\R,p}^\B\varphi_{\z,q}^\B\rangle,
    &\quad
    S|\varphi_{\R,p}^\B\phi_{\z,q}^\F\rangle &=
    A^{\R\z}_{pq}|\varphi_{\R,p}^\B\phi_{\z,q}^\F\rangle + B^{\R\z}_{pq}|\phi_{\R,p}^\F\varphi_{\z,q}^\B\rangle,\\
    S|\phi_{\R,p}^\F\phi_{\z,q}^\F\rangle&=
    D^{\R\z}_{pq}|\phi_{\R,p}^\F\phi_{\z,q}^\F\rangle,&\quad
    S|\phi_{\R,p}^\F\varphi_{\z,q}^\B\rangle &=
    E^{\R\z}_{pq}|\phi_{\R,p}^\F\varphi_{\z,q}^\B\rangle + F^{\R\z}_{pq}|\varphi_{\R,p}^\B\phi_{\z,q}^\F\rangle,
\end{aligned}
\end{equation}
and 
\begin{equation}
\begin{aligned}
    S|\varphi_{\z,p}^\B\varphi_{\R,q}^\B\rangle &= D^{\z\R}_{pq}\,|\varphi_{\z,p}^\B\varphi_{\R,q}^\B\rangle,&\quad
    S|\varphi_{\z,p}^\B\phi_{\R,q}^\F\rangle&= E^{\z\R}_{pq}|\varphi_{\z,p}^\B\phi_{\R,q}^\F\rangle - F^{\z\R}_{pq}|\phi_{\z,p}^\F\varphi_{\R,q}^\B\rangle,\\
    S|\phi_{\z,p}^\F\phi_{\R,q}^\F\rangle &= C^{\z\R}_{pq}\,|\phi_{\z,p}^\F\phi_{\R,q}^\F\rangle,&\quad
    S|\phi_{\z,p}^\F\varphi_{\R,q}^\B\rangle&= A^{\z\R}_{pq}|\phi_{\z,p}^\F\varphi_{\R,q}^\B\rangle - B^{\z\R}_{pq}|\varphi_{\z,p}^\B\phi_{\R,q}^\F\rangle.
\end{aligned}
\end{equation}
As the notation suggests, the S-matrix elements of the type $A^{\star \z}_{pq}$ and $A^{\z \star}_{pq}$ with $\star=\L,\R$ are obtained from $A^{\star \L}_{pq}$ and $A^{ \L \star}_{pq}$ through the replacement $x^\pm_{\L,q} \rightarrow x^\pm_{\z,q}$ and $x^\pm_{\L,p} \rightarrow x^\pm_{\z,p}$ respectively.

\subsubsection{The full S matrix}\label{fiona:subsec:fullSmatrix}
Having worked out the ``factorised'' S matrix we can now turn to the full S matrix. It can be obtained by taking the graded tensor product of two copies of the $\mathcal B$-invariant S matrix constructed above
\begin{equation} \label{eq:fullSsim}
    \mathbf{S} \sim S \hat{\otimes} \acute{S} \,,
\end{equation}
which can be defined in terms of the matrix elements by
\begin{equation}
    (\mathcal{M} \hat{\otimes} \acute{\mathcal{M}})^{I \acute{I},J\acute{J}}_{K \acute{K},L\acute{L}} = (-1)^{F_{\acute{K}}F_L+F_J F_{\acute{I}}} \mathcal{M}^{IJ}_{KL} \acute{\mathcal{M}}^{\acute{I}\acute{J}}_{\acute{K}\acute{L}} \,.
\end{equation}

Notice that we did not use an equality sign in \eqref{eq:fullSsim}. This is because the symmetries only fix the S matrix up to 16 functions (one for each 2-particle representation scattered), the so-called dressing phases. For instance, in the massive sector, there are in principle 4 dressing phases: $\sigma_{\L \L}$ when the 2-particle representation scattered is $\varrho_\L \otimes \varrho_\L$, then $\sigma_{\L \R}$ when the 2-particle representation scattered is $\varrho_\L \otimes \varrho_\R$, and also $\sigma_{\R \L}$ and $\sigma_{\R \R}$.

We do expect some additional discrete symmetries to reduce the number of these dressing factors. Due to left-right symmetry, we expect that $\sigma_{\L \L}$ and $\sigma_{\R \R}$ should be related, and might be expressed in terms of a single function $\sigma^{\bullet \bullet}$. Similarly, $\sigma_{\R \L}$ and $\sigma_{\L \R}$ should be expressible in terms of a single function $\tilde{\sigma}^{\bullet \bullet}$. Moreover, also using the fact that the massless modes transform as a doublet under $\alg{su}(2)_\circ$ we are left with six functions. We shall call these functions $\sigma^{\bullet \bullet}$ for the LL and RR dressing phases, $\tilde{\sigma}^{\bullet \bullet}$ for the LR and RL dressing phases, $\sigma^{\circ \bullet }$ for the massless-L and massless-R phases, $\sigma^{\bullet \circ}$ for the L-massless and R-massless phases, $\sigma^{\circ \circ}$ if the massless modes are of the same chirality and finally $\tilde{\sigma}^{\circ \circ}$ if the massless modes are of opposite chirality.

It is customary and convenient to normalise the S-matrix blocks so that the dressing factors themselves have no poles for physical value of the momenta. Such poles may (and in this case, do) exist, and we highlight them explicitly \textit{e.g.}~in terms of rational expressions in $x^\pm$, in such a way as to make more transparent the bound-state structure of the model. However, how to do so in detail is only understood in the case of the pure-RR background (or pure-NSNS, which is however much simpler~\cite{Dei:2018mfl}). The case of mixed-flux backgrounds is apparently more subtle, see~\cite{Frolov:2023lwd}. Hence, for the remainder of this section, let us specialise to the case
\begin{equation}
    \text{RR~flux~only}:\qquad
    x^\pm(p)\equiv x^\pm_{\L}(p)=x^\pm_{\R}(p)\,.
\end{equation}
% \begin{equation}
%     \bra{Y_q Y_p} \gen{S} \ket{Y_p Y_q} = \frac{x^+_p}{x^-_p} \frac{x^-_q}{x^+_q} \frac{x^-_p - x^+_q}{x^+_p-x^-_q} \frac{1-\frac{1}{x^-_p x^+_q}}{1-\frac{1}{x^+_p x^-_q}} \frac{1}{(\sigma_{pq}^{\bullet \bullet})^2}\,.
% \end{equation}
% \begin{equation}
%     \bra{\bar{Y}_q Y_p} \gen{S} \ket{Y_p \bar{Y}_q} = \frac{x_p^+}{x_p^-} \frac{x^-_q}{x^+_q} \frac{1-\frac{1}{x^+_p x^-_q}}{1-\frac{1}{x^+_p x^+_q}} \frac{1-\frac{1}{x^-_p x^+_q}}{1-\frac{1}{x^-_p x^-_q}} \frac{1}{(\tilde{\sigma}_{pq}^{\bullet \bullet})^2}\,. 
% \end{equation}
This choice has a further advantage: the relation between LL and RR dressing factors is straightforward --- not only they are related, but they are \textit{the same function} of $x^\pm$; the same holds for other phases related by L-R symmetry. We have 
\begin{equation}
\label{eq:massivenorm}
    \begin{aligned}
    \mathbf{S}\,\big|Y_{\L,p}Y_{\L,q}\big\rangle&=&
    e^{+i p}e^{-i q}
    \frac{x^-_p-x^+_q}{x^+_p-x^-_q}
    \frac{1-\frac{1}{x^-_px^+_q}}{1-\frac{1}{x^+_px^-_q}}\big(\sigma^{\bullet\bullet}_{pq}\big)^{-2}\,
    \big|Y_{\L,p}Y_{\L,q}\big\rangle,\\
    \mathbf{S}\,\big|Y_{\L,p}Z_{\R,q}\big\rangle&=&
    e^{-i p}
    \frac{1-\frac{1}{x^-_px^-_q}}{1-\frac{1}{x^+_px^+_q}}
    \frac{1-\frac{1}{x^-_px^+_q}}{1-\frac{1}{x^+_px^-_q}}\big(\tilde{\sigma}^{\bullet\bullet}_{pq}\big)^{-2}\,
    \big|Y_{\L,p}Z_{\R,q}\big\rangle,\\
    \mathbf{S}\,\big|Z_{\R,p}Y_{\L,q}\big\rangle&=&
    e^{+ip}
    \frac{1-\frac{1}{x^+_px^+_q}}{1-\frac{1}{x^-_px^-_q}}
    \frac{1-\frac{1}{x^-_px^+_q}}{1-\frac{1}{x^+_px^-_q}}\big(\widetilde{\sigma}^{\bullet\bullet}_{pq}\big)^{-2}\,
    \big|Z_{\R, p}Y_{\L,q}\big\rangle,\\
    \mathbf{S}\,\big|Z_{\R,p}Z_{\R,q}\big\rangle&=&
    \frac{x^+_p-x^-_q}{x^-_p-x^+_q}
    \frac{1-\frac{1}{x^-_px^+_q}}{1-\frac{1}{x^+_px^-_q}}\big(\sigma^{\bullet\bullet}_{pq}\big)^{-2}\,
    \big|Z_{\R,p}Z_{\R,q}\big\rangle,
    \end{aligned}
\end{equation}
\begin{equation}
\label{eq:mixednormalisation}
    \begin{aligned}
    \mathbf{S}\,\big|Y_{\L,p}\chi^{\dot{\alpha}}_{q}\big\rangle&=&
    e^{+\frac{i}{2} p}e^{-i q}
    \frac{x^-_p-x_q}{1-x^+_px_q}\big(\sigma^{\bullet-}_{pq}\big)^{-2}\,
    \big|Y_{\L,p}\chi^{\dot{\alpha}}_{q}\big\rangle,\\
    \mathbf{S}\,\big|Z_{\R,p}\chi^{\dot{\alpha}}_{q}\big\rangle&=&
    e^{-\frac{i}{2} p}e^{-i q}
    \frac{1-x^+_px_q}{x^-_p-x_q}\big(\sigma^{\bullet-}_{pq}\big)^{-2}\,
    \big|Z_{\R,p}\chi^{\dot{\alpha}}_{q}\big\rangle,\\
    \mathbf{S}\,\big|\chi^{\dot{\alpha}}_{p}Y_{\L,q}\big\rangle&=&
    e^{+i p}e^{-\frac{i}{2} q}
    \frac{1-x_px^+_q}{x_p-x^-_q}\big(\sigma^{+\bullet}_{pq}\big)^{-2}\,
    \big|\chi^{\dot{\alpha}}_{p}Y_{\L,q}\big\rangle,\\
    \mathbf{S}\,\big|\chi^{\dot{\alpha}}_{p}Z_{\R,q}\big\rangle&=&
    e^{+i p}e^{+\frac{i}{2} q}
    \frac{x_p-x^-_q}{1-x_px^+_q}\big(\sigma^{+\bullet}_{pq}\big)^{-2}\,
    \big|\chi^{\dot{\alpha}}_{p}Z_{\R,q}\big\rangle,
    \end{aligned}
\end{equation}
\begin{equation}
\label{eq:masslessnorm}
    \mathbf{S}\,\big|\chi^{\dot{\alpha}}_{p}\chi^{\dot{\beta}}_{q}\big\rangle=
    \big(\sigma^{+-}_{pq}\big)^{-2}\,
    \big|\chi^{\dot{\alpha}}_{p}\chi^{\dot{\beta}}_{q}\big\rangle\,.
\end{equation}
We use the shorthand notation $\sigma_{pq} \equiv \sigma(p,q)$, which can be decorated by bullets and circles depending on the precise dressing phase we are considering.
For the mixed mass and massless scattering events we have chosen a chirality of the massless particles by imposing that $0 \leq p \leq \pi$ and $-\pi \leq q \leq 0$. In the dressing phases this is denoted with a $+$ for chiral and $-$ for anti-chiral massless particles.
Looking at the last line, the reader might be baffled by the lack of an $\su(2)$-invariant tensor structure which rotates the dotted indices, of the form~$R^{\dot{\alpha}\dot{\beta}}_{\dot{\gamma}\dot{\delta}}(p,q)$. In fact, such a factor would be allowed by symmetries and integrability. It is however apparently ruled out by the comparison with perturbative computations which, along with crossing symmetry, would make it non-perturbative~\cite{Borsato:2014hja}.

% \paragraph{Special points.}
% When the momenta are equal, the S matrix is proportional to the identity in each sector. 

% In the left-left sector the S matrix has a pole when $x^+_p = x^-_q$ and a zero when $x_q^- = x^+_p$. At these points the representation becomes reducible. 

% {\color{purple} To finish: discussion of poles and zeroes, bound states. Singlet state.}

\paragraph{Properties.}
The S matrix satisfies the quantum Yang-Baxter equation
% ~\footnote{\color{purple} If you have followed the pre-school lectures then you have seen another form of the Yang-Baxter equation, namely
% \begin{equation}
%  S_{12}(p_2,p_3) S_{23}(p_1,p_3) S_{12}(p_1,p_2) = S_{23}(p_1,p_2) S_{12}(p_1,p_3) S_{23}(p_2,p_3)\,.
% \end{equation}
% This is because another convention for the S matrix was used, with the relation
% \begin{equation}
%     \gen{S} = \Pi S\,,
% \end{equation}
% where $\Pi$ denotes the graded permutation operator. A free theory has $S=1$ but $\gen{S} = \Pi$.  
% }
\begin{equation}
    \gen{S}_{12}(p_1,p_2) \gen{S}_{13}(p_1,p_3) \gen{S}_{23}(p_2,p_3) = \gen{S}_{23}(p_2,p_3) \gen{S}_{13}(p_1,p_3) \gen{S}_{12}(p_1,p_2)\,. 
\end{equation}
The two arguments in the brackets are the two momenta of the particles scattered. The two indices of the S matrix on the other hand denote the spaces in the tensor product on which the S matrix acts. This equation is the hallmark of an integrable model. It is necessary to have consistent factorisation: the two expressions on the left and right hand sides of the equality sign are the two ways in which the $3 \rightarrow 3$ scattering event can be decomposed into a product of $2 \rightarrow 2$ scattering events, and consistency requires both decompositions to give the same result. 
%Graphically, the quantum Yang-Baxter equation takes the form
% \begin{center}
% \begin{tikzpicture}
% \def\xl{4}
% \def\xr{-4}
% \def\xc{0}
% \def\shift{1}
% \def\d{2}
% \def\h{2}
% \def\lw{1.5}
% \def\cA{blue}
% \def\cB{red}
% \def\cC{green}
% \draw[line width=\lw, color=\cA, ->] (\xl-\d,-\h) node[below, color=\cA] {$p$}  -- (\xl+\d,\h);
% \draw[line width=\lw, color=\cB, ->] (\xl-\shift,-\h) node[below, color=\cB] {$q$} -- (\xl-\shift,\h);
% \draw[line width=\lw, color=\cC, ->] (\xl+\d,-\h) node[below, color=\cC] {$r$} -- (\xl-\d,\h);
% \draw[line width=\lw, fill=black] (\xl,0) circle (0.1) node[right] {$\quad \gen{S}_{23}(p,r)$};
% \draw[line width=\lw, fill=black] (\xl-\shift,-1) circle (0.1) node[left] {$\gen{S}_{12}(p,q) \quad $};
% \draw[line width=\lw, fill=black] (\xl-\shift,1) circle (0.1) node[left] {$\gen{S}_{12}(q,r) \quad $};
% \draw (0,0) node[]{$=$};
% \draw[line width=\lw, color=\cA, ->] (\xr-\d,-\h) node[below, color=\cA] {$p$}  -- (\xr+\d,\h);
% \draw[line width=\lw, color=\cB, ->] (\xr+\shift,-\h) node[below, color=\cB] {$q$} -- (\xr+\shift,\h);
% \draw[line width=\lw, color=\cC, ->] (\xr+\d,-\h) node[below, color=\cC] {$r$} -- (\xr-\d,\h);
% \draw[line width=\lw, fill=black] (\xr,0) circle (0.1) node[left] {$\gen{S}_{12}(p,r)\quad$};
% \draw[line width=\lw, fill=black] (\xr+\shift,-1) circle (0.1) node[right] {$\quad \gen{S}_{23}(q,r)  $};
% \draw[line width=\lw, fill=black] (\xr+\shift,1) circle (0.1) node[right] {$\quad \gen{S}_{23}(p,q) $};
% \end{tikzpicture}
% \end{center}
 This equation is satisfied no matter what the dressing phases are, since they appear as the same scalar factor on both sides of the above equation. Other desirable properties of the S matrix will however constrain the dressing phases. In particular, the S matrix also needs to satisfy braiding unitarity
\begin{equation}
    \gen{S}_{12} (p,q) \gen{S}_{21}(q,p)=1\,,
\end{equation}
as well as physical unitarity,
\begin{equation}
    \gen{S}_{12} (p,q) \left( \gen{S}_{12}(p,q) \right)^\dagger =1\,, \qquad p, q \in \mathbb{R}\,.
\end{equation}
This imposes some constraints on the scalar factors, namely
\begin{equation} \begin{gathered}
\sigma_{qp}^{\bullet \bullet} = (\sigma_{pq}^{\bullet \bullet})^* = \frac{1}{\sigma^{\bullet \bullet}_{pq}}\,, \qquad \tilde{\sigma}_{qp}^{\bullet \bullet} = (\tilde{\sigma}_{pq}^{\bullet \bullet})^* = \frac{1}{\tilde{\sigma}_{pq}^{\bullet \bullet}}\,, \qquad \sigma^{\circ \circ}_{qp} = (\sigma^{\circ \circ}_{pq})^* = \frac{1}{\sigma^{\circ \circ}_{pq}}\,, \\
\sigma^{\bullet \circ}_{qp} = (\sigma^{\circ \bullet}_{pq})^* = \frac{1}{\sigma^{\circ \bullet}_{pq}}\,, \qquad \sigma^{\circ \bullet}_{qp} = (\sigma^{\bullet \circ}_{qp})^* = \frac{1}{\sigma^{\bullet \circ}_{pq}}\,.
\end{gathered}
\end{equation}
What do we deduce from these relations? First of all, from the first line it follows that the dressing factors in the massive and massless sectors can be written as exponentials of antisymmetric functions in the momenta, and for real momenta the dressing phases take values in the unit circle. The second line tells us that the two dressing phases in the mixed mass sector (massive-massless and massless-massive) are coupled to each other. Therefore we only have four independent dressing phases $\sigma^{\bullet \bullet}$, $\tilde{\sigma}^{\bullet \bullet}$, $\sigma^{\circ \circ}$ and, for instance, $\sigma^{\bullet \circ}$. More constraints can be imposed on the dressing phases assuming crossing symmetry. Let us see this in the next section.

\subsubsection{Crossing equations for the dressing factors}

The shortening condition \eqref{eq:shortening} involves the square of the Hamiltonian and hence there exist two branches for the dispersion relation. The $H >0$ branch is related to unitary representations. Indeed, recall that the BPS bound $\gen{H} \geq 0$ was inferred using the reality condition on the fermionic generators $\gen{Q}$ and $\gen{S}$. On the other hand, the $H<0$ branch is related to anti-unitary representations, to which we can associate anti-particles. One can go from one representation to the other by using the antipode map. This is a map  $S: \mathcal A \rightarrow \mathcal A$ that acts on the algebra by changing the sign of the Hamiltonian and angular momentum
\begin{equation}
    S(\gen{H}) = - \gen{H}\,, \qquad S(\gen{M}) = - \gen{M}\,.
\end{equation}
For the raising and lowering operators, because of the central extension, on top of changing the sign we also need to add the braiding factor $\gen{U} = e^{i \gen{p}}$,
\begin{equation} \begin{aligned}
    S(\gen{Q}_{\L}{}^A) &= - \gen{U}^{-1} \gen{Q}_{\L}{}^{A}\,, &\qquad S(\gen{Q}_{\R A}) &= - \gen{U}^{-1} \gen{Q}_{\R A}\,, \\
    S(\gen{S}_{\L A}) &=- \gen{U}^{+1} \gen{S}_{\L A}\,, &\qquad S(\gen{S}_\R{}^A) &= - \gen{U}^{+1} \gen{S}_\R{}^A.
    \end{aligned}
\end{equation}
Consistency with the commutation relation implies that
\begin{equation}
    S(\gen{C}) = - \gen{U}^2 \gen{C} = - e^{i \gen{p}} \gen{C}\,.
\end{equation}
On the momentum we have $S(\gen{p}) = - \gen{p}$.
For the $\alg{su}(2)$ generators we have simply
\begin{equation}
    S(\gen{J}_{\bullet \dot{A}}{}^{\dot{B}}) = - \gen{J}_{\bullet \dot{B}}{}^{\dot{A}}\,, \qquad S(\gen{J}_{\circ A}{}^B) = - \gen{J}_{\circ B}{}^A\,.
\end{equation}
The antipode map can be used to define the charge conjugation matrix at the level of the representations,
\begin{equation}
    S(\gen{J}) \cong \mathcal C^{-1} \bar{\gen{J}}^{st} \mathcal C\,,
\end{equation}
where the supertransposition is defined on a matrix realisation as $M_{jk}^{st} = (-1)^{(F_j+1)F_k} M_{kj}$. Solving the above equation for all the generators of the algebra then gives both the explicit expression of the charge conjugation matrix and the antipode representation parameters (that we denote with a bar). In the massive sectors, the charge conjugation matrix exchanges left and right particles. Moreover, it exchanges highest and lowest weight under $\alg{su}(2)_\bullet$. It takes the form
\begin{equation} \begin{aligned}
    \mathcal C \ket{Y} &= \ket{\bar{Y}}\,, &\qquad \mathcal C \ket{\Psi^1} &= -i \ket{\bar{\Psi}^2}\,, &\qquad \mathcal C \ket{\Psi^2} &= +i \ket{\bar{\Psi}^1}\,, &\qquad \mathcal C \ket{Z} &= \ket{\bar{Z}}\,, \\
    \mathcal C \ket{\bar{Y}} &= \ket{Y}\,, &\qquad \mathcal C \ket{\bar{\Psi}^1} &= +i \ket{\Psi^2}\,, &\qquad \mathcal C \ket{\bar{\Psi}^2} &= -i \ket{\Psi^1}\,, &\qquad \mathcal C \ket{\bar{Z}} &= \ket{Z}\,.
    \end{aligned}
\end{equation}
On the massless bosons it acts as
\begin{equation}
    \mathcal C \ket{T^{11}} = \ket{T^{22}}\,, \qquad \mathcal C \ket{T^{12}} = - \ket{T^{21}}\,, \qquad \mathcal C \ket{T^{21}} = -\ket{T^{12}}\,, \qquad \mathcal C \ket{T^{22}} = \ket{T^{11}}\,.
\end{equation}
For the massless fermions we have that
\begin{equation} \begin{aligned}
    \mathcal C \ket{\bar{\chi}^1} &= -i c(p) \ket{\chi^2} \,, &\qquad \mathcal C \ket{\chi^1} &= + i c(p) \ket{\bar{\chi}^2}\,, \\
    \mathcal C \ket{\bar{\chi}^2} &= +i c(p) \ket{\chi^1}\,, &\qquad \mathcal C \ket{\chi^2} &= -i c(p) \ket{\bar{\chi}^1}\,,
    \end{aligned}
\end{equation}
where $c(p) = \frac{a_\L(p)}{b_\R(p)}$. For the antipode representation parameters, there is a simple transformation of the Zhukovski variables, which is actually easy to write in the general mixed-flux case 
\begin{equation}
    x^\pm_\L \rightarrow \bar{x}^\pm_\L = \frac{1}{x^\pm_\R}\,, \qquad x^\pm_\R \rightarrow \bar{x}^\pm_\R = \frac{1}{x^\pm_\R}, \qquad x_\circ^\pm \rightarrow \bar{x}_\circ^{\pm} = \frac{1}{x_\circ^\pm}\,.
\end{equation}
Notice that $\bar{\bar{x}}_\L^\pm = x^\pm_\L$ and $\bar{\bar{x}}_\R^\pm = x^\pm_\R$, so that crossing twice gives back the same expression.
The quantities $\eta_\L$ and $\eta_\R$ involve a square root and are hence not meromorphic functions on the complex plane. As a consequence, we need to resolve some ambiguity when doing the crossing. One choice is such that
\begin{equation}
    \bar{\eta}_\L = \frac{i}{x^+_\R} \eta_\R\,, \qquad \bar{\eta}_\R = \frac{i}{x^+_\L} \eta_\L\,.
\end{equation}

In the context of relativistic quantum field theories, the crossing equation is a constraint arising from requiring that a given scattering process involving particles of momenta $p_j$ can equivalently be seen as a scattering process with corresponding anti-particles of momenta $-p_j$. Here we assume that a similar constraint should hold for non-relativistic theories.
The crossing equations read
\begin{equation} \begin{aligned}
    (\mathcal C^{-1} \otimes 1) \gen{S}^{st_1}(\bar{x}_1, x_2) (\mathcal C \otimes 1) \gen{S}(x_1,x_2) &= 1 \otimes 1\,, \\
    (1 \otimes \mathcal C^{-1}) \gen{S}^{st_2}(x_1, \bar{x}_2) (1 \otimes \mathcal C) \gen{S}(x_1,x_2) &= 1 \otimes 1\,,
    \end{aligned}
\end{equation}
where $^{st_n}$ denotes the supertranspose in the $n$-th factor of the tensor product. From the S matrix structure it follows that the left hand side is automatically proportional to the identity, and one is left with equations for the dressing factors.

Let us now specialise the form of these equations to the pure-RR dressing factors discussed above. We have
\begin{equation}
\begin{aligned}
\left(\sigma^{\bullet\bullet} (x_1^\pm,x_2^\pm)\right)^2\left(\tilde\sigma^{\bullet\bullet} (\bar{x}_1^\pm,x_2^\pm)\right)^2&=
\left(\frac{x_2^-}{x_2^+}\right)^2
\frac{(x_1^- - x_2^+)^2}{(x_1^- - x_2^-)(x_1^+ - x_2^+)}\frac{1-\frac{1}{x_1^-x_2^+}}{1-\frac{1}{x_1^+x_2^-}}\,,\\
\left(\sigma^{\bullet\bullet} (\bar{x}_1^\pm,x_2^\pm)\right)^2\left(\tilde\sigma^{\bullet\bullet} (x_1^\pm,x_2^\pm)\right)^2&=
\left(\frac{x_2^-}{x_2^+}\right)^2
\frac{\left(1-\frac{1}{x^+_1x^+_2}\right)\left(1-\frac{1}{x^-_1x^-_2}\right)}{\left(1-\frac{1}{x^+_1x^-_2}\right)^2}\frac{x_1^--x_2^+}{x_1^+-x_2^-}\,,
\end{aligned}
\end{equation}
for the massive sector,
\begin{equation}
\begin{aligned}
\left(\sigma^{\bullet\circ} (x_1^\pm,x_2)\right)^2\left(\sigma^{\bullet\circ} (\bar{x}_1^\pm,x_2)\right)^2&={%\color{red}
\frac{1}{(x_2)^4}}\frac{f(x_1^+,x_2)}{f(x_1^-,x_2)}\,,\\
\left(\sigma^{\circ\bullet} (x_1,x_2^\pm)\right)^2\left(\sigma^{\circ\bullet} (\bar{x}_1,x_2^\pm)\right)^2&=\frac{f(x_1,x_2^+)}{f(x_1,x_2^-)}\,,
\end{aligned}
\end{equation}
for the mixed-mass sector and
\begin{equation}
\begin{aligned}
    &\big(\sigma^{\circ\circ} (x_1,x_2)\big)^2\big(\sigma^{\circ\circ} (\bar{x}_1,x_2)\big)^2&=&-f(x_1,x_2)^2\,,\\
    &\big(\widetilde{\sigma}^{\circ\circ} (x_1,x_2)\big)^2\big(\widetilde{\sigma}^{\circ\circ} (\bar{x}_1,x_2)\big)^2&=&-f(x_1,x_2)^2\,,
\end{aligned}
\end{equation}
for the massless sector. The function $f(x,y)  = i\frac{1- xy}{x-y}$.

\begin{centering}
\begin{tcolorbox}
\begin{exercise}
  Rederive the crossing equations by requiring that the S matrix acts trivially on the singlet state.
\end{exercise}
\end{tcolorbox}
\end{centering}

Solving these equations is far from straightforward. Even if we have determined that $\bar{\bar{x}}^=x^{\pm}$, it is easy to check that it must be
\begin{equation}
    \sigma^{\bullet\bullet} (\bar{\bar{x}}_1^\pm,x_2^\pm)\neq\sigma^{\bullet\bullet} (x_1^\pm,x_2^\pm)\,,
\end{equation}
and similarly for the other dressing factors. In other words, the dressing factors must have cuts on the $x^\pm$ planes.%
\footnote{This statement is reminiscent of the fact that, for relativistic models with $H=m\cosh\theta$ and $p=m\sinh\theta$, the matrix part of the S-matrix is $2\pi i$-periodic, but the dressing factor is not.}
It is crucial to identify a physical region on the $x^\pm$ plane, and make suitable assumptions on the on the analytic continuation to the ``crossed'' region as well as to other regions of the momentum space. This is currently the main challenge for the mixed-flux case. However, in the pure-RR case, a proposal for the dressing factors has been recently put forward~\cite{Frolov:2021fmj}, correcting a previous guess~\cite{Borsato:2013hoa}. This proposal is what allowed to construct and study the mirror model which is crucial to extract the finite-volume spectrum of the model, as we will see in section~\ref{ana:sec}.

\subsection{Summary and concluding remarks}

In this section we showed how to compute the perturbative and exact S-matrix describing the scattering of excitations on the two-dimensional worldsheet of the string. To achieve this we fixed uniform lightcone gauge so that the worldsheet becomes a cylinder, and took the decompactification limit to have well-defined scattering states. Applying this procedure to strings propagating in an $AdS_3 \times S^3 \times T^4$ background we found that the original $\alg{psu}(1,1|2)_\L \oplus \alg{psu}(1,1|2)_\R$ isometry algebra of the sigma model describing the curved part of the geometry gets broken to a centrally-extended $\mathcal A = \left[ \alg{su}(1|1)_\L \oplus \alg{su}(1|1)_\R \right]_{c.e}^{\oplus 2}$ algebra upon lightcone gauge fixing. The excitations then transform in four different respresentations of $\mathcal A$: two massive respresentations (left and right) and two massless ones (chiral and anti-chiral). Requiring that the scattering respects the symmetry algebra $\mathcal A$ then fixes the S-matrix up to the dressing phases. 

Finding the exact S-matrix is an important step to solve the spectrum of an integrable model. In section \ref{ana:sec} we will see how to compute physical quantities through a technique called the Thermodynamic Bethe Ansatz. Before that, let us take stock of the state of the art and outline some open problems.

\paragraph{Pure RR backgrounds.}
The case of RR backgrounds is the best understood. In that case, the integrability construction of the S-matrix was initiated in~\cite{Borsato:2012ud,Borsato:2013qpa} and completed in~\cite{Borsato:2014hja,Frolov:2021zyc} up to the dressing factors. The latter have been recently proposed in~\cite{Frolov:2021fmj}. The complete control over the S-matrix allowed to derive the equations which describe the spectrum, and to quantitatively investigate the dimensions of string states in the $k=0, h\ll1$ regime (where supergravity or semiclassical techniques cannot be used). It is worth noting that currently there is no other worldsheet approach to tackle this corner of the parameter space. Little is known about the dual theory; a proposal was given in~\cite{OhlssonSax:2014jtq}, but it is not clear if it matches with the worldsheet integrability description.

\paragraph{Pure NSNS and relation to $T\bar{T}$ deformations.} From the mixed-flux S-matrix it is formally easy to obtain the pure RR S-matrix by sending $k \rightarrow 0$, at least in the matrix part.%
\footnote{The dressing factor might change in a more subtle way, reflecting that in principle only integer values of~$k$ are allowed.}
The pure NSNS limit, when the amount of RR flux $h \rightarrow 0$ is however not as straightforward. This is because in that limit the dispersion relation becomes chiral --- \textit{i.e.}, particles move at the ``speed of light'' $\pm k/(2\pi)$. Taking the limit in the S-matrix then correctly describes the head-on scattering processes (when the two particles move in opposite direction), but it also gives a non-trivial result for the collinear scattering events (when the two particles move in the same direction). The latter should not be present in a correct analysis. The S-matrix for strings in pure NSNS background is in fact given by a simple CDD factor~\cite{Baggio:2018gct}. This simple structure takes its origin from the fact that in the pure NSNS case the worldsheet Hamiltonian in uniform lightcone gauge is given by an integrable $T\bar{T}$ deformation~\cite{Cavaglia:2016oda} of a theory of free bosons.

\paragraph{Mixed-flux backgrounds.}
In this case, the integrability construction is still incomplete. The S~matrix was fixed in~\cite{Lloyd:2014bsa} (building on~\cite{Hoare:2013pma,Hoare:2013lja}), up to the dressing factors. The latter have not yet been fixed, due to the rather intricate kinematics; the crossing equations have however been solved in a relativistic limit~\cite{Frolov:2023lwd}. It is worth noting that, despite these issues, the integrability approach appears to be the most promising route to understand mixed-flux backgrounds; at least currently, it is unclear how to use the hybrid approach to compute the spectrum, even at $h\ll1$. A different approach would be to start from the holographic duals, and in particular from the perturbed symmetric-product orbifold theory corresponding to $k=1$ and $h\ll1$. There have been many efforts in this direction, in particular in~\cite{Pakman:2009mi} (in a spin-chain language) and in~\cite{David:2010yg} (for the purposes of bootstrapping an S-matrix along the lines discussed above). The latter idea was recently revisited in~\cite{Gaberdiel:2023lco}, where the symmetry algebra of the deformed symmetric-product orbifold was worked out at first order in conformal perturbation theory. As explained in~\cite{Frolov:2023pjw}, the result for the algebra and representations (as well as the matrix part of the S~matrix which they determine) precisely match with the $k=1,h\ll1$ expansion of the one from~\cite{Lloyd:2014bsa}.

\paragraph{Other AdS${}_{\mathbf{3}}$ spaces.}
This analysis can also be applied to $AdS_3 \times S^3 \times S^3 \times S^1$. In fact, historically the investigation of integrability initially focused on this background, both on the classical~\cite{Babichenko:2009dk} and quantum side~\cite{Borsato:2012ud,Borsato:2012ss}. The main difference is that the geodesics used for gauge fixing is at most $\tfrac{1}{4}$-BPS (see~\cite{Dei:2018yth} for a discussion of the various gauge-fixings), so that the residual algebra is only $\psu(1|1)^{\oplus2}_{\text{c.e.}}$, that is half of that of $AdS_3\times S^3\times T^4$ case (in the language used in this section, it is $\mathcal{B}$ and not~$\mathcal{A}$). As a result, the short representations are two-dimensional, and there are eight of them, with dispersion~\cite{Borsato:2015mma}
\begin{equation}
    H(p)=\sqrt{\left(m+\frac{k}{2\pi}p\right)^2+4h^2\sin^2\left(\frac{p}{2}\right)},\qquad m=\pm0,\, \pm \alpha,\, \pm (1-\alpha),\, \pm1\,,
\end{equation}
where the $\pm0$ denotes that there are two representations with $m=0$. The parameter~$\alpha$ characterises the background --- the radii of $AdS$ and of the spheres satisfy $1/(R_{AdS})^2=1/(R_{S})^2+1/(R_{S}')^2$, so that $\alpha=(R_{AdS}/R_{S})^2$ and $1-\alpha=(R_{AdS}'/R_{S})^2$; it is also the parameter that appears in the superisometry algebra, which is given by two copies of the exceptional Lie superalgebra~$\mathfrak{d}(2,1;\alpha)_{\L}\oplus\mathfrak{d}(2,1;\alpha)_{\R}$. To obtain the~$T^4$ one formally takes $\alpha\to1$ or $\alpha\to0$, whereby the algebra contracts.
The matrix part of the the S~matrix was fixed in~\cite{Borsato:2015mma} from the symmetries, but currently the dressing factor is not known --- due to more complicated mass spectrum, it is more involved than the~$T^4$ case.

\paragraph{Integrable deformations.}
Strings on $AdS_3 \times S^3 \times T^4$ admit a very rich space of integrable deformations. We have already seen that it is possible to add a WZ term to the action while preserving integrability, giving rise to a model with mixed RR and NSNS flux. One can also deform the symmetry algebra into a quantum group $\mathcal U_{q_\L}(\alg{psu}(1,1|2)) \oplus \mathcal U_{q_\R}(\alg{psu}(1,1|2))$, where $q_\L$ and $q_\R$ are two real deformation parameters, which results in a theory where strings are propagating in a deformed $AdS_3 \times S^3 \times T^4$ background. In the pure RR case an exact S-matrix describing the scattering in the massive sector was conjectured in~\cite{Hoare:2014oua} and matched with perturbative calculations in~\cite{Seibold:2021lju}. Also in that case it is possible to add a WZ term while preserving integrability~\cite{Delduc:2018xug}. The tree-level S-matrix for strings propagating in such a three-parameter deformed background was computed in~\cite{Bocconcello:2020qkt} but the exact S-matrix is yet to be found. More recently, an integrable elliptic deformation of the $AdS_3 \times S^3 \times T^4$ string with three deformation parameters was constructed, and its perturbative S-matrix was obtained to tree-level (in the bosonic truncation)~\cite{Hoare:2023zti}. The theory was also embedded into supergravity, but whether the theory remains integrable upon including the RR-fluxes remains an open question. The symmetries of the deformed theory have also not been analysed yet, and the exact S-matrix remains unknown.

\newpage

\section{String Spectrum from the Bethe ansatz}\label{ana:sec}

%\subsection{Introduction}\label{ana:subsec:introduction}

The presence of integrability in a given model is a remarkable advantage. It provides mechanisms to address problems whose solution would otherwise be very difficult or even unattainable.  In sections \ref{s:sibylle}-\ref{s:fiona} important consequences and simplifications due to this property were presented in the context of AdS/CFT, more specifically strings in AdS$ _5 $ and AdS$ _3 $  backgrounds. 

This section continues on this path by focusing on one more aspect of integrability: the computation of the string spectrum by using the so-called \textbf{Thermodynamic Bethe ansatz} (TBA) method.

The TBA is a technique to unravel the thermodynamics of integrable quantum theories. It was first developed \cite{Yang:1968rm} to understand the equilibrium thermodynamics of a bosonic system with repulsive delta interaction in a one-dimensional periodic box. The generalisation to 2D integrable relativistic field theories was presented in \cite{Zamolodchikov:1989cf}, where the idea of what is now known as mirror theory was introduced.  Although the standard procedure leads to the computation of only the ground state energy, excited states can be obtained by analytic continuation \cite{Dorey:1996re}.

The Thermodynamic Bethe ansatz is an effective procedure to construct the free-energy $f$ of an integrable theory. For this computation, the information about the density of particles of each type that contribute to the entropy of the system comes from the Bethe ansatz equations. The relation between the energy $e$ and the density of particles of each type is then obtained by computing the free energy stationary ``points'' (requiring $\delta f$=0). This technique generates a set of coupled nonlinear integral equations, one equation for each type of particle. In most of the cases this system of equations is solved numerically. For ealier reviews on the TBA see \cite{vanTongeren:2016hhc} and the book \cite{essler2005one}. 

In the context of AdS/CFT, the TBA was remarkably successful in the computation of the spectrum of AdS$_5\times$S$^5$ superstring \cite{Beisert:2005tm,Ambjorn:2005wa,Janik:2006dc,Beisert:2006ez,Arutyunov:2007tc,Gromov:2009tv,Bombardelli:2009ns,Arutyunov:2009kf,Arutyunov:2009zu,Arutyunov:2009ur,Gromov:2009bc,Arutyunov:2009iq,Arutyunov:2009mi,Arutyunov:2009ux,Bajnok:2009vm,Arutyunov:2009ax,Arutyunov:2010gb,Cavaglia:2010nm,Gromov:2010km,Ahn:2011xq} (see also the review \cite{Beisert:2010jr}, especially chapter \cite{Bajnok:2010ke}). Furthermore, the TBA was also computed in the quantum deformed case \cite{Arutyunov:2012zt,Arutyunov:2012ai}. In \cite{Gromov:2013pga}, the so-called quantum spectral curve (QSC), consisting of a simplified and compact set of equations, was introduced (for a review see \cite{Gromov:2017blm}). Recently, a QSC solver was also constructed \cite{Gromov:2023hzc}. Progress was also done in AdS$_4\times$CP$^{3}$ with the TBA in \cite{Bombardelli:2009xz} and \cite{Gromov:2009at} for example (for a review see \cite{Klose:2010ki}) and the QSC in \cite{Bombardelli:2017vhk}.

Inspired by the success of the Thermodynamic Bethe ansatz in higher dimensional AdS backgrounds, this programme has been extended to   $\text{AdS}_3/\text{CFT}_2$ \cite{Borsato:2013hoa,Borsato:2016xns,Bombardelli:2018jkj,Baggio:2018gct,Dei:2018mfl,Sfondrini:2020ovj,Frolov:2021fmj,Frolov:2021bwp,Seibold:2022mgg,Frolov:2023wji,Frolov:2023lwd,Baglioni:2023zsf}. In particular, this method was responsible for the computation of the finite-volume (and zero-temperature) ground state energy in both pure NS-NS \cite{Baggio:2018gct,Dei:2018mfl,Sfondrini:2020ovj} and more recently, on pure-Ramond-Ramond \cite{Frolov:2021fmj,Frolov:2021bwp,Frolov:2023wji} backgrounds. Regarding the mixed-flux case, the S-matrix of the string model is understood up to dressing factors~\cite{Hoare:2013pma,Lloyd:2014bsa}. More recently, the complete S-matrix, including the dressing factors, has been fixed in a special relativistic limit~\cite{Frolov:2023lwd}. The mirror S-matrix is expected to be quite unusual for the mixed-flux model, as it can be seen already at tree-level~\cite{Baglioni:2023zsf}. As we will see, the computation of the Thermodynamic Bethe ansatz for the full mixed-flux theory, remains an open question.  For the pure-RR theory, in addition to the TBA a Quantum spectral curve was proposed by two different research groups  \cite{Cavaglia:2021eqr,Ekhammar:2021pys,Cavaglia:2022xld}. But contrary to AdS$_5\times$S$^5$, these were not constructed from the TBA. They were conjectured using a bootstrap approach. The compatibility of these QSC proposals with the recently derived TBA remains to be checked. 

Before explaining the construction of the Thermodynamic Bethe ansatz for AdS$_3$,  we will introduce several basic concepts and definitions on a simpler model: the XXX Heisenberg spin chain. This will allow us, when finally presenting the AdS$_3$ case,  to focus only on the important differences and new features, hopefully making the presentation clearer.

The first step on the TBA construction is to have an exact (or asymptotic) Bethe ansatz (BA). The BA is a technique created with the objective of diagonalize the conserved charges of an integrable model. In fact, integrable models are characterized by the presence of an infinite number of hidden symmetries which generate infinitely many commuting conserved charges $ \mathbb{Q}_j $
\begin{equation*}
	\left[\mathbb{Q}_j,\mathbb{Q}_k\right]=0, \quad j,k=1,2,...
\end{equation*}
The direct diagonalization of these conserved charges is  in principle a complicated task, except on three situations: when the system has very small volume, when it contains very few particles, or both. This is true even for systems whose local Hilbert space is very small (like the $\alg{su}(2)$ XXX spin-1/2 chain). But the higher the Hilbert space, more difficult the direct diagonalization becomes. The BA provides an alternative to perform such a diagonalization.

The Coordinate Bethe ansatz \cite{Bethe:1931hc}, (see also \cite{Reffertln2014} for a review), allows one to write an exact expression for the energy in integrable models. More ``advanced'' versions, like the Algebraic Bethe ansatz (for reviews applied to standard spin chains see \cite{Faddeev:1996iy,Nepomechie:1998jf,Retore:2021wwh}; and applied to AdS$_3$ see \cite{Seibold:2022mgg}) provide ways to diagonalize all the conserved charges and are easier (than the Coordinate one) to apply for more complicated models. 

Independently of the type of BA used, its main advantage is the construction of algebraic expressions for the eigenvalues in terms of quantities called Bethe roots. The importance of such explicit formulas lies on the possibility of taking limits, especially to the case where the number of particles and the size of the system are very large. As we will see, the TBA is obtained through one of these limits.  

The BA is exact in the case of spin chains. For 2D field theories some complications arise, especially related to the so-called wrapping effects. In the context of AdS/CFT the BA is usually a good description of the theory only asymptotically. In such case, the BA is called Asymptotic Bethe ansatz and the Bethe equations are called Bethe-Yang equations \cite{Yang:1968rm}. We expect the details will be clear by the end of this section.

\paragraph{Plan:} The remaining of this section will be divided in three main parts. First, in subsection \ref{ana:subsec:TBAforXXX} we explain the main concepts in the TBA, introducing them with the XXX Heisenberg spin chain as example. Next, in subsection \ref{ana:subsec:TBAforAdS3} we focus on AdS$_3$/CFT$_2$. In particular, in this part we introduce the concept of mirror theory and use it to construct the ground-state energy for finite-volume (and zero temperature) for the case of pure-Ramond-Ramond background. Finally, subsection \ref{ana:subsubsec:summaryTBAAdS3} contains a brief summary of the TBA-related open problems in the context of AdS$_3$/CFT$_2$.

Given that this method involves several technical details, in subsection \ref{ana:subsec:TBAforXXX} in addition to explain the main ideas,  we will also try to be as explicit as possible with the calculations including examples. We also present a summary of the main ideas in subsection \ref{ana:subsubsec:summaryTBAXXX}, which the reader can use to more easily keep track of the steps and the logic involved.

\subsection{Thermodynamic Bethe ansatz for the XXX model}\label{ana:subsec:TBAforXXX}

The thermodynamic limit corresponds to consider large volume and large number of particles while keeping their ratio finite
\begin{equation}
	L\rightarrow \infty,\quad N\rightarrow \infty, \quad \frac{L}{N}\rightarrow \text{ fixed}.
	\label{eq:thermodynamiclimitXXX}
\end{equation}
In this regime, fluctuations are not large enough to be relevant and therefore thermodynamics allows the computation of several quantities including the Gibbs free energy and the chemical potential. 

The TBA consists in taking the thermodynamic limit in the Bethe ansatz and will result in integral equations. This technique plays an important role in computing the string spectrum in AdS$_3$/CFT$_2$. We now proceed to explain each step of this technique using the XXX Heisenberg spin chain as example. 

\subsubsection{The Bethe ansatz}\label{ana:subsubsec:IntroductionTBAXXX}

The XXX spin-$\frac{1}{2}$ chain Hamiltonian is given by
\begin{equation}
    \mathbf{H}= -\frac{J}{2} \sum_{j=1}^{L}\left(\vec{\sigma}_j.\vec{\sigma}_{j+1}-\mathbb{1}_{j,j+1}\right), \qquad \vec{\sigma}_{L+1}\equiv \vec{\sigma}_1, \qquad \vec{\sigma}=\{\sigma^x,\sigma^y,\sigma^z\},
    \label{eq:HamiltonianXXX}
\end{equation}
where $\sigma^a$, $a=1,2,3$ are the Pauli matrices\footnote{The representation we used is given by
\begin{equation}
    \sigma^1=\begin{pmatrix}
        0 & 1\\
        1 & 0
    \end{pmatrix}, \qquad \sigma^2=\begin{pmatrix}
        0 & -i\\
        i & 0
    \end{pmatrix} \qquad \text{and} \qquad \sigma^3=\begin{pmatrix}
        1 & 0\\
        0 & -1
    \end{pmatrix}.
\end{equation}} and $\mathbb{1}$ is the identity matrix. This Hamiltonian acts on $L$ copies of $\mathbb{C}^2$. In other words, in each site of the chain we can put a spin up or a spin down. For $J>0$ the model is ferromagnetic, so the vacuum has all spins aligned, while for $J<0$ the model is antiferromagnetic, and the vacuum has spins maximally anti-aligned. Importantly, $\mathbf{H}$ has a total $\alg{su}(2)$ symmetry
\begin{equation}
    [\gen{H}, \gen{J}^a] =0, \qquad \gen{J}^a = \sum_{j=1}^L \sigma^a_j.
\end{equation}
In particular, the Hamiltonian commutes with the Cartan generator $\gen{J}^3$ of this algebra. As a consequence, the number of overturned spins is conserved and the model has closed sectors.
\begin{itemize}
    \item $\mathbf{N=0:}$ Let us assume that the vacuum is the case with all spin-up
\begin{equation}
    |0\rangle=|\uparrow\uparrow\uparrow...\uparrow\rangle. 
    \label{eq:vac}
\end{equation}
We think of this as a state representing no excitations and write this case as $N=0$. The state $|0\rangle$ is itself an eigenstate of the Hamiltonian (with energy $H=0$). This sector has size one, since there is only one such a state. 
 \item $\mathbf{N=1:}$ The next sector is the one made of 1-particle states. We represent the state with a spin down in the $n_1$-th position as
 \begin{equation}
     |n_1\rangle= |\uparrow\uparrow..\uparrow \underbrace{\downarrow}_{\text{$n_1$-th  site}}\uparrow...\uparrow\rangle. 
 \end{equation}
 If we now act on this state with the Hamiltonian, we obtain a combination of 1-particles states. This sector has size $L$, since there are $L$ different sites that the flipped spin could occupy. Using the periodicity of the chain one can show that the eigenstates of $\mathbf{H}$ are given by 
 \begin{equation} 
     |\psi_p\rangle=\sum_{n_1=1}^Le^{ipn_1}|n_1\rangle \quad \text{if $p$ satisfies} \quad e^{ipL}=1,
 \end{equation}
 with energy $H=J(1-\cos p)$. This is interpreted as a pseudo-particle moving in the chain with quantised momentum $p$. Such excitation is usually called a \textbf{magnon}.
 \item $\mathbf{N=2:}$ Two-particles states, with spins flipped in sites $n_1$ and $n_2$ (with $n_2>n_1$) are represented by $|n_1,n_2\rangle$. There are $L!/[2!(L-2)!]$ possible such a states. 
 When we had only one excitation, the only thing that a magnon could do was to move in the chain. However, if we have two magnons, in addition to move, they can also scatter each other. So, in this case the two-particles eigenstates are written as
 \begin{equation} |\psi_{p_1,p_2}\rangle=\sum_{n_2>n_1}\left(e^{ip_1n_1+ip_2n_2}+S(p_1,p_2)e^{ip_1n_2+ip_2n_1}\right)|n_1,n_2\rangle.
 \label{eq:eigenstateN2}
 \end{equation}
 The corresponding energy and momentum are given by 
 \begin{equation}
 H=2J\sum_{k=1}^{2}\frac{1}{u_k^2+1},    \qquad P=\sum_{k=1}^2p_k, \qquad p_k=-i\log\left(\frac{u_k+i}{u_k-i}\right).
 \label{eq:energyN2}
 \end{equation}
 But \eqref{eq:eigenstateN2} and \eqref{eq:energyN2} are only the eigenstates and eigenvalues, respectively, if the so-called \textbf{Bethe roots} $u_k$ satisfy the following equations
 \begin{equation}
     e^{ip_1L}S(u_1,u_2)=1, \qquad e^{ip_2L}S(u_2,u_1)=1.
 \end{equation}
 These are called \textbf{Bethe-equations}, with the S-matrix $S(u_1,u_2)$ given by
 \begin{equation}
     S(u_1,u_2)=\frac{u_1-u_2-2i}{u_1-u_2+2i}.
 \end{equation}
 These expressions are found by using the periodicity of the chain and of the wave-functions. 
\item \textbf{generic} $\mathbf{N:}$ By continuing this procedure one finds that for a case with $N$-excitations, the energy and momentum are given by
\begin{equation}
    H=2J\sum_{k=1}^{N}\frac{1}{u_k^2+1},    \qquad P=\sum_{k=1}^Np_k, \qquad p_k=-i\log\left(\frac{u_k+i}{u_k-i}\right),
    \label{eq:energyN}
\end{equation}
as long as $\{u_k\}$ and $\{p_j\}$ satisfy the Bethe equations
\begin{equation}
	\left(\frac{u_k-i}{ u_k+i}\right)^L=\prod_{j\neq k}^{N}\frac{u_k-u_j-2i}{u_k-u_j+2i}, \quad \text{for} \quad k=1,...,N.
	\label{eq:BetheEquations}
\end{equation}
This sector contains $L!/[N!(L-N)!]$ eigenstates. 
\end{itemize}
The procedure above is called \textbf{Coordinate Bethe ansatz (CBA)} \cite{Bethe:1931hc}  and is very effective in diagonalizing several different integrable Hamiltonians. For more details on the procedure see \cite{Reffertln2014}.  

We see in the Bethe equations \eqref{eq:BetheEquations} a very characteristic feature of integrability: the fact that the $N\rightarrow N$ particles scattering decomposes in a sequence of $2\rightarrow2$ particles scatterings. This property already appeared in this review in section \ref{fiona:subsec:fullSmatrix} and as discussed there is directly related to the Yang-Baxter equation.

\paragraph{Values of N.} This model has $\alg{su}(2)$ symmetry and for this reason there is a symmetry $ N \leftrightarrow L-N$. As a consequence we don't need to solve the Bethe equations for all $ N=1,...,L $ we can instead solve them only for 

\begin{equation}
N=\begin{cases}
1,...,\frac{L}{2} & \text{ for even } L\\
1,...,\frac{L-1}{2} & \text{ for odd } L
\end{cases},
\label{eq:valuesofN}
\end{equation}
and we will find all the eigenvalues.

\paragraph{On alternatives.} The Coordinate Bethe ansatz is an effective and intuitive way to think about this problem. However, there exist alternative methods that, despite having a less intuitive physical interpretation, are more powerful. In particular, for more complicated spin chains and if we are interested in not only the Hamiltonian, but also higher conserved charges, the so-called Algebraic Bethe ansatz is a more suitable (and easier to apply) technique (for reviews see \cite{Faddeev:1996iy,Nepomechie:1998jf,Retore:2021wwh}). This involves constructing an object called transfer matrix $t(u)$, which is the generating function of infinitely many conserved charges
	\begin{equation}
	\mathbb{Q}_{i}\propto\frac{d^{(i-1)}\left(\text{log}\, t(u)\right)}{du^{i-1}}\Big|_{u=i},\quad  i=1,2,...
	\label{eq:conservedcharges}
\end{equation}
In this tower of conserved charges, $ \mathbb{Q}_{1} $ is the momentum and $ \mathbb{Q}_{2} $ corresponds to the Hamiltonian. The method involves the application of the so-called Quantum Inverse Scattering method \cite{Baxter:1982zz,Korepin:1993kvr,Faddeev:1996iy}. In particular, one defines certain creation and annihilation operators, which are used to generate all the states from a (pseudo-) vacuum. The Yang-Baxter equation is again a fundamental piece in this procedure.

By applying the algebraic Bethe ansatz method we find that the eigenvalues of the transfer matrix are described by
\begin{equation}
	\Lambda(u)\equiv\Lambda(u,\{u_1,...,u_N\})=(u+i)^L\prod_{j=1}^{N}\frac{u-u_j-2i}{u-u_j}+(u-i)^L\prod_{j=1}^{N}\frac{u-u_j+2i}{u-u_j},
	\label{eq:eigenvalue}
\end{equation}
as long as $ \{u_k\} $, known as Bethe roots, satisfy the Bethe equations \eqref{eq:BetheEquations}. In this way, the eigenvalues of any of the conserved charges \eqref{eq:conservedcharges} can be obtained by the logarithmic derivative of $\Lambda(u)$. The total momentum for a spin chain with $ N $ magnons and $L$ sites is given for example by
\begin{equation}
	P=-i\log\left((2i)^{-L}\Lambda(i)\right)=-i\sum_{k=1}^{N}\log\left(\frac{u_k+i}{u_k-i}\right),
 \label{eq:momentumBA}
\end{equation}
while its energy is
\begin{align}
	H(\{u_k\})=- iJ\frac{d}{du}\log (\Lambda(u))\Big|_{u=i}+\frac{JL}{2}=2J\sum_{k=1}^{N}\frac{1}{u_k^2+1}.
	\label{eq:energyBA}
\end{align}
Both quantities match with the ones obtained via CBA.

\paragraph{On solving the Bethe equations.} In order to compute the momentum and the energy in the XXX model we need to solve the Bethe equations \eqref{eq:BetheEquations} and then plug the Bethe roots in equations \eqref{eq:momentumBA} and \eqref{eq:energyBA}. 
The Bethe equations are easy to solve numerically for small number of sites and small number of magnons. For small number of sites most of the solutions are real, and therefore relatively easy to find.
As the number of sites and magnons increase, however, not only the Bethe equations become more numerous and of higher polynomial degree, but more and more of the Bethe roots are actually complex.

Just to give an idea of how quickly the complex roots become important, it is useful to define 
\begin{equation}
	\chi_N=\frac{\# \text{  of different eigenvalues with at least two complex Bethe roots for a given } N}{ \# \text{ of different eigenvalues for a given } N }.
\end{equation}
Using $\chi_N$ we can see in Table \ref{table:XXXspectrum} that the percentage of states described by complex Bethe roots increases as $ L $ and $ N $ increase.

\begin{table}[ht]
	\setlength{\tabcolsep}{0.5em} % for the horizontal padding
	{\renewcommand{\arraystretch}{0.4}
		{\renewcommand{\arraystretch}{0.4}
			\begin{center}
				\begin{tabular}{|c|c|c|c|c|}
					\hline
					& & & & \\
					$ \mathbf{L} $ & \textbf{degeneracies} & $ \mathbf{N} $'s & $ \mathbf{\chi_N} $ & $ \% $\\
					& & & & \\ \hline
					& & & & \\
					1 & 2(1) & 0 & - & - \\
					& & & & \\ \hline
					& & & & \\
					\multirow{2}{*}{2} & \multirow{2}{*}{1$ \oplus $3} & 0 & - & \multirow{2}{*}{0$ \% $}\\ 
					& & 1 & 0/1 & \\ & & & & \\ \hline & & & & \\
					\multirow{2}{*}{3} & \multirow{2}{*}{2(2)$ \oplus $4} & 0 & - &  \multirow{2}{*}{0$ \% $}\\
					& & 1 & 0/2 & \\& & & & \\ \hline& & & & \\
					\multirow{3}{*}{4} & \multirow{3}{*}{2(1)$ \oplus $3(3)$ \oplus $5} & 0 & - & \multirow{3}{*}{17$ \% $}\\
					& & 1 & 0/3 & \\ 
					& & 2 & 1/2 & \\ & & & & \\ \hline& & & & \\
					\multirow{3}{*}{5} & \multirow{3}{*}{5(2)$ \oplus $4(4)$ \oplus $6} & 0 & -& \multirow{3}{*}{20$ \% $}\\
					& & 1 & 0/4 & \\
					& & 2 & 2/5 & \\ & & & & \\ \hline & & & & \\
					\multirow{4}{*}{6} & \multirow{4}{*}{5(1)$ \oplus $9(3)$ \oplus $5(5)$ \oplus $7} & 0 & - & \multirow{4}{*}{35$ \% $}\\
					& & 1 & 0/5 & \\
					& & 2 & 3/9 &\\
					& & 3 & 4/5 & \\ & & & & \\ \hline & & & & \\
					\multirow{4}{*}{7} & \multirow{4}{*}{14(2)$ \oplus $14(4)$ \oplus $6(6)$ \oplus $8} & 0 & -& \multirow{4}{*}{37$ \% $} \\
					& & 1 & 0/6 & \\
					& & 2 & 3/14 & \\
					& & 3 & 10/14 &\\ & & & & \\ \hline
				\end{tabular}
	\end{center}}}
	\caption{The last column shows for a given $L$ the percentage of states that depend on complex Bethe roots. Notice that due to the symmetries, it is enough to consider \eqref{eq:valuesofN}. These results were obtained by solving the BA and comparing its eigenvalues with the ones from direct diagonalization. The column ``degeneracies'' just shows that this model indeed has $\alg{su}(2)$ symmetry. When thinking in the eigenvalues of the transfer matrix, something like $n_A(A)\oplus n_B(B)$ just means that we found $n_A$ eigenvalues with degeneracy $A$ and $n_B$ eigenvalues with degeneracy $B$.}
		\label{table:XXXspectrum}
\end{table}

Table \ref{table:XXXspectrum} indicates that, as mentioned before, if the volume of the system grows, but the number of excitations is kept small, it is still relatively easy to solve (in other words, almost all roots are still real). But  when the number of particles grows, the complexity rapidly increases. \footnote{If one just naively tries to solve the Bethe equations directly, one cannot go much further than what is shown in Table \ref{table:XXXspectrum} for the XXX model. For models whose symmetry is an algebra of higher rank is even more difficult. However, there are more effective approaches to do this. The fast solver in \cite{Marboe:2016yyn}, which is applicable to rational functions and based on the so-called Baxter Q-functions, and the technique based on algebraic varieties \cite{Gainutdinov:2015vba} applied to quantum deformed models, are two examples. } 

Therefore, we can expect that for $ L\rightarrow \infty $ and $ N\rightarrow \infty $ a large number of the Bethe roots will be complex. But for $ \frac{L}{N}= $ fixed, this corresponds exactly to the thermodynamic limit and we need a different strategy to work with such equations.

In the next section, we will see that these complex Bethe roots will arrange themselves in patterns that make possible the computation of several quantities at this regime.
The strategy involving these patterns is called \textbf{String Hypothesis} and will provide a way to write effective Bethe equations which will depend on a real ``rapidity'' $ u $. These patterns will form bound states and can then be each treated as a particle. 
To start, we assume $ L\rightarrow \infty $ but no constraints about $ N $, and only later we approach the fact that $N$ is also infinitely large and $ \frac{L}{N} $ is fixed. 
Please also notice that the word ``string'' here does not carry the same meaning as in previous sections of this review. The name comes from the fact that the roots arrange themselves in towers of discrete points in the complex plane. 

\paragraph{On Integrable 2D Quantum field theories (IQFTs).} Although for spin chains the Bethe ansatz is an exact procedure, the same is not the case for IQFTs. In the latter we define asymptotic states, and it is convenient to consider very large systems. We can create \textit{in} (for $t\rightarrow -\infty$) and \textit{out} (for $t\rightarrow +\infty$) asymptotic states with arbitrary number of magnons,
\begin{equation}
    |\psi_{p_1,p_2,...,p_N}\rangle^{\text{in,out}},
\end{equation}
by starting with a vaccum $|0\rangle$ and acting on it with $N$ creation operators. One can then define the S-matrix as the operator that maps \textit{in} states into \textit{out} states. 

%Since we are interested in physical states, we need to satisfy the level-matching condition\footnote{This statement is made assuming no-winding.} 
%\begin{equation}
%    \mathbf{P}|\psi_{p_1,p_2,...,p_N}\rangle^{\text{in,out}}=(p_1+p_2+...+p_N)|\psi_{p_1,p_2,...,p_N}\rangle^{\text{in,out}}=0
%\end{equation}
%which indeed put extra constraints in the allowed values of momentum. 

As seen in section \ref{fiona:subsec:fullSmatrix}, the presence of integrability in the theory has dramatic consequences (see also \cite{Zamolodchikov:1978xm}  and for reviews in QFT see \cite{Dorey:1996gd,Bombardelli:2016scq,Levkovich-Maslyuk:2016kfv}, and in the context of AdS$_3$ see \cite{Sfondrini:2014via}). Using the points discussed in \ref{fiona:subsec:fullSmatrix}, together with the periodicity of the wave-functions in space leads to the so-called Bethe-Yang equations. 

Despite the important differences, the procedure itself is very similar to the exact one and it is called Asymptotic Bethe ansatz. However, it raises some questions, with one in particular being the difficulty in describing the spectrum of such a QFT at finite-volume. In fact, this difficulty motivated the introduction of an important technique in \cite{Zamolodchikov:1989cf}, which will be discussed in section \ref{ana:subsec:TBAforAdS3}.

\

Let us leave this problem for now and return to the discussion of spin chains in order to introduce the basic concepts of the Thermodynamic Bethe ansatz method. For the Heisenberg spin chain, the TBA was first performed in \cite{Takahashi:1972zza}. 

\subsubsection{Large L limit and String configurations}\label{ana:subsubsec:LargeLXXX}

We would like to consider the Bethe equations \eqref{eq:BetheEquations} in the limit $ L\rightarrow \infty $. Before starting, it is useful to rewrite equation \eqref{eq:BetheEquations} using the definition of momentum \eqref{eq:energyN} such that

\begin{equation}
e^{i p_k L}\prod_{j\neq k}^{N}\frac{u_k-u_j-2i}{u_k-u_j+2i}=1 \quad \text{for} \quad k=1,...,N.
\label{eq:BEnewversion}
\end{equation}

We would like to carefully take the infinite-volume limit to avoid divergences. The process of doing this will lead us to the so-called \textbf{string-configurations}.  In order to achieve this, it is important to notice that although the total momentum and total energy are expected to be real, the same is not necessarily true for these quantities for an individual fundamental particle. So, in order to take the limit we will need to systematically study the analytic structure of the Bethe equations. Additionally, it is important to remember that for many models, the complex Bethe roots appear in complex conjugate pairs. 

We start by checking whether $p_1$ is real or complex. These two choices generate different outcomes. When $p_1$ is real, the exponential $e^{ip_1L}$  oscillates, while if $p_1$ is complex the exponential either diverges or goes to zero (depending on the imaginary part of $p_1$). When the momentum is complex, we will adopt as convention that its imaginary part is always positive. With this choice, the exponential always vanishes when $L\rightarrow\infty$. 

\paragraph{A: $ \mathbf{p_1\in \mathbb{R} }$}: 

For $ p_1\in \mathbb{R} $, $ e^{i p_1 L} $ in equation \eqref{eq:BEnewversion} oscillates, and therefore there are no problems when we take the limit $ L\rightarrow \infty $. What happens is only that as larger the $ L $ more and more solutions will appear (because the degree of the polynomial on $u$ grows as $L$ increases). Notice that from \eqref{eq:energyN}, real momenta implies real rapidity and vice-versa. For reasons that become clear soon, let us call this case a \textbf{1-string} and write $u_1=u$, with $u\in \mathbb{R}$.

\paragraph{B: $ \mathbf{p_1\in \mathbb{C} } $ with $ \text{Im} \mathbf{(p_1)>0} $}:

It is convenient to explicitly write equations \eqref{eq:BEnewversion} for $ k=1,2 $ as
\begin{align}
e^{i p_1 L}\frac{u_1-u_2-2i}{u_1-u_2+2i}\prod_{j\neq 1,2}^{N}\frac{u_1-u_j-2i}{u_1-u_j+2i}& =1,\label{eq:k1}\\
e^{i p_2 L}\frac{u_2-u_1-2i}{u_2-u_1+2i}\prod_{j\neq 1,2}^{N}\frac{u_2-u_j-2i}{u_2-u_j+2i}& =1.\label{eq:k2}
\end{align}

For $ L\rightarrow \infty $, we have $ e^{i p_1 L}\rightarrow 0 $. So, in order for \eqref{eq:k1} to be satisfied we need to have $ u_1-u_2+2i=0 $ as well. This gives 
\begin{equation}
u_2=u_1+2i.
\label{eq:u2}
\end{equation} By choosing \eqref{eq:u2} we avoided all problems in equation \eqref{eq:k1}. But now we need to check if we created any problems in equation \eqref{eq:k2}. In order to do that, let us multiply equations \eqref{eq:k1} and \eqref{eq:k2}  by each other
\begin{equation}
e^{i(p_1+p_2)L}\prod_{j\neq 1,2}^{N}\left(\frac{u_1-u_j-2i}{u_1-u_j+2i}\frac{u_2-u_j-2i}{u_2-u_j+2i}\right)=1,
\label{eq:k1k2}
\end{equation}
and then we repeat the steps done for $p_1$, but now for $p_1+p_2$.

\paragraph{B.1: $ \mathbf{(p_1+p_2)\in \mathbb{R}} $\textbf{:}}

For this case $ e^{i(p_1+p_2)L} $ oscillates and therefore there are no problems. The solution can be written as	
\begin{equation}
u_1=u-i\quad \text{ and } \quad u_2=u+i, \quad \text{with} \quad u\in\mathbb{R}.
\label{eq:stringQ2XXX}
\end{equation} 
This solution can be called \textbf{2-complexes}, a \textbf{2-string}, or \textbf{string of length two}. 

\paragraph{B.2: $ \mathbf{(p_1+p_2)\in \mathbb{C} } $ \textbf{with Im}$  \mathbf{(p_1+p_2)>0} $ }

In this situation, we have that $ e^{i(p_1+p_2)L}\rightarrow 0 $ and therefore we need a pole at $ u_2-u_3+2i=0 $ in order to equation \eqref{eq:k1k2} be satisfied. Consequently we have another particle now with rapidity
\begin{equation}
u_3=u_2+2i.
\end{equation}
We now need to write the equation \eqref{eq:BEnewversion} for $ k=1,2,3 $,
\begin{align}
	e^{i p_1 L}\frac{u_1-u_2-2i}{u_1-u_2+2i}\frac{u_1-u_3-2i}{u_1-u_3+2i}\prod_{j\neq 1,2,3}^{N}\frac{u_1-u_j-2i}{u_1-u_j+2i}& =1,\label{eq:kk1}\\
	e^{i p_2 L}\frac{u_2-u_1-2i}{u_2-u_1+2i}\frac{u_2-u_3-2i}{u_2-u_3+2i}\prod_{j\neq 1,2,3}^{N}\frac{u_2-u_j-2i}{u_2-u_j+2i}& =1,\label{eq:kk2}\\
	e^{i p_3 L}\frac{u_3-u_1-2i}{u_3-u_1+2i}\frac{u_3-u_2-2i}{u_3-u_2+2i}\prod_{j\neq 1,2,3}^{N}\frac{u_3-u_j-2i}{u_3-u_j+2i}& =1,\label{eq:kk3}
\end{align}
respectively; and multiply all of them together
 \begin{equation}
 	e^{i(p_1+p_2+p_3)L}\prod_{j\neq 1,2,3}^{N}\left(\frac{u_1-u_j-2i}{u_1-u_j+2i}\frac{u_2-u_j-2i}{u_2-u_j+2i}\frac{u_3-u_j-2i}{u_3-u_j+2i}\right)=1.
 	\label{eq:k1k3}
 \end{equation}
Next, we check whether the total momentum $ p_1+p_2+p_3 $ is real or not. If it is real  we stop and have a system with 	
\begin{equation}
\{u_2=u_1+2i, \,u_3=u_2+2i\}.
\end{equation}
Given that we had Im$(p_1)>0$ (which implies Im$(u_1)<0$) a natural choice is to write Im$ u_1 = -2$, such that 
\begin{equation}
u_1=u-2i, \quad  u_2=u\quad \text{and} \quad u_3=u+2i, \quad \text{with} \quad u\in\mathbb{R}.
\label{eq:3-complexes}
\end{equation}
This solution can be called \textbf{3-complexes}. 

However, if the total momentum is complex we continue applying the procedure.
In this way we can construct strings of any length. In particular, we obtain

\begin{equation}
u_{j+1}-u_j=2i, \quad \text{for} \quad j=1,...,Q-1
\end{equation}
whose solution is a \textbf{string of length} $ \mathbf{Q} $ given by
\begin{equation}
u_j=u-(Q+1-2j)i,\quad \text{with }\quad u\in\mathbb{R} \quad \text{and} \quad j=1,...,Q.
\label{eq:QstringsXXX}
\end{equation}
The $ u $ is usually called the \textbf{center} of the string complex.

\begin{figure}[ht]
\begin{centering}
    \begin{tcolorbox}
\includegraphics[width=13cm]{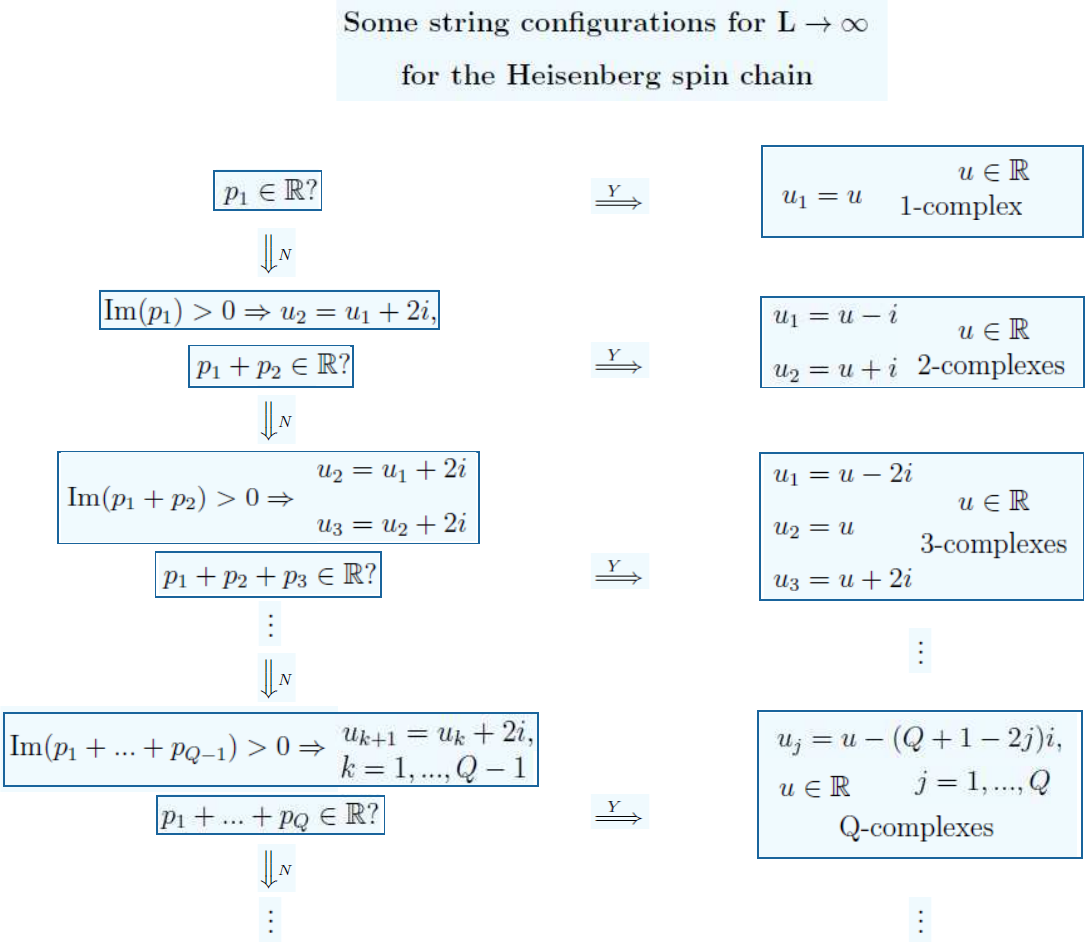}
    \end{tcolorbox}
\end{centering}
	\caption{A summary of the string configurations construction when $ L\rightarrow \infty $. In the figure $Y$ means ``Yes''  and $N$ means ``No'' }
 \label{fig:stringconfig}
\end{figure}

\begin{figure}[ht]
\begin{centering}
\hspace{3cm}\includegraphics[scale=0.8]{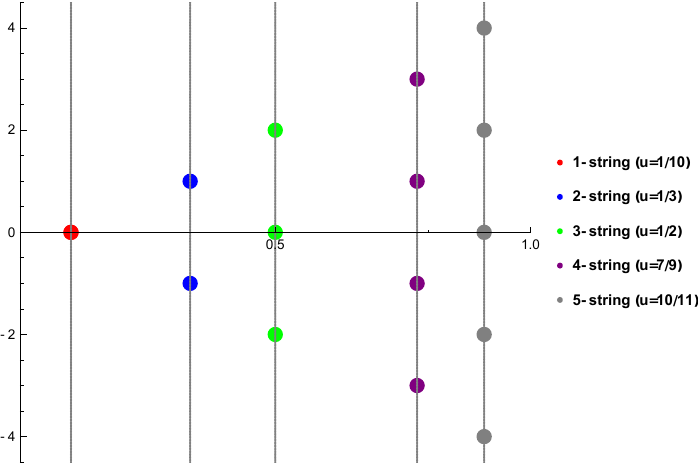}
\end{centering}
	\caption{Q-Strings for $Q=1,2,3,4,5$, centered at different values of $u\in \mathbb{R}$. }
 \label{fig:String-configurationsplot}
\end{figure}

A summary of the construction of the first $Q$-complexes can be seen in Figure \ref{fig:stringconfig}. We also plot the first five possible string configurations on Figure \ref{fig:String-configurationsplot}.

Once we find the first $Q$-string, we can continue the analysis. If the momenta $p_{Q+1}$ is real, the exponential $e^{i p_{Q+1} L}$ oscillates as $L\rightarrow\infty$ and again there are no problems. If it is complex we continue the analysis like before to find a string of length $Q^\prime$, now centered in a real value $u^\prime$ and given by\vspace{-0.2cm}
\begin{equation}
u_{Q+j}=u^\prime-(Q^\prime+1-2j)i, \qquad u^\prime\in \mathbb{R}, \qquad j=1,...,Q^\prime.
\end{equation}
In order to obtain all possible strings, one just keeps applying this procedure, always remembering that each $Q_k$-complexes should in principle be centered in a different real value (named as $u$ and $u^\prime$ in our example of $Q$-string and $Q^\prime$-string, respectively). In general, when we study a system with already $a-1$ strings (including $1$-strings), the $Q_a$-string is obtained by
\begin{equation}
    u_{q_a}=u_a-(Q_a+1-2j)i, \qquad u_a\in\mathbb{R}, \qquad j=1,...,Q_a,
    \label{eq:stringconfiggenXXX}
\end{equation}
where $q_a=Q_1+Q_2+...+Q_{a-1}$.

Notice that this equation and the ones for the energy and momentum are invariant under swapping any pair  $\{u_{j_1},u_{j_2}\}$. Therefore, although we started the analysis by $p_1$, any other starting point would generate the same total energy and momentum.

The discussion on the construction of the string configurations presented above was based on the one presented in the lecture notes by van Tongeren \cite{vanTongeren:2016hhc} in 2016.

\begin{centering}
    \begin{tcolorbox}
     \textbf{Exercise 4.1)}: Consider the system with \\
     $\bullet$ Im$(p_1)>0$, $(p_1+p_2)\in\mathbb{R}$, \\
     $\bullet$ $p_3\in\mathbb{R}$,\\
     $\bullet \text{ and } $ Im$(p_4)>0$, Im$(p_4+p_5)>0$, $(p_4+p_5+p_6)\in\mathbb{R}$. \\
     All the remaining $p_k$ are real. \\
     a) This system has two different strings $Q$ and $Q^\prime$. Compute them and tell their length.\\
     b) If we had instead a system with \\
     $\bullet$ Im$(p_1)>0$, $(p_1+p_2)\in\mathbb{R}$, \\
     $\bullet \text{ and } $ Im$(p_3)>0$, Im$(p_3+p_4)>0$, $(p_3+p_4+p_5)\in\mathbb{R}$. \\
     Would that be fundamentally different from the case with with $p_3\in\mathbb{R}$ above?
    \end{tcolorbox}
\end{centering}

\begin{centering}
    \begin{tcolorbox}
     \textbf{Exercise 4.2}: Consider a system that has a string of length three and a string of length four. Assume that $p_{k}\in\mathbb{R}$ for $k>9$. With that in mind, provide two sets of conditions on $p_j$, for $1\le j\le 9$ that would make that such strings possible. Are these the only possibilities? 
    \end{tcolorbox}
\end{centering}

\begin{centering}
\begin{tcolorbox}
\textbf{Exercise 4.3:} For the Hubbard model \cite{essler2005one}, the analogous of the Bethe equations for a state with a spin-up  and one spin-down are given by
\begin{align} 
e^{ip_1L}&=\frac{\lambda-\sin{p_1}-iu}{\lambda-\sin{p_1}+iu}\\
e^{ip_2L}&=\frac{\lambda-\sin{p_2}-iu}{\lambda-\sin{p_2}+iu}\\
1&=\prod_{j=1}^{2}\frac{\lambda-\sin{p_j}-iu}{\lambda-\sin{p_j}+iu}.
\end{align}
Assuming that 
\begin{equation}
p_1=q-i\xi 
\end{equation}
with $p,\,\xi\in\mathbb{R}$ and $\xi>0$, compute the string configuration for $L\rightarrow\infty$.
\end{tcolorbox}
\end{centering}

It is important to highlight that the complex solutions that arrange themselves in string configurations are solutions of the system only when  $L\rightarrow \infty$. They are not the same complex solutions that one finds when solving the Bethe equations for finite $L$ (like when constructing Table \ref{table:XXXspectrum}).

Some of the finite complex solutions behave like
\begin{equation}
u_j=u+\delta_0-(Q+1-2j)i+\delta_1 i,\quad \text{with } u\in\mathbb{R} \quad \text{and} \quad j=1,...,Q,
\end{equation}
\textit{i.e.}~like a string but with extra contributions to both the real and imaginary parts of the Bethe roots. These contributions decrease as the length of the chain grows, such that
\begin{equation}
    \delta_0\rightarrow 0, \qquad \delta_1\rightarrow 0 \qquad \text{when} \qquad L\rightarrow \infty.
\end{equation}

The concepts and discussion presented in this subsection are focused on the XXX spin-1/2 chain. In this model only one type of fundamental particle is present - the magnon. A  similar story happens when we have more excitations' species. Each different type of particle will have their own string configuration, and one needs to construct each of them. This will be exactly the case in AdS$_3$ as we will see in section \ref{ana:subsec:TBAforAdS3}.

\subsubsection{Bound States}\label{ana:subsubsec:boundstates}

At this point the reader could be wondering what is the interpretation of the center of string or of the string itself. It happens that each string behaves like a bound state with rapidity $u$, \textit{i.e.}~the energy of the string is smaller than the energy of the individual real magnons. 

\paragraph{Energy and momentum.}
 In particular, the momentum of a Q-string is given by

\begin{equation}
p^Q(u)=-i\log\left(\frac{u+iQ}{u-iQ}\right),
\label{eq:pforstringXXX}
\end{equation}
and its energy is

\begin{equation}
H^Q(u)=\frac{2JQ}{Q^2+u^2}.
\label{eq:EforstringXXX}
\end{equation}

Notice that real momentum $p$ implies real rapidity $u$, and vice-versa, for any $Q\in\mathbb{Z}$.

\begin{centering}
\begin{tcolorbox}
\textbf{Exercise 4.4:} Given that 
\begin{equation}
p^Q(u)=\sum_{k=1}^{Q}p(u_k),
\end{equation}
and 
\begin{equation}
H^Q(u)=\sum_{k=1}^{Q}H(u_k),
\end{equation}
with $ u_k $ being the rapidity of the $Q$-string described in equation \eqref{eq:QstringsXXX}, check equations \eqref{eq:pforstringXXX} and \eqref{eq:EforstringXXX} for $ Q=2,3,4 $.
\end{tcolorbox}
\end{centering}

\paragraph{Bethe equations for the center of the bound state.}

We promised above that one can use the string configurations to write effective Bethe equations. On these equations the center $u$ behaves as the rapidity of the bound-state formed by the string-complexes.  The first step is to write equations that describe the scattering between a fundamental excitation (magnon) and a $Q$-string. 

Let us start by rewriting equation \eqref{eq:BEnewversion} as 

\begin{equation}
e^{ip_kL}\prod_{j\neq k}^{N}S^{11}(u_k-u_j)=1\quad \text{where} \quad S^{11}(u_k-u_j)=\frac{u_k-u_j-2i}{u_k-u_j+2i},
\label{eq:BESmatrix}
\end{equation}
and then focus first on a string with $ Q=2 $, \textit{i.e.}~the case with equation \eqref{eq:k1k2} and \eqref{eq:stringQ2XXX}. In equation \eqref{eq:BESmatrix}, $S^{11}(u_k-u_j)$ describes the scattering matrix between particles with rapidities $u_k$ and $u_j$, with $j\neq k$.

Now, let us substitute the $ 2 $-string \eqref{eq:stringQ2XXX} in equation \eqref{eq:k1k2}. We obtain 
\begin{equation}
e^{i(p_1+p_2)L}\prod_{j\neq 1,2}^{N}\frac{u-u_j-3i}{u-u_j+3i}\frac{u-u_j-i}{u-u_j+i}=1,
\end{equation} 
which can be rewritten as
\begin{equation}
e^{i(p_1+p_2)L}\prod_{j\neq 1,2}^{N}S^{12}(u_j-u)=1.
\end{equation}
The S-matrix $ S^{12}(u_j-u)=S^{21}(u-u_j) $ with
\begin{equation}
S^{12}(u_j-u)S^{21}(u-u_j)=S^{11}(u_1-u_j)S^{11}(u_2-u_j)=\frac{u-u_j-3i}{u-u_j+3i}\frac{u-u_j-i}{u-u_j+i},
\end{equation}
can be interpreted as the scattering between a particle with rapidity $ u_j $ and a 2-string with rapidity $ u $.

Similarly we could do the procedure for the $ 3 $-string \eqref{eq:3-complexes}, by substituting the 3-string in equation \eqref{eq:k1k3} and we would obtain 
\begin{equation}
e^{i(p_1+p_2+p_3)L}\prod_{j\neq 1,2,3}^{N}S^{13}(u_j-u)=1,
\label{eq:BES13XXX}
\end{equation}
with $S^{13}(u_j-u)=S^{31}(u-u_j)$ and
\begin{equation}
S^{31}(u-u_j)=S^{11}(u_1-u_j)S^{11}(u_2-u_j)S^{11}(u_3-u_j)=\frac{u-u_j-4i}{u-u_j+4i}\frac{u-u_j-2i}{u-u_j+2i}.
\label{eq:S13XXX}
\end{equation}
The $ S^{13}(u-u_j) $ can be interpreted as describing the scattering between a particle with rapidity $ u_j $ and a 3-string with rapidity $ u $.

We could continue this process and we would find
\begin{equation}
e^{iL\sum_{j=1}^{Q}p_j}\prod_{j\neq 1,..,Q}^{N}S^{1Q}(u_j-u)=1,
\label{eq:BE1Q}
\end{equation}
with 
\begin{equation}
\begin{aligned}
S^{Q1}(u-u_j)&=\prod_{a=1}^{Q}S^{11}(u_a-u_j)=\frac{u-u_j-(Q+1)i}{u-u_j+(Q+1)i}\frac{u-u_j-(Q-1)i}{u-u_j+(Q-1)i},\\
&=S^{1Q}(u_j-u).
\label{eq:SmatrixstringQXXX}
\end{aligned}
\end{equation}

This idea is very convenient when writing the Bethe equations on the thermodynamic limit, as will be clear in section \eqref{ana:subsubsec:BElargeLXXX}.

\begin{centering}
\begin{tcolorbox}
\textbf{Exercise 4.5:} a) Check equations \eqref{eq:BES13XXX} and \eqref{eq:S13XXX}. \\
b) Compute the case with $Q=4$ and check that the cancellations indeed happen such that equation \eqref{eq:SmatrixstringQXXX} is satisfied.
\end{tcolorbox}
\end{centering} 

For  a system with a finite number of magnons, described uniquely by string configurations (including $1$-string's), and whose number of strings of length $Q$ is given by $N^Q$, it is clear that 
\begin{equation}
\sum_{Q\in \text{lengths}}Q N^Q=N,
\label{eq:totalnumberofmagnons}
\end{equation}
where ``lengths'' is a set including the different lengths of the strings.
Consider for instance, a system with $N=16$ magnons, distributed in one 1-string, three 2-strings, one 3-string and one 6-string. This means that we would have lengths$=\{1,2,3,6\}$ (corresponding to the four different lengths) and that 
\begin{align}
    & Q=1,\, N^{(1)}=1,\\
    & Q=2,\, N^{(2)}=3,\\
    & Q=3,\, N^{(3)}=1,\\
    & Q=6,\, N^{(6)}=1,
\end{align}
which using \eqref{eq:totalnumberofmagnons} gives $N=16$, as expected. We are showing this simple thought now because when introducing \textbf{fusion} we will have to make a similar assumption. This example is not realistic because we do not expect for finite $N$ to have a system described \textit{only} by string configurations\footnote{Actually, this assumption is not true even for $N\rightarrow \infty$. But in that case, as we will discuss, it corresponds to an excellent approximation when the quantity of interest is the free energy.}. Nonetheless, we expect that this example will give the reader the intuition behind the less-intuitive assumption \eqref{eq:conditiononQ1} soon to be made for $N\rightarrow\infty$. 

\subsubsection{String Hypothesis}\label{ana:subsubsec:stringhypothesis}

We have seen that for $L\rightarrow\infty$, the complex solutions arrange themselves in patterns called strings. The string hypothesis is the assumption that,\textit{ for $L\rightarrow \infty$, $ N\rightarrow \infty $ and $ \frac{L}{N}\rightarrow  $ fixed, all relevant solutions for the computation of the free energy can be written as string configurations.} This is arguably a very good hypothesis when the quantity of interest is the free energy.

The string hypothesis is focused on the thermodynamics and it is a good approximation only in that limit. As mentioned above, it works remarkably well to compute the free energy, but for several other physical quantities the contribution of the Bethe roots that are not arranged in string configurations becomes important. In our case, the free-energy is exactly the quantity we are are interested in, and therefore we expect that the string hypothesis will provide an accurate description.

\subsubsection{Fusion: Bethe equations in the Thermodynamic limit}\label{ana:subsubsec:BElargeLXXX}

Now we have all the elements to write the Bethe equations in the thermodynamic limit. First of all let us assume that we are in the regime where the number of magnons $N$ also goes to infinity and the string hypothesis is valid.  In that case, extending the argument on \eqref{eq:totalnumberofmagnons}, one can make the following assumption
\begin{equation}
\sum_{Q=1}^{\infty}QN^Q=N.
\label{eq:conditiononQ1}
\end{equation}
This assumption allows one to write the following expression for the Bethe equations of the string configurations

\begin{equation}
e^{ip_jL}\prod_{Q=1}^{\infty}\prod_{b=1}^{N_Q}S^{1Q}(u_j-u_b)=1,
\label{eq:BE1QXXX}
\end{equation}
where $ S^{1Q}(u_j-u_b) $ is defined as in \eqref{eq:SmatrixstringQXXX} and $ u_j $ satisfies \eqref{eq:QstringsXXX}

In the previous section we used products of $ S^{11}(u_k-u_j) $ to construct $ S^{1Q}(u_j-u) $ which can be interpreted as the scattering between a magnon and a string of length $ Q $. In the same way we can construct now an object  $ S^{Q_aQ_b}(u_a-u_b) $ by products of $ S^{1Q_b}(u_k-u_b) $. This object can be interpreted as describing the scattering between a $ Q_a $-string centered on real $ u_a $  and a $ Q_b $-string with real center $ u_b $. With this in mind we can write the following Bethe equations
\begin{equation}
e^{ip_a^{Q_a}L}\prod_{Q_b=1}^{\infty}\prod_{b\neq a}^{N_{Q_b}}S^{Q_aQ_b}(u_a-u_b)=1,
\label{eq:BEforPQXXX}
\end{equation}
where $ u_a ,\, u_b $ belong to \eqref{eq:stringconfiggenXXX}. Also
\begin{align}
S^{Q_aQ_b}&(u_a-u_b)=\prod_{j=1}^{Q_a}S^{1Q_b}(u_j-u_b)\label{eq:SQaQb1}\\ 
&=\prod_{j=1}^{Q_a}\left(\frac{u_b-u_j-(Q_b+1)i}{u_b-u_j+(Q_b+1)i}\frac{u_b-u_j-(Q_b-1)i}{u_b-u_j+(Q_b-1)i}\right)\nonumber\\
&=\frac{u_b-u_a-(Q_b+Q_a)i}{u_b-u_a+(Q_b+Q_a)i}\frac{u_b-u_a-(Q_b-Q_a)i}{u_b-u_a+(Q_b-Q_a)i}\prod_{k=1}^{Q_a-1}\left(\frac{u_b-u_a+(Q_b-Q_a+2k)i}{u_b-u_a-(Q_b-Q_a+2k)i}\right)^2.\label{eq:SQaQb2}
\end{align}

\begin{centering}
\begin{tcolorbox}
\textbf{Exercise 4.6:} Assuming \eqref{eq:SmatrixstringQXXX} and \eqref{eq:stringconfiggenXXX}, compute $ S^{Q_aQ_b}(u_a-u_b)$ using equation \eqref{eq:SQaQb1} for $Q_a=2,3,4$ and check that they can be written as in expression \eqref{eq:SQaQb2}. 
\end{tcolorbox}
\end{centering}

\subsubsection{Densities and the counting function}\label{ana:subsubsec:countingfunctionXXX}

\

We are now almost ready to compute the free energy. In order to do that, it is important to notice that in the thermodynamic limit we are working with a large volume and a large number of particles. For this reason, it makes sense to think in terms of densities, instead of individual ``particles''. Additionally, the introduction of a new  quantum number called  counting function is very enlightning. Let us now see how this happens.

By taking the logarithm of equation \eqref{eq:BEforPQXXX} we obtain 
\begin{equation}
2\pi i c^{Q_a}L+ip^{Q_a}(u_a)L+ \log\left(\prod_{Q_b=1}^{\infty}\prod_{b\neq a}^{N_{Q_b}}S^{Q_aQ_b}(u_a-u_b)\right)=0,
\end{equation}
such that
\begin{equation}
    c^Q(u)=-\frac{p^Q(u)}{2\pi}-\frac{1}{2\pi  i L}\sum_{Q_b=1}^{\infty}\sum_{b=1}^{N_{Q_b}}\log S^{QQ_b}(u-u_b).
\end{equation}
The function $c^{Q_a}$ is called counting function. In particular $L c^{Q_a}(u)\equiv I^{Q_a}$ assumes integer values for specific values of the rapidity and it is thought as a $Q_a$-particle quantum number. One can imagine, as proposed by Yang-Yang \cite{Yang:1968rm}, that we can have not only particles, but also gaps when the particle is not occupying an available slot. These gaps are called holes.

For $L\rightarrow \infty$ we can have a large number of these particles and holes. So, it makes sense to work with densities of particles and densities of holes, which can be defined respectively as
\begin{equation}
    \rho(u)=\frac{\Delta n}{L\Delta u} \qquad \text{and} \qquad \bar{\rho}(u)=\frac{\Delta \bar{n}}{L\Delta u},
    \label{eq:densities}
\end{equation}
where $\Delta n $ is the number of particles whose rapidity is in an interval $\Delta u$ and $\Delta \bar{n} $ is the number of holes in that interval.
The total density and total number of particles in that rapidity interval is given by
\begin{equation}
\rho_t=\rho+\bar{\rho} \quad \text{and} \quad \Delta n_t=\Delta n+\Delta\bar{n}.
\end{equation}
Having the definitions of density, we can rewrite the counting function as 
\begin{align}
c^Q(u)&=-\frac{p^Q(u)}{2\pi}-\frac{1}{2\pi i}\sum_{Q_b=1}^{\infty}\sum_{b=1}^{N_{Q_b}}\log S^{QQ_b}(u-u_b)\frac{(u_b-u_{b+1})}{L(u_b-u_{b+1})},\\
&\rightarrow -\frac{p^Q(u)}{2\pi}-\frac{1}{2\pi i}\sum_{Q^\prime=1}^{\infty}\int_{-\infty}^{\infty}du^{\prime}\log S^{QQ^\prime}(u-u^{\prime})\rho^{Q^\prime}(u^\prime).
\label{eq:countingfunction2}
\end{align}
The function $c^{Q}(u)$ has to be monotonic increasing, but proving that this in fact happens is a model dependent task and not necessarily straightforward (for a clear discussion on this point see \cite{vanTongeren:2016hhc}).

From the discussion above, it is clear that the counting function keeps track of the number of particles and holes in a given rapidity range such that
\begin{align}
    & L(\rho(u)+\bar{\rho}(u))du=L dc(u),\nonumber\\
    & \Rightarrow \quad \rho(u)+\bar{\rho}(u)=\frac{dc(u)}{du}.
\end{align}
This relation when used together with \eqref{eq:countingfunction2} provides a clear relation between the momentum of the bound-states and their densities. It will play an important role in the next sections.

\subsubsection{The TBA equations and the free-energy}\label{ana:subsubsec:TBAXXX}

\paragraph{Free-energy - The basics:}
The free energy per site $f$ is given by 

\begin{equation}
f=e-\mathcal{T}s,
\end{equation}
where  $\mathcal{T}$ is the temperature.

In order to compute $f$ we need to determine the energy per site $e$ and the entropy per site $s$. These quantities are directly connected with the number of available states and therefore the density of particles and holes in the model. 
The entropy is defined as the logarithm of the available states, \textit{i.e.}~the number of possibilities to distribute $ \Delta n $ particles among $ \Delta n_t $ vacancies, and it is given by
\begin{equation}
\begin{aligned}
\Delta \mathcal{S}&=\log\left(\frac{\Delta n_t!}{\Delta n!\Delta\bar{n}!}\right),\\
&=\log\left(\frac{\left(L\Delta u\rho_t(u)\right)!}{\left(L\Delta u\rho(u)\right)!\left(L\Delta u\bar{\rho}\right)!}\right),\\
&=\log\left(L\Delta u\rho_t(u)\right)!-\log\left(L\Delta u\rho(u)\right)!-\log\left(L\Delta u\bar{\rho}(u)\right)!\\
&\sim L \Delta u\left(\rho_t\log\rho_t-\rho\log\rho-\bar{\rho}\log\bar{\rho}\right),
\end{aligned}
\end{equation}
where in the last step we used the Stirling formula ($\log m!\sim m\,\log m-m$).

The entropy per site is therefore given by

\begin{equation}
s=\int_{-\infty}^{\infty}du\left(\rho_t\log\rho_t-\rho\log\rho-\bar{\rho}\log\bar{\rho}\right).
\end{equation}
So, in order to compute the free energy we need to learn more about the densities for our model. As we have seen in the previous subsection, this is exactly the information coming from the Bethe equations. 

The expression above assumes only one type of excitation, but notice that we are now dealing with bound states generated by string configurations. As a consequence, the entropy will depend on the density $\rho^{Q}(u)$ of such strings-complexes and we have to sum over all possible values of $Q$
\begin{equation}
    s=\sum_{Q=1}^{\infty}\int_{-\infty}^{\infty}du\left(\rho^Q_t\log\rho^Q_t-\rho^Q\log\rho^Q-\bar{\rho}^Q\log\bar{\rho}^Q\right).
\end{equation}
If we have more types of particles (like massive and massless ones, or different spins, for instance), the resulting entropy will be the sum of several expressions analog to this one.

\paragraph{TBA equations.}

Let us focus on a $ Q $-string, starting by explicitly writing $\rho^Q(u)+\bar{\rho}^Q(u) =\frac{dc^Q(u)}{du}$ for this model as
\begin{align}
\rho^Q(u)+\bar{\rho}^Q(u)& =\frac{dc^Q(u)}{du},\label{eq:summingrhos1}\\
& =-\frac{1}{2\pi}\frac{dp^Q}{du}-\frac{1}{2\pi i}\sum_{Q^\prime}\int_{-\infty}^{\infty}du^{\prime}\frac{d}{du}\left(\log S^{QQ^\prime}(u-u^{\prime})\right)\rho^{Q^\prime}(u^\prime),\label{eq:summingrhos2}\\
&=-\frac{1}{2\pi}\frac{dp^Q}{du}-\sum_{Q^\prime}\int_{-\infty}^{\infty}du^{\prime}K^{QQ^\prime}(u-u^\prime)\rho^{Q^\prime}(u^\prime),\label{eq:summingrhos3}\\
&=-\frac{1}{2\pi}\frac{dp^Q(u)}{du}-\sum_{Q^\prime}K^{QQ^\prime}\star \rho^{Q^\prime}(u),\label{eq:summingrhos4}
\end{align}
where from \eqref{eq:summingrhos2} to \eqref{eq:summingrhos3} we defined
\begin{equation}
K^{QQ^\prime}(u-u^\prime)=\frac{1}{2\pi i}\frac{d}{du}\left(\log S^{QQ\prime}(u-u^{\prime})\right).
\end{equation}
The $K$ is called Kernel and it is a positive quantity.
To pass from equation \eqref{eq:summingrhos3} to equation \eqref{eq:summingrhos4} we defined \textbf{convolution} as 
\begin{equation}
f\star g(u)=\int_{-\infty}^{\infty}du^\prime f(u-u^\prime)g(u^\prime).
\label{eq:convolution}
\end{equation}
The free energy per site for a  $ Q $-string is given by 
\begin{align}
f&=\sum_Q\int_{-\infty}^{\infty}du\left(H^Q\rho^Q-\mathcal{T}\left(\rho_t^Q\log\rho_t^Q-\rho^Q\log\rho^Q-\bar{\rho}^Q\log\bar{\rho}^Q\right)\right),\\
&=\sum_Q\int_{-\infty}^{\infty}du\left(H^Q\rho^Q-\mathcal{T}\left(\rho^Q\log\left(\frac{\rho_t^Q}{\rho^Q}\right)+\bar{\rho}^Q\log\left(\frac{\rho_t^Q}{\bar{\rho}^Q}\right)\right)\right).
\label{eq:freeenergybeforeTBA}
\end{align}
The thermodynamic equilibrium happens at its stationary point, \textit{i.e.}~at $ \delta f=0 $ where we variate $ f $ with respect to both $ \rho $ and $\bar{\rho}$
\begin{equation}
\delta f=\sum_Q\int_{-\infty}^{\infty}du\left(H^Q\delta\rho^Q-\mathcal{T}\left(\delta\rho^Q\log\left(\frac{\rho_t^Q}{\rho^Q}\right)+\delta\bar{\rho}^Q\log\left(\frac{\rho_t^Q}{\bar{\rho}^Q}\right)\right)\right).
\label{eq:deltaf}
\end{equation}
\begin{centering}
\begin{tcolorbox}
\textbf{Exercise 4.7:} Variate $f$ in equation \eqref{eq:freeenergybeforeTBA} with respect to $\rho^Q$ and $\bar{\rho}^Q$ and show that the variation of the logarithms cancel and only equation \eqref{eq:deltaf} remain.
\end{tcolorbox}
\end{centering} 
We can now variate equation \eqref{eq:summingrhos4}, isolate $ \delta\bar{\rho}^Q $ and put it back in $ \delta f $ obtaining
\begin{align}
\delta f& =\sum_Q\int_{-\infty}^{\infty}du\left(H^Q\delta\rho^Q-\mathcal{T}\delta\rho^Q\log\left(\frac{\bar{\rho}^Q}{\rho^Q}\right)+\mathcal{T}K^{PQ}\star \delta\rho^{Q^\prime}\log\left(1+\frac{\rho^Q}{\bar{\rho}^Q}\right)\right),\\
&=\sum_Q\int_{-\infty}^{\infty}du\left(H^Q\delta\rho^Q-\mathcal{T}\delta\rho^Q\log\left(\frac{\bar{\rho}^Q}{\rho^Q}\right)\right)+\nonumber\\
&\quad +\mathcal{T}\sum_{Q,Q^{\prime}}\int_{-\infty}^{\infty}du\int_{-\infty}^{\infty}du^\prime K^{QQ^\prime}(u-u^\prime)\delta\rho^{Q^\prime}(u^\prime)\log\left(1+\frac{\rho^Q(u)}{\bar{\rho}^Q(u)}\right),\\
&=\sum_Q\int_{-\infty}^{\infty}du\left(H^Q\delta\rho^Q-\mathcal{T}\delta\rho^Q\log\left(\frac{\bar{\rho}^Q}{\rho^Q}\right)\right)+\nonumber\\
&\quad +\mathcal{T}\sum_{Q,Q^{\prime}}\int_{-\infty}^{\infty}du^\prime\int_{-\infty}^{\infty}du K^{Q^\prime Q}(u^\prime-u)\delta\rho^Q(u)\log\left(1+\frac{\rho^{Q^\prime}(u^\prime)}{\bar{\rho}^{Q^\prime}(u^\prime)}\right),\\
&=\sum_Q\int_{-\infty}^{\infty}du\left(H^Q(u)-\mathcal{T}\log\left(\frac{\bar{\rho}^Q}{\rho^Q}\right)+\mathcal{T}\sum_{Q^{\prime}}\log\left(1+\frac{\rho^{Q^\prime}(u)}{\bar{\rho}^{Q^\prime}(u)}\right)\widetilde{\star} K^{Q^\prime Q}(u)\right)\delta\rho^Q(u),
\end{align}
where 
\begin{equation}
f\widetilde{\star} g(u)=\int_{-\infty}^{\infty}du^\prime f(u^\prime)g(u^\prime-u),
\label{eq:rightconvolution}
\end{equation}
is the convolution from the right.
So, by requiring $ \delta f=0 $ we obtain the \textbf{Thermodynamic Bethe ansatz equations}

\begin{equation}
\log\left(\frac{\bar{\rho}^Q}{\rho^Q}\right)=\frac{H^Q(u)}{\mathcal{T}}+\sum_{Q^\prime=1}^{\infty}\log\left(1+\frac{\rho^{Q^\prime}(u)}{\bar{\rho}^{Q^\prime}(u)}\right)\widetilde{\star} K^{Q^\prime Q}(u).
\label{eq:TBAXXX}
\end{equation}
\begin{centering}
\begin{tcolorbox}
\textbf{Exercise 4.8:} Use the results above to prove that
\begin{equation}
f=\frac{\mathcal{T}}{2\pi}\sum_Q \int_{-\infty}^{\infty}du\frac{dp^Q}{du}\log\left(1+\frac{\rho^Q}{\bar{\rho}^Q}\right).
\label{eq:freeenergyTBAXXX}
\end{equation}
\end{tcolorbox}
\end{centering}

\paragraph{Y-functions.}

We can define the \textbf{Y-functions} as 
\begin{equation}
Y^{Q^\prime}(u)=\frac{\bar{\rho}^{Q^\prime}}{\rho^{Q^\prime}}
\label{eq:Yfunctions}
\end{equation}
and as a consequence rewrite the TBA equations as
\begin{equation}
\log Y^Q=\frac{H^Q}{\mathcal{T}}+\sum_{Q^\prime=1}^{\infty}\log\left(1+\frac{1}{Y^{Q^\prime}}\right)\widetilde{\star} K^{Q^\prime Q}.
\label{eq:TBAXXXwithY}
\end{equation}

\paragraph{Free energy - Final form.}

In terms of the $Y$-function the free energy \eqref{eq:freeenergyTBAXXX} becomes
\begin{equation}
f=\frac{\mathcal{T}}{2\pi}\sum_{Q=1}^{\infty} \int_{-\infty}^{\infty}du\frac{dp^Q}{du}\log\left(1+\frac{1}{Y^Q}\right).
\label{eq:freeenergyTBAXXXintermsofY}
\end{equation}
In principle, we now have the free-energy of the model. It is enough to solve the TBA equations \eqref{eq:TBAXXXwithY} and substitute the result in \eqref{eq:freeenergyTBAXXXintermsofY}. However, this is not such an easy task. Notice that the sum in $Q$ goes up to infinity. As a consequence, we have infinitely many equations, and infinitely many Kernels, $\rho^{Q}(u)$'s, etc. In addition, the limits of integration are also infinite. As a consequence, solving these equations numerically requires some approximations. For example, one has to consider a finite number of them instead, by using the assumption that as larger the bound state, more it contributes to $Y^Q$ functions, and less to the free energy (since the integrand is of the form $\log\left(1+\frac{1}{Y^Q}\right)$). 

In addition, excited states can also be obtained. They are usually constructed using an analytical continuation strategy, first introduced in \cite{Dorey:1996re} (for a pedagogical introduction see section 3.2 in \cite{vanTongeren:2016hhc}).

In the context of standard integrable spin chains the TBA has been applied to several different models (to cite a few \cite{Babelon:1982mc,deVega:1993sw,Mezincescu:1993pk,Woynarovich:1981ca}, see also the reviews \cite{Mezincescu:1992yt,vanTongeren:2016hhc} and the book \cite{essler2005one}). The string configurations are not always as simple as the ones for the $\alg{su}(2)$ invariant XXX spin chain. But in general the procedure works well and the computation of the TBA equations and corresponding ground-state energy (and excited states) can be successfully achieved.

\subsubsection{Summary}\label{ana:subsubsec:summaryTBAXXX}

Let us now summarize the TBA approach for integrable spin chains:

\begin{itemize}
\item The first step consists in constructing the Bethe ansatz itself (either via coordinate or algebraic approach) for the finite periodic spin chain.
\item Then for $ L\rightarrow \infty $ study the poles of the Bethe equations and construct string complexes. Verify that these string configurations form bound states.
\item Obtain the Bethe equations in terms of the center of these complexes. This will teach you how fundamental particles scatter these bound states.
\item Assume that all configurations relevant for the free-energy are of string-type when the number of sites and magnons go to infinity. This comes by the name of String Hypothesis. Use this to take the thermodynamic limit.
\item Use fusion to write the Bethe equations relating $ Q_a $-strings with $ Q_b $-strings. This will tell you, in particular, how these $Q$-excitations scatter each other.
\item Proceed with the computation of the counting function by using the new Bethe equations. If possible make sure this function increases monotonically with momentum (this can be hard to check and is not at all obvious for some models).
\item Compute $ \delta f=0  $ by using that the sum of the density of ``particles'' and the density of holes correspond to the derivative of the counting function with respect to the rapidity.
\item Simplify and write the TBA equations.
\item Substitute the TBA equations in the $ f $ and obtain the final form of the free energy.
\item Write the final results in terms of Y-functions.
\end{itemize}

Additionally, after having the Y-system it is important to

\begin{itemize}
    \item Solve the equations numerically, in order to obtain a quantitative analysis of the spectrum of the model.
    \item If possible, investigate how good the string hypothesis is for your model of interest. This is made using some numerical analysis for large but finite $L$.
\end{itemize}
\subsection{Mirror TBA for \texorpdfstring{AdS$_3$/CFT$_2$}{AdS3/CFT2} }\label{ana:subsec:TBAforAdS3}

\subsubsection{Introduction}\label{ana:subsubsec:IntroductionAdS3}

We are now ready to discuss the Thermodynamic Bethe ansatz for AdS$ _3$/CFT$_2 $. On the one hand, we will see that the procedure here will be very similar to what we just applied for the Heisenberg model. In particular, we will look for string configurations, formulate a string hypothesis, compute densities using the counting functions and look for the stationary points in the free energy in order to compute the TBA equations. 

On the other hand, there are fundamental differences. $AdS_3\times S^3 \times T^4$ has some interesting properties that were absent in the XXX spin chain. First of all, instead of only one type of fundamental excitation (magnon), we now have three (two massive and one massless). Furthermore, due to the analytic properties of the S-matrix, not all integrals will run along the same interval. 

Additionally, we are now interested in the spectrum of the string (and therefore in the asymptotic Bethe ansatz) instead of a spin chain.  In principle the two problems look very similar. However, when looking closer this is not exactly the case. We can start by taking the decompactification limit $R\rightarrow \infty$, where the worldsheet cylinder can be thought of as a plane. As already discussed in section \ref{s:fiona}, this allows us to have well defined asymptotic states, compute the scattering matrix and the Bethe-Yang equations. The problem is that the asymptotic Bethe equations are valid only up to the so-called wrapping corrections. In $AdS_3$ we have that particles can wrap around the worldsheet cylinder and disregarding such effects leads to an incorrect energy spectrum. The problem is aggravated by the presence of massless modes which make the wrapping corrections become important already at order $1/R$ \cite{Abbott:2015pps}, instead of the exponential suppression usually present for massive modes. 

So, the question is: \textit{is there a way to compute the string spectrum for the theory at finite volume instead?} The answer to this question is yes and its solution relies on a trick first introduced by Zamolodchikov in  \cite{Zamolodchikov:1989cf}. The idea consists of a double Wick rotation that ``exchanges'' the ground state energy of a theory with finite volume at zero temperature, by the free-energy of a theory whose volume is infinitely large at finite temperature. The latter theory is usually called \textbf{mirror theory} and its Bethe ansatz can be computed in this regime. 

In relativistic theories, the original and mirror theories are basically the same. In particular, their dispersion relations coincide. 
However, in $AdS$ backgrounds the original theory and its mirror counterpart are generally very different. Nonetheless, the strategy can remarkably still be applied\footnote{But it contains additional steps, including the computation of the mirror S-matrix and dressing phases.} and the finite volume spectrum of the original theory can be successfully obtained \cite{Arutyunov:2007tc,Frolov:2021bwp}.

Another subtlety of this model when compared with the XXX chain is its richer analytic structure. In particular, it requires the analysis not only of the poles, but also of the zeros of the S-matrices when taking the infinite volume limit on the mirror Bethe-Yang equations.

\subsubsection{Double Wick-rotation and Mirror model}

The mirror model mentioned above will allow us to indirectly construct the ground state energy for the finite-size theory, at zero temperature. Let us now understand why. Along this section, unless stated otherwise, we write the variables of the mirror theory with the same name as in the original theory, but with an extra tilde. Please also notice that in this review we consider that the volume of the original theory is $R$ and the volume of the mirror theory is $L$, which is the opposite of the convention used in \cite{Frolov:2021bwp}.

\paragraph{Partition Function: original theory \textit{versus} mirror theory.}\

\begin{figure}[ht]
\begin{centering}
\hspace{1cm}\includegraphics[width=13cm]{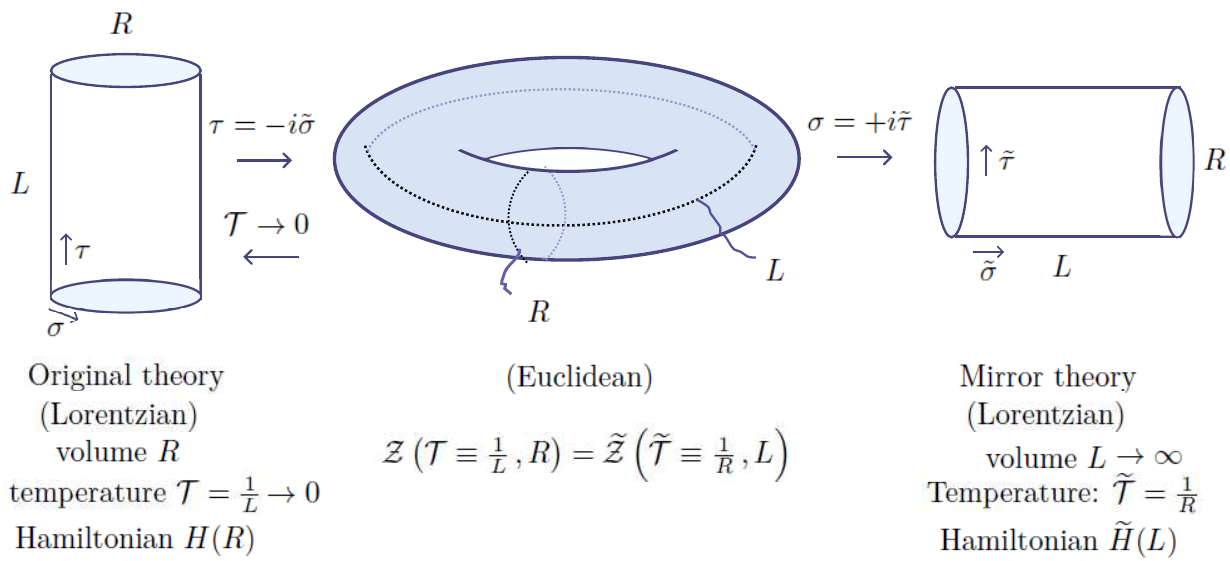}
\end{centering}
	\caption{We can interpret this as the time evolution of the system happening in two different cycles on the torus. In particular, the Euclidean partition function can be computed from two different theories (with Lorentzian signature): the original and the mirror theories.}
 \label{fig:torus}
\end{figure}
The double Wick rotation corresponds to 
\begin{equation}
    \tau\rightarrow \tilde{\sigma}=i\tau \qquad \text{and}\qquad \sigma\rightarrow \tilde{\tau}=-i \sigma.
\end{equation}
The theory after the first Wick rotation is the Euclidean version of the original theory, and it is described by the following partition function 
\begin{equation}
    \mathcal{Z}\left(\mathcal{T}\equiv \frac{1}{L},R\right)=\sum_n e^{-\frac{E_n(R)}{\mathcal{T}}}=\sum_n e^{-L\,E_n(R)}.
    \label{eq:originalpartition}
\end{equation}
This can be thought as putting the theory on a torus. For $L\rightarrow \infty$ (\textit{i.e.}~zero temperature) the torus becomes again the original cylinder. For a schematic representation of this and other details related to the mirror theory, see Figure \ref{fig:torus}.

After the second Wick rotation, we can define the partition function for the mirror theory as
\begin{equation}
    \widetilde{\mathcal{Z}}\left(\widetilde{\mathcal{T}}\equiv \frac{1}{R},L\right)=\sum_n e^{-\frac{\tilde{E}_n(L)}{\widetilde{\mathcal{T}}}}=\sum_n e^{-R\,\tilde{E}_n(L)}.
    \label{eq:mirrorpartition}
\end{equation}
These two partition functions have to be equal.\footnote{An explicit argument about why this is the case can be found in \cite{Arutyunov:2007tc}.}
\begin{equation}
        \mathcal{Z}\left(L,R\right)=\widetilde{\mathcal{Z}}\left(R,L\right).
        \label{eq:equalpartition}
\end{equation}
As we are interested in the original theory, with finite volume $R$ and zero temperature $\mathcal{T}=\frac{1}{L}\rightarrow 0$, we would like to take the limit $L\rightarrow\infty$ in \eqref{eq:originalpartition}. This limit generates $\mathcal{Z}\sim e^{-LE_0(R)}$. But given \eqref{eq:equalpartition}, we need to take the same $L\rightarrow\infty$ limit on \eqref{eq:mirrorpartition}. Therefore, at $L\rightarrow\infty$ 
\begin{align}
    E_0(R)&=\frac{R}{L}\tilde{E}_0(L)\equiv R \tilde{f}.
\end{align}
Hence, by computing the free-energy for the mirror theory $\tilde{f}$ at finite temperature, which is a well-defined problem using the Thermodynamic Bethe ansatz, we will obtain as a consequence the ground-state energy of the original theory at finite volume. 

%We have changed the notation to $\tilde{f}$ to match with the notation used for the XXX model. 

Let us now see how some other quantities are affected by the mirror transformation.

\paragraph{Dispersion relation: original theory \textit{versus} mirror theory.}
Given that the roles of space and time are swapped due to the mirror transformations, naturally we can expect the same  for momentum and energy. Indeed the momentum $p$ and the energy $H$ of the original theory transform as \footnote{Notice that sometimes people use the parity-reverse of the mirror model. So the reader, will find in the literature expressions that differ from \eqref{eq:mirror}} 

\begin{equation}
    p=i \tilde{H}, \qquad H=i\tilde{p},
    \label{eq:mirror}
\end{equation}
where $\tilde{p}$ and $\tilde{H}$ are their counterparts in the mirror theory.

For a relativistic model, the dispersion relation remains invariant
\begin{equation}
    H^2=p^2+m^2 \rightarrow \tilde{H}^2=\tilde{p}^2+m^2, 
\end{equation}
under the double Wick rotation. Therefore, the original theory and the mirror theory are actually the same. What about AdS$_3$?

As discussed in section \eqref{sec:one-particle}, the dispersion relation for AdS$_3$ is given by
\begin{equation}
    H=\sqrt{\left(\mu+\frac{kp}{2\pi}\right)^2+4h^2\sin^2\left(\frac{p}{2}\right)},
\end{equation}
where $k$ is related to the amount of NSNS flux, while $h$ is related to the amount of RR flux. In particular, we have
\begin{equation}
    H=\begin{cases}
        \sqrt{\mu^2+4h^2\sin^2\left(\frac{p}{2}\right)} & \text{for pure-RR},\\
        \hfil \Big|\mu+\frac{kp}{2\pi}\Big| & \text{for pure-NSNS},
    \end{cases}
\end{equation}
which under the transformation \eqref{eq:mirror} becomes
\begin{equation}
    \tilde{H}=\begin{cases}       2\text{arcsinh}\left(\frac{\sqrt{\mu^2+\tilde{p}^2}}{2h}\right) & \text{for pure-RR},\\
        \hfil \frac{2\pi}{k}(|\tilde{p}|+i\mu)& \text{for pure-NSNS}.
    \end{cases}
\end{equation}
We immediately see that, contrary to the relativistic case, the original theory and the mirror theory in AdS$_3$ are very different. Moreover, in the pure-NSNS case, \textit{the mirror energy is not real}. In fact, while the mirror dispersion relation cannot be written in terms of elementary functions for $h\neq0\neq k$, it is easy to see that in that case too, the mirror energy is not real. We might well worry that this signals a breakdown of our construction. However, for $h=0$ and $k>0$, the mirror TBA can be worked out despite the presence of complex energy levels~\cite{Dei:2018mfl}. It has been argued that this can be done even for generic $h,k$~\cite{Baglioni:2023zsf}. In both cases, the crux of the argument is that all states' energies in the mirror partition function must come in pairs with complex-conjugate momenta. This is because the values of $\mu$ come in pairs of opposite signs, which is itself a consequence of crossing symmetry of the original (string) model.  

In what follows, we focus on the pure-RR  case. Notice that the sign in the mirror dispersion relation was chosen such that a real mirror particle with $\tilde{p}\in\mathbb{R}$ has mirror energy $\tilde{H}\ge 0$. It is interesting to note that the mirror dispersion relation can be seen as the analytic continuation of the string one, by introducing a suitable (elliptic) parametrisation. In fact, this can be done for the whole representation of the symmetry algebra. This fact was first observed in~\cite{Arutyunov:2007tc} and then extended to the case of massless particles in~\cite{Frolov:2021zyc}. As a consequence, because the matrix part of the S matrix is fixed by representation theory, \textit{the mirror S matrix can be obtained from the string one by analytic continuation}. It is worth noting that this is not automatically true for the dressing factors; rather, we can \textit{impose} as one of the requirements of the bootstrap procedure that the dressing factors should enjoy a meaningful (unitary, analytic, etc.) continuation to the mirror kinematics. In this way we can think of having a ``\"uber-model'' which encompasses the mirror and string kinematics.%
\footnote{%
Indeed, even if we will not see this, to fully describe the spectrum using the mirror TBA it is sometimes necessary to consider an S~matrix with one ``leg'' in the string kinematics and another in the mirror one.  
}
It is a remarkable fact that such dressing factors could be constructed first for $AdS_5\times S^5$ and then, building on their properties~\cite{Arutyunov:2009kf}, for $AdS_3\times S^3\times T^4$~\cite{Frolov:2021fmj}. This was the key ingredient that was needed to derive the mirror TBA equations of~\cite{Frolov:2021bwp}.

\paragraph{Zhukovsky variables.} Let us use this opportunity to remind the reader about a very useful set of variables, since they make several expressions, including the S-matrix itself, look much simpler. They are  $x^\pm$, $x^\pm=x\left(u\pm \frac{i}{h}\right)$ with
\begin{equation}
x(u)=\frac{1}{2}\left(u-i\sqrt{4-u^2}\right)
\label{eq:Zhukovskyvar}
\end{equation}
and are called Zhukovsky variables.

Using them, let us see how the energy and the momentum of massive particles are affected by the double Wick rotation. 

As we have seen in \eqref{eq:energymomentum}, the energy and momentum of the massive particles can be written as 
\begin{equation}
    \frac{2iH}{h}=x^+-\frac{1}{x^+}-x^-+\frac{1}{x^-}, \qquad \text{and} \qquad e^{ip}=\frac{x^+}{x^-},
\end{equation}
respectively. Under the mirror transformation \eqref{eq:mirror} they become
\begin{equation}
    \tilde{H}=\log\left(\frac{x^-}{x^+}\right), \qquad \text{and} \qquad \tilde{p}=-\frac{h}{2}\left(x^+-\frac{1}{x^+}-x^-+\frac{1}{x^-}\right),
\end{equation}
and can be rewritten as
\begin{equation}
    \tilde{H}=\log\frac{x\left(u-\frac{i}{h}\right)}{x\left(u+\frac{i}{h}\right)}, \qquad \text{and} \qquad \tilde{p}=hx\left(u-\frac{i}{h}\right)-hx\left(u+\frac{i}{h}\right)+i.
\end{equation}
So, again very distinct from the original theory.

\subsubsection{Excitations contributing to the Mirror Bethe-Yang equations}

In the XXX model we had only one type of fundamental particle. Similarly, in $AdS_5\times S^5$ there was only one type of momentum carrying particle. However, in the case of $AdS_3$ strings in the pure-RR case, we have three distinct types of excitations, corresponding to $\mu=0$, $\pm1$. This creates an additional difficulty, since we cannot write both the Bethe-Yang equations associated to $\mu=+1$ and to $\mu=-1$ in a simple way. We are required to choose one of them to be simple and as a consequence will have little control about that happens to the other type.

\paragraph{$\mathbf{\mu=+1}$:} When associated to $\mu=+1$ the particles are called  ``left''-excitations. There are $N_1$ of them. Following \cite{Frolov:2021bwp}, we choose the Bethe-Yang equations for left-momentum carrying modes to have a simple form. As a result, the study of string configurations for these equations will be very simple. In particular, only the poles of the S-matrix will play a role in the construction of these patterns. For a better understanding of the reasons behind this, the reader can see \cite{Seibold:2022mgg}, focusing especially on the discussion about grading choices.

\paragraph{$\mathbf{\mu=-1}$:} There are $N_{\bar{1}}$ ``right''-momentum carrying modes, related to $\mu=-1$. Given the choice of making the left-particles' Bethe-Yang equations simpler, the computation of the string complexes for the \textit{right} ones will be a lot more involved. In particular, it will be necessary to analyse both the poles and zeros of the Bethe-Yang equations in this case.

\paragraph{$\mathbf{\mu=0}$:} Finally, there are the massless excitations due to $\mu=0$. There are $ N_0=N_0^{(1)}+N_0^{(2)} $ modes. These are in a doubet of $\alg{su}(2)$ algebra (called $\alg{su}(2)_\circ$), so the upper indices correspond to a label $\dot{\alpha}=1,2$. 

\paragraph{Auxiliary roots:} In addition to the fundamental excitations, there is one more type of particles appearing in the Bethe-Yang equations.
There are $N_y=N_y^{(1)}+N_y^{(2)}$ auxiliary modes, with rapidity $y_k$. These roots transform in the fundamental representation of a second $\alg{su}(2)$ algebra (called $\alg{su}(2)_\bullet$), for which we introduce the index $\alpha=1,2$. This type of root does not carry momenta and therefore will not contribute for the mirror energy.

\medskip
From these excitations, the ones carrying momentum can be described by a mirror momentum $ \tilde{p}_k $ or by a rapidity, $ u_k $. In several cases it is convenient to use both at the same time. The rapidity for the auxiliary roots is denoted by $ y_k $.

\subsubsection{The Bethe-Yang equations and their string configurations}\label{ana:subsubsec:StringsAdS3}

\

Let us now see which type of string we find when we take the $ L\rightarrow \infty $ limit. Although the dressing factors  $\sigma_{ij}$ are very important (see for example section \ref{fiona:subsec:fullSmatrix}), we will not enter in detail about them. They are constructed in \cite{Frolov:2021fmj} and summarised in Appendix C of \cite{Frolov:2021bwp}.
The relevant information for our computation is the fact that they have a nice analytic structure, in particular, containing no poles that will contribute to any of the bound states \cite{Frolov:2021fmj}.

In principle, given the number of different fundamental excitations one could perhaps expect to have several types of corresponding string-configurations. This is actually not the case, since both massless particles and auxiliary particles do not form bound states. 

Let us start by investigating the ``left''-excitations.

\paragraph{Massive ``left''-modes.}
The Bethe-Yang equations for the ``left'' excitations are given by
\begin{equation}
e^{i\tilde{p}_kL}\prod_{j\neq k}^{N_1}S_{\alg{sl}}^{11}(u_k,u_j)\prod_{j=1}^{N_{\bar{1}}}\tilde{S}^{11}_{\alg{sl}}(u_k,u_j)\prod_{\dot{\alpha}=1}^{2}\prod_{j=1}^{N_{0}^{(\dot{\alpha})}}S^{10}(u_k,u_j^{(\dot{\alpha})})\prod_{\alpha=1}^{2}\prod_{j=1}^{N_{y}^{(\alpha)}}S^{1y}(u_k,y_j^{(\alpha)})=1,
\end{equation}
where $ k=1,...,N_1 $ and the S-matrices are given by
\begin{align}
&S_{\alg{sl}}^{11}(u_k,u_j)=\frac{x_k^+-x_j^-}{x_k^--x_j^+}\frac{1-\frac{1}{x_k^-x_j^+}}{1-\frac{1}{x_k^+x_j^-}}\left(\sigma_{kj}^{\bullet\bullet}\right)^{-2}, \quad \tilde{S}_{\alg{sl}}^{11}(u_k,u_j)=e^{ip_k}\frac{1-\frac{1}{x_k^+x_j^+}}{1-\frac{1}{x_k^-x_j^-}}\frac{1-\frac{1}{x_k^-x_j^+}}{1-\frac{1}{x_k^+x_j^-}}\left(\tilde{\sigma}_{kj}^{\bullet\bullet}\right)^{-2},\nonumber\\
&S^{10}(u_k,u_j)=e^{-\frac{i}{2}p_k}e^{-ip_j}\frac{1-x_k^+x_j}{x_k^--x_j}\left(\tilde{\sigma}_{kj}^{\bullet\circ}\right)^{-2}, \quad S^{1y}(u_k,y_j)=e^{\frac{i}{2}p_k}\frac{x_k^--y_j}{x_k^+-y_j}.
\label{eq:SmatrixleftAdS3}
\end{align}

Let us now think about the string configurations for $ L\rightarrow \infty $ assuming that Im$(\tilde{p_1})>0  $. The equation for $ k=1 $ is given by
\begin{align}
e^{i\tilde{p}_1L}\textcolor{blue}{\prod_{j\neq k}^{N_1}S_{\alg{sl}}^{11}(u_1,u_j)}\prod_{j=1}^{N_{\bar{1}}}\tilde{S}^{11}_{\alg{sl}}(u_1,u_j)\prod_{\dot{\alpha}=1}^{2}\prod_{j=1}^{N_{0}^{\dot{\alpha}}}S^{10}(u_1,u_j^{(\dot{\alpha})})\prod_{\alpha=1}^{2}\prod_{j=1}^{N_{y}^{\alpha}}S^{1y}(u_1,y_j^{(\alpha)})=1.
\label{eq:BEleftbeforefusion}
\end{align}
In this limit we have that $e^{i\tilde{p}_1L}\rightarrow 0$, therefore we need to have a pole in one of the S-matrices. By looking at \eqref{eq:SmatrixleftAdS3}  we notice that using the term in blue we can generate such a pole. Consequently, from $ S_{\alg{sl}}^{11}(u_1,u_2) $ we see that there should be a pole at \begin{equation}
x_1^-=x_2^+.
\end{equation}
If we now multiply the equation for $ k=1 $ by the one for $ k=2 $ and continue the procedure for a few times in a similar way as done for XXX chain, we see that
\begin{equation}
x_1^-=x_2^+, \quad x_2^-=x_3^+, \quad x_3^-=x_4^+, \quad ...\\
\end{equation}
such that
\begin{equation}
x_j^-=x_{j+1}^+, \quad j=1,...,Q.
\label{eq:xjpm}
\end{equation}
Notice that we can write
\begin{equation}
x_j^\pm=x\left(u_j\pm\frac{i}{h}\right)=\frac{1}{2}\left(u_j\pm \frac{i}{h}-i\sqrt{4-\left(u_j\pm \frac{i}{h}\right)^2}\right).
\label{eq:xjpmintermsofu}
\end{equation}
If we substitute \eqref{eq:xjpm} in \eqref{eq:xjpm} it becomes
\begin{equation}
u_{j}-\frac{i}{h}-i\sqrt{4-\left(u_j- \frac{i}{h}\right)^2}=u_{j+1}+\frac{i}{h}-i\sqrt{4-\left(u_{j+1}+ \frac{i}{h}\right)^2},
\end{equation}
whose string complex solutions is given by 
\begin{equation}
u_j=u+\frac{Q+1-2j}{h}i, \quad j=1,...Q.
\end{equation}
So we have bound states with $ Q $ left momentum-carrying particles\footnote{Notice that this is basically the same string pattern we found in the XXX spin chain (exactly the same if we put $h=-1$) .}. Their bound states have $\alg{u}(1) $ charge $ M $ given by a positive $ Q\in \mathbb{Z} $.

\paragraph{Massive ``right''-modes.}
The Bethe-Yang equation for the ``right'' particles are given by
\begin{equation} \label{eq:BErightbeforefusion}
\textcolor{red}{e^{i\tilde{p}_kL}}\prod_{j= 1}^{N_1}\tilde{S}^{11}_{\alg{su}}(u_k,u_j)\textcolor{red}{\prod_{j\neq k}^{N_{\bar{1}}}S_{\alg{su}}^{11}(u_k,u_j)}\prod_{\dot{\alpha}=1}^{2}\prod_{j=1}^{N_{0}^{\dot{\alpha}}}\bar{S}^{10}(u_k,u_j^{(\dot{\alpha})})\textcolor{blue}{\prod_{\alpha=1}^{2}\prod_{j=1}^{N_{y}^{\alpha}}\bar{S}^{1y}(u_k,y_j^{(\alpha)})}=1
\end{equation}
where $ k=1,...,N_{\bar{1}} $ with S-matrices
\begin{align}
&S_{\alg{su}}^{11}(u_k,u_j)=e^{ip_k-ip_j}\frac{x_k^--x_j^+}{x_k^+-x_j^-}\frac{1-\frac{1}{x_k^+x_j^-}}{1-\frac{1}{x_k^-x_j^+}}\left(\sigma_{kj}^{\bullet\bullet}\right)^{-2}, \, \bar{S}^{1y}(u_k,y_j)=\frac{1}{S^{1y}(u_k,\frac{1}{y_j})},\\
&\tilde{S}_{\alg{su}}^{11}(u_k,u_j)=e^{-ip_j}\frac{1-\frac{1}{x_k^-x_j^-}}{1-\frac{1}{x_k^+x_j^=}}\frac{1-\frac{1}{x_k^-x_j^+}}{1-\frac{1}{x_k^+x_j^-}}\left(\tilde{\sigma}_{kj}^{\bullet\bullet}\right)^{-2}, \, \bar{S}^{10}(u_k,u_j)=e^{\frac{i}{2}p_k}e^{-ip_j}\frac{x_k^--x_j}{1-x_k^+x_j}\left(\tilde{\sigma}_{kj}^{\bullet\circ}\right)^{-2}.\nonumber
\label{eq:BErightbeforefusion}
\end{align}
We will see that here we have $ \bar{Q} $-right momentum carrying roots plus $2(\bar{Q}-1)$ auxiliary roots.
So, we again have

\begin{equation}
x_j^-=x_{j+1}^+, \quad j=1,...,Q-1,
\end{equation}
which in $ u $ variables translates as
\begin{equation}
u_j=u+\frac{\bar{Q}+1-2j}{h}i, \quad j=1,...\bar{Q}.
\label{eq:stringforQbar}
\end{equation}
However, notice that this produces a zero in $ S_{\alg{su}}^{11}(u_k,u_j) $ instead of a pole. This means that we have a zero coming from $ e^{i\tilde{p}_kL}\rightarrow 0 $ and another from $ S_{\alg{su}}^{11}(u_k,u_j) $ in \eqref{eq:BErightbeforefusion} in red. We need then two poles to cancel these two zeros. They come from the term in blue. The first terms give for instance for $ k=1 $
\begin{equation}
\prod_{\alpha=1}^{2}\bar{S}^{1y}(u_1,y_1^{(\alpha)})=e^{-\frac{ip_1}{2}}\frac{x_1^+-\frac{1}{y_1^{(1)}}}{x_1^--\frac{1}{y_1^{(1)}}}e^{-\frac{ip_1}{2}}\frac{x_1^+-\frac{1}{y_1^{(2)}}}{x_1^--\frac{1}{y_1^{(2)}}}.
\end{equation}
Therefore, we have a pole at $ y_1^{(1)}=\frac{1}{x_1^-} $ and another at  $ y_1^{(2)}=\frac{1}{x_1^-} $.
If we continue with this procedure what we obtain is 
\begin{equation}
y_j^{(1)}=y_j^{(2)}=\frac{1}{x_j^-}, 
\end{equation}
which using $ x_j^-=x(u_j-\frac{i}{h}) $ and equation \eqref{eq:stringforQbar}
lead us to 
\begin{equation}
y_j^{(1)}=y_j^{(2)}=\frac{1}{x_j^-}=\frac{1}{x\left(u+\frac{(\bar{Q}-2j)i}{h}\right)}, \quad j=1,...,\bar{Q}-1.
\end{equation}

\paragraph{Massless modes.} The Bethe equations for the massless case are given by
\begin{align}
&e^{i\tilde{p}_kL}\prod_{j= 1}^{N_1}S^{01}(u_k^{(1)},u_j)\prod_{j=1}^{N_{\bar{1}}}\bar{S}^{01}(u_k^{(1)},u_j)\prod_{j=1}^{N_{0}^{(1)}}\bar{S}^{00}(u_k^{(1)},u_j^{(1)})\prod_{j=1}^{N_{0}^{(2)}}\bar{S}^{00}(u_k^{(1)},u_j^{(2)})\times\nonumber\\
&\hspace{3cm}\times\prod_{\alpha=1}^{2}\prod_{j=1}^{N_{y}^{(\alpha)}}\bar{S}^{0y}(u_k^{(1)},y_j^{(\alpha)})=-1.
\label{eq:BEmasslessbefore}
\end{align}
In particular, the scattering matrices between massless particles and the other types of excitation are given by
\begin{align}
    &S^{00}(u_k,u_j)=\left(\sigma_{kj}^{\circ \circ}\right)^{-2}, \quad && S^{01}(u_k,u_j)=\frac{1}{S^{10}(u_j,u_k)},\\
    &\bar{S}^{01}(u_k,u_j)=\frac{1}{\bar{S}^{10}(u_j,u_k)}, \quad && S^{0y}(u_k,u_j)=\frac{1}{S^{0y}\left(u_k,\frac{1}{y_j}\right)}.
\end{align}

Massless excitations do not admit bound states, since their momenta and rapidities are real. In particular $-1<x^{\dot{\alpha}}<1$. The thermodynamic limit of equation \eqref{eq:BEmasslessbefore} can be taken directly without any problems. 

\paragraph{Auxiliary equations.} The auxiliary roots satisfy the following Bethe equations:
\begin{equation}
\prod_{j= 1}^{N_1}S^{y1}(y_k,u_j)\prod_{j=1}^{N_{\bar{1}}}\bar{S}^{y1}(y_k,u_j)\prod_{\alpha=1}^{2}\prod_{j=1}^{N_{y}^{\dot{\alpha}}}\bar{S}^{y0}(y_k,y_j^{(\dot{\alpha})})=-1.
\end{equation}
They do not carry momentum and therefore it does not make sense to create strings from them.

Let us now summarize the existing string configurations.

\begin{centering}
\begin{tcolorbox}
\begin{itemize}
    \item $Q$-particles: 
\begin{equation}
u_j=u+\frac{Q+1-2j}{h}i, \quad j=1,...Q.
\end{equation}
We will represent the density of $Q$-particles as $\rho^{Q}$ and the density of $Q$-holes as $\bar{\rho}^Q$ such that they satisfy $\rho_t^{Q}=\rho^{Q}+\bar{\rho}^Q$.
\item $\bar{Q}$-particles:
\begin{equation}
u_j=u+\frac{\bar{Q}+1-2j}{h}i, \quad j=1,...\bar{Q}.
\end{equation}
and additionally
\begin{equation}
y_j^{(1)}=y_j^{(2)}=\frac{1}{x_j^-}=\frac{1}{x\left(u+\frac{(\bar{Q}-2j)i}{h}\right)}, \quad j=1,...,\bar{Q}-1.
\end{equation}
Again, the density of $\bar{Q}$-particles as $\rho^{\bar{Q}}$ and the density of $\bar{Q}$-holes as $\bar{\rho}^{\bar{Q}}$ such that they satisfy $\rho_t^{\bar{Q}}=\rho^{\bar{Q}}+\bar{\rho}^{\bar{Q}}$.
\item Massless particles: 
They do not form string configurations. They appear as excitations of real mirror momentum and real energy. The rapidities are such that $|u|>2$. The density of massless particles can be written as $\rho^{(\dot{\alpha})}$ while massless holes have density $\bar{\rho}^{(\dot{\alpha})}$, with $\rho_t^{(\dot{\alpha})}=\rho^{(\dot{\alpha})}+\bar{\rho}^{(\dot{\alpha})}$ and $\dot{\alpha}=1,2$.
\item Auxiliary particles: they do not carry momentum and cannot generate string complexes. They satisfy $(y^{(\alpha)})^*=(y^{(\alpha)})^{-1}$. Similarly to the massless case, we can also define densities for the auxiliary particles. In addition to defining particles and holes we also divide these roots and corresponding densities into $y^+$ (when Im $y>0$) and $y^-$ (when Im $y<0$).
\end{itemize}
\end{tcolorbox}
\end{centering}

\subsubsection{String Hypothesis}\label{ana:subsubsec:StringHypothesisAdS3}

As we saw for the XXX model, we can write the Bethe equations completely in terms of the centers of strings. In the next subsection we will do this for AdS$ _3 $. The equations written in that way are very convenient for the thermodynamic limit, \textit{i.e.}

\begin{equation}
L\rightarrow \infty, \quad  N\rightarrow \infty, \quad \text{with } \quad \frac{L}{N}=\text{fixed}.
\end{equation}
In particular, in these new equations the momentum (rapidity) is always real.

In AdS$ _3 $ this idea will result in four types of objects, each connected to one type of excitation. 
Namely, $ N_L^{Q} $ and $ N_R^{Q} $ satisfying 
\begin{equation}
\sum_{Q=1}^{\infty}QN_L^{(Q)}=N_L \quad \text{and} \quad \sum_{\bar{Q}=1}^{\infty}\bar{Q}N_R^{(\bar{Q})}=N_R,
\label{eq:firststepstringhypAdS3}
\end{equation}
two sets of $ N_0^{(\dot{\alpha})} $ massless particles, with $ p_k\in \mathbb{R} $ and $ |u|>2 $, and finally the two sets of auxiliary particles corresponding to the auxiliary roots $ y^{(\alpha)} $.

It is important to highlight that the auxiliary roots do not carry momentum and therefore the mirror energy is the sum of the contributions of the massive and massless particles only.

\subsubsection{Free-energy}

The free energy of the mirror theory is made of contributions from $Q$-particles, $\bar{Q}$-particles, massless particles and auxiliary particles. It is given by
\begin{equation}
\tilde{f}=\tilde{f}^Q+\tilde{f}^{\bar{Q}}+\tilde{f}^{(0)}+\tilde{f}^{y^-}+\tilde{f}^{y^+}
\label{eq:freenergyAdS3beforeTBA}
\end{equation}
where
\begin{equation}
\begin{aligned}
    & \tilde{f}^{Q}=\sum_{Q=1}^{\infty}\int_{-\infty}^{+\infty}du\left(\tilde{H}^Q(u)\rho^Q(u)-\widetilde{\mathcal{T}}\left(\rho^{Q}(u)\log\frac{\rho_t^Q(u)}{\rho^Q(u)}+\bar{\rho}^{Q}(u)\log\frac{\rho_t^Q(u)}{\bar{\rho}^Q(u)}\right)\right),\\
    & \tilde{f}^{\bar{Q}}=\sum_{\bar{Q}=1}^{\infty}\int_{-\infty}^{+\infty}du\left(\tilde{H}^{\bar{Q}}(u)\rho^{\bar{Q}}(u)-\widetilde{\mathcal{T}}\left(\rho^{\bar{Q}}(u)\log\frac{\rho_t^{\bar{Q}}(u)}{\rho^{\bar{Q}}(u)}+\bar{\rho}^{\bar{Q}}(u)\log\frac{\rho_t^{\bar{Q}}(u)}{\bar{\rho}^{\bar{Q}}(u)}\right)\right),\\    &\tilde{f}^{(0)}=\sum_{\dot{\alpha}=1}^{2}\int_{|u|>2}du\left(\tilde{H}^{(0)}(u)\rho^{(\dot{\alpha})}(u)-\widetilde{\mathcal{T}}\left(\rho^{(\dot{\alpha})}(u)\log \frac{\rho_t^{(\dot{\alpha})}(u)}{\rho^{(\dot{\alpha})}(u)}+\bar{\rho}^{(\dot{\alpha})}(u) \log \frac{\rho_t^{(\dot{\alpha})}(u)}{\bar{\rho}^{(\dot{\alpha})}(u)}\right)\right),\\    &\tilde{f}^{(y^\pm)}=-\widetilde{\mathcal{T}}\sum_{\alpha=1}^{2}\int_{-2}^{+2}du\left(\rho_{y^{\pm}}^{(\alpha)}(u)\log \frac{\rho_{y^{\pm},t}^{(\alpha)}(u)}{\rho_{y^{\pm}}^{(\alpha)}(u)}+\bar{\rho}_{y^{\pm}}^{(\alpha)}(u) \log \frac{\rho_{y^{\pm},t}^{(\alpha)}(u)}{\bar{\rho}_{y^{\pm}}^{(\alpha)}(u)}\right).
    \end{aligned}
\end{equation}
Remember that $\widetilde{\mathcal{T}}=1/R$. Notice also that as expected there is no energy contribution of type $\tilde{H}^{y^\pm}$.

We would like to compute the value of $\tilde{f}$ when $\delta\tilde{f}=0$, where $\delta$ is the variation with respect to the densities of all types of particles. 
In order to do this computation, we need more information about the densities. This information comes from the fused mirror Bethe equations, which are the focus of the next subsection.

\subsubsection{Fusion of the Bethe-Yang equations}\label{ana:subsubsec:fusionAdS3}

As discussed above, the string complexes act as bound states. It is therefore useful to discover how such bound states scatter each other and how they scatter massless and auxiliary excitations. This information will play an important role in the computation of the mirror theory free energy.

The idea for fusion here is the same as in the Heisenberg model. We can build S-matrices describing the scattering of a particle with a Q-string first by multiplying  $ S^{11}(u_k-u_j) $'s (see for example \eqref{eq:SmatrixstringQXXX} and the equations leading to it ). This results in 
\begin{equation}
S^{Q1}(u,u_j)=\prod_{a=1}^{Q}S^{11}(u_a,u_j)=\frac{u-u_j-(Q+1)\frac{i}{h}}{u-u_j+(Q+1)\frac{i}{h}}\frac{u-u_j-(Q-1)\frac{i}{h}}{u-u_j+(Q-1)\frac{i}{h}}.
\end{equation}
In the same way we can build scatterings of a $ Q $-string with a $ Q^\prime $-string, by multiplying several $ S^{Q1}(u,u_j) $. This results in 
\begin{align}
S^{QQ^\prime}(u,u^\prime)&=\frac{u-u^{\prime}-(Q+Q^{\prime})\frac{i}{h}}{u-u^{\prime}+(Q+Q^{\prime})\frac{i}{h}}\frac{u-u^{\prime}-(Q^{\prime}-Q)\frac{i}{h}}{u-u^{\prime}+(Q^\prime-Q)\frac{i}{h}}\times\nonumber\\
& \hspace{1cm}\times \prod_{j=1}^{Q-1}\left(\frac{u-u^\prime-(Q^\prime-Q+2j)\frac{i}{h}}{u-u^\prime+(Q^\prime-Q+2j)\frac{i}{h}}\right)^2.
\end{align}
These expressions are again very similar to the XXX ones. Nevertheless, in this case, they do not describe the full picture. The reason for that is that we need to also know how the grading affects the expressions, take into account the dressing factor, as well as how the $\{Q,\bar{Q}\}$-excitations scatter massless and auxiliary particles, for example.

Despite that, although technically more challenging, all the procedure is very similar to what was performed before, so we will not repeat the calculations. Therefore, in order to make the presentation more fluid we will skip the details and write explicitly all the fused S-matrices directly as 
\begin{align}
& S_{\alg{sl}}^{Q_aQ_b}(u_a,u_b)=\frac{S^{Q_aQ_b}(u_a-u_b)^{-1}}{\left(\Sigma_{ab}^{Q_aQ_b}\right)^2}, \quad \tilde{S}_{\alg{sl}}^{Q_a\bar{Q}_b}(u_a,u_b) \frac{e^{ip_a}}{\left(\tilde{\Sigma}_{ab}^{Q_a\bar{Q}_b}\right)^2}\frac{1-\frac{1}{x_a^+x_b^+}}{1-\frac{1}{x_a^-x_b^-}}\frac{1-\frac{1}{x_a^+x_b^-}}{1-\frac{1}{x_a^-x_b^+}},\nonumber\\
&S^{Q_a0}(u_a,x_j)=ie^{-\frac{i}{2}p_a}\frac{x_a^+x_j-1}{x_a^--x_j}\frac{\left(\Sigma_{\text{BES}}^{Q_a0}(x_a^{\pm},x_j)\right)^{-2}}{\Phi(\gamma_{aj}^{+\circ})\Phi(\gamma_{aj}^{-\circ})},\quad S^{Q_ay}(u_a,y_b)=e^{\frac{i}{2}p_a}\frac{x_a^--y_b}{x_a^+-y_b},\nonumber\\
&S_{\alg{su}}^{\bar{Q}_a\bar{Q}_b}(u_a,u_b)=\left(\frac{x_a^--x_b^+}{x_a^+-x_b^-}\right)^2\frac{S^{\bar{Q}_a\bar{Q}_b}(u_a-u_b)^{-1}}{e^{-ip_a+ip_b}\left(\Sigma_{ab}^{\bar{Q}_a\bar{Q}_b}\right)^2},\quad S^{00}(u_j,u_k)=\frac{a(\gamma_{jk)}\Phi(\gamma_{jk})^2}{\left(\Sigma_{\text{BES}}^{00}(x_j,x_k)\right)^2},\nonumber\\
& \tilde{S}_{\alg{su}}^{\bar{Q}_aQ_b}(u_a,u_b)=\frac{e^{-ip_b}}{\left(\tilde{\Sigma}^{\bar{Q}_aQ_b}(u_a,u_b)\right)^2}\frac{1-\frac{1}{x_a^-x_b^-}}{1-\frac{1}{x_a^+x_b^+}}\frac{1-\frac{1}{x_a^+x_b^-}}{1-\frac{1}{x_a^-x_b^+}}, \quad S^{0Q_b}(x_b,u_j)=\frac{1}{S^{Q_b0}(u_j,x_b)},\nonumber\\
&\bar{S}^{\bar{Q}_a0}(u_a,x_j)=ie^{+\frac{i}{2}p_a}\frac{x_a^--x_j}{x_a^+x_j-1}\frac{\left(\Sigma_{\text{BES}}^{\bar{Q}_a0}(x_a^{\pm},x_j)\right)^{-2}}{\Phi(\gamma_{aj}^{+\circ}\Phi(\gamma_{aj}^{-\circ})}, \quad \bar{S}^{0\bar{Q}_b}(x_b,u_j)=\frac{1}{\bar{S}^{\bar{Q}_b0}(u_j,x_b)},\nonumber\\
&\bar{S}^{Q_ay}(u_a,y_b)=e^{-\frac{i}{2}p_a}\frac{x_a^+-\frac{1}{y_b}}{x_a^--\frac{1}{y_b}}=\frac{1}{S^{Q_ay}(u_a,\frac{1}{y_b})},\quad S^{0y}(x_k,y_j)=e^{+\frac{i}{2}p_k}\frac{\frac{1}{x_k}-y_j}{x_k-y_j}=\frac{1}{S^{0y}(x_k,\frac{1}{y_j})},\nonumber\\
&S^{yQ}(y,u)=\frac{1}{S^{Qy}(u,y)},\quad \bar{S}^{yQ}(y,u)=\frac{1}{\bar{S}^{Qy}(u,y)},\quad \bar{S}^{y0}(y,u)=\frac{1}{\bar{S}^{0y}(u,y)}.
\end{align}
For more details on the expressions above see \cite{Frolov:2023wji}.

Proceeding in a similar way as done in the spin chain case, we can write the fused Bethe-Yang equations as 
\begin{align}
&e^{i\tilde{p}_aL}\prod_{b\neq a}^{N_L}S_{\alg{sl}}^{Q_aQ_b}(u_a,u_b)\prod_{b=1}^{N_R}\tilde{S}^{Q_a\bar{Q}_b}_{\alg{sl}}(u_a,u_b)\prod_{\dot{\alpha}=1}^{2}\prod_{j=1}^{N_{0}^{\dot{\alpha}}}S^{Q_a0}(u_a,u_j^{(\dot{\alpha})})\times\nonumber\\
&\hspace{1cm}\prod_{\alpha=1}^{2}\prod_{b=1}^{N_{+}^{(\alpha)}}S_+^{Q_ay}(u_a,y_b^{(\alpha)})\prod_{b=1}^{N_{-}^{(\alpha)}}S_-^{Qy}(u_a,y_b^{(\alpha)})=1,
\label{eq:fusedQparticle}\\
&e^{i\tilde{p}_aL}\prod_{b=1}^{N_L}\tilde{S}_{\alg{su}}^{\bar{Q}_aQ_b}(u_a,u_b)\prod_{b\neq a}^{N_R}S^{\bar{Q}_a\bar{Q}_b}_{\alg{su}}(u_a,u_b)\prod_{\dot{\alpha}=1}^{2}\prod_{j=1}^{N_{0}^{\dot{\alpha}}}\bar{S}^{\bar{Q}_a0}(u_a,u_j^{(\dot{\alpha})})\times\nonumber\\
&\hspace{1cm}\prod_{\alpha=1}^{2}\prod_{b=1}^{N_{+}^{(\alpha)}}\frac{1}{S_-^{\bar{Q}_ay}(u_a,y_b^{(\alpha)})}\prod_{b=1}^{N_{-}^{(\alpha)}}\frac{1}{S_+^{\bar{Q}_ay}(u_a,y_b^{(\alpha)})}=1,
\label{eq:fusedQbarparticle}\\
&e^{i\tilde{p}_kL}\prod_{b=1}^{N_L}S^{0Q_b}(u_k,u_b)\prod_{b=1}^{N_R}\bar{S}^{0\bar{Q}_b}(u_k,u_b)\prod_{j\neq k}^{N_{0}^{(1)}}S^{00}(u_k,u_j)\prod_{j=1}^{N_{0}^{(2)}}S^{00}(u_k,u_j)\times\nonumber\\
&\hspace{1cm}\prod_{\alpha=1}^{2}\prod_{b=1}^{N_{+}^{(\alpha)}}S^{0y}(u_k,u_b)\prod_{b=1}^{N_{-}^{(\alpha)}}\frac{1}{S^{0y}(u_k,u_b)}=1,\label{eq:fusedmassless1}\\
&e^{i\tilde{p}_kL}\prod_{b=1}^{N_L}S^{0Q_b}(u_k,u_b)\prod_{b=1}^{N_R}\bar{S}^{0\bar{Q}_b}(u_k,u_b)\prod_{j=1}^{N_{0}^{(1)}}S^{00}(u_k,u_j)\prod_{j\neq k}^{N_{0}^{(2)}}S^{00}(u_k,u_j)\times\nonumber\\
&\hspace{1cm}\prod_{\alpha=1}^{2}\prod_{b=1}^{N_{+}^{(\alpha)}}S^{0y}(u_k,u_b)\prod_{b=1}^{N_{-}^{(\alpha)}}\frac{1}{S^{0y}(u_k,u_b)}=1,\label{eq:fusedmassless2}\\
&\prod_{b=1}^{N_L}S_{-}^{yQ}\left(u_k^{(\alpha)},u_b\right)\prod_{b=1}^{N_R}S_{+}^{y\bar{Q}}\left(u_k^{(\alpha)},u_b\right)\prod_{j=1}^{N_0}S^{y0}(u_k^{(\alpha)},u_j)=-1,\label{eq:fusedyminus}\\
&\prod_{b=1}^{N_L}S_{+}^{yQ}\left(u_k^{(\alpha)},u_b\right)\prod_{b=1}^{N_R}S_{-}^{y\bar{Q}}\left(u_k^{(\alpha)},u_b\right)\prod_{j=1}^{N_0}S^{y0}(u_k^{(\alpha)},u_j)=-1,\label{eq:fusedyplus}
\end{align}
where the auxiliary Bethe roots were separated depending on whether Im$(y)>0$ (with density $\rho^{(\alpha)}_{y^+}$ and $y=x(u)$) or Im$(y)<0$ (with density $\rho^{(\alpha)}_{y^-}$ and $y=\frac{1}{x(u)}$). In the same way, the index ``$\pm$'' in $S^{Q_a,y}$ indicates if that S-matrix depends on a $y^+$ or $y^-$.  With these expressions we can now construct the counting functions.
Notice that we are considering $L\rightarrow \infty$, $N\rightarrow \infty$ (with $\frac{L}{N}$=fixed), and therefore using the string hypothesis (see \eqref{eq:firststepstringhypAdS3}). 
With this mind, in the equations above \eqref{eq:fusedQparticle}-\eqref{eq:fusedyplus} we implicitly assumed that
\begin{equation}
    \prod_{b=1}^{N_L} \quad \text{corresponds to } \prod_{Q_b=1}^{\infty}\prod_{b=1}^{N_L^{Q}}.
    \label{eq:implicit}
\end{equation}
Following the strategy in \cite{Frolov:2023wji}, we have decided not to explicitly write this on the Bethe-Yang equations at this stage, to avoid bulky expressions. An analogous expression to \eqref{eq:implicit} is true for the $\bar{Q}$-particles.

\subsubsection{Counting Functions}\label{ana:subsubsec:Countingfuncions}

By taking the logarithm of equations \eqref{eq:fusedQparticle}-\eqref{eq:fusedyplus} we can define the counting function for each type of particle. They will be very important to write the free-energy. Like in the XXX model, the counting functions $c(u)$ will be obtained by taking the logarithm of the Bethe-Yang equations and then performing a step similar to \eqref{eq:countingfunction2}. In the thermodynamic limit, for $Q$-particles, $\bar{Q}$-particles, massless particles as well as $y^{\pm}$ ``particles'', they are given by
\begin{align}
&c^{Q}(u)=\frac{\tilde{p}^Q(u)}{2\pi}+\frac{1}{2\pi i}\int_{-\infty}^{\infty}du^\prime\left(\sum_{Q^\prime=1}^{\infty}\log S_{\alg{sl}}^{QQ^\prime}(u,u^\prime)\rho^{Q^\prime}(u^\prime)+\sum_{\bar{Q}^\prime=1}^{\infty}\log S_{\alg{sl}}^{Q\bar{Q}^\prime}(u,u^\prime)\rho^{\bar{Q}^\prime}(u^\prime)\right)\nonumber\\
&\hspace{1cm}+\frac{1}{2\pi i}\sum_{\alpha=1}^{2}\int_{-2}^{2}du^\prime\left(\log S_+^{Qy}(u,u^{\prime^{(\alpha)}})\rho_{y^+}^{(\alpha)}(u^\prime)+\log S_-^{Qy}(u,u^{\prime^{(\alpha)}})\rho_{y^-}^{(\alpha)}(u^\prime)\right)\nonumber\\
&\hspace{1cm}+\frac{1}{2\pi i}\sum_{\dot{\alpha}=1}^{2}\int_{|u^{\prime}|>2}du^\prime\log S^{Q0}(u,u^{\prime\dot{\alpha}})\rho_0^{(\dot{\alpha})}(u^\prime),\\
&c^{\bar{Q}}(u)=\frac{\tilde{p}^{\bar{Q}}(u)}{2\pi}+\frac{1}{2\pi i}\int_{-\infty}^{\infty}du^\prime\left(\sum_{Q^\prime=1}^{\infty}\log S_{\alg{su}}^{\bar{Q}Q^\prime}(u,u^\prime)\rho^{Q^\prime}(u^\prime)+\sum_{\bar{Q}^\prime=1}^{\infty}\log S_{\alg{su}}^{\bar{Q}\bar{Q}^\prime}(u,u^\prime)\rho^{\bar{Q}^\prime}(u^\prime)\right)\nonumber\\
&\hspace{1cm}+\frac{1}{2\pi i}\sum_{\alpha=1}^{2}\int_{-2}^{2}du^\prime\left(\log S_+^{\bar{Q}y}(u,u^{\prime^{(\alpha)}})\rho_{y^+}^{(\alpha)}(u^\prime)+\log S_-^{\bar{Q}y}(u,u^{\prime^{(\alpha)}})\rho_{y^-}^{(\alpha)}(u^\prime)\right)\nonumber\\
&\hspace{1cm}+\frac{1}{2\pi i}\sum_{\dot{\alpha}=1}^{2}\int_{|u^\prime|>2}du^\prime\log S^{\bar{Q}0}(u,u^{\prime\dot{\alpha}})\rho_0^{(\dot{\alpha})}(u^\prime),\\
& c^{(0)}(u)=\frac{\tilde{p}^{(0)}(u)}{2\pi}+\frac{1}{2\pi i}\int_{-\infty}^{\infty}\left(\sum_{Q^{\prime}=1}^{\infty}\log S^{0Q^\prime}(u,u^{\prime})\rho^{Q^\prime}(u^\prime)+\sum_{\bar{Q}^{\prime}=1}^{\infty}\log S^{0\bar{Q}^\prime}(u,u^{\prime})\rho^{\bar{Q}^\prime}(u^\prime)\right)\nonumber\\
&\hspace{1cm}+\frac{1}{2\pi i}\sum_{\alpha=1}^{2}\int_{-2}^{2}du^\prime\log S^{0y}(u,u^{\prime})(\rho_{y^+}^{(\alpha)}(u^\prime)-\rho_{y^-}^{(\alpha)}(u^\prime))\nonumber\\
&\hspace{1cm}+\frac{1}{2\pi i}\sum_{\dot{\alpha}=1}^{2}\int_{|u^\prime|>2}\log S^{00}(u,u^\prime)\rho_0^{(\dot{\alpha})}(u^\prime),\\
& c_{y^-}^{(\alpha)}(u)=+\frac{1}{2\pi i}\int_{-\infty}^{\infty}du^\prime\left(\sum_{Q=1}^{\infty}\log S_{-}^{yQ}(u,u^\prime)\rho^Q(u^\prime)+\sum_{\bar{Q}=1}^{\infty}\log S_{+}^{y\bar{Q}}(u,u^\prime)\rho^{\bar{Q}}(u^\prime)\right)\nonumber\\
&\hspace{1cm}+\frac{1}{2\pi i}\sum_{\dot{\alpha}=1}^{2}\int_{|u^\prime|>2}\log S^{y0}(u,u^\prime)\rho_0^{(\dot{\alpha})}(u^\prime),\\
& c_{y^+}^{(\alpha)}(u)=+\frac{1}{2\pi i}\int_{-\infty}^{\infty}du^\prime\left(\sum_{Q=1}^{\infty}\log S_{+}^{yQ}(u,u^\prime)\rho^Q(u^\prime)+\sum_{\bar{Q}=1}^{\infty}\log S_{-}^{y\bar{Q}}(u,u^\prime)\rho^{\bar{Q}}(u^\prime)\right)\nonumber\\
&\hspace{1cm}+\frac{1}{2\pi i}\sum_{\dot{\alpha}=1}^{2}\int_{|u^\prime|>2}\log S^{y0}(u,u^\prime)\rho_0^{(\dot{\alpha})}(u^\prime).
\end{align}

\subsubsection{Densities}\label{ana:subsubsec:CountingfuncionsAdS3}

We can now compute the densities using the counting functions in the following way 
\begin{equation}
\rho^{(A)}(u)+\bar{\rho}^{(A)}(u)=\frac{dc^{(A)}(u)}{du},
\label{eq:densitiesAdS3}
\end{equation}
where $A=Q,\bar{Q}, 0, y^-,y^+$. We can also define the Kernel as
\begin{equation}
    K^{AB}(u,u^\prime)=\frac{1}{2\pi i}\frac{d}{du}\left(\log S^{AB}(u,u^{\prime})\right).
    \label{eq:kernelAB}
\end{equation}
Given that the limits of integration change depending on which type of particle we are referring to, it is convenient to define three types of convolution. The first type we will call \textit{massive}, (represented by ``$\,\star\,$'')
\begin{equation}
K^{AB}\star\rho^B(u)=\int_{-\infty}^{\infty}du^\prime         K^{AB}(u,u^\prime)\rho^B(u^\prime),     
\end{equation}
whose integral is related to $\{Q,\bar{Q}\}$-particles. The second type is the
\textit{massless}  convolution (represented by ``$\,\check{\star}$\,''). 
\begin{equation}
K^{AB}\,\check{\star}\,\rho^B(u)=\int_{|u^{\prime}|>2}du^\prime         K^{AB}(u,u^\prime)\rho^B(u^\prime) .    
\end{equation}
As the name indicates, this case involves integrals whose limit of integration are given by $|u|>2$ like the massless particles. Finally, the third type is given by
\begin{equation}
    K^{AB}\,\hat{\star}\,\rho^B(u)=\int_{-2}^{2}du^\prime K^{AB}(u,u^\prime)\rho^B(u^\prime) ,
\end{equation}
and can be called \textit{auxiliary} convolution and is represented by ``$\,\hat{\star}\,$''.

With all these definitions in mind we can write the densities of the bound-states, massless and auxiliary particles as given by
\begin{align}
&\rho^Q(u)+\bar{\rho}^{Q}(u)=\frac{1}{2\pi}\frac{d\tilde{p}^Q}{du}+\sum_{Q^\prime=1}^{\infty} K_{\alg{sl}}^{QQ^\prime}\star\rho^{Q^\prime}(u)+\sum_{\bar{Q}^\prime=1}^{\infty} \tilde{K}_{\alg{sl}}^{Q\bar{Q}^\prime}\star\rho^{\bar{Q}^\prime}(u)+\nonumber\\
&\hspace{1cm}+\sum_{\dot{\alpha}=1}^{2} K^{Q0}\,\check{\star}\,\rho_0^{(\dot{\alpha})}(u)+\sum_{\alpha=1}^{2} K_+^{Qy}\,\hat{\star}\,\rho_{y^+}^{(\alpha)}(u)+\sum_{\alpha=1}^{2} K_-^{Qy}\,\hat{\star}\,\rho_{y^-}^{(\alpha)}(u),\label{eq:densitiesQ}\\
&\rho^{\bar{Q}}(u)+\bar{\rho}^{\bar{Q}}(u)=\frac{1}{2\pi}\frac{d\tilde{p}^{\bar{Q}}}{du}+\sum_{Q^\prime=1}^{\infty} K_{\alg{su}}^{\bar{Q}Q^\prime}\star\rho^{Q^\prime}(u)+\sum_{\bar{Q}^\prime=1}^{\infty} \tilde{K}_{\alg{su}}^{\bar{Q}\bar{Q}^\prime}\star\rho^{\bar{Q}^\prime}(u)+\nonumber\\
&\hspace{1cm}+\sum_{\dot{\alpha}=1}^{2} \tilde{K}^{\bar{Q}0}\,\check{\star}\,\rho_0^{(\dot{\alpha})}(u)+\sum_{\alpha=1}^{2} K_+^{\bar{Q}y}\,\hat{\star}\,\rho_{y^-}^{(\alpha)}(u)+\sum_{\alpha=1}^{2} K_-^{\bar{Q}y}\,\hat{\star}\,\rho_{y^+}^{(\alpha)}(u),\label{eq:densitiesQbar}\\
&\rho^{(\dot{\alpha})}(u)+\bar{\rho}^{(\dot{\alpha})}(u)=\frac{1}{2\pi}\frac{d\tilde{p}^{(\dot{\alpha})}}{du}+\sum_{Q^\prime=1}^{\infty} K^{0Q^\prime}\star\rho^{Q^\prime}(u)+\sum_{\bar{Q}^\prime=1}^{\infty} \tilde{K}^{0\bar{Q}^\prime}\star\rho^{\bar{Q}^\prime}(u)+\nonumber\\
&\hspace{1cm}+\sum_{\dot{\alpha}=1}^{2} K^{00}\,\check{\star}\,\rho_0^{(\dot{\alpha})}(u)+\sum_{\alpha=1}^{2} K^{0y}\,\hat{\star}\,\rho_{y^+}^{(\alpha)}(u)-\sum_{\alpha=1}^{2} K^{0y}\,\hat{\star}\,\rho_{y^-}^{(\alpha)}(u),\label{eq:densitiesmassless}
\end{align}
\begin{align}
&\rho_{y^-}^{(\alpha)}(u)+\bar{\rho}_{y^-}^{(\alpha)}(u)=\sum_{Q^\prime=1}^{\infty} K_{-}^{yQ^\prime}\star\rho^{Q^\prime}(u)+\sum_{\bar{Q}^\prime=1}^{\infty} K_{+}^{0\bar{Q}^\prime}\star\rho^{\bar{Q}^\prime}(u)+\sum_{\dot{\alpha}=1}^{2} K^{y0}\,\check{\star}\,\rho_0^{(\dot{\alpha})}(u),\label{eq:densitiesyminus}\\
&\rho_{y^+}^{(\alpha)}(u)+\bar{\rho}_{y^+}^{(\alpha)}(u)=\sum_{Q^\prime=1}^{\infty} K_{+}^{yQ^\prime}\star\rho^{Q^\prime}(u)+\sum_{\bar{Q}^\prime=1}^{\infty} K_{-}^{0\bar{Q}^\prime}\star\rho^{\bar{Q}^\prime}(u)+\sum_{\dot{\alpha}=1}^{2} K^{y0}\,\check{\star}\,\rho_0^{(\dot{\alpha})}(u).\label{eq:densitiesyplus}
\end{align}

\subsubsection{The TBA equations}\label{ana:subsubsec:TBA-AdS3}

In order to obtain the ground-state energy, we now need to first compute $\delta\tilde{f}=0$. This will give us a relation between the energies $\{\tilde{H}^Q,\tilde{H}^{\bar{Q}},\tilde{H}^{(0)}\}$ and the densities. These results can be plugged in equation \eqref{eq:freenergyAdS3beforeTBA}, together with \eqref{eq:densitiesQ}-\eqref{eq:densitiesyplus} in order to obtain $\tilde{f}$.
To start we need to require
\begin{equation}
\delta\tilde{f}=\delta\tilde{f}^Q+\delta\tilde{f}^{\bar{Q}}+\delta\tilde{f}^{(0)}+\delta\tilde{f}^{y^-}+\delta\tilde{f}^{y^+}=0,
\label{eq:freenergyAdS3beforeTBAvar}
\end{equation}
where
\begin{equation}
\begin{aligned}
    & \delta\tilde{f}^{Q}=\sum_{Q=1}^{\infty}\int_{-\infty}^{+\infty}du\left(\tilde{H}^Q(u)\delta\rho^Q(u)-\widetilde{\mathcal{T}}\left(\delta\rho^{Q}(u)\log\frac{\rho_t^Q(u)}{\rho^Q(u)}+\delta\bar{\rho}^{Q}(u)\log\frac{\rho_t^Q(u)}{\bar{\rho}^Q(u)}\right)\right),\\
    & \delta\tilde{f}^{\bar{Q}}=\sum_{\bar{Q}=1}^{\infty}\int_{-\infty}^{+\infty}du\left(\tilde{H}^{\bar{Q}}(u)\delta\rho^{\bar{Q}}(u)-\widetilde{\mathcal{T}}\left(\delta\rho^{\bar{Q}}(u)\log\frac{\rho_t^{\bar{Q}}(u)}{\rho^{\bar{Q}}(u)}+\delta\bar{\rho}^{\bar{Q}}(u)\log\frac{\rho_t^{\bar{Q}}(u)}{\bar{\rho}^{\bar{Q}}(u)}\right)\right),\\    &\delta\tilde{f}^{(0)}=\sum_{\dot{\alpha}=1}^{2}\int_{|u|>2}du\left(\tilde{H}^{(0)}(u)\delta\rho^{(\dot{\alpha})}(u)-\widetilde{\mathcal{T}}\left(\delta\rho^{(\dot{\alpha})}(u)\log \frac{\rho_t^{(\dot{\alpha})}(u)}{\rho^{(\dot{\alpha})}(u)}+\delta\bar{\rho}^{(\dot{\alpha})}(u) \log \frac{\rho_t^{(\dot{\alpha})}(u)}{\bar{\rho}^{(\dot{\alpha})}(u)}\right)\right),\\    
    &\delta\tilde{f}^{(y^\pm)}=-\widetilde{\mathcal{T}}\sum_{\alpha=1}^{2}\int_{-2}^{+2}du\left(\delta\rho_{y^{\pm}}^{(\alpha)}(u)\log \frac{\rho_{y^{\pm},t}^{(\alpha)}(u)}{\rho_{y^{\pm}}^{(\alpha)}(u)}+\delta\bar{\rho}_{y^{\pm}}^{(\alpha)}(u) \log \frac{\rho_{y^{\pm},t}^{(\alpha)}(u)}{\bar{\rho}_{y^{\pm}}^{(\alpha)}(u)}\right).
    \end{aligned}
    \label{eq:deltafs}
\end{equation}
Notice that, like in the Heisenberg spin chain, terms of the type
\begin{equation}
    \int du\left(\rho^A\,\delta\log\frac{\rho_t^A}{\rho^A}+\bar{\rho}^A\,\delta\log\frac{\rho_t^A}{\bar{\rho}^A}\right),
\end{equation}
vanish for any $A$. This is why we do not write these terms.

We now variate equations \eqref{eq:densitiesQ}-\eqref{eq:densitiesyplus}
and substitute the results in \eqref{eq:deltafs}. With this we obtain a large expression for $\delta\tilde{f}$, containing terms depending on $\delta\rho^{Q}(u)$, while other terms depend on $\delta\rho^{Q^\prime}(u^\prime)$, etc. One can rewrite these expressions by swapping 
$\{u,u^{\prime}\}$ (since they are integrated over the same interval) in some terms as well as $Q\leftrightarrow Q^{\prime}$. After a carefully rewriting one obtains that
\begin{align}
    &\delta\tilde{f}=\sum_{Q=1}^{\infty}\int_{-\infty}^{\infty}du(\text{EqQ})\delta\rho^{Q}(u)+\sum_{\bar{Q}=1}^{\infty}\int_{-\infty}^{\infty}du(\text{Eq}\bar{\text{Q}})\delta\rho^{\bar{Q}}(u)+\sum_{\dot{\alpha}=1}^{2}\int_{|u|>2}du(\text{Eq0})\delta\rho^{(\dot{\alpha})}(u)\nonumber\\    &\hspace{1cm}+\sum_{\alpha=1}^{2}\int_{-2}^{2}du(\text{Eqy}^+)\delta\rho^{(\alpha)}_{y^+}(u)+\sum_{\alpha=1}^{2}\int_{-2}^{2}du(\text{Eqy}^-)\delta\rho^{(\alpha)}_{y^-}(u)=0,
\end{align}
\begin{equation}
    \Rightarrow \quad \text{EqQ}=0,\quad \text{Eq}\bar{\text{Q}}=0,\quad \text{Eq0}=0, \quad \text{and} \quad \text{Eqy}^{\pm}=0.
\end{equation}

If we define the Y-functions as 
\begin{equation}
Y^Q=\frac{\rho^Q}{\bar{\rho}^Q}, \quad \bar{Y}^{\bar{Q}}=\frac{\rho^{\bar{Q}}}{\bar{\rho}^{\bar{Q}}}, \quad Y_0^{(\dot{\alpha})}=\frac{\rho_0^{(\dot{\alpha})}}{\bar{\rho}_0^{(\dot{\alpha})}}, \quad Y_{\pm}^{(\alpha)}=-e^{i\mu_\alpha}\frac{\bar{\rho}_{y^\pm}^{(\alpha)}}{\rho_{y^\pm}^{(\alpha)}},
\label{eq:Ys}
\end{equation}
with $\mu_{\alpha}=(-1)^{\alpha}\mu$, and use them, we find that the TBA equations are given by
\begin{align}
\text{EqQ}=0:\hspace{1cm}&\nonumber\\
\Rightarrow\quad -\log Y^Q &=R\tilde{H}^Q-\sum_{Q^\prime=1}^{\infty}\log\left(1+Y^{Q^\prime}\right)\star K_{\alg{sl}}^{Q^\prime Q}-\sum_{\bar{Q}^\prime=1}^{\infty}\log\left(1+\bar{Y}^{\bar{Q}^\prime}\right)\star \tilde{K}_{\alg{su}}^{\bar{Q}^\prime Q}\nonumber\\
&-\sum_{\alpha=1}^{2}\log\left(1-\frac{e^{i\mu_\alpha}}{Y_+^{(\alpha)}}\right)\widehat{\star}K_+^{yQ}-\sum_{\alpha=1}^{2}\log\left(1-\frac{e^{i\mu_\alpha}}{Y_-^{(\alpha)}}\right)\widehat{\star}K_-^{yQ}\nonumber\\
&- \sum_{\dot{\alpha}=1}^{2}\log\left(1+Y_0^{(\dot{\alpha})}\right)\check{\star}K^{0Q},\label{eq:eqQ}\\
\text{Eq}\bar{\text{Q}}=0:\hspace{1cm}&\nonumber\\
\Rightarrow\quad -\log \bar{Y}^{\bar{Q}} &=R\tilde{H}^{\bar{Q}}-\sum_{Q^{\prime}=1}^{\infty}\log\left(1+Y^{Q^{\prime}}\right)\star \tilde{K}_{\alg{sl}}^{Q^\prime \bar{Q}}-\sum_{\bar{Q}^\prime=1}^{\infty}\log\left(1+\bar{Y}^{\bar{Q}^\prime}\right)\star K_{\alg{su}}^{\bar{Q}^\prime \bar{Q}}\nonumber\\
&-\sum_{\alpha=1}^{2}\log\left(1-\frac{e^{i\mu_\alpha}}{Y_+^{(\alpha)}}\right)\widehat{\star}K_+^{y\bar{Q}}-\sum_{\alpha=1}^{2}\log\left(1-\frac{e^{i\mu_\alpha}}{Y_-^{(\alpha)}}\right)\widehat{\star}K_-^{y\bar{Q}}\nonumber\\
&- \sum_{\dot{\alpha}=1}^{2}\log\left(1+Y_0^{(\dot{\alpha})}\right)\check{\star}K^{0\bar{Q}},\label{eq:eqQbar}\\
\text{Eq0}=0:\hspace{1cm}&\nonumber\\
\Rightarrow\quad -\log Y^{(\dot{\alpha})} &=R\tilde{H}^{(\dot{\alpha})}-\sum_{Q^\prime=1}^{\infty}\log\left(1+Y^{Q^\prime}\right)\star K^{Q^\prime 0}-\sum_{\bar{Q}^\prime=1}^{\infty}\log\left(1+\bar{Y}^{\bar{Q}^\prime}\right)\star \tilde{K}^{\bar{Q}^\prime 0}\nonumber\\
&-\sum_{\alpha=1}^{2}\log\left(1-\frac{e^{i\mu_\alpha}}{Y_+^{(\alpha)}}\right)\widehat{\star}K^{y0}-\sum_{\alpha=1}^{2}\log\left(1-\frac{e^{i\mu_\alpha}}{Y_-^{(\alpha)}}\right)\widehat{\star}K^{y0}\nonumber\\
&- \sum_{\dot{\alpha}=1}^{2}\log\left(1+Y_0^{(\dot{\alpha})}\right)\check{\star}K^{00},\label{eq:eq0}\\
\text{Eqy}^{\pm}=0:\hspace{1cm}&\nonumber\\
\Rightarrow\quad \log Y_{\pm}^{(\alpha)} &=-\sum_{Q^\prime=1}^{\infty}\log\left(1+Y^{Q^\prime}\right)\star K_{\pm}^{Q^\prime y}-\sum_{\bar{Q}^\prime=1}^{\infty}\log\left(1+\bar{Y}^{\bar{Q}^\prime}\right)\star K_{\mp}^{\bar{Q}^\prime y}\nonumber\\
&\mp \sum_{\dot{\alpha}=1}^{2}\log\left(1+Y_0^{(\dot{\alpha})}\right)\check{\star}K^{0y},\label{eq:eqy}
\end{align} 
where the convolutions presented here are from the right, namely
\begin{align}
& \rho^A\star K^{AB}(u)=\int_{-\infty}^{\infty}du^\prime        \rho^A(u^\prime) K^{AB}(u^\prime,u),     \\
& \rho^A\,\check{\star}\,K^{AB}(u)=\int_{|u^{\prime}|>2}du^\prime         \rho^A(u^\prime) K^{AB}(u^\prime,u), \\
& \rho^A\,\hat{\star}\,K^{AB}(u)=\int_{-2}^{2}du^\prime \rho^A(u^\prime) K^{AB}(u^\prime,u).
\end{align}
Additionally, we also used $\widetilde{\mathcal{T}}=\frac{1}{R}$.

Notice also the introduction of $\mu$ via a twist on the TBA equations.  
For $\mu= 0$ the equations above describe even-winding number and supersymmetric vaccuum. On the other hand, the choice $\mu\neq 0$ breaks supersymmetry.

The equations \eqref{eq:eqQ}-\eqref{eq:eqy} are called TBA equations. They are very coupled and are in principle not easy to solve even numerically. Nonetheless, very importantly, they are non-perturbative and contain all the information necessary for the computation of the ground-state energy. 
This means that we can in principle study them in several regimes. One does not need to stop, however, on the ground state. At this stage we could proceed and compute excited states using analytical continuation \cite{Dorey:1996re} (see \cite{vanTongeren:2016hhc} for a review).

Additionally, systems with auxiliary excitations can be usually simplified, since not all equations are independent (see for example section 2.5 in \cite{vanTongeren:2016hhc} for a pedagogical explanation, and \cite{Arutyunov:2009ux} for this applied to $AdS_5 \times S^{5}$). For this model, the authors of \cite{Frolov:2023wji} performed a first simplification on the TBA equations. It would be interesting to continue towards this direction, and further understand the Y-system. 

\begin{centering}
	\begin{tcolorbox}
		\textbf{Exercise 4.9}: Prove that \eqref{eq:eqQ}-\eqref{eq:eqy} are the conditions required to obtain $\delta\tilde{f}=0$. In order to do that just perform the few steps skipped in section \eqref{ana:subsubsec:TBA-AdS3}.
	\end{tcolorbox}
\end{centering}

\subsubsection{The ground-state energy}\label{ana:subsubsection:groundstate}

As discussed earlier, the ground-state energy of the original theory (with zero temperature and finite volume) is given by $E_0(R)=R \tilde{f}$. By plugging the previous results back into the mirror free-energy we find
\begin{align}
    E_0(R)&=-\int_{-\infty}^{\infty}\frac{du}{2\pi}\frac{d\tilde{p}^Q}{du}\log\left((1+Y^{Q})(1+\bar{Y}^{\bar{Q}})\right)\nonumber\\
    &\qquad-\int_{|u|>2}\frac{du}{2\pi}\frac{d\tilde{p}^{(0)}}{du}\log\left((1+Y_0^{(1)})(1+Y_0^{(2)})\right).
    \label{eq:groundstate}
\end{align}

\begin{centering}
	\begin{tcolorbox}
		\textbf{Exercise 4.10}: Construct the mirror free-energy by replacing the TBA equations as well as \eqref{eq:densitiesQ}-\eqref{eq:densitiesyplus} in \eqref{eq:freeenergybeforeTBA}. Check that it is consistent with expression \eqref{eq:groundstate}.
	\end{tcolorbox}
\end{centering}

\subsection{Conclusions}\label{ana:subsubsec:summaryTBAAdS3}

\subsubsection{Summary}

The main steps in the construction of the  Thermodynamic Bethe ansatz for the pure-Ramond-Ramond $AdS_3\times S^3\times T^4$ are:

\begin{itemize}
\item Introduce the mirror model and S~matrix, which can be done through analytic continuation of the original S~matrix; this requires understanding in detail the rather intricate analytic structure of the dressing factors.
\item Construct the asymptotic Bethe ansatz for the mirror model, which is again related to the one of the original model by analytic continuation.
\item Formally solve the mirror Bethe-Yang equations for $L\rightarrow \infty$, \textit{i.e.}\ formulate the ``string hypothesis'' for the model. In this case we have bound states and auxiliary particles (we do not have, instead, complexes of several auxiliary particles). Both left- and right-particles can form bound-states, and we can write the Bethe equations in a way (or more precisely, in a \textit{grading}) which makes either family of bound states simple, but not both. We choose ``left'' bound states to have a straightforward form, and ``right'' ones to involve auxiliary particles. The massless modes cannot form bound states or strings.
\item Use fusion to construct the S-matrices for bound-states depending only on the center of the strings. This includes the cases where the first particle is of ``left'' type and the second is of any type, then the first of ``right'' type and the second of any type, and so on and so forth;
\item Compute the counting function for each type of excitations and then use them to compute the sum of the densities;
\item Keeping in mind that  $\widetilde{\mathcal{T}}=\frac{1}{R}$, compute the TBA equations by requiring $\delta\tilde{f}=0$;
\item Use the TBA equations and the densities to compute the mirror free-energy and obtain the ground-state energy of the original model as a consequence;
\item Simplify the TBA equations. In principle this step can be refined to give a Y-system, a T-system and eventually the quantum spectral curve; this has not been done for this model. 
\end{itemize}

\subsubsection{Open questions}\label{ana:subsec:conclusions}

There are some very interesting points still to be discussed and further computed:

\paragraph{Simplifications and numerical results.} The derivation of the mirror TBA yields the equations in a ``canonical'' form, which involves sums over infinite types of excitations. This obscures the symmetry structure of the model and makes numerical computations quite challenging. In general, we expect that the canonical TBA equations can be simplified by exploiting suitable identities between the kernels, as it was done in~\cite{Frolov:2021bwp}. However, it is possible that further simplifications can be engineered. This would be necessary in order to rewrite the equations as a Y-system, meaning a set of functional equations supplemented by suitable discontinuity conditions. This is the first step in the derivation of the Quantum Spectral Curve from the TBA.

\paragraph{Computing physical observables.}
The TBA equations which we have discussed only describe the ground-state energy of the model. Because we are dealing with a supersymmetric theory, this is zero, so there is nothing interesting about it! However, from the ground-state equations it is possible to derive interesting predictions, by some minor modifications. Firstly, we may twist the boundary conditions of certain fields in such a way to break (some) supersymmetry (see \eqref{eq:Ys}, for example). Then, the ground-state energy will not vanish, and instead it would be a function of the twist and of the volume. This was studied in~\cite{Frolov:2023wji}, and it is interesting to note that the contribution of the massless modes to the ground-state energy presents some puzzling discrepancies with respect to what is expected from semiclassical arguments.
Another important application is to compute the spectrum of generic (non-protected) states. It is believed~\cite{Dorey:1996re} that the equations describing such states follow from the ground-state ones by a suitable analytic continuation.%
\footnote{It should be noted that this procedure can be especially subtle for certain ``exceptional'' set of excitations~\cite{Arutyunov:2012tx}; this has not yet been investigated in this context.}
This route was taken in~\cite{Brollo:2023pkl,Brollo:2023rgp} where the TBA equations for excited states were written down, and solved numerically in the small-tension limit. Interestingly, at small tension, the most important contribution to the energy comes from massless modes. However, unlike the case of the NSNS models, the tensionless spectrum is not given by the symmetric-product orbifold CFT of $T^4$, but by some yet-to-be-identified interacting theory.

\paragraph{Quantum Spectral Curve (QSC).} As mentioned earlier two proposals for a QSC were put forward recently \cite{Cavaglia:2021eqr,Ekhammar:2021pys}. Unlike what happened for $AdS_5\times S^5$ and $AdS_4\times \mathbb{C}P^3$, however, these proposal are not derived from the mirror TBA --- in fact, they \textit{predate} it! They were derived by imposing that the QSC equations have the correct symmetries as well as some suitable analytic properties (this last point is especially subtle in a model as complicated as this one). Then, the equations were studied in~\cite{Cavaglia:2022xld} where they have been used to predict the small tension energy of certain excited states. However, in the QSC formalism it is currently not clear how to describe massless states, so the authors considered states containing only massive excitations.
Hence, the results of~\cite{Cavaglia:2022xld} and those of~\cite{Brollo:2023pkl,Brollo:2023rgp}  cannot be compared. It would be very desirable to either derive the QSC from the TBA (or viceversa), or carefully compare the numerical prediction for a given observable.

\paragraph{Mixed-flux.} A more challenging but certainly more physically insightful question is \textit{what is the spectrum of mixed-flux theories}. To this end, it is necessary to construct the  TBA (or QSC) for the mixed-flux model. This is a very hard question due to the difficulty in constructing the dressing phases, or in any case the difficulty in understanding the analytic structure of mixed-flux models, especially in the mirror kinematics where they appear to be non-unitary. Recent computations \cite{Frolov:2023lwd,Baglioni:2023zsf} shed some light on this matter, but a full answer is still to be found.

\newpage

\section{Introduction to the hybrid superstring} \label{s:bob}

%\textit{Current author: Bob Knighton. } For questions, comments, typos, \textit{etc.}~on this section, feel free to write to %\href{mailto:rik23@cam.ac.uk}{rik23@cam.ac.uk}.

%\subsection{Introduction}\label{sec:bob:introduction}

The two major approaches to superstring theory are the RNS formalism and the Green-Schwarz formalism. The RNS (also sometimes NSR) formalism \cite{Ramond:1971gb,Neveu:1971rx} places \textit{worldsheet supersymmetry} on the center stage. Although worldsheet supersymmetry does not manifest as a supersymmetry in the effective field theory of the target space, it has proved a powerful tool in the description of superstrings, primarily due to the covariant quantisation it allows on the worldsheet. Indeed, worldsheet observables can be written in a completely covariant way via the BRST procedure that gauge-fixes the worldsheet (super-)Diffeomorphism$\times$Weyl symmetry. On the other hand, the Green-Schwarz formalism \cite{Green:1983wt}, reviewed in detail in Sections \ref{s:sibylle} and \ref{s:saskia}, emphasizes manifest \textit{spacetime/target space supersymmetry}. However, it pays the price of not admitting a known covariant quantisation, and instead is typically quantised in the lightcone gauge.

In these notes, we introduce a third, less known, formalism for superstring theory, known as the \textit{hybrid} formalism \cite{Berkovits:1994wr,Berkovits:1996bf,Berkovits:1999im} (sometimes referred to in the literature as the Berkovits superstring). The basic idea of the hybrid formalism is to consider spacetimes which factorise into the form
\begin{equation}
X\times \mathcal{M}\,,
\end{equation}
where $X$ is, typically, some non-compact spacetime while $\mathcal{M}$ is typically taken to be compact. Furthermore, we assume that the sigma models on $X$ and $\mathcal{M}$ are \textit{independently} supersymmetric. The hybrid formalism can be thought of as a quantisation of the worldsheet theory which makes manifest the spacetime supersymmetries of $X$ and which only makes manifest the worldsheet supersymmetries of $\mathcal{M}$. In other words, the hybrid formalism treats the non-compact dimensions of the target in a Green-Schwarz-like approach, while the compact dimensions are taken in an RNS-like description. The benefit of the hybrid formalism, as we will see throughout these notes, is that, while the spacetime supersymmetries of $X$ are manifest, the resulting worldsheet theory still retains a covariant quantisation, unlike traditional Green-Schwarz approaches.\footnote{Yet another approach to covariant superstring theory is the \textit{pure-spinor} formalism, also developed by Berkovits \cite{Berkovits:2000fe}, which has been an extremely powerful tool in the computation of string amplitudes and treating backgrounds with RR flux. While we will not discuss pure spinors here, we direct the interested reader to the pedagogical introduction \cite{Berkovits:2017ldz}.}

This section is structured as follows. In Section \ref{sec:bob-rns}, we provide a pedagogical review of the covariant quantisation of the bosonic string and RNS superstring. We focus in particular on BRST quantisation and picture changing. In Section \ref{sec:bob-spacetime-susy}, we discuss the role of spacetime supersymmetry in the RNS superstring and develop notions of off-shell spacetime supersymmetry, largely following the treatment of \cite{Berkovits:1996bf}. In Section \ref{sec:bob-4d}, we introduce the hybrid string in four dimensions, starting from RNS string theory compactified on a Calabi-Yau manifold. Special care is given to the set of complicated field redefinitions taken to get to the hybrid description. In Section \ref{sec:bob-6d}, we discuss the hybrid string on a the background $AdS_3 \times S^3$, again starting from the RNS description. Along the way, we briefly introduce (supersymmetric) WZW models. In Section \ref{sec:bob-applications}, we discuss some applications of the hybrid formalism, specifically the conceptual ease of turning on Ramond-Ramond flux in the $AdS_3 \times S^3$ background, and the application of the hybrid formalism to the `tensionless' limit of IIB string theory on $AdS_3 \times S^3\times K3$, which is conjectured to be holographically dual to the symmetric orbifold CFT $\text{Sym}(K3)$.

\subsection{Review of the RNS superstring}\label{sec:bob-rns}

In this section we provide a lightning review of the bosonic string and the RNS superstring. We attempt to keep everything self-contained, while not going beyond the scope of what is needed in later sections. For a more complete description, see, for example, \cite{Polchinski:1998rq,Polchinski:1998rr,Blumenhagen:2013fgp}

\subsubsection{Review of the bosonic string}

We will begin with the simplest string theory that can be written down: the bosonic string in $D$ dimensions. The fields of this theory are $D$ scalars $X:\Sigma\to\mathbb{R}^D$ which parametrize the coordinates of the worldsheet $\Sigma$ in the $D$-dimensional ambient space $\mathbb{R}^D$. Taking the target space metric to be $g_{\mu\nu}=\delta_{\mu\nu}$,\footnote{Throughout this section the signature will play a minor role, so we will stick to Euclidean signature for convenience.} the dynamics of the worldsheet theory are governed by the Polyakov action\footnote{Relative to Section \ref{s:sibylle} we have set $\alpha'=1$, which we will do for the rest of this section.}
\begin{equation}
S_\text{P}=\frac{1}{4\pi}\int_{\Sigma}\mathrm{d}^2\sigma\,\sqrt{h}\,h^{\alpha\beta}\delta_{\mu\nu}\partial_{\alpha}X^{\mu}\partial_{\beta}X^{\nu}\,.
\end{equation}
Here, $h$ is a metric tensor on the worldsheet, which, as we will see, can be (almost) entirely gauged away. The above action is invariant under the following local symmetries:
\begin{itemize}

    \item Reparametrization $\sigma\to\widetilde{\sigma}(\sigma)$.

    \item Weyl transformations $h_{\alpha\beta}\to e^{\omega}h_{\alpha\beta}$.

\end{itemize}
Using these two symmetries, it is possible to (locally) transform $h$ into the flat metric $\delta$. Thus, the action becomes
\begin{equation}
S_p=\frac{1}{4\pi}\int_{\Sigma}\mathrm{d}^2\sigma\,\delta^{\alpha\beta}\delta_{\mu\nu}\partial_{\alpha}X^{\mu}\partial_{\beta}X^{\nu}\,.
\end{equation}
However, upon gauging $h\to\delta$, there are still combinations of diffeomorphisms and Weyl transformations which leave the metric invariant. These are the \textit{conformal transformations}, which are a mixture of coordinate transformations $\sigma\to\widetilde{\sigma}(\sigma)$ which act as $h(\sigma)\to e^{\phi}h(\tilde{\sigma})$, and then a Weyl transformation to remove the $e^{\phi}$ factor. Thus, the gauge-fixed worldsheet theory will have a residual conformal invariance, and thus will constitute a conformal field theory, which will be extremely useful in calculations. We will find it convenient to use complex coordinates $z,\bar{z}$ on the worldsheet. In these coordinates,
\begin{equation}
S=\frac{1}{4\pi}\int\mathrm{d}^2z\,\partial X^{\mu}\bar{\partial}X_{\mu}\,,
\end{equation}
where $\partial=\partial_z$ and $\bar{\partial}=\partial_{\bar{z}}$. The equations of motion are
\begin{equation}
\partial\bar{\partial}X=0\,,
\end{equation}
and the most general solution is
\begin{equation}
X(z,\bar{z})=X_L(z)+X_R(\bar{z})\,.
\end{equation}

In an appropriate quantum treatment of the string, as we know from gauge theory, we need to be careful about gauge transformations. Just as a standard gauge theory with connection $A$ requires integrating over \textit{all} connections $A$ and dividing by the volume of the gauge group, in string theory we integrate over all metrics $h$ moduli the group of local symmetries $\text{Diff}\times\text{Weyl}$. That is, the path integral takes the schematic form
\begin{equation}\label{eq:bob-gauge-fixed-polyakov}
Z=\int\frac{\mathcal{D}h\,\mathcal{D}X}{\text{Diff}\times\text{Weyl}}e^{-S_p[X,h]}\,.
\end{equation}
Of course, the volume of diffeomorphisms and Weyl transformations on $\Sigma$ is infinite, and so this expression needs regulation. The standard trick is to introduce Fadeev-Popov ghosts whose path integral formally computes the (inverse) volume. This is achieved in the case of $\text{Diff}\times\text{Weyl}$ by introducing a pair of anti-commuting ghosts $b,c$ with scaling dimensions $\text{dim}(b)=2$ and $\text{dim}(c)=-1$ and action
\begin{equation}
S_{\text{ghost}}=\frac{1}{2\pi}\int\mathrm{d}^2z\,b\bar{\partial} c\,.
\end{equation}
The equations of motion demand that $b,c$ are holomorphic. To be complete, we have to include an anti-holomorphic pair $(\bar{b},\bar{c})$ and we have
\begin{equation}
S_{\bar{b}\bar{c}}=\frac{1}{2}\int\mathrm{d}^2z\,\bar{b}\partial\bar{c}\,.
\end{equation}
The full gauge-fixed string theory is then given by
\begin{equation}
Z=\int\mathcal{D}X\,\mathcal{D}b\,\mathcal{D}c\,e^{-S[X,b,c]}
\end{equation}
with
\begin{equation}\label{eq:bob-polyakov-and-ghosts}
S[X,b,c]=\frac{1}{2\pi}\int\mathrm{d}^2z\,\left(\frac{1}{2}\partial X^{\mu}\,\bar{\partial}X_{\mu}+b\bar{\partial}c+\bar{b}\partial\bar{c}\right)\,.
\end{equation}

Given the above action, we can write down its stress tensor in the usual way: either by re-introducing a worldsheet metric $h$ and taking the variation of the action $T_{\mu\nu}\sim\delta S/\delta h^{\mu\nu}$, or as the conserved current under coordinate translations $z\to z+a$. In either case, the stress tensor can be derived. In complex coordinates, it is given by
\begin{equation}
T_{zz}(z,\bar{z})=\frac{1}{2}\partial X^{\mu}(z,\bar{z})\partial X_{\mu}(z,\bar{z})+2(\partial c)b(z)+c(\partial b)(z)\,.
\end{equation}
The component $T_{\bar{z}\bar{z}}$ is given by the same expression but with the `right-moving' fields, and the mixed components $T_{z\bar{z}}$ and $T_{\bar{z}z}$ vanish identically as a consequence of conformal symmetry. By the equations of motion,
\begin{equation}
\bar{\partial}T_{zz}=0\,,\quad\partial T_{\bar{z}\bar{z}}=0\,,
\end{equation}
\textit{i.e.}~$T_{zz}$ is holomorphic and $T_{\bar{z}\bar{z}}$ is anti-holomorphic. To simplify notation, we denote
\begin{equation}
T_{zz}(z,\bar{z}):=T(z)\,,\quad T_{\bar{z}\bar{z}}(z,\bar{z}):=\bar{T}(\bar{z})\,.
\end{equation}
We emphasize that $T(z)$ is not necessarily the complex conjugate of $\bar{T}(\bar{z})$.

\subsubsection*{Quantisation and OPEs}

The above discussion has given us a (classical) conformal field theory whose fundamental fields are the coordinates $X^{\mu}$ and the ghosts $b,c$ (as well as their right-moving counterparts). We will now discuss how to promote this theory to a proper quantum string theory.

In standard QFT, we quantise by identifying the canonical momentum $\pi_{\phi}$ conjugate to a field $\phi$ and then impose the equal-time commutation relations
\begin{equation}
[\phi(\textbf{x},t),\pi_{\phi}(\textbf{y},t)]=i\delta(\textbf{x}-\textbf{y})\,.
\end{equation}
These commutation relations tell us the algebraic properties that the operators $\phi$ and $\pi_{\phi}$ satisfy. We then find representations of this algebra, and identify the states in these representations as allowed physical states.\footnote{In the case of gauge theories, we also must impose the BRST conditions, \textit{i.e.}~that the physical states lie in the cohomology of the BRST charge $Q_{\text{BRST}}$.} Of course, the choice of canonical conjugation relations requires a choice of specific time direction $t$.

We can continue with this route in 2D conformal field theory, but due to the holomorphicity of the operators we're considering, it is more useful to phrase the above commutation relations in terms of `operator product expansions' or OPEs. We will briefly review OPEs here, but see for example \cite{Blumenhagen:2013fgp} for a complete treatment.

Given a holomorphic field $\phi(z)$, we can find its conjugate momemtum $\pi_{\phi}(z)$ via the variation
\begin{equation}
\pi_\phi(z)=2\pi\frac{\delta S}{\delta\bar{\partial}\phi(z)}\,.
\end{equation}
The factor of $2\pi$ is customary. In 2D conformal field theory, we implement quantisation with the following recipe:
\begin{itemize}

    \item \textbf{Radial quantisation:} In order to quantise a QFT, we have to pick a time direction. However, the complex plane has Euclidean signature, and so there is no canonical choice. We will artificially pick the `time' coordinate to the the radial coordinate $|z|$. Specifically, we let $z=e^{\tau+i\sigma}$, where $\tau$ is the `time' coordinate. Thus, $z=0$ corresponds to the infinite past $\tau=-\infty$, and is therefore the place where we prepare `asymptotic' states. Slices of constant time are circles centered at the origin. 

    Similar to time ordering in standard QFT, we define the radial ordering of two fields $\Phi_1(z)$ and $\Phi_2(w)$ to be
    \begin{equation}
    R(\Phi_1(z)\Phi_2(w))=
    \begin{cases}
    \Phi_1(z)\Phi_2(w) & |z|>|w|\,,\\
    \Phi_2(w)\Phi_1(z) & |z|<|w|\,.
    \end{cases}
    \end{equation}
    From here forward, we assume that all products of operators are radially ordered.

    \item \textbf{Operator product expansion:} In general, given two local operators $A(z)$ and $B(w)$, their product will be singular as $z\to w$. This is captured in the \textit{operator product expansion} (OPE). Quantisation of a 2D CFT is achieved by imposing that $\phi(z)$ and $\pi_{\phi}(w)$ satisfy the OPE
    \begin{equation}\label{eq:bob-fundamental-OPE}
    \phi(z)\pi_{\phi}(w)\sim\frac{1}{z-w}+(\text{finite as }z\to w)\,,
    \end{equation}
    where, as always, the above expression is radially-ordered. To see how this relates to the standard canonical commutation relations, we define the equal time commutation relation to be obtained by the following setup:
    \begin{equation}
    [\phi(z),\pi_{\phi}(w)]_{|z|=|w|}=\lim_{\delta\to 0^+}\left(\phi(z)\pi_{\phi}(w)\big|_{|z|=|w|+\delta}-\pi_{\phi}(w)\phi(z)\big|_{|z|=|w|-\delta}\right)\,.
    \end{equation}
    Using the OPE, we have
    \begin{equation}
    [\phi(z),\pi_{\phi}(w)]_{|z|=|w|}=\lim_{\delta\to 0^+}\left(\frac{1}{z-w}\bigg|_{|z|=|w|+\delta}-\frac{1}{z-w}\bigg|_{|z|=|w|+\delta}\right)\,.
    \end{equation}
    This expression vanishes when $z\neq w$ and must be regulated when $z=w$, but the end result is
    \begin{equation}
    [\phi(z),\pi_{\phi}(w)]_{|z|=|w|}\propto\delta(z-w)\big|_{|z|=|w|}\,,
    \end{equation}
    \textit{i.e.}~the OPE is equivalent to the canonical commutation relations in QFT. The extra factor of $2\pi$ comes from the factor of $2\pi$ in the definition of $\pi_{\phi}$.

\end{itemize}
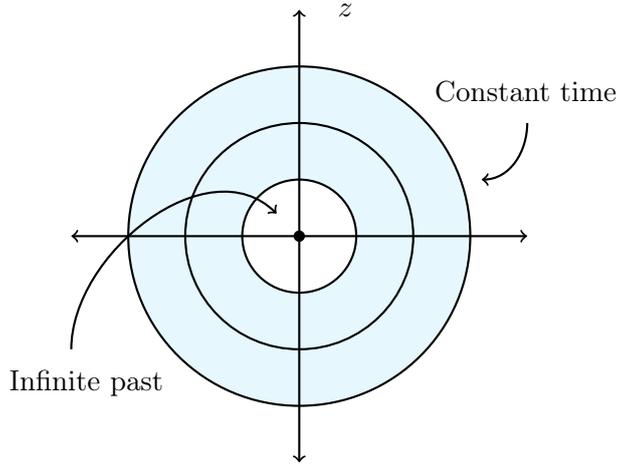
\begin{figure}
\centering
\begin{tikzpicture}[scale = 1.5]
\begin{scope}[xshift = -3cm]
\fill[cyan, opacity = 0.1] (0,0) circle (1.5);
\fill[white] (0,0) circle (0.5);
\draw[thick, <->] (-2,0) -- (2,0);
\draw[thick, <->] (0,-2) -- (0,2);
\draw[thick] (0,0) circle (0.5);
\draw[thick] (0,0) circle (1.0);
\draw[thick] (0,0) circle (1.5);
\node[above right] at (1.1,1.1) {Constant time};
\draw[thick, ->] (2,1) to[out = -90, in = 0] (1.6,0.5);
\node[below left] at (-1.1,-1.1) {Infinite past};
\draw[thick, ->] (-2,-1) to[out = 90, in = 135] (-0.2,0.2);
\fill (0,0) circle (0.05);
\node[right] at (0.25,2) {$z$};
\end{scope}
\end{tikzpicture}
\caption{The geometry of radial quantisation: we let the coordinate $z\in\mathbb{C}$ parametrize our worldsheet. Circles of constant radius are thought of as circles of constant time, and $z=0$ corresponds to the infinite past, where asymptotic states are prepared.}
\end{figure}

\noindent We can use this prescription to derive the OPEs of the various fields appearing in the bosonic string:
\begin{itemize}

    \item The scalars $X$ have conjugate momentum $\partial X$. We have the OPE
    \begin{equation}
    X^{\mu}(z)\partial X^{\nu}(w)\sim\frac{\delta^{\mu\nu}}{z-w}+\cdots\,.
    \end{equation}
    Since $X$ isn't strictly holomorphic, it is usually more convenient to work with the OPE
    \begin{equation}
    \partial X^{\mu}(z)\,\partial X^{\nu}(w)\sim-\frac{\delta^{\mu\nu}}{(z-w)^2}+\cdots\,.
    \end{equation}

    \item The ghost field $c$ has conjugate momentum $b$. Thus, we have the OPE
    \begin{equation}
    c(z)b(w)\sim\frac{1}{z-w}+\cdots\,.
    \end{equation}
    Note that $b$ and $c$ are \textit{anti-commuting}, so this should really be thought of as an anti-commutation relation. We can also swap the order and we find
    \begin{equation}
    b(z)c(w)\sim\frac{1}{z-w}+\cdots\,.
    \end{equation}

\end{itemize}
OPEs are incredibly useful tools since they package the commutation relations of quantum fields in a way that is compatible with the powerful techniques of complex analysis.

\subsubsection*{States, operators, and charges}

In conformal field theory, there is very little difference between talking about \textit{states} and \textit{local operators}, and there is in fact a one-to-one correspondence between these two concepts. This is the \textit{state-operator correspondence}:
\begin{itemize}

    \item The CFT contains a vacuum state $\ket{\Omega}$. Given a local field $\Phi(z)$, you can define a state $\ket{\varphi}$ by taking the limit
    \begin{equation*}
    \ket{\varphi}:=\lim_{z\to 0}\Phi(z)\ket{\Omega}
    \end{equation*}
    (This may need regularized.) Since $z=0$ is the `infinite past', this is like preparing an asymptotic state in QFT.

    \item This can also be done in reverse! Given a state $\ket{\varphi}\in\mathcal{H}$, you can construct a local operator $\Phi(z)$ at any point $z$ in the plane (c.f. Polchinski \cite{Polchinski:1998rq,Polchinski:1998rr}).

    \item Given a (not necessarily local) operator $\mathcal{O}$, we can think of $\mathcal{O}$ acting on a state $\ket{\varphi}$, or we can think of it acting on a local operator $\Phi$ via the commutator $[\mathcal{O},\Phi(z)]$:
    \begin{equation*}
    \mathcal{O}\ket{\varphi}\Longleftrightarrow [\mathcal{O},\Phi(z)]\,.
    \end{equation*}

\end{itemize}
While we will not prove the operator-state correspondence here, we will use it very often, and sometimes not make a distinction between the concept of an operator and a state.

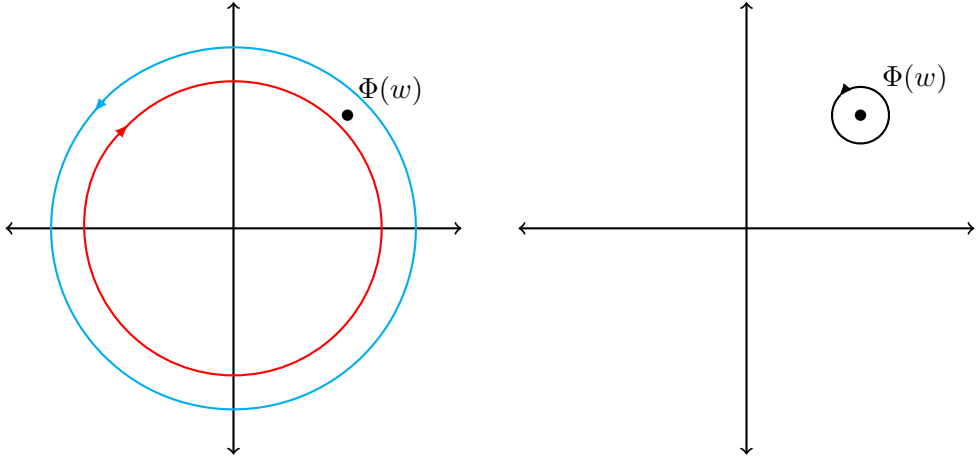
\begin{figure}[!ht]
\centering
\begin{tikzpicture}[scale = 1.5]
\begin{scope}[xshift = -2.25cm]
\draw[thick, <->] (-2,0) -- (2,0);
\draw[thick, <->] (0,-2) -- (0,2);
\fill (1,1) circle (0.05);
\node[above right] at (1,1) {$\Phi(w)$};
\draw[thick, red, latex-] (0,0) [partial ellipse = 135:500:1.3 and 1.3];
\draw[thick, cyan, -latex] (0,0) [partial ellipse = 135:500:1.6 and 1.6];
\end{scope}
\begin{scope}[xshift = 2.25cm]
\draw[thick, <->] (-2,0) -- (2,0);
\draw[thick, <->] (0,-2) -- (0,2);
\fill (1,1) circle (0.05);
\node[above right] at (1.1,1.1) {$\Phi(w)$};
\draw[thick, -latex] (1,1) [partial ellipse = 135:500:0.25 and 0.25];
\end{scope}
\end{tikzpicture}
\caption{The contour integrals for computing the commutator $[Q,\Phi(w)]$ of a charge acting on a state $\Phi$.}
\end{figure}

As an application, let us see how we can read off charges under conserved currents by representing states as local operators and using the OPE. Let $J(z)$ be a conserved current on the worldsheet (for example $\partial X$). Then the conserved charge associated to $J$ is given by the integral of $J$ over a spatial slice. However, since we are working in radial quantisation, `spatial slices' are just circles of constant radius $|z|=R$. Thus, we can write the conserved charge $Q$ as
\begin{equation}
Q=\oint_{|z|=R}\frac{\mathrm{d}z}{2\pi i}J(z)\,.
\end{equation}
The factor of $2\pi i$ is again customary. Now, given a local operator $\Phi(z)$ dual to a state $\ket{\Phi}$ with charge $Q\ket{\Phi}=q\ket{\Phi}$, by the state-operator correspondence, we should have
\begin{equation}
[Q,\Phi(z)]=q\Phi(z)\,.
\end{equation}
We can also evaluate this commutator in terms of the OPE $J(z)\Phi(w)$. Writing $Q$ as a contour integral, we can write this commutator as
\begin{equation}
[Q,\Phi(w)]=\oint\frac{\mathrm{d}z}{2\pi i}\left(J(z)\Phi(w)-\Phi(w)J(z)\right)\,.
\end{equation}
Due to radial ordering, this commutator must be defined by putting $|z|>|w|$ in the first component and $|z|<|w|$ in the second, \textit{i.e.}
\begin{equation}
[Q,\Phi(w)]=\left(\oint_{|z|>|w|}\frac{\mathrm{d}z}{2\pi i}-\int_{|z|<|w|}\frac{\mathrm{d}z}{2\pi i}\right)J(z)\Phi(w)\,,
\end{equation}
where the integrand is radially-ordered. First computing the integral around the contour $|z|>|w|$ and then subtracting the integral around the contour $|z|<|w|$ is equivalent to simply integrating in a small circle around $w$, so that
\begin{equation}
[Q,\Phi(w)]=\oint_{w}\frac{\mathrm{d}z}{2\pi i}J(z)\Phi(w)\,.
\end{equation}

Now, by the Cauchy theorem, we know that the result of the integral of a meromorphic function of $z$ over a small circle around $w$ is just the residue of that function at $w$, \textit{i.e.}~the coefficient of the simple pole in its Laurent expansion around $w$. That is, the charge $q$ is just the coefficient of the $1/(z-w)$ term in the OPE of $J$ with $\Phi$. Specifically,
\begin{equation}\label{eq:bob-jphi-ope}
J(z)\Phi(w)\sim\cdots+\frac{q}{z-w}+\cdots\,.
\end{equation}
Thus, OPEs can be used to read off charges of fields under conserved currents. As a piece of terminology, we define a field $\Phi$ to be a \textit{highest-weight state of charge $q$} under $J$ if the higher-order poles in the OPE \eqref{eq:bob-jphi-ope} vanish.

Finally, we define a type of normal ordering which can fuse two local operators into a third local operator. Since the product of two fields can have bad short distance behaviour as in equation \eqref{eq:bob-jphi-ope}, one cannot simply multiply two fields at the same point and expect a well-defined result. As such, given two fields $\Phi_1$ and $\Phi_2$, it is natural to define their normal-ordered product $(\Phi_1\Phi_2)$ to be their product minus the divergent parts. Specifically,
\begin{equation}
(\Phi_1\Phi_2)(w)=\lim_{z\to w}\left(\Phi_1(z)\Phi_2(w)-\text{OPE}\right)\,,
\end{equation}
where `OPE' stands for the divergent parts of the operator-product expansion. Another way to calculate this is to take the integral
\begin{equation}
(\Phi_1\Phi_2)(w)=\oint_{w}\frac{\mathrm{d}z}{2\pi i}\frac{\Phi_1(z)\Phi_2(w)}{z-w}\,.
\end{equation}

\vspace{0.25cm}

\begin{centering}
\begin{tcolorbox}
\textbf{Exercise:} \textit{Show that these two definitions of normal-ordering are equivalent.}
\end{tcolorbox}
\end{centering}

\vspace{0.25cm}

% \begin{centering}
% \begin{tcolorbox}
% \textbf{Solution:} The general form of the OPE between two fields will be
% \begin{equation*}
% \Phi_1(z)\Phi_2(w)\sim\sum_{n=\infty}^{1}\frac{A_n(w)}{(z-w)^n}+A_0(w)+\cdots\,,
% \end{equation*}
% where the $\cdots$ terms vanish as $z\to w$. The first sum gives the divergent part, and so upon subtracting it away, we find
% \begin{equation*}
% A_0(w)=(\Phi_1\Phi_2)(w)\,.
% \end{equation*}
% Now, to show that this can be obtained by a contour integral, we divide the OPE by $(z-w)$. This gives
% \begin{equation*}
% \frac{\Phi_1(z)\Phi_2(w)}{z-w}\sim\sum_{n=\infty}^{1}\frac{A_n(w)}{(z-w)^{n+1}}+\frac{(\Phi_1\Phi_2)(w)}{z-w}+\text{finite}\,.
% \end{equation*}
% Taking the contour integral of $z$ along a small circle centered at $w$, we can apply Cauchy's theorem, which tells us that only the residue at $z=w$ will contribute. Since the residue is precisely the coefficient of $1/(z-w)$ in the above expansion, we find
% \begin{equation*}
% \oint_{w}\frac{\mathrm{d}z}{2\pi i}\frac{\Phi_1(z)\Phi_2(w)}{z-w}=(\Phi_1\Phi_2)(w)\,.
% \end{equation*}
% \end{tcolorbox}
% \end{centering}

\subsubsection*{The stress tensor and critical dimension}

An important consequence of the worldsheet conformal symmetry is that the (left-moving) stress tensor $T(z)$ is holomorphic. As a consequence, if we take any function $f(z)$, then the current $f(z)T(z)$ is also conserved, since
\begin{equation}
\bar{\partial}(fT)=0\,.
\end{equation}
Taking $f(z)=z^{n+1}$, we have the series of conserved currents
\begin{equation}
J^{(n)}(z)=z^{n+1}T(z)\,,
\end{equation}
and the conserved currents associated to $J^{(n)}$ are the \textit{Virasoro modes}
\begin{equation}
L_n=\oint\frac{\mathrm{d}z}{2\pi i}J^{(n)}(z)=\oint\frac{\mathrm{d}z}{2\pi i}z^{n+1}T(z)\,.
\end{equation}
The different modes generate infinitesimal conformal transformations on the worldsheet. Among them, the mode $L_0$ is special, as it generates transformations of the form $z\to\lambda z$, \textit{i.e.}~scaling transformations. We say that a field $\Phi$ has scaling dimension $h$ if
\begin{equation}
[L_0,\Phi(w)]=h\Phi(w)\,.
\end{equation}
Equivalently,
\begin{equation}
J^{(0)}(z)\Phi(w)\sim\cdots+\frac{h}{z-w}+\cdots\,.
\end{equation}
Furthermore, since $T(z)$ itself is the generator of translations $z\to z+a$, and $L_{-1}$ is the conserved charge of $T(z)$, we have
\begin{equation}
[L_{-1},\Phi(w)]=\partial\Phi(w)\,,
\end{equation}
or
\begin{equation}
T(z)\Phi(w)\sim\cdots+\frac{\partial\Phi(w)}{z-w}+\cdots\,.
\end{equation}
Putting these two OPEs together, we have
\begin{equation}
T(z)\Phi(w)=\cdots+\frac{h}{(z-w)^2}+\frac{\partial\Phi(w)}{z-w}+\cdots\,.
\end{equation}
We say that $\Phi$ is a primary if the higher-order divergent terms in the OPE vanish.

For the bosonic string, we can write down the stress tensor
\begin{equation}
T=\frac{1}{2}\partial X^{\mu}\partial X_{\mu}+2\partial c\,b+c\partial b\,,
\end{equation}
where all products are assumed to be normal-ordered. Since we know the OPEs of $\partial X$ and of $b$ and $c$, we can derive the OPE of $T$ with itself. We find
\begin{equation}
T(z)T(w)\sim\frac{D-26}{2(z-w)^4}+\frac{2T(w)}{(z-w)^2}+\frac{\partial T(w)}{z-w}+\cdots\,.
\end{equation}
This OPE is known as the \textit{Virasoro algera} with central charge $c=D-26$, and it is a characteristic of 2D CFTs that the stress tensor obeys this algebra.

\vspace{0.25cm}

\begin{centering}
\begin{tcolorbox}
\noindent\textbf{Exercise:} Compute the OPE $T(z)T(w)$ from the above definitions.
\end{tcolorbox}
\end{centering}

\vspace{0.25cm}

% \begin{centering}
% \begin{tcolorbox}
% \noindent\textbf{Solution:} 
% \end{tcolorbox}
% \end{centering}

% \vspace{0.25cm}

The $1/(z-w)^4$ term tells us that $T$ itself is not a primary field unless $D=26$. The coefficient of $1/2(z-w)^4$ in the $TT$ OPE is called the `central charge' of the theory, and it turns out that the worldsheet path integral \eqref{eq:bob-gauge-fixed-polyakov} is plagued with anomalies unless this coefficient vanishes. We thus conclude that bosonic string theory must have
\begin{equation}
D=26\,.
\end{equation}

\subsubsection*{BRST quantisation}

The action \eqref{eq:bob-polyakov-and-ghosts} on its own defines a quantum field theory whose states are excitations of some vacuum state $\ket{\Omega}$ by the modes of the fields $\partial X$, $b$, and $c$. However, we originally introduced the ghosts $b,c$ in order to fix a gauge symmetry on the worldsheet, namely the diffeomorphism and Weyl symmetry. Just as in the quantisation of Yang-Mills theories, once we introduce the ghosts, the gauge symmetry is not completely lost. The worldsheet is still invariant under the fermionic transformations:
\begin{equation}
\begin{split}
\delta_{\epsilon}X^{\mu}&=\epsilon\,c\,\partial X\\
\delta_{\epsilon}c&=\epsilon \, c\,\partial c\\
\delta_{\epsilon}b&=\epsilon\,T\,.
\end{split}
\end{equation}
This is the left-over BRST symmetry of the original diffeomorphism symmetry on the worldsheet. The conserved current associated to it is
\begin{equation}
J_{\text{BRST}}=c\,T_{X}+\frac{1}{2}c\,T_{b,c}\,,
\end{equation}
where the products are assumed to be normal-ordered, and $T_X$ and $T_{b,c}$ are the components of the stress-tensor made of $X$ and of $b,c$, respectively. The conserved charge associated to this current is the BRST charge
\begin{equation}
Q_{\text{BRST}}=\oint\frac{\mathrm{d}z}{2\pi i}J_{\text{BRST}}(z)\,,
\end{equation}
where, as usual, the integral is taken around a small circle centered at the origin.

The BRST charge $Q_{\text{BRST}}$ is used to identify physical states of the theory. One can check that the square of the BRST charge vanishes if $D=26$, which, as we mentioned, is required for a consistent worldsheet theory. Thus, $Q_{\text{BRST}}$ can be used to define a cohomology on the Hilbert space of the theory. Physical states are those which lie in those cohomology, \textit{i.e.}~physical states are those which are annihilated by the BRST charge
\begin{equation}
Q_{\text{BRST}}\ket{\psi}=0\,,
\end{equation}
and we identify two states if they differ by a $Q_{\text{BRST}}$-exact term:
\begin{equation}
\ket{\psi}\sim \ket{\psi}+Q_{\text{BRST}}\ket{\varphi}\,.
\end{equation}

By the state-operator-correspondence, we can also simply consider local operators $\psi(z)$ instead of states $\ket{\psi}$. In this case, the action of $Q_{\text{BRST}}$ on the operator $\psi(z)$ is given by
\begin{equation}
[Q_{\text{BRST}},\psi(z)\}\,,
\end{equation}
where the bracket $[\cdot,\cdot\}$ is an anticommutator if both operators are anticommuting and a commutator otherwise. The BRST cohomology is then given by the equivalence relation
\begin{equation}
\psi(z)\sim\psi(z)+[Q_{\text{BRST}},\varphi(z)\}\,.
\end{equation}

Furthermore, given the existence of the $b,c$ ghost system, we define one more constraint for the worldsheet theory, namely that physical states have no excitations in the $b,c$ fields. We define the ghost number current to be
\begin{equation}
J_{bc}=bc\,.
\end{equation}
A physical state is said to have ghost number zero if the charge of the state under this current vanishes. We demand that physical states have ghost number zero.\footnote{There is an ambiguity in the definition of ghost number due to the normal-ordering procedure. Some authors demand that the ghost number is $+1$, others $+\frac{1}{2}$, and others zero. Since we don't deal with the details of this constraint too much, we will keep with the latter convention.}

\subsubsection{The RNS superstring}

We now turn our attention to the $\mathcal{N}=1$ superstring in flat space. It is obtained by adding a worldsheet fermion $\psi^{\mu}$ which is the supersymmetric partner to the worldsheet scalar coordinate $X^{\mu}$. The action for this theory is given by
\begin{equation}\label{eq:bob-rns-string}
S=\frac{1}{4\pi}\int\mathrm{d}^2z\,\partial X^{\mu}\bar{\partial}X_{\mu}+\frac{i}{2\pi}\int\mathrm{d}^2z\,\left(\psi^{\mu}\bar{\partial}\psi_{\mu}+\bar{\psi}^{\mu}\partial\bar{\psi}_{\mu}\right)\,.
\end{equation}
Here, we have taken the worldsheet metric to be flat ($h_{\alpha\beta}=\delta_{\alpha\beta}$). Just as in the bosonic string, this is only a well-defined gauge if the total central charge of the worldsheet theory has central charge $c=0$, so that the Weyl transformations $h\to e^{\omega}h$ is truly a quantum symmetry of the theory. We will revisit this point later when we discuss BRST quantisation.

The equations of motion are simply
\begin{equation}
\partial\bar{\partial}X^{\mu}=0\,,\quad\partial\bar{\psi}^{\mu}=0\,,\quad\bar{\partial}\psi^{\mu}=0\,.
\end{equation}
As before, the general solution for $X$ is a sum of left-movers and right-movers, whereas $\psi$ ($\bar{\psi}$) is a holomorphic (anti-holomorphic) function. That is,
\begin{equation}
X^{\mu}(z,\bar{z})=X^{\mu}_L(z)+X^{\mu}_R(\bar{z})\,,\quad\psi^{\mu}(z,\bar{z})=\psi^{\mu}(z)\,,\quad\bar{\psi}^{\mu}(z,\bar{z})=\bar{\psi}^{\mu}(\bar{z})\,.
\end{equation}
From now on, we will focus \textit{only} on the left-moving (holomorphic) part of the theory. The right-moving part will behave in precisely the same way. The RNS superstring action is invariant under $X\to X+a$. The conserved currents $J^{\mu}$ associated to this symmetry are simply given by
\begin{equation}
J^{\mu}=\partial X^{\mu}\,.
\end{equation}
Since we can obtain $X^{\mu}$ (up to a constant term) by integrating $J^{\mu}$, we will work with $J^{\mu}$ instead of $X^{\mu}$ whenever possible. Thus, in the left-moving sector, we consider the pair $(\psi,J)$ to be the set of fundamental fields of the theory.

The action \eqref{eq:bob-rns-string} has a further symmetry given by supersymmetry transformations on the worldsheet. Let $\varepsilon$ be a Grassmann-odd variable. Then the transformations
\begin{equation}\label{eq:bob-rns-susy-transformations}
\delta_{\varepsilon}X^{\mu}=\varepsilon\psi^{\mu}\,,\quad\delta_{\varepsilon}\psi^{\mu}=\frac{i}{2}\varepsilon \partial X^{\mu}\,,
\end{equation}
leave the action invariant. The conserved current defined by this symmetry is the \textit{supercurrent} and is given by
\begin{equation}
G(z)=J^{\mu}(z)\psi_{\mu}(z)\,.
\end{equation}
Finally, the action \eqref{eq:bob-rns-string} is symmetric under worldsheet translations $z\to z+a$. The conserved current associated to this symmetry is the stress-tensor, and is given by
\begin{equation}\label{eq:bob-rns-stress-tensor}
T(z)=\frac{1}{2}J^{\mu}(z)J_{\mu}(z)+\frac{i}{2}\psi^{\mu}(z)\partial\psi^{\mu}(z)\,.
\end{equation}
Together, $J^{\mu}$, $G$, and $T$ make up the conserved currents of the worldsheet theory, and thus are the fundamental objects we consider during quantisation.

The above supersymmetry transformation \eqref{eq:bob-rns-susy-transformations} defines what is known as $\mathcal{N}=(1,0)$ supersymmetry, since it acts only on the left-moving worldsheet fermion $\psi$. An independent supersymmetry transformation is given by
\begin{equation}\label{eq:bob-01-susy}
\delta_{\overline{\varepsilon}}X^{\mu}=\overline{\varepsilon}\overline{\psi}^{\mu}\,,\quad\delta_{\overline{\varepsilon}}\overline{\psi}^{\mu}=\frac{i}{2}\overline{\varepsilon}\overline{\partial}X^{\mu}\,.
\end{equation}
This generates so-called $\mathcal{N}=(0,1)$ supersymmetry, and the conserved current is the superpartner $\overline{G}$ of the right-moving stress tensor $\overline{T}$. Together \eqref{eq:bob-rns-susy-transformations} and \eqref{eq:bob-01-susy} generate the full worldsheet $\mathcal{N}=(1,1)$ supersymmetry.

Finally, we should emphasize that, a priori, \textit{worldsheet} supersymmetry has nothing to do with \textit{spacetime} supersymmetry. The former is a symmetry under the exchange of worldsheet quantities, whereas spacetime supersymmetry is either \textit{i)} a symmetry between spacetime bosonic and fermionic fields or, alternatively, \textit{ii)} isometries of the \textit{superspace} describing the spacetime. As it stands, the worldsheet fermions $\psi^{\mu}$ have no spacetime geometric interpretation, but are rather just degrees of freedom we add to the worldsheet (in particular, the worldsheet fermions are not spacetime superspace coordinates). We will see later that the RNS string indeed has spacetime supersymmetry, but it is not obvious. There are alternate (and equivalent) formulations of superstring theory which make spacetime supersymmetry manifest. Two examples are:
\begin{itemize}

    \item The Green-Schwarz (GS) superstring. The action of this string theory quantifies the `area' of the worldsheet in the spacetime superspace. The GS string has the advantage of making spacetime supersymmetry manifest, but the disadvantage of being difficult to quantise.

    \item The \textit{hybrid} string, the subject of these lectures. The hybrid string makes (some) spacetime supersymmetry manifest while also keeping manifest conformal symmetry on the worldsheet. However, although the hybrid string allows for `covariant' quantisation, it can only be formulated on special backgrounds.

\end{itemize}

\vspace{0.25cm}

\begin{centering}
\begin{tcolorbox}
\noindent\textbf{Exercise:} \textit{Show that \eqref{eq:bob-rns-string} is invariant under the supersymmetry transformations \eqref{eq:bob-rns-susy-transformations}. Derive the conserved current $G(z)$ from these transforamtions using the Noether procedure.}
\end{tcolorbox}
\end{centering}

\vspace{0.25cm}

% \begin{centering}
% \begin{tcolorbox}
% \noindent\textbf{Solution:} The variation of the action under $\delta_{\varepsilon}$ can be broken down into two pieces:
% \begin{equation*}
% \begin{split}
% \delta_{\varepsilon}(\partial X\overline{\partial}X)&=\varepsilon(\partial\psi\overline{\partial}X+\partial X\overline{\partial}\psi)\\
% \delta_{\varepsilon}(\psi\partial\psi)&=\frac{i}{2}\varepsilon(\partial X\overline{\partial}\psi-\psi\overline{\partial}\partial X)\,.
% \end{split}
% \end{equation*}
% Invariance of the RNS action \eqref{eq:bob-rns-string} under these SUSY transformations is obtained upon plugging the above transformations into the action and integrating by parts.
% \end{tcolorbox}
% \end{centering}

\subsubsection*{Canonical quantisation, OPEs, and the superconformal algebra}

Just as in the bosonic string, we can quantise the theory by promoting $\partial X,\psi$ to operators and imposing the canonical OPEs with their conjugate momenta. We note that $\psi$ is its own conjugate momentum, and so it satisfies the OPE:
\begin{equation}
\psi^{\mu}(z)\psi^{\nu}(w)\sim\frac{\delta^{\mu\nu}}{(z-w)}+\cdots\,.
\end{equation}
The $JJ$ OPE was derived above and is given by
\begin{equation}
\partial X^{\mu}(z)\partial X^{\nu}(w)\sim-\frac{\delta^{\mu\nu}}{(z-w)^2}
\end{equation}

The conserved supercurrent $G(z)$ and stress tensor $T(z)$ are expressed in terms of $J$ and $\psi$ as
\begin{equation}
G(z)=(\partial X^{\mu}\psi_{\mu})(z)\,,\quad T(z)=\frac{1}{2}(\partial X^{\mu}\partial X_{\mu})(z)+\frac{i}{2}(\psi^{\mu}\partial\psi_{\mu})(z)\,,
\end{equation}
where, as always, we take the products to be normal-ordered. One can compute the OPEs among the fields $G$ and $T$. The details of the computation are a bit tendious, but the result is
\begin{equation}
\begin{split}
T(z)T(w)&\sim\frac{3D}{4(z-w)^4}+\frac{2T(w)}{(z-w)^2}+\frac{\partial T(w)}{z-w}\,,\\
T(z)G(w)&\sim\frac{3G(w)}{2(z-w)^2}+\frac{\partial G(w)}{z-w}\,,\\
G(z)G(w)&\sim\frac{D}{(z-w)^3}+\frac{2T(w)}{z-w}\,.
\end{split}
\end{equation}
This is the so-called $\mathcal{N}=(1,0)$ superconformal algebra, which is an extension of the Virasoro algebra we saw in the bosonic string. The central charge of this algebra can be read off by the $TT$ OPE and is given by
\begin{equation}
c=\frac{3D}{2}\,.
\end{equation}
Quantisation of the worldsheet theory requires understanding representations of this algebra, as they form the symmetry algebra of the worldsheet.

\vspace{0.25cm}

\begin{centering}
\begin{tcolorbox}
\noindent\textbf{Exercise:} \textit{Show that $T(z)$ and $G(z)$ satisfy the above algebra.}
\end{tcolorbox}
\end{centering}

\subsubsection*{Ramond and Neveu-Schwarz sectors}

One fundamental difference between the worldsheet theory of the RNS string and that of the bosonic string is the existence of fields with half-integer conformal weight. We will call such fields `spinors' (and use the term `fermion' to refer to anything which obeys fermionc, \textit{i.e.}~anticommuting, statistics).

A special feature of spinors is that, just as in standar QFT, they are only defined up to an overall minus sign. Thus, if we consider, say, the OPE $\psi(z)\Phi(w)$ as a function of $w$, it is entirely feasible that, as $w$ traverses a small circle around $z$, their product can pick up a minus sign, which can be re-absorbed into $\psi$. This defines a \textit{branch cut} in the OPE of $\psi$ with some state $\Phi$. Specifically, if such a branch cut is present, we have
\begin{equation}
\psi(z)\Phi(w)\propto(z-w)^{n+\frac{1}{2}}
\end{equation}
for some $n\in\mathbb{Z}$. Of course, we can also have a state $\Phi$ whose OPE with $\psi$ is globally defined, \textit{i.e.}~has no branch cut.

It is useful to divide states into the \textit{Ramond} (R) sector and the \textit{Neveu-Schwarz} (NS) sector based on whether the OPE has a branch cut or not:
\begin{equation*}
\Phi\in
\begin{cases}
\text{R sector}\,, & \psi(z)\Phi(w)\text{ has a branch cut}\\
\text{NS sector}\,, & \psi(z)\Phi(w)\text{ otherwise}
\end{cases}
\end{equation*}
Note that, because of the superconformal symmetry, if a state is to be in a definite sector of $G$, then it must be in the \textit{same} sector for all of the $\psi^{\mu}$.\footnote{However, a state can be in a different sector for the left- and right-moving components.} As a general rule, R sector states generate spacetime fermions while NS sector states generate spacetime bosons.

\subsubsection*{BRST quantisation and the superconformal ghost system}

Now that we have postulated commutation relations (OPEs) on the worldsheet, we procede to actually quantising the theory. Since worldsheet string theory is a gauge theory (the gauge symmetries being worldsheet (super-)diffeomorphisms and Weyl transformations), a proper treatment of its quantisation requires the introduction of ghost fields conjugate to these gauge transformations, as well as a BRST charge to define an appropriate cohomology.

Within the bosonic string this was accomplished by introducing two new fields $(b,c)$ to the worldsheet with anti-commuting statistics and weights $h(b)=2$, $h(c)=-1$ with action
\begin{equation}
S_{bc}=\frac{1}{2\pi}\int\mathrm{d}^2z\,b\bar{\partial}c\,.
\end{equation}
Upon quantisation, the conjugate momentum to $b$ is $c$, and so we impose the OPEs
\begin{equation}
c(z)b(w)\sim\frac{1}{z-w}\,.
\end{equation}
The $bc$ ghost system is required to fix worldsheet reparametrization symmetry.

Similarly, in the superstring, we not only have to fix the reparametrization invariance, but also the choice of spin connection used to define the fermions on the curved worldsheet. This is accomplished by introducing a commuting pair of ghosts $(\beta,\gamma)$ with conformal weights $h(\beta)=\frac{3}{2}$ and $h(\gamma)=-\frac{1}{2}$. The action is given by
\begin{equation}
S_{\beta\gamma}=\frac{1}{2\pi}\int\mathrm{d}^2z\,\beta\bar{\partial}\gamma\,.
\end{equation}
Similarly to the $bc$ ghost system, canonical quantisation of the $\beta\gamma$ system takes the form of the OPE
\begin{equation}
\gamma(z)\beta(w)\sim\frac{1}{z-w}\,.
\end{equation}
After gauge-fixing, the full action of the RNS string theory is given by
\begin{equation}
S=S_{\text{RNS}}+S_{bc}+S_{\beta\gamma}\,.
\end{equation}
The full stress tensor of this theory is given by
\begin{equation}
T=T_{\text{RNS}}+2\partial c\,b+c\,\partial b-\frac{3}{2}\beta\,\partial\gamma-\frac{1}{2}\partial\beta\,\gamma\,.
\end{equation}
The central charge of this stress tensor is tedious to calculate, but it can be done. It is broken up into three pieces: the piece coming from the RNS string, the piece from the $bc$ system, and the piece from the $\beta\gamma$ system. The full result is
\begin{equation}
c=c_{\text{RNS}}+c_{bc}+c_{\beta\gamma}=\frac{3D}{2}-26+11=\frac{3(D-10)}{2}\,.
\end{equation}
Since the worldsheet theory is only consistent if the full central charge vanishes, we demand $c=0$, or $D=10$. This is the origin of the critical dimension in the RNS superstring.

We can also define a supercharge $G$ of the full RNS string with ghosts, and it is given by
\begin{equation}
G=(J^{\mu}\psi_{\mu})+(\partial\beta)c+\frac{3}{2}\beta(\partial c)-\frac{1}{2}b\gamma\,.
\end{equation}
Together, the `full' stress tensor $T$ and supercharge $G$ satisfy the $\mathcal{N}=1$ superconformal algbera with central charge $c=0$, \textit{i.e.}
\begin{equation}
\begin{split}
T(z)T(w)&\sim\frac{2T(w)}{(z-w)^2}+\frac{\partial T(w)}{z-w}+\cdots\,,\\
T(z)G(w)&\sim\frac{3G(w)}{2(z-w)^2}+\frac{\partial G(w)}{z-w}+\cdots\,,\\
G(z)G(w)&\sim\frac{2T(w)}{z-w}+\cdots\,.
\end{split}
\end{equation}

\subsubsection*{BRST quantisation}

Finally, just as in the bosonic string, we need to construct a BRST charge and impose that physical states live in the BRST cohomology of this charge. First, let us break up $T$ and $G$ into the contributions from the RNS fields $(J,\psi)$ and from the ghosts $(b,c,\beta,\gamma)$. That is, we define
\begin{equation}
T=T_{J,\psi}+T_{\text{gh}}\,,\quad G=G_{J,\psi}+G_{\text{gh}}\,.
\end{equation}
Then we can define a BRST current on the worldsheet as
\begin{equation}
J_{\text{BRST}}=c\left(T_{J,\psi}+\frac{1}{2}T_{\text{gh}}\right)+\gamma\left(G_{J,\psi}+\frac{1}{2}G_{\text{gh}}\right)\,,
\end{equation}
and a BRST charge as
\begin{equation}
Q_{\text{BRST}}=\oint\frac{\mathrm{d}z}{2\pi i}J_{\text{BRST}}(z)\,.
\end{equation}
Just as in the case of the bosonic string, one can show that, as long as $D=10$, the BRST charge satisfies
\begin{equation}
Q_{\text{BRST}}^2=0\,,
\end{equation}
and thus can be used to define a cohomology. Physical states are now those which satisfy
\begin{equation}
Q_{\text{BRST}}\ket{\varphi}=0\,,\quad\ket{\varphi}\sim\ket{\varphi'}+Q_{\text{BRST}}\ket{\psi}\,,
\end{equation}
or, in the language of local operators
\begin{equation}
[Q_{\text{BRST}},\Phi\}=0\,,\quad\Phi\sim\Phi'+[Q_{\text{BRST}},\Psi\}\,.
\end{equation}

\subsubsection{Bosonisation}

The RNS string as described above is composed of fields $(X,\psi,b,c,\beta,\gamma)$. The theory also has many conserved currents, such as the spacetime momentum $-i\partial X$, the ghost numbers $bc$ and $\beta\gamma$, as well as a set of conserved currents for the fermioins $\psi^{\mu}$ that we will discuss shortly. These currents commute amongst themselves, and thus their eigenstates should form a basis of the Hilbert space of states. By the state-operator correspondence, we should be able to write down local operators which have specific eigenvalues under these currents. For the scalar $X$, this is actually straightforward, and we have
\begin{equation}
\ket{k}\Longleftrightarrow e^{ik\cdot X}\,.
\end{equation}
This state has spacetime momentum $k$. For the other currents, however, it is somewhat tricky to write down states that have the appropriate quantum numbers.

For the fields $\psi,b,c$ we can try to do exactly this, using what is known as the \textit{fermion/boson duality} in two-dimensions. A classic example is the duality between the `Sine-Gordon' model and the `Thirring' model \cite{Thirring:1958in}:
\begin{equation}
\begin{split}
S_{\text{Sine-Gordon}}&=\frac{1}{2\pi}\int\left(\frac{1}{2}\partial_{\alpha}\varphi\partial^{\alpha}\varphi-g\cos{\beta\varphi}\right)\,,\\
S_{\text{Thirring}}&=\frac{1}{2\pi}\int\left(\bar{\psi}(i\slashed{\partial}-m)\psi-g'(\bar{\psi}\gamma^{\mu}\psi)(\bar{\psi}\gamma_{\mu}\psi)\right)\,.
\end{split}
\end{equation}
Although the fundamental fields and Lagrangians of these theories are different, Coleman showed that they are equivalent at the quantum level \cite{Coleman:1974bu}. Intuitively, the fermions $\psi$ of the Thirring model are thought to condensates or coherent states of the Sine-Gordon field, while the scalars in the Sine-Gordon theory can be thought of as bound states of the fermions on the Thirring model.

In our case, the fermionic degrees of freedom in the RNS are essentially described by the \textit{free} Thirring model with $m=g'=0$. This is dual to the free sine-Gordon model with $g=\beta=0$. Thus, we should, in a sense, be able to replace the fermions $\psi,b,c$ with scalars. We will show how to do this below.

\noindent\textbf{Note:}
\begin{itemize}

    \item A \textit{boson} is any local operator which satisfies \textit{commutation} relations.

    \item A \textit{fermion} is any local operator which satisfies \textit{anti-commutation} relations.

\end{itemize}
\textit{Example:} The conformal ghosts have $h(b)=2$ and $h(c)=-1$ but are fermions. The superconformal ghosts have $h(\beta)=3/2$ and $h(\gamma)=-1/2$ but are bosons.

\subsubsection*{\boldmath Bosonising the \texorpdfstring{$bc$}{bc} system}

Consider as a prototypical example the $bc$ conformal ghost system. This system consists of two (chiral) anticommuting fields with action
\begin{equation}
S_{bc}=\frac{1}{2\pi}\int\mathrm{d}^2z\,b(z)\bar{\partial}c(z)\,.
\end{equation}
This theory has a conserved current:
\begin{equation}
J_{bc}=bc
\end{equation}
which generates the $\text{U}(1)$ transformation $b\to e^{i\alpha}b$, $c\to e^{-i\alpha}c$. A simple computation shows that the OPE of $J_{bc}$ with itself is given by
\begin{equation}\label{eq:bob-bc-current-ope-bosonisation}
J_{bc}(z)J_{bc}(w)\sim\frac{1}{(z-w)^2}\,.
\end{equation}

Recall that in the bosonic string the current $J_X$ associated to shifts in $X$ was defined by
\begin{equation}
J_{X}=\partial X\,,
\end{equation}
and satisfied the OPE
\begin{equation}\label{eq:bob-x-current-ope-bosonisation}
J_X(z)J_X(w)\sim-\frac{1}{(z-w)^2}\,.
\end{equation}
Given the formal resemblance between \eqref{eq:bob-bc-current-ope-bosonisation} and \eqref{eq:bob-x-current-ope-bosonisation}, one might be tempted to postulate the existence of a scalar field $\sigma$ such that
\begin{equation}
J_{bc}=\partial\sigma\,.
\end{equation}
If such a scalar field exists, then its chiral half should satisfy the OPE
\begin{equation}
\sigma(z)\sigma(w)\sim\log(z-w)\,.
\end{equation}

Given the fields $bc$, it is clear how to construct such a scalar: simply take the normal-ordered product $bc$ and integrate it to get
\begin{equation}
\sigma(z)=\int^{z}bc\,.
\end{equation}
However, a remarkable property about 2D conformal field theories is that this process is invertible. If we define
\begin{equation}
b=e^{-\sigma}\,,\quad c=e^{\sigma}\,,
\end{equation}
it is possible to show that the OPE between $b$ and $c$ is
\begin{equation}
b(z)c(w)=e^{-\sigma}(z)e^{\sigma}(w)\sim\frac{1}{z-w}\,.
\end{equation}
Furthermore, the normal ordered product of $e^{-\sigma}$ and $e^{\sigma}$ is $\partial\sigma$. Thus, by considering the exponentials of $\sigma$, we can recover fields $b,c$ which satisfy anticommuting statistics. These fields satisfy the correct OPEs for a $b,c$ system, and the product $bc$ is simply $\partial\sigma$. This hints at an invertible relationship
\begin{equation}
(b,c)\longleftrightarrow\sigma\,,
\end{equation}
and we can consider the scalar $\sigma$ instead of the two fermions $b,c$.

Naively, however, this cannot be true, since we know that the central charge of the $b,c$ system is $c=-26$, while that of a free scalar is $c=1$. However, this is reconciled by noting that the stress tensor of the $b,c$ system
\begin{equation}
T=\frac{1}{2}(\partial b)c+\frac{1}{2}(\partial c)b-\frac{3}{2}\partial(bc)
\end{equation}
written in terms of the scalar $\sigma$ is
\begin{equation}
T=\frac{1}{2}\partial\sigma\,\partial\sigma-\frac{3}{2}\partial^2\sigma\,.
\end{equation}
Thus, the stress tensor is not that of a standard scalar field, but is `twisted' by a term $-3\partial^2\sigma/2$. The central charge is computed by noting the leading term in the $TT$ OPE, and we indeed verify that the central charge of the scalar $\sigma$ is $c=-26$.

\subsubsection*{Anomalous conservation}

The extra term $-\frac{3}{2}\partial^2\sigma$ has an interpretation in terms of conservation laws. Let $J=\partial\sigma$. We can rewrite the stress tensor as
\begin{equation}
T=\frac{1}{2}JJ-\frac{3}{2}\partial J\,.
\end{equation}
The OPE of $T$ with $J$ is given by
\begin{equation}
T(z)J(w)\sim-\frac{3}{(z-w)^3}+\frac{J(w)}{(z-w)^2}+\frac{\partial J(w)}{z-w}\,.
\end{equation}
The presence of the cubic pole signals that $J$ is not a primary, and thus, as a quantum current, is not conserved. This is not visible from the conformal gauge ($h_{\alpha\beta}=\delta_{\alpha\beta}$), but becomes clear once we couple the worldsheet fields back to a curved worldsheet metric. It can be shown that the above OPE is equivalent to the non-conservation law:
\begin{equation}\label{eq:bob-non-conservation}
\nabla^zJ_z=\frac{3}{4}R\,,
\end{equation}
where $R$ is the curvature on the worldsheet. Although we typically choose the worldsheet metric to be locally flat, this cannot always be done globally. If the worldsheet is a compact Riemann surface with genus $g\neq 1$, then the curvature must be non-zero somewhere, because of the Gauss-Bonet theorem, which states that the integral of the scalar curvature $R$ over the worldsheet is a topological invariant and therefore depends only on the genus $g$:
\begin{equation}
\frac{1}{4\pi}\int_{\Sigma}\mathrm{d}^2z\sqrt{h}\,R=2-2g\,.
\end{equation}
The non-conservation law \eqref{eq:bob-non-conservation} is a signal that there is a quantum anomaly in the ghost number symmetry on the worldsheet.

A consequence of this non-conservation is that correlation functions of fields $\Phi_i$ vanish unless their charges under $J$ sum up to $3-3g$. To see this, note that
\begin{equation}
\begin{split}
\frac{1}{2\pi}\int_{\Sigma}\mathrm{d}^2z\sqrt{h}\left\langle\Delta^zJ_z(z)\prod_{i=1}^{n}\Phi_i(z_i)\right\rangle&=-\sum_{j=1}^{n}\mathop{\mathrm{Res}}_{z = z_j}\left\langle J(z)\prod_{i=1}^{n}\Phi_i(z_i)\right\rangle\\
&=-\left(\sum_{j=1}^{n}Q_j\right)\left\langle\prod_{i=1}^{n}\Phi_i(z_i)\right\rangle\,,
\end{split}
\end{equation}
where we have used the OPE \eqref{eq:bob-jphi-ope} in passing from the first to second line, as well as the identity 
\begin{equation}
\bar\partial\left(\frac{1}{z-z_i}\right)=-2\pi\delta^{(2)}(z-z_i)\,.
\end{equation}
However, by \eqref{eq:bob-non-conservation} and the Gauss-Bonet theorem, we have
\begin{equation}
\begin{split}
\frac{1}{2\pi}\int_{\Sigma}\mathrm{d}^2z\sqrt{h}\left\langle\Delta^zJ_z(z)\prod_{i=1}^{n}\Phi_i(z_i)\right\rangle&=\frac{3}{8\pi}\int_{\Sigma}\mathrm{d}^2z\sqrt{h}\,R\left\langle\prod_{i=1}^{n}\Phi_i(z_i)\right\rangle\\
&=(3-3g)\left\langle\prod_{i=1}^{n}\Phi_i(z_i)\right\rangle\,.
\end{split}
\end{equation}
Thus, we conclude that either the correlator vanishes, or
\begin{equation}
\sum_{i=1}^{n}Q_i=3g-3\,.
\end{equation}

\subsubsection*{Bosonising the worldsheet fermions}

Now that we have bosonised the $bc$ ghost system, we can also consider the worldsheet fermions $\psi^{\mu}$. In \eqref{eq:bob-rns-string} they don't obviously have $\text{U}(1)$ conserved currents, but if we define the complex linear combinations
\begin{equation}
\Psi^{\pm i}=\frac{1}{\sqrt{2}}\left(\psi^{2i-1}\pm i\,\psi^{2i}\right)\,,
\end{equation}
then the fermion action can be written as
\begin{equation}
S_{\psi}=\frac{i}{2\pi}\int\mathrm{d}^2z\,\psi^{\mu}\bar{\partial}\psi_{\mu}=\frac{i}{2\pi}\int\mathrm{d}^2z\,(\Psi^{+i}\bar{\partial}\Psi^{-i}+\Psi^{-i}\bar{\partial}\Psi^{+i})\,.
\end{equation}
This action now looks roughly like $D/2$ copies of the $bc$ system. In particular, there are $D/2$ $\text{U}(1)$ symmetries
\begin{equation}
\Psi^{\pm i}\to e^{\pm i\alpha}\Psi^{\pm i}\,,
\end{equation}
and the conserved currents are
\begin{equation}
J^{i}=\Psi^{+i}\Psi^{-i}\,.
\end{equation}
If $D=10$, there are $5$ such conserved currents.\footnote{In reality, the fermionic action $S_{\psi}$ is invariant under a full set of $\text{SO}(10)$ transformations. The above $\text{U}(1)$ generators are the \textit{Cartan} generators of the (complexified) Lie algebra $\mathfrak{so}(10)$.} Just as for the $bc$ system, we can introduce scalars $\sigma_i$ such that
\begin{equation}
J^i=\Psi^{+i}\Psi^{-i}=\partial\sigma^i\,.
\end{equation}
Given the scalars $\sigma^i$, we can also write the fermions $\Psi^{\pm i}$ as\footnote{Strictly speaking, in defining $\Psi^{\pm i}$ in terms of the scalars $\sigma^i$, we need to introduce phases called \textit{cocycle factors}, which ensure that $\Psi^{\pm i}$ and $\Psi^{\pm j}$ anticommute for $i\neq j$. See \cite{Blumenhagen:2013fgp} for a more careful treatment. We will generally drop cocycle factors wherever they should appear.}
\begin{equation}
\Psi^{\pm i}=e^{\pm\sigma^i}\,.
\end{equation}

Unlike in the case of the $bc$ system, the currents $J^i$ are truly conserved quantities in the quantum theory. This is because the stress tensor \eqref{eq:bob-rns-stress-tensor} has no term analogous to the $\partial(bc)$ term in the $b,c$ system stress tensor.

\subsubsection{Vertex operators}

The benefit of bosonisation is that it provides a direct way to construct vertex operators with specified values under conserved currents. Assume we have a current $J$ on the worldsheet which has conformal weight $h(J)=1$. As we discussed above, a state $\ket{q}$ with charge $q$ is equivalent to a local operator $\Phi_q$ such that
\begin{equation}
J(z)\Phi_q(w)\sim\cdots+\frac{q\,\Phi_q(w)}{z-w}+\cdots\,.
\end{equation}
Rather than postulating the existence of such a state/operator, we can attempt to construct it. Let us assume that a scalar $\sigma$ exists which bosonises the above current, \textit{i.e.}~such that $J=\partial\sigma$. Then by the above discussion, the state $e^{q\sigma}$ satisfies precisely the correct requirements, since
\begin{equation}
J(z)\,e^{q\sigma}(w)\sim\frac{q}{z-w}+\cdots\,.
\end{equation}
The exponential $e^{q\sigma}$ is therefore not only a state with the desired charge $q$, but moreover is a \textit{highest-weight} state for the current $J$ (one for which the $J\Phi$ OPE has no higher-order poles).

We can apply the above construction to vertex operators in the RNS string: so far, we have 16 conserved currents that we know how to bosonise:
\begin{itemize}

    \item The $D=10$ spacetime momentum current $J_X=\partial X$, bosonised trivially though $X$.

    \item The $D/2=5$ fermion number currents $J^i=\Psi^{+i}\Psi^{-i}$, bosonised through the scalars $\boldsymbol{\sigma}=\{\sigma_1,\ldots,\sigma_5\}$ (with background charge $Q=0$).

    \item The $bc$ ghost number current $J_{bc}=bc$ bosonised through the scalar $\sigma$ (with background charge $Q=3$).

\end{itemize}
Using these currents, we can immediately write down a generic highest-weight state of these currents:
\begin{equation}
e^{ik\cdot X}e^{\boldsymbol{\lambda}\cdot\boldsymbol{\sigma}}e^{q\sigma}\,,
\end{equation}
where $k^{\mu}$ labels the spacetime momentum of the state and $q$ labels the $bc$ ghost number. The 5-component vector $\boldsymbol{\lambda}$ labels the 'occupation numbers' of the worldsheet fermions, such that an excitation in $\Psi^{+i}$ counts for $\lambda_i\to\lambda_i+1$ and $\Psi^{-i}$ counts for $\lambda_i\to\lambda_i-1$.

Another advantage of the description of vertex operators via bosonisation is that it is immediately clear which states live in the Ramond (R) sector and which live in the Neveu-Schwarz (NS) sector. Recall that we defined a state $\Phi$ to be in the R sector if the $\Psi^{\pm i}\Phi$ OPE has square-root branch cuts and in the NS sector otherwise. Writing the fermions $\Psi^{\pm i}$ as exponentials, we can read off the $\Psi^{\pm i}$ OPEs with a state of the form $e^{\boldsymbol{\lambda}\cdot\boldsymbol{\sigma}}$ as
\begin{equation}
\Psi^{\pm i}(z)\,e^{\boldsymbol{\lambda}\cdot\boldsymbol{\sigma}}(w)\sim(z-w)^{\pm\lambda_i}e^{\boldsymbol{\lambda}_{\pm i}\cdot\boldsymbol{\sigma}}\,,
\end{equation}
where $\boldsymbol{\lambda}_{\pm i}$ is the original weight vector $\boldsymbol{\lambda}$ with the replacement $\lambda_i\to\lambda_i\pm 1$\,. Thus, we see that the state $e^{\boldsymbol{\lambda}\cdot\boldsymbol{\sigma}}$ is in the R sector when the $\lambda_i$'s are half-integer and in the NS sector when the $\lambda_i$'s are integer. Two special sets of choices of the weight vectors $\boldsymbol{\lambda}$ are:
\begin{itemize}

    \item $\boldsymbol{\lambda}=(0,0,0,0,0)$, so that $e^{\boldsymbol{\lambda}\cdot\boldsymbol{\sigma}}=\textbf{1}$ is the identity operator. This state is also referred to as the `NS vacuum', since it is the NS state with smallest conformal weight with $h=0$.

    \item $\boldsymbol{\lambda}=(\pm\frac{1}{2},\pm\frac{1}{2},\pm\frac{1}{2},\pm\frac{1}{2},\pm\frac{1}{2})$. These are the R sector states with the lowest conformal weight, namely $h=\frac{5}{8}$, and have a 32-fold degeneracy. This is also the dimension $2^{D/2}=32$ of the Dirac algebra in $D=10$ dimensions. This is not a coincidence and, as we will discuss below, these states are spinors in ten dimensions.\footnote{The fundamental reason for this is that the 5 scalars $\sigma_i$ essentially label the Cartan subalgebra of the Lie group $\mathfrak{so}(10)$, the Lorentz algebra in ten (Euclidean) dimensions. The weight vectors $\boldsymbol{\lambda}$ represent labels on the Dynkin diagram, and thus correspond to representations of $\mathfrak{so}(10)$. Weight vectors with half-integer coefficients correspond to the states in the spinor representation of $\mathfrak{so}(10)$.}. We label the states $e^{\boldsymbol{\lambda}\cdot\boldsymbol{\sigma}}$ by $S_{\alpha}$ if there are an even number of $+$ signs in the exponential and $S_{\dot\alpha}$ if there are an odd number of $+$ signs in the exponential, where $\alpha=\{1,\ldots,8\}$ and $\dot\alpha=\{\dot{1},\ldots,\dot{8}\}$. We will show later that these form a representation of the 10 dimensional Clifford algebra.

\end{itemize}

\subsubsection{Picture number and picture changing}

We are finally ready to discuss the last ingredient necessary to define BRST-invariant vertex operators in the RNS string: the bosonisation of the superconformal $\beta,\gamma$ system.

As a reminder, the $\beta,\gamma$ system is a CFT with conformal weights $h(\beta)=\frac{3}{2},h(\gamma)=-\frac{1}{2}$, and is the superpartner of the $b,c$ system. The action of this system is
\begin{equation}
S_{\beta\gamma}=\frac{1}{2\pi}\int\mathrm{d}^2z\,\beta\overline{\partial}\gamma\,.
\end{equation}
Naturally, there is a $\text{U}(1)$ symmetry $\beta\to e^{i\alpha}\beta$, $\gamma\to e^{-i\alpha}\gamma$, and the conserved current is given by
\begin{equation}
J_{\beta\gamma}=\beta\gamma\,,
\end{equation}
which satisfies the OPE
\begin{equation}
J_{\beta\gamma}(z)J_{\beta\gamma}(w)\sim-\frac{1}{(z-w)^2}\,.
\end{equation}
This again is precisely the form of the OPE $\partial\varphi(z)\partial\varphi(w)$ for a scalar which satisfies
\begin{equation}
\varphi(z)\varphi(w)\sim-\log(z-w)\,.
\end{equation}
Therefore, we \textit{define} the scalar $\varphi$ such that
\begin{equation}
J_{\beta\gamma}=-\partial\varphi\,.
\end{equation}
(The minus sign is arbitrary but customary.) One might be tempted to jump the gun and right down an ansatz for $\beta$ and $\gamma$ in terms of $\varphi$, \textit{i.e.}
\begin{equation}
\beta\stackrel{?}{=}e^{-\varphi}\,,\quad\gamma\stackrel{?}{=}e^{\varphi}\,,
\end{equation}
however this cannot be the correct answer for two reasons:
\begin{itemize}

    \item The vertex operators $e^{\pm\varphi}$ obey \textit{fermionic} statistics, while the ghosts $\beta,\gamma$ obey bosonic statistics.

    \item In order for $e^{\pm\varphi}$ to have the correct conformal weights, the stress tensor for the scalar has to be
    \begin{equation*}
    T_{\varphi}=-\frac{1}{2}\partial\varphi\,\partial\varphi+\partial^2\varphi\,.
    \end{equation*}
    However, upon computing the $T_{\varphi}T_{\varphi}$, we can derive the central charge of this stress tensor to be
    \begin{equation*}
    c(\varphi)=13\,,
    \end{equation*}
    whereas we know that the central charge of the $\beta,\gamma$ system is
    \begin{equation*}
    c(\beta,\gamma)=11\,.
    \end{equation*}
    Thus, the scalar $\varphi$ cannot be the full description of the theory.

\end{itemize}

The trick is to define a \textit{fermionic} ghost system $\eta,\xi$ with central charge $c(\eta,\xi)=-2$, so that the scalar $\varphi$ together with the $\eta,\xi$ system has central charge
\begin{equation}
c(\varphi)+c(\xi,\eta)=13-2=11=c(\beta,\gamma)\,.
\end{equation}
Such a ghost system needs conformal weights $h(\eta)=1$ and $h(\xi)=0$, an action
\begin{equation}
S_{\eta\xi}=\frac{1}{2\pi}\int\mathrm{d}^2z\,\eta\overline{\partial}\xi
\end{equation}
and should satisfy the OPEs
\begin{equation}
\eta(z)\xi(w)\sim\frac{1}{z-w}\,,\quad\xi(z)\eta(w)\sim\frac{1}{z-w}
\end{equation}
Once this system is defined, we can use it to `bosonise' the $\beta,\gamma$ system. It turns out that the correct definition is
\begin{equation}
\beta=e^{-\varphi}\partial\xi\,,\quad\gamma=\eta e^{\varphi}\,.
\end{equation}

We should also note that once the $\eta,\xi$ system has been introduced, it contains its own `accidental' $\text{U}(1)$ symmetry $\eta\to e^{i\alpha}\eta$, $\xi\to e^{-i\alpha}\xi$ with current
\begin{equation}
J_{\eta\xi}=\eta\xi\,,
\end{equation}
which can itself be bosonised by introducing a scalar $\chi$ with
\begin{equation}
J_{\eta\xi}=\partial\chi\,,\quad \chi(z)\chi(w)\sim\log(z-w)\,.
\end{equation}
In terms of $\chi$, we can write
\begin{equation}
\eta=e^{-\chi}\,,\quad\xi=e^{\chi}\,,
\end{equation}
and thus
\begin{equation}
\beta=e^{-\varphi}e^{\chi}\partial\chi\,,\quad\gamma=e^{-\chi}e^{\varphi}\,.
\end{equation}
To summarize, the bosonisation of the $\beta,\gamma$ system requires the introduction of \textit{two} scalar fields $\varphi,\chi$. Alternatively, it requires the introduction of one scalar field $\varphi$ and a fermionic ghost system $\eta,\xi$.

\subsubsection*{Picture number and picture changing}

There is a subtlety of the above construction: the $\beta,\gamma$ ghost system is written only in terms of the fields $\varphi,\eta,\partial\xi$ and knows nothing about the `zero mode' (constant of integration) of $\xi$. Specifically, the shift
\begin{equation}
\xi\to\xi+\zeta\,,\quad\partial\zeta=0
\end{equation}
leaves the $\beta,\gamma$ system invariant. That is, the algebra of observables generated by $(\varphi,\eta,\xi)$ is much larger than the original algebra of the $\beta,\gamma$ ghost system. We refer to the former as the \textit{large} algebra and the latter as the \textit{small} algebra. One can check that the conserved current associated to the shift $\xi\to\xi+\zeta$ is just $\eta$, and so operators in the small algebra are required to satisfy
\begin{equation}
\left[\oint\frac{\mathrm{d}z}{2\pi i}\eta,\Phi\right]=0\,,
\end{equation}
\textit{i.e.}~that $\Phi$ is invariant under the $\xi\to\xi+\zeta$ shift.

Another feature of the $\beta\gamma$ system is that it has an infinite number of `ground states' that one could choose for quantisation. Given a state $\Phi_q$ with OPEs
\begin{equation}
\beta(z)\Phi_q(w)\sim\mathcal{O}\left((z-w)^{-3/2+q}\right)\,,\quad\gamma(z)\Phi_q(w)\sim\mathcal{O}\left((z-w)^{1/2+q}\right)\,,
\end{equation}
we say that the state $\Phi_q$ is a ground state of the $\beta,\gamma$ system with \textit{picture} $q$. This state will still have an OPE of with the current $J_{\beta\gamma}=\beta\gamma$ of the form
\begin{equation}
J_{\beta\gamma}(z)\Phi(w)\sim\mathcal{O}((z-w)^{-1})\,,
\end{equation}
and thus still be highest-weight with respect to this current.

Upon bosonising the $\beta\gamma$ system, the above vacua are manifested as different charges under the so-called `picture counting' operator:
\begin{equation}
N_p=\oint\frac{\mathrm{d}z}{2\pi i}(J_{\eta\xi}-J_{\beta\gamma})=\oint\frac{\mathrm{d}z}{2\pi i}(\partial\chi-\partial\varphi)\,.
\end{equation}
This is the difference of the $\beta\gamma$ ghost number and the $\eta\xi$ ghost number. The $\beta,\gamma$ ghosts have picture number $N_p=0$, while the fields $(\varphi,\eta,\xi)$ have
\begin{equation}
N_p(e^{q\varphi})=q\,,\quad N_p(\eta)=-1\,,\quad N_p(\xi)=1\,.
\end{equation}
Any state which is built from the $q$-picture vacuum $\Phi_q$ with only operators built from $\beta,\gamma$ will stay in the $q$-picture sector.

As it turns out, the sectors with picture $q$ and picture $q'$ are isomorphic to each other if $q-q'\in\mathbb{Z}$.\footnote{Since $\beta,\gamma$ are spinor fields, states are either in the R or NS sector. The R sector must always have $q$ half-integer, and the NS sector must always have $q$ integer.} Indeed, given a state $\Phi_{q}$ with picture number $q$, we can always construct new vertex operators with other pictures $\widetilde{q}$. A systematic way to construct such vertex operators is to use the `picture-raising' operator
\begin{equation}
Z\cdot \Phi_q:=[Q_{\text{BRST}},(\xi \Phi_q)\}\,,
\end{equation}
\textit{i.e.}~one takes the normal-ordered product of $\xi$ with $V$ and then commutes the result with the BRST operator. Since $\xi$ has picture number $N_p(\xi)=+1$, the resulting state has picture number $q+1$.

One may object that the state $Z\cdot \Phi_q$ is BRST-exact and thus trivial in the Hilbert space of states. However, this is not quite true, since the BRST cohomology is defined in the \textit{small Hilbert space}, \textit{i.e.}~on those states for which
\begin{equation}
\left[\oint\frac{\mathrm{d}z}{2\pi i}\eta(z),\Phi\right]=0\,.
\end{equation}
However, the state $\xi \Phi_q$ is a member of the \textit{large} Hilbert space, and so $[Q_{\text{BRST}},\xi \Phi_q\}$ is not an exact state from the point of view of the BRST cohomology, which is only defined on the \textit{small} Hilbert space.

The picture-changing operator $Z$ is a useful operator to consider since it allows for changes in the picture number while mapping physical states to physical states. Indeed, if $V_q$ is in the BRST cohomology, then so is $Z\cdot V_q$ since
\begin{equation}
[Q_{\text{BRST}},[Q_{\text{BRST}},\xi V_q\}\}\propto [\xi V_{q},[Q_{\text{BRST}},Q_{\text{BRST}}\}\}=0\,,
\end{equation}
by the (graded) Jacobi identity. Finally, we mention that the inverse of the picture raising operator $Z$ is the picture lowering operator
\begin{equation}
Y=c\,\partial\xi\,e^{-2\varphi}\,,
\end{equation}
which has picture number $N_p(Y)=-1$. The existence of $Y$ shows that the map $Z$ between BRST cohomologies of different picture charge is truly an isomorphism.

\begin{figure}
\centering
\begin{tikzpicture}
\draw[latex-] (0,-1.5) -- (0,-0.5);
% \node[right] at (0,-1) {$Q_{\text{BRST}}$};
\node at (0,0) {$\mathcal{H}_{q-1}$};
\draw[latex-] (0,0.5) -- (0,1.5);
\node[right] at (0,1) {$Q_{\text{BRST}}$};
\node at (0,2) {$\mathcal{H}_{q-1}$};
\draw[latex-] (0,2.5) -- (0,3.5);
\node[right] at (0,3) {$Q_{\text{BRST}}$};
\node at (0,4) {$\mathcal{H}_{q-1}$};
\draw[latex-] (0,4.5) -- (0,5.5);
% \node[right] at (0,5) {$Q_{\text{BRST}}$};
\draw[latex-] (3,-1.5) -- (3,-0.5);
% \node[right] at (3,-1) {$Q_{\text{BRST}}$};
\node at (3,0) {$\mathcal{H}_{q}$};
\draw[latex-] (3,0.5) -- (3,1.5);
\node[right] at (3,1) {$Q_{\text{BRST}}$};
\node at (3,2) {$\mathcal{H}_{q}$};
\draw[latex-] (3,2.5) -- (3,3.5);
\node[right] at (3,3) {$Q_{\text{BRST}}$};
\node at (3,4) {$\mathcal{H}_{q}$};
\draw[latex-] (3,4.5) -- (3,5.5);
% \node[right] at (3,5) {$Q_{\text{BRST}}$};
\draw[latex-] (6,-1.5) -- (6,-0.5);
% \node[right] at (6,-1) {$Q_{\text{BRST}}$};
\node at (6,0) {$\mathcal{H}_{q+1}$};
\draw[latex-] (6,0.5) -- (6,1.5);
\node[right] at (6,1) {$Q_{\text{BRST}}$};
\node at (6,2) {$\mathcal{H}_{q+1}$};
\draw[latex-] (6,2.5) -- (6,3.5);
\node[right] at (6,3) {$Q_{\text{BRST}}$};
\node at (6,4) {$\mathcal{H}_{q+1}$};
\draw[latex-] (6,4.5) -- (6,5.5);
% \node[right] at (6,5) {$Q_{\text{BRST}}$};
\draw[-latex] (-1.5,0) -- (-0.5,0);
\draw[-latex] (0.5,0) -- (2.5,0);
\node[above] at (1.5,0) {$Z$};
\draw[-latex] (3.5,0) -- (5.5,0);
\node[above] at (4.5,0) {$Z$};
\draw[-latex] (6.5,0) -- (7.5,0);
\draw[-latex] (-1.5,2) -- (-0.5,2);
\draw[-latex] (0.5,2) -- (2.5,2);
\node[above] at (1.5,2) {$Z$};
\draw[-latex] (3.5,2) -- (5.5,2);
\node[above] at (4.5,2) {$Z$};
\draw[-latex] (6.5,2) -- (7.5,2);
\draw[-latex] (-1.5,4) -- (-0.5,4);
\draw[-latex] (0.5,4) -- (2.5,4);
\node[above] at (1.5,4) {$Z$};
\draw[-latex] (3.5,4) -- (5.5,4);
\node[above] at (4.5,4) {$Z$};
\draw[-latex] (6.5,4) -- (7.5,4);
\end{tikzpicture}
\caption{The physical Hilbert space in different `pictures'. The picture changing operator $Z$ maps physical states to physical states, and thus defines an isomorphism $Z:H(Q_{\text{BRST}}:\mathcal{H}_q\to\mathcal{H}_q)\to H(Q_{\text{BRST}}:\mathcal{H}_{q+1}\to\mathcal{H}_{q+1})$ between BRST cohomologies at different pictures.}
\end{figure}

\subsubsection*{Physical states and `canonical' picture}

Now that we have bosonised the $\beta,\gamma$ system, we can write down vertex operators (states) which satisfy the BRST condition. A simple guess would be
\begin{equation}\label{eq:bob-rns-vertex-operator}
V_{q,\boldsymbol{\lambda},k}:=e^{q\varphi}e^{\boldsymbol{\lambda}\cdot\boldsymbol{\sigma}}e^{ik\cdot X}\,.
\end{equation}
These states have definite charges under spacetime momentum, have definite fermion occupation numbers, and definite picture number. Moreover, one can compute the conformal weight of this state and we find
\begin{equation}
h(V_{q,\boldsymbol{\lambda},k})=-\frac{k^2}{2}+\frac{\boldsymbol{\lambda}^2}{2}-\frac{q^2}{2}-q\,.
\end{equation}
Of course, for states in the Ramond sector, any half-integer picture $q$ is allowed, and for states in the NS sector, any integer picture $q$ is allowed, but the following choice, dubbed the `canonical' picture, can be chosen such that states of the form \eqref{eq:bob-rns-vertex-operator} are physical without extra ingredients. The canonical picture is defined as
\begin{equation}
q=
\begin{cases}
-1\,,\quad\text{NS sector}\\
-\frac{1}{2}\,,\quad\text{R sector}
\end{cases}
\end{equation}

Recall that a necessary (but not sufficient) condition for a state to be BRST closed is that its conformal weight is equal to one. Thus, if the above state has any chance at being physical, we must have
\begin{equation}
-\frac{k^2}{2}+\frac{\boldsymbol{\lambda}^2}{2}-\frac{q^2}{2}-q=1\,.
\end{equation}
Let's look at a few examples:
\begin{itemize}

    \item $\boldsymbol{\lambda}=\{0,0,0,0,0\}$: This is in the NS sector, so we take $q=-1$, and the physical state condition $h=1$ becomes
    \begin{equation*}
    \frac{k^2}{2}=-\frac{1}{2}\,.
    \end{equation*}
    This is the so-called \textit{tachyon} state and its mass is $m^2=k^2=-1$ (in Planck units).

    \item $\boldsymbol{\lambda}=\{0,\ldots,\pm 1,\ldots,0\}$. This is also in the NS sector, so we take $q=1$, and the physical state condition $h=1$ becomes
    \begin{equation*}
    -\frac{k^2}{2}+\frac{1}{2}-\frac{1}{2}+1=1\,,
    \end{equation*}
    and so $k^2=0$. These state is massless and there are ten of them, one for each vector index in $\mu=0,\ldots,9$.

    \item $\boldsymbol{\lambda}=\{\pm\frac{1}{2},\pm\frac{1}{2},\pm\frac{1}{2},\pm\frac{1}{2},\pm\frac{1}{2}\}$: This is in the R sector, so we take $q=-\frac{1}{2}$, and the physical state condition is
    \begin{equation*}
    -\frac{k^2}{2}+\frac{5}{8}-\frac{1}{4}+\frac{1}{2}=1\,,
    \end{equation*}
    or $k^2=0$. These states will be spinors and also massless.

\end{itemize}
The above states of the superstring constitute its \textit{low-lying spectrum}, \textit{i.e.}~the physical states of smallest possible mass. An infinite tower of other physical states can be constructed, and their masses will be strictly positive $M^2\geq 1$. Since mass is measured in Planck units\footnote{More specifically, units of $\hbar c/\ell_s$, where $\ell_s$ is the string length}, these states will be completely inaccessible to low energy physics.

\begin{table}
\begin{center}
\begin{tabular}{ | c | c | c | c | }
    \hline
     & & \\
    {\textbf{State}} & {\textbf{Sector}} & {\textbf{Mass}} \\
     & & \\
    \hline
     & & \\
    {$e^{-\varphi}e^{ik\cdot X}$} & {Neveu-Schwarz} & {$M^2=-1$} \\
     & & \\
    \hline
     & & \\
    {\hspace{0.5cm}$e^{-\varphi/2}e^{\pm\sigma_1/2\pm\cdots\pm\sigma_5/2}e^{ik\cdot X}$\hspace{0.5cm}} & {\hspace{1cm}Ramond}\hspace{1cm} & {\hspace{0.75cm}$M^2=0$\hspace{0.75cm}} \\
     & & \\
    \hline
     & & \\
    {$e^{-\varphi}e^{\pm\sigma_i}e^{ik\cdot X}$} & {Neveau-Schwarz} & {$M^2=0$} \\
     & & \\
    \hline
\end{tabular}
\end{center}
\caption{The first low-lying physical states of the (left-moving part of the) RNS string.}
\end{table}

\subsubsection{The GSO projection}

The above low-lying spectrum for the superstring theory has two primary problems.
\begin{itemize}

    \item Most obviously, it contains a non-physical state with mass $M^2=-1$ (in Planck units). In any physically sensible theory, such states need to be taken care of.

    \item Less obviously, once one puts together the left- and right-moving parts of the string spectrum, the number of fermionic and bosonic degrees of freedom in the spacetime theory are not equal. If we want to construct a theory which has spacetime supersymmetry at high energies, then we need a way to reconcile this mis-match of the counting of degrees of freedom.

\end{itemize}
It turns out that both of these problems can be fixed in one fell swoop via the \textit{Gliozzi-Scherk-Olive (GSO) projection} \cite{Gliozzi:1976qd}.

The first step in the GSO projection is to define a quantum number $(-1)^F$ which, for a given state, counts the number of fermionic modes acting on it. In practical terms, given a state of the form
\begin{equation}
e^{q\varphi}e^{\boldsymbol{\lambda}\cdot\boldsymbol{\sigma}}\,,
\end{equation}
the fermion number $(-1)^F$ is defined by
\begin{equation}
(-1)^F=(-1)^{q+\sum_{i=1}^{5}\lambda_i}\,.
\end{equation}
The \textit{GSO projection} is the following recipe:
\begin{itemize}

    \item In the NS sector, demand $(-1)^F=1$.

    \item In the Ramond sector, there are two choices: either demand $(-1)^F=1$ or $(-1)^F=-1$. These choices differ by a choice of normalization, and the identification $\sigma_i\to-\sigma_i$ swaps them. We will take $(-1)^F=-1$ for convenience.

\end{itemize}
States which do not satisfy the above criteria are said to be \textit{projected out} by the GSO projection. We define the physical states of the RNS string to be those which survive the GSO projection. Within the low-lying states, the Tachyon $e^{-\varphi}$ has $(-1)^F=-1$, and is projected out. In the R sector, the states
\begin{equation}
e^{-\varphi/2}e^{\pm\sigma_1/2\pm\cdots\pm\sigma_5/2}
\end{equation}
have
\begin{equation}
(-1)^F=(-1)^{\#\text{plus signs}-3}\,,
\end{equation}
which is $+1$ for an odd number of plus signs and $-1$ for an even number of plus signs. Given our above convention, we keep the states with an even number of plus signs. The physical spectrum after GSO projection is listed in Table \ref{tab:bob-gso-spectrum}.

\begin{table}
\begin{center}
\begin{tabular}{ | c | c | c | c | } 
    \hline
     & & \\
    { \textbf{State}} & { \textbf{Sector}} & { \textbf{Mass}} \\
     & & \\
    \hline
     & & \\
    {\textcolor{black}{$e^{-\varphi}e^{ik\cdot X}$}} & { Neveu-Schwarz} & { $M^2=-1$} \\
     & & \\
    \hline
     & & \\
    {\hspace{0.5cm}\textcolor{black}{$e^{-\varphi/2}\underbrace{e^{\pm\sigma_1/2\pm\cdots\pm\sigma_5/2}}_{\text{odd number of + signs}}e^{ik\cdot X}$}\hspace{0.5cm}} & {\hspace{1.5cm} Ramond\hspace{1.5cm}} & {\hspace{1cm} $M^2=0$\hspace{1cm}} \\
     & & \\
    \hline
     & & \\
    {\textcolor{black}{$e^{-\varphi/2}\underbrace{e^{\pm\sigma_1/2\pm\cdots\pm\sigma_5/2}}_{\text{even number of + signs}}e^{ik\cdot X}$}} & { Ramond} & { $M^2=0$} \\
     & & \\
    \hline
     & & \\
    { \textcolor{black}{$e^{-\varphi}e^{\pm\sigma_i}e^{ik\cdot X}$}} & { Neveu-Schwarz} & { $M^2=0$} \\
     & & \\
    \hline
\end{tabular}
\end{center}
\caption{The spectrum of the (left-moving part) of the RNS superstring after GSO projection. The tachyonic state of the NS sector gets projected out, along with half of the R sector ground states.}
\label{tab:bob-gso-spectrum}
\end{table}

\subsection{Spacetime supersymmetry in the RNS string}\label{sec:bob-spacetime-susy}

In the previous lectures, we introduced the superstring in the RNS formalism. Let us briefly review the salient features:
\begin{itemize}

    \item The RNS string is based on the field content $(J^{\mu}=\partial X^{\mu},\psi^{\mu})$ as well as the superconformal ghost system $(b,c,\beta,\gamma)$. Together, if we assume $\mu=1,\ldots,10$, these fields form an $\mathcal{N}=1$ superconformal field theory on the worldsheet with vanishing central charge.

    \item The physical states are identified with the cohomology of the BRST charge
    \begin{equation*}
    Q_{\text{BRST}}=\oint\frac{\mathrm{d}z}{2\pi i}\left(c\left(T_{J,\psi}+\frac{1}{2}T_{\text{gh}}\right)+\gamma\left(G_{J,\psi}+\frac{1}{2}G_{\text{gh}}\right)\right)\,.
    \end{equation*}

    \item Physical states must have eigenvalue $+1$ under the ghost number charge
    \begin{equation*}
    Q_{\text{ghost}}=\oint\frac{\mathrm{d}z}{2\pi i}\left(bc+\beta\gamma\right)\,.
    \end{equation*}

    \item After bosonising the $\beta,\gamma$ ghost system, we require that vertex operators live in the `small' Hilbert space of the $\eta,\xi$ system, \textit{i.e.}
    \begin{equation}
    \left[\oint\frac{\mathrm{d}z}{2\pi i}\eta(z),\Phi\right]=0\,.
    \end{equation}

    \item Physical states must satisfy the GSO projection, which amounts to summing over spin structures on the worldsheet. This ensures a consistent spacetime spectrum with spacetime supersymmetry.

    \item Upon bosonising the ghost system, the Hilbert space contains infinitely physically equivalent copies of the same vertex operator which lie in the BRST cohomology. These vertex operators are differentiated by their `picture number' $q$, and different picture numbers are related to each other by the `picture raising' and `picture lowering' operators
    \begin{equation*}
    Z=[Q,\xi\,\cdot\}\,,\quad Y=c\,\partial\xi\,e^{-2\varphi}\,.
    \end{equation*}

\end{itemize}
The RNS formalism is incredibly powerful: it allows one to write vertex operators on the worldsheet for every spacetime state generated by a single string, and can be used to compute correlation functions in a conceptually straightforward manner. The drawback of the RNS formalism, however, is that the symmetries of the spacetime are not always manifest. In particular, looking at the form of the action of the RNS string, it is not at all obvious that the target space enjoys spacetime supersymmetry.

One way to make spacetime supersymmetry manifest is to abandon the RNS string altogether and define a new worldsheet theory whose fundamental fields are the coordinates $X$, along with anticommuting superspace coordinates $\theta,\bar{\theta}$. This leads to the \textit{Green-Schwarz} formalism which we explored in Section~\ref{s:sibylle}. An alternative route is to try to redefine the fields in the RNS formalism such that they manifestly demonstrate spacetime supersymmetry. This will lead to the hybrid formalism, which is the main topic of this section. As mentioned in above, we will closely follow the treatment of \cite{Berkovits:1996bf}.

In order to understand the hybrid formalism, we must first understand how spacetime supersymmetry is obtained from the RNS formalism in the first place.

\subsubsection{Spacetime supersymmetry generators on the worldsheet}

Recall that, in bosonic string theory in flat space, there is a global symmetry corresponding to constant shifts $X^{\mu}\to X^{\mu}+a^{\mu}$ in the spacetime coordinates. The conserved current corresponding to this shift is simply $J^{\mu}=\partial X^{\mu}$, and the conserved charge is computed by integrating $J^{\mu}$ over a constant time slice on the worldsheet. In radial quantisation, a constant time slice is a circle encircling the origin, and so the conserved charge is given by
\begin{equation}\label{eq:bob-spacetime-momentum}
P^{\mu}:=\oint\frac{\mathrm{d}z}{2\pi}\partial X^{\mu}\,.
\end{equation}
Since this is the conserved charge associated to shifts in the spacetime coordinate, we are justified in calling this conserved charge the `spacetime momentum'.\footnote{Recall, again, that we supress right-moving contributions in these notes. The actual spacetime momentum includes \eqref{eq:bob-spacetime-momentum} together with its Hermitian conjugate.} We have defined $P^{\mu}$ with a relative factor if $i$ to the standard definition of the conserved charge associated to $J$, because we want states of the form $e^{ik\cdot X}$ to have eigenvalue $k^{\mu}$ under $P^{\mu}$\,.

A string theory which describes a spacetime that has supersymmetry should have a conserved charge on the worldsheet which generates this supersymmetry. While there is no obvious symmetry on the worldsheet which corresponds to spacetime supersymmetry, we can try to hunt for such an operator regardless and hope it corresponds to the spacetime supersymmetry transformations. Since all charges we have seen so far have been constructed as line integrals of weight-1 fields on the worldsheet, we are searching for a worldsheet field $q(z)$ which would have to satisfy the following criteria:
\begin{itemize}

    \item It would have to live in the Ramond sector, so that $q(z)$ represents a spacetime fermion.

    \item It would have to transform in the spinor representation of the ten-dimensional Lorentz group.

    \item They should generate the ten-dimensional supersymmetry algebra. Let $\alpha$ be a spinor index in ten-dimensions. Then if $Q_{\alpha}$ is the conserved charge associated to $q_{\alpha}$, then we should have
    \begin{equation*}
    \{Q_{\alpha},Q_{\beta}\}=\Gamma_{\alpha\beta}^{\mu}P_{\mu}\,.
    \end{equation*}

\end{itemize}

A particularly simple set of vertex operators in the RNS formalism satisfying the above requirements are given by
\begin{equation}
V_{q,\boldsymbol{\lambda},k}=e^{q\varphi}e^{\boldsymbol{\lambda}\cdot\boldsymbol{\sigma}}e^{ik\cdot X}\,,
\end{equation}
where $q=-1$ if $\boldsymbol{\lambda}$ has integer entries (NS sector) and $q=-\frac{1}{2}$ if $\boldsymbol{\lambda}$ has half-integer entries (R sector). Let us take $k=0$ for simplicity. The lowest-energy states in the Ramond sector are given by $\boldsymbol{\lambda}=\{\pm\frac{1}{2},\ldots,\pm\frac{1}{2}\}$ and have conformal weight
\begin{equation}
h=\frac{\boldsymbol{\lambda}^2}{2}-\frac{q^2}{2}-q=1\,,
\end{equation}
and are physical states, \textit{i.e.}~in the BRST cohomology. We will not show it here, since we have not discussed the Lorentz symmetry generators on the worldsheet, but it can also been shown that the above states transform in the spinor representation of $\text{SO}(10)$.

There are $2^5=32$ such states. The GSO projection tells us to only keep half of them, for example only those with an even number of plus signs in $\boldsymbol{\lambda}$. Thus, we collectively define
\begin{equation}
S_{\alpha}=e^{\pm\frac{\sigma_1}{2}\pm\frac{\sigma_2}{2}\pm\frac{\sigma_3}{2}\pm\frac{\sigma_4}{2}\pm\frac{\sigma_5}{2}}\,,
\end{equation}
where $\alpha=1,\ldots,16$. It can be shown that the OPE between these $S_{\alpha}$ fields is given by
\begin{equation}
S_{\alpha}(z)S_{\beta}(w)\sim\frac{\Gamma_{\alpha\beta}^{\mu}\psi_{\mu}(w)}{(z-w)^{3/4}}+\cdots\,,
\end{equation}
where $\Gamma^{\mu}$ are the $10$-dimensional Dirac matrices.\footnote{The full spinor algebra in 10 dimensions in 32 dimensional. The GSO projection is equivalent to decomposing the $\mathfrak{so}(10)$ representations as $\textbf{32}=\textbf{16}\oplus\textbf{16}'$, where $\textbf{16}$ and $\textbf{16}'$ are the Weyl representations. Here, we take spinors to be in the $\textbf{16}$ and the gamma matrices are the ones projected onto this subalgebra.}

We define the `spacetime supersymmetry' generators $Q_{\alpha}$ to be the integral of these vertex operators at $k=0$ around a circle centered at the origin.\footnote{Since supersymmetry generators satisfy $[P_{\mu},Q_{\alpha}\}=0$, they should carry no momentum.} That is, we define $q_{\alpha}=e^{-\varphi/2}S_{\alpha}$ and
\begin{equation}
Q_{\alpha}=\oint\frac{\mathrm{d}z}{2\pi i}q_{\alpha}(z)=\oint\frac{\mathrm{d}z}{2\pi i} S_{\alpha}e^{-\varphi/2}\,.
\end{equation}

So far, we have simply defined a set of operators which we call the sapcetime supersymmetry generators. For the operators $Q_{\alpha}$ to truly be the supersymmetry generators in spacetime, then they must satisfy the supersymmetry algebra
\begin{equation}
\{Q_{\alpha},Q_{\beta}\}=\Gamma^{\mu}_{\alpha\beta}P_{\mu}=\oint\frac{\mathrm{d}z}{2\pi}\Gamma^{\mu}_{\alpha\beta}\partial X_{\mu}\,,
\end{equation}
where in the second step we used that the momentum operator $P_{\mu}$ is the integral of the conserved current $\partial X_{\mu}$. However, we can compute the anticommutator of $Q_{\alpha}$ with $Q_{\beta}$ and we find (see the exercise below)
\begin{equation}
\{Q_{\alpha},Q_{\beta}\}=\oint\frac{\mathrm{d}z}{2\pi i}e^{-\varphi}\psi_{\mu}\Gamma^{\mu}_{\alpha\beta}\,.
\end{equation}
This is, of course, \textit{not} the usual supersymmetry algebra. But why?

The answer lies in picture number. The gravitino vertex operator lives in the $q=-\frac{1}{2}$ picture (since it is defined with a factor of $e^{-\varphi/2}$), and so the anticommutator $\{Q_{\alpha},Q_{\beta}\}$ has picture number $q=-1$. However, the momentum operator clearly has picture number $q=0$, since there are no exponentials of $\varphi$ in its definition. Thus, the anticommutator of the supercharges never had a chance of reproducing the desired supersymmetry algebra. However, we can act on the anticommutator $\{Q_{\alpha},Q_{\beta}\}$ with the picture changing operator $Z$, and we find
\begin{equation}
Z\cdot\{Q_{\alpha},Q_{\beta}\}=\oint\frac{\mathrm{d}z}{2\pi i}\Gamma^{\mu}_{\alpha\beta}Z\cdot\left(e^{-\varphi}\psi_{\mu}\right)=\oint\frac{\mathrm{d}z}{2\pi}\Gamma^{\mu}_{\alpha\beta}\partial X^{\mu}=\Gamma^{\mu}_{\alpha\beta}P^{\mu}\,.
\end{equation}
Thus, the spacetime supersymmetry algebra is satisfied \textit{up to picture-changing}.

\begin{centering}
\begin{tcolorbox}
\textbf{Exercise:} \textit{Given the form of the supersymmetry generators $Q_{\alpha}$ above, show that}
\begin{equation*}
\{Q_{\alpha},Q_{\beta}\}=\oint\frac{\mathrm{d}z}{2\pi i}e^{-\varphi}\psi_{\mu}\Gamma^{\mu}_{\alpha\beta}\,.
\end{equation*}
\textbf{Hint:} \textit{Use the OPEs}
\begin{align}
S_{\alpha}(z)S_{\beta}(w)&\sim\frac{\Gamma^{\mu}_{\alpha\beta}\psi^{\mu}(w)}{(z-w)^{3/4}}+\cdots\,,\\
e^{q\varphi}(z)e^{q'\varphi}(w)&\sim\frac{e^{(q+q')\varphi}(w)}{(z-w)^{qq'}}+\cdots
\end{align}
\textit{to figure out the OPE $Q_{\alpha}(z)Q_{\beta}(w)$, and then perform contour integrals to extract the anticommutator $\{Q_{\alpha},Q_{\beta}\}$.}
\end{tcolorbox}
\end{centering}

% \begin{centering}
% \begin{tcolorbox}
% \textbf{Solution:} Letting $Q_{\alpha}(z)=e^{-\varphi/2}(z)S_{\alpha}(z)$, since $e^{-\varphi/2}$ and $S_{\alpha}$ commute, we can compute the OPEs of the $Q_{\alpha}$'s by the product of $e^{-\varphi/2}(z)e^{-\varphi/2}(w)$ and $S_{\alpha}(z)S_{\beta}(w)$. Using the hint in the exercise, this gives
% \begin{equation*}
% Q_{\alpha}(z)Q_{\beta}(w)=\left(e^{-\varphi/2}(z)e^{-\varphi/2}(w)\right)\left(S_{\alpha}(z)S_{\beta}(w)\right)\sim\frac{\Gamma^{\mu}_{\alpha\beta}e^{-\varphi}(w)\psi^{\mu}(w)}{z-w}+\cdots\,.
% \end{equation*}
% Now, realising that $Q_{\alpha}=\oint\frac{\mathrm{d}z}{2\pi i}Q_{\alpha}(z)$, we can take contour integrals of both sides and we find
% \begin{align}
% \{Q_{\alpha},Q_{\beta}\}&=\oint_{w}\frac{\mathrm{d}z}{2\pi i}\oint_{0}\frac{\mathrm{d}w}{2\pi i}\frac{\Gamma^{\mu}_{\alpha\beta}e^{-\varphi}(w)\psi^{\mu}(w)}{z-w}\\
% &=\int\frac{\mathrm{d}w}{2\pi i}\Gamma^{\mu}_{\alpha\beta}(e^{-\varphi}\psi^{\mu})(w)\,.
% \end{align}
% \end{tcolorbox}
% \end{centering}

\begin{centering}
\begin{tcolorbox}
\textbf{Exercise:} \textit{Let $Z$ be the picture-changing operator. Show that the spacetime supersymmetry algebra is preserved up to picture-changing, \textit{i.e.}}
\begin{equation*}
Z\cdot\{Q_{\alpha},Q_{\beta}\}=\Gamma^{\mu}_{\alpha\beta}P_{\mu}\,.
\end{equation*}
\end{tcolorbox}
\end{centering}

% \begin{centering}
% \begin{tcolorbox}
% \textbf{Solution:}
% \end{tcolorbox}
% \end{centering}

\subsubsection{Off-shell spacetime supersymmetry?}

In the previous section we showed that the spacetime supersymmetry generators $Q_{\alpha}$ of the ten-dimensional superstring do not satisfy the usual spacetime supersymmetry algebra, but instead satisfy the a modified version
\begin{equation}
Z\cdot\{Q_{\alpha},Q_{\beta}\}=\Gamma^{\mu}_{\alpha\beta}P_{\mu}\,,
\end{equation}
\textit{i.e.}~the generators obey the supersymmetry algebra up to the inclusion of the picture-changing operator $Z$. This means that the spacetime supersymmetry algebra is only satisfied \textit{on-shell}.\footnote{The idea of two states being `equivalent' under picture changing is only true for states which satisfy the BRST conditions.} In order to define a superstring theory for which the spacetime supersymmetry is manifest, we need a supersymmetry algebra which closes \textit{off-shell}.

Recall that the fundamental reason for the failure of the closure of the supersymmetry algebra is that the operators $Q_{\alpha}$ have picture number $q=-\frac{1}{2}$, while the momentum operator $P_{\mu}$ has picture number $q=0$. One possible way to get a sensible supersymmetry algebra would be to introduce supersymmetry generators $\widetilde{Q}_{\alpha}$ with picture number $q=+\frac{1}{2}$. This would allow the anticommutator $\{Q_{\alpha},\widetilde{Q}_{\beta}\}$ to have picture number $q=0$, and thus have a chance at reproducing the spacetime SUSY algebra off-shell. A natural guess for $\widetilde{Q}_{\alpha}$ would be simply $Z\cdot Q_{\alpha}$. This operator can be calculated, and we have the resulting supersymmetry algebra is
\begin{equation}
q_{\alpha}(z)\widetilde{q}_{\beta}(w)\sim\frac{\Gamma_{\alpha\beta}^{\mu}\partial X_{\mu}}{z-w}\implies \{Q_{\alpha},\widetilde{Q}_{\beta}\}=\Gamma^{\mu}_{\alpha\beta}P_{\mu}\,,
\end{equation}
where $\widetilde{q}_{\alpha}=Z\cdot q_{\alpha}$. That is, $\{Q_{\alpha},\widetilde{Q}_{\beta}\}$ indeed computes the correct supersymmetry algebra!

However, there is now a problem, namely that we have twice as many supersymmetry generators as we should. Thus, in order to have a well-defined set of supersymmetry generators, we need to find a way to choose half of the original generators to be promoted to generators with picture number $q=+\frac{1}{2}$. In ten dimensions, there is no Lorentz-invariant choice of half of the supersymmetry generators. However, as we will see later, this is possible in four dimensions, due to the chiral decomposition of the Lorentz algebra $\mathfrak{so}(4)\cong\mathfrak{su}(2)\oplus\mathfrak{su}(2)$.

\subsection{The hybrid string in four dimensions}\label{sec:bob-4d}

In the previous section, we introduced the problems of spacetime supersymmetry in the RNS string. Briefly restated, they are:
\begin{itemize}

    \item The supersymmetry generators $Q_{\alpha}$ on the worldsheet do not satisfy the supersymmetry algebra (off shell).

    \item Spacetime supersymmetry is only preserved when the vertex operators are GSO projected.

    \item One can define new supersymmetry generators $\widetilde{Q}_{\alpha}$ such that $\{Q_{\alpha},\widetilde{Q}_{\alpha}\}=\Gamma^{\mu}_{\alpha\beta}P_{\mu}$ is the off-shell supersymmetry algebra. However, there is no Lorentz-invariant choice of which supercharges to keep as $Q_{\alpha}$ and which to replace with $\widetilde{Q}_{\alpha}$.

\end{itemize}
As we will see in this section, all of these problems can be avoided in the context of compactifications to four dimensions.

\subsubsection{The four-dimensional SUSY algebra}

The magical feature of supersymmetry in four dimensions is that the Lorentz algebra $\mathfrak{so}(3,1)$ has a \textit{chiral splitting} as $\mathfrak{su}(2)\oplus\mathfrak{su}(2)$. The spinor representation of the Lorentz group in terms of the $\mathfrak{su}(2)$ algebras is
\begin{equation}
\text{spinor of }\mathfrak{so}(3,1)\cong(\tfrac{1}{2},0)\oplus(0,\tfrac{1}{2})\,.
\end{equation}
Thus, every spinor in four-dimensions can be written as the direct sum of two $\mathfrak{su}(2)$ spinors transforming under the two $\mathfrak{su}(2)$ algebras.

This chiral splitting allows us to split the supersymmetry generators in four-dimensions into two sets of two: the two generators $Q_{\alpha}$ form a doublet of the first $\mathfrak{su}(2)$ (\textit{i.e.}~transform in the $(\frac{1}{2},0)$ of $\mathfrak{so}(3,1)$) while the other two generators $Q_{\dot\alpha}$ form a doublet of the second $\mathfrak{su}(2)$ (\textit{i.e.}~transform in the $(0,\frac{1}{2})$ of $\mathfrak{so}(3,1)$). The supersymmetry algebra, in an appropriate basis, is then
\begin{equation}
\{Q_{\alpha},Q_{\dot\beta}\}=2\sigma^{\mu}_{\alpha\dot\beta}P_{\mu}\,,\quad\{Q_\alpha,Q_\beta\}=\{Q_{\dot\alpha},Q_{\dot\beta}\}=0\,.
\end{equation}
Crutially, since $Q_{\alpha}$ (resp. $Q_{\dot\alpha}$) anticommute among each other, when defining string theory in four dimensions, we can choose only one-half of the supercharges (say $Q_{\dot\alpha}$) to modify so that the spacetime supersymmetry algebra can close off-shell on the worldsheet. First, however, we need to review the compactification of superstrings to four-dimensions.

\subsubsection{Calabi-Yau compactifications}

In order to describe a consistent superstring theory in four-dimensions, we need to `compactify' six of the original ten spacetime dimensions. This is implemented by splitting the original superstring action into a four-dimensional part and a six-dimensional `compactified' part, \textit{i.e.}
\begin{equation}
S=\frac{1}{2\pi}\int\mathrm{d}^2z\,\left(\frac{1}{2}\partial X^{\mu}\bar{\partial}X_{\mu}+i\psi^{\mu}\bar{\partial}\psi_{\mu}+i\bar{\psi}^{\mu}\partial\bar{\psi}_{\mu}\right)+S_{C}\,,
\end{equation}
where $S_C$ is the action of the six compactified directions, and now the index $\mu$ runs from $0$ to $3$.

Upon including the $b,c$ conformal and the $\beta,\gamma$ superconformal ghosts, the full central charge of the theory should be $0$. We know
\begin{equation}
c(X)=4\cdot 1\,,\quad c(\psi)=4\cdot\frac{1}{2}\,,\quad c(b,c)=-26\,,\quad c(\beta,\gamma)=11\,,
\end{equation}
and so the total central charge of the $X,\psi$ fields along with the ghosts is $c(X,\psi,b,c,\beta,\gamma)=-9$. Thus, in order for the ten-dimensional theory to have a vanishing Weyl anomaly, we need
\begin{equation}
c(\text{compactification})=9\,.
\end{equation}
There are several possible ways to achieve this. For example, the RNS string on $\mathbb{R}^6/\Lambda$, where $\Lambda$ is some lattice, is an example of such a theory.

For our purposes, we will consider a special class of string compactifications -- so-called \textit{Calabi-Yau} compactifications. There are many ways to define a Calabi-Yau compactification, see for example \cite{Polchinski:1998rr}. However, in the following we will only need to know that Calabi-Yau manifolds define worldsheet CFTs with extended supersymmetry. Specifically:
\begin{RCText}[2cm]
\textit{A Calabi-Yau compactification defines a conformal field theory with central charge $c=9$ which possesses an $\mathcal{N}=(2,2)$ superconformal algebra.}
\end{RCText}
By an $\mathcal{N}=(2,2)$ superconformal algebra, we mean an extension of the standard $\mathcal{N}=(1,1)$ superconformal algebra which has two supercharges $G_C^{\pm}$ in both the left- and right-moving sectors. Furthermore, there is a conserved $\text{U}(1)$ current $J_C$ under which $G^{\pm}_C$ carry charge $\pm 1$. Together with the stress tensor $T_C$, the $\mathcal{N}=(2,0)$ (that is, the left-moving part of the full $\mathcal{N}=(2,2)$) algebra is given by
\begin{equation}
\begin{split}
T_C(z)T_C(w)&\sim\frac{c}{2(z-w)^4}+\frac{2T_C(w)}{(z-w)^2}+\frac{\partial T_C(w)}{z-w}\,,\\
T_C(z)G^{\pm}_C(w)&\sim\frac{3G_C^{\pm}(w)}{2(z-w)^2}+\frac{\partial G_C^{\pm}(w)}{z-w}\,,\\
G_C^+(z)G_C^-(w)&\sim\frac{c}{6(z-w)^3}+\frac{J_C(w)}{2(z-w)^2}+\frac{2T_C(w)+\partial J_C(w)}{z-w}\,,\\
J_C(z)G_C^{\pm}(w)&\sim\pm\frac{G^{\pm}_C(w)}{z-w}\,,\\
J_C(z)J_C(w)&\sim\frac{c}{3(z-w)^3}\,,
\end{split}
\end{equation}
where, for our case, $c=9$. We need no information about the structure of the six-dimensional compactified theory except for the existence of the generators $(T_C,G^{\pm}_C,J_C)$.

The stress tensor and supersymmetry current of the full theory are given by
\begin{equation}
\begin{split}
T&=\frac{1}{2}\partial X^{\mu}\partial X_{\mu}+\frac{i}{2}\psi^{\mu}\partial\psi_{\mu}+T_C+T_{\text{ghosts}}\\
G&=\psi^{\mu}\partial X_{\mu}+G_C^++G_C^-+G_{\text{ghosts}}\,.
\end{split}
\end{equation}
Together, $T,G$ generate the usual $c=0$ $\mathcal{N}=(1,0)$ superconformal algebra.

\subsubsection{Off-shell supersymmetry in four dimensions}

The vertex operators which generate the four-dimensional spacetime supersymmetry are given, as usual, by exponentials of the bosonised fermions $\psi^{\mu}$. Let us define
\begin{equation}
\Psi^{\pm 1}=\psi^1\pm i\psi^2\,,\quad\Psi^{\pm2}=\psi^3\pm i\psi^4\,.
\end{equation}
Then we can define the fermion number currents
\begin{equation}
J_i=\Psi^{+i}\Psi^{-i}
\end{equation}
and, as before, we bosonise this current by introducing scalars $\sigma_i$ such that
\begin{equation}
J_i=\partial\sigma_i\,.
\end{equation}
In terms of these two scalars, we can write down fields in the NS sector of the four-dimensional CFT as
\begin{equation}
S_{\alpha}=e^{\frac{\alpha}{2}(\sigma_1+\sigma_2)}\,,\quad S_{\dot\alpha}=e^{\frac{\dot\alpha}{2}(\sigma_1-\sigma_2)}\,,
\end{equation}
where $\alpha,\dot{\alpha}=\{+,-\}$. Replicating what we did in ten dimensions, we can try to write down a physical state corresponding to these fields by putting them into the `canonical' $q=-\frac{1}{2}$ picture, \textit{i.e.}~we consider the states
\begin{equation}
e^{-\varphi/2}S^{\alpha}\,,\quad e^{-\varphi/2}S^{\dot\alpha}\,.
\end{equation}
However, these cannot be physical states. To see why, we can just compute their conformal dimension. Each $e^{-\varphi/2}$ contributes $-\frac{q^2}{2}-q=\frac{3}{8}$, while the each term of the form $e^{\pm \sigma_1/2\pm\sigma_2/2}$ contributes $1/4$, and so the total conformal weight is
\begin{equation}
h(e^{-\varphi/2}S^{\alpha})=e^{-\varphi/2}S^{\dot\alpha}=\frac{3}{8}+\frac{1}{4}=\frac{5}{8}\,,
\end{equation}
and in order to be a physical state, we require $h=1$, and so something is missing -- something with conformal weight $3/8$. 

Fortunately, since we have a compactified theory with $\mathcal{N}=(2,2)$ superconformal symmetry, we have access to another scalar that we can use to construct states, namely the bosonisation $H_C$ of the current $J_C$. If we define
\begin{equation}
J_C=\partial H_C\,,
\end{equation}
then by the OPE of $H_C$ is determined by the self-OPE of $J_C$, namely
\begin{equation}
J_C(z)J_C(w)\sim\frac{3}{(z-w)^2}\implies H_C(z)H_C(w)\sim-3\log(z-w)\,.
\end{equation}
The exponential of $H_C$ has conformal weight
\begin{equation}
h(e^{qH_C})=\frac{3q^2}{2}\,.
\end{equation}
Thus, vertex operators of the form $e^{\pm H_C/2}$ have conformal weight $3/8$. Inspired by this, we define states of the form
\begin{equation}
q_{\alpha}=e^{-\varphi/2}S^{\alpha}e^{-H_C/2}\,,\quad q_{\dot\alpha}=e^{-\varphi/2}S^{\dot\alpha}e^{H_C/2}\,,
\end{equation}
which have conformal weight $h=1$. The plus and minus signs in the exponentials of $H_C$ are chosen so that the full vertex operators $q^{\alpha},q^{\dot\alpha}$ have an even number of plus signs in the exponents for the scalars $(\sigma_1,\sigma_2,H_C)$, and thus survive the GSO projection.

Now we have a candidate for the generators of four-dimensional supersymmetry. Indeed, we can show that the spinor fields $S^{\alpha},S^{\dot\alpha}$ satisfy the algebra
\begin{equation}
S_{\alpha}(z)S_{\dot\beta}(w)\sim\sigma^{\mu}_{\alpha\dot\beta}\psi_{\mu}(w)\,,
\end{equation}
and so, using
\begin{equation}
e^{-H_C/2}(z)e^{H_C/2}(w)\sim\frac{1}{(z-w)^{3/4}}\,,\quad e^{-\varphi/2}(z)e^{\varphi/2}(w)\sim\frac{e^{-\varphi}(w)}{(z-w)^{1/4}}\,,
\end{equation}
we have
\begin{equation}
q_{\alpha}(z)q_{\dot\beta}(w)\sim\frac{e^{-\varphi}\sigma^{\mu}_{\alpha\dot\beta}\psi_{\mu}(w)}{z-w}\,.
\end{equation}

This is, of course, not precisely the supersymmetry algebra in four dimensions, and in particular it has the wrong picture number $N_p=-1$. Just as in the ten-dimensional case, we can hit the left-hand-side with the picture-raising operator $Z$, and we find
\begin{equation}
Z\cdot(q_{\alpha}(z)q_{\dot\beta}(w))\sim\frac{\sigma^{\mu}_{\alpha\dot\beta}\partial X_{\mu}}{z-w}\,,
\end{equation}
which is the supersymmetry algebra in four-dimensions.

As we mentioned in the ten-dimensional discussion, we can attempt to form a manifest 4D supersymmetry algebra by picking half of the supersymmetry generators to lie in the $q=+\frac{1}{2}$ picture. This is possible in four-dimensions, since the supersymmetry algebra splits into two pieces. Let us pick the dotted supersymmetry generators. We define the `new' supersymmetry generator to be
\begin{equation}\label{eq:bob-new-susy-4d}
\begin{split}
\widetilde{q}_{\dot\alpha}&=Z\cdot q_{\dot\alpha}\\
&=e^{\varphi}Gq_{\alpha}+b\eta e^{2\varphi}q_{\alpha}\\
&=Ge^{\varphi/2+\frac{\dot\alpha}{2}(\sigma_1-\sigma_2)+H_C/2}+b\eta e^{3\varphi/2+\frac{\dot\alpha}{2}(\sigma_1-\sigma_2)+H_C/2}\,.
\end{split}
\end{equation}
We can compute the OPE between the new supersymmetry generator and $q_{\alpha}$, and we find
\begin{equation}
q_{\alpha}(z)\widetilde{q}_{\dot\beta}(w)\sim\frac{\sigma^{\mu}_{\alpha\dot\beta}\partial X_{\mu}(w)}{z-w}\,,
\end{equation}
\textit{i.e.}~they satisfy the standard 4D supersymmetry algebra. Note that the $q_{\alpha}q_{\beta}$ and $\widetilde{q}_{\dot\alpha}\widetilde{q}_{\dot\beta}$ OPEs are regular.

With the above discussion in mind, we define the 4D supersymmetry generators to be:
\begin{equation}
\begin{split}
Q_{\alpha}&=\oint\frac{\mathrm{d}z}{2\pi i}q_{\alpha}\,,\\
\quad Q_{\dot\alpha}&=\oint\frac{\mathrm{d}z}{2\pi i}\left(Ge^{\varphi/2+\frac{\dot\alpha}{2}(\sigma_1-\sigma_2)+H_C/2}+b\eta e^{3\varphi/2+\frac{\dot\alpha}{2}(\sigma_1-\sigma_2)+H_C/2}\right)\,.
\end{split}
\end{equation}
The expression for $Q_{\dot\alpha}$ in the RNS formalism is indeed a bit messy, but as we have seen it is required for the supercharges to satisfy the 4D supersymmetry algebra without the need to invoke picture changing.

\subsubsection{The hybrid fields}

Having constructed supersymmetry generators which satisfy the off-shell supersymmetry algebra is not enough to say that we have `manifest' spacetime supersymmetry on the worldsheet. For that, we would need fields on the worldsheet which act as odd coordinates in superspace. For example, we say that the bosonic string theory has `manifest' Poincar\'e symmetry since there are fields $X^{\mu}$ which represent the spacetime coordinates on which the Poincar\'e transformations form a global worldsheet symmetry. Given four-dimensional supersymmetry generators $Q_{\alpha},Q_{\dot\alpha}$, superspace coordinates $\theta^{\alpha},\theta^{\dot\alpha}$ must satisfy the anticommutation relations:
\begin{equation}
\{Q_{\alpha},\theta^{\alpha}\}=\delta\indices{_\alpha^\beta}\,,\quad\{Q_{\dot\alpha},\theta^{\dot\alpha}\}=\delta\indices{_{\dot\alpha}^{\dot\beta}}\,.
\end{equation}
On the worldsheet, we are thus searching for fields $\theta^{\alpha},\theta^{\dot\alpha}$ whose OPEs with $q_{\alpha},\widetilde{q}_{\dot\alpha}$ are given by
\begin{equation}
q_{\alpha}(z)\theta^{\beta}(w)\sim\frac{\delta\indices{_{\alpha}^{\beta}}}{z-w}\,,\quad \widetilde{q}_{\dot\alpha}(z)\theta^{\dot\beta}(w)\sim\frac{\delta\indices{_{\dot\alpha}^{\dot\beta}}}{z-w}\,.
\end{equation}
For $\theta^{\alpha}$, there is a natural guess, given by taking the expression for $q_{\alpha}$ and negating all of the $\text{U}(1)$ charges. Indeed, the operator
\begin{equation}
\theta^{\alpha}=e^{\varphi/2}e^{-\frac{\alpha}{2}(\sigma_1+\sigma_2)}e^{H_C/2}
\end{equation}
satisfies the desired OPEs. Similarly, we can write an ansatz for $\theta^{\dot\beta}$ by similarly proposing a state with opposite quantum numbers to $\widetilde{q}_{\dot\alpha}$. The appropriate guess is
\begin{equation}
\theta^{\dot\alpha}=c\,\xi\,e^{-3\varphi/2}e^{-\frac{\dot\alpha}{2}(\sigma_1-\sigma_2)}e^{-H_C/2}\,,
\end{equation}
where we think of the $c\xi$ prefactor as being conjugate to the $b\eta$ factor in \eqref{eq:bob-new-susy-4d}. This field alone is conjugate to $q_{\dot\alpha}$, and there is no need to introduce a separate term conjugate to the first term in the second line of \eqref{eq:bob-new-susy-4d}.

Now we have candidate fields $\theta^{\alpha},\theta^{\dot\alpha}$ behave precisely like superspace variables under supersymmetry transformations. Furthermore, $\theta^{\alpha}$ and $\theta^{\dot\alpha}$ have trivial OPEs with each other. We would like to write a field theory on the worldsheet in terms of these fields, so that the $\theta$'s become fundamental fields in a sigma model on superspace. A natural guess for such a theory would be to introduce a conjugate field $p^{\alpha}$ for $\theta_{\alpha}$ and $p^{\dot\alpha}$ for $\theta_{\dot\alpha}$ such that $(p^{\alpha},\theta_{\alpha})$ and $(p^{\dot\alpha},\theta_{\dot\alpha})$ form two fermionic systems. For such a conjugate field $p$ to exist for each $\theta$, we would demand the OPEs:
\begin{equation}
p_{\alpha}(z)\theta^{\beta}(w)\sim\frac{\delta\indices{_{\alpha}^{\beta}}}{z-w}\,,\quad p_{\dot\alpha}(z)\theta^{\dot\beta}(w)\sim\frac{\delta\indices{_{\dot\alpha}^{\dot\beta}}}{z-w}
\end{equation}
A natural candidate for the $p$ fields are then just the supersymmetry generators $q_{\alpha},\widetilde{q}_{\dot\alpha}$. However, the supersymmetry generators have the problem that they have a non-trivial OPE with each other, namely the supersymmetry algebra. We would like to have fields $p_{\alpha}$, $p_{\dot\alpha}$ which are conjugate to the $\theta$'s and have \textit{trivial} OPEs among themselves, so that they decouple. A natural guess would be to take the expressions above for the $\theta$'s and just reverse the quantum number of every component. That is, we define
\begin{equation}
p_{\alpha}=e^{-\varphi/2}e^{\alpha(\sigma_1+\sigma_2)/2}e^{-H_C/2}\,,\quad p_{\dot\alpha}=b\eta\,e^{3\varphi/2}e^{\dot\alpha(\sigma_1-\sigma_2)/2}e^{H_C/2}\,.
\end{equation}
The field $p_{\alpha}$ is just the original supersymmetry generator $q_{\alpha}$, while the field $p_{\dot\alpha}$ is the second term in the third line of \eqref{eq:bob-new-susy-4d}.

After defining the fields $\theta^{\alpha},p_{\alpha}$ and $\theta^{\dot\alpha},p_{\dot\alpha}$, by construction they satisfy the OPEs
\begin{equation}
p_{\alpha}(z)\theta^{\beta}(w)\sim\frac{\delta\indices{_{\alpha}^{\beta}}}{z-w}\,,\quad p_{\dot\alpha}(z)\theta^{\dot\beta}(w)\sim\frac{\delta\indices{_{\dot\alpha}^{\dot\beta}}}{z-w}
\end{equation}
and are both anti-commuting fields. Crucially, the fields $(p_{\alpha},\theta^{\alpha}$ decouple from $(p_{\dot\alpha},\theta^{\dot\alpha})$ (\textit{i.e.}~their OPEs between each other are regular) This is precisely the OPE of four $b,c$ ghost systems described by the action
\begin{equation}
S_{p,\theta}=\int\mathrm{d}^2z\left(p_{\alpha}\bar{\partial}\theta^{\alpha}+p_{\dot\alpha}\bar{\partial}\theta^{\dot\alpha}\right)\,.
\end{equation}
We call the fields $(\theta^{\alpha},p_{\alpha})$ and $(\theta^{\dot\alpha},p_{\dot\alpha})$ 
the \textit{hybrid fields}. Together with the scalars $X^{\mu}$, they describe a sigma model on superspace in four dimensions.

\subsubsection{Decoupling the internal CFT}

In the original RNS description of the 10D superstring compactified to 4 dimensions, the 4D CFT described by $X^{\mu},\psi^{\mu}$ was completely decoupled from the internal 6D CFT with action $S_C$. The only assumptions we made on the form of the internal field theory was that it possessed an $\mathcal{N}=(2,2)$ worldsheet supersymmetry and that the action $S_C$ did not have any couplings to the 4D fields $X^{\mu},\psi^{\mu}$ or the ghosts. Other than those two assumptions we have been completely agnostic of the details of this CFT.

So far, while trying to introduce a manifest supersymmetric 4D string theory, we have been able to keep the compact CFT unaffected. However, the cost we paid is that the fields $\theta^{\alpha},p_{\alpha}$ and $\theta^{\dot\alpha},p_{\dot\alpha}$ were defined via the R-symmetry current $J_C=\partial H_C$. This was required so that, for example, the supersymmetry generators $q_{\alpha},q_{\dot\alpha}$ obeyed the physical state conditions ($h=1$). A consequence of the use of $H_C$ in the definition of the hybrid fields, however, is that they can no longer be said to decouple from the fields of the internal CFT. Indeed, we have the OPEs
\begin{equation}
\begin{split}
J_C(z)\theta^{\alpha}(w)\sim\frac{3\theta^{\alpha}(w)}{2(z-w)}\,,&\quad J_C(z)\theta^{\dot\alpha}(w)\sim-\frac{3\theta^{\dot\alpha}(w)}{2(z-w)}\\
J_C(z)p_{\alpha}(w)\sim-\frac{3p_{\alpha}(w)}{2(z-w)}\,,&\quad J_C(z)p_{\dot\alpha}(w)\sim\frac{3p_{\dot\alpha}(w)}{2(z-w)}\,.
\end{split}
\end{equation}
Since $J_C$ is built from fields in the internal CFT, this shows that the hybrid fields cannot be thought of as decoupled from the 6D fields.

One way around this is to \textit{change the definition of the internal fields}. The trick is to notice that all of the hybrid fields obey the constraint $Q_C-3N_p=0$ where $Q_C$ is the charge under $J_C$ and $N_p$ is their picture number. Indeed:
\begin{equation}
\begin{gathered}
Q_C(\theta^{\alpha})=\frac{3}{2}\,,\quad Q_C(\theta^{\dot\alpha})=-\frac{3}{2}\,,\quad Q_C(p_{\alpha})=-\frac{3}{2}\,,\quad Q_C(p_{\dot\alpha})=\frac{3}{2}\,,\\
N_p(\theta^{\alpha})=\frac{1}{2}\,,\quad N_p(\theta^{\dot\alpha})=-\frac{1}{2}\,,\quad N_p(p_{\alpha})=-\frac{1}{2}\,,\quad N_p(p_{\dot\alpha})=\frac{1}{2}\,.
\end{gathered}
\end{equation}
Thus, we can define the new current
\begin{equation} \label{eq:jcnewxyz}
J_C^{\text{new}}:=J_C-3\mathcal{P}\,,
\end{equation}
where
\begin{equation}
\mathcal{P}=\eta\xi-\partial\varphi
\end{equation}
is the picture-number current. The `new' current $J_C^{\text{new}}$ has the property that its OPE with all of the hybrid fields $(X,\theta,p)$ vanishes. Thus, we would like to `redefine' the internal CFT such that $J_C^{\text{new}}$ is the new R-symmetry current.

This can indeed be done. We consider the operator
\begin{equation}
\mathcal{W}=\oint\frac{\mathrm{d}z}{2\pi i}(\varphi-\chi)J_C=\oint\frac{\mathrm{d}z}{2\pi i}\mathcal{P}H_C\,.
\end{equation}
The `new' internal R-symmetry current $J_C^{\text{new}}$ can be shown to be related to the old R-symmetry current $J_C$ by the following similarity relation:
\begin{equation}
J_C^{\text{new}}=e^{\mathcal{W}}J_Ce^{-\mathcal{W}}.
\end{equation}
The result is a current which has regular OPEs with all of the hybrid variables $(X,p,\theta)$. One can show that we can do the same for \textit{all} of the fields $\Phi_C$ of the compact CFT, that is we define
\begin{equation}
\Phi_C^{\text{new}}=e^{\mathcal{W}}\Phi_Ce^{-\mathcal{W}}\,,
\end{equation}
and the resulting fields will all have regular OPEs with the hybrid fields \cite{Berkovits:1996bf}. The effect of this operation on the $\mathcal{N}=(2,0)$ generators in the compactified CFT is given by
\begin{equation}
\begin{gathered}
G_C^{+,\text{new}}=e^{\varphi-\chi}G_C^+\,,\quad G_C^{-,\text{new}}=e^{-\varphi+\chi}G_C^{-}\,,\\
T_C^{\text{new}}=T_C+\partial(\varphi-\chi)J_C+\frac{3}{2}(\partial\varphi-\partial\chi)^2\,.
\end{gathered}
\end{equation}
These operators still satisfy the same $c=9$ $\mathcal{N}=(2,0)$ supersymmetry algebra.

\vspace{0.25cm}

\begin{centering}
\begin{tcolorbox}
\textbf{Exercise:} \textit{Show that the new and old R-symmetry currents of the compact CFT are indeed related to each other by conjugation with $e^{\mathcal{W}}$. That is, show that}
\begin{equation*}
e^{\mathcal{W}}J_Ce^{-\mathcal{W}}=J_C-3\mathcal{P}\,.
\end{equation*}
\textbf{Hint:} Consider $e^{\alpha\mathcal{W}}J_Ce^{-\alpha\mathcal{W}}$ as a function of $\alpha$.
\end{tcolorbox}
\end{centering}

\vspace{0.25cm}

% \begin{centering}
% \begin{tcolorbox}
% [breakable, enhanced]
% \textbf{Solution:} We start by considering the function
% \begin{equation*}
% f_{\alpha}(w)=e^{\alpha\mathcal{W}}J_C(w)e^{-\alpha\mathcal{W}}\,.
% \end{equation*}
% Taking the derivative with respect to $\alpha$ gives
% \begin{equation*}
% f_{\alpha}(w)=e^{\alpha\mathcal{W}}[\mathcal{W},J_C(w)]e^{-\alpha\mathcal{W}}\,.
% \end{equation*}
% However, we know that the commutator $[\mathcal{W},J_C(w)]$ can be computed by the simple pole in the OPE of $\mathcal{P}H_C$ with $J_C$. Since $\mathcal{P}$ has a regular OPE with $J_C$, we have
% \begin{equation*}
% (\mathcal{P}H_C)(z)J_C(w)\sim-\frac{3\mathcal{P}(w)}{z-w}\,,
% \end{equation*}
% where we have used $H_C(z)J_C(w)\sim -3/(z-w)$. Thus, $[\mathcal{W},J_C(w)]=-3\mathcal{P}(w)$. Thus,
% \begin{equation*}
% \frac{\mathrm{d}}{\mathrm{d}\alpha}f_{\alpha}(w)=-3e^{\alpha\mathcal{W}}\mathcal{P}(w)e^{-\alpha\mathcal{W}}\,.
% \end{equation*}
% Now, taking a second derivative gives a vanishing result, since $[\mathcal{W},\mathcal{P}(w)]=0$. Indeed, the OPE of $\mathcal{P}H_C$ and $\mathcal{P}$ has no simple pole. Thus, we have a second-order differential equation with initial conditions:
% \begin{equation*}
% \frac{\mathrm{d}^2}{\mathrm{d}\alpha^2}f_{\alpha}(w)=0\,,\quad f_{0}(w)=J_C(w)\,,\quad\frac{\mathrm{d}}{\mathrm{d}\alpha}f_{\alpha}(w)\bigg|_{\alpha=0}=-3\mathcal{P}\,.
% \end{equation*}
% The general solution is
% \begin{equation*}
% f_{\alpha}(w)=J_C(w)-3\alpha\mathcal{P}(w)\,,
% \end{equation*}
% and the desired result is obtained after setting $\alpha=1$.
% \end{tcolorbox}
% \end{centering}

\subsubsection[The \texorpdfstring{$\rho$}{rho}-ghost and the hybrid action]{\boldmath The \texorpdfstring{$\rho$}{rho}-ghost and the hybrid action}

So far, in defining a worldsheet string theory in four dimensions, we have taken the original RNS fields are replaced them with four fermionic systems $(p,\theta)$. However, it is not clear that this field redefinition is invertible. Let us take a look at the bosonisation of each of the $(p,\theta)$ systems:
\begin{equation}
\begin{split}
\theta^\alpha&\implies\frac{\varphi}{2}-\frac{\alpha}{2}(\sigma_1+\sigma_2)+\frac{H_C}{2}\,,\\
p_{\alpha}&\implies-\frac{\varphi}{2}+\frac{\alpha}{2}(\sigma_1+\sigma_2)-\frac{H_C}{2}\,,\\
\theta^{\dot\alpha}&\implies\sigma+\chi-\frac{3\varphi}{2}-\frac{\dot\alpha}{2}(\sigma_1-\sigma_2)-\frac{H_C}{2}\,,\\
p_{\dot\alpha}&\implies-\sigma-\chi+\frac{3\varphi}{2}+\frac{\dot\alpha}{2}(\sigma_1-\sigma_2)+\frac{H_C}{2}\,.
\end{split}
\end{equation}
This defines the bosonisation of the four $(p,\theta)$ systems in terms of six scalars. The $H_C$ belongs to the compactified theory, and so to invert the field redefinitions, we need to make 5 independent scalars out of the four $(p,\theta)$ systems. This yields four equations for five unknowns, and thus is not an invertible transformation. We can fix this by introducing a new scalar $\rho$ made up of $(\sigma,\chi,\varphi,\sigma_1,\sigma_2,H_C)$ such that the field redefinitions $\text{RNS}\to\text{hybrid}$ is invertible. The appropriate scalar is defined as
\begin{equation}
\rho=-3\varphi+\sigma+2\chi-H_C\,.
\end{equation}
This is the so-called $\rho$ ghost and it defines the full content of the hybrid formalism. It satisfies the OPE
\begin{equation}
\rho(z)\rho(w)=-\log(z-w)\,.
\end{equation}

\vspace{0.25cm}

\begin{centering}
\begin{tcolorbox}
\textbf{Exercise:} \textit{Show that $\rho$ as defined above is the unique scalar which commutes with the bosonisations of $(p_{\alpha},\theta^{\alpha})$, $(p_{\dot\alpha},\theta^{\dot\alpha})$, and $J_C^{\text{new}}$ and which also satisfies}
\begin{equation*}
\rho(z)\rho(w)\sim-\log(z-w)\,.
\end{equation*}
\end{tcolorbox}
\end{centering}

\vspace{0.25cm}

\noindent After the introduction of the $\rho$ scalar, the action of the hybrid string is given by
\begin{equation}
S=\frac{1}{2\pi}\int\mathrm{d}^2z\left(\frac{1}{2}\partial X^{\mu}\bar{\partial}X_{\mu}+p_{\alpha}\bar{\partial}\theta^{\alpha}+p_{\dot\alpha}\bar{\partial}\theta^{\dot\alpha}+\bar{p}_{\alpha}\partial\bar{\theta}^{\alpha}+\bar{p}_{\dot\alpha}\partial\bar{\theta}^{\dot\alpha}\right)+S_{C}^{\text{new}}[\Phi_C^{\text{new}}]+S_{\text{ghost}}[\rho]\,,
\end{equation}
where now $S_{\text{ghost}}$ includes contributions both from the $b,c$ ghost system (equivalently the scalar $\sigma$) and the $\rho$ ghost.

\subsubsection{The topological twist}

In order for the hybrid string to be truly quantum equivalent to the RNS string, we need the stress tensors of the two theories to be equal, so that the conformal weights of all states agree. The stress tensor of the hybrid string is given by
\begin{equation}
T_{\text{hybrid}}=\frac{1}{2}\partial X^{\mu}\partial X_{\mu}-p_{\alpha}\partial\theta^{\alpha}-p_{\dot\alpha}\partial\theta^{\dot\alpha}-\frac{1}{2}\partial\rho\,\partial\rho-\frac{1}{2}\partial^2\rho+T^{\text{(new)}}_C\,.
\end{equation}
Now, we can compare this stress tensor by writing each of the hybrid variables $(p,\theta,\rho)$ in terms of the original scalars of the RNS formalism. After a rather tedious calculation, we find
\begin{equation}
T_{\text{hybrid}}=T_{\text{RNS}}-\frac{1}{2}\partial J_C^{\text{new}}\,,
\end{equation}
and so there is a discrepancy between the stress tensors of the two theories.

To fix this discrepancy, we have to modify the internal CFT one more time by performing a so-called `topological twist'
\begin{equation}
T_C^{\text{new}}\to T_C^{\text{new}}+\frac{1}{2}J_C^{\text{new}}\,.
\end{equation}
Under this twist, we have $T_{\text{hybrid}}\to T_{\text{hybrid}}+\frac{1}{2}J_C^{\text{new}}=T_{\text{RNS}}$. Thus, we have a quantum equivalence:
\begin{equation}
\begin{gathered}
\text{RNS strings on }\mathbb{R}^4\times\mathcal{M}_6\\
\Longleftrightarrow\\
\text{hybrid strings on }\mathbb{R}^4\times\text{topologically twisted }\mathcal{M}_6\,.
\end{gathered}
\end{equation}

As a final sanity check, we can calculate the central charge of the hybrid stress tensor. The central charge of the topologically twisted internal CFT is just $c=0$. Indeed, we have
\begin{equation}
J_C^{\text{new}}(z)J_{C}^{\text{new}}(w)\sim\frac{3}{(z-w)^2}\implies\partial J_C^{\text{new}}(z)\partial J_{c}^{\text{new}}(w)\sim-\frac{18}{(z-w)^4}\,,
\end{equation}
which shows that the $1/(z-w)^4$ term in the topologically twisted OPE of $T_C^{\text{new}}(z)T_C^{\text{new}}(w)$ vanishes. Furthermore, we know the central charges of the $p,\theta$ systems
\begin{equation}
c(p_{\alpha},\theta^{\alpha})=-4\,,\quad c(p_{\dot\alpha},\theta^{\dot\alpha})=-4\,,
\end{equation}
and the central charge of the $\rho$ system is given by $c(\rho)=4$. Thus, the full central charge of the theory is
\begin{equation}
c(X^{\mu})+c(p_{\alpha},\theta^{\alpha})+c(p_{\dot\alpha},\theta^{\dot\alpha})+c(\rho)=0\,,
\end{equation}
and so the hybrid formalism describes a consistent string theory with vanishing Weyl anomaly.

\vspace{0.25cm}

\begin{centering}
\begin{tcolorbox}
Show that, within the RNS formalism, the operators
\begin{equation*}
G^{\pm,\text{new}}_{C}=e^{\pm(\varphi-\chi)}G_C^{\pm}
\end{equation*}
have conformal weight $h(G_C^{+,\text{new}})=2$, $h(G_C^{-,\text{new}})=1$. Show that, within the hybrid formalism, this is consistent with the topological twist
\begin{equation*}
T_{C}^{\text{new}}\to T_C^{\text{new}}+\frac{1}{2}J_C^{\text{new}}\,.
\end{equation*}
\end{tcolorbox}
\end{centering}

% \vspace{0.25cm}

% Finally, we note the difference between the hybrid formalism in type IIB and type IIA string theory. In IIB, we take the GSO projection to be the same in the left and right. The end effect is that, once moving to the hybrid formalism, we take the topological twisting to be the same in the left- and right-moving sectors. This defines the so-called \textit{topological A-twist} on $\mathcal{M}_6$. In type IIA, we take the opposite GSO projection in the left and right. This can be achieved by making the replacement $\overline{H}_C\to-\overline{H}_C$ compared to type IIB. This has the effect of doing the opposite topological twist in the left- and right-moving sectors, \textit{i.e.}
% \begin{equation}
% T_C^{\text{new}}\to T_C^{\text{new}}+\frac{1}{2}\partial J_C^{\text{new}}\,,\quad \overline{T}_C^{\text{new}}\to \overline{T}_C^{\text{new}}-\frac{1}{2}\bar{\partial}\bar{J}_C^{\text{new}}\,.
% \end{equation}
% This defines the so-called \textit{topological B-twist} on $\mathcal{M}_6$.\footnote{The fact that the type IIA/IIB hybrid superstrings require a topological B/A-twist, respectively, is an unfortunate consequence of naming conventions.}

\subsubsection{The physical state conditions}

So far we have described in detail how to pass from the RNS variables to the hybrid variables of \cite{Berkovits:1996bf}. However, a string theory is not described only by its field content and stress tensor, but one also has to specify a guage-fixing procedure. In the RNS formalism, this gauge-fixing was performed by demanding that physical states lie in the BRST cohomology, i.e
\begin{equation}
[Q_{\text{BRST}},\Phi\}=0\,,\quad\Phi\sim\Phi'+[Q_{\text{BRST}},\Psi\}\,.
\end{equation}
Furthermore, since we are working in the bosonised description of the superconformal ghosts $\beta,\gamma$, we have to restrict to the small Hilbert space, \textit{i.e.}~the states which are invariant under the shift $\xi\to \xi+\varepsilon$. This was imposed in the RNS formalism by demanding that we only keep states in the kernel of $\eta_0$, \textit{i.e.}
\begin{equation}
\left[\oint\frac{\mathrm{d}z}{2\pi i}\eta(z),\Phi\right\}=0\,.
\end{equation}

In terms of the hybrid fields $x$, $(p_{\alpha},\theta^{\alpha})$, and $\rho$, the physical state conditions become quite complicated \cite{Berkovits:1994wr,Berkovits:1996bf}. The BRST charge can be expressed in terms of the hybrid fields as
\begin{equation}
J_{\text{BRST}}=e^{\rho}d_{\alpha}d^{\alpha}+G^+_C\,,
\end{equation}
with
\begin{equation}
d_{\alpha}=p_{\alpha}+\frac{i}{2}\sigma^{\mu}_{\dot\alpha\,\alpha}\theta^{\dot\alpha}\partial X_{\mu}-\frac{1}{8}(\theta^{\dot\alpha}\theta_{\dot\alpha})\partial\theta_{\alpha}+\frac{1}{16}\theta_{\alpha}\partial(\theta^{\dot\alpha}\theta_{\dot\alpha})\,.
\end{equation}
Moreover, the fermion $\eta$ can be written as
\begin{equation}
\eta=e^{-2\rho+H_C}d^{\dot\alpha}d_{\dot\alpha}+G^-_C\,e^{-\rho+H_C}\,,
\end{equation}
with
\begin{equation}
d_{\dot\alpha}=p_{\dot\alpha}+\frac{i}{2}\sigma^{\mu}_{\alpha\dot\alpha}\theta^{\alpha}\partial X_{\mu}-\frac{1}{8}(\theta^{\alpha}\theta_{\alpha})\partial\theta_{\dot\alpha}+\frac{1}{16}\theta_{\dot\alpha}\partial(\theta^{\alpha}\theta_{\alpha})\,.
\end{equation}
Note that we are raising/lowering spinor indices with the epsilon tensors $\varepsilon_{\alpha\beta},\varepsilon_{\dot\alpha\dot\beta}$.

The physical state conditions in the hybrid formalism, while complicated, also have an interpretation in terms of a twisted topological $\mathcal{N}=4$ algebra. Out of the hybrid variables, one can define worldsheet generators
\begin{equation}
T,G^{\pm},\widetilde{G}^{\pm},J,J^{\pm\pm}
\end{equation}
which satisfy the `small' twisted $\mathcal{N}=4$ algebra on the worldsheet. Two of the supercharges can then be identified with the BRST current and $\eta$, namely
\begin{equation}
G^+=J_{\text{BRST}}\,,\quad\widetilde{G}^+=\eta\,.
\end{equation}
Thus, the physical state conditions can be written in the form:
\begin{equation}
[G^+_0,\Phi]=[\widetilde{G}^+_0,\Phi]=0\,,\quad \Phi\sim\Phi'+[G^+_0,\Psi]\,,
\end{equation}
for any state $\Psi$ satisfying $[\widetilde{G}^+_0,\Psi]=0$. However, $\widetilde{G}^+_0$ has trivial cohomology, since if $[\widetilde{G}^+_0,\Psi]=0$ then $\Psi=[\widetilde{G}^+_0,\xi\Psi]$. Thus, the physical state conditions can be brought into the more symmetric form
\begin{equation}
[G^+_0,\Phi]=[\widetilde{G}^+_0,\Phi]=0\,,\quad \Phi\sim\Phi'+[G^+_0,[\widetilde{G}^+_0,\Psi]]\,,
\end{equation}
for any state $\Psi$.

The above discussion identifies physical states $\Phi$ in the hybrid formalism with the elements of a \textit{double-cohomology} of an $\mathcal{N}=4$ topological algebra. We will not go into more detail of this point, as it is beyond the scope of these lectures, but see \cite{Berkovits:1994vy,Berkovits:1994wr,Berkovits:1996bf,Berkovits:1999im} for more details.

\subsubsection{Summary}

The full theory (the GS-like $(X,p,\theta)$ variables in four-dimensions, the topologically twisted Calabi-Yau sigma model, and the $\rho$ ghost) defines the full content of the hybrid formalism in four dimensions. This theory has the following properties:

\begin{itemize}

    \item It has manifest four-dimensional spacetime supersymmetry, since it describes a sigma-model on superspace.

    \item The compactified CFT $S_C$ is modified to a new internal CFT $S_{C}^{\text{new}}$, which is then topologically twisted.

    \item Since the operators $p,\theta$ and all integer powers $e^{n\rho}$ of the $\rho$ ghosts automatically have fermion number $(-1)^F=1$, these operators are automatically GSO projected! That is, any composite operator which can be constructed from products of the $p$'s and $\theta$'s, as well as integer powers $e^{n\rho}$ and exponentials $e^{ik\cdot X}$ automatically satisfies the GSO projection, and thus the above theory automatically has spacetime supersymmetry in its spectrum by construction.

    \item The hybrid string action bares a large formal resemblance to the Green-Schwarz action. Indeed, the hybrid formalism as it is defined above can be defined in a completely independent way by starting with the Green-Schwarz superstring and quantising it in a covariant fashion. This requires adding a ghost field $\rho$ and `twisting' the internal manifold, just as we did in order to get from the RNS string to the hybrid string. In fact, this is how the hybrid string was originally discovered \cite{Berkovits:1994wr}.

\end{itemize}

\subsection[The hybrid string on \texorpdfstring{$AdS_3 \times S^3$}{AdS3xS3}]{\boldmath The hybrid string on \texorpdfstring{$AdS_3 \times S^3$}{AdS3xS3}}\label{sec:bob-6d}

Now that we have a basic understanding of the hybrid formalism in the example of a Calabi-Yau compactification from ten to four dimensions, we can now turn to a more exciting example: string theory on $AdS_3 \times S^3\times\mathcal{M}_4$. This background is of particular interest due to the AdS/CFT correspondence: string theory on this background is supposed to be dual to a 2D conformal field theory which lives on the boundary of $\text{AdS}_3$. The dual 2D CFT has all of the (super)-isometries of $AdS_3 \times S^3$ as its (global) super-conformal symmetries. Thus, a formalism of string theory which makes the super-isometries of $AdS_3 \times S^3$ manifest is invaluable for understanding the qualitative features of the AdS$_3$/CFT$_2$ correspondence. The hybrid formalism on allows for such a description. Surprisingly, the hybrid formalism also offers a strength over the traditional RNS background in that, once one has defined the hybrid string on $AdS_3 \times S^3$, it is conceptually straightforward to add \textit{Ramond-Ramond flux} \cite{Berkovits:1999im}.

In this section we will review the hybrid formalism on this background, starting with the basics of strings on $AdS_3 \times S^3$, adding supersymmetry, then redefining to hybrid variables. The result is a so-called Wess-Zumino-Witten model on the supergroup $\text{PSU}(1,1|2)$, which is the group of super-isometries of $AdS_3 \times S^3$, as well as the group of super-conformal transformations in two dimensions.

\subsubsection[The geometry of \texorpdfstring{$AdS_3 \times S^3$}{AdS3xS3}]{\boldmath The geometry of \texorpdfstring{$AdS_3 \times S^3$}{AdS3xS3}}\label{sec:bob-ads3-geometry}

The bosonic isometries of $AdS_3 \times S^3$ are given by $\text{SL}(2,\mathbb{R})$ transformations acting on $\text{AdS}_3$ and $\text{SU}(2)$ transformations acting on $\text{S}^3$. Since these transformations are non-abelian, their generators ($J^a$ for $\mathfrak{sl}(2,\mathbb{R})$ and $K^a$ for $\mathfrak{su}(2)$) will have non-trivial self-commutators, and satisfy the algebra:
\begin{equation}
\begin{split}
[J^+,J^-]=-2J^3\,,&\quad [J^3,J^{\pm}]=\pm J^{\pm}\,,\\
[K^+,K^-]=2K^3\,,&\quad [K^3,K^{\pm}]=\pm K^{\pm}\,,\\
\end{split}
\end{equation}
Spacetime supersymmetry is added to this background in the same way as in flat space: we introduce supercharges which transform as spinors under the bosonic isometries, and whose anticommutators return the bosonic generators. It turns out that the appropriate (anti)-commutation relations are satisfied by introducing $8$ super-charges $Q^{\alpha\beta\gamma}$ with $\alpha,\beta,\gamma\in\{+,-\}$ which satisfy the algebra:
\begin{equation}\label{eq:bob-psu112-global}
\begin{gathered}
\big[J^a,Q^{\alpha\beta\gamma}\big]=\frac{1}{2}(\tilde{\sigma}^a)\indices{^\alpha_{\delta}}Q^{\delta\beta\gamma}\,,\\
[K^a,Q^{\alpha\beta\gamma}]=\frac{1}{2}(\sigma^a)\indices{^\beta_{\delta}}Q^{\alpha\delta\gamma}\,,\\
\{Q^{\alpha\beta+},Q^{\gamma\delta-}\}=-\varepsilon^{\beta\delta}(\tilde{\sigma}_a)\indices{^\alpha^\gamma}J^a+\varepsilon^{\alpha\gamma}(\sigma_a)^{\beta\delta}K^a\,.
\end{gathered}
\end{equation}
with all other commutators vanishing. Here $\sigma^a$ are the usual $\sigma$-matrices which generate the two-dimensional spinor representation of $\mathfrak{su}(2)$, while $\tilde{\sigma}^a$ are the $\mathfrak{sl}(2,\mathbb{R})$ sigma matrices, which differ from $\sigma$ only by a sign, and which form the $\mathfrak{sl}(2,\mathbb{R})$ spinor representation. Thus, the supercharges $Q^{\alpha\beta}$ and $\widetilde{Q}^{\alpha\beta}$ are spinors of $\mathfrak{sl}(2,\mathbb{R})$ in the first index and of $\mathfrak{su}(2)$ in the second index.

The above algebra, including bosonic and fermionic transformations, is the spacetime supersymmetry algebra of the $AdS_3 \times S^3$ background. The above algebra has a name: $\mathfrak{psu}(1,1|2)$, and it is the algebra of super-isometries of $AdS_3 \times S^3$.\footnote{The algebra $\mathfrak{psu}(1,1|2)$ is also, very critically, the chiral part of the global $\mathcal{N}=4$ superconformal algebra in two-dimensions. This is a very important observation in holography, since the super-isometry group of $AdS_3 \times S^3$ becomes the superconformal group of an $\mathcal{N}=4$ superconformal field theory living on the boundary.} As a supergroup, $\text{PSU}(1,1|2)$ is defined to have a block decomposition, where the diagonal blocks generate $\text{SL}(2,\mathbb{R})$ and $\text{SU}(2)$, whereas the off-diagonal blocks contain the supersymmetry generators:
\begin{equation}
\text{PSU}(1,1|2)\cong
\left( \begin{array}{c|c}
   \text{SL}(2,\mathbb{R}) & \text{supercharges} \\
   \midrule
   \text{supercharges} & \text{SU}(2) \\
\end{array}\right)
\end{equation}
In fact, not only is $\text{PSU}(1,1|2)$ the superisometry algebra of $AdS_3 \times S^3$, but it is possible to write the superspace of $AdS_3 \times S^3$ as a quotient
\begin{equation}
\text{Super}(AdS_3 \times S^3)\cong\frac{\text{PSU}(1,1|2)_{L}\times\text{PSU}(1,1|2)_R}{\text{SL}(2,\mathbb{R})\times\text{SU}(2)}\,.
\end{equation}

The group $\text{PSU}(1,1|2)$ will play a very important role when we discuss the hybrid formalism on $AdS_3 \times S^3$, where the currents $J^a,K^a,Q^{\alpha\beta\gamma}$ will be promoted to worldsheet fields which generate a Wess-Zumino-Witten (WZW) model on $\text{PSU}(1,1|2)$. We will review how this works in the RNS description in the next section, starting with the bosonic description and then introducing worldsheet supersymmetry.

\subsubsection{Bosonic WZW models}

As was explained in detail in Section \ref{s:saskia}, string theory on $AdS_3 \times S^3$ is rare among superstring backgrounds in that it has a relatively simple worldsheet description. This is because both $\text{AdS}_3$ and $\text{S}^3$ with their constant curvature metrics are isometric to Lie groups:
\begin{equation}
\text{AdS}_3\cong\text{SL}(2,\mathbb{R})\,,\quad\text{S}^3\cong\text{SU}(2)\,,
\end{equation}
where the metric on the group is taken to be the Killing form on the Lie algebra.\footnote{Since the Killing form is a non-degenerate bilinear form $\mathfrak{g}\times\mathfrak{g}\to\mathbb{R}$, it defines a bilinear form on the space of tangent vectors to any Lie group $G$, and thus defines a canonical $G$-invariant metric on $G$.} String theory backgrounds on group manifolds are particularly special due to the high amount of symmetry, and the string theory on these backgrounds can be understood largely by studying the representation theory of so-called \textit{affine algebras }$\mathfrak{g}_k$, which for our purposes are the OPE algebras of $\mathfrak{g}$-valued conserved currents $J$ on the worldsheet.

The appropriate language to describe such backgrounds is in terms of so-called \textit{Wess-Zumino-Witten} models, which were also discussed in Sections \ref{s:sibylle} and \ref{s:saskia} in the context of Green-Schwarz superstrings. We will briefly discuss them and their quantisation here. As this topic has already been discussed in previous sections, this will be a lightning review: for more details about WZW models, we direct the reader to the textbook \cite{DiFrancesco:1997nk} as well as the lecture notes \cite{Eberhardt_wzw_notes}.

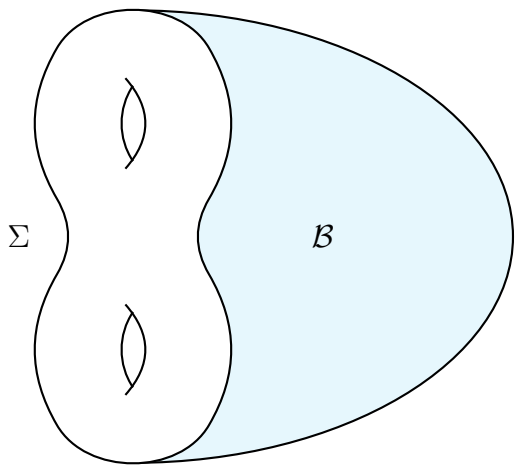
\begin{figure}
\centering
\begin{tikzpicture}
\fill[cyan, opacity = 0.1] (0,2.5) [partial ellipse = -90:90:5 and 3];
\draw[thick] (0,2.5) [partial ellipse = -90:90:5 and 3];
\begin{scope}[rotate = 90]
\fill[white] (0,1) to[out=30,in=150] (2,1) to[out=-30,in=210] (3,1) to[out=30,in=150] (5,1) to[out=-30,in=30] (5,-1) to[out=210,in=-30] (3,-1) to[out=150,in=30] (2,-1) to[out=210,in=-30] (0,-1) to[out=150,in=-150] (0,1);
\draw[smooth, thick] (0,1) to[out=30,in=150] (2,1) to[out=-30,in=210] (3,1) to[out=30,in=150] (5,1) to[out=-30,in=30] (5,-1) to[out=210,in=-30] (3,-1) to[out=150,in=30] (2,-1) to[out=210,in=-30] (0,-1) to[out=150,in=-150] (0,1);
\draw[smooth, thick] (0.4,0.1) .. controls (0.8,-0.25) and (1.2,-0.25) .. (1.6,0.1);
\draw[smooth, thick] (0.5,0) .. controls (0.8,0.2) and (1.2,0.2) .. (1.5,0);
\draw[smooth, thick] (3.4,0.1) .. controls (3.8,-0.25) and (4.2,-0.25) .. (4.6,0.1);
\draw[smooth, thick] (3.5,0) .. controls (3.8,0.2) and (4.2,0.2) .. (4.5,0);
\end{scope}
\node at (2.5,2.5) {\large $\mathcal{B}$};
\node at (-1.5,2.5) {\large $\Sigma$};
\end{tikzpicture}
\caption{The three-manifold $\mathcal{B}$ is constructed so that its boundary is the worldshet $\Sigma$. Different choices of continuation of $g$ as a function on $\mathcal{B}$ should be physically equivalent.}
\label{fig:bob-filled-worldsheet}
\end{figure}

Given a compact group $G$, we can describe the motion of a string worldsheet $\Sigma$ in $G$ by a map $g:\Sigma\to G$. To define a field theory with $g$ as the fundamental field, we start by writing down a kinetic term:
\begin{equation}\label{eq:bob-principal-chiral-model}
S_{\text{kin}}[g]=\frac{1}{4\pi f^2}\int_{\Sigma}\text{Tr}[g^{-1}\partial g\,g^{-1}\overline{\partial}g]\,.
\end{equation}
The fields $g^{-1}\partial g$ and $g^{-1}\overline{\partial}g$ define $\mathfrak{g}$-valued one-forms on $\Sigma$, and the trace $\text{Tr}$ is taken in the adjoint representation of $\mathfrak{g}$. As a (familiar) example, let us take $G=\text{U}(1)$. Any point in $G$ is expressed as $e^{i\alpha}$ for some real $\alpha$. Taking the derivative of $g$ gives
\begin{equation}
g^{-1}\partial g=i\partial\alpha\,,\quad g^{-1}\overline{\partial}g=i\overline{\partial}\alpha\,.
\end{equation}
And so the kinetic term for $G=\text{U}(1)$ can be written in terms of the real field $\alpha$ as
\begin{equation}
S_{\text{kin}}[\alpha]=-\frac{1}{4\pi f^2}\int_{\Sigma}\partial\alpha\overline{\partial}\alpha\,,
\end{equation}
whic is just the kinetic term of a string in one flat dimension.

The equations of motion of the above action are of the form
\begin{equation}\label{eq:bob-wzw-conservation}
\overline{\partial}\left(g^{-1}\partial g\right)+\partial\left(g^{-1}\overline{\partial}g\right)=0\,.
\end{equation}
This is equivalent to the conservation of the current
\begin{equation}
J_z=g^{-1}\partial g\,,\quad J_{\bar{z}}=g^{-1}\overline{\partial}g\,.
\end{equation}
This current is associated to the global symmetry of the theory associated to left-translations $g\to gg_R^{-1}$. In the language of differential forms, $J$ is a $\mathfrak{g}$-valued one-form on the worldsheet $\Sigma$, which is the pullback of the so-called Maurer-Cartan form:
\begin{equation}
J=g^{-1}\mathrm{d}g\in\Omega^1(\Sigma)\otimes\mathfrak{g}\,.
\end{equation}

In a conformal field theory, we want currents $J$ for which the components are (anti)-holomorphic.\footnote{Strictly speaking, we only need the stress-tensor of the theory to be holomorphic, via traclessness and conservation. However, as we will discuss below, the stress-tensor is built out of bilinears in the conserved currents, \textit{i.e.}~$T\sim\text{Tr}[JJ]$. In most cases, then, we require $J$ to be holomorphic for $T$ to be holomorphic. If one allows $G$ to be a supergroup, then there are exceptions to this rule, see the discussion of Section \ref{sec:bob-applications}.} For this to be the case, we would like, for example, that $J_z$ is holomorphic and $J_{\bar{z}}$ is anti-holomorphic. However, for non-abelian groups $G$, we cannot have both $\partial J_{\bar{z}}=0$ and $\overline{\partial}J_z=0$. To see this, we note that the Maurer-Cartan form $J=g^{-1}\mathrm{d}g$ satisfies the flatness condition
\begin{equation}
\mathrm{d}J+J\wedge J=0\,.
\end{equation}
However, if $\overline{\partial}J_z=\partial J_{\overline{z}}=0$, then expanding in components $J=J_z\mathrm{d}z+J_{\bar{z}}\mathrm{d}\bar{z}$, we have
\begin{equation}
\mathrm{d}J=\overline{\partial}J_z\,\mathrm{d}\bar{z}\wedge\mathrm{d}z+\partial J_{\bar{z}}\,\mathrm{d}\bar{z}\wedge\mathrm{d}z=0\,,
\end{equation}
since both terms vanish individually. However, we know that
\begin{equation}
\mathrm{d}J=-J\wedge J
\end{equation}
does not vanish unless $G$ is abelian. Thus, the components of $J$ cannot be (anti)holomorphic unless $G$ is abelian.
In order to improve the kinetic term so that the conserved currents are (anti)holomorphic, we introduce the so-called \textit{Wess-Zumino} term
\begin{equation}
S_{\text{WZ}}[g]=-\frac{ik}{2\pi}\int_{\mathcal{B}}\text{Tr}[g^{-1}\mathrm{d}g\wedge g^{-1}\mathrm{d}g\wedge g^{-1}\mathrm{d}g]\,.
\end{equation}
Here, $\mathcal{B}$ is a 3-manifold with boundary $\partial\mathcal{B}=\Sigma$. For abelian groups, this term identically vanishes due to the anti-symmetry of the wedge product. For non-abelian groups, it is necessary for the existence of a holomorphic stress tensor on the worldsheet. The equations of motion for the total action $S_{\text{kin}}+S_{\text{WZ}}$ are given by
\begin{equation}
(1+f^2k)\partial\left(g^{-1}\overline{\partial}g\right)+(1-f^2k)\overline{\partial}(g^{-1}\partial g)=0\,.
\end{equation}
Thus, for $f^2=1/k$, we have that the right-moving current $J=g^{-1}\overline{\partial} g$ is anti-holomorphic. One can also show that the current $\mathrm{d}g\,g^{-1}$ associated to left-translations $g\to g_Lg$ has a holomorphic conserved component if $f^2=1/k$. We call this choice of parameters the \textit{Wess-Zumino-Witten point} and, for most choices of group $G$ the WZW model is not conformal unless we are at the WZW point. Thus, the action of the WZW model is
\begin{equation}
S_{\text{WZW}}[g]=\frac{k}{4\pi}\int_{\Sigma}\text{Tr}[g^{-1}\mathrm{d}g\wedge\star g^{-1}\mathrm{d}g]-\frac{ik}{2\pi}\int_{\mathcal{B}}\text{Tr}[g^{-1}\mathrm{d}g\wedge g^{-1}\mathrm{d}g\wedge g^{-1}\mathrm{d}g]\,.
\end{equation}

The above construction should not depend on how we choose to extend the field $g:\Sigma\to G$ to a field $g:\mathcal{B}\to G$. It turns out that, if $g,\tilde{g}$ are two separate such extensions, then
\begin{equation}
S_{\text{WZ}}[g]-S_{\text{WZ}}[\tilde{g}]\in 2\pi k\mathbb{Z}\,.
\end{equation}
Thus, in order for the path integrand $e^{iS_{\text{WZ}}}$ to be independent of the choice $g:\Sigma\to\mathcal{B}$, we demand that $k$ is an integer.\footnote{Here, it is important that $G$ is compact, for which the above statement is equivalent to $H_3(G,\mathbb{Z})\cong\mathbb{Z}$. If $G$ is non-compact, then the action is truly independent of the continuation $g:\mathcal{B}\to G$, and there is no need to quantise $k$.}

From now on, we focus only on the left-moving part of the theory. The current $J$, as Lie-algebra valued one-forms, admits an expansion in the basis of the Lie algebra $\mathfrak{g}$. Let us take such a basis $T_a$ to have
\begin{equation}
\text{Tr}(T_aT_b)=\kappa_{ab}\,,
\end{equation}
where $\kappa_{ab}$ is the Killing form on $\mathfrak{g}$, whose inverse we denote by $\kappa^{ab}$. Expanding in this basis, we define the currents $J^a$ as
\begin{equation}
J(z)=J^a(z)T_a\,.
\end{equation}
As an example, let us take $g:\Sigma\to G$ close to the identity in $G$. Then
\begin{equation}
g=e^{T_ax^a}\implies J=g^{-1}\partial g=T_a\partial x^a
\end{equation}
so that $J^a$ is the holomorphic derivative of the local coordinate $x^a$ on $G$ around the identity.

Upon quantisation of the theory, we impose canonical OPEs between fields and their conjugate momenta. Doing this for the WZW model, we find the following OPEs between the currents $J^a$:
\begin{equation}
J^a(z)J^b(w)\sim\frac{k\,\kappa^{ab}}{(z-w)^2}+\frac{f\indices{^a^b_c}J^c(w)}{z-w}\,.
\end{equation}
The above OPE algebra is actually the most general algebra that can be satisfied between two fields of conformal weight $h=1$ in a unitary CFT, and is known as a \textit{Kac-Moody} algebra, and is denoted by $\mathfrak{g}_k$ or sometimes $\widehat{\mathfrak{g}}_k$. The current algebra is the primary object of study in Wess-Zumino-Witten models, and an understanding of the unitary representations of $\mathfrak{g}_k$ is paramount to understanding the spectrum of the above CFT.

\subsubsection*{The Sugawara construction}

Given the WZW model, there is a standard method of constructing a stress tensor on the worldsheet given only the currents $J^a$. Using the flat space string as an example, we had $J^{\mu}=\partial X^{\mu}$ and the stress tensor was simply
\begin{equation}
T=\frac{1}{2}\partial X^{\mu}\partial X_{\mu}=\frac{1}{2}\delta_{\mu\nu}J^{\mu}J^{\nu}\,.
\end{equation}
That is, the stress tensor is built purely as a bilinear in the holomorphic conserved currents $J^{\mu}$.

A natural guess for the stress tensor of an arbitrary WZW model with current algebra $\mathfrak{g}_k$ would be
\begin{equation}
T\stackrel{?}{=}\gamma\kappa_{ab}J^aJ^b\,,
\end{equation}
\textit{i.e.}~as the natural bilinear combination of the currents $J^a$. Since we know that the stress tensor has to be of weight $h=1$, we can fix the coefficient $\gamma$ by demanding the OPE
\begin{equation}
T(z)J^a(w)\sim\frac{J^a(w)}{(z-w)^2}+\frac{\partial J^a(w)}{z-w}\,.
\end{equation}
This calculation yields
\begin{equation}
\gamma=\frac{1}{2(k+h^{\vee})}\,,
\end{equation}
where $h^{\vee}$ is the \textit{dual Coxeter number} of the Lie algebra $\mathfrak{g}$, defined by
\begin{equation}
f\indices{^d_b_c}f\indices{^a^b^c}=h^{\vee}\kappa^{ad}\,.
\end{equation}
Two examples that we will use later are $\mathfrak{g}=\mathfrak{sl}(2,\mathbb{R})$ and $\mathfrak{g}=\mathfrak{su}(2)$, for which
\begin{equation}
h^{\vee}(\mathfrak{sl}(2,\mathbb{R}))=-2\,,\quad h^{\vee}(\mathfrak{su}(2))=2\,.
\end{equation}

Once we have constructed a stress tensor $T$ for the WZW model, we can compute its central charge by demanding
\begin{equation}
T(z)T(w)\sim\frac{c}{2(z-w)^4}+\frac{T(w)}{(z-w)^2}+\frac{\partial T(w)}{z-w}\,.
\end{equation}
This calculation is again quite tedious, but can be done, and at the end of the day the WZW model on the Lie group $G$ with level $k$ is given by
\begin{equation}
c(\mathfrak{g}_k)=\frac{k\,\text{dim}(\mathfrak{g})}{k+h^{\vee}}\,.
\end{equation}
For example,
\begin{equation}
c(\mathfrak{sl}(2,\mathbb{R})_k)=\frac{3k}{k-2}\,,\quad c(\mathfrak{su}(2)_k)=\frac{3k}{k+2}\,.
\end{equation}

\subsubsection*{As a string theory}

The above discussion has so far treated WZW models only as a 2D conformal field theory. In order to promote them to a proper string theory, we need to couple them to a dynamical worldsheet metric $h$ which we then gauge away. The procedure is precisely analogous to that of the flat space string, and the result is that we have to supplement the WZW model with a $b,c$ ghost system with central charge $c_{b,c}=-26$.

For the resulting string theory to be consistent, the total central charge has to vanish. If we consider the WZW model on $G^1\times\cdots\times G^n$, we can in principle pick different levels for each simple factor, and the total central charge will be the sum of the central charges on the individual factors. Thus, if our worldsheet theory is described entirely by a product of WZW models, we require
\begin{equation}
\sum_{i=1}^{n}c(\mathfrak{g}^i_{k_i})=\sum_{i=1}^{n}\frac{k_i\text{dim}(\mathfrak{g}^i)}{k_i+h^{\vee}(\mathfrak{g}^i)}=26\,.
\end{equation}
Alternatively, we can also balance the central charge by considering a model of the form $G^1\times\cdots\times G^n\times\mathcal{M}$, where $\mathcal{M}$ is a 2D CFT of appropriate central charge which is not necessarily a WZW model. For example, if we consider bosonic string theory on $AdS_3 \times S^3\times\mathcal{M}$, we need
\begin{equation}
\frac{3k}{k-2}+\frac{3k'}{k'+2}+c(\mathcal{M})=26\,,
\end{equation}
where $k$ and $k'$ are the levels of the $\text{SL}(2,\mathbb{R})$ and $\text{SU}(2)$ WZW models, respectively.

Finally, we comment on the physical meaning of the level $k$. For a spacetime like $\text{AdS}_3$ or $\text{S}^3$ with constant curvature, we can associate a typical length scale (\textit{e.g.} the inverse curvature) $L$ with the spacetime. We can also associate a length scale to the string itself, which we denote by $\ell_s$. The level $k$ is related to these scales by
\begin{equation}
k=\left(\frac{L}{\ell_s}\right)^2\,.
\end{equation}
In practice, we take the string length $\ell_s$ to be a fixed length (say the Planck length), and so we interpret $k$ as describing the size of the target space in string length units. The interesting limits are
\begin{equation}
k\gg 1\,,\quad k\to 1\,.
\end{equation}
The first limit describes a small string propagating in a curved background, and thus has an interpretation of a `semi-classical' limit. The second limit, on the other hand, describes a string which is as large as the spacetime itself. This is the `tensionless' limit, and we will discuss it later at the end of these lectures in the case of $AdS_3 \times S^3$.

\subsubsection{Supersymmetric WZW models}

In flat space, we can construct a worldsheet theory with $\mathcal{N}=1$ worldsheet supersymmetry by introducing spin-$\frac{1}{2}$ Fermions $\psi^{\mu}$ which are the superpartners of the bosonic currents $J^{\mu}=\partial X^{\mu}$. The supersymmetry transformations are (focusing only on the left-moving sector) given by
\begin{equation}
\delta_{\varepsilon}\psi^{\mu}=\varepsilon J^{\mu}\,,\quad \delta_{\varepsilon}J^{\mu}=\varepsilon\partial\psi^{\mu}\,,
\end{equation}
so that the doublet
\begin{equation}
\begin{pmatrix}
\psi^{\mu}\\ J^{\mu}
\end{pmatrix}
\end{equation}
is an $\mathcal{N}=1$ SUSY multiplet.

This $\mathcal{N}=1$ worldsheet construction generalises immediately to non-abelian WZW models. In the bosonic theory, one has (holomorphic) currents $J^{a}$ transforming in the adjoint representation of the affine Lie algebra $\mathfrak{g}_k$. In order to add worldsheet supersymmetry, one then simply introduces worldsheet fermions $\psi^a$. The appropriate supersymmetry transformations are the same as those in flat space, namely
\begin{equation}
\delta_{\varepsilon}\psi^{a}=\varepsilon J^{a}\,,\quad \delta_{\varepsilon}J^{a}=\varepsilon\partial\psi^{a}\,.
\end{equation}
The fields now satisfy the OPEs
\begin{equation}\label{eq:bob-n1-wzw}
\begin{split}
J^a(z)J^b(w)&\sim\frac{k\,\kappa^{ab}}{(z-w)^2}+\frac{f\indices{^a^b_c}J^c(w)}{z-w}+\cdots\,,\\
J^a(z)\psi^b(w)&\sim\frac{f\indices{^a^b_c}\psi^c(w)}{z-w}+\cdots\,,\\
\psi^a(z)\psi^b(w)&\sim\frac{k\,\kappa^{ab}}{z-w}+\cdots\,,
\end{split}
\end{equation}
where we recall that $k$ is the level of the WZW model and $\kappa^{ab}$ is the Killing form on the Lie algebra $\mathfrak{g}$. We denote the $\mathcal{N}=1$ Wess-Zumino-Witten model on the group $G$ with level $k$ by the symbol $\mathfrak{g}_k^{(1)}$. We refer to the above OPE algebra by the same name.

\subsubsection*{Decoupling the fermions}

The above construction of an $\mathcal{N}=1$ WZW model as a doublet of fields $(\psi^a,J^a)$ with $\psi^a$ transformin in the adjoint of $\mathfrak{g}$ is convenient in that it makes the $\mathcal{N}=1$ supersymmetry on the worldsheet manifest. However, it is inconvenient from the point-of-view of quantisation.

Recall that when we quantise a WZW model, we consider states $\ket{\psi}$ which lie in highest-weight representations of the current algebra. This ensures that these states have energies which are bounded from below. However, the algebra \eqref{eq:bob-n1-wzw} is rather complicated since the bosonic currents $J^a$ and the fermionic superpartners $\psi^a$ have a non-zero OPE with each other, and thus constructing its highest-weight representations is a rather difficult task.

Thankfully, there is a simple way around this difficulty. We simply define new bosonic currents $\mathcal{J}^a$ by
\begin{equation}\label{eq:bob-decoupled-currents}
\mathcal{J}^a=J^a+\frac{1}{2k}f\indices{^a_b_c}(\psi^b\psi^c)\,.
\end{equation}
Given the definition of $\mathcal{J}^a$, one can check that the OPE of $\mathcal{J}^a$ with the fermions is regular, meaning that the new bosonic currents and the worldsheet fermions `decouple'. Similarly, one can show that $(\mathcal{J}^a,\psi^a)$ satisfy the algebra
\begin{equation}
\begin{split}
\mathcal{J}^a(z)\mathcal{J}^b(w)&\sim\frac{(k-h^{\vee})\kappa^{ab}}{(z-w)^2}+\frac{f\indices{^a^b_c}\mathcal{J}^c(w)}{z-w}\,,\\
\psi^a(z)\psi^b(w)&\sim\frac{k\kappa^{ab}}{z-w}\,,\\
\mathcal{J}^a(z)\psi^b(w)&\sim\cdots\,.
\end{split}
\end{equation}
That is, the currents $\mathcal{J}^a$ satisfy the algebra $\mathfrak{g}_{k-h^{\vee}}$. Therefore, if we chose to describe the worldsheet theory with respect to the fields $(\psi^a,\mathcal{J}^a)$, the algebra they satisfy is simply a bosonic Ka\v{c}-Moody algebra $\mathfrak{g}_{k-h^{\vee}}$ and the algebra of $\text{dim}(\mathfrak{g})$ \textit{free} fermions. That is,
\begin{equation}
\mathfrak{g}_{k}^{(1)}\cong\mathfrak{g}_{k-h^{\vee}}\oplus(\text{dim}(\mathfrak{g})\text{ free fermions}).
\end{equation}
From this, we can immediately calculate the central charge of the $\mathcal{N}=1$ WZW model, and we find
\begin{equation}
\begin{split}
c(\mathfrak{g}_k^{(1)})&=c(\mathfrak{g}_{k-h^{\vee}})+\frac{1}{2}\text{dim}(\mathfrak{g})\\
&=\frac{(3k-2h^{\vee})\text{dim}(\mathfrak{g})}{2k}\,.
\end{split}
\end{equation}

\vspace{0.25cm}

\begin{centering}
\begin{tcolorbox}
\textbf{Exercise:} \textit{Show that the currents $\mathcal{J}$ defined in \eqref{eq:bob-decoupled-currents} satisfy the algbra $\mathfrak{g}_{k-h^{\vee}}$.}
\end{tcolorbox}
\end{centering}

\vspace{0.25cm}

% \begin{centering}
% \begin{tcolorbox}
% \textbf{Solution:}
% \begin{enumerate}[1)]

%   \item 

%   \item 

% \end{enumerate}
% \end{tcolorbox}
% \end{centering}

\subsubsection[The RNS string on \texorpdfstring{$AdS_3 \times S^3$}{AdS3xS3}]{\boldmath The RNS string on \texorpdfstring{$AdS_3 \times S^3$}{AdS3xS3}}

Now that we have reviewed Wess-Zumino-Witten models in general, let us turn to the problem of quantising strings on $AdS_3 \times S^3$. The key observation is that the Riemannian manifolds $\text{AdS}_3$ and $\text{S}^3$ are isometric to the group manifolds $\text{SL}(2,\mathbb{R})$ and $\text{SU}(2)$, respectively.\footnote{Strictly speaking, since $\text{AdS}_3$ is simply connected and $\pi_1(\text{SL}(2,\mathbb{R}))=\mathbb{Z}$, we identify $\text{AdS}_3$ with the universal cover $\widetilde{\text{SL}(2,\mathbb{R})}$. This will largely play no role in the following.} Thus, string theory on the background $AdS_3 \times S^3$ with pure NS-NS flux is equivalent to the Wess-Zumino-Witten model
\begin{equation}
\mathfrak{sl}(2,\mathbb{R})_k^{(1)}\oplus\mathfrak{su}(2)_{k'}^{(1)}\,.
\end{equation}
Recall that the central charge of an $\mathcal{N}=1$ WZW model $\mathfrak{g}_k^{(1)}$ with dual coxeter number $h^{\vee}(\mathfrak{g})$ is given by 
\begin{equation}
c(\mathfrak{g}_k^{(1)})=\frac{(3k/2-h^{\vee})\text{dim}(\mathfrak{g})}{k}\,,
\end{equation}
we can use $h^{\vee}(\mathfrak{sl}(2,\mathbb{R}))=-2$ and $h^{\vee}(\mathfrak{su}(2))=2$ and we have
\begin{equation}
c(\mathfrak{sl}(2,\mathbb{R})_k^{(1)}\oplus\mathfrak{su}(2)_{k'}^{(1)})=\frac{3(3k/2+2)}{k}+\frac{3(3k'/2-2)}{k}\,.
\end{equation}
In order to have a well-defined superstring theory, we need the full worldsheet field content (minus the ghosts) to have $c=15$. We can add a four-dimensional compact theory $\mathcal{M}_4$ with central charge $c=6$ and we find that the total central charge is
\begin{equation}
c_{\text{tot}}=3\left(\frac{3k/2+2}{k}+\frac{3k'/2-2}{k'}+2\right)\,.
\end{equation}
Demanding that this combination gives the critical central charge $c_{\text{tot}}=15$ leads to the Diophantine equation
\begin{equation}
\frac{3k+4}{k}+\frac{3k'-4}{k'}=6\implies\frac{1}{k}-\frac{1}{k'}=0\,.
\end{equation}
Thus, in order for $AdS_3 \times S^3\times\mathcal{M}_4$ to be a consistent superstring background, we need the levels of the $\text{SL}(2,\mathbb{R})$ and $\text{SU}(2)$ WZW models to be the same. In other words, the radius of $\text{AdS}_3$ and the radius of $\text{S}^3$ must match.

As a side note, we also comment on the case of $AdS_3 \times S^3\times\text{S}^3\times\text{S}^1$, which is an interesting holographic background in its own right, but not one we will study in detail here. We can model this as
\begin{equation}
\mathfrak{sl}(2,\mathbb{R})^{(1)}_{k}\oplus\mathfrak{su}(2)_{k_1}^{(1)}\oplus\mathfrak{su}(2)_{k_2}^{(1)}\oplus\mathfrak{u}(1)^{(1)}\,.
\end{equation}
Note that the `level' of a $\text{U}(1)$ WZW model is not defined, since $\text{U}(1)$ is abelian, and so there is no Wess-Zumino term. The central charge of this theory is
\begin{equation}
\begin{split}
c\Big(\mathfrak{sl}(2,\mathbb{R})^{(1)}_{k}&\oplus\mathfrak{su}(2)_{k_1}^{(1)}\oplus\mathfrak{su}(2)_{k_2}^{(1)}\oplus\mathfrak{u}(1)^{(1)}\Big)\\
&=\frac{3}{2}\left(\frac{3k+4}{k}+\frac{3k_1-4}{k_1}+\frac{3k_2-4}{k_2}\right)+\frac{3}{2}\\
&=\frac{3}{2}\left(10+\frac{4}{k}-\frac{4}{k_1}-\frac{4}{k_2}\right)\,,
\end{split}
\end{equation}
which we demand is equal to $15$ as to cancel the superconformal ghosts. This gives
\begin{equation}
\frac{1}{k}=\frac{1}{k_1}+\frac{1}{k_2}\,.
\end{equation}
Recalling that the radius $L$ of a spacetime is related to the string coupling by $k=L^2/\ell_s^2$, we see that the radii of the factors in $AdS_3 \times S^3\times\text{S}^3\times\text{S}^1$ satisfy
\begin{equation}
R^2=R_1^2+R_2^2\,,
\end{equation}
\textit{i.e.}~they satisfy a Pythagorean relation. Defining $R_1=R\sin\alpha$ and $R_2=R\cos{\alpha}$ defines a two-parameter family of $\text{AdS}_3$ backgrounds parametrized by the level $k$ of the $\text{AdS}_3$ model and an angle $\alpha$.\footnote{We emphasize that the levels $k_1,k_2$ are required by quantum consistency to be integers, while the level $k$ can be arbitrary (but positive). For example, taking $\alpha=\pi/4$ gives $k_1=k_2$ and $k=k_1/2$.}

In the last subsection we argued that an $\mathcal{N}=1$ supersymmetric WZW model can be `decoupled' in the sense that we can redefine the bosonic currents and arrive at an algebra for which the bosonic and fermionic currents decouple. Specifically, given a WZW model $\mathfrak{g}_k^{(1)}$, we have
\begin{equation}
\mathfrak{g}_k^{(1)}\cong\mathfrak{g}_{k-h^{\vee}}\oplus(\text{dim}(\mathfrak{g})\text{ free fermions})\,.
\end{equation}
Applying this to the case of strings on $AdS_3 \times S^3$, and setting the level $k$ of the two algebras to be the same, we have
\begin{equation}
\mathfrak{sl}(2,\mathbb{R})_k^{(1)}\oplus\mathfrak{su}(2)_k^{(1)}\cong\mathfrak{sl}(2,\mathbb{R})_{k+2}\oplus\mathfrak{su}(2)_{k-2}\oplus(6\text{ free fermions})\,.
\end{equation}
To summarize, the field content of the RNS formalism of string theory on $AdS_3 \times S^3\times\mathcal{M}_4$ is given by
\begin{itemize}

    \item Bosonic $\mathfrak{sl}(2,\mathbb{R})$ currents $\mathcal{J}^a$ which form the algebra $\mathfrak{sl}(2,\mathbb{R})_{k+2}$.

    \item Bosonic $\mathfrak{su}(2)$ currents $\mathcal{K}^a$ which form the algebra $\mathfrak{su}(2)_{k-2}$.

    \item 3 free fermions $\psi^a$ associated to the $\mathfrak{sl}(2,\mathbb{R})$ algebra, and 3 free fermions $\chi^a$ associated to the $\mathfrak{su}(2)$ algebra.

    \item The field content of the $c=6$ compactified theory $\mathcal{M}_4$.

    \item The usual $bc$ and $\beta\gamma$ ghost systems.

\end{itemize}
The stress tensor of this theory is complicated but straightforward to write down, and reads
\begin{equation}
\begin{split}
T=&\frac{1}{2k}\left(-2\mathcal{J}^3\mathcal{J}^3+\mathcal{J}^+\mathcal{J}^-+\mathcal{J}^-\mathcal{J}^++2\psi^3\partial\psi^3-\psi^+\partial\psi^--\psi^-\partial\psi^+\right)\\
&+\frac{1}{2k}\left(2\mathcal{K}^3\mathcal{K}^3+\mathcal{K}^+\mathcal{K}^-+\mathcal{K}^-\mathcal{K}^+-2\chi^3\partial\chi^3-\chi^+\partial\chi^--\chi^-\partial\chi^+\right)\\
&+\frac{1}{2}(\partial b)c+\frac{1}{2}(\partial c)b-\frac{3}{2}\partial(bc)-\frac{1}{2}(\partial\beta)\gamma+\frac{1}{2}(\partial\gamma)\beta-\partial(\beta\gamma)
\end{split}
\end{equation}

\subsubsection[The hybrid string on \texorpdfstring{$AdS_3 \times S^3$}{AdS3xS3}]{\boldmath The hybrid string on \texorpdfstring{$AdS_3 \times S^3$}{AdS3xS3}}

We are now ready to introduce the hybrid formalism on $AdS_3 \times S^3\times\mathcal{M}_4$. We recall briefly that the field content of the string theory in the RNS formalism is given by 
\begin{equation}
\underbrace{\mathcal{J}^a,\psi^a}_{\mathfrak{sl}(2,\mathbb{R})_{k}^{(1)}}\quad\underbrace{\mathcal{K}^a,\chi^a}_{\mathfrak{su}(2)_k^{(1)}}\quad\underbrace{b,c,\beta,\gamma}_{\text{ghosts}}\,.
\end{equation}
and the fields of the internal manifold $\mathcal{M}_4$.

In the flat space superstring, we found it convenient to bosonise the worldsheet fermions as well as the $bc$ and $\beta\gamma$ ghost system. For the worldsheet fermions $\psi^a$ and $\chi^a$, a convenient basis for bosonisation is given by defining scalars $\sigma_1,\sigma_2,\sigma_3$ such that
\begin{equation}
\frac{1}{k}\psi^+\psi^-=\partial\sigma_1\,,\quad\frac{1}{k}\chi^+\chi^-=\partial \sigma_2\,,\quad\frac{2}{k}\psi^3\chi^3=\partial\sigma_3\,.
\end{equation}
Given the OPEs between the worldsheet fermions, one can readily calculate the OPEs of the scalars $\sigma_i$ and find
\begin{equation}
\sigma_i(z)\sigma_j(w)\sim\delta_{ij}\log(z-w)\,.
\end{equation}
From the scalars $H_i$, it is possible to recover the original RNS fermions via
\begin{equation}
\psi^{\pm}=\sqrt{k}\,e^{\pm\sigma_1}\,,\quad\chi^{\pm}=\sqrt{k}\,e^{\pm\sigma_2}\,,\quad\psi^3\mp\chi^3=\sqrt{k}\,e^{\pm\sigma_3}\,.
\end{equation}
We also include the bosonised scalar $\varphi$ of the $\beta\gamma$ system whose charge gives the picture number. 

Finally, we assume that the internal manifold $\mathcal{M}_4$ has at least $\mathcal{N}=(2,2)$ worldsheet supersymmetry and has a central charge $c=6$ (as is the case for $\mathcal{M}_4=\mathbb{T}^4,\text{K3}$). If this is the case, the internal $\mathcal{M}_4$ theory contains a $\text{U}(1)$ current $J_C$ which generates the R-symmetry of the $\mathcal{N}=2$ algebra. We also bosonise this current by including a scalar $H$ such that
\begin{equation}
\partial H_C=J_C\,.
\end{equation}
Here, $J_C$ is normalized so that its self-OPE is
\begin{equation}
J_C(z)J_C(w)\sim\frac{c}{3(z-w)^2}=\frac{2}{(z-w)^2}\,,
\end{equation}
so that $H$ satisfies
\begin{equation}
H_C(z)H_C(w)\sim 2\log(z-w)\,.
\end{equation}
Thus, we have a set $\{\sigma_i,\varphi,H_C\}$ of scalars which we can use to generate vertex operators.

In the flat space superstring, the scalars introduced to bosonise the fermions had the advantage that they allowed a simple description of both the R/NS-sector vertex operators, but also of the spacetime supersymmetry generators. Similarly, we can define fields on the worldsheet which generate spacetime supersymmetry on $AdS_3 \times S^3$ in a similar fashion. In particular, we define a set of $8$ fermionic generators by 
\begin{equation}
q^{\alpha\beta\gamma}=\exp\left(\frac{\alpha}{2}\sigma_1+\frac{\beta}{2}\sigma_2+\frac{\alpha\beta\gamma}{2}\sigma_3+\frac{\gamma}{2}H_C-\frac{1}{2}\varphi\right)\,,
\end{equation}
where $\alpha,\beta,\gamma\in\{+,-\}$. The $e^{-\varphi/2}$ is again chosen so that the operator $q^{\alpha\beta\gamma}$ is in the canonical $q=-\frac{1}{2}$ picture of R-sector operators. The conformal weight of this state is given by
\begin{equation}
h(q^{\alpha\beta\gamma})=\frac{\alpha^2+\beta^2+2\gamma^2+\alpha^2\beta^2\gamma^2}{8}-\frac{1}{8}+\frac{1}{2}=\frac{5}{8}-\frac{1}{8}+\frac{1}{2}=1\,.
\end{equation}
Similar to in the flat space case, these generators are designed so that the number of minus signs is even (excluding the $-\frac{1}{2}\varphi$ term), which is why the scalar $\sigma_3$ has a funny coefficient of $\alpha\beta\gamma$. Also like in the flat space case, all of these operators are in the $q=-\frac{1}{2}$ picture, and thus there is no chance of their anti-commutators satisfying the supersymmetry algebra.

The trick to move from the RNS formalism to the hybrid formalism in $AdS_3 \times S^3\times\mathcal{M}_4$ is now the same as in four-dimensions: we define half of the supersymmetry generators to live in the $q=-\frac{1}{2}$ picture, while the other half is defined through picture changing to be in the $q=\frac{1}{2}$. Let us pick $q^{\alpha\beta +}$ to be in the $q=-\frac{1}{2}$, while we picture-change $q^{\alpha\beta -}$. That is, we define
\begin{equation}
S^{\alpha\beta+}=q^{\alpha\beta+}\,,\quad S^{\alpha\beta-}=Z\cdot q^{\alpha\beta -}\,.
\end{equation}
The naming $S^{\alpha\beta\gamma}$ is conventional.

Now we can search for superspace variables conjugate to the above supersymmetry generators. For the supercharges $S^{\alpha\beta+}$ this is achieved by just reversing the quantum numbers and defining
\begin{equation}
\theta_{\alpha\beta}=\exp\left(-\frac{\alpha}{2}\sigma_1-\frac{\beta}{2}\sigma_2-\frac{\alpha\beta}{2}\sigma_3-\frac{1}{2}H_C+\frac{1}{2}\varphi\right)
\end{equation}
This field has weight $h=0$ and satisfies
\begin{equation}
\theta_{\alpha\beta-}(z)S^{\gamma\delta+}(w)\sim\frac{\delta\indices{_\alpha^\beta}\delta\indices{_\beta^\delta}}{z-w}\,.
\end{equation}
Thus, the supersymmetry generators $Q^{\alpha\beta +}$ act on the $\theta$ coordinates geometrically by translations. Writing the explicit form of $S^{\alpha\beta-}$, we might also look for coordinates $\widetilde{\theta}_{\alpha\beta}$ which transform geometrically under $Q^{\alpha\beta-}$. However, a careful analysis shows that one cannot introduce $\theta_{\alpha\beta+}$ in a way that is independent of $\theta_{\alpha\beta}$, and in particular we cannot choose both to be free fields. This is in stark contrast to the case of four flat spacetime dimensions, where all of the superspace variables could be introduced and chosen independently to form free field theories.

The (unfortunate) conclusion of the above discussion is that the hybrid formalism on $AdS_3 \times S^3$ can only make half of the spacetime supersymmetry completely manifest. Letting $p^{\alpha\beta}$ be the conjugate momenta of $\theta_{\alpha\beta}$, we can immediately write down
\begin{equation}
\begin{split}
p^{\alpha\beta}&=\exp\left(\frac{\alpha}{2}\sigma_1+\frac{\beta}{2}\sigma_2+\frac{\alpha\beta}{2}\sigma_3+\frac{1}{2}H_C-\frac{1}{2}\varphi\right)\,.
\end{split}
\end{equation}
Together, the fields $\theta,p$ form a set of four first-order fermionic systems with OPEs
\begin{equation}
\theta_{\alpha\beta}(z)p^{\gamma\delta}(w)\sim\frac{\delta\indices{_\alpha^\gamma}\delta\indices{_\beta^\delta}}{z-w}\,.
\end{equation}
These four systems form four manifest super-coordinates in the target space. Bosonising gives four scalars. Since we started with six scalars (the three bosonised fermion systems $\sigma_i$, the $\varphi$ of the $\beta\gamma$ system, and the R-symmetry $H_C$ of the compactified theory), we simply do not have enough scalars to generate a second set of super-coordiantes which make $Q^{\alpha\beta-}$ manifest.

Having introduced the system $p,\theta$, however, we can write all of the supersymmetry generators purely in terms of these free fields and $\mathcal{J}$,$\mathcal{K}$. First, we define the `fermionic' currents
\begin{equation}
J_{(\text{f})}^a=\frac{1}{2}(\widetilde{\sigma}^a)\indices{_{\alpha}^{\beta}}(p^{\alpha\gamma}\theta_{\beta\gamma})\,,\quad K^a_{(\text{f})}=\frac{1}{2}(\sigma^a)\indices{_{\alpha}^{\beta}}(p^{\gamma\alpha}\theta_{\gamma\beta})\,.
\end{equation}
It can be checked that these satisfy the algebras $\mathfrak{sl}(2,\mathbb{R})_{-2}$ and $\mathfrak{su}(2)_2$, respectively. In fact, these are simply the fermion bilinears we subtracted from $J^a,K^a$ in order to define the `decoupled' currents $\mathcal{J}^a,\mathcal{K}^a$, just expressed in the hybrid variables $p,\theta$. Thus, we have
\begin{equation}
J^a=\mathcal{J}^a+J^a_{(\text{f})}\,,\quad K^a=\mathcal{K}^a+K^a_{(\text{f})}\,.
\end{equation}
Second, we can use $\theta,p$ to write down the supersymmetry generators $S^{\alpha\beta\gamma}$. By the definition of $p^{\alpha\beta}$, we have
\begin{equation}
S^{\alpha\beta+}=p^{\alpha\beta}\,.
\end{equation}
Far less obviously, however, is the following expression for $S^{\alpha\beta-}$, namely
\begin{equation}
S^{\alpha\beta-}=k\partial\theta^{\alpha\beta}+(\widetilde{\sigma}_a)\indices{^\alpha_\gamma}\left(\mathcal{J}^a+\frac{1}{2}J_{(\text{f})}^a\right)\theta^{\gamma\beta}-(\sigma_a)\indices{^\beta_{\gamma}}\left(\mathcal{K}^a+\frac{1}{2}K^{a}_{(\text{f})}\right)\theta^{\alpha\gamma}\,,
\end{equation}
where we have raised the indices on $\theta$ using the usual $\varepsilon$ symbol. Defining the supersymmetry charges
\begin{equation}
Q^{\alpha\beta\gamma}=\oint\frac{\mathrm{d}z}{2\pi i}S^{\alpha\beta\gamma}(z)\,,
\end{equation}
a direct computation shows that the supercharges $Q^{\alpha\beta\gamma}$, along with the charges
\begin{equation}
J_0^a=\oint\frac{\mathrm{d}z}{2\pi i}J^a(z)\,,\quad K^a_0=\oint\frac{\mathrm{d}z}{2\pi i}K^a(z)
\end{equation}
satisfy the supersymmetry algebra $\mathfrak{psu}(1,1|2)$ in equation \eqref{eq:bob-psu112-global}.

In fact, we can do better. Taking the fields $J^a,K^a,S^{\alpha\beta\gamma}$ as being currents on the worldsheet, we can determine their current algebra. They satisfy the current algebra
\begin{equation}
\begin{split}
J^a(z)J^b(w)&\sim\frac{k\widetilde{\kappa}^{ab}}{(z-w)^2}+\frac{\widetilde{f}\indices{^a^b_c}J^b(w)}{z-w}\,,\\
K^a(z)K^b(w)&\sim\frac{k\kappa^{ab}}{(z-w)^2}+\frac{f\indices{^a^b_c}K^c(w)}{z-w}\,,\\
J^a(z)S^{\alpha\beta\gamma}(w)&\sim\frac{(\widetilde{\sigma}^a)\indices{^\alpha_\delta}S^{\delta\beta\gamma}(w)}{z-w}\,,\\
K^a(z)S^{\alpha\beta\gamma}(w)&\sim\frac{(\sigma^a)\indices{^\beta_\delta}S^{\alpha\delta\gamma}(w)}{z-w}\,,\\
S^{\alpha\beta+}(z)S^{\gamma\delta+}(w)&\sim\frac{k\,\varepsilon^{\alpha\gamma}\varepsilon^{\beta\delta}}{(z-w)^2}+\frac{\varepsilon^{\alpha\gamma}(\sigma_a)\indices{^\beta^\delta}K^a(w)-\varepsilon^{\beta\delta}(\widetilde{\sigma}_a)^{\alpha\gamma}J^a(w)}{z-w}\,,
\end{split}
\end{equation}
where $\kappa,f,\sigma$ (resp. $\widetilde{\kappa},\widetilde{f},\widetilde{\sigma}$) are the Killing form, structure constants, and spinor generators of $\mathfrak{su}(2)$ (resp. $\mathfrak{sl}(2,\mathbb{R})$). This is the current algebra $\mathfrak{psu}(1,1|2)_k$ which would be obtained from a Wess-Zumino-Witten model on the supergroup $\text{PSU}(1,1|2)$ at level $k$. A more careful analysis of the worldsheet fields shows that the action of the hybrid variables $p,\theta$ along with $J,K$ indeed describes a WZW model on $\text{PSU}(1,1|2)$ \cite{Berkovits:1999im}.

As a quick aside, we mention some of the group theoretic properties of the supergroup $\text{PSU}(1,1|2)$.\footnote{The WZW model on $\text{PSU}(1,1|2)$ was studied in detail in \cite{Gotz:2006qp}. See also \cite{Gerigk:2012lqa}.} As was mentioned in Section \ref{sec:bob-ads3-geometry}, $\text{PSU}(1,1|2)$ is a supergroup with bosonic subgroup $\text{SL}(2,\mathbb{R})\times\text{SU}(2)$ and with eight odd dimensions, corresponding to the supercharges of $AdS_3 \times S^3$. As such, we can assign its `superdimension' to be
\begin{equation}
\text{sdim}(\text{PSU}(1,1|2))=6-8=-2\,,
\end{equation}
where the superdimension counts the number of even dimensions minus the number of odd dimensions of the supergroup. Moreover, the superalgebra $\mathfrak{psu}(1,1|2)$ has the surprising property that its dual Coxeter number $h^{\vee}$ vanishes.\footnote{The supergroups with vanishing dual Coxeter number have been classified, and they are $\mathfrak{psl}(n|n)$, $\mathfrak{ops}(2n+2|2n)$, and $\mathfrak{d}(2,1;\alpha)$.} Thus, a naive calculation for the central charge of the $\mathfrak{psu}(1,1|2)_k$ WZW model is
\begin{equation}
c(\mathfrak{psu}(1,1|2)_k)=\frac{k\,\text{sdim}(\mathfrak{psu}(1,1|2))}{k+h^{\vee}}=-2\,.
\end{equation}
This agrees with the counting of central charges from the bosonic $\mathfrak{sl}(2,\mathbb{R})_{k+2}\oplus\mathfrak{su}(2)_{k-2}$ WZW models and the four first-order $(p,\theta)$ systems with $c=-2$ each:
\begin{equation}
c(\mathfrak{sl}(2,\mathbb{R})_{k+2})+c(\mathfrak{su}(2)_{k-2})-4\cdot 2=\frac{3(k+2)}{k}+\frac{3(k-2)}{k}-8=-2\,.
\end{equation}
In order to quantise the worldsheet theory, we need something of central charge $+2$ to cancel the Weyl anomaly of the $\mathfrak{psu}(1,1|2)_k$ theory. As we will see, this will be the total central charge of the left-over ghost system in the hybrid formalism.

\subsubsection{Decoupling the internal CFT}

Since used the R-symmetry current $J_C$ of $\mathcal{M}_4$ to define the hybrid variables in $AdS_3 \times S^3$, there is no way that the hybrid variables can decouple from the RNS variables on $\mathcal{M}_4$. Just like in the flat space example, however, we can perform a similarity transformation
\begin{equation}
\Phi_C^{\text{new}}=e^{\mathcal{W}}\Phi_Ce^{-\mathcal{W}}\,,
\end{equation}
where
\begin{equation}
\mathcal{W}=\oint\frac{\mathrm{d}z}{2\pi i}(H_C\mathcal{P})(z)\,,
\end{equation}
where $\mathcal{P}=\partial\chi-\partial\varphi$ is the picture current. The effect on the R-symmetry current is
\begin{equation}
J_C^{\text{new}}=J_C-2\mathcal{P}\,,
\end{equation}
which decouples from the hybrid fields. Just like in the flat space case, the $c=9$ $\mathcal{N}=(2,0)$ generators are modified to
\begin{equation}
G^{\pm,\text{new}}_C=e^{\pm(\varphi-\chi)}G_C^{\pm}\,,\quad T_C^{\text{new}}=T_C-\mathcal{P}J_C+\frac{3}{2}\mathcal{P}^2\,,
\end{equation}
which, together with $J_C^{\text{new}}$, still satisfy an (untwisted) $c=6$ $\mathcal{N}=(2,0)$ superconformal algebra.

\subsubsection[The \texorpdfstring{$\sigma,\rho$}{sigma,rho} ghosts]{\boldmath The \texorpdfstring{$\sigma,\rho$}{sigma,rho} ghosts}

In the RNS formalism, we started with the six scalars
\begin{equation}
\varphi,\chi,\sigma_1,\sigma_2,\sigma_3,H_C
\end{equation}
from which we defined four first order systems $(p,\theta)$ and a new R-symmetry current $J_C^{\text{new}}$. Thus, a scalar is missing. The $p,\theta$ system and the new R-symmetry are bosonised by the five scalars
\begin{equation}
\frac{\alpha}{2}\sigma_1+\frac{\beta}{2}\sigma_2+\frac{\alpha\beta}{2}\sigma_3+\frac{H_C}{2}-\frac{\varphi}{2}\,,\quad H_C-2\chi+2\varphi\,.
\end{equation}
The unique linear combination which is orthogonal to all five of these currents and has OPE $\rho(z)\rho(w)=-\log(z-w)$ is given by
\begin{equation}
\rho=2\varphi-H_C-\chi\,.
\end{equation}
This scalar has background charge $Q_{\rho}=3$ and central charge
\begin{equation}
c(\rho)=(1+3Q_{\rho}^2)=28\,.
\end{equation}
Furthermore, we never used the $b,c$ system when constructing the hybrid fields, and so we still have the $\sigma$ ghost with $c=-26$. Together, $\rho$ and $\sigma$ define the full ghost content of the hybrid formalism on $AdS_3 \times S^3\times\mathcal{M}_4$, and we refer to them collectively as the $(\rho,\sigma)$ system, which has central charge
\begin{equation}
c(\rho,\sigma)=28-26=2\,.
\end{equation}

\subsubsection{The topological twist}

In the process of decoupling the internal CFT, we wound up introducing supercharges $G^{\pm,\text{new}}_C$ and R-symmetry $J^{\text{new}}_C$ which, together with $T_C^{\text{new}}$ still satisfy the $c=6$ $\mathcal{N}=(2,0)$ superconformal algebra. However, in the full theory, these redefined generators have the following two properties:
\begin{itemize}

    \item $G^{\pm,\text{new}}_{C}$ have conformal weights $h=\frac{3}{2}\pm\frac{1}{2}$ with respect to the stress tensor $T$ of the full 10-dimensional theory.

    \item $J^{\text{new}}_C$ has a background charge of $Q=2$.

\end{itemize}
As in the flat space case, this indicates that the stress tensor of the compactified theory has been `twisted'. Indeed, a direct check gives
\begin{equation}
T_{\text{RNS}}(\mathcal{J},\psi,\mathcal{K},\chi,b,c,\beta,\gamma,\Phi_C)=\underbrace{T_{\text{hybrid}}(J,K,S)}_{\mathfrak{psu}(1,1|2)_k}+T_{\text{ghosts}}(\rho,\sigma)+\underbrace{T_{C}^{\text{new}}(\Phi_C^{\text{new}})+\frac{1}{2}\partial J_C^{\text{new}}}_{\text{topological twist}}\,.
\end{equation}
That is, in order for the hybrid theory to be equivalent to the original RNS description, we need to include the topological twisting term to the compact theory $\mathcal{M}_4$. Therefore, the full hybrid formalism is given by the quantum equivalence:
\begin{equation}
\begin{gathered}
\text{hybrid strings on }AdS_3 \times S^3\times\mathcal{M}_4\\
\Longleftrightarrow\\
\mathfrak{psu}(1,1|2)_k\oplus[(\rho,\sigma)\text{ ghosts}]\oplus[\text{topologically twisted }\mathcal{M}_4]\,.
\end{gathered}
\end{equation}
As a sanity check, we can calculate the full central charge of the theory. Since topologically twisted theories have vanishing central charge, we have
\begin{equation}
c_{\text{tot}}=c(\mathfrak{psu}(1,1|2)_k)+c(\sigma,\rho)=-2+2=0\,,
\end{equation}
and so the hybrid string on $AdS_3 \times S^3\times\mathcal{M}_4$ defines a consistent string theory.

\subsection{Applications}\label{sec:bob-applications}

The hybrid formalism, as discussed above, is built from the RNS formalism from a complicated series of field redefinitions. The conceptual advantage in the end is that one is left with a covariant way to quantise the worldsheet theory while also having manifest super-Poincare covariance in some component of the target space.

Despite this nice conceptual feature, practical computations in the hybrid formalism are generically technically difficult. One reason for this is that, upon performing all of the necessary field redefinitions, the BRST operator of the theory becomes massively more complicated than its RNS cousin. Moreover, the prescription for calculating correlation functions (which we have not discussed here) also becomes much more complicated than the usual definitions in the RNS formalism. Although both of these problems (the BRST cohomology and the definition of correlation functions) can be given a natural interpretation in terms of an $\mathcal{N}=4$ topological string  \cite{Berkovits:1994vy,Berkovits:1994wr,Berkovits:1996bf,Berkovits:1999im,Kappeli:2006fj}, this reinterpretation does very little in the way of simplifying actual calculations. (This is not to say that calculations in the hybrid formalism are entirely impossible, see for example \cite{Berkovits:2001nv,Bobkov:2002bx,Gaberdiel:2011vf,Gerigk:2012cq} and references therein).

However, as we will briefly review below, there are two things that the hybrid formalism does extremely well. Focusing on the case of the $AdS_3 \times S^3\times\mathcal{M}_4$ superstring, the hybrid formalism allows for 1) a conceptually straightforward way to quantise strings on backgrounds with nonzero RR-flux and 2) allows, in a very special limit, exact calculations of string correlation functions at all orders in string perturbation theory.

\subsubsection{Adding Ramond-Ramond flux}

In the previous section we defined the hybrid string on $AdS_3 \times S^3$ in terms of a WZW model on the supergroup $\text{PSU}(1,1|2)$ at level $k$. The level $k$ of the WZW model (and in turn of the current algebra $\mathfrak{psu}(1,1|2)_k$) is determined by the amount of NS-NS flux present in the string background. However, NS-NS flux is not the only type of flux which can be present in a background of the form $AdS_3 \times S^3\times\mathcal{M}_4$. Indeed, as discussed in Section \ref{s:saskia}, $\text{AdS}_3$ string backgrounds are special among holographic string theories in that they can be supported by a mixture of both NS-NS flux and Ramond-Ramond (RR) flux. Mixed flux backgrounds (backgrounds with both NS-NS and RR fluxes) are notoriously difficult to quantise from a worldsheet perspective, particularly in the RNS formalism. However, in this section we will show that within the hybrid formalism it is (conceptually) straightforward to include RR flux on the worldsheet. In fact, this was one of the original motivators of formulating the hybrid superstring on $AdS_3 \times S^3\times\mathcal{M}_4$ \cite{Berkovits:1999im,Berkovits:1999xv}.

Let us consider a WZW model whose target is some (super)group $G$. The fundamental field is a map $g:\Sigma\to G$ from the worldsheet into the $G$, and the action is given by
\begin{equation}
S_{\text{WZW}}=\frac{1}{4\pi f^2}\int\mathrm{d}^2z\,\text{STr}\left(g^{-1}\partial g\,g^{-1}\bar{\partial}g\right)+k S_{\text{WZ}}[g]\,,
\end{equation}
where $S_{\text{WZ}}$ is the Wess-Zumino term
\begin{equation}
S_{\text{WZ}}=-\frac{i}{2\pi}\int_{\mathcal{B}}\text{STr}[g^{-1}\mathrm{d}g\wedge g^{-1}\mathrm{d}g\wedge g^{-1}\mathrm{d}g]\,.
\end{equation}
In a standard WZW model, consistent quantisation requires that $f^2k=1$.\footnote{Technically, we could also have $f^2k=-1$, but for the sake of discussion we will stick with $f^2k=1$. These two cases are related to each other by a reversal of worldsheet orientation.} This is the requirement for conformal symmetry to be preserved on the worldsheet at the quantum level. The reasoning is that if $f^2k\neq 1$, then the conserved current $J$ is not holomorphic (equivalently, $\bar{J}$ is not anti-holomorphic). This naively spoils worldsheet conformal invariance since the stress tensor
\begin{equation}
T=\frac{1}{2(k+h^{\vee})}\kappa_{ab}J^aJ^b
\end{equation}
will fail to be holomorphic.

As it turns out, $\mathfrak{psu}(1,1|2)$ belongs to a class of Lie superalgebras for which the requirement $f^2k=1$ can be relaxed. Specifically, this is due to the fact that $\mathfrak{psu}(1,1|2)$ has vanishing dual Coxeter number, \textit{i.e.}~$h^{\vee}=0$. Supergroup WZW models with vanishing dual Coxeter number were studied in detail in \cite{Ashok:2009xx,Benichou:2010rk} One can show that this property implies that the stress tensor $T$, built from the Sugawara construction, is still holomorphic despite the fact that $J$ itself is not. Furthermore, the holomorphicity of $T$ does not receive quantum corrections \cite{Berkovits:1999im,Bershadsky:1999hk}. This implies that the WZW model on $\text{PSU}(1,1|2)$ is still a consistent quantum theory away from the WZW point!

In the hybrid superstring on $AdS_3 \times S^3\times\mathcal{M}_4$, relaxing the condition $f^2k=1$ has the physical interpretation of adding Ramond-Ramond (RR) flux to the background. In terms of the geometry of $\text{AdS}_3$, the first coefficient of the WZW model action is given by the radius of $\text{AdS}_3$ in units of the string length. The level $k$ is associated to the amount of NS-NS flux. A supergravity analysis \cite{Maldacena:1998bw} shows that, if one adds $Q_5^{\text{RR}}$ D5 branes to the NS5-F1 system, then the radius of the near-horizon $\text{AdS}_3$ geometry is given by (see the discussion around equation \eqref{eq:string-tension})
\begin{equation}
\frac{1}{f^2}=\frac{R^2_{\text{AdS}}}{\ell_s^2}=\sqrt{(Q^{\text{NS}}_5)^2+g_s^2(Q^{RR}_5)^2}=\sqrt{k^2+g_s^2(Q^{\text{RR}}_5)^2}\,.
\end{equation}
Here, $g_s$ is the \textit{string coupling constant}, which we take to be small. In terms of the value $k$ of NS-NS and $Q_{5}^{\text{RR}}$ of RR-flux, the quantity $f^2k$ can be expressed as
\begin{equation}
f^2k=\frac{k}{\sqrt{k^2+g_s^2(Q^{\text{RR}}_5)^2}}=\left(1+g_s^2\frac{(Q^{\text{RR}}_5)^2}{k^2}\right)^{-1/2}\,.
\end{equation}
If we take $g_s^2$ to be small, we find that, although $Q^{\text{RR}}_5\in\mathbb{Z}_{\geq 0}$ the quantity $f^2k$ is essentially continuous on the interval $0\leq f^2k\leq 1$.\footnote{Without loss of generality, we have chosen the positive branch of the square root.} We should note that, in the notation of Section \ref{s:saskia}, we have
\begin{equation}
q=f^2k\,.
\end{equation}
Indeed, $q=1$ is the condition of having a background with pure NS-NS flux, while $q\to 0$ is given by the limit $g_sQ_{5}^{\text{RR}}\gg k$, \textit{i.e.}~the limit in which the background is entirely dominated by RR flux.

While the addition of Ramond-Ramond flux by choosing $f^2k\neq\pm1$ is conceptually straightforward, the quantisation of such a theory is a complete mess. For one, the introduction of the deformation term actually introduces nonlinear couplings between the fields of the WZW model and the $\rho,\sigma$ ghost, which makes the BRST quantisation of the theory nearly completely intractable. Second, even at the level of the WZW model itself (\textit{i.e.}~ignoring the ghosts and BRST quantisation), the non-holomorphicity of the current $J$ leads to an OPE algebra whose representation theory is not well understood. The current algebra between the non-holomorphic conserved currents can be derived as \cite{Benichou:2010rk}
\begin{equation}\label{eq:bob-psu-rr-flux-currents}
\begin{split}
J^a_z(z)J^b_z(w)&\sim\frac{(1+kf^2)\kappa^{ab}}{4f^2(z-w)^2}+\frac{f\indices{^a^b_c}}{4}\left(\frac{(3-kf^2)(1+kf^2)J^c_z(w)}{z-w}+\frac{(1-kf^2)^2J^a_{\bar{z}}(w)}{(z-w)^2}\right)\,,\\
J^a_{\bar{z}}(z)J^b_{\bar{z}}(w)&\sim\frac{(1-kf^2)\kappa^{ab}}{4f^2(\bar{z}-\bar{w})^2}+\frac{f\indices{^a^b_c}}{4}\left(\frac{(3+kf^2)(1-kf^2)J^c_{\bar{z}}(w)}{\bar{z}-\bar{w}}+\frac{(1-kf^2)^2J^a_{z}(w)}{(\bar{z}-\bar{w})^2}\right)\,,\\
J^a_z(z)J^b_{\bar{z}}(w)&\sim(1-kf^2)^2f\indices{^a^b_c}\left(\frac{J^c_z(w)}{\bar{z}-\bar{w}}+\frac{J^c_{\bar{z}}(w)}{z-w}\right)\,.
\end{split}
\end{equation}
The quantisation of this current algebra is extremely complicated. Here, by $\sim$ we are excluding terms in the OPE which are less divergent than $1/(z-w)$, such as terms like $\log(z-w)$. 

Despite the complexity of the above algebra, certain properties of string theory in mixed-flux backgrounds can be deduced from its representation theory. For example, for vanishing RR flux, the worldsheet theory contains a continuum of states known as `long strings' which can propagate toward the boundary of $\text{AdS}_3$ with finite cost in energy \cite{Maldacena:2000hw}. However, an analysis of the current algebra \eqref{eq:bob-psu-rr-flux-currents} shows that as soon as the background includes any RR flux, i.e. as soon as one perturbs $kf^2$ away from $1$, the worldsheet conformal dimensions of these long string states acquires a nonzero imaginary part, and thus cannot be part of the physical spectrum of the worldsheet theory \cite{Eberhardt:2018vho}.

\subsubsection{Applications to holography}

So far, we have introduced the hybrid formalism as a formal redefinition of the RNS formalism for which allows for manifest spacetime supersymmetry without the need to artificially impose the GSO projection. As we argued above, the hybrid formalism on $AdS_3 \times S^3$ can also be employed to study stringy backgrounds with non-vanishing RR flux, even if it is technically challenging in practice. Here, we will explain another advantage of the hybrid formalism which has only become clear in recent years: it allows the study of $AdS_3 \times S^3$ backgrounds in the deeply stringy regime.

Recall that, in the RNS formalism, we can describe the $\mathcal{N}=(1,1)$ WZW model $\mathfrak{g}^{(1)}_k$ by a purely bosonic WZW model at level $k-h^{\vee}$ and $\text{dim}(\mathfrak{g})$ free, decoupled fermions. That is,
\begin{equation}
\mathfrak{g}^{(1)}_{k-h^{\vee}}\cong\mathfrak{g}_{k-h^{\vee}}\oplus(\text{dim}(\mathfrak{g})\text{ free fermions})\,.
\end{equation}
Applied to the background $AdS_3 \times S^3$ (with pure NS-NS flux), we have
\begin{equation}
\text{RNS strings on AdS}_3\times\text{S}^3\cong\mathfrak{sl}(2,\mathbb{R})_{k+2}\oplus\mathfrak{su}(2)_{k-2}\oplus(6\text{ free fermions})\,.
\end{equation}
This is precisely the model we began with when defining the hybrid string on $AdS_3 \times S^3$.

In the brane construction of $AdS_3 \times S^3$ backgrounds, $k$ represents the number of units of NS-NS flux sourced by the stack of NS5 branes or, equivalently, the number of NS5 branes. Clearly, the smallest value of $k$ allowed, then, would be $k=1$ (at $k=0$ we no longer have an $AdS_3 \times S^3$ near-horizon geometry). However, precisely at $k=1$ the worldsheet theory becomes
\begin{equation}
\mathfrak{sl}(2,\mathbb{R})_{3}\oplus\mathfrak{su}(2)_{-1}\oplus(6\text{ free fermions})\,.
\end{equation}
The negative level in the $\mathfrak{su}(2)_{-1}$ factor is a problem and this factor alone is non-unitary. Specifically, it has central charge
\begin{equation}
c(\mathfrak{su}(2)_{-1})=\frac{3\cdot(-1)}{-1+h^{\vee}(\mathfrak{su}(2))}=-3\,.
\end{equation}
It is still feasible to formally treat the $\mathfrak{su}(2)_{-1}$ factor as a `ghost' system, but beyond the counting of states, it is not very clear how to do this.

An alternative approach is to simply start with the hybrid formalism. Since the hybrid formalism on $AdS_3 \times S^3$ is a WZW model on the supergroup $\text{PSU}(1,1|2)$ with level $k$, we have\footnote{We also have to topologically twist the internal manifold $\mathcal{M}_4$ and include the $\rho,\sigma$ ghosts.}
\begin{equation}
\text{hybrid strings on AdS}_3\times\text{S}^3\text{ at }k=1\cong\mathfrak{psu}(1,1|2)_1\,.
\end{equation}
The affine Ka\v{c}-Moody superalgebra $\mathfrak{psu}(1,1|2)_1$ is perfectly well-defined and admits unitary representations, and thus allows for a conceptually straightforward analysis of the worldsheet (although the physical state conditions become rather complicated).

Typically, exact calculations in string theory on curved spacetimes are completely out of reach. However, the key feature of the $k=1$ worldsheet theory lies in the fact that, even though the worldsheet lies on a highly curved target space, it can be described by a \textit{free worldsheet theory}. In particular, the WZW model $\mathfrak{psu}(1,1|2)_1$ can be realised by eight free fields \cite{Eberhardt:2018ouy,Dei:2020zui}:
\begin{equation}
\underbrace{(\lambda,\mu^{\dagger})\,,\quad(\mu,\lambda^{\dagger})}_{\text{four }\beta\gamma\text{ systems with }\lambda=1/2}\quad\underbrace{(\psi^a,\psi^{\dagger}_a)}_{\text{four }bc\text{ systems with }\lambda=1/2}
\end{equation}
To see this, we group the free fields into two supervectors of dimension $2|2$:
\begin{equation}
Y=
\begin{pmatrix}
\mu^{\dagger} & \lambda^{\dagger} & \psi_1^{\dagger} & \psi_2^{\dagger}
\end{pmatrix}\,,\quad
Z=
\begin{pmatrix}
\lambda \\ \mu \\ \psi^1 \\ \psi^2
\end{pmatrix}\,,
\end{equation}
for which the action of the fields reads
\begin{equation}
S=\frac{1}{2\pi}\int_{\Sigma}Y\bar\partial Z\,.
\end{equation}
This action contains a set of linear symmetries generated by any invertible supermatrix of dimension $2|2$ acting as
\begin{equation}
Z\to MZ\,,\quad Y\to YM^{-1}\,.
\end{equation}
This symmetry generates the supergroup $\text{GL}(2|2)$, and the conserved currents, which are generated by bilinears of the form $J\indices{_a^b}=Y_aZ^b$, generate the algebra $\mathfrak{gl}(2|2)_1$. If one also gauges the current
\begin{equation}
\mathcal{Z}=\frac{1}{2}Y\cdot Z\,,
\end{equation}
which generates the diagonal $\text{U}(1)$ subgroup $Z\to\alpha Z$ $Y\to\alpha^{-1}Y$, one reduces the symmetry group from $\text{GL}(2|2)$ to $\text{PSU}(1,1|2)$. See \cite{Dei:2020zui,Gaberdiel:2022als} for more details. The advantage of this free field realisation is that much of the difficulty of string theory on $AdS_3 \times S^3$ reduces to computing quantities in a free CFT (although a slightly unconventional one).

Holographically, string theory on $AdS_3 \times S^3\times\mathcal{M}$ has long believed to be dual to a CFT in the moduli space of so-called \textit{symmetric orbifold theories} \cite{Dijkgraaf:1998gf,deBoer:1998us,Maldacena:1999bp,Larsen:1999uk,Seiberg:1999xz,Argurio:2000tb,Gaberdiel:2007vu} (see also \cite{David:2002wn} for a review). These theories are CFTs obtained by starting with a `seed' theory $X$ of central charge $c$ and tensoring it with itself $N$ times to obtain the CFT $X^{\otimes N}$. One then gauges or `orbifolds' the $S_N$ symmetry of permuting the copies of $X$ to obtain the CFT
\begin{equation}
\text{Sym}^N(X):=X^{\otimes N}/S_N\,,
\end{equation}
which has central charge $c_N=cN$. Taking $N$ large then gives a large-$c$ CFT, which potentially has a hologrpahic dual. When we say that string theory on $AdS_3 \times S^3\times\mathcal{M}$ lives in the same moduli space as $\text{Sym}^N(X)$, what is meant is that the dual theory is given by deforming $\text{Sym}^N(X)$ by an exactly marginal operator $\mathcal{O}$ which breaks the orbifold structure.\,\footnote{By this, we mean that the deformed theory can no longer be written as $\text{Sym}^N(Y)$ for some CFT $Y$.}

The strength of the deformation $\mathcal{O}$ determines how far away one is from the so-called `orbifold point', \textit{i.e.}~the point in the moduli space for which the dual CFT is \textit{exactly} a symmetric orbifold. Since the spectrum of symmetric orbifold theories does not meet the requirement to be described by an effective theory of $\text{AdS}_3$ supergravity (specifically, symmetric orbifolds do not have sparse spectra), they cannot be dual to the supergravity limit of string theory. It is thus believed that the orbifold point is a string theory for which the spacetime curvature is very large. 

If the background has pure units of NS/NS flux, this means that the level $k$ of the $\mathfrak{psu}(1,1|2)_k$ WZW model should be small. However, since it is quantised, there is a smallest value it can reach while still describing a consistent theory, namely $k=1$. In recent years, using the above free field realisation of $\mathfrak{psu}(1,1|2)_1$, a large amount of evidence has been gathered supporting the fact that the $k=1$ limit of $\text{AdS}_3$ string theory is \textit{exactly} dual to a symmetric orbifold. Specifically,
\begin{equation}
\begin{gathered}
\text{pure NS-NS strings on AdS}_3\times\text{S}^3\times\mathcal{M}_4\text{ at }k=1\\
\Longleftrightarrow\\
\text{the symmetric product CFT Sym}^N(\mathcal{M}_4)\,.
\end{gathered}
\end{equation}
Evidence for this proposal includes:
\begin{itemize}

    \item A full matching of the physical spectra \cite{Eberhardt:2018ouy,Eberhardt:2020bgq,Naderi:2022bus}.

    \item A full matching of correlation functions of twisted-sector ground states at tree-level \cite{Eberhardt:2019ywk,Dei:2020zui,Dei:2023ivl}.

    \item A full matching of correlation functions of twisted-sector ground states at all loop-level \cite{Eberhardt:2020akk,Knighton:2020kuh}.

    \item A matching of the first nontrivial terms in the perturbative series as one deforms away from the orbifold point \cite{Fiset:2022erp}.

\end{itemize}
All of these calculations were performed within the framework of the hybrid formalism. In fact, it is not currently known how to treat the $k=1$ theory in the RNS or Green-Schwarz formalisms, so for now the only method we have of exploring this region in the parameter space of superstring theories is with the hybrid formalism.

\subsection{Outlook}

In the previous sections, we presented in detail the hybrid formalism of superstrings both in four dimensional flat space and on the bacground $AdS_3\times S^3\times\mathcal{M}_4$. In an attempt to be completely self-contained, there were unfortunately aspects of the theory which could not be addressed. Most notably, we only made a passing reference to the physical state conditions one must impose on worldsheet states, and we have not made any mention of exactly how to compute correlation functions of the theory. For both of these topics, we encourage the interested reader to consult \cite{Berkovits:1996bf} and \cite{Berkovits:1999im}.

Before we conclude our discussion of the hybrid formalism, let us briefly discuss two points that we believe to be important gaps in our knowledge, which will hopefully be better understood.

\paragraph{The hybrid formalism in other backgrounds:} Since the hybrid formalism is such a powerful tool for analysing backgrounds like $\mathbb{R}^4$ and $AdS_3\times S^3$ in a way that is SUSY-covariant, one might ask whether the formalism can be readily generalised to other backgrounds. The formulation of the hybrid superstring on $\mathbb{R}^4$ required the background to be obtained by a Calabi-Yau compactification. Similarly, defining the hybrid string on $AdS_3\times S^3\times\mathcal{M}_4$ required us to use the fact that $\mathcal{M}_4$ is either $\mathbb{T}^4$ or $K3$. In both of these cases, the existence of $\mathcal{N}=(2,2)$ supersymmetry in the compactified directions was a crucial ingredient in the field redefinitions in passing from the RNS formalism to the hybrid formalism. Thus, it seems that the ability to write down a hybrid superstring on a background is dependent on the details of that background.

Interestingly, there are other backgrounds on which the hybrid formalism can be defined. For example, the hybrid formalism can be described $\mathbb{R}^2\times\mathcal{M}_8$, where $\mathcal{M}_8$ is a Calabi-Yau four-fold \cite{Berkovits:2001tg}, as well as on $AdS_2\times S^2\times \mathcal{M}_6$, where $\mathcal{M}_6$ is a Calabi-Yau three-fold \cite{Berkovits:1999zq}. More recently, it has been shown that the background $AdS_3\times S^3\times S^3\times S^1$ also admits a hybrid description, which is based on a WZW model with group $\text{D}(2,1;\alpha)$ \cite{Eberhardt:2019niq}, where $\alpha$ controls the relationship between the radii of the two $S^3$'s (see Section \ref{s:saskia}). 

Despite this, the holy grail of holographic backgrounds, namely type IIB superstrings on $AdS_5\times S^5$, has no known hybrid description. There are hints \cite{Berkovits:1999zq,Gaberdiel:2021qbb,Gaberdiel:2021jrv} that the appropriate description should be based on a WZW model with group $\text{PSU}(2,2|4)$. However, a consistent hybrid description with this group is not currently known.

\paragraph{The geometric meaning of the hybrid formalism:} The RNS formalism is at its heart a geometric theory of supergravity in two dimensions. Indeed, the $bc$ and $\beta\gamma$ ghost systems of the RNS formalism are nothing more than the Fadeev-Popov ghosts needed to gauge the diffeomorphism and super-diffeomorphism symmetry on the worldsheet. When passing to the hybrid formalism on, say, $AdS_3\times S^3$, the $\beta\gamma$ system gets lost in all of the field redefinitions. What is left is the $\text{PSU}(1,1|2)$ WZW model, the compact CFT $\mathcal{M}_4$, the original $bc$ diffeomorphism ghosts, and a new scalar dubbed $\rho$.

The existence of the extra scalar left over after field redefinitions is a smoking gun of the hybrid formalism. Despite usually being referred to as a `ghost', it is not clear what, if anything, $\rho$ is actually gauging. It might be natural to guess that $\rho$, similar to the $bc$ and $\beta\gamma$ ghosts, is gauging some sort of geometric structure on the worldsheet, but there are very few hints as to what this kind of structure would be.

While the hybrid formalism as presented in this section is obtained from the RNS formalism via a series of field redefinitions, one would hope that there would be a kind of `intrinsic' derivation of the hybrid formalism that is independent of the RNS or Green-Schwarz formalisms. Such a derivation could potentially shed light on the meaning of the hybrid ghost $\rho$, as well as the complexity of the physical state conditions in the hybrid formalism.\footnote{See \cite{McStay:2023thk} for recent progress in this direction in the context of the $AdS_3\times S^3\times\mathbb{T}^4$ hybrid string.}

\newpage

%%%%%%%%%%%%%%%%%%%%%%%%%%%%%%%%%%%%%%%%%%%%%%%%%%%%%%%
%%%%%%%%%%%%%%%%%%%%%%%%%%%%%%%%%%%%%%%%%%%%%%%%%%%%%%%
%%%%%%%%%%%%%%%%%%%%%%%%%%%%%%%%%%%%%%%%%%%%%%%%%%%%%%%
\section{Matrix theory and the string worldsheet}
%%%%%%%%%%%%%%%%%%%%%%%%%%%%%%%%%%%%%%%%%%%%%%%%%%%%%%%
%%%%%%%%%%%%%%%%%%%%%%%%%%%%%%%%%%%%%%%%%%%%%%%%%%%%%%%
%%%%%%%%%%%%%%%%%%%%%%%%%%%%%%%%%%%%%%%%%%%%%%%%%%%%%%%

%\noindent \emph{Author: Ziqi Yan}

%\noindent \emph{Email:} \href{ziqi.yan@su.se}{ziqi.yan@su.se}

%%%%%%%%%%%%%%%%%%%%%%%%%%%%%%%%%%%%%%%%%%%%%%%%%%%%%%%
%\subsection{Introduction}
%%%%%%%%%%%%%%%%%%%%%%%%%%%%%%%%%%%%%%%%%%%%%%%%%%%%%%%

We give a pedagogical review of aspects of nonperturbative string theory in connection with Matrix theory. This section has a different but closely related motivation compared to the rest of the review: we would like to develop tools to understand how string theory behaves when the string coupling is very large, where conventional perturbative treatments fail to apply. A thorough understanding of nonperturbative phenomena in string theory is indispensable before one can justly determine how it should be applied to Nature~\cite{Polchinski:1998rr}. For example, in the Standard Model, one would have falsely concluded that strong and weak forces were short ranged without understanding the nonperturbative phenomena such as the Higgs mechanism and quark confinement. From a broader perspective, viewing string theory as a mathematical framework that provides useful techniques and insights for other fields, it is equally important to map out various nonperturbative aspects in order to reveal its full power for solving related problems in other disciplines of theoretical physics.  

It is well known that the nature of ten-dimensional type IIA (and $E_8 \times E_8$ heterotic) superstring theory changes in the strongly coupled regime, where an eleventh dimension emerges and the fundamental role played by the string is replaced with the membrane~\cite{Bergshoeff:1987cm, Townsend:1995kk, Witten:1995ex, Horava:1995qa}. This eleven-dimensional theory of the supposedly fundamental membranes is conjectured to be ``M-theory''~\cite{Horava:1995qa}. One important lesson that we have learned from the second superstring revolution is that the hypothetical M-theory unifies different perturbative superstring theories, namely, type I, IIA, IIB, and heterotic superstring theory, which describe remote corners in the web of solutions of the single eleven-dimensional theory. Although we have acquired a fairly comprehensive understanding of perturbative string theory, our knowledge of the fundamental nature of M-theory, \emph{\textit{i.e.}}~nonperturbative string theory and quantum gravity, is still rather limited. One major technical difficulty is that, viewing M-theory as a UV-complete theory, the dynamics of the supermembrane is described by a nonrenormalizable sigma model and nonperturbative methods are required for its quantisation~\cite{Horava:1995qa}.\,\footnote{See~\cite{Horava:2008ih, Yan:2022dqk} for a candidate UV-completion of the supermembrane using the theory of quantum critical membrane, whose associated sigma model is power-counting renormalizable and it flows back to the the conventional supermembrane sigma model at low energies, at least at the classical level.} Moreover, due to its intrinsic instability~\cite{deWit:1988xki}, the supermembrane probably does not even exist as a fundamental object and tends to be dissolved into dynamical bits. This begs for a multiparticle state (and membrane field-theoretical) interpretation~\cite{Banks:1996vh}. In this sense, the approaches that we will review in this section are ``exact'' in nature, as they are beyond perturbative string theory. Moreover, to be distinguished from a standard review on M-theory, and in line with the title of this current review, we focus on the recent progress of related string worldsheet approaches~\cite{Gomis:2023eav}, which appear to be surprisingly powerful in terms of mapping out various nonpeturbative corners in string and M-theory. 

Even though the fundamental principles of M-theory remain mysterious, we are still able to learn a great deal about different facets of it via studying certain decoupling limits of string and M-theory.
In such decoupling limits, one zooms in on self-consistent corners where certain states become inaccessible. These corners are often significantly simpler than the original theory.
We have learned that perturbative superstring theories are corners of M-theory, which are achieved by compactifying M-theory over the eleventh dimension with a very small radius, such that the membrane-related excitations are decoupled. Field-theoretical limits of string and M-theory give rise to ten- and eleven-dimensional supergravities, where the excitations associated with any extended objects, such as strings and D-branes, are decoupled. The renowned AdS/CFT correspondence also arises from a decoupling limit, where the bulk IIB supergravity modes are decoupled from the near-horizon AdS geometry, and the latter is then identified with $\mathcal{N} = 4$ super Yang-Mills (SYM) theory on the asymptotic boundary~\cite{Maldacena:1997re, Klebanov:1997kc}. 

In this section, we focus on the decoupling limits in type II superstring theories that are related to Matrix theory~\cite{Banks:1996vh}, which describes M-theory in the Discrete Light Cone Quantisation (DLCQ)~\cite{Susskind:1997cw, Seiberg:1997ad, Sen:1997we}. DLCQ M-theory is usually defined via a limiting procedure: we start with M-theory compactified over a spacelike circle, and then perform an infinite momentum limit along this circle, which can be heuristically thought of as an infinite boost transformation. Effectively, this procedure turns the originally spacelike circle \emph{lightlike}. In this sense, DLCQ M-theory in practice means that we compactify M-theory over a lightlike circle. In the infinite momentum limit, almost all light excitations except the Kaluza-Klein particle states in the lightlike compactification are decoupled. The associated particle dynamics is described by Banks-Fischler-Shenker-Susskind (BFSS) Matrix theory~\cite{Banks:1996vh}, which is a nonrelativistic quantum mechanical system of nine $N \! \times \! N$ matrices. Here, $N$ is associated with the Kaluza-Klein excitation number. In the language of type IIA superstring theory, such a Kaluza-Klein state with momentum number $N$ corresponds to a bound state of $N$ D0-particles. It is conjectured that BFSS Matrix theory may describe the full M-theory at large $N$, where the lightlike circle decompactifies. 

\begin{figure}[t]
	\centering
	\includegraphics[scale=0.48]{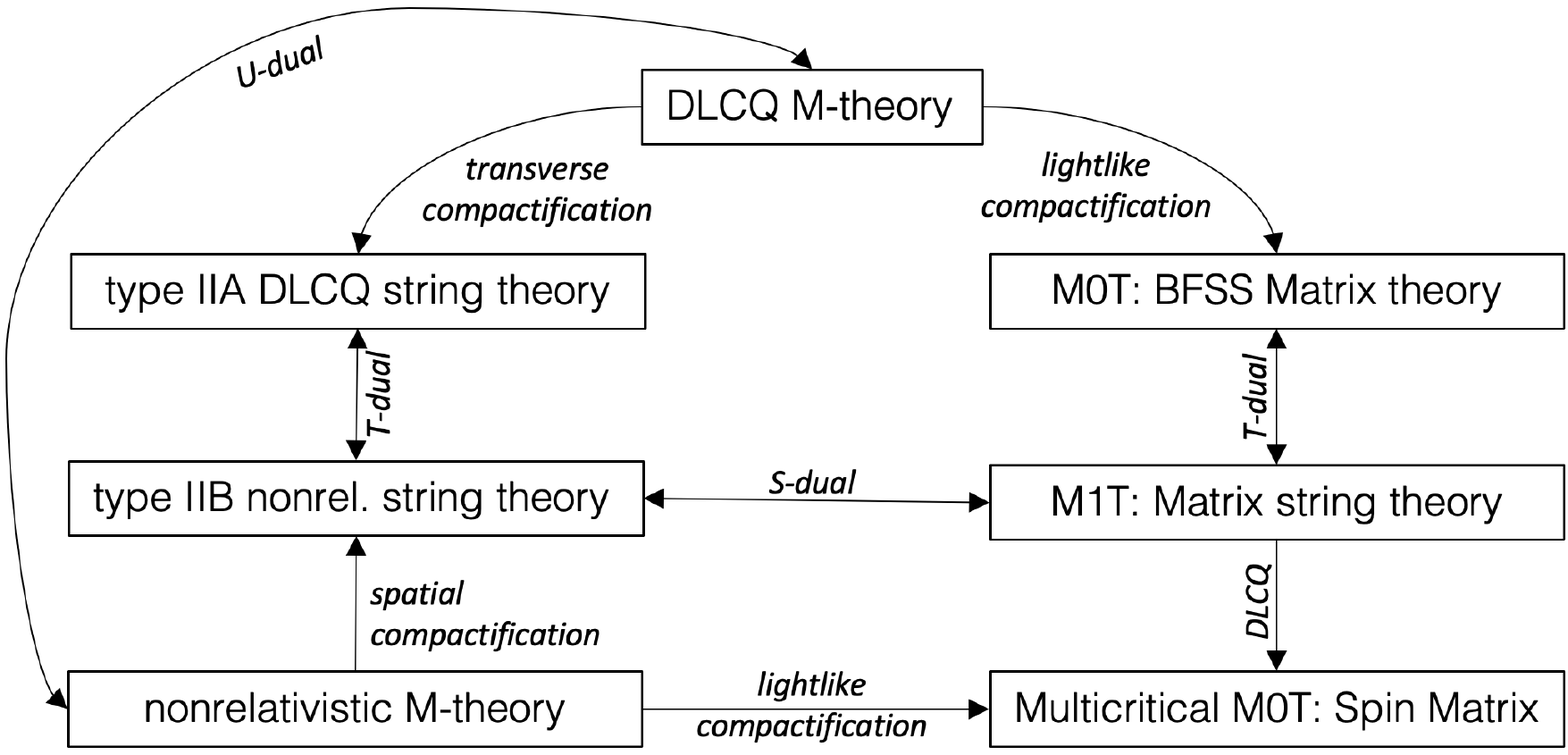}
    \vspace{-1cm}
	\caption{Duality web of decoupling limits that are discussed in the review. Here, ``DLCQ'' stands for Discrete Light Cone Quantisation, \emph{\textit{i.e.}}~the theory is compactified over a lightlike circle, and ``M$p$T'' stands for Matrix $p$-brane Theory.}
	\label{fig:roadmap}
\end{figure}

In a spacetime with a lightlike compactification, the system exhibits nonrelativistic behaviors~\cite{KOGUT197375}. This might be counter-intuitive as one typically expects nonrelativistic physics to appear when the speeds of the physical contents are small,\,\footnote{Along different lines, it is also possible to consider a more general nonrelativistic framework that provides candidate UV-completions of relativistic systems~\cite{Horava:2008ih, Horava:2009uw}.} while at large boost everything moves almost at the speed of light. We dedicate Section~\ref{sec:intro} as a pedagogical review to explain this exotic phenomenon. In Section~\ref{sec:mtmqm}, we review the basic ingredients of Matrix theory and discuss the related decoupling limit in type IIA superstring theory. In Section~\ref{sec:swdw}, we discuss the worldsheet theory of the fundamental string in this IIA corner and review its T-duality transformations, which allow us to make contact with the worldsheet description of Spin Matrix Theory (SMT) in Section~\ref{sec:smtl}. SMT refers to certain near BPS limits of $\mathcal{N} = 4$ SYM and the AdS/CFT correspondence, which lead to a class of integrable models including the Landau-Lifshitz theory~\cite{Harmark:2017rpg}. In Section~\ref{sec:sdnrs}, we review a further extension of the duality web via S-duality and the corner of nonrelativistic string theory, which is a self-contained, perturbative string theory with a Galilean invariant string spectrum~\cite{Gomis:2000bd}.\,\footnote{There is a recent review~\cite{Oling:2022fft} that focuses on the geometric aspects of nonrelativistic string theory, which is complementary to the current review.} Section~\ref{sec:dlcqnst} further elaborates the basics of nonrelativistic string theory, which plays an anchoring role in the duality web and in the original derivation of the SMT string. We will review the relation between the decoupling limit that leads to nonrelativistic string theory and the $T\bar{T}$ deformation in Section~\ref{sec:ttbar}. In Section~\ref{sec:udbdlcq}, we discuss the M-theory uplift of nonrelativistic string theory and its U-dual relation to DLCQ M-theory. See Figure~\ref{fig:roadmap} for a road map of the duality web to be detailed in this section, and see~\cite{Blair:2023noj, Gomis:2023eav} for recent advances and further extensions of this duality web. We will focus on the bosonic contents throughout this section, which is sufficient for conveying the central ideas.

%%%%%%%%%%%%%%%%%%%%%%%
%%%%%%%%%%%%%%%%%%%%%%%
\subsection{Nonrelativistic physics from an infinite boost} \label{sec:intro}
%%%%%%%%%%%%%%%%%%%%%%%
%%%%%%%%%%%%%%%%%%%%%%%

We know that Newtonian physics is valid for describing the macroscopic world where objects move at a relatively much smaller speed compared to the speed of light $c$\,. 
More explicitly, consider a relativistic particle with a rest mass $m_0$ that satisfies the dispersion relation
\be \label{eq:reldr}
	E^2 - |\mathbf{p}|^2 \, c^2 = m^2_0 \,c^4\,,
\ee
where $E$ is the energy and the spatial momentum $\mathbf{p}$ is related to the velocity $\mathbf{u}$ of the particle via
\be \label{eq:momexp}
	\mathbf{p} = \frac{m_0 \, \mathbf{u}}{\sqrt{1 - |\mathbf{u}|^2 / c^2}}\,.
\ee
In the regime where $|\mathbf{u}| / c \ll 1$\,, we find
\be
	E - m_0 \, c^2 = \frac{1}{2} \, m_0 \, |\mathbf{u}|^2 + O \bigl( |\mathbf{u}| / c \bigr)^4\,.
\ee
Define $\varepsilon \sim E - m_0 \, c^2$ to be the effective energy. In the low-speed regime, we obtain the following quadratic dispersion relation:
\be
	\varepsilon = \frac{|\mathbf{p}|^2}{2 \, m}\,.
\ee
This is the standard nonrelativistic limit that we usually consider.

Intriguingly, there is another, less intuitive, way to take a limit of special relativity such that nonrelativistic behaviors arise. Instead of keeping the speed of the massive particle much smaller than the speed of light, we now take an infinite boost limit such that the massive particle moves almost at the speed of light in a particular direction that we call $x^1$ \cite{KOGUT197375}. 
In the infinitely boosted frame, the velocity components of the particle in all the other spatial directions approach zero. This observation can be made manifest in the Lorentz transformation of velocity. Let the massive particle move in the original frame $S$ with a velocity of $\mathbf{u} = (u^{}_1\,, \mathbf{w})$\,, where $u_1$ is along $x^1$\,. Consider a boosted frame $S'$ moving with a relative velocity of $\mathbf{v} = (- v^{}_1\,, \mathbf{0})$ with respect to $S$\,. For an observer in $S'$, the particle moves with a velocity of $\mathbf{u}' = (u'_1\,, \mathbf{w}')$\,, where
\begin{align}
	u'_1 = \frac{u^{}_1 + v^{}_1}{1 + u^{}_1 \, v^{}_1 / c^2}\,, 
		\qquad%
	\mathbf{w}' = \mathbf{w} \, \frac{\sqrt{1 - v^2_1 / c^2}}{1 + u^{}_1 \, v^{}_1 / c^2}\,.
\end{align}
After a large boost in $x^1$\,, with $\delta \equiv 1 - v^{}_1 / c \rightarrow 0^+$\,, we find
$u'_1 = c + O \bigl( \delta \bigr)$ and $\mathbf{w}' = O \bigl( \delta^{1/2} \bigr)$\,.
In $S'$, denote the spatial momentum as $\mathbf{p}' = (p'_1\,, \mathbf{k}')$\,, where $p'_1$ is the momentum in $x^{\prime1}$ and $\mathbf{k}'$ is the transverse momentum.
Using Eq.~\eqref{eq:momexp}\,, we find 
\begin{align}
	p'_1 = m \, c \, \delta^{-1/2} + O(\delta)\,, 
        \qquad%
	\mathbf{k}' = \frac{m_0 \, \mathbf{w}}{\sqrt{1 - |\mathbf{u}|^2 / c^2}} + O\bigl(\delta^2\bigr)\,,
\end{align}
where 
\be
	m = \frac{1}{\sqrt{2}} \frac{\bigl( 1 + u^{}_1 / c \bigr)}{\sqrt{1 - |\mathbf{u}|^2 / c^2}} \, m^{}_0\,.
\ee
Effectively, we are taking an infinite momentum limit in $x^1$ while keeping the momentum components in the other spatial directions finite. Expanding with respect to a small $\delta$\,, %
we find that the relativistic dispersion relation \eqref{eq:reldr} becomes
\be
	\delta^{-1/2} \Bigl( E' - |p'_1| \, c \Bigr) = \frac{1}{2 \, m} \Bigl( |\mathbf{k}'|^2 + m_0^2 \, c^2 \Bigr) + O \bigl( \delta^2 \bigr)\,.
\ee
Further define
the effective energy to be $\varepsilon \sim \delta^{-1/2} \, (E' - |p'_1| \, c)$\,. In the zero $\delta$ limit, the effective dispersion relation is
\be \label{eq:imfee}
	\varepsilon = \frac{1}{2 \, m} \Bigl( |\mathbf{k}'|^2 + m_0^2 \, c^2 \Bigr)\,,
\ee
which, at a fixed $m$\,, can be interpreted as nonrelativistic.

This infinite-momentum frame can be equivalently obtained by going to the lightcone spacetime coordinates $(x^+, \, x^-, \, x^{i})$ with $i = 1, \cdots\,, d-2$\,, which are related to $x = (x^0,\, x^{d-1},\, x^{i})$ via
\be
	x^+ = \frac{1}{\sqrt{2}} \bigl( x^0 + x^{d-1} \bigr)\,,
		\qquad%
	x^- = \frac{1}{\sqrt{2}} \bigl( x^0 - x^{d-1} \bigr)\,.
\ee
Here, $x^0$ is the time direction and $d$ refers to the spacetime dimension.
For simplicity, we take $d=4$\,, such that $i = 1\,, \, 2$\,. From now on, we also take the natural unit such that $c=1$\,. The dispersion relation in the lightcone frame is
$- 2 \, p^{}_- \, p^{}_+ + |\mathbf{k}|^2 + m_0^2 = 0$\,.
In terms of the effective energy $\varepsilon \equiv p^{}_+$\,, we find the following dispersion relation at a fixed lightcone momentum $p^{}_-$\,:
\be
	\varepsilon = \frac{1}{2 \, p^{}_-} \Bigl( |\mathbf{k}|^2 + m_0^2 \Bigr)\,,
\ee
which resembles Eq.~\eqref{eq:imfee}. 
This procedure of fixing the lightcone momentum allows us to restrict to the \emph{Bargmann} subgroup of the Poincar\'{e} group. To make this manifest, we start with the Poincar\'{e} algebra with the non-vanishing commutators
\begin{subequations}
\label{eq:poincare-algebra}
\begin{align}
	\bigl[ {M}^{}_{\mu\nu}\,, {P}^{}_\rho \bigr] & =  \eta^{}_{\nu\rho} \, {P}^{}_\mu - \eta^{}_{\mu\rho} \, {P}^{}_\nu\,, \\[4pt]
	\bigl[ {M}^{}_{\mu\nu}\,, {M}^{}_{\rho\sigma} \bigr] & = \eta^{}_{\mu\sigma} \, {M}_{\nu\rho} + \eta_{\nu\rho} \, M_{\mu\sigma} - \eta_{\mu\rho} \, {M}_{\nu\sigma} - \eta_{\nu\sigma} \, {M}_{\mu\rho} \, .
\end{align}
\end{subequations}
Perform the following changes of variables \cite{PhysRevD.1.2901}:
\begin{subequations}
\begin{align}
	H & = \frac{1}{\sqrt{2}} \, \Bigl( {P}_0 - {P}_3 \Bigr)\,, 
		&%
	G_2 & = \frac{1}{\sqrt{2}} \, \Bigl( {M}_{10} + {M}_{13} \Bigr)\,,
		&%
	S_1 & = \frac{1}{\sqrt{2}} \, \Bigl( {M}_{10} - {M}_{12} \Bigr)\,, \\[2pt]
	N & = \frac{1}{\sqrt{2}} \, \Bigl( {P}_0 + {P}_3 \Bigr)\,,
		&%
	G_3 & = \frac{1}{\sqrt{2}} \, \Bigl( {M}_{20} - {M}_{23} \Bigr)\,,
		&%
	S_3 & = \frac{1}{\sqrt{2}} \, \Bigl( {M}_{20} + {M}_{23} \Bigr)\,,
\end{align}
\end{subequations}
and define $J = {M}_{12}$\,. The generators $H$, $P_i$\,, $G_i$\,, $J$, and $N$ form the Bargmann algebra, defined by the following non-vanishing commutators:
\begin{subequations} \label{eq:bargalg}
\begin{align}
	\bigl[ H\,, \,G^{}_i \bigr] & = P^{}_{i}\,,
		&%
	\bigl[ P^{}_i\,, \, J \bigr] & = \epsilon^{}_{ij} \, P^{}_{j}\,, \\[2pt]
	\bigl[ P^{}_{i}\,, \,G_{j} \bigr] & = \delta^{}_{ij} \, N\,,
		&%
	\bigl[ G^{}_{i}\,, \, J \bigr] & = \epsilon^{}_{ij} \, G^{}_{j}\,. 
\end{align}
\end{subequations}
Here, $H$ is associated with the time translation, $P_{i}$ the spatial translations, $G_{i}$ the Galilei boosts, $J$ the spatial rotation, and $N$ the central extension. The central extension $N$ is associated with the conservation of particle number. 

Going to an infinite momentum frame in both QFTs and string/M-theory has the benefit of zooming in on a much simpler vacuum, which has Galilean symmetry. The associated dynamics can be described by a Hamiltonian for a quantum mechanical system with conserved particle number. 

%%%%%%%%%%%%%%%%%%%%%%%%%%%%%%%%%%%%%%%%%%%%%%%%%%%%%%%
\subsection{M-theory as a Matrix quantum mechanics} \label{sec:mtmqm}
%%%%%%%%%%%%%%%%%%%%%%%%%%%%%%%%%%%%%%%%%%%%%%%%%%%%%%%

The infinite momentum frame turns out to be useful for understanding M-theory and is closely related to its Matrix theory description. We now give a brief introduction to this topic. See \cite{Taylor:2001vb} for a comprehensive review. Part of the current review also follows the textbook~\cite{Kiritsis:2019npv} by Kiritsis. 

%%%%%%%%%%%%%%%%%%%%%%%%%%%%%%%%%%%%%%%%%%%%%%%%%%%%%%%
\subsubsection{Quantisation of the membrane}
%%%%%%%%%%%%%%%%%%%%%%%%%%%%%%%%%%%%%%%%%%%%%%%%%%%%%%%

We will follow the historical route that led to the Matrix theory description of M-theory. It is natural to start with the supermembrane in order to construct such a microscopic interpretation. This is because we would na\"{i}vely expect that the fundamental role played by the string in superstring theory should now be replaced by the supermembrane. For simplicity, we will only focus on the bosonic contents of the supermembrane in the following discussions. We start with the Nambu-Goto action for the membrane in 11D flat spacetime~\cite{Bergshoeff:1987cm},
\be \label{eq:mtba}
	S = - T \int \de^3 \sigma \, \sqrt{-\det \lr \p_\alpha X^\text{M} \, \p_\beta X_\text{M} \rr}\,,
		\qquad%
	\alpha = 0\,,\, 1\,,\, 2\,, 
		\qquad%
	\text{M} = 0\,,\, 1\,,\, \cdots\,, 10\,.
\ee
This action can be rewritten in the Polyakov formulation by introducing an auxiliary worldvolume metric $\gamma^{}_{\alpha\beta}$\,, such that
\be \label{eq:mapol}
	S = - \frac{T}{2} \int \de^3 \sigma \, \sqrt{-\gamma} \, \Bigl( \gamma^{\alpha\beta} \, \p_\alpha X^\text{M} \, \p_\beta X_\text{M} - 1 \Bigr)\,,
\ee
where $\gamma^{\alpha\beta}$ is the inverse of $\gamma_{\alpha\beta}$ and $\gamma = \det ( \gamma_{\alpha\beta} )$\,. 
Varying the action with respect to $\gamma^{}_{\alpha\beta}$ leads to the following equation of motion:
\be
	\Bigl( - \gamma^{\alpha\gamma} \, \gamma^{\beta\delta} + \tfrac{1}{2} \, \gamma^{\alpha\beta} \, \gamma^{\gamma\delta} \Bigr) \, \p_\gamma X^\text{M} \, \p_\delta X_\text{M} - \tfrac{1}{2} \, \gamma^{\alpha\beta} = 0\,,
\ee
which is solved by
$\gamma^{}_{\alpha\beta} = \p_\alpha X^\text{M} \, \p_\beta X_\text{M}$\,.
As in bosonic string theory, we can use the diffeomorphisms to fix some of the components of the worldvolume metric $\gamma_{\alpha\beta}$\,. On the three-dimensional worldvolume, there are three diffeomorphism symmetries, which we use to perform the following gauge fixing:
\be \label{eq:gchoice}
	\gamma^{}_{0a} = 0\,,
		\qquad%
	\gamma^{}_{00} = - \det \Bigl( \p_a X^\text{M} \, \p_b X_\text{M} \Bigr)\,.
\ee
We have taken the split of the worldvolume index $\alpha = (0\,, a)$\,, with $a = 1\,, 2$\,. However, unlike string theory, we are unable to fix all the metric components up to a conformal factor. Namely, $\gamma^{}_{ab}$ is not fixed. The membrane action then becomes
\be \label{eq:magf}
	S = \frac{T}{2} \int \de^3 \sigma \, \Bigl[ \p_\tau X^\text{M} \, \p_\tau X_\text{M} - \det \bigl( \p_\alpha X^\text{M} \, \p_\beta X_\text{M} \bigr) \Bigr]\,.
\ee
Note that $\sigma^\alpha = (\tau\,, \, \sigma^a)$\,. Define the Poisson bracket,
\be
	\lc f\,,\, g \rc \equiv \epsilon^{ab} \, \frac{\p f}{\p \sigma^a} \, \frac{\p g}{\p \sigma^b}\,.
\ee
Then, the gauge-fixed action \eqref{eq:magf} can be rewritten as
\be \label{eq:memaction}
	S = \frac{T}{2} \int \de^3 \sigma \, \Bigl[ \p_\tau X^\text{M} \, \p_\tau X_\text{M} - \tfrac{1}{2} \, \bigl\{ X^\text{M},\, X^\text{N} \bigr\} \, \bigl\{ X_\text{M}\,,\, X_\text{N} \bigr\} \Bigr]\,,
\ee
where we have integrated out $\gamma_{ab}$\,. This needs to be supplemented with the constraints from varying $\gamma^{}_{00}$ and $\gamma^{}_{0a}$ in Eq.~\eqref{eq:mapol},  before fixing them as in Eq.~\eqref{eq:gchoice}. Varying $\gamma^{}_{00}$ gives the constraint
\be \label{eq:const1}
	\p_\tau X^\text{M} \, \p_\tau X_\text{M} = - \frac{1}{2} \, \bigl\{ X^\text{M},\, X^\text{N} \bigr\} \bigl\{ X_\text{M}\,,\, X_\text{N} \bigr\}\,.
\ee
Varying $\gamma^{}_{0a}$ gives
\be \label{eq:secondcon}
	\p_a X^\text{M} \, \Bigl[ \p_1 X_\text{M} \, \p_\tau X^\text{N} \, \p_2 X_\text{N} - \bigl( 1 \leftrightarrow 2\bigr) \Bigr] = 0\,.
\ee
The constraint~\eqref{eq:secondcon} is equivalent to
$\p_\tau X^\text{M} \, \p_a X_\text{M} = 0$\,,  which implies $\lc \p_\tau X^\text{M}, X_\text{M} \rc = 0$\,.

In order to solve the above constraints, we define the lightcone coordinates $X^\pm = X^0 \pm X^{10}$ in the target space and consider the lightcone gauge 
$X^- (\tau\,, \sigma^a) = \tau$\,. 
The constraints now become
\begin{subequations}
\begin{align}
	\p_\tau X^+ & = \frac{1}{4} \, \p_\tau X^{i} \, \p_\tau X^{i} + \frac{1}{8} \, \lc X^{i}, X^{j} \rc \lc X^{i}, X^{j} \rc, \\[2pt]
	\p_a X^+ & = \frac{1}{2} \, \p_\tau X^{i} \, \p_a X^{i}, \\[4pt]
	0 & = \lc \p_\tau X^{i}, X^{j} \rc,
\end{align}
\end{subequations}
where $i = 1\,,\, \cdots, \, 9$\,. The lightlike momentum $p = \frac{1}{2} \lr p^{}_0 + p^{}_1 \rr$ is then fixed to be 
$p = V \, T$\,,
with $V = 2 \pi$ the spatial volume of the membrane.
The lightcone Hamiltonian is
\be \label{eq:lchmltn}
	H = \frac{T}{2} \int \de^2 \sigma \lr \p_\tau X^{i} \, \p_\tau X^{i} + \frac{1}{2} \lc X^{i}, X^{j} \rc \lc X^{i}, X^{j} \rc \rr.
\ee

Even though the constraints in Eqs.~\eqref{eq:const1} can be solved by going to lightcone gauge just like in string theory, the equations of motion for the membrane are not linear as they are for the string. Namely, the equations of motion from varying $X^{i}$ in the membrane action \eqref{eq:memaction} are
\be \label{eq:mbeom}
	\ddot{X}^{i} = \Bigl\{ \bigl\{ X^{i}, X^{j} \bigl\}\,, X_{j} \Bigr\}\,,
\ee
which is nonlinear and thus hard to solve. Unlike string theory, it is still very difficult to quantise the membrane in lightcone gauge.

It is possible to quantise a regularized version of the membrane by discretizing the two-dimensional spatial manifold of the worldvolume~\cite{Goldstone:1982, hoppe1987phd}. For simplicity, we assume that the membrane surface is a sphere, in which case the worldvolume has the topology $\mathbb{R} \times S^2$\,. We then take both $\sigma^1$ and $\sigma^2$ to have $N$ lattice points. As a result, the embedding coordinates $X^{i}$ now become Hermitian $N \times N$ matrices, 
\be
	X^{i} \bigl( \tau\,, \, \sigma^1, \, \sigma^2 \bigr) \rightarrow X^{i}_\text{I\,J} (\tau)\,,
		\qquad%
	\text{I},\, \text{J} = 1, \cdots, N\,. 
\ee
Meanwhile, the Poisson brackets become matrix commutators,
\be
	\bigl\{ f, \, g \bigr\} \rightarrow - \frac{i}{2} \, \bigl[ f, \, g \bigr]\,, 
\ee
and the integral over the membrane surface is replaced with
\be
	\int \de^2 \sigma \, F(\tau, \sigma^a) \rightarrow \frac{V}{N} \, \sum_{\text{I}=1}^N F^{}_\text{I\,I} (\tau)\,.
\ee
The matrix Hamiltonian associated with Eq.~\eqref{eq:lchmltn} is~\cite{deWit:1988wri} (see also~\cite{baake1985fierz, flume1985quantum, Claudson:1984th})
\be \label{eq:mthmil}
	H = \frac{T}{8} \, \tr \! \lr \p_\tau X^{i} (\tau) \, \p_\tau X^{i} (\tau) - \frac{1}{2} \, \bigl[ X^{i} (\tau)\,, X^{j} (\tau) \bigr] \, \bigl[ X^{i} (\tau)\,, X^{j} (\tau) \bigr] \rr,
\ee
which is supplemented by the constraint 
$\bigl[ \p_\tau X^{i}, \, X^{i} \bigr] = 0$\,.
Moreover, the nonlinear equation of motion \eqref{eq:mbeom} gives rise to the matrix equations of motion,
\be \label{eq:meom}
	\p_\tau^2 X^{i} = - \Bigl[ \bigl[ X^{i}, X^{j} \bigr]\,, X^{j} \Bigr].
\ee
This is a quantum mechanical system described by nine $N \times N$ matrices. This theory has a symmetry group $U(N)$\,, under which the matrices $X^{i}$ are in the adjoint representation. This is true even when the membrane surface is a general Riemann surface of any genus. The $U(N)$ symmetry can be traced to the symmetry group of the so-called \emph{area-preserving diffeomorphisms} satisfied by the membrane. The fact that the $U(N)$ Matrix theory is not associated with a particular topology implies that it can probably approximate the system of multiple membranes with arbitrary topologies. Even though we started with the first-quantised theory of the membrane, it seems that we are led to a Matrix-theory description of the second-quantised theory! In retrospective, this is somewhat expected: the continuum limit of the discretized membrane is not guaranteed to only give a single first-quantised membrane, but instead it leads to a multi-membrane scenario where a second quantisation interpretation is required. 

%%%%%%%%%%%%%%%%%%%%%%%%%%%%%%%%%%%%%%%%%%%%%%%%%%%%%%%
\subsubsection{The Matrix theory interpretation} \label{sec:mti}
%%%%%%%%%%%%%%%%%%%%%%%%%%%%%%%%%%%%%%%%%%%%%%%%%%%%%%%

In string theory, it is important that there is a discrete spectrum of states, resulting in massless states in the string Hilbert space that are in one-to-one correspondence to particle states in the target space. This property is key to the interpretation that there is a massless graviton state separated from massive excitations. 

In contrast, the membrane spectrum turns out to be continuous and is plagued by intrinsic instability~\cite{deWit:1988xki}. This can be seen already at the classical level. Let us create a spike that is roughly cylindrical on the membrane surface, which costs energy $\sim T \, A \, L$\,, where $A$ is the area of the cross section of the cylinder and $L$ is the length. Recall that $T$ is the membrane tension. Consider a very narrow and long spike, \emph{\textit{i.e.}}~with a small $A$ but a large $L$\,. The energy required for creating such a configuration can be arcitrarily low. This suggests that a membrane tends to have many long-spike fluctuations, which makes it unlikely to define the membrane as a localized object. 

The instability of the membrane is associated with a set of flat directions in Matrix theory. It is cured in the quantum bosonic Matrix theory, where an effective confining potential emerges and the flat directions of the classical theory are removed. However, such instability reappears in the supersymmetric theory, indicating that there is \emph{no} simple interpretation of the supermembrane states as a discrete particle-like spectrum.   

The problem of continuous spectrum finds its resolution in the Banks-Fischler-Shenker-Susskind (BFSS) proposal of the Matrix theory conjecture \cite{Banks:1996vh}, which suggests that the large $N$ limit of supersymmetric Matrix theory might describe the full M-theory in lightcone coordinates. The BFSS argument involves initially compactifying M-theory over a spacelike circle, which leads to type IIA superstring theory in ten dimensions. The Kaluza-Klein momentum states in the compact eleventh dimension correspond to the D0-branes in the IIA theory, with mass
\be
	\tau^{}_0 = \frac{1}{g^{}_s \, \alpha'{}^{1/2}}\,,
\ee
where $g^{}_s$ is the string coupling and $\alpha'$ the Regge slope. For any $N$ D0-branes there is a ultrashort multiplet of bound states that has mass $N \, \tau^{}_0$\,. 
In the context of M-theory, this mass formula can be interpreted as the Kaluza-Klein momentum in the compact direction, with the radius of the circle being
\be \label{eq:rogsa}
	R^{}_0 = g^{}_s \, \alpha'{}^{1/2}\,,
\ee
which implies that a single D0-brane state corresponds to a graviton in M-theory with a unit longitudinal momentum.
In the strongly coupled limit $g^{}_s \rightarrow \infty$\,, the M-theory circle decompactifies.

Next, we take a further limit where the momentum in the compact circle is taken to be infinite, such that all light excitations except the D0-branes are suppressed. Therefore, the Matrix theory Hamiltonian \eqref{eq:mthmil} now gains an alternative interpretation as the low-energy effective action of D0-branes in the infinite momentum limit. We will discuss in the next subsection how to make this description more precise, using the discrete lightcone quantisation (DLCQ) of M-theory.

We now discuss the implications of the D0-brane interpretation for the continuous spectrum of the membrane. Consider a set of nine block-diagonal matrices,
\be \label{eq:bdxap}
	X^{i} = 
	\begin{pmatrix}
		x^{i}_1 &\,\, 0 &\,\, 0 &\,\, 0\\[2pt]
		0 &\,\, x^{i}_2 &\,\, 0 &\,\, 0 \\[2pt]
		0 & \,\, 0 &\,\, \cdots &\,\, 0 \\[2pt]
		0 &\,\, 0 &\,\, 0 & \,\, x^{i}_k
	\end{pmatrix}\,,
\ee
where $x^{i}_r$ are $m_r \times m_r$ matrices with $\sum_r m_r = N$\,. Plugging Eq.~\eqref{eq:bdxap} into the Matrix equation of motion \eqref{eq:meom}, we find the following decoupled equations:
\be
	\p_\tau^2 x_r^{i} = - \Bigl[ \bigl[ x_r^{i}\,, \, x_r^{j} \bigr]\,, \, x_r^{j} \Bigr]\,,
		\qquad%
	r = 1\,, \, \cdots, \, k\,.
\ee 
These classically independent equations of motion govern the dynamics of $k$ Matrix theory objects. For example, these can be bound D0-brane states in type IIA superstring theory that correspond to (super)gravitons in M-theory.  
In this case, the matrix configuration in Eq.~\eqref{eq:bdxap} describes a multi-particle state of independent, classical gravitons in M-theory, while the off-diagonal entries encode interactions between these gravitons. From the string theoretical perspective, these interactions are mediated by the ground-state open strings connecting the D0-branes. Given that Matrix theory should describe a second quantised theory, the puzzle of the continuous spectrum is resolved. The long, narrow spikes that seemed to cause instability of the membrane correspond to a configuration in the target space with multiple macroscopic membranes connected by narrow tubes of infinitesimally small energies.
The above observation implies that Matrix theory is already beyond perturbative string theory, since the latter describes a first-quantised theory in the target space. 

%%%%%%%%%%%%%%%%%%%%%%%%%%%%%%%%%%%%%%%%%%%%%%%%%%%%%%%
\subsubsection{M-theory in the DLCQ} \label{sec:mtitdlcq}
%%%%%%%%%%%%%%%%%%%%%%%%%%%%%%%%%%%%%%%%%%%%%%%%%%%%%%%

We now study further the compactification of M-theory on a circle, which gives rise to the type IIA superstring description. We will provide detailed formulae that further support the observations in Section~\ref{sec:mti} and review how M-theory in the discrete lightcone quantisation (DLCQ), \emph{\textit{i.e.}}~M-theory in spacetime with a lightlike compactification, corresponds to BFSS Matrix theory~\cite{Susskind:1997cw, Seiberg:1997ad, Sen:1997we}. 

The desired lightlike circle is commonly defined via a subtle infinite boost limit of a spacelike circle. Boosting along a compact direction might be somewhat deceptive as the compactification already breaks Lorentz invariance, but this way of thinking at least makes some sense in the case where the circle is very large, especially bearing in mind that we are ultimately interested in the decompactification limit where M-theory in eleven dimensions is recovered. Even though we are really dealing with a decoupling limit, this heuristic argument using the infinite boost still provides valuable intuitions. 

To be concrete, we denote the target-space coordinates as $X^\text{M}$, $\text{M} = 0\,, 1\,, \cdots, 10$ and consider a large boost transformation along $X^{10}$, \emph{\textit{i.e.}}, 
\begin{align} \label{eq:lrzbst}
	X'{}^0 & = \gamma \, \bigl( X^0 + v \, X^{10} \bigr)\,, 
        \qquad%
	X'{}^{10} = \gamma \, \bigl( X^{10} + v \, X^{0} \bigr)\,,
\end{align}
together with the periodic boundary condition,
$X^{10} \sim X^{10} + 2 \pi \, R_0$\,,
which implies that the $x^{10}$ direction is compactified over a circle of radius $R_0$\,. Here, $\gamma = 1 / \sqrt{1 - v^2}$ is the Lorentz factor. We have set $c=1$\,. Define the change of variables in the boosted frame,
\begin{align} \label{eq:lccoords}
    X'{}^+ & = X'{}^0 + X'{}^{10}\,, 
		\qquad%
	X'{}^- = X'{}^0 - v \, X'{}^{10}\,,
\end{align}
such that $X'{}^- = X^0 / \gamma$
does not contain any periodicity and therefore can be identified with the effective time direction in a lightlike coordinate system. Indeed, in the ``boosted frame'' with a large $\gamma$\,, we find
\be \label{eq:xpmrs}
	X'{}^+ = 2 \, \gamma \, \bigl( X^0 + X^1 \bigr) + O (\gamma^{-1})\,,
		\qquad%
	X'{}^- = \frac{X^0}{\gamma}\,, 
\ee
and
the boundary conditions are
$X'{}^+ \! \sim \! X'{}^+ + 2 \pi \, \bigl( 2 \gamma R_0 \bigr) + O(\gamma^{-1})$ and $X'{}^- \! \sim \! X'{}^-$\,. 
Also note that, in the infinitely boosted frame where $v$ is almost the speed of light, Eq.~\eqref{eq:lccoords} does give the lightcone coordinates.
In the double scaling limit~\cite{Seiberg:1997ad}, 
\be \label{eq:dsl}
	\gamma \rightarrow \infty\,,
		\qquad%
	R_0 \rightarrow 0\,,
		\qquad%
	R \equiv 2 \, \gamma \, R_0 \text{ is fixed},
\ee
we find 
\be
    X'{}^+ \sim X'{}^+ + 2 \pi R\,,
        \qquad%
    X'{}^- \sim X'{}^-\,,
\ee
which defines a lightlike circle. The above procedure defines DLCQ M-theory. 

We have discussed that BFSS Matrix theory describes certain D0-brane excitations in type IIA superstring theory. We have learned that DLCQ M-theory is defined via the double scaling limit~\eqref{eq:dsl}. From the string theoretical perspective, this double scaling limit must correspond to a limit of the IIA theory. The resulting string theory in this limit is precisely the one that arises from compactifying M-theory over the lightlike circle. %\SIB{<- strange sentence?} 
In order to understand how this string theory looks like, we start with M-theory and consider the probe membrane described by the action~\eqref{eq:mtba}. 
Wrap the membrane around the target space lightlike circle, such that~$X^+ = \sigma^2$\,. It is not difficult to show that the membrane action~\eqref{eq:mtba} now reduces to the following string worldsheet action~\cite{Gomis:2023eav}:\,\footnote{See also~\cite{Batlle:2016iel} for related studies of the same Nambu-Goto action. The compactification of the M2-brane over a lightlike circle has also been studied in~\cite{Kluson:2021pux}, where Eq.~\eqref{eq:nvs} appeared as an intermediate step in Eq.~(3.3). However, the last equality in Eq.~(3.3) does not hold unless it is modified to be
\[
    \left| 
    \begin{tabular}{ll}
        \!\!0 &\!\! $\tau^{}_\beta$\!\!\! \\[4pt]
        \!\!$\tau^{}_\alpha$ &\!\! $h^{}_{\alpha\beta}$\!\!\!
    \end{tabular}
    \right| = \det \bigl( - \tau^{}_\alpha \, \tau^{}_\beta + h^{}_{\alpha\beta} \bigr) - \det \bigl( h^{}_{\alpha\beta} \bigr)\,.
\]}
\begin{align} \label{eq:nvs}
    S_\text{string} & = - T \int \de^2 \sigma \, \sqrt{-\det \! 
    \begin{pmatrix}
        0 &\, \p_\beta X^0 \\[4pt]
        \p_\alpha X^0 &\, \p_\alpha X^i \, \p_\beta X^i 
    \end{pmatrix}}\,,
        \quad%
    \alpha = 0\,, \, 1\,,
        \quad%
    i = 1\,, \cdots, \, 9\,.
\end{align}
Here, $X^0 = - X^-$ and $X^i = x^i$. We also defined $\sigma^\alpha = (\tau, \sigma)$\,. The string action~\eqref{eq:nvs} also arises from a nonrelativistic limit of the conventional Nambu-Goto action
\be \label{eq:sng}
    S_\text{NG} = - T \int \de^2 \sigma \sqrt{-\det \Bigl( \p^{}_\alpha X^\mu_{\phantom{h}} \, \p^{}_\beta X^{}_\mu \Bigr)}\,,
        \qquad%
    \mu = 0\,, \cdots, \, 9\,.
\ee
Here, $T = (2\pi\alpha')^{-1}$ is the string tension. Perform the rescaling
\be \label{eq:scl}
    X^0 \rightarrow \sqrt{\omega} \, X^0,
        \qquad%
    X^i \rightarrow \frac{1}{\sqrt{\omega}} \, X^i,
        \qquad%
    \alpha' \rightarrow \alpha'
\ee
in Eq.~\eqref{eq:sng}, we find that the action~\eqref{eq:nvs} is recovered at $\omega \rightarrow \infty$\,. Note that, up to subleading orders in large $\omega$\,, we have $\omega \propto \gamma$\,, where $\gamma$ is the Lorentz factor introduced in Eq.~\eqref{eq:lrzbst}. Therefore, the infinite-boost limit $\gamma \rightarrow \infty$ in M-theory translates to the $\omega \rightarrow \infty$ limit in type IIA superstring theory. 

We now discuss how BFSS Matrix theory arises from the $\omega \rightarrow \infty$ limit of the D0-branes. For pedagogical reason, we first focus on a single D0-brane, so we expect to reproduce the free action in BFSS Matrix theory~\eqref{eq:mthmil}, with the action
\be \label{eq:fbfss}
    S_\text{BFSS} \propto \int \de^2 \tau \, \p_\tau X^i \, \p_\tau X^i\,. 
\ee
Before the $\omega \rightarrow \infty$ limit is performed, the effective action for a single D0-brane describes a relativistic particle,
\be \label{eq:sdz}
    S_\text{D0} = - \frac{1}{g^{}_s \, \alpha'{}^{1/2}} \int \de\tau \, \sqrt{- \p_\tau X^\mu_{\phantom{\dagger}} \, \p_\tau X_\mu}\,.
\ee
Here, $g_s$ is the string coupling. 
Plug Eq.~\eqref{eq:scl} into the D0-brane action~\eqref{eq:sdz}, and expand with respect to a large $\omega$\,, we find
\be
    S_\text{D0} = \frac{\omega^{-3/2}}{g^{}_s \, \alpha'^{1/2}} \int \de\tau \, \Bigl[ - \omega^2 + \tfrac{1}{2} \, \p_\tau X^i \, \p_\tau X^i + O (\omega^{-2}) \Bigr]\,. 
\ee
We have chosen the static gauge $X^0 = \tau$\,.
In order for the desired term $\p_\tau X^\mu \, \p_\tau X_\mu$ to survive at $\omega \rightarrow \infty$\,, we are required to rescale the string coupling $g_s$ as
\be \label{eq:rsgs}
    g_s \rightarrow \omega^{-3/2} \, g_s\,. 
\ee
However, there still remains a divergent term in Eq.~\eqref{eq:sdz}, which can be canceled upon the introduction of the Ramond-Ramond (RR) 1-form. The RR term in the D0-brane action is in form the same as an electromagnetic gauge potential coupled to the D0-particle,
\be \label{eq:dzcs}
    S_\text{CS} = \int \de\tau \, \p_\tau X^\mu \, C_\mu\,.
\ee
Setting 
\be \label{eq:cc1rr}
    C^{(1)} = \omega^2 \, g^{-1}_s \, dX^0\,, 
\ee
we find that the $\omega$ divergence in $S_\text{D0} + S_\text{CS}$ cancel and the $\omega \rightarrow \infty$ limit leads to a finite action that recovers the free part~\eqref{eq:fbfss} of BFSS Matrix theory. 

When a stack of D0-branes are considered, the same limiting prescription gives rise to the full (bosonic) BFSS Matrix theory~\eqref{eq:mthmil}~\cite{Blair:2023noj}. The bosonic sector of the non-abelian action describing $N$ coinciding D0-branes is given by~\cite{Myers:1999ps}
\be \label{eq:nonab}
    S_\text{D0} = - \frac{1}{g^{}_s \, \alpha'{}^{1/2}} \int \de \tau \, \tr \sqrt{\Bigl( - \p_\tau X^\mu_{\phantom{\dagger}} \, \p_\tau X_\mu \Bigr) \, \det \Bigl( \delta^i_j + i \, \bigl[ X^i \,, \, X_j \bigr] \Bigr)}\,, 
\ee
where $X^i$ are scalars in the adjoint representation of $U(N)$\,. The appearance of the commutator $[X^i, X^j]$ can be understood from compactifying ten-dimensional $\mathcal{N} = 1$ super Yang-Mills theory all the way to (0+1)-dimension. Plugging Eqs.~\eqref{eq:scl} and~\eqref{eq:rsgs} into the non-abelian action~\eqref{eq:nonab}, and further taking into account the $\omega^2$ term from the RR 1-form in Eq.~\eqref{eq:cc1rr},  the action principle associated with the Hamiltonian~\eqref{eq:mthmil} describing the bosonic sector of BFSS Matrix theory is recovered in the $\omega \rightarrow \infty$ limit. 

The above $\omega \rightarrow \infty$ limit defines a decoupling limit of type IIA superstring theory, which we refer to as \emph{Matrix 0-brane theory} (M0T). In M0T, the D0-branes are the light excitations, which coexist with a whole range of extended objects, such as all the other D$p$-branes and the fundamental string. Importantly, this decoupling limit is associated with a nontrivial RR 1-form~\eqref{eq:cc1rr} that is tuned to its critical value, such that it cancels the background D0-brane tension. We therefore refer to such an $\omega \rightarrow \infty$ limit as the critical RR 1-form limit~\cite{Blair:2023noj, Gomis:2023eav}. See also a historical account of related limits in \textit{e.g.}~\cite{Gopakumar:2000ep,Harmark:2000ff} focusing on the ``open string'' decoupling limits and \cite{Gomis:2000bd, Danielsson:2000gi} for the ``closed string'' limits without the necessity of introducing any D-branes. 

%%%%%%%%%%%%%%%%%%%%%%%%%%%%%%%%%%%%%%%%%%%%%%%%%%%%%%%
\subsection{Worldsheet formalism and duality web from two DLCQs} \label{sec:swdw}
%%%%%%%%%%%%%%%%%%%%%%%%%%%%%%%%%%%%%%%%%%%%%%%%%%%%%%%

We now turn our attention to the string worldsheet under the critical RR 1-form limit and study its T-dual properties. We have derived the associated Nambu-Goto action~\eqref{eq:sng} from compactifying the M2-brane over a lightlike circle. We start with reviewing the Polyakov formulation of the same fundamental string~\cite{Gomis:2023eav}, which will prepare us to consider the T-duality transformations later in this subsection. 

%%%%%%%%%%%%%%%%%%%%%%%%%%%%%%%%%%%%%%%%%%%%%%%%%%%%%%%
\subsubsection{Polyakov action}
%%%%%%%%%%%%%%%%%%%%%%%%%%%%%%%%%%%%%%%%%%%%%%%%%%%%%%%

The Polyakov formulation for the Nambu-Goto action~\eqref{eq:sng} is given by~\cite{Gomis:2023eav}:\,\footnote{See also~\cite{Albrychiewicz:2023ngk} where an in-form similar string sigma model is considered in the context of tropological sigma models.}
\be \label{eq:mztfscw}
	S_\text{P} = - \frac{T}{2} \int \de^2 \sigma \, e \, \biggl[ \Bigl( e^\alpha{}^{}_0 \, \p^{}_\alpha X^0 \Bigr)^2 + e^\alpha{}^{}_1 \, e^\beta{}^{}_1 \, \p_\alpha X^{i} \, \p_\beta X^{i} + \lambda \, e^\alpha{}^{}_1 \, \p_\alpha X^0 \biggr]\,.
\ee
We have introduced the zweibein field $e_\alpha{}^a$ with $a = 0\,, 1$ being the frame index, together with its inverse $e^\alpha{}_a = e^{-1} \, \epsilon^{\alpha\beta} \, e^{}_\beta{}^b \, \epsilon^{}_{ab}$ and $e = \epsilon^{\alpha\beta} \, e^{}_\alpha{}^0 \, e^{}_\beta{}^1$\,.
Here, $\lambda$ is a Lagrange multiplier that imposes the constraint $e^\alpha{}_1 \, \p_\alpha X^0 = 0$\,, which is solved by $e^\alpha{}_1 = \Omega \, \epsilon^{\alpha\beta} \p_\beta X^0$\,. Plugging this solution back into Eq.~\eqref{eq:mztfscw}, we find
\be \label{eq:sptng}
	S_\text{P} \rightarrow \frac{T}{2} \int \de^2 \sigma \, \ls \frac{1}{\Omega^2 \, e} - \Omega^2 \, e \, \det \! 		\begin{pmatrix}
        		0 &\, \p_\beta X^0 \\[4pt]
        		\p_\alpha X^0 &\, \p_\alpha X^i \, \p_\beta X^j 
    \end{pmatrix} \rs\,,
\ee
where now $e = \bigl( \Omega \, e^\alpha{}_0 \, \p_\alpha X^0 \bigr)^{-1}$\,. Integrating out $e$ in Eq.~\eqref{eq:sptng} leads to the Nambu-Goto action~\eqref{eq:nvs}. Eq.~\eqref{eq:mztfscw} describes the fundamental string theory in M0T. 

In flat gauge with $e_\alpha{}^a = \delta_\alpha^a$\,, the Polyakov action~\eqref{eq:mztfscw} becomes
\be \label{eq:mzts}
	S_\text{P} = - \frac{T}{2} \int \de^2 \sigma \Bigl( \p_\tau X^0 \, \p_\tau X^0 + \p_\sigma X^{i} \, \p_\sigma X^{i} + \lambda \, \p_\sigma X^0 \Bigr)\,.
\ee
This action exhibits non-Lorentzian structures in both the target space and worldsheet: it is invariant under the target space Galilean boost transformation
\be
	\delta_\text{G} X^0 = 0\,,
		\qquad%
	\delta_\text{G} X^{i} = \Lambda^{i} \, X^0,
		\qquad%
	\delta_\text{G} \lambda = - 2 \, \Lambda_{i} \, \p_\sigma X^{i},
\ee
as well as the worldsheet Galilean boost transformation\,\footnote{Note that the worldsheet is Carrollian instead of Galilean in~\cite{Gomis:2023eav}. In Carrollian spacetime, it is the space instead of the time that is absolute. The Carrollian and Galilean worldsheets are formally equivalent to each other in Wick rotated time. We stick to the Galilean worldsheet in order to be aligned with the convention in Section~\ref{sec:smtl}.}
\be
	\delta_\text{g} \tau = 0\,,
		\qquad%
	\delta_\text{g} \sigma = v \, \tau\,,
		\qquad%
	\delta_\text{g} \lambda = 2 \, v \, \p_\tau X^0\,.
\ee
It is expected that the target space develops a non-Lorentzian behavior, as the space and time are rescaled differently in Eq.~\eqref{eq:scl}, followed by setting $\omega$ to infinity. However, it might be surprising that the worldsheet geometry also becomes non-Lorentzian. In fact, in order to derive Eq.~\eqref{eq:mzts} from a decoupling limit of the conventional string theory action
\be \label{eq:relsa}
	S = - \frac{T}{2} \int \de^2 \sigma \, \p_\alpha X^\mu \, \p^\alpha X_\mu\,,
\ee
we have to supplement the anisotropic rescalings of the embedding coordinates $X^\mu$ in Eq.~\eqref{eq:scl} with the following rescaling of the worldsheet coordinates:
\be \label{eq:tos}
	\tau \rightarrow \omega \, \tau\,,
		\qquad%
	\sigma \rightarrow \sigma\,.
\ee
Under these rescalings of both the embedding coordinates and worldsheet coordinates, we expand Eq.~\eqref{eq:relsa} with respect to a large $\omega$ as
\be
	S = - \frac{T}{2} \int \de^2 \sigma \, \Bigl[ - \omega^2 \, \p_\sigma X^0 \, \p_\sigma X^0 + \p_\tau X^0 \, \p_\tau X^0 + \p_\sigma X^i \, \p_\sigma X^i + O(\omega^{-2}) \Bigr]\,.
\ee
Rewrite the $\omega^2$ term using the Hubbard-Stratonovich transformation,
\be
	- \omega^2 \, \p_\sigma X^0 \, \p_\sigma X^0 \rightarrow \lambda \, \p_\sigma X^0 + \frac{\lambda^2}{4 \, \omega^2} \,,
\ee
Eq.~\eqref{eq:mzts} is recovered at $\omega \rightarrow \infty$\,. It is shown in~\cite{Gomis:2023eav} that, upon Wick rotation, the topology of the nonrelativistic worldsheet is described by the nodal Riemann spheres, akin to the case in ambitwistor string theory~\cite{Geyer:2015bja, Geyer:2018xwu}. This relation to ambitwistor string theory~\cite{Mason:2013sva} is expected: Matrix 0-brane theory is dual to tensionless string theory~\cite{Lindstrom:1990qb, Isberg:1993av} via a timelike T-duality transformation~\cite{Blair:2023noj, Gomis:2023eav},\,\footnote{It is also shown in \cite{Blair:2023noj, Gomis:2023eav} that further dualising tensionless string theory leads to string theory coupled to Carroll-like geometries in the target space.} where the latter is closely related to ambitwistor string theory when a ``flipped'' vacuum is chosen~\cite{Siegel:2015axg, Casali:2016atr, Bagchi:2020fpr}.   

In the rest of this section, we will review different T-duality transformations of the string action~\eqref{eq:mzts}. As a preparation, we start with a review of conventional T-duality transformations in string theory. 

\subsubsection{Introduction to T-duality}

For an extensive review of T-duality, see \textit{e.g.}~\cite{Alvarez:1994dn}. 
Consider conventional string theory in $d$-dimensional flat spacetime. In conformal gauge, the relevant string action is given by Eq.~\eqref{eq:relsa}. 
Compactify the spatial $X^1$ direction over a circle of radius $R$  
by imposing the boundary condition
\be \label{eq:bdrycond}
	X^1 (\sigma + 2\pi) = X^1 (\sigma) +2 \pi R \, w \,,
		\qquad%
	w \in \mathbb{Z}\,,
\ee 
where $w$ describes how many times that the string wraps around the $X^1$ direction.
To T-dualise $X^1$, we start with rewriting the action~\eqref{eq:relsa} equivalently as
\be \label{eq:paction}
	S_\text{parent} = - \frac{T}{2} \int \de^2 \sigma \, \Bigl( \p_\alpha X^m \, \p^\alpha X_m + V^\alpha \, V_\alpha + 2 \, \tilde{X}^1 \, \epsilon^{\alpha\beta} \, \p_\alpha V_\beta \Bigr)\,,
\ee
where $m = (0\,, \, 2\,, \, \cdots\,, \, d-1)$\,. Integrating out the Lagrange multiplier $X^1$ in the associated path integral leads to the constraint, $\epsilon^{\alpha\beta} \, \p_\alpha V_\beta = 0$\,,
which, by Poincar\'{e}'s lemma, can be solved locally by
\be \label{eq:valapha}
	V_\alpha = \p_\alpha \Theta\,.
\ee
Identifying $\Theta = Y^1$\,, we recover the original action~\eqref{eq:relsa} from the ``parent'' action \eqref{eq:paction}. 
Instead, in order to derive the T-dual action, we integrate out $V_\alpha$ in Eq.~\eqref{eq:paction}. Varying the action \eqref{eq:paction} with respect to $V_\alpha$ gives the equation of motion,
\be \label{eq:solv}
	V^{}_\alpha = - \epsilon^{}_\alpha{}^\beta \, \p^{}_\beta \tilde{X}^1\,.
\ee
Plugging Eq.~\eqref{eq:solv} into Eq.~\eqref{eq:paction} gives the dual action,
\be
	S_\text{dual} = - \frac{T}{2} \int \de^2 \sigma \, \Bigl( \p_\alpha \tilde{X}^1 \, \p^\alpha \tilde{X}^1 + \p_\alpha X^m \, \p^\alpha X_m \Bigr)\,. 
\ee
This action describes another string theory with the dual worldsheet field $(\tilde{X}^1, X^m)$\,. 

The worldsheet field $X^1$ in Eq.~\eqref{eq:relsa} satisfies the equation of motion, $\p^\alpha \p_\alpha X^1 = 0$\,.
Taking into account the boundary condition \eqref{eq:bdrycond}, this equation of motion is solved by
\be \label{eq:x1me}
	X^1 = x^1 + \alpha' \, p^{}_1 \, \tau + R \, w \, \sigma + \text{oscillations}.
\ee
For the physical vertex operator 
$\exp \bigl( i \, p^{}_\mu X^\mu \bigr)$ 
to be single-valued, the boundary condition \eqref{eq:bdrycond} implies (see \textit{e.g.}~\cite{Polchinski:1998rq})
\be
	\exp \Bigl( 2 \pi i \, R \, w \, p^{}_1 \Bigr) = 1
		\qquad \implies \qquad%
	p^{}_1 = \frac{n}{R}\,,
		\quad%
	n \in \mathbb{Z}\,.
\ee
Here, $n$ is the Kaluza-Klein (KK) number. Similarly, we have the following generic solution to $\tilde{X}^1$\,:
\be \label{eq:yme}
	\tilde{X}^1 = \tilde{x}^1 + \alpha' \, \tilde{p}^{}_1 \, \tau + \tilde{R} \, \tilde{w} \, \sigma + \text{oscillations}.
\ee
We have assumed that $\tilde{X}^1$ is compactified over the dual circle with a radius $\tilde{R}$\,. Over this compactification, we have
\be
	\tilde{p}^{}_1 = \frac{\tilde{n}}{\tilde{R}}\,.
\ee
Using Eq.~\eqref{eq:valapha} with $\Theta = X^1$ together with Eq.~\eqref{eq:solv}, we find the duality mapping,
\be \label{eq:dm}
	\p^{}_\alpha X^1 = - \epsilon^{}_\alpha{}^\beta \, \p_\beta \tilde{X}^1\,,
\ee
\emph{\textit{i.e.}}, $\p_\tau X^1 = \p_\sigma \tilde{X}^1$ and $\p_\sigma X^1 = \p_\tau \tilde{X}^1$. Plugging Eqs.~\eqref{eq:x1me} and \eqref{eq:yme} into Eq.~\eqref{eq:dm} gives
$\alpha' \, R^{\,\text{-}1} \, n = \tilde{R} \, \tilde{w}$ and $R \, w = \alpha' \, \tilde{R}^{\,\text{-}1} \, \tilde{n}$\,.
We therefore find the T-dual dictionary,
\be
	\tilde{n} = w\,,
		\qquad%
	\tilde{w} = n\,,
		\qquad%
	\tilde{R} = \frac{\alpha'}{R}\,.
\ee
In the T-dual frame, the theory is compactified over a dual circle of radius $\alpha' / R$\,, and the KK and winding numbers are swapped. This T-duality symmetry is manifestly reflected in the closed string dispersion relation,
\be \label{eq:sdp}
	-p^m \, p^{}_m = \frac{n^2}{R^2} + \frac{w^2 \, R^2}{\alpha'{}^2} + \frac{1}{\alpha'} \bigl( N + \bar{N} - 2 \bigr)\,, 
\ee
where $N$ and $\bar{N}$ are the string excitation numbers. This dispersion relation is supplemented with the level-matching condition,
$N - \bar{N} = n \, w$\,.
In the T-dual frame, the dispersion relation and the level-matching condition take the same form, with 
\be
	-\tilde{p}^{\,m} \, \tilde{p}^{}_m = \frac{\tilde{n}^2}{\tilde{R}^2} + \frac{\tilde{w}^2 \, \tilde{R}^2}{\alpha'{}^2} + \frac{1}{\alpha'} \bigl( N + \bar{N} - 2 \bigr)\,, 
		\qquad%
	N - \bar{N} = \tilde{n} \, \tilde{w}\,.
\ee
Therefore, T-duality is a $\mathbb{Z}_2$ symmetry in the target space, where ``T'' is for ``Target.''

%%%%%%%%%%%%%%%%%%%%%%%%%%%%%%%%%%%%%%%%%%%%%%%%%%%%%%%
\subsubsection{T-dual string worldsheet in Matrix 0-brane theory}
%%%%%%%%%%%%%%%%%%%%%%%%%%%%%%%%%%%%%%%%%%%%%%%%%%%%%%%

We are now ready to discuss the T-duality transformations of the M0T string action~\eqref{eq:mzts}. We first consider the T-duality transformations along $p$ spatial directions, which requires us to rewrite the M0T action~\eqref{eq:mzts} as~\cite{Gomis:2023eav}
\begin{align} \label{eq:mztspa}
\begin{split}
	S_\text{parent} = - \frac{T}{2} \int \de^2 \sigma \Bigl( \p_\tau X^0 \, \p_\tau X^0 & + \p_\sigma X^{A'} \p_\sigma X^{A'} \! + \lambda \, \p_\sigma X^0 \\[4pt]
	& + V^{u}_\sigma \, V^u_\sigma + 2 \, \tilde{X}^u \, \epsilon^{\alpha\beta} \, \p_\alpha V^u_\beta \Bigr).
\end{split}
\end{align}
We have split the transverse index $i = 1\,, \, \cdots, \, 9$ to be $i = (u\,,  A')$\,, with $u = 1\,, \, \cdots, \, p$ and $A' = p+1\,, \, \cdots, \, 9$\,. Integrating out $\tilde{X}^u$ imposes the constraint $\epsilon^{\alpha\beta} \, \p_\alpha V^u_\beta = 0$\,, which is solved by $V^u_\alpha = \p_\alpha X^u$\,. Plugging this solution back into Eq.~\eqref{eq:mztspa} gives back the original M0T string action~\eqref{eq:mzts}. Instead, integrating out $V^u_\sigma$ in the parent action~\eqref{eq:mztspa} gives the dual action~\cite{Gomis:2023eav}
\be \label{eq:mpts}
	S_\text{M$p$T} = - \frac{T}{2} \int \de^2 \sigma \Bigl( - \p_\tau \tilde{X}^A \, \p_\tau \tilde{X}_A + \p_\sigma X^{A'} \p_\sigma X^{A'} \! + \tilde{\lambda}_A \, \p_\sigma \tilde{X}^A \Bigr)\,,
\ee
where $\tilde{X}^A = (X^0, \, \tilde{X}^u)$ and we have defined $\tilde{\lambda}_u = (\lambda\,, \, 2 \, V_\tau^u)$\,, $A = 0\,, 1\,, \cdots, \, p$\,. For simplicity, we drop the tildes on $\tilde{X}^A$ and $\tilde{\lambda}_u$ below. This dual action arises from a decoupling limit of type II superstring theories that naturally generalises the M0T limit in Section~\ref{sec:mtitdlcq}. This generalised decoupling limit is defined by the following reparametrizations~\cite{Gomis:2023eav, Gopakumar:2000ep, Harmark:2000ff, Blair:2023noj}:
\begin{subequations} \label{eq:mptopre}
\begin{align}
	X^A & \rightarrow \sqrt{\omega} \, X^A\,,
		&%
	X^{A'} & \rightarrow \frac{X^{A'}}{\sqrt{\omega}}\,, \\[4pt]
	g_s & \rightarrow \omega^{\frac{p-3}{2}} \, g_s\,,
		&
	C^{(p+1)} & \rightarrow \frac{\omega^2}{g_s} \, dX^0 \wedge \cdots \wedge dX^{p+1}\,.
\end{align}
\end{subequations}
Here, $C^{(p+1)}$ is an RR ($p$\,+1)-form. The $\omega \rightarrow \infty$ limit of type II superstring theory leads to the \emph{Matrix $p$-brane theory} (M$p$T), where the light excitations are the D$p$-branes that are described by Matrix gauge theory~\cite{Blair:2023noj}.\,\footnote{M$p$T has been generalised to the cases where $p < 0$~\cite{Blair:2023noj, Gomis:2023eav}: while M(-1)T is related to tensionless and ambitwistor string theroy and IKKT Matrix theory, M$p$Ts with $p < -1$ are associated with strings in Carroll-like target space. See also~\cite{Cardona:2016ytk} for previous discussions on Carroll strings.} For examples, M$1$T is associated with Matrix string theory~\cite{Dijkgraaf:1997vv} and M$3$T is associated with $\mathcal{N} = 4$ super Yang-Mills theory. 

We have seen in Section~\ref{sec:mtitdlcq} that M0T arises from the DLCQ of M-theory. An interesting observation is that it is now possible to consider a second DLCQ in M$p$T with $p \neq 0$ by forming a lightlike circle. In this review, we focus on the simplest case of DLCQ M1T, where the fundamental string action is given by 
\be \label{eq:dlcqmot}
	S^\text{DLCQ}_\text{M1T} = - \frac{T}{2} \int \de^2 \sigma \Bigl( - 2 \, \p_\tau X^+ \, \p_\tau X^- + \p_\sigma X^{A'} \, \p_\sigma X^{A'} \! + \lambda_+ \, \p_\sigma X^+ + \lambda_- \, \p_\sigma X^- \Bigr)\,,
\ee
where $X^\pm = (X^0 \pm X^1) / \sqrt{2}$\,, $\lambda_\pm = (\lambda_0 \pm \lambda_1) / \sqrt{2}$\,, and the lightlike direction $X^+$ is compactified over a circle. In order to understand better what this DLCQ M1T means, we perform a T-duality transformation in $X^+$ and show that it maps the lightlike $X^+$ circle to a spatial circle. We start with introducing the parent action,
\begin{align}
\begin{split}
	S_\text{parent} = - \frac{T}{2} \int \de^2 \sigma \Bigl( \p_\sigma X^{A'} \, \p_\sigma X^{A'} & + \lambda_- \, \p_\sigma X^- \\[2pt]
	& - 2 \, V_\tau \, \p_\tau X^- + \lambda_+ \, V_\sigma + 2 \, \tilde{X} \, \epsilon^{\alpha\beta} \, \p_\alpha V_\beta \Bigr)\,,
\end{split}
\end{align}
Integrating out the Lagrange multiplier $\tilde{X}$ gives back the DLCQ M0T string action~\eqref{eq:dlcqmot}. Instead, integrating out $V_\alpha$ leads to the dual action,
\be \label{eq:sdualmmpt}
	S_\text{dual} = - \frac{T}{2} \int \de^2 \sigma \Bigl( \p_\sigma X^{A'} \, \p_\sigma X^{A'} + \lambda_- \, \p_\sigma X^- \Bigr)\,,
		\qquad%
	\p_\sigma \tilde{X} = \p_\tau X^-\,. 
\ee
This dual action is invariant under the transformations when it is on shell,
\be \label{eq:mmztbt}
	\delta X^{A'} = \Lambda^{A'} X^-\,,
		\qquad%
	\delta \tilde{X} = v \, X^-\,, 
		\qquad%
	\delta X^- = 0\,.
\ee
Interpreting $X^-$ as the time direction, we define $\tilde{X}^0 = X^-$ and $\tilde{X}^1 = \tilde{X}$, we find that $v$ is the Galilean boost velocity in the $\tilde{X}^0$--$\tilde{X}^1$ sector while $\Lambda^{A'}$ parametrizes the Galilean boost in the $\tilde{X}^0$--$\tilde{X}^{A'}$ sector.\,\footnote{Interpreting $X^-$ as a spatial direction and $\tilde{X}$ as the target space time, these two directions are related to each other via a Carrollian boost, which results from a zero (instead of infinite for the Galilean boost) speed-of-light limit of Lorentzian boost. The resulting string action is then identified with the longitudinal Carrollian string in~\cite{Bidussi:2023rfs} (see also~\cite{Gomis:2023eav}).} Also note that the original lightlike circle along $X^+$ in the DLCQ M1T string action~\eqref{eq:dlcqmot} becomes spacelike along $\tilde{X}^1$ in the T-dual frame. Further define $\tilde{\lambda}_0 = \lambda_-$\,, we rewrite Eq.~\eqref{eq:sdualmmpt} as~\cite{Blair:2023noj, Gomis:2023eav}
\be \label{eq:mm0t}
	S_\text{dual} = - \frac{T}{2} \int \de^2 \sigma \Bigl( \p_\sigma X^{A'} \, \p_\sigma X^{A'} + \tilde{\lambda}_A \, \p_\sigma \tilde{X}^A - \tilde{\lambda}_1 \, \p_\tau X^0 \Bigr)\,. 
\ee
Here, $A = 0\,, 1$\,. We have introduced the Lagrange multiplier $\tilde{\lambda}_1$ to incorporate the constraint from Eq.~\eqref{eq:sdualmmpt}. 
Note that the new action~\eqref{eq:mm0t} is invariant under the boost transformations in Eq.~\eqref{eq:mmztbt} even when it is off shell, as long as we introduce the additional transformation 
\be
    \delta \tilde{\lambda}_0 = - 2 \, \Lambda^{A'} \p_\sigma X^{A'} - v \, \tilde{\lambda}_1 \,.
\ee
This dual string action~\eqref{eq:mm0t} is closely related to the Spin Matrix Theory (SMT) discussed in Section~\ref{sec:smtl} (see Eq.~(A.15) in~\cite{Harmark:2018cdl}). This fundamental string lives in \emph{Multicritical Matrix 0-brane Theory} (MM0T), which arises from a limit of type IIA superstring theory with multiple background fields taken to be ``critical''~\cite{Blair:2023noj}.

We now review the ``multicritical" decoupling limit of type IIA superstring theory that leads to MM0T~\cite{Blair:2023noj, Gomis:2023eav}. We again start with the conventional string action~\eqref{eq:relsa}, but now supplemented with an electric $B$-field term, such that Eq.~\eqref{eq:relsa} becomes
\be \label{eq:relsabf}
	S = - \frac{T}{2} \int \de^2 \sigma \, \Bigl( \p_\alpha X^\mu \, \p^\alpha X_\mu + \epsilon^{\alpha\beta} \, \p_\alpha X^\mu \, \p_\beta X^\nu \, B_{\mu\nu} \Bigr)\,.
\ee
Next, we perform the reparametrization in Eq.~\eqref{eq:relsabf},
\be \label{eq:zoab}
	X^0 \rightarrow \omega \, X^0\,,
		\qquad%
	X^1 \rightarrow X^1\,,
		\qquad%
	X^{A'} \rightarrow \frac{X^{A'}}{\sqrt{\omega}}\,,
		\qquad%
	B \rightarrow - \omega \, dX^0 \wedge dX^1\,,
\ee
together with the rescaling of the worldsheet coordinates in Eq.~\eqref{eq:tos}. Note that we have taken $B_{01}$ to be constant here, in which case the $B$-field term in Eq.~\eqref{eq:relsabf} is a boundary term. The inclusion of this term is important for the self-consistency of the theory, particularly when a more general $B$-field or compactifications are considered.  
The string action~\eqref{eq:relsabf} now becomes
\begin{align} \label{eq:divo}
\begin{split}
	S = - \frac{T}{2} \int \de^2 \sigma \, \Bigl[ - \omega^3 \, \bigl( \p_\sigma X^0 & - \omega^{-2} \, \p_\tau X^1 \bigr)^2 + \omega \, \bigl( \p_\sigma X^1 - \p_\tau X^0 \bigr)^2 \\[4pt]
	& + \p_\sigma X^{A'} \p_\sigma X^{A'} - \omega^{-2} \, \p_\tau X^{A'} \p_\tau X^{A'} \Bigr]\,.
\end{split}
\end{align}
Using the Hubbard-Stratonovich method to integrate in the auxiliary fields $\lambda_A$\,, $A = 0 \,, 1$\,, we rewrite Eq.~\eqref{eq:divo} as
\begin{align}
\begin{split}
	S = - \frac{T}{2} \int \de^2 \sigma \, \biggl[ \p_\sigma X^{A'} \, \p_\sigma X^{A'} & + \lambda_0 \, \Bigl( \p_\sigma X^0 - \omega^{-2} \, \p_\tau X^1 \Bigr) + \lambda_1 \Bigl( \p_\sigma X^1 - \p_\tau X^0 \Bigr) \\[4pt]
	& - \omega^{-2} \, \p_\tau X^{A'} \p_\tau X^{A'} + \frac{\lambda_0^2}{4 \, \omega^3} - \frac{\lambda_1^2}{4 \, \omega} \biggr]\,.
\end{split}
\end{align}
In the $\omega \rightarrow \infty$ limit, we recover the MM0T string action~\eqref{eq:mm0t}. 

In addition to the prescriptions in Eq.~\eqref{eq:zoab}, the T-dual of Eq.~\eqref{eq:mptopre} also implies that we are required to take into account the following reparametrizations of the background string coupling and RR 0-form~\cite{Blair:2023noj}:
\be \label{eq:gsc1}
	g_s \rightarrow \omega^{-1} \, g_s\,,
		\qquad%
	C^{(1)} \rightarrow \omega^2 \, g^{-1}_s \, d X^0\,.
\ee
In order to gain some intuition about why the prescription~\eqref{eq:gsc1} is necessary, we apply the MM0T limit defined above to the conventional D0-brane action that is a sum of Eqs.~\eqref{eq:sdz} and \eqref{eq:dzcs}, which gives
\begin{subequations}
\begin{align}
	S_\text{D0} & \rightarrow - \frac{1}{g_s \, \alpha'{}^{1/2}} \int \de\tau \, \Big[ \omega^2 - \frac{1}{2} \, \p_\tau X^1 \, \p_\tau X^1 + O(\omega^{-2}) \Bigr]\,, \\[4pt]
	S_\text{CS} & \rightarrow \frac{\omega^2}{g_s \, \alpha'{}^{1/2}} \int \de\tau\,.
\end{align}
\end{subequations}
Note that we have taken the static gauge $X^0 = \tau$\,. 
The $\omega \rightarrow \infty$ limit of $S_\text{D0} + S_\text{CS}$ gives a finite D0-brane action in MM0T, which describes a particle moving along the background string associated with the critical background $B$-field. As there are two different background fields, $B$ and $C^{(1)}$\,, that become critical (such that they cancel respectively the string and D0-brane tension in the background bound F1-D0 configuration), it is natural to refer to this decoupling limit of type IIA superstring theory as \emph{multicritical} M0T. It is also possible to generalise MM0T to MM$p$T, where a background bound F1-D$p$ configuration is taken to be critical. See~\cite{Blair:2023noj, Gomis:2023eav} for further discussions.  

%%%%%%%%%%%%%%%%%%%%%%%%%%%%%%%%%%%%%%%%%%%%%%%%%%%%%%%
\subsubsection{Lorentzian worldsheet from S-duality} \label{sec:sdnrs}
%%%%%%%%%%%%%%%%%%%%%%%%%%%%%%%%%%%%%%%%%%%%%%%%%%%%%%%

So far, we have only considered T-duality transformations of the fundamental strings associated with various Matrix theories. T-dualities are perturbative and do \emph{not} alter the nature of the string worldsheet. We have learned that the string worldsheets in these theories are non-Lorentzian, which makes it somewhat exotic to consider their quantisation. In fact, the light excitations in these decoupling limits of type II string theories are described by various Matrix theories, which supposedly encode the dynamics analogous to how the fundamental string encodes the dynamics in perturbative string theories. Intriguingly, before this connection to Matrix theories was established, there has been already abundant literature discussing the (perturbative) quantisation of related string theories with similar non-Lorentzian worldsheet in the context of tensionless~\cite{Isberg:1993av} and ambitwistor~\cite{Mason:2013sva, Geyer:2022cey} strings. All these worldsheet theories are related to the M0T string sigma model~\eqref{eq:mztfscw} via T-duality transformations~\cite{Gomis:2023eav}. 

We now review a decoupling limit of type IIB superstring theory that is related to Matrix 1-brane theory (M1T) via an S-duality transformation~\cite{Dijkgraaf:1997vv, Gomis:2000bd, Danielsson:2000gi, Ebert:2023hba}.\,\footnote{See~\cite{Bergshoeff:2022iss, Bergshoeff:2023ogz, Ebert:2023hba} for the SL($2\,, \mathbb{Z}$) generalisation, which exhibits novel branching behaviors and reveals a polynomial realisation of SL($2\,, \mathbb{Z}$).} 
The M1T string sigma model is given by Eq.~\eqref{eq:mpts}. We will see that the S-duality transformation maps the non-Lorentzian M1T string worldsheet to the conventional Lorentzian worldsheet. In this S-dual frame, standard conformal field theoretical techniques become available again on the Lorentzian string worldsheet. 

We start with constructing the D1-brane action in M1T. A single D1-brane in type IIB superstring theory is described by the effective action,
\be \label{eq:doba}
	S^{}_\text{D1} = - \frac{1}{g^{}_s \alpha'} \int \de^2 \sigma \, \sqrt{-\det \Bigl( \p_\alpha X^\mu_{\phantom{\dagger}} \, \p_\beta X_\mu + F_{\alpha\beta} \Bigr)} + \frac{1}{g^{}_s \, \alpha'} \int C^{(2)}. 
\ee
We have set the $B$-field to zero and $F = dA$ is the U(1) gauge strength on the D1-brane. Using the prescription~\eqref{eq:mptopre}, the $\omega \rightarrow \infty$ limit of the D1-brane action~\eqref{eq:doba} gives rise to the following D1-brane action in M1T:
\be \label{eq:sdomot}
	S^\text{M1T}_\text{D1} = - \frac{1}{2 \, g^{}_s \alpha'} \int \de^2 \sigma \, \sqrt{-\tau} \, \Bigl( \tau^{\alpha\beta} \, \p_\alpha X^{A'} \, \p_\beta X^{A'} - \tfrac{1}{2} \, \tau^{\alpha\gamma} \, \tau^{\beta\delta} \, F_{\alpha\beta} \, F_{\gamma\delta} \Bigr)\,,
\ee
where $\tau = \det \tau_{\alpha\beta}$ and $\tau^{\alpha\beta}$ is the inverse of $\tau_{\alpha\beta}$\,, with $\tau_{\alpha\beta} = \p_\alpha X^{\!A} \, \p_\beta X_{\!A}$\,. We recall that $A = 0\,, \, 1$ and $A' = 2\,, \, \cdots, \, 9$\,. Now, we are ready to consider the S-dual of the D1-brane action~\eqref{eq:sdomot} in M1T, which can be implemented as a magnetic duality of the U(1) gauge potential $A_\alpha$\,. Therefore, we rewrite Eq.~\eqref{eq:sdomot} as
\begin{align} \label{eq:spsd}
\begin{split}
	S_\text{parent} = - \frac{1}{2 \, g^{}_s \alpha'} \int \de^2 \sigma \, \sqrt{-\tau} \, \Bigl[ \tau^{\alpha\beta} \, \p_\alpha X^{A'} \, \p_\beta X^{A'} & - \tfrac{1}{2} \, \tau^{\alpha\gamma} \, \tau^{\beta\delta} \, F_{\alpha\beta} \, F_{\gamma\delta} \\[4pt]
	& - \Theta^{\alpha\beta} \bigl( F_{\alpha\beta} - 2 \, \p_{[\alpha} A_{\beta]} \bigr) \Bigr]\,,
\end{split}
\end{align}
where $F_{\alpha\beta}$ is treated as an independent two-form instead of an exact form. Integrating out the anti-symmetric two-form $\Theta^{\alpha\beta}$ in Eq.~\eqref{eq:spsd} imposes that $F = dA$\,, under which Eq.~\eqref{eq:spsd} gives back Eq.~\eqref{eq:sdomot}. Instead, integrating out $A_\alpha$ in Eq.~\eqref{eq:spsd} imposes the condition $d \Theta = 0$\,, which is solved locally by $\Theta^{\alpha\beta} = \theta \, \epsilon^{\alpha\beta}$\,, with $\theta$ being constant. Further integrating out $F_{\alpha\beta}$ in Eq.~\eqref{eq:spsd} gives the S-dual action,
\be \label{eq:ngafs}
	S_\text{parent} \rightarrow - \frac{1}{2 \, g^{}_s \alpha'} \int \de^2 \sigma \, \sqrt{-\tau} \, \tau^{\alpha\beta} \, \p_\alpha X^{A'} \, \p_\beta X^{A'} - 2 \, g_s \, \theta^2 \, \int dX^0 \wedge dX^1\,.
\ee
In the case where $\theta = 0$\,, we find the Nambu-Goto action~\cite{Andringa:2012uz},
\be \label{eq:ngns}
	S_\text{NG} = - \frac{T}{2} \int \de^2 \sigma \, \sqrt{-\tau} \, \tau^{\alpha\beta} \, \p_\alpha X^{A'} \, \p_\beta X^{A'}\,.
\ee
In terms of the auxiliary metric $h^{}_{\alpha\beta} = e^{}_\alpha{}^a \, e^{}_\beta{}^b \, \eta^{}_{ab}$\,, $a = 0\,, 1$\,, the associated Polyakov formulation is~\cite{Gomis:2000bd, Gomis:2005pg, Bergshoeff:2018yvt}
\be \label{eq:spns}
	S_\text{P} = - \frac{T}{2} \int \de^2 \sigma \, \sqrt{-h} \, \Bigl( h^{\alpha\beta} \, \p_\alpha X^{A'} \, \p_\beta X^{A'} + \lambda \, \bar{e}^\alpha \, \p_\alpha X + \bar{\lambda} \, e^\alpha \, \p_\alpha \overline{X} \Bigr)\,,
\ee
where  
\begin{subequations}
\begin{align}
	e^\alpha & = e^\alpha{}_0 + e^\alpha{}_1\,,
		&%
	e_\alpha & = e_\alpha{}^0 + e_\alpha{}^1\,, 
		&%
	X & = X^0 + X^1\,, \\[4pt]
	\bar{e}^\alpha & = - e^\alpha{}_0 + e^\alpha{}_1\,,
		&%
	\bar{e}_\alpha & = - e_\alpha{}^0 + e_\alpha{}^1\,, 
		&%
	\overline{X} & = X^0 - X^1\,.
\end{align}
\end{subequations}
Moreover, $h^{\alpha\beta}$ is the inverse worldsheet metric, $e^\alpha{}^{}_a = e^{-1} \, \epsilon^{\alpha\beta} \, e^{}_\beta{}^b \, \epsilon^{}_{ab}$ is the inverse zweibein field, $e = \sqrt{-h}$\,, and $h = \det (h_{\alpha\beta})$\,. Integrating out the Lagrange multipliers $\lambda$ and $\bar{\lambda}$ in Eq.~\eqref{eq:spns} imposes the constraints  $\epsilon^{\alpha\beta} \, e^{}_\alpha \, \p^{}_\beta X = 0$ and $\epsilon^{\alpha\beta} \, \bar{e}^{}_\alpha \, \p^{}_\beta \overline{X} = 0$\,, which are solved locally by $e_\alpha \propto \p_\alpha X$ and $\bar{e}_\alpha \propto \p_\alpha \overline{X}$\,. Plugging these solutions back into Eq.~\eqref{eq:spns} recovers the Nambu-Goto action~\eqref{eq:ngns}. The string action~\eqref{eq:spns} has a conventional Lorentzian worldsheet and defines \emph{nonrelativistic string theory}~\cite{Klebanov:2000pp, Gomis:2000bd, Danielsson:2000gi}, which has a (string) Galilean invariant string spectrum that we will review in the next subsection. 

%%%%%%%%%%%%%%%%%%%%%%%%%%%%%%%%%%%%%%%%%%%%%%%%%%%%%%%
\subsection{From DLCQ to nonrelativistic string theory} \label{sec:dlcqnst}
%%%%%%%%%%%%%%%%%%%%%%%%%%%%%%%%%%%%%%%%%%%%%%%%%%%%%%%

At the end of the previous subsection, we introduced type IIB nonrelativistic superstring theory via the S-duality transformation of Matrix 1-brane theory (M1T). We have learned that the light excitations in M1T are bound states of D1-branes that are described by Matrix string theory~\cite{Dijkgraaf:1998gf}, and the fundamental string in M1T is defined on a non-Lorentzian worldsheet. Meanwhile, the light excitations in nonrelativistic string theory are the fundamental strings, whose string worldsheet is Lorentzian. Therefore, the dynamics of the nonrelativistic string can be described by conformal field theory, which allows us to quantise it as in conventional perturbative string theory. This observation shows that nonrelativistic string theory plays an anchoring role in the duality web of decoupling limits in string theory, in view of that the target space physics can be derived in nonrelativistic string theory from first principles. 
The physical quantities~\footnote{See \emph{e.g.}~\cite{Gomis:2000bd, Danielsson:2000gi, Yan:2021hte} for discussions on string amplitudes in nonrelativistic string theory.} in nonrelativistic string theory can then be mapped to other corners in the duality web that are nonperturbative from the string perspective. It is therefore valuable to develop a comprehensive understanding of nonrelativistic string theory, which we review in this subsection. 

As M1T is related to DLCQ M-theory, nonrelativistic string theory, which is S-dual to M1T, is also related to DLCQ M-theory. This relation via Matrix $p$-brane theories (M$p$Ts), although concrete, is somewhat a roundabout. At the beginning of this subsection we will review a more direct relation to DLCQ M-theory: nonrelativistic string theory arises from T-dualising the lightlike circle in the DLCQ of conventional string theory~\cite{Gomis:2000bd, Danielsson:2000gi}, while the latter arises from compactifying DLCQ M-theory over a spatial circle. This procedure recovers the same string sigma model~\eqref{eq:spns} that describes nonrelativistic string theory, which is unitary, UV-complete and its string spectrum and S-matrix enjoy nonrelativistic symmetry. In this T-dual frame described by nonrelativistic string theory, the lightlike circle in DLCQ string theory maps to a regular spatial circle, and the non-Lorentzian nature of the spacetime geometry becomes manifest. In this sense, nonrelativistic string theory provides a first principles definition of the DLCQ of relativistic string theory~\cite{Bergshoeff:2018yvt}.  

%%%%%%%%%%%%%%%%%%%%%%%%%%%%%%%%%%%%%%%%%%%%%%%%%%%%%%%
\subsubsection{Lightlike T-dual of DLCQ string theory} \label{sec:lltddlcqs}
%%%%%%%%%%%%%%%%%%%%%%%%%%%%%%%%%%%%%%%%%%%%%%%%%%%%%%%

We now present a more direct route from DLCQ M-theory to nonrelativistic string theory, without relying on the S-duality transformation in Section~\ref{sec:sdnrs}.
We start with compactifying DLCQ M-theory over a transverse, spatial circle. This extra compactification leads to the DLCQ of type IIA superstring theory, where the target space lightlike circle from the DLCQ in M-theory still survives in the ten-dimensional theory. The embedding coordinates in DLCQ string theory are $X^\mu$, $\mu = 0\,, \, \cdots, \, 9$\,, with the lightlike directions $X^\pm = X^0 \pm X^1$\,. We require that $X^+$ be compactified, \emph{\textit{i.e.}}
\be
	X^+ (\sigma + 2\pi) = X^+ (\sigma) + 2 \pi R \, w\,,
		\qquad%
	w \in \mathbf{Z}\,,
\ee
with $w$ the winding number in $X^+$\,. 
It might be strange to think about windings around a lightlike circle, but we will soon find a simple physical interpretation in the T-dual frame. 
In terms of the lightlike coordinates $X^\pm$ and the transverse coordinates $X^{A'}$, where $A' = 2, \cdots, 9$\,, we rewrite the conventional string action~\eqref{eq:relsa} as 
\be \label{eq:llcompaction}
	S = - \frac{T}{2} \int \de^2 \sigma \, \Bigl( \p_\alpha X^{A'} \, \p^\alpha X^{A'} - \p_\alpha X^+ \, \p^\alpha X^- \Bigr)\,.
\ee
The KK momentum in $X^+$ is quantised, with
\be \label{eq:quantisedp}
	p_+ = \frac{n}{R}\,,
		\qquad%
	n \in \mathbf{Z}\,.
\ee
Note that $p_+ \sim \int d\sigma \, \p_\tau X^+$  
is the momentum conjugate to $Y^+$. 

In order to perform a lightlike T-duality transformation in $X^+$, we rewrite Eq.~\eqref{eq:llcompaction} equivalently as
\be \label{eq:llpa}
	S_\text{parent} = - \frac{T}{2} \int \de^2 \sigma \, \Bigl( \p_\alpha X^{A'} \, \p^\alpha X^{A'} - V_\alpha \, \p^\alpha X^- - \tilde{X}^1 \, \epsilon^{\alpha\beta} \, \p_\alpha V_\beta \Bigr)\,.
\ee
Upon integrating out $\tilde{X}^1$, we find the local solution,
\be \label{eq:vapy}
	V_\alpha = \p_\alpha X^+\,,
\ee
plugging this back into the parent action \eqref{eq:llpa}, the original action \eqref{eq:llcompaction} is recovered. Instead, integrating out $V_\alpha$ gives rise to the action
\be \label{eq:sdual}
	S_\text{dual} = - \frac{T}{2} \int \de^2 \sigma \, \p_\alpha X^{A'} \, \p^\alpha X^{A'}\,, 
\ee
which is supplemented by the constraint,
\be \label{eq:constraint}
	\p_\alpha \tilde{X}^0 = - \epsilon_\alpha{}^\beta \, \p_\beta \tilde{X}^1\,,
		\qquad%
	\tilde{X}^0 \equiv X^-\,.
\ee
In terms of the lightlike coordinates on the worldsheet and target space,
\begin{subequations} \label{eq:defccs}
\begin{align}
	\p & = \p_\tau + \p_\sigma
		&
	X & = \tilde{X}^0 + \tilde{X}^1\,, \\[4pt]
	\bar{\p} & = - \p_\tau + \p_\sigma\,,
		&%
	\overline{X} & = \tilde{X}^0 - \tilde{X}^1\,,
\end{align}
\end{subequations}
the constraint \eqref{eq:constraint} becomes
\be \label{eq:constraintpm}
	\p \overline{X} = \bar{\p} X = 0\,,
\ee
According to Eq.~\eqref{eq:constraintpm}, we have
\be \label{eq:holoantiholo}
	X = X (\tau + \sigma)\,,
		\qquad%
	\overline{X} = \overline{X} (\tau - \sigma)\,.
\ee
We have used the indicative notation $\tilde{X}^0$ and $\tilde{X}^1$ in Eq.~\eqref{eq:defccs}, implying that $\tilde{X}^0$ is the time direction and $\tilde{X}^1$ a spatial direction in the T-dual frame. It then follows that $X$ and $\overline{X}$ are lightlike coordinates in the dual theory. This is demanded by the target space symmetries, which we demonstrate now. We consider the dual theory in the decompactification limit, where the Lorentz boost in the longitudinal sector is recovered. The chiral conditions in Eq.~\eqref{eq:holoantiholo} demand that the most general symmetry transformations for $X$ and $\overline{X}$ are
$X \rightarrow f (X)$ and $\overline{X} \rightarrow \bar{f}(\overline{X})$\,,
where $f$ and $\bar{f}$ are arbitrary functions. Infinitesimally, we find
$\delta X = \bigl( \Theta + \Lambda \bigr) \, X$ and $\delta \overline{X} = \bigl( \Theta - \Lambda \bigr) \, \overline{X}$\,.
Here, $\Theta$ and $\Lambda$ receive the interpretation as the parameters for the dilatation and Lorentz boost transformation, respectively, only if $X$ and $\overline{X}$ are interpreted as the target space lightlike coordinates. See~\cite{Bergshoeff:2018yvt} for further details of this T-duality transformation that maps between DLCQ and nonrelativistic string theory. See~\cite{Bergshoeff:2019pij} for the same duality map from the limiting procedure and see~\cite{Harmark:2017rpg, Kluson:2018egd, Harmark:2018cdl, Harmark:2019upf} using the method of null reduction. 

To further understand the physics in the dual frame, we study the mode expansions. We start with considering the mode expansions of $X^\pm$ in DLCQ string theory,
\begin{subequations} \label{eq:ybarymexp}
\begin{align}
	X^+ & = x^+ + \alpha' \, p_- \, \tau + R \, w \, \sigma + \text{oscillations}\,, \\[2pt]
    X^- & = x^- + \alpha' \, p_+ \, \tau + \text{oscillations}\,. 
\end{align}
\end{subequations}
Recall that $p_+$ is defined in Eq.~\eqref{eq:quantisedp}, which is quantised. 
Using Eqs.~\eqref{eq:constraint}, we find
\begin{subequations}
\begin{align}
	\tilde{X}^0 & = \tilde{x}^0 + \tilde{w} \, \tilde{R} \, \tau + \text{oscillations}\,, \\[2pt]
	\tilde{X}^1 & = \tilde{x}^1 + \tilde{w} \, \tilde{R} \, \sigma + \text{oscillations}\,, \label{eq:x1winding}
\end{align}
\end{subequations}
where 
\be \label{eq:wpn}
	\tilde{w} = n \,,
\ee 
is the winding number in $\tilde{X}^1$ in the dual frame, and the dual circle is now spacelike in $\tilde{X}^1$, with a radius $\tilde{R} = \alpha'/R$\,. This is rather surprising: T-duality maps the spacetime with a lightlike compactification to a regular spacetime with a spacelike compactification!

As a final crosscheck, Eq.~\eqref{eq:defccs} implies
\begin{subequations} \label{eq:mexpzm}
\begin{align}
	X & = x + \tilde{w} \, \tilde{R} \, \bigl( \tau + \sigma \bigr) + \text{oscillations}\,, \\[2pt]
	\overline{X} & = \bar{x} + \tilde{w} \, \tilde{R} \, \bigl( \tau - \sigma \bigr) + \text{oscillations}\,,
\end{align}
\end{subequations}
which indeed satisfy the constraints in Eq.~\eqref{eq:constraintpm}.  

%%%%%%%%%%%%%%%%%%%%%%%%%%%%%%%%%%%%%%%%%%%%%%%%%%%%%%% 
\subsubsection{Nonrelativistic closed strings} \label{sec:ncs}
%%%%%%%%%%%%%%%%%%%%%%%%%%%%%%%%%%%%%%%%%%%%%%%%%%%%%%%

Next, we will show that the above lightlike T-duality leads us to nonrelativistic string theory, which has a Galilean-invariant string spectrum~\cite{Gomis:2023eav}. We are therefore trading exotic physics in lightlike compactifications with more familiar nonrelativistic physics in spacelike compactifications.

We have seen in Eq.~\eqref{eq:x1winding} that the KK number $n$ in string theory over a lightlike circle is mapped to winding along the spatial circle in the dual theory. The winding along the lightlike circle should be mapped to momentum in the dual spatial circle. However, there is no dual KK momentum in the zero modes of the mode expansions of $X$ and $\overline{X}$ in Eq.~\eqref{eq:mexpzm}. Where is the KK momentum hiding?

The dual action~\eqref{eq:sdual} can be rewritten in a way such that the constraints~\eqref{eq:constraintpm} are incorporated via Lagrange multipliers. This essentially gives back the ``parent'' action~\eqref{eq:llpa}, which we rewrite by defining
\be \label{eq:eqlbl}
	\lambda = - \frac{1}{2} \, \bigl( V_\tau + V_\sigma \bigr)\,,
		\qquad%
	\bar{\lambda} = - \frac{1}{2} \, \bigl( - V_\tau + V_\sigma \bigr)\,. 
\ee
The resulting dual action is
\be \label{eq:nstaction}
	S = - \frac{T}{2} \int \de^2 \sigma \, \Bigl( \p_\alpha X^{A'} \, \p^\alpha X^{A'} + \lambda \, \bar{\p} X + \bar{\lambda} \, \p \overline{X} \Bigr)\,.
\ee 
This is referred to as the \emph{Gomis-Ooguri theory} in the literature \cite{Gomis:2000bd}, which is identical to the action~\eqref{eq:spns} in conformal gauge. Here, $\lambda$ and $\bar{\lambda}$ are one-form fields that play the role of Lagrange multipliers. Integrating out these Lagrange multipliers recovers the constraints in Eq.~\eqref{eq:constraintpm}. One intriguing consequence of these constraints is that the moduli space is now localized to be a submanifold of the one in relativistic string theory. For example, in the calculation of the one-loop free energy where the worldsheet torus wraps over a spacetime torus that involves a Euclidean time direction, the integral over the moduli space of the fundamental domain associated with SL(2,\,$\mathbb{Z}$) is now localized to be a sum over a set of discrete points within the fundamental domain. Moreover, it turns out that the critical dimensions for the bosonic and supersymmetric strings are the same as in the standard case, which are 26 and 10, respectively. This string theory is known to be unitary and ultra-violet complete and can be studied independently of the full relativistic string theory. 

According to the duality map \eqref{eq:vapy}, we find~\cite{Yan:2021lbe}
\be
	\lambda = - \p X^+\,,
		\qquad%
	\bar{\lambda} = - \bar{\p} X^+\,.
\ee
These one-form fields are associated with the conjugate momenta of $X$ and $\overline{X}$\,, with
\be
	\lambda = \frac{2}{T} \frac{\p \CL}{\p \bigl( \p_\tau X \bigr)}\,,
		\qquad%
	\bar{\lambda} = - \frac{2}{T} \frac{\p \CL}{\p \bigl( \p_\tau \overline{X} \bigr)}\,,
\ee
with $\CL$ the Lagrangian density associated with the action \eqref{eq:nstaction}. Note that the following variables with tildes match the ones in Section~\ref{sec:lltddlcqs}. Define the energy (conjugate to $\tilde{X}^0$) to be $\tilde{p}^{}_0$ and the longitudinal spatial momentum (conjugate to $\tilde{X}^1$) to be $\tilde{p}^{}_1$\,, we then have the following mode expansions:
\be
	\lambda = - \alpha' \, \bigl( \tilde{p}^{}_0 + \tilde{p}^{}_1 \bigr) + \text{oscillations}\,, 
		\qquad%
	\bar{\lambda} = - \alpha' \, \bigl( \tilde{p}^{}_0 - \tilde{p}^{}_1 \bigr) + \text{oscillations}\,.
\ee
Since the dual coordinate $X^1$ is compactified over a circle of radius $\tilde{R} = \alpha' / R$\,, we have
\be
	\tilde{p}^{}_1 = \frac{\tilde{n}}{\tilde{R}}\,,
\ee
where $\tilde{n}$ is the KK number in the T-dual frame described by nonrelativistic string theory. 
Comparing Eqs.~\eqref{eq:vapy}, \eqref{eq:ybarymexp} and \eqref{eq:eqlbl}, we find the following dictionary between the two mutually dual frames:
\be
	\tilde{p}^{}_0 = p_-\,,
		\qquad%
	\tilde{n} = w\,.
\ee
This completes the duality map between $n$ and $\tilde{w}$ in Eq.~\eqref{eq:wpn}. Therefore, we still have the common lore that T-duality maps the KK and winding number to each other, however, in the Gomis-Ooguri theory described by Eq.~\eqref{eq:nstaction}, the KK and winding number are not on the same footing anymore. 

Back in the DLCQ of relativistic string theory, and in the case of zero winding, we have the dispersion relation,
$2 \, p^{}_+ \, p^{}_- = p^{}_{A'} \, p^{}_{A'} + \cdots$\,.
Using Eq.~\eqref{eq:quantisedp} with $p^{}_+ = n / R$\,, we find
$p^{}_- = p^{}_{A'} \, p^{}_{A'} \, R/ (2 \, n) + \cdots$\,.
In the T-dual frame, we have
\be \label{eq:p0p}
	\tilde{p}^{\phantom{\dagger}}_0 = \frac{\alpha'}{2 \, \tilde{w} \, \tilde{R}} \, p^{\phantom{\dagger}}_{\!A'} \, p^{\phantom{\dagger}}_{\!A'} + \cdots\,.
\ee
Intriguingly, this implies that the only asymptotic states in the string spectrum of the Gomis-Ooguri theory \eqref{eq:nstaction} are those with nonzero windings, and the dispersion relation is Galilei invariant. This is the reason why we refer to the Gomis-Ooguri theory as \emph{nonrelativistic string theory}.\,\footnote{Nonrelativistic string theory is also referred to as ``wound string theory" in~\cite{Danielsson:2000gi}.}

%%%%%%%%%%%%%%%%%%%%%%%%%%%%%%%%%%%%%%%%%%%%%%%%%%%%%%%
\subsubsection{Critical \texorpdfstring{$B$}{B}-field limit} \label{sec:bfb}
%%%%%%%%%%%%%%%%%%%%%%%%%%%%%%%%%%%%%%%%%%%%%%%%%%%%%%%

Instead of viewing nonrelativistic string theory as a T-dual of the DLCQ of relativistic string theory, we now review how to derive the Gomis-Ooguri action~\eqref{eq:nstaction} from a decoupling limit, where a background critical $B$-field is fine tuned to cancel the background fundamental string tension~\cite{Gomis:2000bd}. For simplicity, we drop the tildes over various variables in nonrelativistic string theory that we have used in Sections~\ref{sec:lltddlcqs} and \ref{sec:ncs}. 

We start with the sigma model~\eqref{eq:relsabf} describing relativistic strings in a nontrivial $B$-field, and perform the following reparametrization that replaces the one in Eq.~\eqref{eq:relsabf}:
\be \label{eq:gbconf}
	X^A \rightarrow \omega \, X^A\,,
		\qquad%
	X^{A'} \rightarrow X^{A'}\,,
		\qquad%
	B_{\mu\nu} \rightarrow - \omega^2 \, dX^0 \wedge dX^1\,. 
\ee
Here, $A = 0\,, \, 1$ and $A' = 2\,, \, \cdots, \, 9$\,. 
The action \eqref{eq:relsabf} then becomes
\be
	S \rightarrow - \frac{T}{2} \int \de^2 \sigma \lr - \omega^2 \, \bar{\p} X \, \p \overline{X} + \p_\alpha X^{A'} \, \p^\alpha X^{A'} \rr.
\ee
Superficially, it seems that the first term diverges in the $\omega \rightarrow 0$ limit. However, this divergence disappears upon integrating in the auxiliary fields $\lambda$ and $\bar{\lambda}$\,, such that
\be \label{eq:hst}
	S \rightarrow - \frac{T}{2} \int \de^2 \sigma \lr \p_\alpha X^{A'} \, \p^\alpha X^{A'} + \lambda \, \bar{\p} X + \bar{\lambda} \, \p \overline{X} + \omega^{-2} \, \lambda \, \bar{\lambda} \rr.
\ee 
Now, the $\omega \rightarrow 0$ limit is well defined, which precisely leads to the Gomis-Ooguri action \eqref{eq:nstaction}. In this limiting procedure, the infinite boost in DLCQ string theory is encoded by the fine-tuned critical $B$-field that cancels the background string tension.

In the presence of the $B$-field configuration and the rescalings in Eq.~\eqref{eq:gbconf}, and in the case where the string is wrapped $\tilde{w}$ times around the compactified $X^1$ direction, %\SIB{[are the use of tildes correct here?]} 
the relativistic string dispersion relation \eqref{eq:sdp} is modified to be
\be
	\lr \frac{{p}^{}_0}{\omega} + \frac{{w} \, \omega \, {R}}{\alpha'} \rr^2 - p^{\phantom{\dagger}}_{\!A'} \, p^{\phantom{\dagger}}_{\!A'} = \frac{{n}^2}{\bigl( \omega \, {R} \bigr)^2} + \frac{{w}^2 \, \bigl(\omega \, {R} \bigr)^2}{{\alpha'}^2} + \frac{1}{\alpha'} \lr N + \bar{N} - 2 \rr,
\ee
which in the $\omega \rightarrow 0$ limit gives the dispersion relation in nonrelativistic closed string theory, 
\be
	{p}^{}_0 = \frac{\alpha'}{2 \, {w} \, {R}} \ls p^{\phantom{\dagger}}_{\!A'} \, p^{\phantom{\dagger}}_{\!A'} + \frac{1}{\alpha'} \bigl( N + \bar{N} - 2 \bigr) \rs.
\ee
This dispersion relation reproduces Eq.~\eqref{eq:p0p} derived from dualizing the closed string spectrum in DLCQ string theory.

%%%%%%%%%%%%%%%%%%%%%%%%%%%%%%%%%%%%%%%%%%%%%%%%%%%%%%%
\subsubsection{Symmetries and curved target space} \label{sec:scts}
%%%%%%%%%%%%%%%%%%%%%%%%%%%%%%%%%%%%%%%%%%%%%%%%%%%%%%%

We now examine the symmetries of the nonrelativistic string action \eqref{eq:nstaction}, which is invariant under the following global transformations:
\begin{subequations}
\begin{align}
	\delta X^A & = \Theta^A - \Lambda \, \epsilon^A{}_B \, X^B\,, \\[4pt]
	\delta X^{A'} & = \Theta^{A'} - \Lambda^{A'}{}_{B'} \, X^{B'} + \Lambda_A{}^{A'} \, X^A\,, \\[2pt]
	\delta \lambda_A & = - \Lambda \, \epsilon_A{}^B \, \lambda_B - \Lambda_{AA'} \, \p_\tau X^{A'} - \epsilon_{AB} \, \Lambda^B{}_{A'} \, \p_\sigma X^{A'},
\end{align}
\end{subequations}
where
$\lambda_0 = \frac{1}{2} \lr \lambda - \bar{\lambda} \rr$ and $\lambda_1 = \frac{1}{2} \lr \lambda + \bar{\lambda} \rr$\,. 
The target-space manifold is partitioned into the longitudinal sector with the index $A = 0\,, \, 1$ and the transverse sector with the index $A' = 2\,, \, \cdots, \, 9$\,. 
The Lie algebra parameters $\Theta^A$, $\Theta^{A'}$, $\Lambda$\,, $\Lambda^{A'B'}$, and $\Lambda^{AA'}$ are respectively associated with the following generators:
\begin{align} \label{eq:generators}
\begin{split}
	\text{\emph{longitudinal translations}} & \qquad H^{}_A \\[2pt]
	\text{\emph{transverse translations}} & \qquad P^{}_{A'} \\[2pt]
	\text{\emph{longitudinal Lorentz boost}} & \qquad M \\[2pt]
	\text{\emph{transverse rotations}} & \qquad J^{}_{A'B'} \\[2pt]
	\text{\emph{string Galilean boosts}} & \qquad G^{}_{AA'}
\end{split}
\end{align}
Note that the string Galilean boost
$\delta X^A = 0$ and $\delta X^{A'} = \Lambda_A{}^{A'} \, X^A$ 
is a direct generalisation of the Galilean boost $\delta t = 0$ and $\delta x^{A'} = v^{A'} \, t$ for nonrelativistic particles. 
We then find a stringy generalisation of the Bargmann algebra \eqref{eq:bargalg}, defined by the following non-vanishing commutators~\cite{Brugues:2006yd, Bergshoeff:2019pij}:\,\footnote{Noncentral extensions of the algebra are required for the quantum consistency in nonrelativistic string theory~\cite{Yan:2021lbe}.}
\begin{subequations}
\begin{align}
	[H^{}_A\,, M] & = \epsilon^{}_A{}^B \, H_B\,,  
		&%
	[H^{}_A\,, G^{}_{BA'}] & = \eta^{}_{AB} \, P^{}_{A'}\,, \\[2pt]
	[P^{}_{A'}, J^{}_{B'C'}] & = \delta^{}_{A'B'} \, P^{}_{C'} - \delta^{}_{A'C'} \, P^{}_{B'}\,,
		&%
	[G^{}_{\!AA'}, M] & = \epsilon^{}_A{}^B \, G^{}_{BA'}\,, \\[2pt]
	[G^{}_{AA'}, J^{}_{B'C'}] & = \delta^{}_{A'B'} \, G^{}_{\!AC'} - \delta^{}_{A'C'} \, G^{}_{AB'}\,, 
\end{align}
\vspace{-7mm}
\be
	\!\!\!\!\!\!\![J_{A'B'}\,, J_{C'D'}] = \delta_{B'C'} \, J_{A'D'} - \delta_{A'C'} \, J_{B'D'} + \delta_{A'D'} \, J_{B'C'} - \delta_{B'D'} \, J_{A'C'}\,.
\ee
\end{subequations}
These commutators define the string Galilei algebra in nonrelativistic string theory that replace the Poincar\'{e} algebra in relativistic string theory.~\footnote{See~\cite{Bidussi:2021ujm} for discussions on F-string Galilei algebra, whose gauging leads to the geometry that incorporates the $B$-field~\cite{Bergshoeff:2021bmc}.}

The target space geometry can be constructed by associating spacetime gauge fields to each of the generators in Eq.~\eqref{eq:generators}. In particular, we define $\tau_\mu{}^A$ and $E_\mu{}^{A'}$ to be the vielbein fields encoding the geometry of the longitudinal and transverse sectors, respectively. These vielbein fields are associated with the longitudinal and transverse translational generators $H_A$ and $P_{A'}$\,, respectively.
In curved spacetime, the flat spacetime action \eqref{eq:nstaction} is generalised to the interacting sigma model,
\begin{align} \label{eq:ism}
\begin{split}
	S = - \frac{T}{2} \int \de^2 \sigma \Bigl[ \p_\alpha X^\mu \, \p^\alpha X^\nu \, E_{\mu\nu} (X) & + \lambda \, \bar{\p} X^\mu \, \tau_\mu (X) + \bar{\lambda} \, \p X^\mu \, \bar{\tau}_\mu (X) \\[2pt]
		& + \epsilon^{\alpha\beta} \, \p_\alpha X^\mu \, \p_\beta X^\nu \, B_{\mu\nu} (X) \Bigr]\,,
\end{split}
\end{align}
where, up to a redefinition of $\lambda$ and $\bar{\lambda}$\,, the symmetric two-tensor $E_{\mu\nu}$ can be brought into the form 
$E^{}_{\mu\nu} = E^{}_\mu{}^{A'} \, E^{}_\nu{}^{B'} \, \delta_{A'B'}$\,, which can be regarded as a metric in the transverse sector. 
Note that $E^{}_{\mu\nu}$ is a rank $8$ matrix and does \emph{not} constitute a metric in the ten-dimensional target space.
We also defined
$\tau^{}_\mu = \tau^{}_\mu{}^0 + \tau^{}_\mu{}^1$ and $\bar{\tau}^{}_\mu = \tau^{}_\mu{}^0 - \tau^{}_\mu{}^1$\,.	
The appropriate target-space geometry to which nonrelativistic string theory is coupled is equipped with a codimension-two foliation structure. The longitudinal and transverse sectors are related to each other via a string Galilean boost,
\be
	\delta \tau^{}_\mu{}^A = 0\,,
		\qquad%
	\delta E^{}_\mu{}^{A'} = \Lambda^{}_A{}^{A'} \, \tau^{}_\mu{}^A\,.
\ee
Supplemented with appropriate transformations of $\lambda$\,, $\bar{\lambda}$\,, and $B_{\mu\nu}$, the sigma model action \eqref{eq:ism} is invariant under the string Galilei boost. This non-Lorentzian geometry is referred to as \emph{string Newton-Cartan geometry}, which naturally generalises Newton-Cartan geometry associated with Newtonian gravity. Dynamically, there is \emph{no} graviton in this theory. Instead, the only gravitational interaction in nonrelativistic string theory is the Newton-like force between winding strings.

So far, we have been working with conformal gauge. Undoing this introduces an auxiliary worldsheet zweibein field $e_\alpha{}^a$, $a = 0, 1$ as in Eq.~\eqref{eq:spns}, and
the action describing nonrelativistic string theory with a dynamical worldsheet is then given by \cite{Bergshoeff:2018yvt}
\begin{align}
  \label{eq:snc-pol-action}
    S & = - \frac{T}{2} \int \de^2 \sigma \, e \, \Bigl( h^{\alpha\beta} \, \p_\alpha X^\mu \, \p_\beta X^\nu \, E_{\mu\nu} + \lambda \, \bar{e}^\alpha \, \p_\alpha X^\mu \, \tau_\mu + \bar{\lambda} \, e^\alpha \, \p_\alpha X^\mu \, \bar{\tau}_\mu \Bigr) \notag \\[2pt]
    & \quad - \frac{T}{2} \int \de^2 \sigma \, \epsilon^{\alpha\beta} \, \p_\alpha X^\mu \, \p_\beta X^\nu \, B_{\mu\nu} + \frac{1}{4\pi} \int \de^2 \sigma \sqrt{-h} \, R (h) \, \Phi\,.
\end{align}
Here, $e = \det (e_\alpha{}^a)$\,, $a = 0\,, \, 1$\,, $R(h)$ is the scalar curvature associated with $h_{\alpha\beta}$\,, and $\Phi$ is the dilaton field. Integrating out the non-dynamical worldsheet metric $h_{\alpha\beta}$ gives rise to the following Nambu-Goto action that generalises Eq.~\eqref{eq:ngns} to curved spacetime~\cite{Andringa:2012uz}:
\begin{align}
  \label{eq:snc-ng-action}
\begin{split}
    S_\text{NG} & = - \frac{T}{2} \int \de^2\sigma \lr \sqrt{-\tau} \, \tau^{\alpha\beta} \, \p_\alpha X^\mu \, \p_\beta X^\nu \, E_{\mu\nu} + \epsilon^{\alpha\beta} \, \p_\alpha X^\mu \, \p_\beta X^\nu \, B_{\mu\nu} \rr, %\\[2pt]
        %
    %& \quad + \frac{1}{4\pi} \int \de^2\sigma \sqrt{-\tau} \, R(\tau) \, \Phi\,.
\end{split}
\end{align}
where we have omitted the dilaton term. 

The Weyl anomalies of the interacting sigma model~\eqref{eq:snc-pol-action} have been analysed in~\cite{Gomis:2019zyu, Gallegos:2019icg, Bergshoeff:2019pij, Yan:2019xsf, Yan:2021lbe}. 
We refer the interested readers to the recent review~\cite{Oling:2022fft} on nonrelativistic string theory for further details on its curved target space geometry. Note that the $\lambda \, \bar{\lambda}$ term in~\eqref{eq:hst} would be generated due to quantum corrections, which deforms nonrelativistic string theory towards relativistic string theory. It is shown in~\cite{Gomis:2019zyu, Yan:2019xsf, Yan:2021lbe} that it is possible to impose extra symmetries such that the $\lambda \, \bar{\lambda}$ term is prevented from being generated at the quantum level in the string sigma model. One consequence of these global worldsheet symmetries is that they impose extra geometric constraints on the target space vielbein data: certain components of the intrinsic torsion of the vielbein $\tau_\mu{}^A$ are now set to zero. Note that similar constraint also naturally arises in nonrelativistic supergravity~\cite{Bergshoeff:2021tfn}. For this reason, this $\lambda \, \bar{\lambda}$ term is referred to as the \emph{torsional deformation}~\cite{Yan:2021lbe}. When the torsional deformation is included, we are essentially working with relativistic string theory expanded around the nonrelativistic string vacuum. This deformation plays an important role when it comes to the standard AdS/CFT correspondence. See~\emph{e.g.}~\cite{Danielsson:2000mu, Guijosa:2023qym, Avila:2023aey} in the context of the gravity dual of noncommutative open string theory. 

%%%%%%%%%%%%%%%%%%%%%%%%%%%%%%%%%%%%%%%%%%%%%%%%%%%%%%%
\subsubsection{A first principles definition of DLCQ string theory} \label{sec:fpdl}
%%%%%%%%%%%%%%%%%%%%%%%%%%%%%%%%%%%%%%%%%%%%%%%%%%%%%%%

Now, we are ready to use nonrelativistic string theory described by the action principle \eqref{eq:ism} to provide a first principles definition of the DLCQ of string theory, via a well-defined T-duality transformation in a longitudinal spatial isometry~\cite{Bergshoeff:2018yvt}. See also~\cite{Harmark:2017rpg, Kluson:2018egd, Harmark:2018cdl} for the perspective of null reduction, which will be discussed in Section~\ref{sec:nisnrsb}. 

Assume that there is a Killing vector $k^\mu \p_\mu$ satisfying
\be
	k^\mu \, \tau^{}_\mu{}^0 = 0\,,
		\qquad%
	k^\mu \, \tau^{}_\mu{}^1 \neq 0\,,
		\qquad%
	k^\mu \, E^{}_\mu{}^{A'} = 0\,.
\ee
We then take a coordinate system $X^\mu = (y\,, X^m)$\,, $m = 0\,, \, 2\,, \, \cdots, \, 9$ adapted to $k^\mu$, with $k^\mu \, \p_\mu = \p_y$\,. This abelian isometry is represented by a translation in the longitudinal spatial direction $y$\,. We then write the embedding coordinates as $X^\mu = (y\,, X^i)$\,. %\SIB{[What is the difference with small x and big X?]} 
Rewrite the action \eqref{eq:ism} describing nonrelativistic string theory as follows:
\begin{align} \label{eq:pac}
	S^{}_\text{parent} 
	& = - \frac{T}{2} \int \de^2 \sigma \, \Bigl[ \p_\alpha X^m \, \p_\beta X^n \, E_{mn} + \lambda \, \bigl( \overline{V} \, \tau^{}_y + \bar{\p} X^m \, \tau^{}_m \bigr) + \bar{\lambda} \, \bigl( V \, \bar{\tau}^{}_y + \p X^m \, \bar{\tau}^{}_m \bigr) \Bigr] \notag \\[2pt]
	& \quad - \frac{T}{2} \int \de^2 \sigma \, \epsilon^{\alpha\beta} \, \Bigl( 2 \, V_\alpha \, \p_\beta X^m \, B_{ym} + \p_\alpha X^m \, \p_\beta X^n \, B_{mn} + 2 \, \tilde{Y} \, \p_\alpha V_\beta \Bigr)\,.
\end{align}
Integrating out $\tilde{y}$ imposes the constraint 
$\p_\alpha V_\beta = 0$\,.
Locally, this can be solved by $V_\alpha = \p_\alpha y$. Plugging this into the action \eqref{eq:pac} gives back to the original action \eqref{eq:ism}. Instead, integrating out $V_\alpha$ leads to the T-dual action
\be
	\tilde{S} = - \frac{1}{4\pi\alpha'} \int \de^2 \sigma \, \Bigl( \p_\alpha X^\mu \, \p^\alpha X^\nu \, \widetilde{G}_{\mu\nu} + \epsilon^{\alpha\beta} \, \p_\alpha X^\mu \, \p_\beta X^\nu \, \widetilde{B}_{\mu\nu} \Bigr)\,,
\ee
with $\tilde{X}^\mu = (\tilde{Y}, X^i)$ and the Buscher-like rules
\begin{align}
	\widetilde{G}^{}_{yy} & = 0\,, 
		&%
	\widetilde{G}^{}_{mn} & = E_{mn} + \frac{B^{}_{ym} \, \tau^{}_n{}^0 + B^{}_{n} \, \tau^{}_m{}^0}{\tau^{}_{y}{}^1}\,,
		&%
	\widetilde{G}^{}_{ym} & = \frac{\tau^{}_m{}^0}{\tau^{}_{y}{}^1}\,, \\[2pt]
	\widetilde{B}^{}_{ym} & = \frac{\tau^{}_{ym}}{\tau^{}_{yy}}\,, 
		&%
	\widetilde{B}^{}_{ij} & = B^{}_{mn} + \frac{B^{}_{ym} \, \tau^{}_{n}{}^1 - B^{}_{yn} \, \tau^{}_{m}{}^1}{\tau^{}_{y}{}^1}\,.
\end{align}
Moreover, the dilaton field transforms as
$\widetilde{\Phi} = \Phi - \log \bigl| \tau^{}_{y}{}^1 \bigr|$\,.
This dual theory is coupled to a regular pseudo-Riemannian target-space geometry and describes relativistic string theory. The fact that $\widetilde{G}^{}_{yy} = 0$ implies that the dual $\tilde{y}$ direction is a lightlike isometry. Therefore, this dual theory is relativistic string theory in the DLCQ.

%%%%%%%%%%%%%%%%%%%%%%%%%%%%%%%%%%%%%%%%%%%%%%%%%%%%%%%
\subsubsection{Relation to the \texorpdfstring{$T\bar{T}$}{TTbar} deformation} \label{sec:ttbar}
%%%%%%%%%%%%%%%%%%%%%%%%%%%%%%%%%%%%%%%%%%%%%%%%%%%%%%%
 
Before we end this subsection, we review an intriguing relation between the critical $B$-field limit that leads to nonrelativistic string theory and the $T\bar{T}$ deformation~\cite{Blair:2020ops}, and show how the $T\bar{T}$ deformation is related to the torsional deformation $\lambda\bar{\lambda}$ when the target space is three dimensional. In the $T\bar{T}$ deformation, a two-dimensional quantum field theory (QFT) defined by a Lagrangian $\CL$ is deformed by an irrelevant term that is the determinant of the energy-momentum tensor. To be more specific, we follow a trajectory in the space of field theories parametrized by $t$~\cite{Zamolodchikov:2004ce, Smirnov:2016lqw, Cavaglia:2016oda}. Each point of this trajectory is associated with a 2D QFT, whose Lagrangian we denote as $\CL(t)$\,, such that $\CL(0) = \CL$\,, \emph{\textit{i.e.}} the original theory $\CL$ is recovered at $t=0$\,. Denote the stress energy tensor associated with $\CL(t)$ by $T_{\alpha\beta}(t)$\,. The trajectory that we are interested in is then defined via the flow equation 
\be \label{eq:dldt}
	\frac{d\CL(t)}{dt} = \det \bigl[ T^{}_{\alpha\beta} (t) \bigr]\,. 
\ee
In terms of the lightcone coordinates $z = \frac{1}{2} (\tau + \sigma)$ and $\bar{z} = \frac{1}{2} (- \tau + \sigma)$\,, we find
\be
\label{eq:Ottbar}
	\det \bigl[ T^{}_{\alpha\beta} (t) \bigr] = \frac{1}{4} \, \Bigl[ T^{}_{zz} (t) \, T^{}_{\bar{z}\bar{z}} (t) - T_{z\bar{z}}^2 (t) \Bigr]\,.
\ee
We chose the definition of $z$ and $\bar{z}$ such that
\be
    \p = \frac{\p}{\p z} = \p_\tau + \p_\sigma\,,
        \qquad%
    \bar{\p} = \frac{\p}{\p \bar{z}} = - \p_\tau + \p_\sigma\,,
\ee
which matches the convention in Eq.~\eqref{eq:defccs}. 
If the original theory is conformal, $T_{z\bar{z}} = T^\alpha{}_\alpha (0) = 0$ and we are left with $T \, \bar{T}$ with $T \sim T_{zz}$ and $\bar{T} \sim T_{\bar{z}\bar{z}}$ in Eq.~\eqref{eq:Ottbar}, which motivates the name of the $T\bar{T}$ deformation. 
Despite the deformation being irrelevant, the deformed theory is relatively well-behaved.
In particular, the deformation is solvable: it acts in a simple way on the spectrum of the model, so that the deformed energy levels can be determined from the undeformed ones by an ordinary differential equation~\cite{Smirnov:2016lqw, Cavaglia:2016oda}. It also affects in a simple way the S-matrix of the model~\cite{Dubovsky:2017cnj}. However, it does spoil  locality and, unless we are considering a supersymmetric theory~\cite{Baggio:2018rpv}, some energy levels of the deformed theory become complex. 

\vspace{3mm}

\noindent $\bullet$~\textbf{Nambu-Goto action from $T\bar{T}$ deformation.} We start with a brief review of the $T\bar{T}$ deformation of a free boson before relating it to nonrelativistic strings. This relationship was first noticed in~\cite{Cavaglia:2016oda}.
We start with the free action of a single massless real scalar $\phi$\,,\,\footnote{It is straightforward to generalise this to any number of scalar fields.}
\be \label{eq:freea}
    S = - \frac{1}{2} \int\limits_{-\infty}^{+\infty}\de\tau\int\limits_{0}^{L} \de \sigma \, \p_\alpha \phi \, \p^\alpha \phi. 
\ee 
The deformed Lagrangian $\CL(t)$ has the stress energy tensor
\be \label{eq:defset}
	T^{}_{\alpha\beta} (t) = \frac{\p \CL(t)}{\p (\p^\alpha \phi)} \, \p^{}_\beta \phi - \eta^{}_{\alpha\beta} \, \CL(t)\,.
\ee
Starting from Eq.~\eqref{eq:freea} at $t = 0$\,, the deformed Lagrangian can be constructed order by order. Expanding $\CL(t)$ with respect to $t$ gives
\begin{equation}
    \CL(t) = \sum_{n=0}^\infty t^n \CL^{(n)},
        \qquad% 
    \CL^{(0)} = - \frac{1}{2} \, \p \phi \, \bar{\p} \phi\,.
\end{equation}
The first-order correction given by the $T\bar{T}$ deformation is built out of the stress-energy tensor of the free theory
\begin{equation}
     \CL^{(1)} = \det \bigl[ T^{}_{\alpha\beta} (t) \bigr]_{t=0}
     = \frac{1}{4}\big(\partial \phi\, \bar{\partial} \phi\big)^2\,.
\end{equation}
The subsequent orders would be built from the stress-energy tensor of the deformed theory. This iterative procedure can be resummed to give an analytic function of~$t$ as in~\cite{Cavaglia:2016oda}. A more straightforward and generalisable approach~\cite{Bonelli:2018kik} is to first note that the deformed Lagrangian can only be a function of the dimensionless quantity $t \, \partial \phi \, \bar{\partial} \phi$ (recall that $t$ has mass dimension $[t]=-2$), \emph{\textit{i.e.}}~$\mathcal{L}(t) \sim F(t \, \partial \phi \, \bar{\partial} \phi)$\,. To recover the free theory~\eqref{eq:freea} at small~$t$, we impose the condition
\begin{equation}
\label{eq:TTbinitial}
    F(0) = 0\,,
        \qquad%
    F'(0) = - \frac{1}{2}\,,
\end{equation} 
The deformed action takes the form
\be \label{eq:freeat}
    S(t) = \int\limits_{-\infty}^{+\infty}\de\tau\int\limits_{-R/2}^{R/2}\de\sigma \, \mathcal{L} (t)\,,
        \qquad%
    \mathcal{L} (t) =  \frac{1}{t} \, F \bigl(t \, \partial \phi \, \bar{\partial} \phi \bigr)\,,
\ee 
The stress-energy tensor $T_{\alpha\beta} (t)$ in Eq.~\eqref{eq:defset} implies
\begin{equation}
    \det \bigl[ T^{}_{\alpha\beta} (t) \bigr]=
    %z^2\Big(F'(z)\Big)^2-\Big(zF'(x)-F(x)\Big)^2=
    \frac{1}{t^2}\Bigl[ 2\,x\,F(x)\,F'(x) - F(x)^2 \Bigr]\,,
\end{equation}
where the explicit $t$-dependence signals that the operator is irrelevant with dimension~$4$.
Hence, the flow equation~\eqref{eq:dldt} gives
\begin{equation}
    \left(x \, \frac{\de}{\de x} - 1\right)\Bigl[ F(x) - F(x)^2 \Bigr] = 0\,,
\end{equation}
Imposing the initial conditions in Eq.~\eqref{eq:TTbinitial}, we find
\begin{equation} \label{eq:lt}
    \CL (t) = - \frac{1}{2 \, t} \lr \sqrt{1 + 2 \, t \, \p^\alpha \phi \, \p_\alpha \phi} - 1 \rr\,,
\end{equation}
which takes the form of the Nambu-Goto action that describes the fundamental string in static gauge.

It is worth noting that the energy levels of the theory are also deformed in a simple way. In the undeformed theory, the energy levels $E_n(R)$ of a conformal field theory on a cylinder of circumference $R$ are
\begin{equation} \label{eq:enzl}
    E_n(R)=\frac{2\pi}{R} \lr  n-\frac{c}{24} \rr,\qquad n\in\mathbb{N}\,.
\end{equation}
Here, $n$ and $\bar{n}$ are the eigenvalues of the Virasoro generators and $c$ is the central charge. For simplicity, we focus on the zero-momentum states that satisfy $n=\bar{n}$ in Eq.~\eqref{eq:enzl}. In the special case of Eq.~\eqref{eq:freea}, $c = 1$\,.  
The deformed energy levels $\mathcal{E}_n (t\,, R)$ obey the Burgers equation~\cite{Smirnov:2016lqw, Cavaglia:2016oda},
\begin{equation}
\label{eq:Burgers}
    \frac{\p}{\p t} \mathcal{E}_n(t\,, R) = \mathcal{E}_n (t\,, R) \, \frac{\p}{\p L} \mathcal{E}_n(t\,, R)\,,
\end{equation}
which implies that
\be \label{eq:iden}
    \mathcal{E}_n(t\,,R) = E_n \Bigl( R + t \, \mathcal{E}_n (t\,, R) \Bigr)\,.
\ee
Also note that $\mathcal{E}_n(0\,,L) = E_n(L)$\,. 
Plugging Eq.~\eqref{eq:enzl} into Eq.~\eqref{eq:iden} gives
\begin{equation}
    \mathcal{E}_n(t\,,R) = \frac{R}{2 \, t} \left[ \sqrt{1 + \frac{4 \, t}{R} \, E_n (R)} - 1 \right].
\end{equation}
Note that, when $t$ is sufficiently large, the energy $\mathcal{E}_0$ is imaginary, which resembles the tachyon in bosonic string theory.

\vspace{3mm}

\noindent $\bullet$~\textbf{$T\bar{T}$ as a torsional deformation of nonrelativistic string theory.} We are now ready to identify the relation~\cite{Blair:2020ops} between the $T\bar{T}$ deformation and the critical $B$-field limit discussed in Section~\ref{sec:bfb}, where the latter leads to nonrelativistic string theory. We  start with the conventional Nambu-Goto action in three-dimensional target space, 
\be \label{eq:relnga}
    S_\text{NG} = - T \int \de^2 \sigma \sqrt{ - \det \Bigl( \p_\alpha X^\mu_{\phantom{\dagger}} \, \p_\beta X_\mu \Bigr)} - T \int B\,,
        \qquad%
    \mu = 0\,, \, 1\,,\, 2\,.
\ee
Plugging Eq.~\eqref{eq:gbconf} that parametrizes the critical $B$-field limit into the Nambu-Goto action~\eqref{eq:relnga}, we find in the static gauge $X^0 = \tau$ and $X^1 = \sigma$ that 
\be \label{eq:sngnrelexp}
    S_\text{NG} = - T \int \de^2 \sigma \, \omega^2 \lr \sqrt{1 + \omega^{-2} \, \p X^2 \, \bar{\p} X^2} - 1 \rr. 
\ee
Identifying $X^2 = \phi$ and $\omega^{-2} = 2 \, t$\,, we find that $\CL(t)$ in Eq.~\eqref{eq:lt} is the same as the Lagrangian associated with the Nambu-Goto action~\eqref{eq:sngnrelexp}. Moreover, the $\omega \rightarrow \infty$ limit of Eq.~\eqref{eq:sngnrelexp} leads to the Nambu-Goto formulation of the nonrelativistic string in three-dimensional spacetime. 
Therefore, the $T\bar{T}$ deformation of the free boson receives the interpretation as the deformation from nonrelativistic to relativistic string theory. In the Polyakov formulation (see Eqs.~\eqref{eq:spns}, \eqref{eq:hst} and~\eqref{eq:snc-pol-action}), the $T\bar{T}$ deformed action associated with the Nambu-Goto action~\eqref{eq:sngnrelexp} becomes
\be \label{eq:sptd}
    S_\text{P} = - \frac{T}{2} \int \de^2 \sigma \, e \, \Bigl( h^{\alpha\beta} \, \p_\alpha \phi \, \p_\beta \phi + \lambda \, \bar{e}^\alpha \, \p_\alpha X + \bar{\lambda} \, e^\alpha \, \p_\alpha \overline{X} + 2 \, t \, \lambda \, \bar{\lambda} \Bigr)\,. 
\ee
Here, $X = \tau + \sigma$ and $\overline{X} = \tau - \sigma$\,. Note that we already set $X^2 = \phi$ and $\omega^{-2} = 2 \, t$\,. Integrating out the auxiliary fields $e_\alpha{}^a$\,, $\lambda$\,, and $\bar{\lambda}$ in Eq.~\eqref{eq:sptd} gives the theory defined by the Lagrangian~\eqref{eq:lt}. The $\lambda \bar{\lambda}$ term in Eq.~\eqref{eq:sptd} is precisely the torsional deformation in~\cite{Yan:2021lbe}, which we discussed briefly at the end of Section~\ref{sec:scts}. In this special case with a three-dimensional target space, the torsional deformation of nonrelativistic string theory is the $T\bar{T}$ deformation. 

\vspace{3mm}

\noindent $\bullet$~\textbf{Uniform lightcone gauge.} Finally, we revisit the $T\bar{T}$ deformation by starting with conventional Polyakov action for the free closed string, %\SIB{<- abrupt transition / strange sentence }
\begin{equation}
    S=-\frac{T}{2}\int\limits_{-\infty}^{+\infty}\de\tau\int\limits_{-R/2}^{R/2} \de \sigma \, g^{\alpha\beta} \, \partial_{\alpha}X^\mu \, \partial_{\beta}X_\mu\,,
\end{equation}
where $g^{\alpha\beta}$ is the unit-determinant worldsheet metric. We again take the target space to be three-dimensional, with coordinates $X^\mu=(t\,,\,\varphi\,,\,\phi)$\,. Define the canonical momenta that is conjugate to $X^\mu$, with 
\be
    P_\mu = \frac{\delta S}{\delta \partial_\tau X^\mu} = - T \, g^{\tau\alpha} \, \p_\alpha X_\mu\,.
\ee
We slightly generalise the discussion of Section~\ref{s:strings-lightcone-gauge}, and perform the following change of variables following~\cite{Sfondrini:2019smd}
\begin{subequations}
\begin{align}
    X^+ & = (1-a)\,t + a\,\varphi\,, & P_{+} & = \frac{1}{\Delta_{ab}} \, \Bigl[ \bigl( 1 - b \bigr) \, P_t + b \, P_\varphi \Bigr], \\[4pt]
    X^- & = - b \, t + (1-b) \, \varphi\,, & P_- & = \frac{1}{\Delta_{ab}} \, \Bigl[ - a \, P_t + \bigl( 1 - a \bigr) \, P_\varphi \Bigr],
\end{align}
\end{subequations}
where $0\leq a<1$ and $0\leq b<1$ are parameters, and $\Delta_{ab} = 1 - a - b + 2 \, a \, b \neq 0$\,.
Recall that these conjugate momenta are related to two conserved charges
\begin{equation}
    \gen{E} = - \int\limits_{-R/2}^{R/2} \de \sigma \,P_t\,,
        \qquad%
    \gen{J} = \int\limits_{-R/2}^{R/2} \de \sigma \,P_\varphi\,.
\end{equation}
We choose the uniform lightcone gauge~\cite{Arutyunov:2009ga}
\begin{equation}
    X^+ = \tau\,,
        \qquad% 
    P_- = \frac{1}{1-b}\,.
\end{equation}
As reviewed in section~\ref{s:strings-lightcone-gauge}, the Hamiltonian of the model is one of the lightcone momenta,
\begin{equation} \label{eq:genh}
    \gen{H} = \int\limits_{-R(a)/2}^{R(a)/2} \de \sigma \, P_+ = \frac{1}{\Delta_{ab}} \, \Bigl[ - ( 1 - b ) \, \gen{E} + b \, \gen{J} \Bigr]\,,
\end{equation}
while the other lightcone momentum fixes the volume of the model
\begin{equation} \label{eq:lapm}
    \gen{P}_- = \int\limits_{-R(a)/2}^{R(a)/2} \de \sigma \, P_- = \frac{1}{\Delta_{ab}} \, \Bigl[ a \, \gen{E} + ( 1 - a ) \,  \mathbf{J} \Bigr] = \frac{L(a)}{1-b}\,.
\end{equation}
Using Eq.~\eqref{eq:hden}, we find that the Hamiltonian is (for simplicity, we set $b = 1/2$)
\begin{align}
    \gen{H} = - \int\limits_{-R(a)/2}^{R(a)/2}\de\sigma\,\frac{1}{2 \, a} \left[ \sqrt{\Bigl( 1 - 2 \, a \, P_\phi^2 \Bigr) \Bigl( 1 - 2 \, a \, \p_\sigma \phi \Bigr)} -1 \right]\,, 
\end{align}
The associated Lagrangian is
\be
    \CL = \frac{1}{2 \, a} \Bigl( \sqrt{1 - 2 \, a \, \p \phi \, \bar{\p} \phi} - 1 \Bigr)\,.
\ee
Clearly, this is the same Lagrangian (and Hamiltonian) of the $T\bar{T}$-deformed theory defined by Eq.~\eqref{eq:lt}, upon setting $a = - t$\,.
The Burgers equation~\eqref{eq:Burgers} explains \textit{why} there is a relationship between $T\bar{T}$-deformed theories and strings in the uniform lightcone gauge~\cite{Baggio:2018gct}. Clearly, in string theory the choice of the gauge parameter~$a$ must not have any effect on observables such as the spectrum:
\begin{equation}
    \frac{\de}{\de a}\gen{H}= -\frac{\de}{\de a}\int\limits_{-R(a)/2}^{R(a)/2} \de \sigma \, P_+(a)=0\,.
\end{equation}
Hence, the effect of changing the volume~$R(a)$ of the string should be perfectly compensated by the $a$-dependence of the Hamiltonian density $-P_+(a)$. But the volume dependence from Eq.~\eqref{eq:lapm} is exactly the one predicted by the Burgers equation~\eqref{eq:iden}:
\begin{equation}
    L(a) = \frac{1-b}{\Delta_{ab}} \Bigl[ a \, \gen{E} + ( 1 - a ) \, \mathbf{J} \Bigr] = L(0) - a\,\gen{H}\,.
\end{equation}
We have used Eq.~\eqref{eq:genh} here. 
Hence, the $a$-dependence of the Hamiltonian density $-P_+(a)$ is that of a $T\bar{T}$ deformed theory with parameter $a=-t$. The above structure can be extended to more general actions (also the ones involving fermions) and has been used to generate a plethora of $T\bar{T}$-deformed theories in closed form~\cite{Frolov:2019nrr}.

\vspace{3mm}

\noindent $\bullet$~\textbf{T-duality revisited.}
The uniform lightcone gauge can be related to the (perhaps more straightforward) static gauge in a T-dual frame. %of the type discussed in Section~\ref{sec:lltddlcqs}. 
Define $\widetilde{X}^-$ to be the T-dual coordinate with respect to $X^-$. The gauge-fixing condition now becomes $\widetilde{X}^- \sim \sigma$\,.
In the original theory, the target-space line element is 
\begin{align}
\begin{split}
    \de s^2=
    \frac{1}{(\Delta_{ab})^2} \, \Bigl[ - \bigl( 1 - 2 \, b \bigr) \, \bigl( \de X^+ \bigr)^2 & + 2 \, \bigl( a + b - 2 \, a \, b \bigr) \, \de X^+ \, \de X^- \\[2pt]
    & + \bigr( 1 - 2 \, a \bigr) \, (\de X^-)^2 \Bigr] + \bigl( \de\phi \bigr)^2 \,.
\end{split}
\end{align}
By the Buscher rules, we T-dualise along $X^-$, which generates a constant $B$-field in the T-dual frame with
\begin{equation} \label{eq:tbttb}
    \widetilde{B} = \frac{a + b - 2 \, a \, b}{1 - 2 \, a}\,\de X^+\wedge \de \widetilde{X}^-\,,
\end{equation}
while the T-dual metric is
\begin{equation} \label{eq:gttb}
     \de \tilde{s}^2=
      - \frac{1}{1 - 2 \, a} \, \bigl( \de X^+ \bigr)^2 + \frac{(\Delta_{ab})^2}{1 - 2 \, a} \, \bigl( \de\widetilde{X}^- \bigr)^2 + \bigl( \de\phi \bigr)^2\,.
\end{equation}
The Nambu-Goto action is then
\be \label{eq:relnga2}
    S_\text{NG} = - T \int \de^2 \sigma \ls \sqrt{ - \det \Bigl( \p_\alpha X^\mu_{\phantom{\dagger}} \, \p_\beta X^\nu_{\phantom{\dagger}} \, \tilde{G}_{\mu\nu} \Bigr)} + \epsilon^{\alpha\beta} \, \p_\alpha X^\mu \, \p_\beta X^\nu \, \tilde{B}_{\mu\nu} \rs.
\ee
We are interested in the almost light-like limit, where
$a=\frac{1}{2}+\epsilon$ and $b=\frac{1}{2}+\epsilon$ with a small $\epsilon$\,.
By setting the static gauge conditions $X^+=\tau$, $\widetilde{X}^- = \sigma/\Delta_{ab}$, we find that the gauge fixed action reads, 
\begin{equation} \label{eq:snged}
    S_\text{NG} =T\int\de^2\sigma \, \frac{1}{2 \, \epsilon}\left[1-\sqrt{1-2 \, \epsilon\, \partial \phi \, \bar{\partial} \phi} + O\bigl(\epsilon^2\bigr) \right].
\end{equation}
Upon the identification $\epsilon = - t$\,, Eq.~\eqref{eq:snged} matches Eq.~\eqref{eq:lt} up to subleading terms in $\epsilon$\,. Note that, in terms of $\epsilon$\,, the dual metric~\eqref{eq:gttb} and $B$-field~\eqref{eq:tbttb} become
\be
    d\tilde{s}^2 = - \frac{1}{2 \, \epsilon} \, \Bigl[ - \bigl( \de\widetilde{X}^0 \bigr)^2 + \bigl( \de\widetilde{X}^1 \bigr)^2 \Bigr] + \bigl(\de\phi \bigr)^2 + O(\epsilon)\,,
        \qquad%
    \widetilde{B} = \frac{1}{2 \, \epsilon} \, \de \widetilde{X}^0 \! \wedge \de \widetilde{X}^1 + O(\epsilon)\,.
\ee
Here, $\widetilde{X}^0 = - \widetilde{X}^+$ and $\widetilde{X}^1 = \widetilde{X}^-/2$\,. Identify $2 \, \epsilon = - \omega^{-2}$\,, Eq.~\eqref{eq:nrstp} matches the reparametrization~\eqref{eq:gbconf} that defines the decoupling limit for nonrelativistic string theory (in three dimensions). In the limit $\epsilon \rightarrow 0$\,, the above T-duality is identified with the T-duality relation between DLCQ and nonrelativistic string theory that we discussed in Section~\eqref{sec:lltddlcqs}. See Section~\ref{sec:fpdl} and~\cite{Bergshoeff:2018yvt} for discussions on T-duality transformations of nonrelativistic string theory in curved background fields, and see~\cite{Bergshoeff:2019pij, Ebert:2021mfu} for the associated limiting procedures.\,\footnote{See~\cite{Bergshoeff:2023fcf} for a recent generalisation to nonrelativistic heterotic string theory.} See~\cite{Blair:2020ops} for relations between the TsT transformation~\cite{Lunin:2005jy} (which corresponds to a $T\bar{T}$ deformation) and nonrelativistic T-duality, where the nonrelativistic limit is viewed as a ``reverse $T\bar{T}$ deformation.'' As $T\bar{T}$ deformations preserve integrability, it is curious to think what this may imply for the nonrelativistic string limit of the AdS${}_5 \times S^5$ superstring (see \textit{e.g.}~\cite{Gomis:2005pg} and relevant studies of integrability in~\cite{Roychowdhury:2019vzh, Fontanella:2020eje, Fontanella:2021btt, Fontanella:2022fjd}).  

%%%%%%%%%%%%%%%%%%%%%%%%%%%%%%%%%%%%%%%%%%%%%%%%%%%%%%%
\subsection{Back to M-theory} \label{sec:udbdlcq}
%%%%%%%%%%%%%%%%%%%%%%%%%%%%%%%%%%%%%%%%%%%%%%%%%%%%%%%

In this final subsection, we review nonrelativistic M-theory that uplifts nonrelativistic string theory. A simple way to derive this uplift is by studying nonperturbative dualities of the probe D-branes in nonrelativistic string theory~\cite{Ebert:2021mfu}. We also use this subsection to summarize how the string theories discussed above are defined in curved target space. 

\subsubsection{Decoupling limits in curved spacetimes}
\label{ssec:curved-actions}

Before we consider the D-brane worldvolume actions in nonrelativistic string theory. We first review how the decoupling limits of type II superstring theories that lead to nonrelativistic string theory and (multicritical) Matrix $p$-brane theories are defined in curved backgrounds~\cite{Blair:2023noj, Gomis:2023eav}. Using the curved geometry data from Section~\ref{sec:scts}, the limiting prescriptions can be readily generalised to arbitrary bosonic background fields (see Eq.~\eqref{eq:mptopre} for Matrix $p$-brane theory, Eqs.~\eqref{eq:zoab} and \eqref{eq:gsc1} for multicritical Matrix $p$-brane theories, and Eq.~\eqref{eq:gbconf} for nonrelativistic string theory. % \SIB{<- quite long (and therefore a bit confusing) sentence}. 
The bosonic contents in type II superstring theories include the metric field $\hat{G}_{\mu\nu}$\,, $B$-field $B_{\mu\nu}$\,, dilaton field $\hat{\Phi}$\,, and RR potentials $\hat{C}^{(p)}$\,. 
Define
\be
	\tau^{}_{\mu\nu} = \tau^{}_\mu{}^A \, \tau^{}_\nu{}^B \, \eta^{}_{AB}\,, 
		\qquad%
	E^{}_{\mu\nu} = E^{}_\mu{}^{A'} \, E^{}_\nu{}^{A'}\,.
\ee
We summarize these limiting prescriptions below (the desired corners are defined by sending $\omega$ to infinity; see~\cite{Blair:2023noj, Gomis:2023eav} for further details):
\begin{itemize}

\item

\emph{Matrix $p$-brane theory} ($A = 0\,,\,\cdots,\, p$ and $A' = p+1\,,\, \cdots, 9$):
\begin{subequations} \label{eq:presmpbt}
\begin{align}
	\hat{G}^{}_{\mu\nu} & = \omega \, \tau^{}_{\mu\nu} + \omega^{-1} \, E^{}_{\mu\nu}\,, 
		\qquad%
	\hat{\Phi} = \Phi + \tfrac{1}{2} \, \bigl( p-3 \bigr) \, \ln \omega\,, \\[4pt]
	\hat{C}^{(p+1)} & = \omega^2 \, e^{-\Phi} \, \tau^0 \wedge \cdots \wedge \tau^p\,,
\end{align}
\end{subequations}
while $\hat{B} = B$ and $\hat{C}^{(q)} = C^{(q)}$ for $q \neq p+1$\,. Different Matrix $p$-brane theories are related to each other via T-duality transformations along spatial directions. We apply the M$p$T prescription~\eqref{eq:presmpbt} to the conventional string sigma model in curved background fields,
\be \label{eq:mptscb}
	\hat{S} = - \frac{T}{2} \int \de^2 \sigma \, \Bigl( \sqrt{-\hat{h}} \, \hat{h}^{\alpha\beta} \, \p_\alpha X^\mu \, \p_\beta X^\nu \, \hat{G}^{}_{\mu\nu} + \epsilon^{\alpha\beta} \, \p_\alpha X^\mu \, \p_\beta X^\nu \, \hat{B}^{}_{\mu\nu} \Bigr)\,.
\ee
Together with the reparametrization of the string worldsheet metric,
\be \label{eq:habps}
	\hat{h}^{}_{\alpha\beta} = - \omega^2 \, e^{}_\alpha{}^0 \, e^{}_\beta{}^0 + e^{}_\alpha{}^1 \, e^{}_\beta{}^1 \,,
\ee
that covariantizes Eq.~\eqref{eq:tos}, we find that, in the $\omega \rightarrow \infty$ limit, the action~\eqref{eq:mptscb} becomes the M$p$T string action in curved spacetime,
\begin{align}
	S_\text{M$p$T} & = \frac{T}{2} \int \de^2 \sigma \, e \, \Bigl[ \bigl( e^\alpha{}_0 \, e^\beta{}_0 \, \tau_{\mu\nu} - e^\alpha{}_1 \, e^\beta{}_1 \, E_{\mu\nu} \bigr) \p_\alpha X^\mu \, \p_\beta X^\nu - \lambda_A \, e^\alpha{}_{1} \, \p_\alpha X^\mu \, \tau_\mu{}^A \Bigr] \notag \\[4pt]
	& \quad - \frac{T}{2} \int \de^2 \sigma \, \epsilon^{\alpha\beta} \, \p_\alpha X^\mu \, \p_\beta X^\nu \, B_{\mu\nu}\,,
\end{align}
which naturally generalises the action~\eqref{eq:mpts}.  

\item

\emph{Multicritical Matrix $0$-brane theory} ($A' = 2\,, \, \cdots, \, 9$):
\begin{subequations}
\begin{align}
	\hat{G}^{}_{\mu\nu} & = - \omega^2 \, \tau^{}_\mu{}^0 \, \tau^{}_\nu{}^0 + \tau^{}_\mu{}^1 \, \tau^{}_\nu{}^1 + \omega^{-1} \, E^{}_{\mu\nu}\,,
		&%
	\hat{\Phi} & = \Phi - \ln \omega\,, \\[4pt]
	\hat{C}^{(1)} & = \omega^2 \, e^{-\Phi} \, \tau^0 + C^{(1)}\,,
		&% 	
	\hat{B} & = - \omega \, \tau^0 \wedge \tau^1 + B_{\mu\nu}\,, \\[4pt]
	C^{(q)} & = \omega \, \tau^0 \wedge \tau^1 \wedge C^{(q-2)} + C^{(q)}\,, \quad q \neq 1\,.
\end{align}
\end{subequations}
This $\omega \rightarrow \infty$ corner is T-dual to Matrix ($p$\,+1)-brane theory in the DLCQ. This T-duality transformation maps the $x^1$ isometry direction associated with $\tau^1$ to the lightlike circle in DLCQ M($p$\,+1)T\,. Together with the worldsheet reparametrization~\eqref{eq:habps}, we find the MM0T string action in the $\omega \rightarrow \infty$ limit,
\begin{align}
    \label{eq:mm0t-curved-string-action}
	S_\text{MM0T} & = \frac{T}{2} \! \int \! \de^2 \sigma \, e \, \Bigl( e^\alpha{}_1 \, e^\beta{}_1 \, \p_\alpha X^\mu \, \p_\beta X^\nu \, E_{\mu\nu} + \lambda_A \, e^{\alpha}{}_1 \, \p_\alpha X^\mu \, \tau_\mu{}^A - \lambda_1 \, e^\alpha{}_0 \, \p_\alpha X^\mu \, \tau_\mu{}^0 \Bigr) \notag \\[4pt]
	& \quad - \frac{T}{2} \int \de^2 \sigma \, \epsilon^{\alpha\beta} \, \p_\alpha X^\mu \, \p_\beta X^\nu \, B_{\mu\nu}\,.
\end{align}

\item

\emph{Nonrelativistic string theory} ($A = 0\,, \, 1$ and $A' = 2\,,\, \cdots, 9$):
\begin{subequations} \label{eq:nrstp}
\begin{align}
	\hat{G}^{}_{\mu\nu} & = \omega^2 \, \tau^{}_{\mu\nu} + E^{}_{\mu\nu}\,,
		&%
	\hat{\Phi} & = \Phi + \ln \omega\,, \\[4pt]
	\hat{B}^{}_{\mu\nu} & = - \omega^2 \, \tau^0 \wedge \tau^1 + B^{}_{\mu\nu}\,, 
		&%
	\hat{C}^{(q)} & = \omega^2 \, \tau^0 \wedge \tau^1 \wedge C^{(q-2)} + C^{(q)}\,.
\end{align}
\end{subequations}
This prescription is S-dual to the M1T prescription in Eq.~\eqref{eq:presmpbt}. We have learned that the string worldsheet in nonrelativistic string theory is Lorentzian. The $\omega \rightarrow \infty$ limit of the string action~\eqref{eq:mptscb} in terms of the reparametrization~\eqref{eq:nrstp} reproduces the string sigma model~\eqref{eq:snc-pol-action} in nonrelativistic string theory. 

\end{itemize}

%%%%%%%%%%%%%%%%%%%%%%%%%%%%%%%%%%%%%%%%%%%%%%%%%%%%%%%
\subsubsection{Dual D-branes in nonrelativistic string theory}
%%%%%%%%%%%%%%%%%%%%%%%%%%%%%%%%%%%%%%%%%%%%%%%%%%%%%%%

Equipped with the complete set of the limiting prescription~\eqref{eq:nrstp} that leads to nonrelativistic string theory, we are able to derive the D$p$-brane worldvolume actions in nonrelativistic string theory, by taking the $\omega \rightarrow \infty$ limit of the corresponding D$p$-brane actions in relativistic string theory. 
We start with the effective action of a single D$p$-brane in relativistic string theory, 
\be \label{eq:dpbrel}
	\hat{S}_{\text{D}p} = - T_p \int \de^{p+1} \sigma \, e^{-\hat{\Phi}} \sqrt{- \det \lr \hat{G}_{\alpha\beta} + \hat{\CF}_{\alpha\beta} \rr} + \int \sum_q \hat{C}^{(q)} \wedge e^{\hat{\CF}} \Big|_{p+1}\,.
\ee
Here, $\hat{\CF} = \hat{B} + dA$ and 
\be
	\hat{G}_{\alpha\beta} = \p_\alpha f^\mu (\sigma) \, \p_\beta f^\nu (\sigma) \, \hat{G}_{\mu\nu}\,,
		\qquad%
	\hat{B}_{\alpha\beta} = \p_\alpha f^\mu (\sigma) \, \p_\beta f^\nu (\sigma) \, \hat{B}_{\mu\nu}\,,
\ee
where $f^\mu (\sigma)$ is the embedding function that defines how the D$p$-brane is embedded in the target spacetime, \emph{\textit{i.e.}}, 
\be
	X^\mu \big|_{\Sigma_p} = f^\mu (\sigma)\,,
\ee
with $\Sigma_p$ the D$p$-brane submanifold. It is understood that only $(p+1)$-forms are kept in Eq.~\eqref{eq:dpbrel}.

Taking the $\omega \rightarrow \infty$ limit in Eq.~\eqref{eq:dpbrel}, which we reparametrize using~\eqref{eq:nrstp}, leads to the following D$p$-brane action in nonrelativistic string theory~\cite{Gomis:2020fui, Ebert:2021mfu}: 
\be
	S^{}_{\text{D}p} = - T_p \int \de^{p+1} \sigma \, e^{-\Phi} \, \sqrt{-\det 
	\begin{pmatrix}
		0 & \,\, \tau_\beta \\[2pt]
		\bar{\tau}_\alpha &\,\, E^{}_{\alpha\beta} + \CF^{}_{\alpha\beta}
	\end{pmatrix}}
		+
	\int \sum_q C^{(q)} \wedge e^\CF \Big|_{p+1}\,,
\ee
which can also be derived from the worldsheet perspective by requiring quantum consistency in nonrelativistic open string theory~\cite{Gomis:2020fui} (see~\cite{Danielsson:2000mu, Gomis:2020izd} for closely related studies of nonrelativistic/noncommutative open strings).
This action is in form similar to its counterpart in relativistic string theory, but there is an important distinction: the matrix within the determinant is a $(p+2) \times (p+2)$ matrix instead of a $(p+1) \times (p+1)$ matrix as in Eq.~\eqref{eq:dpbrel}. Moreover, such a D$p$-brane is coupled to the string Newton-Cartan geometry of the closed strings; this non-Lorentzian target-space has a codimension-two foliation structure.

We will mostly focus on type IIA superstring theory in the following discussion, where there are only RR potentials of odd degrees. To probe what kind of M-theory should correspond to type IIA nonrelativistic superstring theory, we study a electromagnetic duality transformation of the associated D2-brane action, 
\be \label{eq:sd2}
	S^{}_{\text{D}2} = - T_2 \int \de^3 \sigma \, e^{-\Phi} \, \sqrt{-\det 
	\begin{pmatrix}
		0 & \,\, \tau_\beta \\[2pt]
		\bar{\tau}_\alpha &\,\, E^{}_{\alpha\beta} + \CF^{}_{\alpha\beta}
	\end{pmatrix}}
		+
	\int \Bigl( C^{(3)} + C^{(1)} \wedge \CF \Bigr)\,.
\ee
We now dualise~\cite{Ebert:2021mfu} the $U(1)$ gauge field $A_\alpha$ on the D2-brane in nonrelativistic string theory by first rewriting the action~\eqref{eq:sd2} equivalently as
\begin{align}
\begin{split}
	S^{}_{\text{D}2} & = - T_2 \int \de^3 \sigma \, e^{-\Phi} \, \sqrt{-\det 
	\begin{pmatrix}
		0 & \,\, \tau_\beta \\[2pt]
		\bar{\tau}_\alpha &\,\, E^{}_{\alpha\beta} + \CF^{}_{\alpha\beta}
	\end{pmatrix}} \\[4pt]
	& \quad +
	\int \Bigl( C^{(3)} + C^{(1)} \wedge \CF \Bigr)
	+ \frac{1}{2} \int \de^3 \sigma \, \widetilde{\Theta}^{\alpha\beta} \Bigl( F_{\alpha\beta} - 2 \, \p_\alpha A_\beta \Bigr)\,,
\end{split}
\end{align}
where $\CF = B + F$ and $F_{\alpha\beta}$ is treated as an independent field. Integrating out $\widetilde{\Theta}^{\alpha\beta}$ imposes the condition $F = dA$ and thus recovers the original D2-brane action \eqref{eq:sd2}. Instead, integrating out $F$ and $A$ leads to 
$\widetilde{\Theta}^{\alpha\beta} = \epsilon^{\alpha\beta\gamma} \, \p_\gamma \Theta$\,,
and furthermore the dual action,
\be \label{eq:nrm2}
	\widetilde{S} = - \frac{1}{2} \int \de^3 \sigma \sqrt{-\gamma} \, \gamma^{\alpha\beta} \, \widetilde{E}_{\alpha\beta} - \int A^{(3)}\,,
\ee
with $\gamma^{}_{\alpha\beta} = \gamma^{}_\alpha{}^u \, \gamma^{}_\beta{}^v \, \eta^{}_{uv}$ and $\gamma^{}_\alpha{}^u = \p_\alpha f^\text{M} \, \gamma^{}_\text{M}{}^u$\,, where $\text{M} = 0\,, \, \cdots, \, 10$\,, $f^{10} = \Theta$\,, and $u\,, \, v = 0\,, \, 1\,, \, 10$\,. The longitudinal vielbein $\gamma^{}_\text{M}{}^u$ is given by
\be
	\gamma^{}_\text{M}{}^u = e^{-\Phi/3}
	\begin{pmatrix}
		\tau^{}_\mu{}^A &\,\, 0 \\[2pt]
		e^\Phi \, C^{(1)}_\mu &\,\, e^\Phi
	\end{pmatrix}.
\ee
We also defined the pullbacks $\widetilde{E}^{}_{\alpha\beta} = \p_\alpha f^\text{M} \, \p_\beta f^\text{N} \, \widetilde{E}^{}_\text{MN}$ and $A^{(3)}_{\alpha\beta\gamma} = \p^{}_\alpha f^\text{M} \, \p^{}_\beta f^\text{N} \, \p^{}_\gamma f^\text{L} \, A^{(3)}_\text{MNL}$\,. The transverse metric $\widetilde{E}^{}_\text{MN}$ and the three-form potential $A^{(3)}$ are given by
\begin{align}
	\widetilde{E}^{}_\text{MN} = e^{2\Phi/3} 
	\begin{pmatrix}
		E^{}_{\mu\nu} &\,\, 0 \\[2pt]
		0 &\,\, 0
	\end{pmatrix}\,, 
		\qquad%
	A^{(3)}_{\mu\nu\rho} = - C^{(3)}_{\mu\nu\rho}\,,
		\qquad%
	A^{(3)}_{\mu\nu10} = B^{}_{\mu\nu}\,. 
\end{align}
The dual action \eqref{eq:nrm2} describes the M2-branes in propagating in eleven-dimensional spacetime, with the dual field $\Theta$ playing the role of the extra eleventh dimension. Note that this M2-brane is coupled to eleven-dimensional \emph{membrane Newton-Cartan} geometry equipped with a codimension-three foliation structure, in contrast to ten-dimensional string Newton-Cartan geometry with a codimension-two foliation to which nonrelativistic string theory is coupled. Unlike DLCQ M-theory, there is \emph{no} lightlike circle the membrane Newton-Cartan geometry. 

%%%%%%%%%%%%%%%%%%%%%%%%%%%%%%%%%%%%%%%%%%%%%%%%%%%%%%%
\subsubsection{U-duality between decoupling limits of M-theory}
%%%%%%%%%%%%%%%%%%%%%%%%%%%%%%%%%%%%%%%%%%%%%%%%%%%%%%%

We now use the probe M2-brane action~\eqref{eq:nrm2} to define the M-theory uplift of nonrelativistic string theory. We start with reviewing how to derive the M2-brane action~\eqref{eq:nrm2} from the standard M2-brane action~\eqref{eq:mtba} supplemented with the Chern-Simons term. We transcribe this conventional M2-brane action in general bosonic background fields below:
\be \label{eq:relmtb}
	\hat{S}_\text{M2} = - T \int \de^3 \sigma \, \sqrt{-\det \Bigl( \p_\alpha X^\text{M} \, \p_\beta X^\text{N} \, \hat{G}_\text{MN} \Bigr)} - T \int \hat{A}^{(3)}\,. 
\ee
Consider the following reparametrizations~\cite{Blair:2021waq, Ebert:2021mfu, Ebert:2023hba} (see also \textit{e.g.}~\cite{Danielsson:2000gi, Gomis:2000bd, Gopakumar:2000ep, Bergshoeff:2000jn, Garcia:2002fa, Kamimura:2005rz} for related decoupling limits):
\be \label{eq:nrmtp}
	\hat{G}^{}_\text{MN} = \omega^{4/3} \, \gamma^{}_\text{MN} + \omega^{-2/3} \, E^{}_\text{MN}\,,
		\qquad%
	\hat{A}^{(3)} = - \omega^2 \, \gamma^0 \wedge \gamma^1 \wedge \gamma^2 + A^{(3)}\,.
\ee
We have defined 
\be
	\gamma^{}_\text{MN} = \gamma^{}_\text{M}{}^u \, \gamma^{}_\text{N}{}^v \, \eta^{}_{uv}\,,\,\,
		u = 0\,, \, 1\,, \, 2\,;
		\qquad
	E_\text{MN} = E^{}_\text{M}{}^{u'} \, E^{}_\text{N}{}^{v'} \, \eta^{}_{u'v'}\,,\,\, 
		u' = 3\,, \, \cdots, \, 10\,.
\ee
In the $\omega \rightarrow \infty$ limit, the M2-brane action~\eqref{eq:relmtb} gives rise to the action~\eqref{eq:nrm2} that is dual to the D2-brane action in nonrelativistic string theory. The decoupling limit defined by taking the $\omega \rightarrow \infty$ of M-theory reparametrized as in Eq.~\eqref{eq:nrmtp} defines \emph{nonrelativistic M-theory}, whose 11D target space membrane Newton-Cartan geometry has a codimension-three foliation structure.  

We now consider U-duality transformation that maps DLCQ to nonrelativistic M-theory. For simplicity, we focus on the flat spacetime limit. We start with a review of the standard U-duality rules in M-theory (see \textit{e.g.}~\cite{Becker:2006dvp}; see also~\cite{Obers:1998fb} for an extensive review of U-duality). Compactify the directions $X^1$, $X^2$, and $X^{10}$ over circles of radii $R_1$\,, $R_2$\,, and $R_{10}$\,, respectively. Regard the $X^{10}$ circle as the M-theory circle. Eq.~\eqref{eq:rogsa} implies $R^{}_{10} = g^{}_s \, \ell^{}_s$\,, where $g_s$ and $\ell_s = \alpha'{}^{1/2}$ are the string coupling and string length, respectively, from the perspective of type IIA nonrelativistic string theory. T-dualising the $X^1$ and $X^2$ circles gives the dual $\widetilde{X}^1$ and $\widetilde{X}^2$ circles of radii $\widetilde{R}_1$ and $\widetilde{R}_2$\,, respectively, with 
$\widetilde{R}_{1,\,2} = {\ell_s^2 / R_{1,\,2}}$\,.
In terms of the eleven-dimensional Planck length $\ell_{11} = g^{1/3}_s \, \ell_s$, we find the generalised T-duality transformations, 
\be \label{eq:trr}
	\widetilde{R}_1 = \frac{\ell_{11}^3}{R_1 \, R_{10}}\,, 
		\qquad%
	\widetilde{R}_2 = \frac{\ell_{11}^3}{R_2 \, R_{10}}\,.
\ee
Moreover, the string coupling $g_s$ transforms under the T-duality as
\be \label{eq:gsrtlt}
	\tilde{g}_s = g_s \,  \frac{\ell_s^p}{R_1 \, R_2} 
		\quad \implies \quad
	\widetilde{R}_{10} = \frac{\ell^{3}_{11}}{R_1 \, R_2}\,,
		\quad%
	\tilde{\ell}^{\,3}_{11} = \frac{\ell_{11}^6}{R_1 \, R_2 \, R_{10}}\,. 
\ee
We have used the fact that the string length $\ell_s$ does not change under the T-duality transformation. We can also apply an S-duality transformation that swaps the $\widetilde{X}^1$ and $\widetilde{X}^2$ circles, in which case the U-duality transformation becomes
\be \label{eq:udual}
	\widetilde{R}_\text{I} = \frac{\ell_{11}^3}{{R}_\text{J} \, {R}_\text{K}}\,,
		\qquad%
	\tilde{\ell}_{11}^3 = \frac{\ell_{11}^6}{R_\text{I} \, R_\text{J} \, R_\text{K}}\,. 
\ee
Here, I\,, J\,, and K take distinct values in $\{1\,, \, 2\,, \, 10 \}$\,. The above U-duality transformations apply to any M-theory, as long as the compact circles are orthogonal to each other. 

We now turn out attention to DLCQ M-theory in flat spacetime. In terms of the lightlike coordinates $X^\pm$ and the transverse coordinates $X^i$\,, $i = 1\,, \cdots\,, 9$\,, the Lorentz boost in the $X^+\!\!-\!\!X^-$ sector takes the form
\be
    X^+ \rightarrow \gamma \, X^+\,,
        \qquad%
    X^- \rightarrow \gamma^{-1} \, X^-\,,
        \qquad%
    X^i \rightarrow X^i\,.
\ee
The $\gamma \rightarrow \infty$ defines DLCQ M-theory. 
Next, we consider the U-duality transformation of DLCQ M-theory compactified over a three-torus along $X^-$\,, $X^1$\,, and $X^2$\,. We assume that $X^-$ is the M-theory circle so that it plays the role of the ``tenth'' spatial direction. Denote the radii associated with the $X^1$\,, $X^2$\,, and $X^-$ circle as $R_1$\,, $R_2$\,, and $R_{10}$\,, which scale under the boost transformation as
\be
    R_{1,\,2} \rightarrow R_{1,\,2}\,,
        \qquad%
    R_{10} \rightarrow \gamma^{-1} \, R_{10}\,.
\ee
According to the U-duality transformation~\eqref{eq:gsrtlt}, we find that the dual radii and the Planck length in eleven dimensions scale as
\be
    \tilde{R}_{1,\,2} \rightarrow \gamma \, \tilde{R}_{1,\,2}\,,
        \qquad%
    \tilde{R}_{10} \rightarrow \tilde{R}_{10}\,,
        \qquad%
    \tilde{\ell}_{11} \rightarrow \gamma^{1/3} \, \tilde{\ell}_{11}\,.
\ee
It then follows that the U-dual coordinates scale as
\be
    \tilde{X}^0 \rightarrow \gamma \, \tilde{X}^0\,,
        \qquad
    \tilde{X}^{1,\,2} \rightarrow \gamma \, \tilde{X}^{1,\,2}\,,
        \qquad%
    \tilde{X}^{u'} \rightarrow \tilde{X}^{u'},
        \qquad%
    u' = 3\,, \, \cdots, \, 10\,.
\ee
We have defined $\tilde{X}^0 = X^+$\,.
In order to measure the dual coordinates with respect to the rescaled Planck length, we define $x^\text{M} = \tilde{X}^\text{M} / \tilde{\ell}_{11}$\,, which acquire the following $\gamma$ scalings:
\be
    x^u \rightarrow \gamma^{2/3} \, x^u\,, 
        \,\,%
    u = 0\,, \, 1\,, \, 2\,;
        \qquad%
    x^{u'} \rightarrow \gamma^{-1/3} \, x^{u'}\,,
        \,\,%
    u' = 3\,, \, \cdots, \, 10\,.
\ee
The associated metric at a finite $\gamma$ takes the form
\be \label{eq:firstfundf}
    \frac{ds^2}{\tilde{\ell}^{\,2}_{11}} = \gamma^{4/3} \, dx^u \, dx^v \, \eta^{}_{uv} + \gamma^{-2/3} \, dx^{u'} \, dx^{u'}\,.
\ee 
Identifying the original Lorentz factor $\gamma$ with $\omega$\,, we observe that the metric~\eqref{eq:firstfundf} matches the nonrelativistic M-theory prescription~\eqref{eq:nrmtp}. A more careful analysis shows that the three-form in Eq.~\eqref{eq:nrmtp} also arises in the U-dual frame. Therefore, in the $\gamma \rightarrow \infty$ limit, we find that DLCQ M-theory is U-dual to nonrelativistic M-theory~\cite{Blair:2023noj}. Intriguingly, under this U-duality transformation, the \emph{lightlike} circle in DLCQ M-theory maps to the transverse $X^{10}$ circle in nonrelativistic M-theory, which becomes \emph{spacelike}.

Note that it is also possible to further consider the U-dual of DLCQ nonrelativistic M-theory, which gives rise to \emph{multicritical M-theory} that uplifts certain MM$p$Ts. We also note that MM0T arises from compactifying nonrelativistic M-theory over a lightlike circle. See~\cite{Blair:2023noj} for relevant discussions. A powerful complementary perspective is studied in~\cite{bpslimits}, where the U-dual orbits that involve one or multiple DLCQs are considered in the context of the BPS mass formulae. 

\newpage

%%%%%%%%%%%%%%%%%%%%%%%%%%%%%%%%%%%%%%%%%%%%%
%%%%%%%%%%%%%%%%%%%%%%%%%%%%%%%%%%%%%%%%%%%%%
%%%%%%%%%%%%%%%%%%%%%%%%%%%%%%%%%%%%%%%%%%%%%

\section{Spin Matrix Theory limits} \label{sec:smtl}
Building on the previous section, we now further explore applications of nonrelativistic strings to decoupling limits in the context of the AdS/CFT correspondence.
In particular, we will study a certain class of decoupling limits of $\mathcal{N}=4$ super-Yang--Mills that describes a small subset of the full spectrum by zooming in on the dynamics close to a BPS bound.
In these limits, which are known as \emph{Spin Matrix} limits~\cite{Harmark:2014mpa}, the remaining spectrum becomes non-relativistic, and there turns out to be a natural connection to the non-relativistic strings discussed in the previous section.
Again, the contents of this section are slightly out of focus with the majority of the rest of this review, as we will work in the context of holography with a five-dimensional AdS bulk.
However, developing similar decoupling limits in the context of string theory on a three-dimensional AdS background and its holographic duals is an important open problem, and we hope that the tools presented in the totality of this review will aid and motivate research in this direction.

%\subsection{Introduction}
To motivate these Spin Matrix decoupling limits from the perspective of holography, recall that it has been extremely fruitful to consider the planar limit $N\to\infty$ of $\mathcal{N}=4$, where $N$ is the rank of the gauge group $SU(N)$ and the 't Hooft coupling $\lambda = g_\text{YM}^2N$ is held fixed.
The planar limit allows us to use integrability to bridge the gap between perturbative computations in field theory at weak 't Hooft coupling and perturbative computations on the string worldsheet, which correspond to strong 't Hooft coupling.
However, it has proven difficult to extend the integrability perspective beyond the planar limit.
Since this prevents us from exploring many interesting aspects of the holographic correspondence, including strong gravity effects and non-perturbative states such as black holes in the bulk,
it is important to explore different approaches to AdS/CFT that have the potential to lead to tractable computations on both sides of the correspondence at finite $N$.

One such approach is represented by the aforementioned Spin Matrix limits, which zoom in on the one-loop dynamics of $\mathcal{N}=4$ close to BPS bounds while keeping $N$ fixed but arbitrary.
As such, this perspective goes beyond the integrable spin chain description that emerges in the planar limit at $N\to\infty$, and $1/N$ corrections allow for splitting and joining of a `gas' of such spin chains.
See the review~\cite{Baiguera:2023fus} for an overview on recent work constructing the explicit Hamiltonians for these limits.
In the current discussion, we will focus mainly on the $N\to\infty$ limit, but even though we are at small 't Hooft coupling, it turns out that the Spin Matrix limits still allow for a string worldsheet description in the bulk.
However, in line with the decoupling limits discussed in the previous section, these string worldsheets will be non-relativistic.

After first introducing the Spin Matrix limits in more detail in Section~\ref{ssec:smt-from-ft} from a field theory perspective, we will focus on the semiclassical large charge regime, where the resulting non-relativistic Hamiltonians can be represented by non-relativistic sigma models.
For the particular Spin Matrix limit where the remaining global symmetry group is $SU(2)$, this gives the Landau--Lifshitz model.
Subsequently, we will translate these limits to the bulk in Section~\ref{ssec:smt-to-st}.
We first focus on constructing appropriate non-relativistic backgrounds from AdS$_5\times S^5$.
Building on the non-relativistic string actions constructed in Section~\ref{sec:scts} we then take the Spin Matrix limit on the worldsheet, where it results in an action similar to the multicritical Matrix 0-brane theory (MM0T) strings discussed in Section~\ref{sec:swdw}.
After gauge fixing, we see that these string actions reproduce the non-relativistic sigma models, and in particular we recover the Landau--Lifshitz model from the $SU(2)$ string limit.

Spin Matrix limits are not the only non-relativistic decoupling limits that can be considered in the context of AdS/CFT.
Following the work of Gomis, Gomis and Kamimura~\cite{Gomis:2005pg}, one can also consider an AdS$_5$ version of the critical $B$-field limit considered in Section~\ref{sec:bfb} above, and several aspects analogous to (but often subtly different from) the relativistic string integrability picture have been developed, see for example~\cite{Fontanella:2023men,Fontanella:2022wfj,Fontanella:2022pbm,Fontanella:2022fjd,Fontanella:2021btt,Fontanella:2021hcb,Roychowdhury:2020dke,Roychowdhury:2020cnj,Roychowdhury:2020yun}.
See also~\cite{Guijosa:2023qym} for a recent discussion of the relation between non-relativistic string limits, non-commutative open string theory and the AdS/CFT correspondence.

\subsection{Spin Matrix decoupling limits in field theory}
\label{ssec:smt-from-ft}
To start, we give a very brief introduction to the field theory origins of these limits, see the reviews~\cite{Harmark:2014mpa,Baiguera:2023fus} for more information.
Consider $\mathcal{N}=4$ Yang--Mills theory with gauge group $SU(N)$ on $\RR \times S^3$.
The $PSU(2,2|4)$ global symmetry of this theory leads to several BPS bounds of the form
\begin{equation}
  \label{eq:bps-bound}
  E \geq Q,
\end{equation}
where $Q = J + S$ is a particular combination of the Cartan generators of the $SU(4)$ R-symmetry, which we denote as $(J_1,J_2,J_3)$, and the Cartan generators of the $SO(4)$ symmetry of the $S^3$, which we denote as $(S_1,S_2)$.
In the following, we will only consider integer combinations.
Using the 't Hooft coupling $\lambda = g_{YM}^2 N$, consider the limit
\begin{equation}
  \label{eq:smt-ft-limit}
  \lambda \to 0,
  \quad
  N \text{ fixed},
  \quad
  \frac{E-Q}{\lambda} \text{ fixed}.
\end{equation}
This limit zooms in on fluctuations directly above the BPS bound~\eqref{eq:bps-bound}, with interactions governed by the one-loop Hamiltonian.
This greatly simplifies the theory.
For example, if we take $Q= J = J_1 + J_2$, the corresponding limit only preserves states that transform under a $SU(2)$ subgroup of the total $PSU(2,2|4)$ global symmetry.
Other BPS bounds lead to different subgroups (as listed in Table~\ref{tab:smt-spin-groups}), which correspond to the `spin group' associated to the remaining excitations.
Since the fields are also matrices in the adjoint of $SU(N)$, these subsectors are called \emph{Spin Matrix} limits.
In principle, we can consider this limit for any $N$, and a general procedure to obtain the corresponding interacting Hamiltonians has recently been developed in~\cite{Harmark:2019zkn,Baiguera:2020jgy,Baiguera:2020mgk,Baiguera:2021hky,Baiguera:2022pll}.

Instead, we will focus on the planar limit $N\to\infty$, which results in a nearest-neighbor spin chain for single-trace operators of length $J$, corresponding to the total R-charge. 
The $1/N$ corrections allow for splitting and joining of the chains, but we will not consider them here.
For the $SU(2)$ case, the $N\to\infty$ Hamiltonian corresponds to the $XXX_{1/2}$ ferromagnetic Heisenberg spin chain.
Its semiclassical large $J$ regime can be described using the Landau-Lifshitz sigma model
\begin{equation}
  \label{eq:ll-sigma-model}
  S_\text{LL}
  = \frac{J}{4\pi} \int d^2 \sigma \left(
    \cos\theta\, \dot\vphi
    - \frac{1}{4} \left[\left(\theta'\right)^2 + \sin^2\theta \left(\vphi'\right)^2\right]
  \right).
\end{equation}
Here, the $S^2$ coordinates $(\theta,\vphi)$ are functions of $(\sigma^0,\sigma^1)$, and we take $\sigma^1$ to be $2\pi$-periodic.
The total momentum along $\sigma^1$ vanishes due to the cyclicity of the trace in the original operators.
Note that the Spin Matrix limit in this regime is similar to the limit considered by Kruczenski in~\cite{Kruczenski:2003gt}, although that limit considers $J\to\infty$ with $\lambda/J^2$ fixed, while the Spin Matrix limit takes $\lambda\to0$ with $J$ fixed.

\begin{table}[ht]
    \centering
    \begin{tabular}{|l|l|}
        \hline
         charge $Q$ & spin group
         \\ \hline
         $J+1 + J_2$ & $SU(2)$
         \\
         $J+1+J_2+J_3$ & $SU(2|3)$
         \\
         $S_1+J_1$ & $SU(1,1)$
         \\
         $S_1+J_1+J_2$ & $PSU(1,1|2)$
         \\
         $S_1+ S_2 + J_1$ & $SU(1,2|2)$
         \\
         $S_1 + S_2 + J_1+ J_2 + J_3$ & $PSU(1,2|3)$
         \\
         \hline
    \end{tabular}
    \caption{Spin groups of Spin Matrix theories obtained from integer BPS bounds $E\geq Q$}
    \label{tab:smt-spin-groups}
\end{table}

\subsection{Translating Spin Matrix limits to string theory}
\label{ssec:smt-to-st}
In the following, we will recover the sigma model~\eqref{eq:ll-sigma-model} as the gauge-fixed worldsheet action of a limit of non-relativistic strings on AdS$_5\times S^5$.
For this, our goal is first to translate the field theory limit~\eqref{eq:smt-ft-limit} to the bulk.
We will investigate how we can choose appropriate bulk coordinates to implement the limit.
We will then see that the limit constrains the dynamics of the string to a particular submanifold of AdS$_5\times S^5$ which depends on what BPS bound we choose.
This discussion largely follows~\cite{Harmark:2017rpg,Harmark:2018cdl,Harmark:2020vll}, see also the review~\cite{Oling:2022fft}.

\subsubsection{Choosing adapted bulk coordinates}
Preparing for the limit, we will now discuss how to construct appropriate bulk coordinates.
Recall that we can see AdS$_5$ and $S^5$ as hypersurfaces in $\RR^{2,4}\simeq \CC^{1,2}$ and $\RR^6\simeq \CC^3$ using
\begin{equation}
  \label{eq:ads5-s5-hypersurfaces}
  - |z_0|^2 + |z_1|^2 + |z_2|^2 = - R^2,
  \qquad
  |w^1|^2 + |w^2|^2 + |w^3|^2 = R^2,
\end{equation}
where $R$ is the AdS$_5$ and $S^5$ radius.
Writing $z_0 = R \cosh\rho\, e^{it}$ and $(z_1,z_2) = R \sinh\rho\,(v_1,v_2)$ where $(v_1,v_2)$ describe a three-sphere $S^3\subset\CC^4$, we obtain the metric for AdS$_5\times S^5$ in global coordinates,
\begin{equation}
  \label{eq:global-ads5-s5-metric}
  ds^2 = R^2 \left[
    -\cosh^2\rho\, dt^2 + d\rho^2 + \sinh^2\rho\, d\Omega_3^2 + d\Omega_5^2
  \right].
\end{equation}
Here, $d\Omega_3^2$ and $d\Omega_5^2$ denote the unit $S^3$ and $S^5$ metrics, respectively.
The AdS$_5$ conformal boundary at $\rho\to\infty$ corresponds to the $\RR\times S^3$ background in the dual field theory.
As a result, the energy $E$ in the field theory is associated to the bulk (asymptotic) isometry $i\pd_t$.
On the other hand, the $SO(4)$ and $SU(4)$ Cartan charges correspond to particular combinations $-i\pd_{\zeta}$ and $-i\pd_\gamma$ of the $S^3$ and $S^5$ isometries given by the rotations of the complex $z_i$ and the $w_i$ coordinates in~\eqref{eq:ads5-s5-hypersurfaces}, respectively.

To take the limit~\eqref{eq:smt-ft-limit}, we then want to combine the coordinates $(t,\zeta,\gamma)$ into new coordinates $(x^0, u)$ such that their conserved charges are
\begin{equation}
  \label{eq:x0-u-charges}
  i \pd_{x^0} = E - Q,
  \qquad
  -i \pd_u = \frac{1}{2} \left(E - S + J\right).
\end{equation}
Now recall from the AdS/CFT dictionary that $\lambda = 4\pi g_s N$.
To take the limit, we rescale
\begin{equation}
  \label{eq:rescale-x0}
  x^0 = \frac{\tilde{x}^0}{4\pi g_s N}
  \qiq
  i \pd_{\tilde{x}^0} = \frac{E-Q}{4\pi g_s N}.
\end{equation}
Taking $\lambda\to0$ with fixed $N$ then corresponds to sending $g_s\to0$, and we see that the limit~\eqref{eq:smt-ft-limit} corresponds to sending $g_s\to0$ with $N$ and the rescaled coordinate $\tilde{x}^0$ fixed.
In this limit, we see that the charge $-i\pd_u$ corresponds to the total angular momentum $J$ along the $S^3$, which is the momentum in the dual spin chain.

Now let us see how this works for the $SU(2)$ case, where $Q=J_1+J_2$.
This charge only contains $S^5$ isometries and we can therefore parametrize it using a coordinate vector for a suitable coordinate $\gamma$ on the $S^5$.
Specifically, the Cartan generators $J_i$ correspond to the isometries $-i\pd_{w^i}$ of the embedding coordinates in Equation~\eqref{eq:ads5-s5-hypersurfaces}.
Now to combine $J_1$ and $J_2$ we parametrize the $S^3\subset S^5$ using Hopf coordinates,
\begin{subequations}
  \begin{align}
    w_1
    &= R \sin(\beta/2) \sin(\theta/2) e^{i(\gamma - \vphi/2)},
    \\
    w_2
    &= R \sin(\beta/2) \sin(\theta/2) e^{i(\gamma + \vphi/2)},
    \\
    w_3
    &= R \cos(\beta/2) e^{i\alpha},
  \end{align}
\end{subequations}
where $\beta$ and $\theta$ run from $0$ to $\pi$ and $\gamma$, $\vphi$ and $\alpha$ are $2\pi$-periodic.
Then $-i\pd_\gamma = J_1 + J_2$.
Using these coordinates, the metric~\eqref{eq:global-ads5-s5-metric} is
\begin{equation}
  \begin{split}
    ds^2 / R^2
    &= - \cosh^2\rho\, dt^2 + d\rho^2 + \sinh^2\rho\, d\bar\Omega_3^2
    \\
    &{}\qquad
    + \frac{1}{4} d\beta^2 + \sin^2(\beta/2)
    \left(d\Sigma_1^2 + \left(d\gamma + A\right)^2\right)
    + \cos^2(\beta/2) d\alpha^2,
  \end{split}
\end{equation}
where the Hopf potential and Fubini-Study metric are
\begin{equation}
  A = \frac{1}{2} \cos\theta\, d\vphi,
  \qquad
  d\Sigma_1^2 = \frac{1}{4} \left(d\theta^2 + \sin^2\theta\, d\vphi^2\right).
\end{equation}
Next, to achieve~\eqref{eq:x0-u-charges} we use the coordinate transformation
\begin{equation}
  t = x^0 - \frac{u}{2},
  \qquad
  \gamma = x^0 + \frac{u}{2}.
\end{equation}
Then the metric becomes
\begin{equation}
  \label{eq:su2-metric-decomposition}
  \begin{split}
    ds^2 / R^2
    &= \sin^2(\beta/2) \left[
      2 \left(dx^0 + \frac{1}{2}A\right) \left(du + A\right)
      + d\Sigma_1^2
    \right]
    \\
    &{}\qquad
    - \left[\sinh^2\rho + \cos^2(\beta/2)\right] \left(dx^0 - \frac{1}{2} du\right)^2
    \\
    &{}\qquad
    + d\rho^2 + \sinh^2\rho\, d\Omega_3^2
    + \frac{1}{4} d\beta^2 + \cos^2(\beta/2) d\alpha^2.
  \end{split}
\end{equation}
If we then rescale $x^0 = \tilde{x}^0/(4\pi g_s N)$ the second line of this metric results in a quadratic potential on the worldsheet that is proportional to $\left[\sinh^2\rho + \cos^2(\beta/2)\right]/g_s$.
As a result, the $g_s\to0$ limit forces the string excitations to a submanifold where $\rho=0$ and $\beta=\pi$.

Thus, in the $SU(2)$ Spin Matrix limit, we zoom in on string configurations that are restricted to lie at the center of AdS$_5$ (since $\rho=0$) and that only probe the $S^3\subset S^5$ corresponding to $w_3=0$ in the embedding coordinates (since $\beta=\pi$).
Together with the global AdS time $t$, the Hopf fiber $\gamma$ of this $S^3$ is combined into the coordinates $x^0$ and $u$, and the base $S^2$ gives rise to the $SU(2)$ global symmetry.
In total, the dynamics is constrained to a four-dimensional submanifold of AdS$_5\times S^5$ which is described by $x^0$, $u$ and the $S^2$ coordinates $\theta$ and $\vphi$.
On this submanifold, both $x^0$ and $u$ are null isometries.
On the other hand, in the $SU(1,2|3)$ case, the corresponding $x^0$ and $u$ coordinates are null isometries across all of AdS$_5\times S^5$, and the full ten-dimensional target space is retained in the limit.
Other limits in Table~\ref{tab:smt-spin-groups} fall in between these two examples.
The interplay between Spin Matrix limits and Penrose limits has also been discussed in~\cite{Harmark:2020vll} and results in non-relativistic versions of pp-wave background metrics.

To understand the effect of the Spin Matrix limit on the string action, it is useful to take this submanifold as an intermediate step.
More precisely, by restricting the string dynamics to this submanifold, we first obtain a non-relativistic string using a null T-duality along the $u$-direction of the above background, along the lines of Section~\ref{sec:lltddlcqs}.
We then implement the Spin Matrix limit on the resulting non-relativistic string action, which will result in a novel sigma model with non-relativistic geometry on the worldsheet.

\subsubsection{Null isometry submanifolds and non-relativistic string backgrounds} \label{sec:nisnrsb}
In the above, we saw that the $SU(2)$ Spin Matrix limit constrains the string dynamics to a submanifold where the appropriate $x^0$ and $u$ coordinates parametrize null isometries.
This happens for any of the Spin Matrix limits in Table~\ref{tab:smt-spin-groups} above, and different limits will keep string excitations along different directions of AdS$_5\times S^5$.
Using the appropriate coordinates, we can write the metric restricted on the resulting submanifold $M$ as
\begin{equation}
  \label{eq:null-isom-submanifold}
  ds^2|_M / R^2
  = 2 \tau{}^0 \left(du - m_i dx^i\right) + E_{ij} dx^i dx^j,
\end{equation}
where the one-form $\tau^0$ has components $\tau^0 = dx^0 + \tau_i{}^0 dx^i$ and none of the components of the metric depend on the $x^0$ and $u$ coordinates.
Here, $i=1,\ldots,2k$ is an appropriate even-numbered set of indices -- in the $SU(2)$ case, we saw that $k=1$ and $x^i=(\theta,\vphi)$.
Less restrictive Spin Matrix limits retain a larger part of the global symmetries and are therefore able to probe more directions of the AdS$_5\times S^5$ geometry.

For simplicity, we will set the AdS radius $R$ to one for now, but we will reintroduce it later.
In the $SU(2)$ case, the decomposition~\eqref{eq:null-isom-submanifold} on the $\rho=0$ and $\beta=\pi$ submanifold corresponds to the top line of~\eqref{eq:su2-metric-decomposition}, which we can write as
\begin{equation}
  \label{eq:su2-tnc-data}
  \begin{gathered}
    \tau{}^0
    = dx^0 + \frac{1}{4} \cos\theta d\vphi,
    \qquad
    m_i dx^i
    = - \frac{1}{2} \cos\theta d\vphi,
    \\
    E_{ij} dx^i dx^j
    =  \frac{1}{4} \left(d\theta^2 + \sin^2\theta d\vphi\right).
  \end{gathered}
\end{equation}
Following the discussion in Section~\ref{sec:lltddlcqs}, we can map the relativistic string action on the Lorentzian background~\eqref{eq:null-isom-submanifold} to a non-relativistic string action coupling to the dual non-relativistic geometry.
Now using $x^M=(x^0,u,x^i)$ to denote the $2(k+1)$-dimensional manifold described by~\eqref{eq:null-isom-submanifold}, we can apply a T-duality along the $u$-direction, which results in the following non-relativistic geometry,
\begin{equation}
  \label{eq:intermediate-snc-backgrounds}
  \begin{gathered}
    \tau_\mu{}^0 dx^\mu = dx^0 + \tau_i{}^0 dx^i,
    \qquad
    \tau_\mu{}^1 dx^\mu = dv,
    \\
    E_{\mu\nu} dx^\mu dx^\nu
    =  E_{ij} dx^i dx^j,
    \qquad
    B_{\mu\nu} dx^\mu \wedge dX^\nu
    = m_i dx^i \wedge dv.
  \end{gathered}
\end{equation}
Note that the T-duality turns on a Kalb--Ramond field corresponding to $g_{ui} = - m_i$ in the original Lorentzian metric.
Additionally, recall that we are considering string configurations that have momentum $J$ along the $u$ direction in the Lorentzian background.
This coordinate gets mapped to a dual coordinate $v$, and the $u$-momentum $J$ gets mapped to a winding mode along this $v$ coordinate.

The non-relativistic string background~\eqref{eq:intermediate-snc-backgrounds} will be modified in the SMT limit.
Since we rescale the $x^0$ coordinate according to~\eqref{eq:rescale-x0}, the one-form $\tau^0$ reduces to $d\tilde{x}^0$.
The other objects do not depend on $x^0$ and their expressions do not change.
However, as we will now see, the Spin Matrix limit changes their coupling to the string action.

\subsubsection{Limit in non-relativistic string action}
Our starting point is the non-relativistic string Polyakov action~\eqref{eq:snc-pol-action},
\begin{equation}
  \label{eq:nrcs-polyakov-action}
  \begin{split}
    S
    &= - \frac{1}{4\pi \alpha'} \int d^2\sigma \left[
      \sqrt{-h} h^{\alpha\beta} \pd_\alpha X^\mu \pd_\beta X^\nu E_{\mu\nu}
      + \epsilon^{\alpha\beta} \pd_\alpha X^\mu \pd_\beta X^n B_{\mu\nu}
    \right]
    \\
    &{}\qquad
    - \frac{1}{4\pi\alpha'}\int d^2\sigma \left[
      \lambda \epsilon^{\alpha\beta} e_\alpha \pd_\beta X^\mu \tau_\mu
      + \bar\lambda \epsilon^{\alpha\beta} \bar{e}_\alpha \pd_\beta X^\mu \bar{\tau}_\mu
    \right],
  \end{split}
\end{equation}
where we have introduced worldsheet vielbeine $e_\alpha{}^a$ such that $h_{\alpha\beta} = \eta_{ab} e_\alpha{}^a e_\beta{}^b$.
Furthermore, we use
$e_\alpha = e_\alpha{}^0 + e_\alpha{}^1$
and
$\bar{e}_\alpha = e_\alpha{}^0 - e_\alpha{}^1$,
as well as
$\tau_\alpha = \tau_\alpha{}^0 + \tau_\alpha{}^1$
and
$\bar{\tau}_\alpha = \tau_\alpha{}^0 - \tau_\alpha{}^1$.
Finally, we decompose
$\lambda_\alpha = \lambda_\alpha{}^0 + \lambda_\alpha{}^1$
ands
$\bar{\lambda}_\alpha = \lambda_\alpha{}^0 - \lambda_\alpha{}^1$.
Following our previous discussion, we will mainly be interested in backgrounds of the form~\eqref{eq:intermediate-snc-backgrounds}, but we will keep them general for now.

We now want to implement the Spin Matrix limit in the non-relativistic string action~\eqref{eq:nrcs-polyakov-action}.
In fact, it is easier to first consider the limit in the corresponding Nambu--Goto action~\eqref{eq:snc-ng-action}.
We will ignore the dilaton term, so the action is
\begin{equation}
  S
  = - \frac{1}{4\pi\alpha'} \int d^2\sigma \left[
    \sqrt{-\tau} \eta^{AB} \tau^\alpha{}_A \tau^\beta{}_B E_{\alpha\beta}
    + \epsilon^{\alpha\beta} B_{\alpha\beta}
  \right].
\end{equation}
Here, $\tau_{\alpha\beta} = \eta_{AB} \tau_\alpha{}^A \tau_\beta{}^B$ is the Lorentzian worldsheet metric that is induced by $\tau_\mu{}^A$, the target space longitudinal one-forms.
We then consider the scaling limit
\begin{equation}
  \label{eq:smt-limit-ng}
  \begin{gathered}
  \tau_\mu{}^0 = \omega^2\, \tilde\tau_\mu{}^0,
  \qquad
  \tau_\mu{}^1 = \omega\, \tilde\tau_\mu{}^1,
  \qquad
  \alpha' = \omega\, \tilde\alpha',
  \\
  E_{\mu\nu} = \tilde{E}_{\mu\nu},
  \qquad
  B_{\mu\nu} = \omega \tilde{B}_{\mu\nu},
  \end{gathered}
\end{equation}
where we have defined $\omega = \sqrt{4\pi g_s N}$ so that $\omega\to\infty$ corresponds to $g_s\to0$ with $N$ fixed.
This results in a non-relativistic Galilean limit on the \emph{worldsheet} of the theory, since we are scaling the induced vielbeine $\tau_\alpha{}^A$ separately.
Dropping the tildes, the resulting action is
\begin{equation}
  S = - \frac{1}{4\pi\alpha'} \int d^2\sigma \left[
    \sqrt{-\tau} \tau^\alpha{}_1 \tau^\alpha{}_1 E_{\alpha\beta}
    + \epsilon^{\alpha\beta} B_{\alpha\beta}
  \right].
\end{equation}
We can equivalently take this limit on the level of the Polyakov action.
Then the worldsheet vielbeine and the Lagrange multipliers scale as follows,
\begin{equation}
  e_\alpha{}^0 = \omega^2 \tilde{e}_\alpha{}^0,
  \qquad
  e_\alpha{}^1 = \omega\, \tilde{e}_\alpha{}^1,
  \qquad
  \lambda^0 = \frac{1}{2\omega^3} \tilde{\lambda}^0,
  \qquad
  \lambda^1 = \frac{1}{2\omega^2} \tilde{\lambda}^1.
\end{equation}
Dropping the tildes, the resulting Polyakov-type action is
\begin{equation}
  \label{eq:smt-polyakov-action}
  \begin{split}
    S
    &= - \frac{1}{4\pi\alpha'} \int d^2\sigma \left[
      e\, e^\alpha{}_1 e^\alpha{}_1 E_{\alpha\beta}
      + \epsilon^{\alpha\beta} B_{\alpha\beta}
    \right]
    \\
    &{}\qquad
    - \frac{1}{4\pi\alpha'} \int d^2\sigma \left[
      \lambda^0 \epsilon^{\alpha\beta} e_\alpha{}^0 \tau_\beta{}^0
      + \lambda^1 \epsilon^{\alpha\beta} \left(
        e_\alpha{}^0 \tau_\beta{}^1 + e_\alpha{}^1 \tau_\beta{}^0
      \right)
    \right].
  \end{split}
\end{equation}
This reproduces the Nambu-Goto above action upon integrating out the constraints.
These actions match the multicritical Matrix 0-brane theory (MM0T) string actions~\eqref{eq:mm0t-curved-string-action} above, see also~\cite{Gomis:2023eav}.
In particular, the $SU(1,2|3)$ Spin Matrix limit can be directly related to the MM0T limit (or equivalently the DLCQ of the Matrix 1-brane theory) applied to AdS$_5\times S^5$.
In the MM0T limit, including the dilaton term in the limit gives rise to nodal Riemann surfaces, which have also been observed in the context of ambitwistor string theory~\cite{Geyer:2015bja}.

We are mainly interested in non-relativistic metric data of the form~\eqref{eq:intermediate-snc-backgrounds}.
In the $\omega\to\infty$ limit, the $\tau^0$ longitudinal one-form simplifies, as we discussed previously, and the resulting backgrounds are
\begin{equation}
  \label{eq:smt-backgrounds}
  \begin{gathered}
    \tau_\mu{}^0 dx^\mu = dx^0
    \qquad
    \tau_\mu{}^1 dx^\mu = dv,
    \\
    E_{\mu\nu} dx^\mu dx^\nu
    =  E_{ij} dx^i dx^j,
    \qquad
    B_{\mu\nu} dx^\mu \wedge dX^\nu
    = m_i dx^i \wedge dv.
  \end{gathered}
\end{equation}
For the $SU(2)$ background~\eqref{eq:su2-tnc-data}, this corresponds to
$\tau^0 = dx^0$ and $\tau^1 = dv$ and
\begin{equation}
  E_{\mu\nu} dx^\mu dx^\nu = \frac{1}{4} \left(d\theta^2 + \sin^2\theta d\vphi^2\right),
  \qquad
  B_{\mu\nu} dx^\mu \wedge dX^\nu
  = \frac{1}{2} \cos\theta dv \wedge d\vphi.
\end{equation}

\subsubsection{Non-Lorentzian worldsheet structure and residual GCA symmetries}
The Polyakov-type action~\eqref{eq:smt-polyakov-action} is still invariant under Weyl transformations,
\begin{equation}
  e^A \to \Omega e^A,
  \qquad
  \lambda^A \to \Omega\inv \lambda^A.
\end{equation}
However, instead of the local Lorentz symmetries of $h_{\alpha\beta} = \eta_{ab} e_\alpha{}^a e_\beta{}^b$, the worldsheet is now invariant under local \emph{Galilean} boosts acting on the vielbeine $e^0$ and $e^1$ (plus appropriate transformations of the Lagrange multipliers),
\begin{equation}
  e^0 \to e^0,
  \qquad
  e^1 \to e^1 + \gamma\, e^0,
  \qquad
  \lambda^0 \to \lambda^0  - \gamma\, \lambda^1,
  \qquad
  \lambda^1 \to \lambda^1.
\end{equation}
This reflects the fact that the worldsheet geometry is no longer Lorentzian but Galilean.

As a result, going to flat gauge, the residual symmetries are also no longer given by the Virasoro algebra.
One way to see this is by solving the constraints in the action~\eqref{eq:smt-polyakov-action} on the general background~\eqref{eq:smt-backgrounds}.
If we use the flat gauge $e^0 = d\sigma^0$ and $e^1 = J d\sigma^1$ so that the total string length is $J$, the constraint multiplying $\lambda^0$ implies
\begin{equation}
  \label{eq:zero-constraint-introd-F}
  0 = X'^M \tau_M{}^0 = X'^0
  \qiq
  X^0 = F(\sigma^0).
\end{equation}
On the other hand, the constraint multiplying $\lambda^1$ implies
\begin{equation}
  \label{eq:one-constraint-introd-G}
  J X'^M \tau_M{}^1 = \dot{X}^M \tau_M{}^0
  \qiq
  X^v = \frac{\dot{F}(\sigma^0)}{J} \sigma^1 + G(\sigma^0).
\end{equation}
We see that after going to flat gauge, the two embedding functions $X^0$ and $X^v$ contain two arbitrary functions of the worldsheet time, which correspond to the coordinate reparametrizations generated by
\begin{equation}
  \xi = f(\sigma^0) \pd_{\sigma^0}
  + \left(
    g(\sigma^0) + \dot{f}(\sigma^0) \sigma^1
  \right) \pd_{\sigma^1}.
\end{equation}
Expanding these generators in the modes $L_m$ and $M_m$ of these functions, we see that they satisfy the \emph{Galilean conformal algebra (GCA)},
\begin{equation}
  \label{eq:gca-alg}
  [L_m, L_n]
  = (m-n) L_{m+n},
  \qquad
  [L_m, M_n]
  = (m-n) M_{m+n},
  \qquad
  [M_m, M_n]
  = 0.
\end{equation}
Note that this algebra is closely related to the BMS$_3$ algebra of three-dimensional asymptotically flat spacetimes~\cite{Bagchi:2010zz}, and it also appears in the context of tensionless limits of strings~\cite{Isberg:1993av,Bagchi:2013bga,Bagchi:2020fpr}.

At this point, the algebra~\eqref{eq:gca-alg} is purely classical, and it is not known if central charges can arise upon quantisation on general backgrounds.
Covariant quantisation of the action~\eqref{eq:smt-polyakov-action} is subtle due to the non-Lorentzian structure encoded in the constraints.
(For a recent discussion on gauge fixing and global symmetries in these actions, see~\cite{Bidussi:2023rfs}.)
However, the classes of backgrounds we've derived in the above are directly motivated by consistent decoupling limits in the boundary field theory, so they should be good candidates for consistent string backgrounds.
As we will now see, fixing the residual GCA symmetry allows us to reproduce the Landau--Lifshitz action for the $SU(2)$ background.

\subsubsection{Gauge-fixed action}
Now let us fix the residual GCA symmetry by setting $F(\sigma^0)=J^2\sigma^0$ and $G(\sigma^0)=0$, so
\begin{equation}
  X^0 = J^2 \sigma^0,
  \qquad
  X^v = J \sigma^1.
\end{equation}
After this Monge-like gauge fixing, we obtain the action
\begin{equation}
  S = - \frac{J R^2}{2\pi\alpha'} \int d^2\sigma \left[
    m_i \dot{X}^i + \frac{1}{2} E_{ij} X'^i X'^j
  \right].
\end{equation}
Here, we have restored the AdS radius $R$ that we previously dropped from the background variables.
The AdS/CFT dictionary then tells us that
$R^2/\alpha' = (R/\ell_s)^2 = \sqrt{\lambda} = \sqrt{4\pi g_s N}$,
which is precisely the parameter $\omega$ we split off from $\alpha'$ in~\eqref{eq:smt-limit-ng}.
Therefore, the effective string tension is just given by $1/(2\pi)$, so that the action is
\begin{equation}
  S = - \frac{J}{2\pi} \int d^2\sigma \left[
    m_i \dot{X}^i + \frac{1}{2} E_{ij} X'^i X'^j
  \right].
\end{equation}
Plugging in the background data that we obtained from the $SU(2)$ limit in Equation~\eqref{eq:su2-tnc-data}, we see that the corresponding gauge-fixed action becomes
\begin{equation}
  S
  = \frac{J}{4\pi} \int d^2 \sigma \left(
    \cos\theta\, \dot\vphi
    - \frac{1}{4} \left[\left(\theta'\right)^2 + \sin^2\theta \left(\vphi'\right)^2\right]
  \right).
\end{equation}
This is precisely the Landau--Lifshitz model in~\eqref{eq:ll-sigma-model} corresponding to the $N\to\infty$ and large $J$ limit of the $SU(2)$ Spin Matrix theory.

\newpage
%%%%%%%%%%%%%%%%%%%%%%%%%%%%%%%%%%%%%%%%%%%
\acknowledgments
We are grateful to the organisers and the students of the Young Researchers Integrability School and Workshop 2022 ``Taming the string worldsheet'' at NORDITA (Stockholm), for which this material was originally conceived. The YRISW 2022 was financially supported by NORDITA and the University of Padova, the GATIS+ network, and the Royal Society \& Science Foundation Ireland.

Sa.D. is grateful to the Azrieli foundation for the award of an Azrieli fellowship. The work of Sa.D. is also partially supported by the Israel Science Foundation (grant No. 1417/21),  by Carole and Marcus Weinstein through the BGU Presidential Faculty Recruitment Fund, the ISF Center of Excellence for theoretical high energy physics and by the Origins Excellence Cluster and by the German-Israel-Project (DIP) on Holography and the Swampland. Sa.D. would also like to thank Anayeli M. Ramirez for helpful comments. The work of Si.D. is  supported by the Swiss National Science Foundation through the NCCR SwissMAP. Si.D. would like to thank Riccardo Borsato for  helpful discussions that aided the preparation of the notes and lectures.
B.K. is grateful to Cassiano Daniel, Kiarash Naderi and Vit Sriprachyakul for helpful discussions. The work of B.K. is supported by STFC consolidated grants ST/T000694/1 and ST/X000664/1.
The work of G.O. is supported in part by the VR project grant 2021-04013 of Niels Obers and by the Royal Society URF of Jelle Hartong through the Research Fellows Enhanced Research Expenses 2022 (RF\textbackslash ERE\textbackslash 221013).
The work of A.L.R. was supported by a UKRI Future Leaders Fellowship (grant number MR/T018909/1) and partially by the grant 18/EPSRC/3590. A.L.R. would like to also thank Rafael Nepomechie, Chiara Paletta and Ben Hoare for useful discussions. F.S. is supported by the European Union’s Horizon 2020 research and innovation programme under the Marie Sk\l{}odowska-Curie grant agreement No 101027251.
A.S.~acknowledged support from the European Union --
NextGenerationEU, from the program STARS\@UNIPD, under project ``Exact-Holography -- A new exact approach to holography: harnessing the power of string theory, conformal field theory, and integrable models'', and from the PRIN Project n.~2022ABPBEY, ``Understanding quantum field theory through its deformations''. A.S.~also thanks the MATRIX Institute in Creswick \& Melbourne, Australia, for support through a MATRIX Simons fellowship during the preparation of this work, and in conjunction with the MATRIX program ``New Deformations of Quantum Field and Gravity Theories''. 
A.L.R., F.S. and A.S. would like to thank the Kavli Institute for Theoretical Physics, where several discussions regarding this review were held. 
Z.Y. is supported by the European Union’s Horizon 2020 research and innovation programme under the Marie Sk\l{}odowska-Curie grant agreement No 31003710. Nordita is supported in part by NordForsk. Z.Y. would like to thank Chris Blair for useful comments. 

%%%%%%%%%%%%%%%%%%%%%%%%%%%%%%%%%%%%%%%%%%%
%%%%%%%%%%%%    Bibliography     %%%%%%%%%%	
%%%%%%%%%%%%%%%%%%%%%%%%%%%%%%%%%%%%%%%%%%%
\bibliographystyle{JHEP}
\bibliography{bib_YRIS.bib}
\end{document}

\end{document}